\documentclass[11pt,a4paper]{book}
\usepackage{epsfig}
\usepackage[T1]{fontenc}    
\usepackage{graphics}
\usepackage{graphicx}
\usepackage{pstricks,pst-coil,pst-fill,pst-plot}
\usepackage[fleqn]{amsmath}    
\usepackage{amssymb}    
\usepackage{amsfonts}   
\usepackage{verbatim}   
\usepackage{mathrsfs}   
\usepackage{dsfont}
\usepackage{euscript}
\usepackage{yfonts}
\usepackage{enumerate}     
\usepackage{txfonts}
\usepackage{marvosym}
\usepackage{vmargin}        
\usepackage{wasysym}		

\setmarginsrb{1.8cm}{2cm}{1.8cm}{2cm}{1cm}{1cm}{1cm}{1.6cm}
 \makeatletter
 \@addtoreset{equation}{section}
 \makeatother

\providecommand{\bysame}{\leavevmode\hbox to3em{\hrulefill}\thinspace}
\providecommand{\MR}{\relax\ifhmode\unskip\space\fi MR }

\providecommand{\href}[2]{#2}

\newcommand{\bra}[1]{\big< \,#1\,|}
\newcommand{\ket}[1]{|\,#1\, \big>}
\newcommand{\braket}[2]{\big<\,#1 \, \big|\,  #2  \big>}

\let\ua=\uparrow
\let\da=\downarrow
\let\tend=\rightarrow

\long\def\symbolfootnote[#1]#2{\begingroup%
\def\thefootnote{\fnsymbol{footnote}}\footnote[#1]{#2}\endgroup}

\newtheorem{theorem}{Theorem}[section]
\newtheorem{prop}[theorem]{Proposition}
\newtheorem{cor}[theorem]{Corollary}
\newtheorem{defin}[theorem]{Definition}

\newtheorem{conj}[theorem]{Conjecture}

\newtheorem{lemme}{Lemma}[section]

\def\Proof{\medskip\noindent {\it Proof --- \ }}

\def\qed{\hfill\rule{2mm}{2mm}}

\newcommand\beq{\begin{equation}}
\newcommand\enq{\end{equation}}
\newcommand\bem{\begin{multline}}
\newcommand\enm{\end{multline}}

\def\beqa{\begin{eqnarray}}
\def\eeqa{\end{eqnarray}}
\def\ba{\begin{array}}
\def\ea{\end{array}}
\def\det{\operatorname{det}}

\newcommand{\f}[2]{{\ensuremath{%
    \mathchoice%
    {\dfrac{#1}{#2}}
    {\dfrac{#1}{#2}}
    {\frac{#1}{#2}}
    {\frac{#1}{#2}}
}}}
\newcommand{\tf}[2]{\ensuremath{#1/#2}}
\newcommand{\pa}[1]{\ensuremath{\left(#1\right)}}
\newcommand{\paa}[1]{\ensuremath{\left\{#1\right\}}}
\newcommand{\pac}[1]{\ensuremath{\left[#1\right]}}
\newcommand{\paf}[2]{\ensuremath{\left(\f{#1}{#2}\right)}}
\newcommand{\pab}[2]{\ensuremath{\pa{\ba{c} #1 \\ #2 \ea }}}
\newcommand{\pabb}[3]{\ensuremath{\pa{ #1 \left| \ba{c} #2 \\ #3 \ea \right .}}  }

\def\a{\alpha}
\def\al{\aleph}
\def\be{\beta}
\def\ga{\gamma}
\def\Ga{\Gamma}

\def\de{\delta}

\def\De{\Delta}
\def\eps{\epsilon}
\def\veps{\varepsilon}
\def\la{\lambda}
\def\La{\Lambda}

\def\sg{\sigma}
\def\vsg{\varsigma}
\def\Sg{\Sigma}

\def\Ups{\Upsilon}

\def\th{\theta}

\def\vth{\vartheta}

\def\Om{\Omega}
\def\om{\omega}
\def\vp{\varphi}

\newcommand{\mc}[1]{\ensuremath{\mathcal{#1}}}
\newcommand{\mf}[1]{\ensuremath{\mathfrak{#1}}}
\newcommand{\msc}[1]{\ensuremath{\mathscr{#1}}}

\newcommand{\bs}[1]{\ensuremath{\boldsymbol{#1}}}
\newcommand{\mbb}[1]{\ensuremath{\mathbb{#1}}}

\DeclareFontFamily{OT1}{pzc}{}
\DeclareFontShape{OT1}{pzc}{m}{it}{<-> s * [1.10] pzcmi7t}{}
\DeclareMathAlphabet{\mathpzc}{OT1}{pzc}{m}{it}

\def \i{ \mathrm i}

\newcommand{\ov}[1]{\ensuremath{\overline{#1}}}
\newcommand{\wt}[1]{\ensuremath{\widetilde{#1}}}
\newcommand{\wh}[1]{\ensuremath{\widehat{#1}}}

\newcommand{\Int}[2]{\ensuremath{\int\limits_{#1}^{#2}}}
\newcommand{\Oint}[2]{\ensuremath{\oint\limits_{#1}^{#2}}}
\newcommand{\Fint}[2]{\ensuremath{\fint\limits_{#1}^{#2}}}

\newcommand{\sul}[2]{\ensuremath{\sum\limits_{#1}^{#2}}}
\newcommand{\pl}[2]{\ensuremath{\prod\limits_{#1}^{#2}}}

\newcommand{\R}{\ensuremath{\mathbb{R}}}
\newcommand{\Cx}{\ensuremath{\mathbb{C}}}

\newcommand{\Dp}[1]{\ensuremath{\partial_{#1}}}

\newcommand{\limit}[2]{\ensuremath{\underset{#1 \tend #2}{\longrightarrow} }}
\newcommand{\J}[1]{\ensuremath{J_{\{#1 \,\}}}  }

\newcommand{\s}[1]{\ensuremath{\sinh\pa{#1}}}
\newcommand{\sd}[1]{\ensuremath{\mathfrak{s}\pa{#1}}}

\newcommand{\ex}[1]{\ensuremath{\e{e}^{#1}}}

\def\tr{\operatorname{tr}}

\newcommand{\op}[1]{ \boldsymbol{ \texttt{#1} } }

\newcommand{\abs}[1]{\ensuremath{\left| #1 \right|}}
\newcommand{\Norm}[1]{\ensuremath{\abs{\abs{#1}} }}

\newcommand{\norm}[1]{\ensuremath{|| #1 ||}}

\newcommand{\moy}[1]{\ensuremath{\langle #1 \rangle}}

\newcommand{\dd}{\mathrm{d}}
\newcommand{\e}[1]{\ensuremath{\mathrm{#1}}}

\newcommand{\intff}[2]{\ensuremath{ [  #1 \,; #2 ] }}
\newcommand{\intfo}[2]{\ensuremath{ [  #1 \,; #2 [ }}

\newcommand{\intoo}[2]{\ensuremath{ ]  #1 \,; #2 [ }}

\newcommand{\intn}[2]{\ensuremath{[\![ \, #1 \,;\, #2 \,]\!]}}

\begin{document}

\begin{titlepage}
\setlength{\topmargin}{-1cm} \enlargethispage{6 cm}

\begin{center}
\begin{tabular}{l p{5 cm} r}
& & Ann\'{e}e: $2015$\\

\end{tabular}

\vspace{1cm}
\textbf{\Large Institut de Math\'{e}matiques de Bourgogne\\ }
\textbf{\Large Universit\'{e} de Bourgogne} 
\vspace{0.3cm}

\begin{figure}[htbp]
     \begin{center}
        \end{center}
\end{figure}

\vspace{0cm}
{\LARGE \bf \scshape{Th\`{e}se d'habilitation \`{a} diriger les recherches}}\\
\medskip
{Sp\'{e}cialit\'{e}: Math\'{e}matiques}\\
\medskip
r\'{e}alis\'{e}e par\\
\medskip
\textbf{\LARGE \scshape{Karol Kajetan Klemens Kozlowski}}\\

\rule{\textwidth}{0.4pt}

\vspace{1 cm}
\textbf{\LARGE \scshape{Asymptotic analysis and quantum integrable models.} }\\
\vspace{1cm}

\begin{figure}[htbp]
\centerline{
        }
\end{figure}

\vspace{7mm}

\medskip

Soutenue publiquement le $19$ juin $2015$\\ devant la commission
d'examen constitu\'{e}e par

\medskip

\sc \bf {%
\begin{tabular}{l l p{1 cm} l}

Anne & Boutet de Monvel && \textnormal{Professeur} (Universit\'{e} Paris Diderot)  \\ 

Jean-S\'ebastien  & Caux & & \textnormal{Professeur} (Universiteit van Amsterdam) \\

 Nikola\"{i} & Kitanine   & & \textnormal{Professeur} (Universit\'{e} de Bourgogne) \\

Andreas & Kl\"{u}mper & & \textnormal{Professeur} (Wuppertal Universit\"{a}t) \\

Barry & McCoy & & \textnormal{Professeur} (Stony Brook University)\\

Fedor   & Smirnov & & \textnormal{Directeur de Recherches} (Universit\'{e} Pierre et Marie Curie)  \\
\end{tabular}%
}

\vspace{7mm}

\textnormal{Rapporteurs}

\vspace{2mm}

\sc \bf {%
\begin{tabular}{l l p{1 cm} l}

Jean-S\'ebastien  & Caux & & \textnormal{Professeur} (Universiteit van Amsterdam) \\

Percy  & Deift   & & \textnormal{Professeur} (New-York University) \\

Barry & McCoy & & \textnormal{Professeur} (Stony Brook University)   \\

Fedor   & Smirnov & & \textnormal{Directeur de Recherches} (Universit\'{e} Pierre et Marie Curie)   \\
\end{tabular}%
}

\end{center}

\end{titlepage}
\setlength{\topmargin}{.75 true cm}

\tableofcontents



\chapter*{Acknowledgments}

This habilitation thesis represents a synthesis of the research that I have carried out over the last six years,  between January 2009 and January 2015. 
During this period, I focused my attention on developing methods allowing one to study various asymptotic regimes of the correlation functions in quantum integrable models. 
Doing so was a source of incredible pleasure and joy. I do have to acknowledge, though, that my research would never 
have been possible if I didn't meet the right persons during my studies nor read the books that woke up my passionate interest in 
mathematical physics.

How could I  not be thankful to my parents? Throughout my childhood and early adult life, they have constantly kept stimulating
my curiosity towards the world and the dazzling amount of tiny details that make life so beautiful. 
I should also thank them for all the efforts they made to stimulate my imagination and for the so numerous encouragements I received 
to push the limits of my interior world beyond any tangible or reasonable boundary that I could be tempted to set. 
It is this curiosity that pushed me towards science, then physics and, more recently, mathematics. 
My close friends also played an important role in helping me to push the boarders of my imagination
and so often encouraged me to satisfy my curiosity.  I am so grateful to the Maripoleona for her constant, everyday, encouragements and support of my work.

I am extremely grateful to all the excellent teachers I had during my studies and whose enthralling lectures  opened me to 
the extreme beauty of physics and mathematics. I also am indebted to all the authors of books and articles whose lecture kept me 
absorbed for countess hours and led me deeper and deeper into the subject.

I consider myself very lucky to have met and interacted  with my collaborators: 
G. Borot, M. Dugave, F. G\"{o}hmann, A. Guionnet, A.R. Its, N. Kitanine, J.-M. Maillet, B. Pozsgay, E.K. Sklyanin, N.A. Slavnov, J. Suzuki, V. Terras and J. Teschner. 
I learned so much form the joint discussions and exchanges of ideas we have had.  My deepest gratitude goes to all of them.

I am also thankful to the administrative staff of the Institut de Mathématiques de Bourgogne. 
I thank L. Paris, the director, for all his efforts that resulted in an efficient and harmonious functioning of the laboratory. 
I thank the secretaries, Anissa Bellaassali, Ibtissam Bourtadi, Nadia Bader,  Caroline Gerin and Magali Crochot for all the help they have provided 
relatively to the various administrative queries I could have.

Last but definitely not least, I would like to express my  gratitude to Jean-S\'{e}bastien Caux, Percy Deift, Barry McCoy and Fedor Smirnov for having accepted to 
referee this manuscript. I am grateful as well to the members of my defence Jury Anne Boutet de Monvel, Jean-S\'{e}bastien Caux, Nikola\"{i} Kitanine, Andreas  Kl\"{u}mper, Barry McCoy
and Fedor Smirnov.

Finally, I am indebted  to all the institutions which supported my research in the last six years: 
the European Science foundation (Marie Curie Excellence Grant MEXT-CT-2006-042695), 
the Burgundy region and its FABER starting-grant program (PARI 2013-2014 "Structures et asymptotiques d'int\'{e}grales multiples"), the agence national pour la recherche (ANR DIADEMS) and the PEPS-PTI grant program of CNRS
("Asymptotique d'int\'{e}grales multiples"). 
I am also grateful to all the institutions that employed me throughout these years: the Deutsches Elektron-Synchrotron, the Indiana University Purdue University Indianapolis
and the Centre National pour la Recherche Scientifique. I am as well indebted to all the institutes which have hosted me, during my travels, and allowed me 
to pursue my work, always  in excellent conditions.

\chapter*{Special functions}

From its very first days, physics has been a tremendous source of inspiration for developing new mathematics.
A previously unknown phenomenon that gets unravelled by an experiment often requires the introduction of new abstract tools
and paradigms allowing one to model it and, more importantly, to gain a predictive power on future experiments of the same kind. First developments in mathematics took their origin in the need of quantifying
very simple observations of every day life. 
Most probably, early forms of trade led to the development of the notion of numbers and basic calculus. 
Trade was intimately related to a change in human behaviour: an evolution from hunter-gatherer societies to the sedenteral ones. 
With trade and sedimentation came wealth; with it emerged  the desire to build even more impressive edifices. All this gave an important impetus to the development of basic geometry, 
at least in what concerns the various forms it took in the ancient world. 
Formulae relating lengths or angles to other lengths, heights or areas of the involved object emerged around 2000 BC in the Egyptian and Babylonian civilisation
with the notion of angle being, however, mainly due to Greeks.
A formula expresses that several quantities are inter-related. Think, for instance, of the area law of a triangle relating  
the area to a base length and to its height relative to this base. As such, a formula offers the possibility of deducing the value of one quantity provided that all the other ingredients of the formula are known. 
In other words, formulae provide one with equations for some unknowns and thus enjoy of a tremendous predictive power... at least in theory. 
In practice, for an equation to be of any use to understanding or solving  a given problem, 
not only one has to be able to construct -whatever this word means- its solutions but one also needs to have at one's disposal tools allowing one to extract all the desired answers out of these. 
For instance, consider $a^2=b$ as an equation for $a$, with $a,b\in \R^+$. Then, the solution is $a=\sqrt{b}$. Still, if one is not able to compute effectively $\sqrt{b}$
then even though one can build the square root function on a purely theoretical ground, knowing that $\sqrt{b}$ solves the problem is useless if one wants to have, say, an explicit
few digit approximant of $a$. 

The equations that were first studied were rather simple from today's perspective: one way or another they boiled down to finding the roots of a polynomial in 
one variable. Despite the apparent simplicity of the problem, it became clear very quickly that the roots of a polynomial are  only seldom  
expressible in terms of objects belonging to the same class, \textit{viz}. some multi-variable polynomial in the coefficients of the original problem. 
In fact, it took quite a while before there arose a satisfactory way of describing the roots
of a generic polynomial equation of low degree. The latter demanded to introduce a new class of functions: the radicals. 
Some further generalisations (radicals of higher order, "imaginary" numbers) allowed to construct explicit solutions to  all  polynomial
equations of the second, third and fourth degree. 

For some time, this gave an impression that, \textit{in fine}, a finite number of algebraic manipulations adjoined to 
taking radicals is all that is needed so as to build the roots associated with an arbitrary polynomial equations. 
Needless to remind that this statement is known to be wrong! Even in the case of such simple equations one needs, as first argued by Galois \cite{GaloisOeuvresMathematiques}, 
to allow for an important change in the perspective. Galois introduced the concept of what one could refer
to as non-constructive existence, namely the fact that an equation may have solutions and yet one is by no means able to construct them explicitly. 
Slowly Galois' ideas steeped throughout the mathematics community and were applied to all types of equations. 

With the appearance of new mathematical structures, be it real-variable differential calculus, integration theory, non-commutative algebra, manifolds, groups,..., there arose a plethora of new types of equations, sometimes taking their origin in some physics related problem and, sometimes, 
solely appearing on purely abstract grounds. Numerous among these equations were then proven to admit solutions. Yet, only in a subset of cases the proofs of existence were constructive and, even in these
special cases, the "constructive" solutions were of no use in the sense that they were too involved so as to extract any useful knowledge or information out of them. 
 
In fact, it became clear that the equations which actually do admit solutions in a closed, explicit and, above all, manageable form are pretty rare. The functions corresponding to such solutions
were then baptised special functions, a concept that emerged in the mid-XVIII century. In the early days of the theory of special functions, this
speciality manifested itself in an overabundance of identities satisfied by the object: summation identities, addition formulae, \textit{etc}. 
(see \textit{e}.\textit{g}.  \cite{BatemanHigherTranscendentalFunctions} for a quite extensive account). 
As it turned out, the very existence of such identities stems from a deep algebraic structure at the root of special functions, the representation theory of groups. This connection was first observed by  
Bragmann \cite{BragmannFirstIntrLinkSpecFctsAndRepTh} in 1947 and systematised by Vilenkin and co-authors \cite{KlimykVilenkinRepsOfLieI}. The matter is that
special functions appear as matrix elements of a representation of a group. The variables and parameters associated with a given special function correspond to the parameters labelling the representation or, simply,
to a parametrisation of the group element $g$ being represented. Summation identities or addition formulae then arise as consequences of the fact that a representation is a group homomorphism
or simply follow from writing down a change of basis in the representation space leading to transformation laws for the matrix elements. Although providing a beautiful structure, this interpretation 
cannot be considered as the end of the story. 
Indeed, as already pointed out, constructing an object, even on the basis of crystal clear algebraic grounds, is not a synonym of "understanding" it! Yet, it is precisely the property 
of being understandable that is required from a special function. To rephrase things, a function -whatever the way of defining it- is "special" if one is able to extract from its definition all the desired information
on the function: local properties, specific values, asymptotic behaviour... 
It is at this stage that analysis kicks into the problem. The very existence of the deep algebraic structure at the root of special functions leaves a hope to actually
be able to develop suitable tools of analysis allowing one to characterise completely, that is to say extract all the desired data out of the special function. 

\vspace{2mm}

An important part of the present thesis is nothing else but the description of tools that I have developed with my collaborators which allow one to analyse, in the above sense, certain specific special functions. 
However, in order to provide a better account of its content, I need to push the discussion further. 

\vspace{2mm}

As the complexity of the equations that were being  manipulated grew, so as to continue being able to solve them explicitly, one had to revisit 
the very idea of what one understands by "explicit solution", and thus by special function.  One of the features of doing so is 
to allow these functions to be built through more and more involved operations. In itself, this poses no problem, as long as one is able to extract from the "explicit" answer all the desired informations on the
object itself. This setting naturally leads to a stratification of special functions. Namely, these are organised in layers, the complexity of a function
increasing with the number of layers lying below the stratum that it belongs to.  Functions belonging to a given layer are expressed through "manageable" manipulations of functions belonging 
to all lower layers. Here "manageable" has a rather loose sense and means the ability to perform an analysis of the expression which defines the function. 
This however does not mean that the analysis is necessarily easy.  I will now be slightly more explicit and provide some examples. 

\vspace{2mm}

The simplest equations are solved through a finite number of algebraic manipulations
(think of solving a system of linear equation: the solution is a rational function in the entries of the matrix associated with the linear system
as inferred from Cramer's rule). 
Yet, very quickly, one stumbles onto a situation where it becomes indispensable to use an infinite number of algebraic manipulations. 
For instance, the solutions to the first order linear differential equation $y^{\prime}(x)=y(x)$ are given by the exponential function

\beq
y(x) \; = \; C \cdot \ex{x} \qquad \e{where} \qquad  \ex{x} \; \equiv \; \sul{n \geq 0}{ + \infty} \f{x^n}{ n!}  \; . 
\nonumber
\enq
In this case, one has to recourse to an infinite summation. The power functions  $x\mapsto (1+x)^{\nu}$ and the logarithms can be constructed in a similar way. The procedure yields 
a new class of special functions generalising polynomials. 
These functions transcendent algebra in the sense that one has to step our of a purely algebraic framework so as to construct them. 
For this reason, they are referred to as the basic transcendental functions. 
A discrete sum, be it finite -as for polynomials- or infinite -as for basic transcendental functions- can be thought of as a specific instance of an integration in respect to a discrete
measure. Clearly, integration theory allows one to construct even more general functions, simply by choosing different types of measures. 
It is thus not a surprise that one-dimensional integrals whose integrands are obtained by compositions and multiplications of basic transcendental functions allow one to construct the next layer 
in the theory of special functions: the higher transcendental functions. 
Probably one of the simplest examples of such a function is the celebrated Euler Gamma function \cite{EulerStudyofGammaFunction} 
which arises as the unique, meromorphic on $\Cx$, solution to the finite difference equation $\Ga(x+1)=x\Ga(x)$ that is log-convex on $\R^+$. The $\Ga$-function admits the integral representation
\beq
\Ga(z) \; = \; \Int{0}{+\infty} t^{z-1} \ex{-t} \dd t \; \qquad \e{valid} \; \e{provided}\; \e{that} \qquad \Re(z)>0  \;. 
\nonumber
\enq

A slightly more involved example is provided by the solutions to the hypergeometric differential equation \cite{GaussEtudeDeFctHypergeometrique}
\beq
z\, (1-z) \cdot \f{ \dd^2 u  }{ \dd^2 z} \; + \; \big( c-(a+b+1) \, z \big) \cdot \f{\dd u }{ \dd z} - ab \cdot u \; = 0 \;. 
\nonumber
\enq
The regular at the origin solution is given in terms of the so-called Gauss one-dimensional integral representation
\beq
u(z) \; = \; \f{ \Ga(c) }{ \Ga(b) \Ga(c-b) } \Int{0}{1} t^{b-1} (1-t)^{c-b-1} (1-t z)^{-a} \cdot \dd t 
\nonumber
\enq
which is valid for $\Re(c) > \Re(b)>0$ and $|\e{arg}(1-z)| <  \pi$ or in terms of the Mellin-Barnes one
\beq
u(z) \; = \; \f{ \Ga(c) }{ \Ga(b) \Ga(a) } \Int{ \i \R }{}  \f{ \Ga(a+s) \Ga(b+s) \Ga(-s) }{  \Ga(c+s) }  (-z)^{s} \cdot \f{ \dd s}{2 \i \pi}
\nonumber
\enq
which is valid for $|\e{arg}(-z)| <  \pi$, $a, b \not\in -\mathbb{N}$. The $\i\R$  path of integration in the Mellin-Barnes representation 
is such that it separates the poles of the integrand belonging to   $ \mathbb{N}$ from those in $\{-a-\mathbb{N} \}\cup \{ -b - \mathbb{N} \}$. 
\textit{Per se}, the hypergeometric function should be considered as a slightly more structured object than the $\Ga$ function
in that its integral representation does involve the $\Ga$-function. In other words, from the hypergeometric function's perspective, the $\Ga$ function is to be considered as a known object. 

It is worth focussing a while on this last example. In itself,  an integral representation for  hypergeometric function, apart from providing a "closed"
well-defined representation, cannot be considered as THE "explicit" form of solutions to the hypergeometric equation. Without tools for analysing it, \textit{i}.\textit{e}. 
extracting all the desired information on the object being represented, the integral representations would only provide one with yet another hardly 
useful formal object. 
Still, the development of complex analysis gave birth to the saddle-point method and to techniques of analysis of integrals
based on contour deformations. It is these techniques that turned one-dimensional integral representations into extremely powerful 
tools. In the case of the hypergeometric function, such representations allow one for an easy access to: \vspace{1mm}
\begin{itemize}
\item[i)] determining the regions of analyticity in the auxiliary parameters ($a, b, c$) and in the principal variable $z$; \vspace{1mm}
\item[ii)] determining the small principal variable expansions; \vspace{1mm}
\item[iii)] determining the value of the function at a few special points so as to get a better insight on its growth and order of magnitude (\textit{e}.\textit{g}. for $z=1$ the Gauss integral representation 
simply reduces to the Euler's beta integral, whereas computing the explicit value of the associated hypergeometric series 
demands a much tougher analysis); \vspace{1mm}
\item[iv)] extracting asymptotic behaviours in the principal variable or in the auxiliary parameters in the vicinity  of the singularities (\textit{e}.\textit{g}. $z \tend \infty$).  \vspace{1mm}
\end{itemize}
We do stress that a specific integral representation is usually only effective for studying a given regime of the function. 
Thus, in a sense, it only provides one with a glimpse at one of the  many faces of the special  function it represents.
For instance, the Gauss one-fold integral representation for the hypergeometric function is particularly well suited for 
obtaining its expansion around $z=0$. However, one has to provide much efforts so as to extract the $z\tend \infty$ asymptotic expansion out it. 
However, the Mellin-Barnes representation is perfectly well suited for this purpose
in that the asymptotic expansion can be \textit{read-off} from it; it is simply enough to  deform the original integration contour to the left
and evaluate the residues of the simple poles of the integrand that were crossed in the process. 
I remind, for that matter, that, when $z\tend\infty$, $u(z)$ admits the asymptotic expansion\symbolfootnote[2]{The latter is, in fact, convergent around $\infty$, but this is a peculiarity of the 
hypergeometric function rather than a general rule}:
\beq
u(z) \; = \; \f{ \Ga(c) \Ga(b-a) }{ \Ga(b) \Ga(c-a) } \big(-z\big)^{-a} \bigg( 1 \; + \; \f{ a(a+1-c) }{ (a+1-b) z } \: + \; \cdots  \bigg) \; + \; 
\f{ \Ga(c) \Ga(a-b) }{ \Ga(a) \Ga(c-b) } \big(-z\big)^{-b} \bigg( 1 \; + \; \f{ b(b+1-c) }{ (b+1-a) z } \: + \; \cdots  \bigg) \;. 
\nonumber
\enq
All building blocks of this asymptotic expansion are given in terms of special functions belonging to "lower" layers. Note that, in this respect, 
it is not surprising that the $\Ga$-function is present, since as mentioned earlier, it belongs to a slightly lower layer. 
The "layer" of higher transcendental functions is now very well understood, see \textit{e.g.} the Bateman project book \cite{BatemanHigherTranscendentalFunctions}.

At this stage it seems natural to wonder what is the stratum lying just above the one of higher transcendental functions. Probably,
the so-called Painlev\'{e} transcendents \cite{FokasItsKapaevNovokshenovPainleveTranscendents} are the best known representatives of this layer. This name refers to  a class of special functions that was discovered in the early XX century 
by Painlev\'{e} on the occasion of classifying all second order non-linear differential equation that have the so-called Painlev\'{e} property: the only singularities of the solution 
which are allowed to depend on the initial data are the \textit{locii} of its poles.
Although it is not a theorem, it appears that these functions do not admit one-fold, or for that matter any \textit{finite} order, integral representations
whose integrands are built from, at most, higher transcendental functions. 
However, should one agree to extend the notion of integral representation to the case of series whose $n^{\e{th}}$ summand is given by a  $k\cdot n$-fold
multiple integral whose integrand is given by certain "simple" combinations of transcendental functions, then indeed one can construct 
such explicit representations for the Painlev\'{e} transcendents. For instance, the solution $\sg$ of the so-called Hirota form of the fifth Painlev\'{e} equation 
\beq
\Big( x \f{ \dd^2 \sg }{ \dd x^2  }  \Big)^2 \; = \; -4 \cdot \bigg(\sg - x \f{ \dd \sg }{ \dd x  } -    \Big( \f{ \dd \sg }{ \dd x  }\Big)^2 \bigg) \cdot 
\Big( \sg - x \f{ \dd \sg }{ \dd x  } \Big)
\nonumber
\enq
that has the $x\tend 0$ expansion
\beq
\sg(x) \; = \; -\f{x}{\pi}-\f{x^2}{\pi^2} \; + \; \e{O}\big( x^3\big)
\nonumber
\enq
can be recast as the logarithmic derivative of a series of multiple integrals which corresponds  to the Fredholm determinant of the so-called sine kernel; namely
\beq
\sg(x) \; = \; x \f{ \dd  }{ \dd x } \ln \det \big[ \e{id}-\op{S}_{x/2}\big]
\nonumber
\enq
where $\op{S}_{x}$ appearing above is the integral operator on $L^2\big(\intff{-1}{1}\big)$ characterised by the integral kernel
\beq
S_x(\la,\mu) \; = \; \f{ \sin[x(\la-\mu)] }{ \pi (\la-\mu) } \;. 
\nonumber
\enq
Its Fredholm determinant admits the below series of multiple integral representation
\beq
\det \big[ \e{id} \, - \,  \op{S}_{x}\big] \; = \; \sul{n \geq 0}{} \f{ (-1)^n }{ n! } \Int{-1}{1} \det_n\big[ S_{x}(\la_a,\la_b) \big] \cdot \dd^n \la \qquad \e{with} \quad  \dd^n \la \; \equiv \; \pl{a=1}{n} \dd \la_a \; . 
\nonumber
\enq
One could immediately wonder if any Fredholm determinant associated with an integral operator $\e{id}+\op{V}_{x}$ on $L^{2}(J)$, $J$ a sufficiently regular curve in $\Cx$, whose integral kernel depends on some auxiliary parameter $x$
deserves to be called  special function of $x$. The answer appears to be negative in that, for a generic integral kernel $V_x(\la,\mu)$, it seems impossible to realise the program i)-iv) relatively to the 
associated Fredholm determinant. 
Thus, in order to embrace the class of special functions that belong to the Painlev\'{e} transcendent layer one should first answer the question of why are the Painlev\'{e}
transcendents so special. To cut the long story short, the answer is that there lies a deep algebraic structure at their root, the Riemann--Hilbert problems \cite{FokasItsKapaevNovokshenovPainleveTranscendents}. 
More precisely, Painlev\'{e} transcendents appear as some of the matrix entries of a solution $\chi$ to a $2\times2$ matrix Riemann--Hilbert problem
whose jump matrix depends parametrically on the auxiliary parameters and the transcendent's principal variable. 
The existence of powerful methods of asymptotic analysis of matrix Riemann--Hilbert problems (the non-linear steepest descent of Deift and Zhou \cite{DeiftZhouSteepestDescentForOscillatoryRHP,DeiftZhouSteepestDescentForOscillatoryRHPmKdVIntroMethod}, 
the concept of local parametrix introduced by Its \cite{ItsDifferentialMethodForParametrix} or the dressing-function trick introduced by Deift, Its and Zhou \cite{DeiftItsZhouSineKernelOnUnionOfIntervals}
that also appeared in a more sophisticated variant adapted for studying orthogonal polynomials \cite{DeiftKriechMcLaughVenakZhouOrthogonalPlyExponWeights}) 
allow one to complete the program i)-iv) for the Painlev\'{e} transcendents, see \textit{e}.\textit{g}. \cite{FokasItsKapaevNovokshenovPainleveTranscendents}. 
Although a generic integral operator $\e{id} +\op{V}_x$ does not admit any connection with a Riemann--Hilbert problem, there exists an algebra of integral operators 
whose Fredholm determinant can be fully characterised in terms of a solution to a matrix Riemann--Hilbert problem: the integrable integral operators. 
These are operators whose integral kernel takes the form
\beq
V_{x}(\la,\mu) \; = \; \f{ \sum_{a=1}^{N} f_a(\la) e_{a}(\mu)  }{ \la - \mu } \qquad \e{with} \qquad \sum_{a=1}^{N} e_a(\la) f_{a}(\la) \;=0
\nonumber
\enq
and $e_a$, $f_a$ are functions whose regularity depends on the given problem\symbolfootnote[2]{In fact, the diagonal vanishing condition can be omitted for the price of working with
a principal value regularisation \cite{DeiftIntegrableOperatorsDiscussion}} and which depend parametrically on the auxiliary variable $x$. 
Note that one can even consider a slightly more general form for these operators by replacing the discrete sum -which can be thought of as an integration
with respect to a discrete compactly supported measure- by an integration in respect to a measure $ \mu$ on a measure space $X$. In that case, the 
index $a$ runs through $X$.

The relation between integrable integral operators and Riemann--Hilbert problems was established in 
the 1990 work of Its, Izergin, Korepin and Slavnov \cite{ItsIzerginKorepinSlavnovDifferentialeqnsforCorrelationfunctions}. However, \textit{per} \textit{se}, the theory of integrable
integral operators definitely takes its roots in the 1980 work of Jimbo, Miwa, Mori and Sato \cite{JimMiwaMoriSatoSineKernelPVForBoseGaz} but 
some elements of the theory were already implicitly present in the 1968 work of Sakhnovich \cite{SakhnovichFirstStepsOfElementsOfIntegrableIntOps}. 
In the case where  the functions $f_a$ and $e_a$ take an explicit and 
sufficiently simple form, one can show, using general arguments from the theory of Riemann--Hilbert problems, that 
the determinants or minors associated with operators $\e{id}+\op{V}_{\{x_i\}}$, where $\op{V}_{\{x_i\}}$ is an integrable integral operator, 
satisfy systems of partial differential equations in respect to the auxiliary parameters -which I have denoted by $\{x_i\}$- entering in the
expression of their kernel or in respect to the endpoints of the support on which the operator acts. In some cases, these equations can be reduced to ordinary differential equation of Painlev\'{e} type. 
Pioneering work on the partial differential equation aspects has been carried out in \cite{ItsIzerginKorepinSlavnovDifferentialeqnsforCorrelationfunctions}. Also, several aspects of these, 
especially in what concerns the study of partial differential equations and Painlev\'{e} equations for specific types of kernels that arise in random matrix theory, have been developed in 1994
by Tracy and Widom in \cite{TracyWidomPDEforFredholms}.

Having climbed that far into the height of strata  of the theory of special functions, one would like to move further and describe the layer of special functions that lies just above the 
Painlev\'{e} transcendent/integrable integral operator class. This class of function has only started to emerge recently, in the 90's,  
and is intimately related to the correlation functions in so-called  interacting quantum integrable models. 
It is the study and description of these functions that is at the heart of the present habilitation thesis. 
I have been developing, over the last six years, methods allowing one to analyse  effectively certain of these functions, that is to say realise the analogue of the program i)-iv)
that was mentioned earlier. 
I have achieved a certain substantial progress on the matter although many open questions remain. In order to describe the problems I was interested in and  
the techniques that I developed to solve them, completely or at least partially, I need to provide some background. This discussion will also 
be an occasion to describe the physics context in which the study is embedded, hence providing another motivation for
my work. Once that I will introduce the general setting, I shall provide an account, which I do not claim to be exhaustive, of the state of the art of the field of quantum integrable models and the theory of its correlation functions. 
This will permit me to delineate the open problems that I faced. 
I will then discuss the solution to these problems in the core of the thesis.

\chapter{Introduction}

\section{Motivations from one-dimensional physics}

\subsection{Some general facts}

A quantum mechanical model in one spatial dimension consists, \textit{primo} in the choice of a Hilbert space $\mf{h}_{\e{phys}}$ where the physical reality of interest will be represented. 
Usually, this Hilbert space admits a tensor product decomposition
\beq
\mf{h}_{\e{phys}} \; = \; \mf{h}_1 \otimes \cdots \otimes \mf{h}_L
\label{ecriture espace Hilbert physique du modele}
\enq
into a product of $L$ "local" spaces. For instance, in the case of lattice models such as spin chains, each space $\mf{h}_a$ occurring in the tensor product decomposition represents a given site of the model and is finite dimensional. 
It represents the spin degrees  of freedom attached to the site. 
In that case, the number of sites $L$ plays the role of the model's volume. For models in the continuum, such as a system of $N$  interacting particles, 
the number of "sites" corresponds to the number of particles in interaction while the space $\mf{h}_{a}$ is attached to the $a^{\e{th}}$ particle and represents its degrees of freedom. 
In this case, the terminology volume rather refers to the physical volume in which the particles evolve. 
One often  considers  a situation where the "local" spaces $\mf{h}_a$ are all isomorphic $\mf{h}_a \simeq \mf{h}$. This will always be the case in this thesis. 
One needs to supplement these informations with a basis $\op{O}^{(k)}$ of operators on $\mf{h}$, $k=1,\dots, K_{\mf{h}}$, where $K_{\mf{h}}$ can be finite or not. This basis provides one with a 
basis of operators on $\mf{h}_{\e{phys}}$ as
\beq
\op{O}^{(k)}_a \; = \; \underbrace{ \e{id} \otimes \cdots \otimes \e{id} }_{a-1} \otimes\,  \op{O}^{(k)} \!\! \otimes  \underbrace{ \e{id} \otimes \cdots \otimes \e{id} }_{L-a} \;. 
\enq
The operators $\op{O}^{(k)}_a$ are called local in that they only act non-trivially only on a single "site" of the model. 

\textit{Secundo}, in order to speak of a quantum-mechanical model, on top of providing a Hilbert space and a basis of operators on this space, one has to specify a privileged operator 
$\op{H}_L$ on  $\mf{h}_{\e{phys}}$ called the Hamiltonian. $\op{H}_L$ is, by requirement, a self-adjoint operator on $\mf{h}_{\e{phys}}$. 
Furthermore, the finiteness of the volume plays the role of a sort of regularisation in that it makes, in concrete situation, the spectrum of $\op{H}_L$
to be discrete\symbolfootnote[4]{Although trivial if the spaces $\mf{h}_a$ in \eqref{ecriture espace Hilbert physique du modele} are finite dimensional, this is a quite non-trivial 
property when some of the $\mf{h}_a$'s are infinite dimensional.}. The eigenvalues and eigenvectors of $\op{H}_L$ contain all the informations on the physics of the model.  
The eigenstate of $\op{H}_L$ associated with the lowest eigenvalue is called the ground state\symbolfootnote[3]{Here, for simplicity of the discussion, I will assume no degeneracy
of the lowest eigenvalue} 
while all other are called the excited states. Unless stated otherwise, I will consider models having translational invariance meaning that there exists an operator 
$\ex{\i \op{P}_L}$ such that $\big[ \op{P}_L, \op{H}_L \big]=0$ and the adjoint action of $\ex{\i \op{P}_L}$  realises a translation along $\mf{h}_{\e{phys}}$:
$\op{O}^{(k)}_{a+1} = \ex{- \i \op{P}_L } \op{O}^{(k)}_a \ex{\i \op{P}_L}$. The operator $\op{P}_L$ is called the momentum operator. 

At this stage, the model is still disconnected from physics. Indeed, in order to compare a theoretical model with experimental data
one isn't that much interested in the spectrum of excitations above the ground state nor in its associated eigenstates but rather in its finite temperature correlation functions. 
For simplicity, assume that the model is defined by the finite volume $L$ Hamiltonian $\op{H}_L$ acting on a finite dimensional Hilbert space, \textit{viz}. a situation where all the local spaces $\mf{h}_a$ are finite dimensional
\symbolfootnote[2]{In such a case the dimension of the space grows exponentially fast in $L$ since, in order for the $a^{\e{th}}$ site to be non-trivial, one has to take $\dim \mf{h}_a \geq 2$}. 
The simplest physically pertinent observable is the \textit{per} site free energy 
\beq
f_L \; = \; -\f{T}{L  } \ln \mc{Z}_T[\op{H}_L] \qquad \e{where}   \qquad  \mc{Z}_T[\op{H}_L]\; = \; \e{tr}_{ \mf{h}_{\e{phys}} }  \big[ \ex{-\frac{1}{T} \op{H}_L } \big] 
\enq
is the model's finite temperature partition function. The finite temperature correlation functions
 correspond to the so-called thermal expectation values of products of local, time evolved, operators:
\beq
\Big< \op{O}^{(k_1)}_{1+x_1}(t_1)\cdots \op{O}^{(k_m)}_{1+x_m}(t_m) \Big>_{T,L} \; \equiv \; 
\e{tr}_{ \mf{h}_{\e{phys}} }\bigg[ \op{O}^{(k_1)}_{1+x_1}(t_1)\cdots \op{O}^{(k_m)}_{1+x_m}(t_m) \cdot  \f{  \ex{-\frac{1}{T} \op{H}_L }  }{ \mc{Z}_T[\op{H}_L] } \bigg] 
\enq
where the temporal evolution\symbolfootnote[3]{Whenever there is no time dependence, we shall drop the principal argument and write $\op{O}^{(k)}_{x}\equiv \op{O}^{(k)}_{x}(0)$ } of an operator is defined as 
\beq
\op{O}^{(k)}_{x}(t) \; = \; \ex{\i t \op{H}_L }  \cdot  \op{O}^{(k)}_{x} \cdot  \ex{  -\i t \op{H}_L } \;. 
\enq
The presence of temperature tremendously complicates the situation. For instance, the zero-temperature analogue of $f_L$ is the per-site ground state energy. 
This follows readily, at least under the above hypothesis, by taking the $T\tend 0^+$ limit of the expression for $f_L$. 
Hence, while only the knowledge of a single eigenvalue is necessary to characterise the zero temperature case, one needs to access to the whole tower
of eigenvalues 
and then be able to sum up the associated series. 
The task is, usually, intractable. In fact, explicit expressions at finite $L$ could only have been obtained for free energies associated with free fermion equivalent models.

When setting the temperature to zero, the situation simplifies. In what concerns the zero-temperature limit of the correlation functions, the thermal averages reduce to 
the ground state expectation values 
\beq
\Big< \op{O}^{(k_1)}_{1+x_1}(t_1)\cdots \op{O}^{(k_m)}_{1+x_m}(t_m)  \Big>_{L} \; \equiv \bra{ \Psi_{GS} } \op{O}^{(k_1)}_{1+x_1}(t_1)\cdots \op{O}^{(k_m)}_{1+x_m}(t_m)  \ket{ \Psi_{GS} } \;, 
\enq
where $\ket{ \Psi_{GS} }$ is the eigenstate associated with the lowest lying eigenvalue of $\op{H}_L$. 

Let $\ket{\Psi_i}$ be an orthonormal basis of $\mf{h}_{\e{phys}}$ built out of the eigenvectors of $\op{H}_L$. 
 Since $\op{P}_L$ and $\op{H}_L$ commute, one can choose the $\ket{\Psi_i}$ as common eigenvectors for these two operators. With each local operator $\op{O}_x^{(a)}$, 
one can associate its form factors, namely its matrix elements relatively to $\op{H}_L$'s eigenbasis, 
\beq
\mc{F}_{ \op{O}_1^{(a)} }\big(\Psi_k,\Psi_{\ell}\big) \, = \, \bra{ \Psi_k } \op{O}_1^{(a)}\ket{ \Psi_{\ell} } \; .
\enq
Due to the translation invariance of the model, the matrix elements of the local operator $\op{O}_{1+x}^{(a)}$
factorise as
\beq
\bra{ \Psi_k }  \op{O}_{1+x}^{(a)} \ket{ \Psi_{\ell} } \; = \; 
\ex{ \i x ( \wh{P}_{\ell} \, - \,  \wh{P}_ {k} ) \, - \,  \i t ( \wh{E}_{\ell} \, - \,  \wh{E}_{k} )  } \cdot  \mc{F}_{ \op{O}_1^{(a)} }\big(\Psi_k,\Psi_{\ell}\big) 
\enq
where $\wh{P}_k$ and $\wh{E}_k$ are the eigenvalues of the momentum $\op{P}_L$ and the Hamiltonian $\op{H}_L$ associated with the eigenstate $\ket{\Psi_k}$ and relatively to the model's ground state.
As a consequence, one can expand an $m$ point function at zero temperature and finite volume $L$ into its so-called form factor series:
\beq
\Big<  \op{O}^{(k_1)}_{1+x_1}(t_1)\cdots \op{O}^{(k_m)}_{1+x_m}(t_m) \Big>_{L} \; = \; \sul{ i_1 , \dots , i_{m-1} }{} \pl{a=1}{m} \Bigg\{ \mc{F}_{ \op{O}_a } \big( \Psi_{ i_{a-1} }, \Psi_{ i_{a} } \big) \Bigg\}
\pl{a=1}{m-1} \bigg\{ \ex{\i (x_{a+1}-x_{a}) \wh{P}_{i_a} - \i (t_{a+1}-t_a) \wh{E}_{i_a} } \bigg\}
\label{ecriture dvpmt fct deux pts cas tres tres general}
\enq
This expansion is quite natural from the physics' point of view. It can be seen as a decomposition of a correlation function into energy-momentum modes in that
the dependence of the individual terms  on the space and time is completely factorised in \eqref{ecriture dvpmt fct deux pts cas tres tres general}. 
In the case of a finite dimensional base space $\mf{h}$ with $\mf{h}_{a}\simeq \mf{h}$ for all $a$ -what is the case of spin lattice models for instance- 
the set $\big\{ \ket{\Psi_k} \big\}$ is finite. The form factor expansion is thus given in terms of a finite sum and thus perfectly well defined. In particular, one can 
take any finite order, space or time, partial derivatives thereof. However, should $\mf{h}$ be infinite dimensional -what is typically the case of interacting systems of particles on the line-
the form factor expansion \textit{a priori} converges (think of Fourier series) in $L^2$, but usually not in a stronger sense. Hence, in general, although useful from the physics' perspective, such expansions should be 
handled with utmost care.

I have, so far, focused the discussion to finite volume  systems. 
Yet, the situation of interest to physics corresponds to a model at finite temperature and infinite volume, \textit{viz}. at $L\tend +\infty$. 
Immediately, then, pops up the question of the well-definiteness of the limit be it on the level 
of the free energy or the correlation function. Surprisingly, the existence of the $L\tend +\infty$ limit can be established on rigorous grounds, at least 
starting from numerous realistic Hamiltonians $\op{H}_L$. I refer to the book of Ruelle \cite{RuelleRigorousResultsForStatisticalMechanics} for more details. 
I will henceforth refer to the $L\tend +\infty$ as the thermodynamic limit. 
This limit does strongly alter the picture in that the volume plays a sort of regularisation parameter. First of all, the spectrum of $\op{H}_L$
will usually exhibit a continuous part which consists of so-called particle-hole excitations and a discrete part which consists of so-called bound states. 
On this level, one distinguishes two cases: \vspace{1mm}
\begin{itemize}
 
 \item[$\bullet$] the tower of states forming the continuous part of the spectrum is directly connected to the ground state. In that case, one speaks of a massless model in that there
 exists an arbitrarily large number of excited states  having a zero energy relatively to the ground state. \vspace{1mm}
 
 \item[$\bullet$] The continuous part of the spectrum is disconnected from the ground state, \textit{viz}. there exists a gap between the ground state\symbolfootnote[3]{Note that, in principle, one can allow 
 for a finite number of eigenstates to crush onto the ground state. In that case, the latter is finitely degenerated in the thermodynamic limit.} and the continuous part of the spectrum.
 In that case, one speaks of a massive model in that an arbitrarily large number of zero energy excitations is \textit{not} possible. \vspace{1mm}

\end{itemize}

These two regimes of the spectrum are believed to have strong implications on the large-distance behaviour of the zero temperature behaviour of a model's correlations functions. 
For instance, consider a two-point function $\big< \op{O}_1^{(\a)}  \op{O}_{1+x}^{(\be)} \big>$ directly in the thermodynamic limit and at zero temperature. 
At infinite spacing, the operators should be blind to each other's presence so that the correlator should go to the product of  expectation values of individual local operators
building up the correlator: $\big< \op{O}_1^{(\a)}  \big> \cdot \big< \op{O}_{1}^{(\be)} \big>$. 
This argument holds independently of the 
massive or massless structure of the spectrum. The massive or massless nature of the spectrum is however believed to influence the speed at which the correlator approaches its limit, or,
in other words the type of large-$x$ decay of the so-called connected correlators
$\big< \op{O}_1^{(\a)} \op{O}_{1+x}^{(\be)} \big>_{\e{c}} \, = \,  \big<\op{O}_1^{(\a)} \op{O}_{1+x}^{(\be)} \big> \, - \, \big< \op{O}_1^{(\a)}  \big> \big<  \op{O}_{1+x}^{(\be)} \big>$. 
Presumably, the connected correlation functions of a massive model decay exponentially fast in the distance of separation between the operators
\beq
 \big< \op{O}_1^{(\a)} \op{O}_{1+x}^{(\be)}\big>_{\e{c}} \; = \; \mc{A}_1^{(\a   \be)} \ex{-\f{ x }{\xi_1} } \; + \; \mc{A}_2^{(\a  \be)} \ex{-\f{ x }{\xi_2} } \; + \; \dots 
\enq
There $\xi_k$ are the so-called correlation lengths while the $\mc{A}_k^{(\a   \be)}$ are the associated amplitudes. The $\dots$ include faster decaying terms, \textit{viz}. those having smaller
correlation lengths. The same structure of the asymptotic expansion is believed to hold true for massless models at finite temperature $T$. In such a case the correlation lengths are functions 
of the temperature which furthermore satisfy $\lim_{T\tend 0^+}\xi_a  = +\infty$.

The large-$x$ asymptotic behaviour changes drastically in the case of a massless model:  the correlation function are expected to approach their infinite distance
limit much slower, algebraically fast in the distance of separation $x$:
\beq
 \big< \op{O}_1^{(\a)} \op{O}_{1+x}^{(\be)}\big>_{\e{c}} \; = \; \f{ \msc{A}_1^{(\a \be)} }{ x^{\nu_1} } \; + \;\f{ \msc{A}_2^{(\a \be)} }{ x^{\nu_2} } \; + \; \dots 
\enq
There, the $\nu_k$ are the critical exponents while $\msc{A}_k^{(\a \be)}$ stand for the associated amplitudes. Note that the massless regime of a model is sometimes referred to as its critical regime. 
Although this has not been written down, both types of asymptotics -be it massive or massless- can posses, on top of a pure decay, vanishing terms that are also oscillatory in respect to the distance.


 \subsection{Approximate methods and predictions}
 
 For a generic realistic, even oversimplified, model it is hopeless to solve the spectral problem associated with $\op{H}_L$, not to mention  calculating its thermal partition or correlation functions, even at zero temperature.
However, one can hope to study certain overall properties of the correlation functions on the basis of the universality principle. 
Models sharing the same value of their zero temperature critical exponents are said to belong to the same universality class. It is believed that 
critical models sharing the same symmetry properties of their Hamiltonian should belong to the same universality class. 
In some situations, the universality principle allows one to go even further and relate the decay properties at large-distances for different massless models at finite, but very small, temperatures. 

The universality principle provides one with powerful means of characterising certain limiting regimes of a massless model's correlation functions. 
Namely, if one is able to obtain the critical exponents of a model, than one is immediately able to predict those of all models belonging to the same universality class. 
In one spatial dimension\symbolfootnote[2]{In the language of classical statistical mechanics, this corresponds to the case of a two-dimensional model \cite{SuzukiCorrespondence(D+1)StatPhysDQuantumHamiltonians}.},
there exist several field theoretic massless models whose spectra and critical exponents can be determined explicitly:
 the Luttinger model (see \cite{HaldaneStudyofLuttingerLiquid} and references therein) and  two-dimensional conformal field theories \cite{BeliavinPolyakovZalmolodchikovCFTin2DQFT,DiFrancescoMathieuSenechalCFTKniga}.
 These can be taken as a basis for accessing to the value of critical exponents for models belonging to their universality classes. I shall now develop 
 on these ideas a bit more.

The  transformation laws for so-called primary operators in a conformal field theory impose 
that correlation function of such operators display an algebraic in the distances between the operators pre-factor. These symmetries are sufficiently constraining 
so as to completely determine the distance dependence of two and three point functions  in a conformal field theory:
it is purely-algebraic. 
In such a situation, the exponents driving the power-law behaviour of a correlator are then expressed in terms of the scaling dimensions of the primary operators that build up the correlation function. 
A given conformal field theory is characterised by its central charge $c$. Different values of the central charge correspond to distinct universality classes: the 
 data issuing from the study of conformal field theories can be exploited so as to predict the critical exponents of a plethora of massless quantum models 
in one spatial dimension. The conformal field theory-based predictions build, in fact, on two pillars. 
 
 Polyakov \cite{PolyakovArgumentAboutCFTInvarianceCriticalCorrelators} argued that the correlation functions of a massless model should  exhibit a conformal invariance in the 
limit where all of the local operators  are far apart, \textit{i}.\textit{e}. in the large-distance limit. Polyakov's argument suggests that the leading contribution to the long-distance asymptotics of a correlator in a massless model 
should be reproduced by correlation functions of appropriate operators in some two-dimensional conformal field theory. Still, the identification of which conformal 
 field theory is to be used and which are the "appropriate" operators  is left open at this stage.

 In 1986 and independently, Bl\"{o}te, Cardy and Nightingale \cite{BloteCardyNightingalePredictionL-1correctionsEnergyAscentralcharge} and Affleck \cite{AffleckCFTPreForLargeSizeCorrPartitionFctonAndLowTBehavior}
 argued that the ground state energy $\wh{E}_{G.S.}$ of a Hamiltonian $\op{H}_L$ having a massless spectrum in the thermodynamic limit takes the form 
\beq
\wh{E}_{G.S.} \; = \; L E_0 \;  - \; \f{ \pi c }{6 L} v_F  \; + \;  \cdots 
\enq
in which $E_0$ in the model's per-site energy in the thermodynamic limit, $v_F$ is the velocity of the excitation on the Fermi boundary while $c$ corresponds to the central charge of the conformal field theory belonging to the model's
universality class. 
Later on, Cardy \cite{CardyConformalDimensionsFromLowLSpectrum} went much further and argued that all the conformal dimensions $\{ \De_{k} \}_{ k \in I}$ of the primary operators, on the 
conformal field theory side, which arise in the modelling of the large-distance behaviour of local operators in the original model can be read-off from the relative to the ground state 
energies $\wh{E}_n$ and momenta $\wh{P}_n$ associated with the model's low-lying excitations, \textit{i}.\textit{e}. those whose excitation energies scale as $1/L$:
\beq
\wh{E}_n  \; \simeq  \; \f{2\pi v_F }{L} \Big(  \De_{k_n}^++\De^{-}_{k_n} \; + \; m_n^+\, + \, m_n^-  \Big) \; + \;  \cdots   \quad \e{and} \quad 
\wh{P}_n  \; \simeq  \; 2 p_F \ell_{k_n} \, + \, \f{2\pi }{L} \Big(  \De_{k_n}^+ - \De^{-}_{k_n} \; + \; m_n^+\, - \, m_n^-  \Big) \; + \;  \cdots   
\nonumber
\enq
The integer $k_n$ picks the scaling dimension of the primary operators associated with the excited state $\wh{E}_n$ while $m_n^{\pm} \in \mathbb{N}$
issues from the fact that the given excited state may also contain contributions of the descendants of the primary operator. Finally, $\ell_{k_n}$ is an integer encoding the 
fact that zero energy excitations in massless models can carry a non-zero momentum due to the possibility of Umklapp excitations.
The momentum of an Umklapp excitation is $2p_F$ where $p_F$ is the so-called Fermi momentum of excitations. 
The presence of such a non-vanishing Umklapp momentum was argued  in 1986
by Bogoliubov, Izergin and Reshetikhin in \cite{BogoluibovIzerginReshetikhinCriticalExponentsforXXZ}. 

With these two pillars available, the choice of the "appropriate" operators, on the conformal field theory side, which should grasp the asymptotic behaviour of a given correlator on the 
physical model side is done by advocating that the conformal operators should inherit of all the symmetries that hold for the operators involved in the correlation function one starts with.
For instance, if a lattice operator $\op{O}^{(k)}_a$ has non-zero form factors between the ground state and an excited state with energy $\wh{E}_n$,
then the primary operators with scaling dimensions $\De_{k_n}^{\pm}$ will be among the appropriate operators, on the conformal field theory side, 
which grasp the asymptotic behaviour of correlators involving the operator $\op{O}^{(k)}_a$.

A similar line of reasoning applies when exploiting the Luttinger liquid model as a tool for providing the critical exponents.  One argues that a model belongs to the Luttinger liquid universality class 
if it has the same form of the low-lying excitations above its ground state. The parameters describing these excitations fix the Luttinger liquid model that 
is pertinent for the model of interest. The critical exponents of the model one starts with are then deduced from the ones computed explicitly 
for the Luttinger liquid model 
\cite{HaldaneCritExponentsAndSpectralPropXXZ,HaldaneLuttingerLiquidCaracterofBASolvableModels,LutherPeschelCriticalExponentsXXZZeroFieldLuttLiquid}. 
It is worth singling out that the Luttinger liquid approach is limited to a smaller amount of models than
the more general, conformal field theory based one.

\subsubsection{The large-distance asymptotics at zero temperature}

To rephrase things, Polyakov's argument justifies why a conformal field theory should emerge as an effective large-distance theory,
while Cardy's observation permits one to extract, from the knowledge of the structure of a model's excitations, the quantities which would characterise
the scaling dimensions of the operators - on the conformal field theory side- which describe the large-distance regime of the model's correlators. 
On the basis of these arguments one predicts that, at zero temperature, a connected two-point function of a model belonging to the universality class
of some two-dimensional conformal field theory will exhibit the large-$x$ asymptotic expansion
\beq
\big< \op{O}_{1+x}^{(\a)}\cdot \op{O}_{1}^{(\be)} \big>_{\e{c}}  \; \simeq \;  \sul{ k\in I^{(\a\be)}  }{} \;  \f{ \mc{A}_k^{(\a \be)} \cdot  \ex{  2 \i x  \ell_{k}  p_F } }{ |x|^{ 2 \De^+_k +   2 \De^-_k } } \; . 
\label{ecriture predictions conformes DA 2 pt fct zero time}
\enq
The sum runs through the set of integers $I^{(\a\be)}$ which contains all the labels of the scaling dimensions that are relevant to the description of the two-point function's asymptotic behaviour while $ \mc{A}_k^{(\a \be)} $
are model dependent amplitudes associated with the contribution of a given conformal dimension.  
Within this approach, one also argues \cite{CardyConformalDimensionsFromLowLSpectrum} that the amplitudes should be given as the limits
\beq
\mc{A}_k^{(\a \be)} \; = \; \lim_{L\tend + \infty}  \bigg\{  L^{ 2 \De^+_k +   2 \De^-_k } \cdot \bra{\Psi_{G.S.} }  \op{O}_{1}^{(\a)} \ket{ \Psi_{\De^{\pm}_{k} } } \cdot \bra{ \Psi_{\De^{\pm}_{k} } }  \op{O}_{1}^{(\be)} \ket{\Psi_{G.S.} } \bigg\}\;
\label{ecriture prediction CFT pour definition amplitude dans fct 2 pts} 
\enq
where $\ket{ \Psi_{\De^{\pm}_{k} } }$ would be the eigenstate of $\op{H}_L$ that is associated, on the conformal field theory side, with the primary operators
having scaling dimensions $\De^{\pm}_k$. 

Note that  expansions of the type \eqref{ecriture predictions conformes DA 2 pt fct zero time}  don't say anything in respect to the 
sub-dominant in $|x|$ corrections to each term of the sum with a fixed exponent $2 \De^+_k +   2 \De^-_k$. This behaviour is expected not to be universal 
and thus not graspable directly within the sole framework of conformal field theory. 

\subsubsection*{The case of time-dependent correlators at zero temperature}

There were a few attempts to extend the predictive arsenal of conformal field theory to the case of the large-distance and long-time asymptotic behaviour 
of time and space dependent correlation functions. Such predictions then take the form: 
\beq
\big< \op{O}_{1+x}^{(\a)}(t) \cdot \op{O}_{1}^{(\be)}(0) \big>_{\e{c}} \; \simeq \;  \sul{ k\in I^{(\a\be)} }{} \; \f{ \mc{A}_k^{(\a \be)} \cdot  \ex{  2 \i x  \ell_{k}  p_F } }{  |x-v_Ft|^{ 2 \De^+_k} \cdot  |x+v_F t|^{ 2 \De^-_k} } \; . 
\label{ecriture conformal prediction DA fct 2 pt time and space}
\enq
The amplitudes are the \textit{same} as in \eqref{ecriture predictions conformes DA 2 pt fct zero time}. 
However, \textit{per se}, these predictions only hold for $|x| >> |v_F t|$. They are thus meaningless in that, when $|x| >> |v_F t|$,   the time dependence
is of the same order of magnitude that the sub-dominant corrections to each terms of the sum. 

\subsubsection*{The response functions at zero temperature}

Even though correlation function carry a lot of deep informations on a given model, it is not possible to measure them directly. 
Experiments such as  neutron scattering, 
Bragg spectroscopy \cite{StamperetalMeasureStructureFactorbyBraggSpectroscopyonBEC}
or even experimental set-ups solely proposed on theoretical grounds such as the Fourier sampling of time of flight images 
\cite{DuanTimeofFlightdetectionmethodforDSFinColdAtoms} or stimulated Raman spectroscopy 
\cite{DaoGeorgesDalibardSalomonCarusottoRamanSpectroscopyTechniqueforMeasureSpectralFunction} 
all measure directly the space and time Fourier transform of the connected two-point functions:
\beq
S^{(\a\be)} (k, \om) \; = \; \Int{ }{} \ex{ \i (\om t - k x) } \big< \op{O}_{1+x}^{(\a)}(t) \cdot \op{O}_{1}^{(\be)}(0) \big> _{\e{c}} \cdot \dd \mu_{\e{spc}}(x) \,  \dd t  \;. 
\enq
The choice $\a, \be$ of operators is specific of the given experiment. Above,  $\dd \mu_{\e{spc}}(x) $ is the Lebesgue measure for models exhibiting a continuous distance variable and it is the discrete measure attached to the sites of the model 
for lattice models such as spin chains. It seems natural to wonder whether there exists some universal  behaviour of the response functions 
at least in some range of the momentum $k$ and the energy $\om$.

The conformal field theory/Luttinger liquid based predictions for the large-distance and long-time asymptotic behaviour 
allow one to predict the behaviour of these Fourier transforms in the limit when the 
frequency $\om$ and the momentum $k$ both go to zero. In order to say anything more about the Fourier transforms one has to go beyond such a regime 
and, in particular, take into account the non-linearities in the excitation spectrum of the model. 
This remained a technical challenge for long, even on a heuristic basis.

\begin{figure}[h]
\begin{center}

\begin{pspicture}(10,5)

\psline(0,0.5)(9.5,0.5)
\psline(0.5,0)(0.5,4.5)
\psline[linewidth=2pt]{->}(9.4,0.5)(9.5,0.5)
\psline[linewidth=2pt]{->}(0.5,4.4)(0.5,4.5)

\rput(9.5,0.3){$k$}
\rput(0.3,4.3){$\om$}

\pscurve[linestyle=dashed, dash=3pt 2pt]{-}(0.5,0.5)(2.2,1.5)(4,2)(6.3,1.5)(8,0.5)

\pscurve{-}(0.5,0.5)(2.5,2.1)(4,4.5)

\psline(8,0.5)(8,0.3)
\rput(8.5,0.2){$2p_F$}

\rput(7.5,1.5){$\om=\veps_h(k)$}

\rput(5,4){$\om=\veps_p(k)$}


\psline[linewidth=2pt]{->}(5.8,2)(5.3,2.7) 

\psline[linewidth=2pt]{<-}(2,3)(1.5,3.3)

\rput( 4,1.5 ){zero}
\rput( 6.6,2.3 ){growing}
\rput( 1.7,2.7 ){growing}

\end{pspicture}
\caption{Typical behaviour of the modulus $|S^{(\rho\rho)} (k, \om)|$ of the density structure factor in the plane $k\geq 0, \om\geq 0$. 
The response function presents a strong divergence close to the particle excitation threshold (black continuous line) and vanishes on the hole excitation threshold
(dashed line). The response function is zero between the dashed line and the $\om=0$ axis.  \label{Graphe de la fonction de reponse} }
\end{center}
\end{figure}
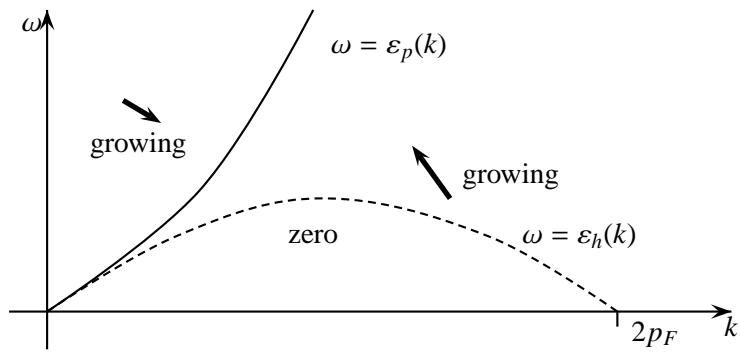

I will now describe the typical behaviour of a specific response function of a system of interacting bosons, the density structure factor $S^{(\rho\rho)} (k, \om)$
( Fourier transform of the time and space dependent density-density correlation function of the model). 
Typically one observes, in the $k\geq 0$, $\om \geq 0$
plane,  a behaviour represented in Figure \ref{Graphe de la fonction de reponse}. The modulus $|S^{(\rho\rho)} (k, \om)|$ of the response function is relatively small 
between the $k=0$ axis and the continuous line (the so-called particle excitation threshold given by the equation $\om=\veps_p(k)$). 
Its value grows sharply when one approaches the particle threshold line. It basically diverges on this line. 
Then, it decays in magnitude. The density structure factor vanishes on the dotted line (the so-called hole excitation threshold $\om=\veps_h(k)$). 
It is strictly zero between the dashed line and the $\om=0$
axis. 

The vanishing of the response functions between the hole excitation threshold and the $\om=0$ line follows, on heuristic grounds, from the 
sub-additivity of the energies of the excitations, that is to say the property that $\veps(k_1+k_2) \leq \veps(k_1)+\veps(k_2)$. This seems to be a general feature of 
many condensed matter physics models. The details of the behaviour of the response function  above the hole excitation threshold and away from the particle excitation threshold 
strongly depend on the model. There is thus no hope to say much about it by means of a universal feature. 
However, the vanishing on the hole excitation threshold $\om=\veps_h(k)$ and the divergence on the particle excitation threshold $\om=\veps_p(k)$ are both universal for integrable models.

Very recently, in 2006, Glazmann, Kamenev, Khodas and Pustilnik \cite{GlazmanKamenevKhodasPustilnikNLLLTheoryAndSpectralFunctionsFremionsFirstAnalysis}
managed to find a way of treating the pertinent effects of non-linearities in the spectrum
of a model of non-interacting fermions what allowed them to argue that the so-called density structure factor 
 exhibits a power-law behaviour in $|\om-\veps_{h,p}(k)|$
at fixed momentum $k$ when $\om$ approaches the particle or hole excitation threshold. The exponent governing this power-law behaviour 
is called the edge exponent. 
In the same year, Pustilnik \cite{PustilnikDSFForCalogero} argued such a power-law behaviour for the density structure factor of the Calogero-Sutherland model, 
this starting from its multiple integral representation. In 2007, Glazmann, Kamenev, Khodas and Pustilnik generalised their approach so as to deal with more complicated 
response functions \cite{GlazmanKamenevKhodasPustilnikNLLLTheoryAndSpectralFunctionsFremionsBetterStudy} as well as with bosonic models \cite{GlazmanKamenevKhodasPustilnikDSFfor1DBosons}.
In 2008, Glazman and Imambekov \cite{GlazmanImambekovComputationEdgeExpExact1DBose} computed the edge exponents describing the power-law vanishing of the 
response functions of a gas of one-dimensional bosons in $\de$-function interactions, the so-called Lieb-Liniger model. 
Their approach was a mixture of the previous handlings and the use of exact expressions for certain thermodynamic quantities that can be deduced from
the quantum integrability of that model. Finally, in 2009, Glazmann and Imambekov brought the theory to a satisfactory level of effectiveness. 
First of all \cite{GlazmanImambekovGeneralRelationEdgeExpForGeneralModel}, they provided a way to express, for Galilean-invariant models, the edge exponents, at momentum $k$, in terms of $\veps_{p,h}(k)$
and certain other natural observables for the model.  Further, they provided \cite{GlazmanImambekovDvPMTCompletTheoryNNLL} a convenient effective model -the non-linear Luttinger liquid-  allowing one to
grasp all the effects of the non-linearities in the original model's spectrum. In that paper they also exhibited some universal feature
of the ratios of response functions $\tf{ S^{(\a\be)}(k,\veps_p(k)+\de \om) }{S^{(\a\be)}(k,\veps_p(k)-\de \om) }$ when $\de \om \tend 0^+$.

\subsubsection*{The case of non-zero temperature}

Although certain experimental situations actually deal with models that are, effectively speaking, at zero temperature, this especially due 
to the development of modern experimental techniques of cold atom physics, most of the measurements take place at finite temperature $T$. 
Going back to the case of massless models at finite temperature, one can argue, within the setting of Cardy's approach, the general form of the low-temperature behaviour 
for the model's free energy in infinite volume and also the general form of the large-distance asymptotic behaviour of its correlation functions. 
Note that these, at finite temperature, are expected to decay exponentially fast in the distance of separation between the operators.  
In general, one expects that the thermodynamic limit $f=\lim_{L\tend+\infty} f_L$ of the \textit{per} site free energy has the low-$T$ behaviour 
\beq
f \; \sim \; \f{E_{0}}{T} - \f{\pi c v_F}{6} T \; + \, \cdots 
\enq
in which $E_0$ is the model's \textit{per} site ground state energy, $c$ is the central charge of the conformal field theory that is supposed to grasp the large-distance behaviour of the model's correlation functions
while $v_F$ is the velocity of the excitation on the Fermi boundary. 

More generally, one argues that the low-temperature behaviour of a model's correlation functions in given by an analogous formula to the zero-temperature case.  
In the case of a two-point function, the expansion takes the form 
\beq
\Big< \op{O}_{1+x}^{(\a)} \cdot \op{O}_{1}^{(\be)} \Big>_{\e{c};T} \; = \;\sul{ k\in I^{(\a\be)} }{} \msc{A}_k^{(\a \be)} \cdot  \ex{  2 \i x  \ell_{k}  p_F } \cdot 
\Bigg( \f{ \pi T/ v_F }{ \sinh \big[ \frac{1}{v_F} \pi T  x \big] }  \Bigg)^{ 2( \De^+_k + \De^-_k ) } \;. 
\enq
in which $\De^{\pm}_k $ is the tower of conformal  dimensions of the primary operators, on the conformal field theory side, representing the given operators $\op{O}^{(\a)} \cdot \op{O}^{(\be)}$
on the finite volume $L$ model side. These conformal dimensions coincide with the ones that appeared in the large-distance asymptotic behaviour 
of the zero temperature correlator 
\newline $\Big< \op{O}_{1+x}^{(\a)}(0) \cdot \op{O}_{1}^{(\be)}(0)\Big>_{\e{c}}$.  Also, the above prediction for the large-distance asymptotic expansion is supposed to be valid only in the regime of low-temperatures, namely $x\tend +\infty$, $T\tend 0^{+}$ with $xT$ being fixed and finite.

\vspace{2mm}

Having at one's disposal such universal predictions for various asymptotic regimes of the correlators, several questions arise: \vspace{1mm}
\begin{itemize}

\item[$\bullet$] can one access, even for some small amount of models, to the finite-size corrections to the ground and excited states energies and momenta and check that these are indeed of conformal type? \vspace{1mm}

\item[$\bullet$] Assuming  that the corrections are of such a form, can one confirm the conformal field theory or Luttinger liquid model based predictions for the large-distance asymptotic behaviour 
of at least one correlation function in this model at zero temperature, this starting from the first principle, \textit{viz}. solely by using the definition of the model and no other heuristics? \vspace{1mm}

\item[$\bullet$] Can a similar program be carried out for some model at finite but small temperature?\vspace{1mm}

\item[$\bullet$] Can anything more be said relatively to the long-distance and large-time asymptotic behaviour? \vspace{1mm}

\item[$\bullet$] Can one reproduce, starting from first principles, the predictions for the response functions and, in particular, test their universal behaviour? \vspace{1mm}

\end{itemize}

As will be discussed in the following, the first point has been answered positively in the 80's. Also, some checks of the second point could have been carried out for models
equivalent to a free fermion model. The first derivation of the second point has been done by myself and collaborators during my PhD studies. 
 
In the core of this manuscript, I will explain how I subsequently managed to provide a rigorous framework, building however on auxiliary hypothesis, that allows one to answer positively to
the second, third and fourth points. I will also discuss a less rigorous, but quite convincing, approach
 that I developed so as to answer positively to the second, fourth and fifth point for a large class of models.

\section{Quantum integrable models}

So far, I have been carrying out a rather general discussion, without making much assumptions on the one-dimensional quantum model. 
It is high time to remind that there exists a specific class of models
in one spatial dimension: the so-called quantum integrable models. These models enjoy of a sufficiently rich underlying algebraic structure that renders possible all of the mentioned computations 
relative to an explicit characterisation of their spectrum, their correlation functions \textit{etc}. 
Due to such properties, the quantum integrable models can naturally serve as a huge laboratory for testing the CFT/Luttinger liquid based predictions 
and, sometimes, even going beyond their scope.

\subsection{The coordinate Bethe Ansatz}

The field of quantum integrable models started to flourish in 1931 with the seminar work of Bethe \cite{BetheSolutionToXXX}. 
In that paper, Bethe proposed an Ansatz for the eigenfunctions of the so-called XXX spin-$1/2$ Hamiltonian $\op{H}_{XXX}$, which is the isotropic limit ($\De=1$) of the XXZ spin-$1/2$ Hamiltonian:
\beq
\op{H}_{XXZ} \; = \; J \sul{a=1}{L} \Big\{ \sg_a^x \sg_{a+1}^x \, + \, \sg_a^y \sg_{a+1}^y \, + \, \De \big( \sg_a^z\sg_{a+1}^z + \e{id}_L \big)   \Big\} \; - \;  \f{ h }{ 2 } \sul{a=1}{L} \sg_a^z  \;. 
\nonumber
\enq
$\op{H}_{XXZ}$ is an operator on the Hilbert space $\mf{h}_{XXZ}=\bigotimes_{a=1}^{L} \mf{h}_{XXZ}^{a}$, where $ \mf{h}_{XXZ}^{a} \simeq \Cx^2$. Further
$\sg^x$, $\sg^y$ and $\sg^z$ are the standard $2\times 2$ Pauli matrices, and the index $a$ in $\sg_a^{w}$, $w=x,y,z$, means that the matrix acts non-trivially solely 
in the $a^{\e{th}}$ Hilbert space $ \mf{h}_{XXZ}^{a} $ as the Pauli matrix $\sg^{w}$. Also $\e{id}_L $ is the identity operator on $\mf{h}_{XXZ}$.
Furthermore, $J>0$ is a coupling constant - the exchange interaction -, $\De$ takes into account the possible anisotropy of the interactions between the spins in the longitudinal and 
transverse directions  while $h$ is an overall magnetic field pointing in the direction of the anisotropy. The model represents an antiferromagnet when $\De \geq -1$. 
For moderate anisotropy $-1\leq \De \leq 1$ and magnetic fields $0\geq |h| < 4J(1+\De)$ the model is a massless antiferromagnet while, for $\De \geq 1$ and magnetic fields $0 \geq |h| < 4 J(\De-1)$ it is 
a massive antiferromagnet.

In 1958, Orbach \cite{OrbachXXZCBASolution} generalised Bethe's construction to the XXZ case. Owing to $\big[ \op{H}_{XXZ} , \sum_{a=1}^{L} \sg^z_a  \big] = 0$, the eigenvectors of $\op{H}_{XXZ}$ belongs to sectors with a fixed eigenvalue $N$ of
the operator $\sum_{a=1}^{L} (\e{id}_L - \sg^z_a)/2$. Bethe's Ansatz provides one with a combinatorial expression for the coordinates of the eigenvectors of $\op{H}_{XXZ}$ in the canonical basis of $\mf{h}_{XXZ}$
graded by $N$. 
These coordinates depend on a set of auxiliary parameters -the rapidities- $\{\la_a\}_1^N$. In order to produce eigenvectors, among other things, these parameters have to satisfy a set of 
algebraic equations of high degree, the so-called Bethe Ansatz equations. For the XXZ Heisenberg magnet at $-1<\De<1$, \textit{viz}. when $\De$ can be parametrised as $\De=\cos(\zeta)$, these take the form:
\beq
\Bigg( \f{ \sinh\big(\la_k-\i\tf{\zeta}{2} \big)  }{  \sinh\big(\la_k+\i\tf{\zeta}{2} \big) } \Bigg)^{L} \cdot 
\pl{ \substack{\ell=1 \\ \not= k}  }{N} \bigg\{ \f{ \sinh\big(\la_{\ell}-\la_k-\i\zeta \big)  }{  \sinh\big(\la_{\ell}-\la_k+\i\zeta \big) } \bigg\} \; = \; 1 \;. 
\label{ecriture eqns de Bethe XXZ Intro}
\enq
The eigenvalues $\mc{E}_{XXZ}\big( \{\la_a\}_1^N\big)$ of $\op{H}_{XXZ}$ are then expressed as simple functions of the rapidities
associated with the given state:
\beq
\mc{E}_{XXZ}\big( \{\la_a\}_1^N\big) \; = \; 2J L \De \, - \, \sul{a=1}{N}  \f{ 2J \sin^2(\zeta) }{  \sinh\big(\la_a+\i \tf{\zeta}{2} \big)  \sinh\big(\la_a - \i \tf{\zeta}{2} \big)  }       - \f{ L-2N }{2} h \; . 
\enq

It is natural to ask about the completeness and the orthogonality of the system of Bethe eigenvectors. This remained an open problem for a long time.  
Bethe gave some counting of solution argument in favour of completeness. 
However, subsequently, the argument was shown to be wrong. The proof of completeness was only given in 1994 by Tarasov and Varchenko for the so-called inhomogeneous XXZ chain and by 
Mukhin, Tarasov and Varchenko \cite{MukhinTarasovVarchenkoCompletenessandSimplicitySpectBA} in 2009 for the homogeneous XXX chains, this provided that one slightly changes the perspective of understanding the Bethe Ansatz equations and its solutions.

The XXZ chain is not the only instance of a quantum integrable model. Numerous other models were studied in the literature, the non-linear Schrödinger model 
or the lattice sine-Gordon model so as to name a few. For instance, the non-linear Schrödinger model is a quantum field theoretical model
of canonical Bose fields $\Phi(x)$, $\Phi^{\dagger}(x)$ living on a circle of length $L$, subject to periodic boundary conditions
and interacting through the Hamiltonian
\beq
\op{H}_{NLS} \; = \; \Int{0}{L} \Big\{ \Dp{x}\Phi^{\dagger}(x) \Dp{x}\Phi(x) + c \Phi^{\dagger}(x)\Phi^{\dagger}(x)\Phi(x)\Phi(x) \Big\} \cdot \dd x \; - \; h \op{N}_{NLS} 
\quad \e{with} \quad
\op{N}_{NLS} \; = \; \Int{0}{L} \Big\{ \Phi^{\dagger}(x)\Phi(x)\Big\}  \cdot \dd x \;. 
\nonumber
\enq
There, $h>0$ plays the role of a chemical potential while $c>0$ is a coupling constant characterising the strength of the repulsive interactions.  
\textit{Per se}, this model goes out of the setting discussed earlier on. In particular, it is ill defined in that one cannot really provide a rigorous construction of 
$\Phi$, $\Phi^{\dagger}$ as operators on some space.  It can nonetheless be understood as an inductive limit, in the weak sense, of a lattice model \textit{c}.\textit{f}. 
Chapter \ref{Chapitre AB grand volume FF dans modeles integrables}. 
Independently of such fine points, in each sector corresponding to a fixed eigenvalue of the number of particle operator $\op{N}_{NLS}$ the model is equivalent to a gas
of $N$ bosonic particles interacting through a $\de$-function pairwise potential with strength $2c$.

The model at $c=+\infty$, the so-called impenetrable bosons,
was introduced in 1960 by Girardeau \cite{GirardeauIntroductionImpBosons} who also obtained its spectrum and eigenfunctions. In 1964, Br\'{e}zin, Pohil and Finkelberg \cite{BrezinPohilFinkelbergFirstIntroBoseGas}
introduced the model in its full generality and proposed a coordinate Bethe Ansatz construction of its eigenfunctions up to $N=3$ particles. During the very same year, 
and probably independently, Lieb and Liniger \cite{LiebLinigerCBAForDeltaBoseGas} solved the model for any value of $N$. In a subsequent publication, 
Lieb \cite{LiebExcitationStructureBoseGas} carried out an extensive analysis of its spectrum. In 1993, Dorlas \cite{DorlasOrthogonalityAndCompletenessNLSE}
proved the completeness of the system of eigenfunctions built through the Bethe Ansatz.

\subsection{Spectrum in the large volume limit}

In the limit where the volume $L$ diverges, one can provide a rather convincing description of the energy levels for Bethe Ansatz solvable models. 
First calculations of the sort have been carried out by H\'{u}lten \cite{HultenGSandEnergyForXXX} in 1938 for the case of the XXX chain at zero magnetic field. The author was able to identify -on a heuristic basis- 
the specific eigenvector constructed by Bethe which gives rise to the model's ground state. He then proposed an approach allowing one to compute the leading in 
the model's volume (number of sites) $L$  behaviour of the ground state energy. H\'{u}lten argued that the parameters $\{\la_a\}_1^N$  describing the model's ground state at zero magnetic field 
are such that $N=L/2$ -for chains of even length- and that, in the $L\tend + \infty$ limit, they form a dense distribution on $\R$ with a density $\rho_{XXX}$. Building on the form of the Bethe equation, 
he argued that $\rho_{XXX}$ satisfies a linear integral equation which he was able to solve. He then used this information to compute the leading in $L$
behaviour of the ground state energy. H\'{u}lten's density approach was generalised to many other models, but always in a rather heuristic setting. 
Orbach \cite{OrbachXXZCBASolution} applied it to the case of the XXZ chain but was not able to solve explicitly the associated linear integral equation. 
A year later, Walker \cite{WalkerFirstExplicitSolToLinIntEqnMassiceXXZ} proposed a change of variable that allowed him to bring Orbach's linear integral equation at $\De>1$ into one 
driven by a $\pi$-periodic Wiener-Hopf operator acting on $L^2(\intff{0}{\pi})$. In this way, he could solve the integral equation by using Fourier 
series expansions. In 1964, Griffiths \cite{GriffithsXXZFirstLinIntEqnFiniteMagField} generalised the approach what allowed him to deal with the ground state of the massless XXZ 
chain in the presence of a finite magnetic field.

The investigation of the excited state was initiated by des Cloizeaux and Pearson \cite{DescloizeauxPearsonExcitationsXXX} in 1962. The authors
derived the form of the dispersion relation of the excitations above the ground state.
Four years later, des Cloizeaux and Gaudin \cite{DescloizeauxGaudinExcitationsXXZ+Gap} extended the analysis to the case of the  XXZ chain. 
Although they did obtain the correct form of the dispersion law, they gave a wrong interpretation of the spin of the excitations. This fact was corrected later, in 1981
by Faddeev and Takhtadzhan \cite{FaddeevTakhtadzhanSpinOfExcitationsInXXX}.
In 1967, Lieb \cite{LiebExcitationStructureBoseGas} provided a very convenient description of the  energies of the excited states
in the non-linear Schr\"{o}dinger model. Namely, he argued that the leading in $L$ behaviour\symbolfootnote[3]{\textit{viz}. when terms vanishing in $L$ are dropped}
of the relative excitation energies in respect to the ground state $\wh{E}_{\e{ex}}$ can be grasped within the particle-hole excitation picture.  
This characterisation of the energies of the excited states was only based on handlings of real valued solutions $\{\la_a\}_1^N$
to the Bethe equation. However most models, the XXZ chain being a paradigmatic example, do have complex valued solutions. 
This fact was already known to Bethe. In fact, starting from Bethe's seminal work and 
until the early '80s, it was widely accepted that such complex-valued solutions, on top of consisting of a certain amount of real roots
also contain strings -sub-sets $\{x_k +\i a \zeta + \eps_{L;k;a}, a=-n,\cdots,n\}$ with the "centre" of the string $x_k$ 
satisfying some modification of the original Bethe equations for the model-. The deviations $\eps_{L;k;a}$ in respect to the ideal strings were believed to be exponentially small in respect to the model's volume $L$.
The so-called string hypothesis has even been used to provide wrong proofs of the completeness of the Bethe Ansatz. 
The breakthrough came in 1982 with the pioneering analysis of Destri and Lowenstein \cite{DestriLowensteinFirstIntroHKBAEAndArgumentForStringIsWrong} 
of the structure of the complex solutions to the Bethe equations 
describing the chiral invariant Gross-Neveu model. These Bethe Ansatz equations are structurally similar to those arising in the XXX spin chain. 
Destri and Lowenstein \cite{DestriLowensteinFirstIntroHKBAEAndArgumentForStringIsWrong} showed that, in the $L\tend +\infty$ limit,  
the complex solutions to the Bethe Ansatz equations describing the low-lying excited states form 2-strings, quartets and wide pairs,
but do not form general, larger, strings.  The centres of these strings were then show to satisfy a set of coupled equations called the higher level Bethe Ansatz equations. 
The work of Destri and Lowenstein was followed, a few months later, by the two independent papers
\cite{BabelondeVegaVialletStringHypothesisWrongXXZ,WoynaorwiczHLBAEMAsslessXXZ0Delta1}. 
Woynarovich \cite{WoynaorwiczHLBAEMAsslessXXZ0Delta1} studied the case of the massless regime of the XXZ chain at anisotropy 
$1>\De>0$ while the analysis by Babelon, Viallet and de Vega \cite{BabelondeVegaVialletStringHypothesisWrongXXZ} was the most complete 
and dealt with all the regimes of the XXZ chain. 
Later, in 1984,  the derivation of the higher level Bethe equations for the massive regime of the XXZ chain was reconsidered by Virosztek-Woynarowich 
\cite{VirosztekWoynarovichStudyofExcitedStatesinXXZHigherLevelBAECalculations}. 
This last paper corrected certain spurious terms present in the higher level Bethe equations for the massive chain obtained earlier. 
All the aforementioned works allowed for a deep understanding of the structure of the large volume limit of solutions to the Bethe equations that describe the low-lying excited states
of the quantum integrable model.
Also, recently in 2007, Caux and Hagemans \cite{CauxHagemansDeformedStringsInXXX} studied the complex solutions to the Bethe equations for the XXX chain and
provided a thorough numerical description of the solutions for 10 sites long chains. This analysis what manifestly showed a strong deviation from the 
"string-hypothesis" form.

The density of Bethe roots approach described above  was rather heuristic. Several elements of rigour have been brought by Yang and Yang \cite{Yang-YangXXZproofofBetheHypothesis,Yang-YangXXZStructureofGS,Yang-YangXXZApplication} 
in 1966. In particular these authors provided a rather good justification, for the XXZ chain, of the specific solution 
to the Bethe equation that gives rise to the ground state. Yang and Yang also gave a firmer ground to why a densification of the ground state Bethe roots should hold 
for the XXZ chain. Furthermore, they provided \cite{Yang-YangXXZStructureofGS} a rather extensive discussion of the properties of the linear integral equations that is satisfied by the 
density of the ground state Bethe roots. However, the proof of the densification, for $-1< \De \leq  0$ and also for $\De>1$ and large enough, has only been 
given recently by Dorlas  and Samsonov \cite{DorlasSamsonovThermoLim6VertexAndConvergceToDensityInSomeCases6VrtX} in 2009. 
The proof of the densification for the ground state Bethe roots as well as for Bethe roots describing a class of particle-hole excited states, this for any $\De >-1$
was given by myself in 2015 \cite{KozProofOfDensityOfBetheRoots}.

Still, the main interest for studying the large-$L$ behaviour of the Bethe roots resided in being able to access  
the $1/L$ corrections to the low-lying excited states' energies, first, so as to test the conformal 
field structure of the excitations in the model and, second, so as to extract from there the critical exponents. 
The first really effective approach allowing one to do so was proposed by de Vega and Woynarovich \cite{DeVegaWoynarowichFiniteSizeCorrections6VertexNLIEmethod} in 1985 and built on an ingenious handling
of densities. The authors treated the case of the massive regime of the XXZ chain. Although not formulated directly in this way, their method built at some point 
on non-linear integral equations. During the late 80's, the method of de Vega and Woynarovich was conformed with success to the calculation of the finite-volume corrections 
to the low-lying excited states' energies of various quantum integrable models. 
The method had, however, certain limits of applicability. In 1990 Batchelor and Kl\"{u}mper \cite{KlumperBatchelorNLIEApproachFiniteSizeCorSpin1XXZIntroMethod} proposed a 
non-linear integral equation based framework that allowed one for a very effective computation of the finite-volume correction. This approach was subsequently developed in 1991
by Kl\"{u}mper, Batchelor and Pearce \cite{KlumperBatchelorPearceCentralChargesfor6And19VertexModelsNLIE} and in 1993 by Kl\"{u}mper, Wehner and Zittartz
\cite{KlumperWehnerZittartzConformalSpectrumofXXZCritExp6Vertex}. 
In 1992, Destri and de Vega \cite{DestriDeVegaAsymptoticAnalysisCountingFunctionAndFiniteSizeCorrectionsinTBAFirstpaper} elaborated further the non-linear 
integral equation based-approach to extracting finite-volume corrections and brought it to a quite satisfactory level of 
 effectiveness in 1995 \cite{DestriDeVegaAsymptoticAnalysisCountingFunctionAndFiniteSizeCorrectionsinTBAFiniteMagField}. 

 These works, and many other that followed, confirmed Cardy's predictions for the large-$L$ behaviour of the ground and low-lying excited states's enegies. 
This allowed for an identification of the central charges and conformal dimensions associated with numerous quantum integrable models.

\subsection{The algebraic Bethe  Ansatz}

During his 1967 analysis \cite{LiebFModelAntiFerro,LiebIceModelEntropyCalculation,LiebFModelFerro,LiebIceModelSolutionAndTranferMatrx/XXZCorresp} of the various phases of the six-vertex model Lieb
 observed \cite{LiebIceModelEntropyCalculation,LiebIceModelSolutionAndTranferMatrx/XXZCorresp} that the transfer matrix associated with this model of two-dimensional statistical physics possesses exactly the same Bethe eigenvectors as the XXZ spin-$1/2$ chain. 
 A similar statement has been made by Sutherland \cite{SutherlandSixVertexSolGenerel}, the same year and, apparently, on independent grounds
In 1968, McCoy and Wu \cite{McCoyWuFirstProofXXZCommutesWith6VTransferMatrix} went one sept further and showed by a direct calculation that the XXZ Hamiltonian commutes with the transfer matrix of the six-vertex model,
hence ensuring that the two object actually share the same eigenvectors, irrespectively of whether the Bethe eigenvectors span the whole Hilbert space of the models. 
In 1970, Sutherland \cite{SutherlandDirectProofThatXYZCommutesWith8vTransferMat} generalised the result of McCoy and Wu and demonstrated that the so-called 
XYZ Hamiltonian\symbolfootnote[3]{This is a generalisation of the XXZ Hamiltonian where all nearest-neighbour spin-spin interaction bare a different weight.}
commutes with the transfer matrix of the eight-vertex model. It was however Baxter that, on the one hand, established a clear connection between the two objects  and, on the other hand, unravelled the deep algebraic structure at their root. 
In his 1972 analysis of the free energy of the eight-vertex model, Baxter \cite{BaxterPartitionfunction8Vertex-FreeEnergy} constructed a one-parameter $\la$ family of transfer matrices $\op{t}_{\e{8V}}(\la)$ associated with this model. 
The parameter $\la$ is usually referred to as the spectral parameter. Baxter represented $\op{t}_{\e{8V}}(\la)$ as the trace of a product of "local" $\op{L}$ matrices. 
This product is called, nowadays, the monodromy matrix whereas the $\op{L}$-matrix
if referred to as the Lax matrix of the model. In the case of the eight vertex model, it is a $2\times 2$ matrix over the ring of linear operators on $\Cx^2$. 
 Baxter showed that these transfer matrices commute at different values of the spectral parameter. 
The commutation follows from a set of constraints that are satisfied by the weights of the eight-vertex model which he referred to as the star-triangle relation. 
In fact, this very relation was already known to Onsager when he solved the two-dimensional Ising model \cite{Onsager2DIsingModel}.  Baxter showed that 
the star-triangle relation can be rewritten as a matrix relation between products of local $\op{L}$ matrices $R_{12}\op{L}_1\op{L}_{2} = \op{L}_{2} \op{L}_1 R_{12}$ where $1,2$
are auxiliary spaces associated with the Lax matrices while  $R$ is a $4\times 4$ matrix acting in the tensor product of the spaces $1$ and $2$. 
A similar equation, has been obtained in 1967 by Yang \cite{YangFactorizingDiffusionWithPermutations} in a completely different setting when he 
studied  the factorisable scattering on the line for the gas of $N$-particles, not necessarily  bosonic, interacting through a $\de$-like repulsive potential. Nowadays, 
the $R_{12}\op{L}_1 \op{L}_{2} = \op{L}_{2} \op{L}_1 R_{12}$ equation is referred to as the Yang-Baxter equation.
On top of proving the commutativity of the transfer matrices, in a subsequent paper, Baxter  \cite{BaxterXYZSolution-GroundStateEnergy} showed that the XYZ Hamiltonian 
can be expressed as $\Dp{\la} \ln \op{t}_{8V}(\la) _{\mid \la=0} $. 
Upon restricting the parameters of the model, one then finds that the XXZ Hamiltonian is given by the logarithmic derivative at zero spectral parameter of the six-vertex model transfer matrix. 
This allowed Baxter to establish a direct connection between transfer matrices of two-dimensional statistical mechanics and quantum integrable models in one dimension. 
I emphasise that Baxter's work added a deep algebraic structure to quantum integrable models: a quantum integrable Hamiltonian is an explicit member of a commutative
algebra of charges generated by a family of commuting transfer matrices. In this respect, one obtains a conceptual bridge between the quantum and the classical integrabilities: 
integrability is the existence of a sufficiently large explicit set of charges in involution. 
At that stage, however, it was not clear how, starting from a given quantum integrable 
Hamiltonian, one can find the commuting transfer matrix that would embed the Hamiltonian into a sufficiently large set of charges in involution.  

In 1979, Faddeev, Sklyanin and Takhtadjan \cite{FaddeevSklyaninTakhtajanSineGordonFieldModel} showed that what Baxter observed was actually just the tip of the iceberg of a deep algebraic structure. 
As pointed out by Baxter, the local RLL relation can be lifted to a global $R_{12}\op{T}_{1}\op{T}_{2}\, =\, \op{T}_{2}\op{T}_{1}R_{12}$ one satisfied by the 
model's monodromy matrix. While Baxter used this relation solely to prove the commutativity of the transfer matrices, it actually provides ones with the algebra  
satisfied by the entries of the model's monodromy matrix. Faddeev, Sklyanin and Takhtadjan \cite{FaddeevSklyaninTakhtajanSineGordonFieldModel} built on these algebraic relations so as to construct the eigenvectors 
of the trace of the  monodromy matrix - the so-called a transfer matrix of the model-. 
They realised this program for the lattice discretisation of the sine-Gordon model. They were able to provide  a  Lax matrix 
satisfying the local Yang-Baxter equation, up to second order corrections in the lattice space. Although their Lax matrix was only approximate, the authors argued the error to be negligible in the continuous limit
when the lattice spacing is sent to zero and the \textit{per} \textit{se} quantum version of the sine-Gordon model formally recovered. 
In the approach of the Leningrad school, the eigenvectors are constructed as products of the $12$ entries of the $2\times 2$ auxiliary space representation of the monodromy matrix:
\beq
\ket{  \{ \la_a\}_1^N }  \; = \; \op{T}_{12}(\la_1)\cdots \op{T}_{12}(\la_N) \ket{0} \;. 
\enq
The operators $\op{T}_{12}(\la)$ form a commutative family and act on the so-called pseudo-vacuum vector $\ket{0}$ which satisfies $\op{T}_{21}(\la) \ket{0}=0$, for any $\la$. 
For $\ket{  \{\la_a\}_1^N  }$ to be an actual eigenvector of the commuting family of transfer matrices,
the parameters $\{\la_a\}_1^N $ have to satisfy  the Bethe Ansatz equations of the model. 
The algebraic approach to quantum integrable models described above leads to tremendous simplifications in the resolution of their spectral problems. 
 Indeed,  the eigenvectors are not expressed as some interlocked combinatorial sum as in the coordinate Bethe Ansatz but, rather as the action of a string of pseudo-creation operators on a reference state. 
One thus recovers, at least in part, the structure of  a Fock space so familiar to free-fermionic models. 
It is important to stress that it was precisely this algebraic construction of the eigenstates that really opened the possibility of an efficient investigation of the 
structure of the space of state (calculation of the norms, the scalar products and, more generally, calculation of the correlation function) of quantum integrable models. 
Furthermore, it was the discovery of this algebraic structure that was at the roots of the theory of quantum groups, formalised 
by Jimbo \cite{JimboQdeformationofU(g)} and Drinfel'd \cite{DrinfeldQuantumGroupsAsHopfAlgebras} in the mid and late 80's. 

In the years that followed its discovery, the algebraic Bethe Ansatz was applied to a plethora of models. The main difficulty was, in fact, to construct the solutions to 
the "local" Yang-Baxter equation defining the fundamental Lax matrix of the model. This could have been done for the XXX chains by 
Faddeev and Takhtadjan \cite{FaddeevTakhtadzhanXXXgeneralOverwiev} or 
the lattice discretisations of the sine-Gordon and non-linear Schr\"{o}dinger model by Izergin and Korepin \cite{IzerginKorepinLatticeVersionsofQFTModelsABANLSEandSineGordon}
so as to name a few. Note that the Lax matrices obtained by Izergin and Korepin were satisfying the Yang-Baxter equation to all order in the
lattice discretisation parameters.  The mentioned models were built over the rank $1$ Lie algebra $\mf{sl}_2$. 
In order to treat more complex models built over of higher rank Lie algebras, it was necessary to generalise the notion of the algebraic Bethe Ansatz 
what led to the so-called nested algebraic Bethe Ansatz developed by Kulish and Reshetikhin \cite{KulishReshetikhinNestedBAFirstIntroduction,KulishReshetikhinNestedBASomeGeneralisationstoGL(N)Reps}
in the early 80's. The Kulish-Reshetikhin procedure is an algebraisation of the nested coordinate Bethe Ansatz approach used earlier by Yang \cite{YangFactorizingDiffusionWithPermutations}
for a multicomponent Bose gas 
and by Sutherland \cite{SutherlandCBAForHigerRankSpinChain} for a multicomponent spin chain. I refer to the 1982 review of Kulish and Sklyanin \cite{KulishSklyaninListOfSolutionsYBEexistentesEn82} for a list - definitely non-exhaustive- of other models
for which Lax matrices were constructed. 

It is also important to mention that there exists a class of models for which the algebraic Bethe Ansatz is not directly applicable since the models 
do not posses a \textit{buona fide} pseudo vacuum but for whose it is possible to reduce the construction of the eigenvectors and eigenvalues
model to the one of an auxiliary model which does posses a pseudo-vacuum. An archetype of such models is the XYZ spin-$\tf{1}{2}$ chain. 
The relationship between the original and auxiliary models is obtained by means of the so-called vertex-IRF\footnote[2]{interaction round-the-face} transformation. 
Such a transformation was first proposed by Baxter \cite{BaxterXYZ/8VExactDZEquivalenceRSOS}, what allowed him to determine the spectrum of the zero field eight vertex transfer
matrix by means of a coordinate Bethe Ansatz \cite{BaxterXYZ/8VExactDZEigenvectors}. Later, Faddeev and Takhtadjan conformed Baxter's trick to the algebraic Bethe Ansatz setting
\cite{FaddeevTakhtadzhanXYZmodelABASolution}. Among models solvable by means of such an extension of the algebraic Bethe Ansatz, one can mention higher spin or rank variants of the XYZ chain.

\subsection{Failure of the algebraic Bethe Ansatz: the quantum separation of variables}
\label{SousSection Intro qSoV}

Despite its tremendous effectiveness, the algebraic Bethe Ansatz cannot always be applied be it directly on through a vertex IRF transformation. Indeed, its implementation heavily relies on the existence of the pseudo-vacuum vector $\ket{0}$. 
The latter, however, does not always exist. One can, in fact, provide many examples of quantum integrable models possessing a local Lax matrix and yet \textit{not} admitting a pseudo-vacuum vector, the simplest one being   the 
quantum Toda chain. It is a quantum mechanical $N+1$-body Hamiltonian   in one spatial dimension defined as 
\beq
\op{ H }_{ \hspace{-2mm} \e{Td}} \; = \; \sul{a=1}{N+1} \f{ \op{p}_{a}^2}{2} \; + \;\kappa \cdot \ex{ \op{x}_{N+1} - \op{x}_1} 
\; + \;  \sul{a=1}{N} \ex{\op{x}_a -\op{x}_{a+1} } \qquad \e{with} \quad \op{p}_n = \f{ \hbar }{ \i } \f{ \Dp{} }{ \Dp{} \op{x}_n } 
\enq
and acting on 
\beq
\mf{h}_{\e{Td}}\,  = \,  L^2\big( \R^{N+1}, \dd^{N+1}x \big) \simeq \bigotimes\limits_{n=1}^{N+1} \mf{h}_{\e{Td}}^n \qquad \e{with} \qquad \mf{h}_{\e{Td}}^n\simeq  L^2\big( \R, \dd x \big) \;. 
\enq
There, $\op{p}_n$ and $\op{x}_n$ are pairs of conjugated variables satisfying to the canonical commutation $\big[ \op{x}_k, \op{p}_{\ell} \big] \; =  \;  \i \hbar$. 
Just as for the XXZ chain, the index $n$ refers to the quantum space $\mf{h}_{\e{Td}}^n$ where these operators act non-trivially. 
When $ \kappa =1$, one speaks of the so-called closed Toda chain whereas, when  $\kappa=0$, the model is referred to
as the open Toda chain. The Lax matrix for the Toda chain takes the form
\beq
\op{L}(\la) \; = \; \left( \ba{cc} \la-\op{p} & \ex{-\op{x}} \\ - \ex{ \op{x} } & 0 \ea \right)  \;.
\label{ecriture matrice de Lax Toda}
\enq
Clearly, its $21$ entry does not annihilate any local pseudo-vacuum, \textit{viz}. there is no solution to $\ex{ \op{x} } \ket{0} $, even in a generalised sense. 
This renders impossible the use of the algebraic Bethe Ansatz -under any of its variants- as a way to solve the spectral problem associated with $\op{H}_{ \e{Td} }$. 

There are, nonetheless, ways of overcoming such issues. In his analysis of the eight vertex model \cite{BaxterPartitionfunction8Vertex-FreeEnergy}, 
Baxter characterised the spectrum of the transfer matrix $\op{t}_{8V}(\la)$  of the eight vertex model in terms of a second order finite-difference and operator valued equation relating
the transfer matrix $\op{t}_{8V}(\la)$  and an auxiliary object, the so-called $Q$-operator $\op{Q}_{8V}(\la)$. 
Although Baxter obtained this equation for the eight vertex model case, I will write directly the one that arises in the context of the closed Toda chain as this
will serve better  the purpose of the discussion:
\beq
\op{t}_{ \e{Td} }(\la) \cdot  \op{Q}_{ \e{Td} } ( \la ) \; = \; (  \i )^{N+1}  \op{Q}_{ \e{Td} } (\la+ \i \hbar) \; + \;  ( -\i )^{N+1}  \op{Q}_{ \e{Td} } (\la - \i \hbar) \;. 
\label{ecriture eqn TQ Toda forme operatorielle}
\enq
The operator valued $\op{t}-\op{Q}$ equation for the Toda chain was obtained, for the first time, by Gaudin and Pasquier \cite{GaudinPasquierQOpConstructionForTodaChain} in 1992. 
The main point is that $\op{t}-\op{Q}$ equations provide one with an alternative, in respect to the Bethe Ansatz, way of characterising a model's spectrum. 
The sole form of the equation is, however, not enough for this purpose. One needs, in addition, to impose some additional  constraints on $\op{t}(\la)$ and  $ \op{Q}( \la )$. 
The usual setting is that \vspace{2mm}
\begin{itemize}
 \item [$\bullet$] $\op{t}(\la)$ and $ \op{Q}( \la )$ are normal operators for real values of the spectral parameter; \vspace{1mm}
\item[$\bullet$] $\op{t}(\la)$ and $ \op{Q}( \la )$ commute at different values of the spectral parameter $\big[ \op{t}(\la) \, , \,  \op{Q}( \mu ) \big]=0$;  \vspace{1mm}
\item[$\bullet$] $\op{t}(\la)$ is a polynomial (be it  a classical one, a trigonometric one or an elliptic one) in the spectral parameter;  

\item[$\bullet$] $ \op{Q}( \la )$ satisfies, at least in a weak sense, some growth 
conditions in $\la$ and admits a meromorphic continuation to a sufficiently large strip around the real axis. \vspace{1mm}
\end{itemize}

\subsubsection*{Spectrum of the Toda chain}

The family 
$\big\{ \op{t}(\la) \, ; \,  \op{Q}( \mu ) \big\}_{\la,\mu}$ being normal and commutative, one can reason, by virtue of the spectral theorem, on the level of the space
where both operators  $\op{t}(\la)$ and $ \op{Q}( \la )$ are realised as multiplication operators by $t(\la)$ and $q_{t}(\la)$. In the case of the closed Toda chain, this leads to the following
second order finite-difference equation for the \textit{two} unknown functions $t_{ \e{Td} }(\la)$ and $q_{t_{ \e{Td} }}(\la)$:
\beq
t_{ \e{Td} }(\la) \cdot  q_{t_{ \e{Td} }}(\la) \; = \; (\i)^{N+1} q_{t_{ \e{Td} }}(\la + \i \hbar) \; + \; \kappa \cdot (-\i)^{N+1} q_{t_{ \e{Td} }}(\la-\i\hbar) \;. 
\label{ecriture eqn TQ scalaire Toda}
\enq
One looks for solutions of this equation under the requirement that $t_{ \e{Td} }(\la)$ is a monoic polynomial in $\la$ of degree $N+1$ while $q_{t_{ \e{Td} }}$ is entire
and subject to the additional constraint on the asymptotic behaviour of $q_{t_{ \e{Td} }}$ (the bounds on the asymptotic behaviour were  found, in the correct form, by Gaudin-Pasquier \cite{GaudinPasquierQOpConstructionForTodaChain}):
\beq
q_{t_{ \e{Td} }}(\la) = \e{O}\Big(  \ex{- (N+1)\f{ \pi}{2\hbar} |\Re(\la) |  } \cdot  |\la |^{\f{N+1}{2\hbar}(2|\Im(\la)|- \hbar) }  \Big)
\qquad \e{uniformly} \;\e{in}  \qquad  |\Im(\la)| \;\leq \; \f{\hbar}{2} \;.
\label{ecriture eqn TQ Toda conditions sur cptm asymptotique}
\enq
The system might seem under-determined, in the sense that it contains too many unknowns. To convince oneself of the contrary, at least heuristically, it is helpful to  make the parallel with the Sturm-Liouville 
spectral problem 
\beq
\e{find} \; \e{all} \quad  \big( E, y \big) \in \R \times H_2(\R) \; \quad \e{such} \; \e{that} \quad - y^{\prime\prime}(x) \, + \,  V(x)\,  y(x)  \; = \; E \cdot y (x) 
\enq
with $V$ sufficiently regular and growing fast enough at infinity and $H_2(\R)$ is the second Sobolev space associated with $L^2(\R)$. Although the above ordinary differential equation admits two linearly independent solutions 
for any value of $E$, only for very specific values of $E$ does one find solutions that are in $H_2(\R)$. 
Regarding to \eqref{ecriture eqn TQ scalaire Toda}, the regularity and growth requirements on $q_{t_{ \e{Td} }}$ play the same role as the $H_{2}(\R)$ space in the Sturm-Liouville problem:
the scalar $\op{t}-\op{Q}$ equation admits solutions $(t_{\e{Td}}, q_{\e{td}} )$
belonging to the desired class only for 
well-tuned monoic polynomials of degree $N+1$. It is this effect that gives rise to so-called quantisation conditions for the Toda chain.

The  scalar $\op{t}-\op{Q}$ equation based description of the spectrum of the Toda chain appeared for the first time in the  early 80's 
work of Gutzwiller \cite{GutzwillerResolutionTodaChainSmallNPaper1,GutzwillerResolutionTodaChainSmallNPaper2} which dealt with 
the closed Toda chain involving a small number  $2,3$ or $4$ of particles. Gutzwiller derived the equation through a reasoning that appeared, at the time, completely independent of the integrable structure of the model. 
He represented the eigenfunctions of the $2,3$ or $4$-particle closed Toda chain in terms of an integral transform
whose integral kernel was given in terms of the eigenfunctions of the $1,2$ or $3$-particle open Toda chain. The integral transform was acting on a product of functions $q$ of one variable
which arose as solutions to a second order finite difference equation which took exactly the form of
the scalar $\op{t}-\op{Q}$ equation given above. As such, Gutzwiller's handlings appeared to be at the root of a new method allowing one to solve spectra
of certain quantum integrable models. However, the real deep connection which allowed for a systematic development of the method is definitely to be attributed to Sklyanin. 
In \cite{SklyaninSoVFirstIntroTodaChain}, by using an analogy with the classical separation of variables,
Sklyanin gave a quantum inverse scattering method-based interpretation of the aforementioned integral transform.
In the case of a quantum integrable model associated with the six-vertex $R$-matrix -such as the quantum Toda chain-, 
the transform corresponds precisely to the map that intertwines the $\op{T}_{12}(\la)$ operator entry of the model's 
monodromy matrix $\op{T}(\la)$ with a multiplication operator. Sklyanin represented the Yang-Baxter algebra of the quantum Toda chain 
on the space $\mf{h}_{\e{sep}}$ where $\op{T}_{12}(\la)$ acts as a diagonal operator. The main simplification stemmed from the fact that $\op{t}_{ \e{Td} }(\la)$
was represented on  $\mf{h}_{\e{sep}}$ as a sum of second order finite-difference operators in \textit{one} variable. As a consequence, the multi-dimensional (partial differential equations on $\R^{N+1}$)
and multi-parameter (the eigenvalues of the operator coefficients in the spectral parameter expansion of the transfer matrix $\op{t}_{ \e{Td} }(\la)$)
spectral problem for  $\op{t}_{ \e{Td} }(\la)$ was reduced to a multi-parameter\symbolfootnote[3]{because one has to keep track, one way or the other, of the different eigenvalues} \textit{one}-dimensional spectral problem 
which took precisely the form of the $\op{t}-\op{Q}$ mentioned earlier. In Sklyanin's approach, the 
eigenfunctions of the transfer matrix $\op{t}_{\e{Td}}$ on $\mf{h}_{\e{sep}}$  were given as products of the function $q_{t_{ \e{Td} }}$. 
In 1992, Gaudin and Pasquier \cite{GaudinPasquierQOpConstructionForTodaChain} realised the $\op{Q}$ operator for the Toda chain as an integral operator on 
$\mf{h}_{\e{Td}}$ and constructed explicitly its integral kernel. This allowed them to show that $\op{t}_{\e{Td}}(\la)$ and $\op{Q}_{\e{Td}}(\la)$
satisfy to the aforementioned operator $\op{t}-\op{Q}$ equation.

Gutzwiller \cite{GutzwillerResolutionTodaChainSmallNPaper2} provided the general form of the solutions to the scalar $\op{t}-\op{Q}$ equation of the model.
On its basis one could then formulae a set of quantisation conditions for the zeroes of the polynomial $t_{\e{Td}}(\la)$. However, these
equation (see Chapter \ref{Chapitre qSoV pour Toda} for more details) were extremely indirect. Solving them, even with the help of a computer
appears to be a hopeless task. A huge surprise relatively to the spectrum of the Toda chain came from the study of four dimensional gauge theories, an apparently completely disconnected subject. 
In 2009, Nekrasov and Shatashvilii \cite{NekrasovShatashviliCorrespSupSymVacuaAndBetheVectors} argued the existence of a correspondence between supersymmetric vacuua in their theories and eigenstates of certain 
quantum integrable models. This framework allowed them to argue \cite{NekrasovShatashviliConjectureTBADescriptionSpectrumIntModels} a relatively simple description, based on non-linear integral equations, for the spectra of various
quantum integrable models, the Toda chain in particular. The Nekrasov and Shatashvilii description of the spectrum of the Toda chain builds on structures and objects 
which appear, at least at first sight, totally disconnected with those attached to the quantum separation of variables.  
It is thus natural to ask: \vspace{1mm}
\begin{itemize}
\item[$\bullet$]  is it possible to recover and prove the description of the spectrum of the Toda chain conjectured by Nekrasov and Shatashvilii?\vspace{1mm}
\end{itemize}
I will provide a positive answer to this question in Chapter \ref{Chapitre qSoV pour Toda},  Section \ref{Section spectrum}.

\subsubsection*{The quantum separation of variables method}

I will now discuss the quantum separation of variables for the Toda chain in slightly more explicit terms. Let $\op{T}(\la)=\op{L}_{1}(\la) \cdots \op{L}_{N+1}(\la) $ 
be the monodromy matrix for the Toda chain.  
The one parameter abelian family of operators $\big[\op{T}(\la)\big]_{12}$ on $\mf{h}_{\e{Td}}$ is intertwined with a multiplication operator on $\mf{h}_{\e{sep}}$ by a unitary operator 
\beq
\msc{U}\; : \; \mf{h}_{ \e{sep} } \, = \, L^2\big( \R^{N+1}, \dd \nu \big)  \qquad \tend \qquad \mf{h}_{ \e{Td} } \, = \, L^2\big( \R^{N+1}, \dd^{N+1} x \big) \;. 
\enq
 In fact, due to the translation invariance of the closed quantum Toda chain, 
the measure $\dd \nu$ on $\mf{h}_{ \e{sep} }$ factorizes $\dd\nu = \dd \mu \otimes  \dd \veps$ into a "trivial" one-dimensional Lebesgue measure $\dd \veps$ that 
takes into account the spectrum $\veps$ of the total momentum operator and a non-trivial part $\dd \mu$ which is absolutely continuous in 
respect to the $N$-dimensional Lebesgue measure $\dd^Ny$ and which will be specified later on.
The operator $\msc{U}$ is realised as an integral transform that allows one to represent functions
$\Phi \in \mf{h}_{\e{Td}}$ as $\Phi(\bs{x}_{N+1})= \msc{U}\big[ \wh{\Phi} \big](\bs{x}_{N+1})$ where, 
for sufficiently well-behaved functions $\wh{\Phi} \in \mf{h}_{\e{sep}}  $,  
\beq
\msc{U}\big[ \wh{\Phi} \big](\bs{x}_{N+1}) \; = \; 
\Int{\R^{N+1} }{}  \vp_{\bs{y}_{N}} (\bs{x}_{N})  \cdot  \ex{\f{ \i }{\hbar}(\veps-\ov{\bs{y}}_N)x_{N+1}}     
\cdot \wh{ \Phi }(\bs{y}_{N}; \veps) \cdot   \f{ \dd\mu(\bs{y}_{N}) }{ \sqrt{N!} } \otimes \dd \veps  \; . 
\label{definition transfo integrale U super cal}
\enq
Here and in the following, I adopt the below notation for $k$-dimensional vectors
\beq
\bs{x}_{k}\; = \; \big(x_1,\dots, x_{k} \big) \;, \qquad \e{similarly} \; \e{for} \; \bs{y}_{k} \qquad  \e{and} \; \e{agree} \; \e{upon} \qquad  \ov{ \bs{y} }_k \; = \; \sul{a=1}{k} y_a \; . 
\label{definition vecteurs multi-dimensionnels}
\enq
Up to some minor modifications, the non-trivial part $\vp_{\bs{y}_{N}} (\bs{x}_{N})$ of $\msc{U}$'s integral kernel  corresponds to the eigenfunctions, in the generalised sense, 
of $\big[\op{T}(\la)\big]_{12}$:
\beq
\big[\op{T}(\la)\big]_{12} \cdot \vp_{\bs{y}_{N}} (\bs{x}_{N}) \; = \; \ex{-x_{N+1}}\pl{a=1}{N} \big(\la-y_a\big) \cdot \vp_{\bs{y}_{N}} (\bs{x}_{N})\;. 
\label{ecriture equation definition heuristique des fct propres B}
\enq
The factorisation of the measure $\dd \nu$ implies that the map $\msc{U}$ factorises as well 
$\msc{U} \; = \; \mc{U}_N \circ \msc{F}$, in which the $\msc{F}$ transform corresponds to taking a Fourier transform in $\veps$ followed by the 
action of a multiplication operator whereas $\mc{U}_N$ constitutes the non-trivial part of $\msc{U}$.
For\symbolfootnote[3]{The subscript $ {}_{\e{sym}}$ occurring in $L^{1}_{\e{sym}}$ indicates that the function $F$ is a symmetric function of its variables.} 
$F \in L^{1}_{\e{sym}}\big(\R^N , \dd \mu(\bs{y}_N)\big)$, it is given by 
\beq
\mc{U}_N[F] (\bs{x}_N) \; = \;\f{1}{ \sqrt{N!} }  \Int{ \R^N }{}  \vp_{\bs{y}_N}(\bs{x}_N) \cdot F(\bs{y}_N)  \cdot \dd \mu(\bs{y}_N)  \;. 
\label{definition transfo U_N} 
\enq
Henceforth the integral transform $\mc{U}_N$ will be referred to as the separation of variables  transform. 
The main advantage of the separation of variables  transform is that it provides one with a very simple form for the 
eigenfunctions of the family of transfer matrices $\op{t}_{\e{Td}}(\la)$. When focusing on a sector with a fixed momentum 
$\veps$, the joint eigenfunction\symbolfootnote[4]{the spectrum is simple, as follows from An's results \cite{AnCompletenessEigenfunctionsTodaPeriodic}.} $\mc{V}_{t_{\e{Td}}}$ 
of  $\op{t}_{\e{Td}}(\la)$ that is associated with the eigenvalue $t_{\e{Td}}(\la)$ admits a factorised representation
in the space $\mf{h}_{\e{sep}}$, in the sense that 
\beq
\wh{\mc{V}_{t_{\e{Td}}}}(\bs{y}_N;\veps) \; = \; \pl{a=1}{N} \Big\{ q_{ t_{\e{Td}} }(y_a)  \Big\} \; . 
\label{definition forme fct propre Toda dans espace separe}
\enq
$q_{ t_{\e{Td}} }$ appearing above solves the scalar form of the $\op{t}-\op{Q}$ equation for the Toda chain \eqref{ecriture eqn TQ scalaire Toda} that involves the 
polynomial $t_{\e{Td}}$.

\vspace{2mm}

To summarise, within the framework of the quantum separation of variables, 
the resolution of the spectral problem for the quantum Toda chain amounts to \vspace{1mm}
\begin{itemize}
\item[i)] building and characterising the  kernel  $\vp_{\bs{y}_N}(\bs{x}_N)$ of the SoV transform ; \vspace{1mm}
\item[ii)] establishing the unitarity of $\mc{U}_N \; : \; L^2\big(\R^N, \dd^N x \big) \;  \tend  \; 
L^2_{\e{sym}}\big(\R^N, \dd \mu(\bs{y}_N) \big) $; \vspace{1mm}
\item[iii)] characterising all of the solutions to the scalar $\op{t}-\op{Q}$ equation 
and proving the equivalence of this spectral problem to the original one for $\op{t}_{\e{Td}}$, \textit{viz}. as it is formulated on $\mf{h}_{\e{Td}}$.\vspace{1mm}
\end{itemize}
  
\noindent Point iii) has been first argued by Sklyanin \cite{SklyaninSoVFirstIntroTodaChain}, the arguments were then slightly improved by Gaudin and Pasquier  \cite{GaudinPasquierQOpConstructionForTodaChain}
and, finally An \cite{AnCompletenessEigenfunctionsTodaPeriodic} proved the equivalence.  The non-trivial part of An's proof consists in establishing that if, upon factorising the movement of the centre of mass, 
$\mc{V}_{t_{\e{Td}}} \in L^2(\R^N,\dd x)$ is a joint eigenfunction of the family $\op{t}_{\e{Td}}(\la)$, then $\wh{\mc{V}_{t_{\e{Td}}}}$ is entire and satisfies to appropriate growth conditions.

The history related to solving points i) and ii) is more sinuous and among others intersects with the theory of Whittaker functions for $GL(N,\R)$. In fact, 
the road to the resolution of point i) takes its roots in the 1979 work of Kostant 
\cite{KostantIdentificationOfEigenfunctionsOpenTodaAndWhittakerVectors}. 
Kostant found a way to quantise the integrals of motion for the open  classical  Toda chain 
hence showing the existence of an abelian ring of operators containing ${ \op{H}_{\e{Td}} }_{\mid_{\kappa=0}}$, \textit{viz}. the quantum open Toda chain Hamiltonian.
Furthermore, he was able \cite{KostantIdentificationOfEigenfunctionsOpenTodaAndWhittakerVectorsMoreDeep} to identify the system of joint generalised eigenfunctions to this ring as the Whittaker functions for 
$GL(N,\R)$. Kostant's approach has been further developed by Goodman and Wallach \cite{GoodmannWallachQuantumTodaI,GoodmannWallachQuantumTodaII,GoodmannWallachQuantumTodaIII} in the mid 80's
what allowed them to treat, among others, the closed Toda chain Hamiltonians ${ \op{H}_{\e{Td}}}_{\mid_{\kappa=1}}$. 
Within this approach, a thorough characterisation  of the relevant Whittaker functions is of prime importance.

The systematic study of Whittaker functions has been initiated by 
Jacquet \cite{JacquetFctWhittakerPrGrpesChevalley} in 1967  and further developed by  Hashizume \cite{HashizumeSomeCharacterizationWhittakerFunctions} and 
Schiffmann \cite{SchiffmannIntegralRepsAndMoreWhittakerFcts}. At the time, the Whittaker function were constructed by purely group theoretical handlings, which allowed to represent them by means of the so-called Jacquet's multiple integral. 
In 1990, Stade \cite{StadeNewIntRepWhittakerFctOutOfJacquetRep} obtained 
another multiple integral representations for the $GL(N,\R)$ Whittaker functions. 
In 1997, Gerasimov, Kharchev, Marshakov, Mironov, Morosov and Olshanetsky  \cite{GerasimovKharchevMarshakovMorozovMironovOlshanetskyGaussDecmpBasedIntRepWhittFcts}
proposed yet another multiple integral representation for these functions 
which was based on the so-called Gauss decomposition\symbolfootnote[2]{Note that the construction of Jacquet's multiple integral representation is based on the so-called Iwasawa decomposition.} of group elements. 
However, for many technical reasons that I will not discuss, all these representations, although explicit, were 
hard to deal with, not even to mention extracting from them some fine property of the object. 
Despite such technical problems, this state of the art was already enough in what concerned applications 
to the quantum Toda chain.

Building on Gutzwiller's construction and implicitly conjecturing that the ring of operators found by 
Kostant actually coincides with 
the quantum inverse scattering method issued integrals of motion for the open $N$-particle Toda chain,
 Kharchev and Lebedev \cite{KharchevLebedevIntRepEigenfctsPeriodicTodaFromWhittakerFctRep} 
wrote down a multiple  integral representation for the eigenfunctions of the closed periodic Toda chain ${ \op{H}_{\e{Td}} }_{\mid_{\kappa=1}}$
in the form $\msc{U}[ \wh{\mc{V}_{t_{\e{Td}}}}  ]$ outlined previously. Their construction worked for any value of $N$. 
The main point of their implicit conjecture relative to the equality between the rings generated by the two families of integrals of motion
was to allow them
to use Kostant's characterisation of the eigenfunctions of the open Toda chain abelian ring of operators 
so as to identify the kernel $\vp_{\bs{y}_N}(\bs{x}_N)$ with the Whittaker functions for $GL(N,\R)$. 
In this way, they provided an explicit construction for the integral kernel of $\mc{U}_N$.  
In the mentioned paper,  Kharchev and Lebedev  used the so-called Gauss decomposition based multiple integral representations for 
these Whittaker functions \cite{GerasimovKharchevMarshakovMorozovMironovOlshanetskyGaussDecmpBasedIntRepWhittFcts}. 
Later, in \cite{KharchevLebedevMellinBarnesIntRepForWhittakerGLN,KharchevLebedevIntRepEigenfctsPeriodicTodaFromRecConstrofEigenFctOfB},
the two authors managed to connect their approach with Sklyanin's quantum separation of variables \cite{SklyaninSoVFirstIntroTodaChain} 
approach to the quantum Toda chain. Indeed, in \cite{SklyaninSoVGeneralOverviewAndConstrRecVectPofB}, Sklyanin proposed a scheme 
based on the recursive construction of the monodromy matrix which allowed one to build  the kernel of the SoV transform through an inductive,
but explicitly solvable, procedure.  Kharchev and Lebedev
\cite{KharchevLebedevMellinBarnesIntRepForWhittakerGLN,KharchevLebedevIntRepEigenfctsPeriodicTodaFromRecConstrofEigenFctOfB}
managed to implement this scheme on the example of the open Toda chain, hence obtaining
a new multiple integral representation for the $GL(N,\R)$ Whittaker functions
which they called the Mellin-Barnes representation. 
Finally, in \cite{GerasimovKharchevLebedevRepThandQISM}, Gerasimov, Kharchev and Lebedev established a clear connection 
between the group theoretical and the quantum inverse scattering method-based approaches to the open 
Toda chain. In particular, that last paper proved the previously used conjecture relative to 
the concurrency between  Kostant's ring of operators on the one hand and the quantum inverse scattering method issued 
conserved charges on the other hand.  

\vspace{2mm} 

There exists one more multiple integral based representation for the generalised eigenfunctions 
of the open Toda chain which was obtained, for the first time, by Givental \cite{GiventalGaussGivIntRepObtainedForEFOfOpenToda} in 1997. 
The group theoretic interpretation of Givental's multiple integral representation has been given by Gerasimov, Kharchev, Lebedev  and Oblezin 
\cite{GerasimovKharchevLebedevOblezinGaussGiventalIntRepFromQISMAndQOp} in 2006. Since the 
corresponding proof built on a specific type of Gauss decomposition for the group elements of $GL(N,\R)$,
this multiple integral representation bears the name Gauss--Givental. 
Furthermore, the work \cite{GerasimovKharchevLebedevOblezinGaussGiventalIntRepFromQISMAndQOp} also contained some comments relative to a connection between the Gauss--Givental representation and the 
$\op{Q}$-operator of the Toda chain constructed earlier by Gaudin and Pasquier 
\cite{GaudinPasquierQOpConstructionForTodaChain}.
In fact, the connection between $\op{Q}$-operators and Gauss-Givental integral representations for the kernels of separation of variables transforms has been established, on a much deeper level of understanding, 
by  Derkachov, Korchemsky and Manashov \cite{DerkachovKorchemskyManashovXXXSoVandQopNewConstEigenfctsBOp}, this a few years 
prior to paper \cite{GerasimovKharchevLebedevOblezinGaussGiventalIntRepFromQISMAndQOp}. 
Derkachov, Korchemsky and Manashov observed that one can build the
eigenfunctions of the $\big[\op{T}(\la)\big]_{12}$ operators arising in the context of the non-compact XXX chain 
out of the building blocks that were used for constructing the integral kernel of the model's $\op{Q}$-operator. This allowed
them to propose a "pyramidal" representation for the kernel of the separation of variables transform
for the non-compact XXX magnet. In
\cite{SilantyevScalarProductFormulaTodaChain},  Silantyev applied the Derkachov, Korchemsky, Manashov  method so as to re-derive the 
 Gauss-Givental representation for the $GL(N,\R)$ Whittaker functions $\vp_{\bs{y}_N}(\bs{x}_N)$. 

The above works,  by-and-large, close the discussion relative to point i). I do stress that Sklyanin's recursive method leading to 
the Mellin-Barnes representation as well as Derkachov, Korchemsky and Manashov's method leading to the Gauss--Givental representation is quite general
in that it solely builds on objects that are "natural" to the quantum inverse scattering method. Therefore, the methods are well entrenched:  
 their implementation to other quantum integrable models solvable by the quantum separation of variables method, 
this in view of solving the analogue of point i) above,  is only a technical question, all conceptual issues being solved. 
In this sense, the state of the art relative to point i) is quite satisfactory.

I will now discuss the unitarity of the transform $\msc{U}$ realising the quantum separation of variables. In virtue of 
the unitarity of the Fourier transform, this property boils down to proving the unitarity of $\mc{U}_N$, \textit{viz}. the point ii) mentioned earlier. 
One can rephrase the unitarity of $\mc{U}_N$ as the property of completeness and orthonormalilty of the system of generalised eigenfunctions 
of the open Toda chain $\{ \big[\op{T}(\la)\big]_{12} \}_{\la \in \R}$ abelian ring of operators. 
I do stress that the construction of the appropriate measure space, \textit{viz}. the space where the variable $\bs{y}_N$ parametrising the eigenvalues of 
$\{ \big[\op{T}(\la)\big]_{12} \}_{\la \in \R}$ evolve adjoined to the explicit form of the measure $\dd \mu$ on this space, is part of the problem stated in point ii). 
For the sake of simplifying the discussion, I already used part of this answer from the very beginning, namely that $\bs{y}_N \in \R^N$. Still, 
I do stress that this property can only be affirmed, \textit{a posteriori}, when completeness is proven. Again, \textit{a posteriori}, it turns out that the measure $\dd \mu$ is continuous 
with respect to Lebesgue's measure on $\R^N$ and takes the explicit form 
\beq
\dd \mu(\bs{y}_N) \; = \; \mu(\bs{y}_N) \cdot \dd^N y 
\quad \e{with} \quad  \mu(\bs{y}_N)  \; = \; 
\f{1}{(2\pi \hbar)^{N}}  \pl{ k \not= p }{ N } \bigg\{ \Ga\Big( \f{ y_k - y_p }{ \i \hbar }  \Big) \bigg\}^{-1}\;. 
\label{Ecriture mesure Sklyanin Partie Intro}
\enq

Both completeness and orthonormality, have   already been established within the framework of the group theoretical 
based approach to the model.  Semenov-Tian-Shansky proved 
\cite{SemenovTianShanskyQuantOpenTodaLatticesProofOrthogonalityFormulaForWhittVectrs} the orthonormalilty 
of the system $\{ \bs{x}_N \mapsto  \vp_{\bs{y}_N}(\bs{x}_N) \}_{ \bs{y}_N \in \R^N }$ which, written formally, takes the form 
\beq
\Int{ \R^{N} }{} \Big( \vp_{\bs{y}_{N}^{\prime}} (\bs{x}_{N}) \Big)^{*}   \cdot 
 \vp_{\bs{y}_{N}} (\bs{x}_{N}) \cdot \dd^{N} x  \; = \; 
 \big[  \mu(\bs{y}_N) \big]^{-1}  
\sul{ \sg \in \mf{S}_N }{}  \pl{a=1}{N} \de\big( y_a - y^{\prime}_{\sg(a)} \big) \;. 
\label{ecriture relation completude par rapport espace originel}
\enq
Also, the completeness
of the system $\{ \bs{y}_N \mapsto \vp_{\bs{y}_N}(\bs{x}_N) \}_{ \bs{x}_N \in \R^N }$, which written formally, takes the form 
\beq
\Int{ \R^{N} }{} \Big( \vp_{\bs{y}_{N}} (\bs{x}_{N}^{\prime}) \Big)^{*}  \vp_{\bs{y}_{N}} (\bs{x}_{N})\; 
\mu(\bs{y}_N) \cdot   \dd^{N} y  \; = \; 
\pl{a=1}{N+1} \de(x_a - x^{\prime}_{a}) \;, 
\label{ecriture relation completude par rapport espace SoV}
\enq
follows from the material that can be found in chapters 15.9.1-15.9.2 and 15.11 of Wallach's book \cite{WallachRealReductiveGroupsII} and appears
as a culminating result of an 878 pages long two volume book developing various tools for constructing and analysing Whittaker vectors.

The proofs \cite{SemenovTianShanskyQuantOpenTodaLatticesProofOrthogonalityFormulaForWhittVectrs,WallachRealReductiveGroupsII}
are technically involved, rather long, and completely disconnected from the quantum inverse scattering method description. 
It is, in fact, not so much the technical part that poses problem but rather their lack of connection to the quantum inverse
scattering method. The matter is that the quantum Toda chain is a very special model that can be solved by means of group theoretical methods. 
This is no longer the case for other quantum integrable models where one has to resort to the more refined algebraic machinery
of quantum groups. One cannot thus build on the existing harmonic analysis on groups machinery so as to study completeness or orthogonality of such more general models. 
It means that if one would like to repeat such a path, then one would need to start developing an analogous theory, presumably from scratch,
in a context that would be applicable to these other models. Taken into account the complexity and length of the aforementioned methods \cite{SemenovTianShanskyQuantOpenTodaLatticesProofOrthogonalityFormulaForWhittVectrs,WallachRealReductiveGroupsII} 
this would be unreasonable. I stress that, as of today, even for the relatively simple case of the lattice discretisation of the Sinh-Gordon model \cite{BytskoTeschnerSinhGordonFunctionalBA}, 
point ii) remains an open question.

One can thus ask whether \vspace{1mm}
\begin{itemize}
\item[$\bullet$] it would possible to set a simple and systematic approach to the resolution of point ii) that would solely build on the objects and structures that are natural to the 
quantum inverse scattering method? \vspace{1mm}
\end{itemize}
I shall answer positively to this question Chapter \ref{Chapitre qSoV pour Toda}. My construction allows, in principle, to extend the proof of unitarity of the SoV transform 
to other, more complex, models such as the lattice discretisation of the Sinh-Gordon model.

\subsection{Quantum integrable models at finite temperature}

The very first approach to the description of quantum integrable models at finite temperatures was developed in 1969 by Yang and Yang 
\cite{Yang-YangNLSEThermodynamics} on the example of the non-linear Schr\"{o}dinger model. These authors, building on the logarithmic Bethe equations describing the 
spectrum of the model \cite{LiebLinigerCBAForDeltaBoseGas}, carried out a large-volume $L$ saddle-point like analysis of the statistical sum defining the model's
partition function and hence its free energy.  The saddle-point analysis allowed Yang and Yang to argue an integral representation for the \textit{per} site free energy $f_{\e{NLS}}$ of the model in the thermodynamic limit: 
\beq
f_{\e{NLS}} \; = \; \Int{ \R }{} \ln \big[ 1\;+ \;  \ex{- \f{ \veps(\la)}{ T } } \big] \cdot \f{ \dd \la  }{ 2\pi } \;,  
\label{ecriture forme energie libre}
\enq
this for any values of the coupling constant $0 < c \leq +\infty$ just as any value of the chemical potential $h$. The dependence of  $f_{\e{NLS}}$ on the coupling constant and on the chemical
potential is encoded in the function $\veps$ which corresponds to the unique solution to 
the non-linear integral equation referred to, nowadays, as the Yang-Yang equation:
\beq
\veps(\la) = \la^2 - h - \f{T}{2\pi} \Int{ \R }{}\f{ 2c }{ (\la-\mu)^2  \; + \; c^2 }  \cdot \ln \big[ 1+ \ex{-\f{\veps(\mu)}{T}}\big] \cdot \dd \mu  \;. 
\label{ecriture eqn YY}
\enq
 Yang and Yang showed that its solution exists and is unique, this for any value of $T$ and $h$. 
In 1971, Yang-Yang's method has been generalised to the study of the thermodynamics 
of the XXZ spin-$\tf{1}{2}$ chain simultaneously and independently by Gaudin \cite{GaudinTBAXXZMassiveInfiniteSetNLIE} and Takahashi
\cite{TakahashiTBAforXXZFiniteTinfiniteNbrNLIE}. Both works built on the so-called string hypothesis
\cite{BetheSolutionToXXX} and characterised the free energy of the model in terms of a solution to an infinite tower of non-linear integral equations. 
Since then, Yang and Yang's approach has been applied with success to many other models and bears the name of the thermodynamic Bethe Ansatz.
Note that the reasoning of Yang and Yang which gave rise to the integral representation for  $f_{\e{NLS}}$ was raised to the level of theorem, only much later, in 1989 by Dorlas, Lewis and Pul\'{e} 
\cite{DorlasLewisPuleRigorousProofYangYangThermoEqnNLSE}. These authors represented the finite temperature and volume partition function of the model at general $c$ in terms of an integral of a function $\ex{L G}$
\textit{versus} the probability measure $\mathbb{P}_{L;+\infty}$ induced by the statistic operator $\ex{-\frac{1}{T}\op{H}_{NLS}}\mid_{c=+\infty}$ of the impenetrable Bose gas ($c=+\infty$)
on an appropriate subspace of $2^{\mathbb{Z}}$.
They subsequently showed that $\mathbb{P}_{L;+\infty}$ satisfies a large deviation principle and that the integrand $G$ is regular enough. In this way   
they could invoke Varadhan's lemma -an infinite dimensional variant of Laplace method- so as to estimate the large-$L$ limit of the
integral they obtained as a representation of the partition function.

A completely different approach to the study of thermodynamics of spin chains has been pioneered by Koma in 1987 on the example of the XXX Hamiltonian \cite{KomaIntroductionQTM6VertexForThermodynamicsOfXXX}. 
Two years later, Koma generalised his method so as to encompass the XXZ \cite{KomaIntroductionQTM6VertexForThermodynamicsOfXXZ} Hamiltonian. Koma's approach takes its roots in Baxter's 
observation on the correspondence between the eight vertex model and the XYZ spin chain.
Building on such ideas, Koma argued that the computation of the thermodynamic limit of the partition function of the XXX and XXZ models in a magnetic field 
is equivalent to obtaining the largest eigenvalue of the transfer matrix associated with a specific inhomogeneous six-vertex model
whose inhomogeneities are functions of the temperature. This transfer matrix is called, nowadays, the quantum transfer matrix
and its size depends on an auxiliary integer called the Trotter number. 
In Koma's approach, the free energy is expressed as the infinite Trotter number limit of the largest eigenvalue of the quantum transfer matrix. 
Koma proposed a characterisation of this largest eigenvalue, at finite Trotter numbers, in terms of a system of logarithmic Bethe Ansatz equations. 
Although Koma could not provide a proper analytic framework for taking the infinite Trotter number limit,
he was able to carry out a numerical analysis at finite Trotter numbers. He was also able to extrapolate his analysis and describe the infinite Trotter number limit 
of the largest eigenvalue on a numerical basis.
In 1991, Takahashi improved Koma's approach what allowed him to compute the infinite Trotter number limit explicitly. 
More precisely, Takahashi found a way to take the infinite Trotter number limit on the level of the Bethe equations what allowed him to 
obtain a characterisation of the thermodynamics of the XYZ and XXZ spin-$\tf{1}{2}$ chains
in terms of an infinite sequence of Bethe roots that ought to be fixed numerically 
\cite{TakahashiThermoXXZInfiniteNbrRootsFromQTM,TakahashiThermoXYZInfiniteNbrRootsFromQTM}. 
In 1992, Kl\"{u}mper \cite{KlumperNLIEfromQTMDescrThermoRSOSOneUnknownFcton} managed to strongly simplify the analysis of the infinite Trotter number limit for the restricted solid-on-solid model.
He then conformed the approach to the case of the XYZ chain in 1993 \cite{KlumperNLIEfromQTMDescrThermoXYZOneUnknownFcton}. To be more specific,
building on the method of non-linear integral equations \cite{BatchelorKlumperFirstIntoNLIEForFiniteSizeCorrectionSpin1XXZAlternativeToRootDensityMethod}, Kl\"{u}mper proposed a 
way for bypassing the problem of sending the Trotter number $N$ to infinity directly on the level of the Bethe equations describing the 
largest eigenvalue of the quantum transfer matrix. He introduced an auxiliary function whose appropriate subset of zeros corresponds to the Bethe roots 
and was able to construct a non-linear integral equation satisfied by this function. The main feature of Kl\"{u}mper's non-linear integral equation 
was that the Trotter number was only appearing parametrically in its driving term. Thus, at least on formal grounds, taking the infinite Trotter number limit
was trivial in that it simply boiled down to replacing the driving term by its limit. The limiting non-linear integral equation could then be easily solved
on a computer by the iteration method. In this way, Kl\"{u}mper obtained the full description of the XYZ 
and XXZ spin-$\tf{1}{2}$ chains at finite temperature solely in terms of a \textit{single} unknown function 
that satisfies a \textit{single} non-linear integral equation. 
This function allows one to compute the free energy at \textit{any} finite temperature $T$ trough a formula that is similar in spirit to \eqref{ecriture forme energie libre}. 
Kl\"{u}mper's method also allows one to build non-linear integral equations 
describing the sub-leading eigenvalues of the quantum transfer matrix and thus gives access to the finite temperature correlation lengths of the model. 
Kl\"{u}mper's approach allowed to confirm the conformal field theory-based predictions \cite{AffleckCFTPreForLargeSizeCorrPartitionFctonAndLowTBehavior} 
for the low-T behaviour of the free energy. The quantum transfer matrix based method has, since then, been applied to many other quantum integrable models.

\subsection{Quantum integrable models with integrable boundary conditions}

Although I have so far only discussed the case of models subject to periodic boundary conditions, there also exist  quantum integrable models which are subject to 
other types of boundary conditions. Apparently, the first instance of such a model was found by McCoy and Wu \cite{McCoyWuFirstProofXXZCommutesWith6VTransferMatrix} in 1967 
who showed that the $XXZ$ Hamiltonian subject to so-called off-diagonal boundary fields commutes with  the transfer matrix of a generalisation of the six-vertex model
studied by Yang \cite{YangGeneralisationOf6VertexCorrespondingtoARotatedinXYplaneXXZChain} where one substitutes the periodic boundary conditions with the free ones. 
The investigation of such models through the Bethe Ansatz was pioneered by Gaudin \cite{GaudinBoseGasBoundary} in 1971. In that paper he was able to implement new types of  boundary conditions for the Lieb-Liniger model 
while still preserving the integrability of the model: Gaudin proposed a modification of the Bethe Ansatz fit for dealing with 
such boundary condidtions. In 1985  Schultz \cite{SchultzHubbardChainWithReflectingEndsCBA}
obtained the spectrum of the Hubbard Hamiltonian subject to the so-called reflecting boundary conditions. Then, two years later,  Alcaraz, Batchelor, Baxter and Quispel \cite{AlcarazBatchelorBaxterQuispelCBAopenXXZ}
gave a description of the spectrum of the XXZ  spin-$1/2$ chain subject to two longitudinal magnetic fields $h_{\pm}$ acting on the first and last site of the chain:
\beq
\op{H}_{XXZ}^{ ( \e{bd} ) } \; =\; J \sul{a=1}{L-1} \Big\{ \sg^x_a \,\sg^x_{a+1} + \sg^y_a\,\sg^y_{a+1} + \De \,(\sg^z_a\,\sg^z_{a+1}+1)\Big\}
\; + \; h_- \, \sg^z_1 \; +\;  h_+ \,\sg^z_L  \;. 
\label{ecriture intro Hamiltonine XXZ boundary}
\enq
The mentioned results all built on the coordinate Bethe Ansatz. In 1988, Sklyanin \cite{SklyaninABAopenmodels} elaborated a version of the 
algebraic Bethe Ansatz approach fit for dealing with models subject to so-called integrable boundary conditions. His approach utilised the results of 
Cheredink \cite{CherednikReflectionEquationFactorisabilityOfScattering} on the factorisability of scattering in the presence of a boundary which relied on the 
so-called reflection equation. Within Sklyanin's approach, the reflection equation constitutes the equivalent of the Yang--Baxter equation. 
Just as it has been discussed earlier on relatively to periodic boundary conditions, not all integrable models are directly amenable to an algebraic Bethe Ansatz resolution. 
For instance, this is usually the case for models subject to the most general boundary conditions\symbolfootnote[3]{In the case of the XXZ spin chain, these correspond to applying boundary
field which also couple with the $\sg^x$ and $\sg^y$ operators, this on both ends of the chain.}. Such models can however be treated by means of first applying a vertex-IRF transformation
and then solving the auxiliary model by means of an algebraic Bethe Ansatz, analogously to the case of the XYZ chain, this provided that the parameters describing the boundary 
interactions satisfy some "weak" constraints.
An archetype of such model, the XXZ chain subject to non-longitudinal boundary fields, was treated in \cite{WangNondiagonalXXZBetheAnsatz,YangZhangSecondReferenceStateOpenXXZ}.
Note that such models can also be solved by means of the quantum separation of variables approach \cite{FaldellaKitanineNiccoliSpectrumAndScalarProductsForOpenXXZNonDiagBC,KitanineMailletNiccoliOpenXXZGenericBCCompletenessofBA}

Starting from the finite-$L$ expression for the energies of the ground and low-lying excited states provided by the Bethe Ansatz, Alcaraz \textit{et al}. extracted their large-$L$
behaviour using  methods analogous to those employed for periodic boundary conditions. As expected, their analysis showed that the presence of boundary interactions generate an additional, constant in $L$, contribution
-the so-called surface free energy- to the ground state's energy. Furthermore, the $\tf{1}{L}$ corrections to the energies of the excited states obtained in \cite{AlcarazBatchelorBaxterQuispelCBAopenXXZ} 
took precisely the form predicted by conformal field theory \cite{AffleckCFTPreForLargeSizeCorrPartitionFctonAndLowTBehavior,BloteCardyNightingalePredictionL-1correctionsEnergyAscentralcharge}. 

One expects that a similar behaviour also holds for the XXZ chain subject to diagonal boundary fields \eqref{ecriture intro Hamiltonine XXZ boundary}
at finite temperature. Namely, the per-site free energy of this model is expected to admit  the large-$L$ expansion 
\beq
L\cdot f^{(\e{bd})}_L \; = \; L \cdot  f^{(\e{per})}_{XXZ}  \; + \; f_{\e{surf}} \; +\; \cdots 
\label{ecriture DA gd L Free energy at boundary model}
\enq
where $f^{(\e{per})}_{XXZ}$ is the thermodynamic limit of the \textit{per site} free energy of the XXZ chain subject to periodic boundary conditions. 
The additional constant term $f_{\e{surf}}$ is called the surface free energy of the model. 

The only integrability-based investigation of the finite temperature surface free energy was carried out by G\"{o}hmann, Bortz and Frahm in \cite{BortzFrahmGohmannSurfaceFreeEnergy}.  
In that paper, the authors provided a representation for the finite Trotter number 
approximant $ f_{\e{surf}}^{(N)}$ of the boundary free energy. In their approach, $f_{\e{surf}}^{(N)}$ was expressed as
an expectation value of a product of $N/2$ local operators forming the so-called finite temperature boundary
operator. This expectation value was computed in respect to the eigenvector associated with the dominant eigenvalue of the
quantum transfer matrix arising in the description of the thermodynamics of the periodic $XXZ$ spin-$\tf{1}{2}$
chain. At the time, it was however not clear how to take the Trotter limit of the formula or even find some manageable, more explicit, expression for the
surface free energy at finite Trotter numbers, \textit{viz}. one that re-casts the expectation value as some explicit function of the parameters
of the model. 

\noindent Thence, the natural question related with this problem reads :\vspace{1mm}
\begin{itemize}
  \item[$\bullet$] it is possible to find a manageable expression for  $ f_{\e{surf}}^{(N)}$ starting from the G\"{o}hmann, Bortz and Frahm formula, this in such a
 way that the Trotter limit can be taken? If yes, can anything be said about limiting regimes of the surface free energy, such as its low-$T$ expansion? \vspace{1mm}
\end{itemize}
It so happens that one can answer positively to the first question and provide an at least partially positive answer to the second one. 
I shall discuss these matters in Chapter \ref{Chapitre QIS a tempe finie et cptmt grde distance de leurs correlateurs}

\section{Correlation function in quantum integrable models}

The observables that are of main interest to the  physics of a model are its correlation functions. I have argued that the correlators which are the easiest to deal with
are the zero temperature ones since then the thermal average boils down to an expectation value in respect to the model's ground state. 
Still, due to the highly combinatorial structure of the eigenvectors built through the Bethe Ansatz, 
obtaining effective expression for the zero temperature correlators of a quantum integrable model is a quite intricate task. 

Below, I will provide an account of the developments that took place relatively to computing the correlators of quantum integrable  models, the XXZ chain and the non-linear Schr\"{o}dinger
model being paradigmatic examples. 

\subsection{The free fermion equivalent models}

In the early days, the attention was focused on computing two-point functions
in the simplest possible quantum integrable models, namely those explicitly equivalent to a model of free fermions. I remind that
Lieb, Mattis and Schultz \cite{LiebMattisSchultzXYSolutionByFreeFermionsCorrFunctsToo}, 
observed that the XX chain, \textit{viz}. the XXZ Hamiltonian at zero anisotropy\symbolfootnote[4]{In fact, this also holds true for the XYZ chain provided that the longitudinal 
coupling constant in front of the $\sg^z-\sg^z$  interactions is set to zero. For obvious reasons, the resulting model is referred to as the XY model.} can be made explicitly equivalent to a systems of free, \textit{viz}. non-interacting fermions. 
The equivalence with free fermions also exists for many other integrable models such as the non-linear Schr\"{o}dinger model at $c=+\infty$
or the two-dimensional Ising model of statistical mechanics. The existence of an underlying free fermionic structure drastically simplifies the 
combinatorics of a model's eigenvectors. Also, the Bethe equations utterly simplify at the free fermion point 
in that these decouple thus becoming \textit{explicitly} solvable (to convince oneself, set $\zeta=\tf{\pi}{2}$ in \eqref{ecriture eqns de Bethe XXZ Intro}).

First results relative to correlators in free fermion equivalent models go back to the 1949 work of Kaufmann and Onsager \cite{KafmanOnsagerFirstIntroDetRepCorrIsing2D}. These authors managed to provide an explicit  
representation for the thermodynamic limit (the limit of an infinite lattice in this case) of the row-to-row spin-spin correlation function $\moy{\sg_{1,1}\sg_{1,N+1} }$ of the two-dimensional Ising model.
The result of \cite{KafmanOnsagerFirstIntroDetRepCorrIsing2D} was given in terms of a sum of two
Toeplitz determinants, each of size $N$. Kaufmann and Onsager also obtained, although did not publish, a simple representation for the diagonal spin-spin correlation function 
$\moy{\sg_{1,1}\sg_{N+1,N+1} }$ in terms of a single Toeplitz determinant (see the excellent review \cite{DeiftItsKrasovskyToeplitzDetsUnderImpetrusIsing})
\beq
\moy{ \sg_{1 \, 1} \cdot  \sg_{ 1+N \,,1+N}  } \; = \; D_N\big[ \vp_{\e{Isg}} \big] \quad \e{with} \quad 
\vp_{\e{Isg}}\big( \ex{ \i \th} \big) \; = \; \bigg( \f{ 1-k \ex{- \i \th} }{ 1-k \ex{i\th} } \bigg)^{\f{1}{2}} \quad \e{and} \quad
k = \f{1}{ \sinh^2\big(  \tf{2J}{T} \big)   }
\nonumber
\enq
is expressed in terms of the temperature and the model's coupling constant. Above, the symbol $D_N\big[ f \big]$ stands for the Toeplitz determinant of the $N\times N$  matrix generated by the symbol $f$:
\beq
D_N\big[ f \big] \; = \; \det_N\Big[ c_{i-j}[f] \Big]  \qquad \e{and} \qquad 
c_{k}[f] \; = \; \Int{-\pi}{\pi} f\big( \ex{i\th} \big) \ex{-\i k\th} \cdot \f{ \dd \th }{ 2\pi } \;. 
\nonumber
\enq
Such a Toeplitz determinant based representation is very effective be it for computing the spontaneous magnetisation  $\msc{M}(T)$ of the model or studying 
the large-distance asymptotic behaviour of the correlator. 

Numerous expressions for other correlations functions in the two-dimensional Ising model and other models equivalent to free fermions were obtained
between the late '60's and the early 80's by various authors (Abraham, Barouch, McCoy, Tracy, Vaida, Wu,...) what allowed to study many of the asymptotic regimes of these correlation functions
\cite{ChengWuMagnetization2DIsingAsymptoticsAtBelowAndAboveTc,McCoySomeAsymptoticsForXYCorrelators,AbrahamBarouchMcCoy4ptcomputationtoGetlongTimeforTwo,McCoyWuSpinSpin2DISingII,
McCoyWuMagnetization2DBoundaryIsingWithMagAtBound,McCoyWuMagnetization2DBoundaryIsingHysteresisPlusWetting,TracyVaidyaRedDensityMatrixSpaceAsymptImpBosonsT=0,
TracyVaidaTimeDependentTransverseIsingCorrelatorsAsymptotics,WuAsymptoticsSpinSpinIsingModel},... 
See also the book \cite{McCoyWu2DIsingModel}. It is important to mention the important progress made by
Wu, McCoy, Tracy  and Barouch in \cite{BarouchTracyMcCOyWuScalingResultsForIsinfPainleveIII}. These authors  found a way to characterise the transition 
between the critical and off-critical regimes  of two-point functions in the two-dimensional Ising model in terms of a Painlev\'{e} III transcendent. 
They were able to study this transitional regime by deriving connection formulae for the third Painlev\'{e} of interest to their study. 
In a subsequent work, McCoy, Tracy and Wu \cite{McCoyTracyWuPIIIProofForDetRep} brought several elements of rigour to the derivation of the connection formulae
obtained in their earlier work with Barouch. 
One of the applications of the results obtained in \cite{BarouchTracyMcCOyWuScalingResultsForIsinfPainleveIII} was the derivation of the first few terms in the $T\tend T_c\pm 0^+$
asymptotic expansion of the zero field susceptibility defined, for the infinite lattice, as
\beq
\chi(T)=\f{1}{T} \sul{M,N \in \mathbb{Z} }{} \Big\{ \moy{ \sg_{1 \, 1} \cdot  \sg_{ 1+M \,,1+N}  } - \msc{M}^2(T)  \bigg\} \;. 
\enq

The analysis of  \cite{BarouchTracyMcCOyWuScalingResultsForIsinfPainleveIII} was building on so-called exponential form factor expansions for the two-point functions, namely representations
given in terms of the exponent of a series of multiple integrals. Over the years, it was found \cite{BoukraaHassaniMaillardMcCoyOrrickZenineHolonomyIsingFormFactors,LybergMcCoyNewExpression2DIsingFormFactors,PalmerTracyFormFactorExpansionbelowTc2DIsing}
how to explicitly take the exponent of this series and obtain so-called form factor expansions. 
These appeared as very effective tools for studying various properties of the zero field susceptibility. Nickel \cite{NickelBoundaryForIsingSusceptibility,NickelBoundaryForIsingSusceptibilityAddendum}
argued the existence and form of the domain of analiticity for the analytic continuation of $\chi(T)$ from real positive temperature to the complex plane. 
The singular structure leading to the domain of analyticity for the diagonal susceptibility has been argued in \cite{BoukraaHassaniMaillardMcCoyZenineDiagIsingSusceptibilitySingAtRootsUnity}
Finally, in Tracy and Widom \cite{TracyWidomDiagSuscept2DIsing} brought the elements of rigour leading to proof of the shape of this domain in the case of the diagonal susceptibility.  
Also, the papers \cite{AssisMaillardMcCoySomeFactorisationPropsIsingFF,BoukraaHassaniMaillardMcCoyOrrickZenineHolonomyIsingFormFactors} exhibited various algebraic properties
of the form factors multiple integrals which indicate the possibility to explicitly 
separate these into sums of products of one-dimensional integrals. 

In fact, a systematic study of the correlation functions in free fermion equivalent quantum integrable models -the impenetrable Bose gas, the XY model or its isotropic version the XX model-  
culminated in representations of these quantities in terms of Fredholm determinants (or their minors) of so-called integrable 
operators \cite{ColomoIzerginKorepinTognettiEFPinXYtransversefield,ColomoIzerginKorepinTognettiTempCorrFctXX,KorepinSlavnovTimeDepCorrImpBoseGas,
LenardBoseImpGasredDensityMatrixFredminorsRepAndProofs,McCoyPerkShrockSpinTimeSpaceAutoCorrAsModSineKernel,McCoyPerkShrockSpinTimeAutoCorrAsModSineKernel,
SchultzFredholmMinorForReduceDensityImpBosons,ForresterWitteFFexpansionDiagDiag2DIsingAsFDetsOfIntegrableIntOps}. The integrable integral operators  $\e{id} + \op{V}_x$ acts
on $L^2(\msc{C})$ with an integral kernel $V_x(\la,\mu)$ depending, on a parameter $x$ in an oscillatory way.
In the case of two-point functions, $x$ plays the role of the spacial and/or temporal separation
between the two operators that build up the correlator. The large-$x$ asymptotic expansion of the 
associated Fredholm determinants then allows one to test the predictions of conformal field theories for the underlying models. 
In fact, the pure sine kernel determinant -or minors thereof- provide one with the simplest possible example of an integrable kernel that  
represents correlators in a free fermion equivalent model. The large-$x$ asymptotic behaviour of $\log \det \big[ \e{id} - \op{S}_x \big]$  was investigated 
starting from 1973 when des Cloizeaux and Mehta \cite{DescloizeauxMethaSineKernelFirstAsympotics} argued the first two terms in its large-$x$ asymptotic expansion. 
Then, in 1976, Dyson \cite{DysonSineKernelInverseScatteringAsymptoticExpansions} argued the expression for the constant term in the asymptotic expansion of the determinant 
and provided a recursive method allowing one to compute the sub-leading corrections in $x$.   
These results were then proven, within the setting of operator methods, by Widom \cite{WidomSinekernelOnSingleIntervals}, Ehrhardt \cite{EhrhardtConstantinPureSinekernelFredholmDeyt}. 
Basor and Widom \cite{BasorWidomTruncatedWienreHopfPiecewiseContinuousSymbols} and, independently, and Budyn and Buslaev \cite{BudynBuslaevPureGammaSineKernelAsympt} 
obtained the large-$x$ expansion for $\log \det \big[ \e{id} - \ga \op{S}_x \big]$ when $|\ga|<1$ and up to a $\e{o}(1)$ remainder. The latter is of a quite
different nature than for $\ga=1$. Recently, Botchner, Deift, Its and Krasovsky \cite{BothnerDeiftItsKrasovskyTransitionAsymptoticsSineKernel}
were able to evince the transition mechanism between the two kinds of asymptotic expansions, thus establishing an earlier conjecture of 
Dyson \cite{DysonSineKernelTransitionAsymptoticsConjecture}. 
I should also mention that the  discovery of the connection between this determinant \cite{JimMiwaMoriSatoSineKernelPVForBoseGaz} and the Painlev\'e V equation 
allowed to access to many terms in the large-$x$ asymptotic expansion of the associated correlation functions 
\cite{JimMiwaMoriSatoSineKernelPVForBoseGaz,McCoyPerkShrockSpinTimeSpaceAutoCorrAsModSineKernel,McCoyPerkShrockSpinTimeAutoCorrAsModSineKernel, 
McCoyTangSineKernelSubleadingFromPainleveV}. 

Still, the systematic and efficient approach to the asymptotic analysis of various quantities related 
to integrable integral operators $\e{id}+\op{V}_x$ -and hence of correlation functions in free fermion model- has been made possible thanks to the observation 
\cite{ItsIzerginKorepinSlavnovDifferentialeqnsforCorrelationfunctions} that their analysis can be reduced to a resolution of an associated  Riemann--Hilbert problem.
The jump contour in this Riemann-Hilbert problem corresponds to the one on which the integral operator acts while the  jump
matrix is expressed in terms of the functions entering in the description of the kernel. 
In the case of interest to the correlation function, the integral kernel $V_x(\la,\mu)$ depends on $x$ in an oscillatory way: as a consequence, one ends-up with an oscillatory Riemann--Hilbert problem.
One can carry out the asymptotic analysis of its solution by some adaptation of the non-linear steepest descent method invented by 
Deift-Zhou \cite{DeiftZhouSteepestDescentForOscillatoryRHP,DeiftZhouSteepestDescentForOscillatoryRHPmKdVIntroMethod} in 1992.
I do stress that it was precisely the Riemann--Hilbert problem setting that allowed for a complete and above all effective characterisation of the leading
asymptotic behaviour of correlation functions in free-fermion equivalent models. The long-distance,
large-time and long-distance at zero but as well non-zero temperature for various models has been carried out in the series of papers
 \cite{CheianovZvonarevZeroTempforFreeFermAndPureSine,DeiftZhouTimeautocorrInTempForXYatCritMag.Field,ItsIzerginKorepinTemperatureLongDistAsympBoseGas,
ItsIzerginKorepinNovokshenovTempeAutoCorrCriticalTransverseIsing,ItsIzerginKorepinSlavnovTempCorrFctSpinsXY,ItsIzerginKorepinVarguzinTimeSpaceAsymptImpBoseGaz}.
Also, transition asymptotics could have been considered recently within the Riemann--Hilbert approach \cite{BothnerDeiftItsKrasovskyTransitionAsymptoticsSineKernel,ClaeysItsKrasovskyEmergenceSingularityToeplitzAndPV}.

\vspace{1mm}
 \noindent The bottom line of all these studies is that \vspace{1mm}
\begin{itemize}
 
 \item[$\bullet$] the correlation functions in free fermion models are described by special functions belonging to the Painlev\'{e} transcendent class; \vspace{1mm}
\item[$\bullet$] one can fully describe the limiting regimes of these objects, this in a quite efficient way. \vspace{1mm}
 
\end{itemize}

To conclude, one can say that the understanding of many aspects of correlation function in quantum integrable models at their free fermion points is, as of today,
quite satisfactory. Still despite the existence of the powerful machinery of Riemann--Hilbert problems, there are still quite a few open problems remaining
such as a detailed characterisation of general two-point functions in the two-dimensional Ising model and of their Fourier transforms and 
a more effective description of the zero field susceptibility, in particular, a better understanding of the algebraic structures at the root of its form factor expansion. 
Some progress has been recently achieved by Forrester and Witte \cite{ForresterWitteFFexpansionDiagDiag2DIsingAsFDetsOfIntegrableIntOps} who related the form factor expansion
of the diagonal susceptibility to a series of Fredholm determinants of integrable integral operators. 

\subsection{Correlation functions in interacting quantum integrable models}

\subsubsection{The space of states}

The computation of correlation functions in interacting quantum integrable models, that is to say those that are \textit{not} equivalent to free fermions was, and still is, a much more challenging task. 
In fact, prior to addressing the \textit{per se} calculation of the correlators, it was first necessary
to obtain a deeper characterisation of the Hilbert space of such models, namely manageable expressions for
\begin{itemize}
\item[$\bullet$] the norms $\braket{  \{\la_a\}_1^N }{  \{\la_a\}_1^N  } $  of on-shell Bethe vectors, namely in the case when $\{\la_a\}_1^N$ solve the Bethe Ansatz equations of the model; \vspace{1mm}
\item[$\bullet$] more generally the scalar products  $\braket{  \{\la_a\}_1^N }{  \{\mu_a\}_1^N  } $  between an off-shell  Bethe vector $\ket{ \{\mu_a\}_1^N  } $ and an on-shell Bethe vector 
$\ket{  \{\la_a\}_1^N } $, \textit{i}.\textit{e}. when the parameters $\{\la_a\}_1^N$ solve the model's Bethe Ansatz equations while the parameters $\{\mu_a\}_1^N$ are kept generic. \vspace{1mm}
\end{itemize}

The first expression for norms has been obtained  within  the framework of the coordinate Bethe Ansatz by Gaudin \cite{GaudinNormNLSE}. In 1971,
he gave strong arguments in favour of a representation for the norm of an $N$ particle eigenstate of the non-linear Schr\"{o}dinger model which was given in terms of ratios of determinants of 
$N\times N$ matrices. Later, in 1981, Gaudin, McCoy and Wu \cite{GaudinMcCoyWuNormXXZ} gave convincing arguments so as to justify an analogous-type of formula for the norms 
of the on-shell Bethe vectors of the XXZ spin-$1/2$ chain.
Then, in 1982, Korepin \cite{KorepinNormBetheStates6-Vertex} built on the operator formalism of the algebraic Bethe Ansatz so as to establish a set of recurrence relations satisfied by the norms 
associated with any integrable model underlying to a six-vertex $R$-matrix. He was able to solve these recurrence relations in terms of ratios of size-$N$ determinants, hence generalising the two previous
results. 
In 1987 Kirillov and Smirnov \cite{KirillovSmirnovCalculationOfkappaScalarProductsInTermsof2Nx2NDeterminants} derived a $2N \times 2N$ determinant representation for the scalar products in the non-linear Schr\"{o}dinger model.
A more convenient representation, valid for all integrable models underlying to a six-vertex $R$-matrix,
was found by Slavonv \cite{SlavnovScalarProductsXXZ} in 1989 by means of solving certain functional equations associated with such
scalar products. Slavnov subsequently provided proofs, building solely on algebraic structures, of his formula \cite{SlavnovScalarProducts6VertexProofBasedOnDualFields,SlavnovScalarProducts6VertexAlgebraicProof}. 
Slavnov's representation for the scalar products was given in terms of ratios of $N\times N$ determinants. 
This determinant representation is, as of today, at the root of most calculations of correlation functions in quantum integrable models. 

So far, I have only mentioned the simplest case of quantum integrable models, namely those built over a rank $1$
Lie algebra. As already mentioned, there also exist quantum integrable models built over higher rank Lie algebras.
There exists some results relative to determinant representations for norms
and scalar products in quantum integrable models associated with  rank $2$ Lie algebras.
In 1989, Reshetikhin \cite{ReshetikhinNormsSU(3)BetheStates} derived a block matrix determinant based representation for the norms of on-shell Bethe vectors 
in quantum integrable models enjoying of a $S\!U(3)$ symmetry. It was only very recently, in 2012, that Belliard, Pakuliak, Ragoucy and Slavnov \cite{BelliardPakuliakRagoucySlavnovNestedBAScalarProductsSU(3)}
obtained a determinant representation for the scalar products in this model.  Slavnov \cite{SlavnovScalarProductsForGL(3)TrigRMatrix} then
generalised such a representation to the case of models having the so-called $G\!L(3)$ trigonometric $R$-matrix. 

The existence of determinant representation for norms and scalar products appears to be one of the  general characteristics of integrable models. 
I will demonstrate in the core of the thesis that such determinant representations are particularly well suited for the calculation of 
form factors, and more generally, for constructing various types of representations for the correlation functions, this both at finite $L$ and in the 
thermodynamic limit. Finally, they also constitute a good starting point to the calculation of the large-distance asymptotic behaviour of two and multi-point 
correlation functions.

\subsubsection{The early approach}

The first activity relative to obtaining expressions for correlation functions in interacting quantum integrable models arose soon after the appearance
of the algebraic Bethe Ansatz. Indeed, the algebraic construction of the Bethe vectors allowed for a tremendous simplification of the underlying combinatorial
structure of the Bethe vectors. This reflected itself in the very possibility to actually prove the norm formula \cite{KorepinNormBetheStates6-Vertex}. 
Building on this determinant representation and on a crafty use of the algebraic structure provided by the algebraic Bethe Ansatz, Izergin and Korepin 
proposed in the mid 80's first representations for the spin-spin correlation functions of the XXZ chain in a magnetic field. 
Their first result consisted in a combinatorial expression for the correlator of the finite-volume $L$ chain \cite{IzerginKorepinQISMApproachToCorrFns2SiteModel}. 
Korepin subsequently obtained a series of multiple integral representation for the time and space dependent density-density correlation function of the non-linear Schrödinger model \cite{KorepinSeriesofMIForCurrentCurrentInNLSM},
this already after taking the model's thermodynamic limit. This constituted an important progress in that it wasn't even clear how to reorganise the representation so that the thermodynamic limit
can be taken, at least on a formal level, \textit{viz}. without discussing any issues related with the convergence of the resulting series.
A year later, Izergin and Korepin \cite{IzerginKorepinQISMApproachToCorrNextDiscussion} derived an analogous type of representation for the thermodynamic limit  of the spin-spin 
two-point function in the XXZ chain. The structure of the answer was quite complicated though. The integrands of the $n^{\e{th}}$ summand of these series were only defined 
implicitly: they were expressed in terms of certain combinatorial sums whose building blocks were not only defined recursively but also involved solutions 
to auxiliary non-linear integral equations. Although constituting an huge step forward, these first results did not allow to obtain any reasonable representation for the correlators. 
It was not clear how one could recast the combinatorial sums into a compact manageable form. 
A way of bypassing the combinatorics was proposed in 1987 by Korepin \cite{KorepinFirstIntroDualFields}, the so-called dual field approach. His method built on the introduction of auxiliary quantum fields 
which allowed him to recast the aforementioned multiple combinatorial sums in terms of expectation values in respect to an auxiliary vacuum (in the space where the dual fields act) of a Fredholm determinant 
of an integrable integral operator. The dual field valued integral kernel of the operator was depending parametrically on the physical parameters describing the correlator: the temperature $T$, 
the chemical potential $h$ as well as the time $t$ and distance $x$. 
The handling of such expressions remained nonetheless quite complex
in that the operator's kernel took values in an infinite dimensional space of unbounded operators. Dual field-based representations for other
correlators were subsequently obtained in the works \cite{KojimaKorepinSlavnovNLSEDeterminatFormFactorAndDualFieldTempeAndTime,KorepinSlavnovApplicationDualFieldsFredDets}. 
Although the large-$x,t$ asymptotic behaviour of the operator valued Fredholm determinants could, in principle, 
 be analysed -on a cavalierly formal level of rigour- \textit{via} Riemann--Hilbert problems as observed in \cite{ItsIzerginKorepinSlavnovDifferentialeqnsforCorrelationfunctions}, 
one meets serious problems, even on a formal level of rigour,  with applying these results to the characterisation of the large-$x,t$ asymptotic behaviour of the correlation functions. 
The main problem was that the formal asymptotic expansion obtained for the dual field valued Fredholm determinant did not commute with the dual vacuum expectation value. 
Namely, computing the dual vacuum expectation value of apparently subdominant terms of the formal asymptotic expansion of the dual field valued Fredholm determinant produces dominant terms!
Despite these problems, building on several ad-hoc hypothesis,  Its and Slavnov \cite{ItsSlavnovNLSTimeAndSpaceCorrDualFields} derived the
large-parameter asymptotics of the dual field valued Fredholm determinant that represents, after evaluating its dual field vacuum average, the space and time dependent field conjugated field 
two-point function in the non-linear Schr\"{o}dinger model at finite temperature. In a subsequent work, Slavnov \cite{SlavnovComputationDualFieldVaccumExpLongTimeDistTempeRedDensNLSE}
developed an efficient framework that allowed him to evaluate the dual field vacuum expectation value of the leading asymptotics that were obtained
in the aforecited paper. However, due to the mixing of orders problem mentioned earlier, Slavnov could only argue, in this way, the leading asymptotic behaviour of the correlator's
logarithm. Still, despite all problems, it is a impressive piece of work.

\subsubsection{The vertex operator and algebraic Bethe Ansatz approaches}

The first truly explicit and well defined representations for the correlation functions in quantum lattice integrable models away from the free fermion point
should be attributed to the Kyoto school.  In 1993, Davis, Foda, Jimbo, Miwa and Nakayashiki \cite{DavisFodaJimboMiwaNakayashikiDiagonalizationXXZinfiniteDelta>1} developed a 
vertex operator based framework so as to construct the eigenstates and characterise the spectrum of the infinite XXZ chain in the massive regime $\De >1$.  
Their construction built on the assumed presence of a $U_{q}\big( \wh{\mf{sl}_2} \big)$ symmetry in the infinite XXZ chain Hamiltonian\symbolfootnote[2]{This symmetry does hold for certain instances of specific boundary conditions
for the chain, see \cite{KulishSklyaninUqSL2InvariantXXZChain}.}. 
Jimbo, Miki, Miwa and Nakayashiki \cite{JimboMikiMiwaNakayashikiElementaryBlocksXXZperiodicDelta>1} conformed this formalism so as to obtain
explicit $m$-fold multiple integral representations for the so-called elementary blocks of length $m$ in the  $\De>1$ zero temperature and zero magnetic field XXZ chain. 
I remind that the elementary blocks correspond to expectation values in respect to the model's ground state
\beq
\Big<  E^{\eps_1 \eps^{\prime}_1}_1 \cdots E^{\eps_m \eps^{\prime}_m}_m \Big> 
\nonumber
\enq
where $E^{\eps \eps^{\prime}}$ is the $2\times 2$ elementary matrix with zero entry everywhere except at the intersection of line $\eps$ and row $\eps^{\prime}$
where it contains $1$. The elementary blocks represent the matrix entries of the model's zero temperature $m$-site reduced density matrix.
As such, they provide one with a basis allowing one to decompose any correlation function of the model.
The result of the Kyoto school was a very important progress in respect to obtaining closed expressions for the correlation functions. For instance, the multiple integral representation
for the elementary blocks allowed to obtain the value of the correlators at neighbouring sites and also to reproduce
the staggered magnetisation obtained previously in 1973 by Baxter \cite{BaxterStaggeredPolarisationInFModel}. 
 Idzumi, Iohara, Jimbo, Miwa, Nakashima and Tokihiro \cite{IdzumiIoharaJimboMiwaNakashimaTokihiroQ-KZInXXZVertexModels} discovered that the  $U_{\e{q}}(\wh{\mf{sl}}_2)$ symmetry of the chain
implies that the multiple integral representations for the  elementary blocks satisfy to the quantum Kniznik-Zalmolodchikov equations \cite{FrenkelResetikinIntroductionOfqKZEquations}.
 In 1994, Jimbo, Kedem, Kojima, Konno, and Miwa \cite{JimboKedemKonnoMiwaXXZChainWithaBoundaryElemBlcks} applied the vertex operator formalism
 so as to obtain multiple integral representation for the elementary blocks of the half-infinite XXZ chain in the massive regime and subject to a longitudinal magnetic field at the boundary.
The vertex operator formalism was also shown to be effective in respect to computing the form factors of local operators in the massive regime 
of the infinite XXZ chain. Multiple integral representations for these objects can be found in the 1995 book of Jimbo and Miwa \cite{JimboMiwaFormFactorsInMassiveXXZ}. 
It is also worth mentioning that the method can be adapted so as to deal with the massless regime of the chain.  Indeed, in  1996, Jimbo and Miwa \cite{JimboMiwaElementaryBlocksXXZperiodicMassless} 
managed to provide solutions to the quantum Kniznik-Zalmolodchikov equations at $\abs{q}=1$, leading to a multiple integral representation for the elementary blocks of the
XXZ chain in the massless regime.

The setting up of an effective, quite systematic, algebraic Bethe Ansatz based approach to the computation of correlation functions in quantum integrable models
has been made possible by the resolution of the quantum inverse scattering problem by Kitanine, Maillet and Terras \cite{KMTFormfactorsperiodicXXZ} in 1999. 
 Building on the Slavnov formula for the scalar products of Bethe states in the XXZ chain, they provided determinant
representation for the form factors of local operators in the finite volume $L$ XXZ chain, this irrespectively of the value taken by $\De$ or the magnetic field $h$. 
A year later, they \cite{KMTElementaryBlocksPeriodicXXZ} reproduced the results for the elementary blocks obtained by the Kyoto school both in the 
massive and massless regimes of the infinite XXZ chain, further generalising them to the case of a finite magnetic field. 
The algebraic Bethe Ansatz based approach to the characterisation of the elementary blocks was generalised by Kitanine \cite{KitanineXXXSpin1MultIntReps} in 2000 so as to encompass the 
case of higher spin XXX magnets -with an explicit application to the spin 1 case- and to the case of higher spin XXZ chain by Deguchi and Matsui \cite{DeguchiMatsuiElementBlocksHigherSpinXXZ} in 2010, 
provided that the anisotropy $\De$ is close enough to the isotropic limit. In 2007, Damerau, G\"{o}hmann, Hansenclever and Kl\"{u}mper \cite{DamerauGohmannHasencleverKlumperDensityMatrixXXZFiniteLength}
generalised the previous setting for computing the elementary blocks so as to encompass the case of finite size XXZ chains.

\subsubsection{Explicit factorisation of the multiple-integrals}

Although explicit, the $m$-fold multiple integral representations for the elementary blocks could not be properly analysed within the setting of the existing methods. In particular, it was not clear how to extract 
from these the large-distance $x$ asymptotic behaviour of the two-point functions\symbolfootnote[2]{For convenience, I label by $x$ the sites of the chain. However, in this context, $x$ is a discrete variable}
$\big< \sg_1^+ \sg_{x+1}^-\big>$ and $\big< \sg_1^z \sg_{x+1}^z\big>$ so as to test the conformal field theory based predictions for quantum integrable models away from their free fermion point. 
This led to various attempts towards a better understanding of these objects. 
 
 A natural way of dealing with these multiple integral representations for the elementary blocks was to attempt computing them explicitly, hence providing explicit and closed expressions for the correlation
 functions separated by a few sites. In fact, the very first explicit expression for the short-range correlators goes back to the 1938 work of  H\'{u}lten \cite{HultenGSandEnergyForXXX}
who obtained -through a completely different reasoning-  a closed expression for the nearest neighbour spin-spin correlator in the XXX chain, a result that was generalised to the 
next-nearest neighbour case by  Takahashi \cite{TakahashiSpinSpinSecondNeighborXXX} in 1977. 
Still, it was the multiple integral representations for the elementary blocks that allowed for a systematic approach to obtaining closed expressions for short range correlators. In 2002 Boos and Korepin \cite{BoosKorepinEvaluationofEFPforXXX4sites} 
managed to separate the integrals representing the simplest possible elementary block, the so-called emptiness formation probability\symbolfootnote[3]{This correlator corresponds to taking $\eps_a=\eps_{a}^{\prime}=2$
for any $a=1,\dots,m$.} \cite{EsslerIzerginKorepinUglovFirstIntroductionoftheEFP},
this up to the case of four sites (\textit{i}.\textit{e}. $m=4$). Then Boos, Korepin and Smirnov \cite{BoosKorepinSmirnovEFPbyQKZforXXX6sites} built on the machinery of  quantum Kniznik-Zalmolodchikov equations 
so as to extend the separation of the integrals 
for the emptiness formation probability up to the case of six sites. Their method appeared fruitful enough so as to allow Boos, Shiroishi and Takahashi to treat the case of four sites elementary blocks 
 \cite{BoosShiroishiTakahashi4sitesSpinSpinXXX,NishiyamaSakaiShiroishiTakahashi4sitesCorrelatorsXXX}.
Later on, Sato, Shiroishi and Takahashi \cite{SatoShiroishiTakahashiGeneratingFunction8sitesXXX,SatoShiroishiTakahashiElementaryBlocks6sites} were able to push the calculations even further
and provided closed expressions for the spin-spin correlators of the XXX chain up to 7 sites and for generic elementary blocks of the XXZ chain up to 6 sites.

The possibility to separate the multiple integrals by means of the quantum Kniznik-Zalmolodchikov equations 
led Boos, Jimbo, Miwa, Smirnov and Takeyama \cite{BoosJimboMiwaSmironovTakeyamaRecusrionFormulaForXXX} to set an approach, based on  
quantum Kniznik-Zalmolodchikov equations, which allowed them to treat the case of general blocks at any $m$. 
This solution allowed them  \cite{BoosJimboMiwaSmironovTakeyamaReducedqKZEqns} to produce new kinds of representations for the $m$-site reduced density matrix.
Their analysis recast the elementary blocks as an expectation value of the exponent of an operator valued two-fold contour integral \cite{BoosJimboMiwaSmironovTakeyamaCompactFormForOmegaOperatorHomogeneousLimit}.
Their result  proved an earlier conjecture \cite{BoosKorepinSmirnovEFPbyQKZforXXX6sites} that the emptiness formation probability can be expressed 
as a rational multi-variable polynomial in $\ln 2$ and $\zeta(2p+1)$, $p\in \mathbb{N}^{*}$ and $\zeta$ being the Riemann $\zeta$ function. 
It also showed that the length-$m$ density matrix can be expressed as some combinatorial sum only involving products of two-auxiliary functions which are defined, at most, by a double integral. 
In this respect, one recovers one of the characteristics of free fermion equivalent models. The only difference is that the complicated sums do not seem
to be expressible in terms of a single determinant, or any structured object for that matter. Nonetheless, the separability led the aforementioned authors to discover a free-fermionic structure 
in the infinite XXZ chain \cite{BoosJimboMiwaSmironovTakeyamaBasisforFermionicOpsXXZ,BoosJimboMiwaSmironovTakeyamaGrassMannStructureinXXZ}. They were able to
construct the fermionic creation and annihilation operators as well as express the average value in the fermionic vacuum of products of such operators in terms 
of determinants \cite{BoosJimboMiwaSmironovTakeyamaFermionicStructureinXXZCreationOps}. 
 This hidden fermionic structure allowed Jimbo, Miwa and Smirnov to provide a characterisation of the one-point functions in the sine-Gordon model 
both in infinite \cite{JimboMiwaSmirnovOnePointfunctionssineGordon} and finite volume \cite{JimboMiwaSmirnovOnePointfunctionssineGordonFiniteVolume}. 
This approach also allowed Negro and Smirnov \cite{NegroSmirnovOnePtFctsSinhGordonDiffrenceEqns} to provide an analogous characterisation of the 
one-point functions in the sinh-Gordon model in finite volume and write down a system of finite-difference equation satisfied by the latter.  
These last results were checked against other predictions first in the same paper and then by Negro \cite{NegroTBAandFermionicBasis}
through numerical methods. 

 \subsubsection{The large-distance asymptotic expansion of the spin-spin correlation functions}
 
Although fruitful, this was not the path I chose for investigating  the correlation functions in quantum integrable models. 
The approach that I have been developing, since my early days in the field,  takes its roots in the works of the Lyon group-
Kitanine, Maillet, Slavnov and Terras- between 1999 and 2005, the date when I joined the group. 
During the mentioned period, the Lyon group built on the machinery of the algebraic Bethe Ansatz approach so as to construct 
various types of series of multiple integral representations for the $\moy{\sg^z_1\sg^z_{x+1}}$
and $\moy{\sg^-_1\sg^+_{x+1}}$ two-point functions  of the XXZ chain in the massless regime \cite{KMNTPeriodicAlternativeRe-Summation,KMNTOriginalSeries,KMNTMasterEquation}.
They also managed to conform one of their multiple integral representations  to the case of space and time dependent spin-spin two-point functions \cite{KMNTDynamicalCorrelationFunctions}. 
The goal of these constructions was to build a representation that would allow one to extract the large-distance $x$ asymptotic behaviour of these
correlators  and compare it with the CFT/Luttinger liquid based predictions and calculations of the conformal dimensions 
by means of extracting of the $1/L$ corrections to the energies of the low-lying excited states.

I joined the group in 2005 and we continued these efforts. Our work culminated in 2006 \cite{KozKitMailSlaTer6VertexRMatrixMasterEquation} with the construction of an $N$-fold 
multiple integral representation for the so-called generating function $\big<\ex{\a \op{Q}_x } \big>$ of longitudinal correlators  in integrable models associated with a six-vertex $R$ matrix. 
This representation was called the master equation in that it allowed to re-derive many of the previously obtained representations for $\big<\ex{\a \op{Q}_x } \big>$ from this unique object. 
When specialised to a given model, this generating function allowed one to compute diagonal two-point functions: the density-density two-point function in the non-linear Schrödinger model
or the spin-spin two-point function in the  XXZ chain. Building on the master equation, we managed to construct new types of series of multiple integral representations 
for the mentioned correlation functions in the non-linear Schr\"{o}dinger model and the XXZ spin-$1/2$ chain. 
In fact, we succeeded to construct one that had structural similarities with a Fredholm series for a Fredholm determinant. 
In particular, at the free fermion point, the series reduced to the Fredholm determinant $\det\big[ I+\op{V}_x^{(0)} \big]$ of an integrable integral operator
closely related with the pure sine kernel mentioned previously
\beq
V_x^{(0)}(\la,\mu) \; = \; \big(\ex{\a}-1 \big) \cdot \f{ \sin\Big(\frac{x}{2}\big[ p_0(\la)-p_0(\mu) \big] \Big) }{ \pi \sinh(\la-\mu) } 
\qquad \e{with} \quad  p_0(\la) \; = \; -\i \ln \bigg( \f{ \sinh\big( \tf{\i\zeta}{2}-\mu \big) }{  \sinh\big( \tf{\i\zeta}{2}+\mu \big) } \bigg)  
\nonumber
\enq
and $0< \zeta <\pi$ parametrising the anisotropy as $\De = \cos(\zeta)$. 
This structural connection allowed us, at the time, to relate the building blocks of our series -the so-called cyclic integrals- to 
 coefficients  $\Dp{\ga}^n \ln \det\big[ \e{id} +\op{V}_x \big]_{\mid \ga=0}$ present in the Taylor series in $\ga$ expansion of the logarithm $\ln \det\big[ \e{id} +\op{V}_x \big]$ of the Fredholm determinant of the so-called generalised sine kernel. 
 The operator $\op{V}_x$ arising in the determinant is understood to act on $L^2(\intff{-q}{q})$ with the integral kernel
\beq
V_x(\la,\mu) \; = \; \ga F(\la) \cdot \f{ \sin\Big[\frac{x}{2}\big( p_0(\la)-p_0(\mu) \big)-\frac{\i}{2}\big(g(\la)-g(\mu)\big) \Big] }{ \pi \sinh(\la-\mu) } \;. 
\nonumber
\enq
This kernel depends on two functions $F$ and $g$  that are holomorphic in a neighbourhood of $\intff{-q}{q}$. 
We were able to extract the  large-$x$ asymptotic behaviour of $ \det\big[ \e{id} + \op{V}_x \big]$ at $\ga$ small enough 
by carrying out \cite{KozKitMailSlaTerRHPapproachtoSuperSineKernel} a Riemann--Hilbert analysis of the associated $2\times 2$ matrix Riemann--Hilbert problem. This provided us 
with the large $x$ asymptotic behaviour of $\Dp{\ga}^n \ln \det\big[  \e{id} + \op{V}_x \big]_{\mid \ga=0}$. In the same paper, we managed to 
build on this data and  prove the large-$x$ asymptotic behaviour of the cyclic integrals. 
In a subsequent publication \cite{KozKitMailSlaTerXXZsgZsgZAsymptotics}, we inserted these large-$x$ expansions into the series for $\big< \ex{\a \op{Q}_x} \big>$  and managed to re-sum them hence obtaining 
the large-$x$ asymptotic behaviour of the generating function. From there, we were able to access to the first terms present in the large-$x$ asymptotic behaviour of the 
longitudinal two-point  functions, this both for the $XXZ$ spin-$1/2$ chain in the massless regime and for the non-linear Schr\"{o}dinger model this for \textit{all}
values of their coupling constants, \textit{viz}. away from these model's respective free fermion points. 
In the case of the XXZ spin-$1/2$ chain, our result took the form:
\beq
\big< \sg_{1}^z \sg_{x+1}^{z} \big> \; = \; \bigg( 1- \f{ 2  }{ \pi } p(q) \bigg)^2 \; - \; \f{ 2 Z^2(q) }{\pi^2 x^2 } \; + \; 
\big| \mc{A}_{z}\big[ Z \big]  \big|^2 \cdot \f{ 2 \cos\big[ 2x p(q) \big]  }{  x^{ 2 Z^2(q) }   }
\; + \; \dots 
\nonumber
\enq
All the building blocks appearing in the above asymptotic expansion are constructed in terms of two auxiliary functions: the dressed momentum $p(\la)$ and the dressed charge $Z(\la)$. 
The coefficients that are explicitly written down only involve the values taken by these functions at $q$, the Fermi boundary of the model. 
The explicit expression for the  amplitude $| \mc{A}_{z}[ Z]  |^2$ in front of the oscillatorily decaying terms
has not been written down in that it is quite bulky. 
$| \mc{A}_{z} [ Z ]  |^2$ is a functional of the dressed charge whose explicit expression is given in terms of ratios of Fredholm determinants of operators
of the type $\e{id}+ \op{U}$. The integral operators $\op{U}$ act on functions supported on a small counter-clockwise loop surrounding $\intff{-q}{q}$.
Their integral kernels are expressed, in particular, in terms of the function $Z$. On top of the Fredholm determinant dependence of $| \mc{A}_{z} [ Z ]  |^2$,
the amplitude also contains exponents of one and two-fold integral transforms involving $Z$. 
The functions $p$ and $Z$ are defined as the unique solutions to the below linear integral equations
\beq
 \left( \ba{cc} Z(\la) \\ p^{\prime}(\la) \ea \right) \; + \; \Int{-q}{q} \f{ \sin(2\zeta) }{ \sinh(\la-\mu-\i \zeta)\sinh(\la-\mu+\i \zeta) }  \left( \ba{cc} Z(\mu) \\ p^{\prime}(\mu) \ea \right) \cdot \f{ \dd \mu }{  2\pi }
\;= \;  \left( \ba{cc} 1 \\ p^{\prime}_0(\la) \ea \right) \qquad \e{and} 
\quad p(\la) \; = \; \Int{0}{\la} p^{\prime}(\la) \cdot \dd \la \;. 
\nonumber
\enq
I stress that the relation between the parameters (amplitudes and exponents) describing the large-$x$ asymptotics and the original parameters of the model $\zeta$ and $p_0(\la)$ is highly non-trivial;
clearly it goes beyond the type of reparametrisation of the constants arising in the original problem as it was the case of the Painlev\'{e} transcendents where, at most, basic transcendental functions of the original 
parameters describe the asymptotic behaviour. It can thus be seen as a first manifestation of the fact that the correlation functions in interacting quantum integrable models do belong to a new layer of special functions that 
lies above the Painlev\'{e} transcendents class. 

The asymptotics  derived in \cite{KozKitMailSlaTerXXZsgZsgZAsymptotics} were the first results obtained starting from the first principles that confirmed the conformal field theoretic
predictions for the large-distance asymptotic behaviour of two-point functions in massless integrable  models \textit{away} from their free fermion point.

\noindent This naturally raises many new questions. 
\begin{itemize}
\item[$\bullet$] First of all, it is important to stress that the asymptotic expansion for $\big< \sg_{1}^z \sg_{x+1}^{z} \big> $ given above was not obtained through a fully rigorous 
reasoning. Indeed, various handlings such as the existence of limits and the exchanges of limits, summations and integrations  symbols, \textit{etc}. were left unjustified.  It is thus natural to wonder whether 
the method of asymptotic analysis can be set in a rigorous framework? \vspace{1mm}
\item[$\bullet$] Second, the expansion contains only the first few terms. Is it possible to obtain the full structure of the asymptotic expansion of the two-point function, namely all the algebraic orders in $x$? \vspace{1mm}
\item[$\bullet$] Third, the constants (amplitudes and exponents) arising in this expansion are expressed in terms of functions known to characterise the large-volume behaviour of the XXZ chain's
spectrum of excitations above the ground state. The critical exponents correspond precisely to the coefficients arising in the $1/L$ part of the energies of the low-lying excited states exactly 
as predicted by the conformal field theoretic methods. Is it possible to test the conformal field theoretic prediction for the large-volume power-law behaviour in the volume of the finite volume amplitudes? \vspace{1mm}
\item[$\bullet$] Fourth, the very fact that is has been possible to extract the large-$x$ behaviour out of the quite intricate series of multiple integrals representing $\big< \ex{\a \op{Q}_x } \big> $
adjoined to a number of quite intriguing algebraic identities that were used in the course of deriving of the series' large-$x$ asymptotics do seem to indicate that there is 
something special about this series. Could it be that the series of multiple integral representation for $\big< \ex{\a \op{Q}_x } \big> $ constitutes an example of a new class of special function lying one layer above the 
Painlev\'{e} transcendent class? If yes, is it possible to characterise this class slightly better and on a more systematic ground? \vspace{1mm}
\item[$\bullet$] Fifth, the method of asymptotic analysis worked for a specific example of a correlation function. Was this an accident? Namely, can the method be adapted to other settings, such as the correlation functions
of other operators or to the time-dependent case or even to models at finite temperature? \vspace{1mm}
\item[$\bullet$] Sixth, although successful, the method of asymptotic analysis of \cite{KozKitMailSlaTerXXZsgZsgZAsymptotics} was, on the one hand, quite involved and time consuming while, on the other hand, indirect in that 
building on representations, structures and objects that are proper to quantum integrable models but quite far away from the objects and quantities usually employed  in physics. 
Is it possible to simplify the method, even if that would mean relaxing its rigour? Is it possible to derive the large-distance asymptotic behaviour of correlation function
solely by building on objects that are "natural" from the perspective of theoretic physics, such as the form factor series? \vspace{1mm}
\end{itemize}

I devoted an important part of my research activity over the last six years to those question. 
I managed, sometimes alone and sometimes with the help of my collaborators, to provide a positive answer to all of them. 
Describing how and to which extent it is possible to answer positively to these questions will occupy a good part of this habilitation thesis,
namely Chapters \ref{Chapitre AB grand volume FF dans modeles integrables} to \ref{Chapitre approche des FF aux asymptotiques des correlateurs}. 
More precisely, the third point will be discussed in Chapter \ref{Chapitre AB grand volume FF dans modeles integrables}, the first, second and fifth points
will be discussed in Chapter \ref{Chapitre AB grand tps et distance via series de Natte}, the fourth point will be discussed throughout Chapters 
\ref{Chapitre AB grand tps et distance via series de Natte}-\ref{Chapitre Solving some c-shifted RHP}. Finally, the sixth point
will be discussed in Chapter \ref{Chapitre approche des FF aux asymptotiques des correlateurs}.

\subsection{Correlation functions at finite temperature}

The investigation of non-trivial properties of the correlation functions of the XXZ spin-$1/2$ chain away from its free fermion point 
and at finite temperature was pioneered by Fabricius and McCoy \cite{FabriciusMcCoyClassicalQuantumCrossOverInTempeXXZ} in 1999 through exact diagonalisation techniques. 
These authors observed that for anisotropies $-1<\De<0$ "close" to the ferromagnetic point, the correlation functions exhibit a quantum/classical crossover 
in the sense that at low temperatures the spin-spin correlation functions are negative while, at sufficiently large temperature, they become positive.  
They interpreted this change of sign as issuing from a competition between quantum mechanical "kinetic" terms (the spin coupling in the transverse direction)
and the potential energy terms (the spin couplings in the longitudinal direction). This transition was further explored on the basis of 
a numerical diagonalisation of the model's quantum transfer matrix by Fabricius, Kl\"{u}mper and McCoy \cite{FabriciusKlumperMcCoyTemperatureDrivenSpaceOscillationsNumerics}
where the effect was attributed to a merging, at some temperature $T_0$, of real  eigenvalues of the quantum transfer matrix 
giving rise to the largest correlation length followed by a creation, for $T>T_0$, of complex conjugated pairs, hence inducing an
oscillatory behaviour in the distance, on top of an exponential decay. Some of these properties when then investigated by means of the non-linear integral
equation describing these eigenvalues \cite{FabriciusKlumperMcCoyTemperatureDrivenSpaceOscillationsStudyOfNLIE}. 
Generalisations of this study to other regimes of the parameters of the XXZ chain or to other models have been carried out in the works
\cite{KlumperMartinezScheerenShiroishiXXZforPositiveDeltaCrossoverAtTNicePictureCondRootsAtFermiPts,KlumperPatuCrossoverinCorrLengthsLiebLiniger,KlumperSikerCrossoverinCorrLengthsTJModelStudyI,KlumperSikerCrossoverinCorrLengthsTJModelMoreExtensiveStudyII}

In 2004 G\"{o}hmann, Kl\"{u}mper and Seel \cite{GohmannKlumperSeelFinieTemperatureCorrelationFunctionsXXZ} managed to conform the quantum transfer matrix approach to the computation of the thermal 
average of the generating function $\moy{ \ex{\a \mc{Q}_x} }_{T}$ of spin-spin correlation functions in the XXZ chain at finite temperature. 
Subsequently, they \cite{GohmannKlumperSeelElementaryBlocksFiniteTXXZ} obtained multiple integral representations for the elementary blocks at finite temperature. 
In 2006, Boos, G\"{o}hmann, Kl\"{u}mper and Suzuki \cite{BoosGohmannKlumperSuzukiFactorisationDensityMatrixXXZFiniteT} conjectured a factorised form 
for the elementary blocks at finite temperature which was in the spirit of the results of \cite{BoosJimboMiwaSmironovTakeyamaCompactFormForOmegaOperatorHomogeneousLimit}. 
The effectiveness of this representation relatively to computing short distance expression -up to four sites- for the two-point functions of the XXZ chain at finite temperature 
was then demonstrated by Boos, Damerau, G\"{o}hmann, Kl\"{u}mper, Suzuki and Wei$\be$e \cite{BoosDamerauGohmannKlumperSuzukiWeisseFactorised2PtFctXXZFiniteT4sites}.
Also, the quantum transfer matrix based approach was shown to be applicable to the case of the finite temperature spin 1 XXX chain by G\"{o}hmann, Seel and Suzuki \cite{GohmannSeelSuzukiSpin1XXXFiniteTElementaryBlocks}. 

These works were focused mainly on the study of the multiple integral based representation for the elementary blocks or for the generating function $\moy{ \ex{\a \mc{Q}_x} }_{T}$. 
All these representations are such that the  distance dependence of the correlation function is intimately intertwined to the model dependent part. However, as discussed earlier on,
the form factor expansions are known to separate the distance-dependence of, say, a two-point function from its model dependent amplitudes. Since, effectively speaking,
the models become massive at finite temperature, such expansion should provide one with a convenient approach to the large-distance asymptotic behaviour of the correlation functions at finite
temperature. 

\vspace{1mm}

\noindent Hence, one can ask \vspace{1mm}
\begin{itemize}
\item[$\bullet$] it is possible to set a quantum transfer matrix formalism allowing one to provide a form factor based description of the 
correlation functions at finite temperature? \vspace{1mm}
\item[$\bullet$] Should this be possible, can one build on such an answer so as to test the conformal field theory predictions at finite but low temperature? \vspace{1mm}
\item[$\bullet$] Independently of the form factor based approach, is it possible to set -analogously to what has been done for the zero temperature case \cite{KozKitMailSlaTerXXZsgZsgZAsymptotics}-
an approach allowing one to extract the large-distance asymptotic behaviour of two-point functions directly from multiple integral based representations such as 
\cite{GohmannKlumperSeelFinieTemperatureCorrelationFunctionsXXZ}?\vspace{1mm}
\end{itemize}
I will show, in Chapter \ref{Chapitre QIS a tempe finie et cptmt grde distance de leurs correlateurs}, how I managed, in collaboration with 
Dugave and G\"{o}hmann, to answer positively to the first two questions, at least provided that certain reasonable hypotheses are satisfied.
Also, I will briefly comment on the positive answer to the third question in Chapter \ref{Chapitre AB grand tps et distance via series de Natte}, which I 
obtained in collaboration with Maillet and Slavnov.

\subsection{Correlation functions within the framework of the quantum separation of variables}

\subsubsection{The quasi-classical approach}

The study of correlation functions in quantum integrable models solvable by the quantum separation of variables was pioneered by Babelon, Bernard and Smirnov \cite{BabelonBernardSmirnovFFForRestrictedSineBySemiClassicalQuatisationSolitons} in 
1996. The authors developed a procedure allowing one to  quantise  the system of classical separated variables used for solving  the 
classical sine-Gordon model. This enabled them to quantise the classical solitons of the theory and argue the
expressions for certain soliton form factors of the quantum model in infinite volume. The form factor they obtained 
coincided with those obtained by means of the bootstrap approach \cite{KarowskiWeiszFormFactorsFromSymetryAndSMatrices,SmirnovFormFactors}.
Then, in 1998, Smirnov \cite{SmirnovQuasiClassicalFFFiniteVolumeIntegrableCFT} proposed a quasi-classical approach to characterising the form factors of local operators 
of a conformal field theory model with central charge $c<1$ and in finite volume. During the same year, 
building again on a semi-classical approach and the classical separation of variables for the model,
Smirnov \cite{SmirnovMatrixElementsforTodaFromSemiClassics} argued the form taken by the form factors of the $N$-particle Toda chain. 
He conjectured that the non-trivial part of the form factor of local operators $\op{O}$ in this model takes the form of the multiple integral
\beq
\Int{\R^N}{}  \pl{a<b}{N} \Big\{ (y_a-y_b)  \sinh\big[ \frac{ \pi }{ \hbar} (y_a-y_b)\big]   \Big\}
\cdot \pl{a=1}{N} \Big\{ \Big( q_{ \e{Td} }^{ (1) }(y_a) \Big)^*\cdot   q_{ \e{Td} }^{ (2) }(y_a)   \Big\} \cdot F_{\op{O}} (y_1,\dots,y_N) \cdot  \dd^N y \;. 
\enq
In such a representation, $ q_{ \e{Td} }^{ (1) },  q_{ \e{Td} }^{ (2) }$ represent the two solutions to the scalar $\op{t}-\op{Q}$ equation for the Toda chain 
which represent the eigenfunctions, in the separated variables, of the two states involved in the form factor. The Vandermonde/hyperbolic Vandermonde
two-body interaction correspond precisely to the density of the measure on the space where the separation of variables occur for the Toda chain, \textit{c.f.} \eqref{Ecriture mesure Sklyanin Partie Intro}. 
Finally, $F_{\op{O}}$ is some symmetric function in the $y_a$'s that is characteristic of the operator $\op{O}$ involved in the form factor. However, Smirnov did not discuss concrete examples 
for which $F_{\op{O}}$ could have been made explicit.

\subsubsection{The inverse problem approach}

The bottom line of the discussion carried out in Subsection \ref{SousSection Intro qSoV} is that 
the quantum separation of variables provides an utterly simple description of the model's eigenfunctions when these are represented on $\mf{h}_{\e{sep}}$, the
space where quantum separation of variables is realised. Therefore, when trying to characterise, \textit{ab initio}, the correlation functions of a model solvable by 
the quantum separation of variables method,  it seems reasonable to attempt bringing the calculations to the sole handling of objects living in $\mf{h}_{\e{sep}}$. 
This requires to find how the local operators of a given model are realised as operators on the space $\mf{h}_{\e{sep}}$. 
Technically speaking, this requires to determine the form taken by the adjoint action of the separation of variables transform on the local operators of the model of interest. 
Such formulae were provided for the first time, for the quantum Toda chain,
by Babelon  \cite{BabelonQuantumInverseProblemConjClosedToda} in 2004. In the mentioned paper Babelon represented
a subclass of local observables associated with the \textit{classical} Toda chain in terms of the classical separated variables of the model. 
By implementing a canonical quantisation of the classical separated variables, Babelon proposed a realisation 
of certain local operators of the closed quantum Toda chain as operators on $L^{2}_{ \e{sym}\times - }\big( \R^N \times \R ,  \dd\mu(\bs{y}_N) \otimes \dd \veps \big)$,
the space where the quantum separation of variables takes place for the quantum Toda chain\symbolfootnote[2]{The index $\e{sym}\times -$  refers to symmetric functions in the 
first $N$ variables}. 
He performed various consistency tests for his reconstruction. 
In a subsequent paper, Babelon proved \cite{BabelonActionPositionOpsWhittakerFctions} 
one set of his formulae by computing the action of the local operators  of interest on the Whittaker functions. More precisely, he obtained a set of equations in dual variables which take the form: 
\beq
\op{O} \cdot \Psi_{\bs{y}_{N};\veps} (\bs{x}_{N+1}) \; = \; \wh{\op{O}} \cdot  \Psi_{\bs{y}_{N};\veps} (\bs{x}_{N+1}) \;. 
\enq
The operator  $\op{O} $ appearing on the \textit{rhs} of this equation represents a certain class of operators acting on the variables  $\bs{x}_{N+1}$ attached to the original space $\mf{h}_{\e{Td}}$. 
The operator  $\wh{\op{O}}$ appearing on the  \textit{lhs} is the dual operator to  $\op{O}$. It acts on the variables attached to the space where the separation of variables occurs, \textit{viz}.
on the dual variables $\bs{y}_N$ and $ \veps$. 
Since $\Psi_{\bs{y}_{N};\veps} (\bs{x}_{N+1})$ corresponds to the integral kernel of the separation of variables transform, such equations
are enough so as to solve the inverse problem. 
Recently, Sklyanin \cite{SklyaninResolutionIPFromQDet} managed to reproduce Babelon's formulae through simple algebraic 
arguments based on the quantum inverse scattering approach to the quantum Toda chain. \vspace{1mm}

Already, at this point, one can ask:\vspace{1mm}
\begin{itemize}
\item[$\bullet$]  it is possible to systematise and generalise the resolution of the inverse problem for the Toda chain to more general operators that those  considered so far? \vspace{1mm}
\end{itemize}
I will answer positively to this question in Chapter \ref{Chapitre qSoV pour Toda}. 

\vspace{2mm}

Some further investigation of the form factors in models solvable by the quantum separation of variables has been carried out recently by 
Grosjean, Maillet and  Niccoli \cite{GrosjeanMailletNiccoliFFofLatticeSineG} and then by Niccoli  \cite{NiccoliCompleteSpectrumAndSomeFormFactorsAntiPeriodicXXZ,NiccoliCompleteSpectrumAndSomeFormFactorsXXXHigherSpin}. 
The mentioned papers all deal with inhomogeneous "spin-chain" lattice models where all the local Hilbert spaces attached to the sites of the model are finite dimensional. 
In this case, the quantum separated variables are discrete and "live" on some finite-set constructed in terms of the inhomogeneity parameters. 
The method heavily relies on the genericity of these inhomogeneity parameters. In particular, the homogeneous limit which corresponds to the 
physically pertinent model is very singular. Some progress in this direction has been achieved in \cite{KitanineMailletNiccoliRewritingScalarProductsXXXSoVIntoSlavnov}. 
In order to compute the form factors, the work \cite{GrosjeanMailletNiccoliFFofLatticeSineG}  built on the solution to the inverse 
problem by the method of direct/reverse projectors introduced by Oota \cite{OotaInverseProblemForFieldTheoriesIntegrability}
while the other two \cite{NiccoliCompleteSpectrumAndSomeFormFactorsAntiPeriodicXXZ,NiccoliCompleteSpectrumAndSomeFormFactorsXXXHigherSpin} 
used the resolution of the inverse problem introduced in \cite{KMTFormfactorsperiodicXXZ}  and further developed in \cite{MailletTerrasGeneralsolutionInverseProblem}. 
All these papers provide finite-size determinant representations for the form factors of a subclass of local operators in the relevant models they study. 
These determinants depend explicitly on the inhomogeneity parameters and have a quite singular homogeneous limit. 
Although this has yet not been written down anywhere,  it is relatively easy by following the ideas of the work \cite{KazamaKomatsuNishimuraSoVLikeMIRepForPartFctRat6Vertex}
to recast such finite-dimensional determinants as $N$-fold integrals of the type
\beq
\mf{z}_N[W_N \mid F ] \; = \; \Int{\msc{C}^N}{} \pl{a<b}{N} \Big\{ \sinh\big[ \f{\pi}{\om_1} (y_a-y_b) \big]  \sinh \big[ \f{\pi}{\om_2} (y_a-y_b) \big] \Big\} 
\cdot \pl{a=1}{N}\Big\{ \ex{-W_N(y_a)} \Big\} \cdot F_{\op{O}}(y_1,\dots,y_N) \cdot  \dd^N y \;. 
\label{ecriture Intro Fct Partition qSoV}
\enq
Above, the parameters $\om_1, \om_2$ are related to the coupling constants of a model. The confining potential $\ex{-W_N} \, = \,  \big( q^{(2)} \big)^* q^{(1)} $ is given by the product of the two 
solutions to the scalar $\op{t}-\op{Q}$ equation of the model which parametrise the two eigenstates that are involved in the form factor of interest. 
The integration contour $\msc{C}$ is strongly model dependent and, for the three cases mentioned above, corresponds to some loop in the complex plane which surrounds 
some of the poles of the potential $\ex{-W_N}$. These poles of $\ex{-W_N}$ are such that their order grows linearly with $N$. 
Finally, the function $F_{\op{O}}$ represents the input of the operator. Its explicit expression may or may not depend on $q^{(1)}$, depending on the model and operator considered.  
The above representation has already a well defined and, above all, easy to take homogeneous limit.

The bottom line of this discussion is that the form factors of quantum integrable models solvable by the quantum separation of variables admit multiple integral representations of the type 
\eqref{ecriture Intro Fct Partition qSoV} or limits thereof (\textit{c.f.} the Toda chain). 
In some cases such as the Toda chain, or the lattice discretisation
of the sinh-Gordon model \cite{BytskoTeschnerSinhGordonFunctionalBA,LukyanovConjectureOfFieldExpValueSinhGAndRenormalization}, the contour $\msc{C}$ simply coincides with $\R$. In other cases, and in particular for models solvable through a discrete version of the quantum separation of variables,
$\msc{C}$ is some loop in $\Cx$. The confining potential is related to the solutions of the $\op{t}-\op{Q}$ equations in the model while $ F_{\op{O}}$ 
is related to the specific operator being considered. In all cases, it is the large-$N$ asymptotic behaviour of these integrals that is of main interest from the point of view of
applications to physics. In the Toda chain case, this limit allows one for a characterisation of the properties of a system of infinitely many quantum particles in interactions.  
It is also important to stress that, at least in what concerns the known explicit examples,
the operator input $F_{\op{O}}$ to the integral represents a "small" perturbation of the $N$-fold integral $\mf{z}_N[W_N \mid F ]$. By this I mean that its presence will not affect the
mechanism which generates the leading large-$N$ asymptotic behaviour of $\ln \mf{z}_N[W_N \mid F ]$ . Since this very mechanism 
drives all the sub-leading terms in the asymptotic, it seems natural to develop methods of asymptotic analysis first for the unperturbed partition functions 
$\ln \mf{z}_N[W_N \mid 1 ]$ and then generalise the techniques to the more complex but physically pertinent cases.

When $\msc{C}=\R$, the $N$-fold integral $\mf{z}_N[W_N \mid 1 ]$  bares a structural similarity with integrals over the spectra of Hermitian random matrices or, more generally,
$\be$-ensemble integrals. Such ensembles, in the case of so-called varying interactions, are given by the multiple integral 
\beq
\mc{Z}_N^{(\be)}\; = \; \Int{ \R^N }{} \pl{a<b}{N} |\la_a-\la_b|^{\be} \cdot \pl{a=1}{N} \Big\{ \ex{-N V(\la_a)} \Big\}  \cdot \dd^N \la  \;. 
\label{int1}
\enq
$\be>0$ is a positive parameter and $V$ a confining potential growing sufficiently fast at infinity for the integral \eqref{int1} to be convergent. 
It was shown by Dimitriu and Edelman \cite{DumitriuEdelmanConstructionTriDiagRMInterpretationForBetaEnsemblesPartFct} that, for general $\beta > 0$ and quadratic $V$, the partition function \eqref{int1}
corresponds to an integration  over the spectrum of a well-tailored family of random tri-diagonal matrices. The result was generalised to 
polynomial $V$ by Krishnapur, Rider and Vir\'{a}g in \cite{KrishnapurRiderViragProofOfBkAndEdgeUniversalityByOpMethods}. 
Also, for general $V$, $\mc{Z}_N^{(\be)}$ can be interpreted as a result of integration over the spectrum of random matrices which are 
drawn from the so-called orthogonal ($\be=1$), unitary ($\be=2$) or symplectic ($\be=4$) ensembles.

$\be$-ensemble integrals, and in particular numerous properties related with their large-$N$ behaviour, have been studied extensively in the literature over the last 40 years,
see \textit{e.g.} the books \cite{AndersonGuionnetZeitouniIntroRandomMatrices,DeiftGoievRandomMatricesUniversalityInDifferentEnsembles,MehtaRandomMatrices,PasturScherbinaEigenvalueDirstRM}.
A good deal of attention has been given to the study of  universality in the local behaviour of the integration variables  in the large-$N$ limit; this behaviour  
should only depend on $\beta$ and on the local environment of the chosen integration variable on $\mathbb{R}$. 
First results relative to the local universality in the bulk where obtained by Shcherbina and Pastur \cite{PasturShcherbinaFirstProofOfBulkUniversalityHermitianRMT}
at $\beta=2$. Such results where then extended to $\be=1,2,4$ within Riemann--Hilbert based techniques, both for the bulk \cite{DeiftGioevProodUniversalityBulkBeta14RHPApproach,DeiftKriechMcLaughVenakZhouOrthogonalPlyVaryingExponWeights}
and the edge universalities \cite{DeiftGioevProodUniversalityEdgeBeta124RHPApproach}. 
Finally, the bulk and edge universality for general $\beta > 0$ was recently established by Bourgade, Erd\"{o}s and Yau
\cite{BourgadeErdosYauUniversalityBulkBetaEnsGnlPot,BourgadeErdosYauUniversalityEdgeBetaEnsGnlPot,BourgadeErdosYauUniversalityBulkBetaEnsConvPot}.

The leading asymptotic behaviour of the partition function $\mc{Z}_{N}^{(\be)}$ takes the form :
\beq
\ln \mc{Z}^{(\be)}_N \, = \; -N^2\Big( \mc{E}^{(\be)}[\mu_{\e{eq}}] \, + \, \e{o}(1) \Big)
\quad \e{with} \quad \mc{E}^{(\be)}[\mu] \; = \; \Int{}{} V(x)\,\dd\mu(x)  \,- \,  \be\Int{x < y}{} \ln|x-y|\,\dd\mu(x) \dd\mu(y) \;. 
\label{ecriture large N asymptotics beta ensembles}
\enq
The leading term is given by the value of the functional $\mc{E}^{(\be)}$ taken at the equilibrium measure $\mu_{\e{eq}}$. 
The latter corresponds to the unique minimised of  $\mc{E}^{(\be)}$ on the space of probability measures on $\R$. 
The properties of the equilibrium measure $\mu_{\e{eq}}$ have been extensively studied 
\cite{DeiftKriechMcLaughEquiMeasureForLogPot,LandkofFoundModPotTheory,SaffTotikLogarithmicPotential}. 
For sufficiently regular potentials, the equilibrium measure is Lebesgue continuous and supported on a finite union of intervals. In such a case, 
its can be expressed in terms of the solution to 
a \textit{scalar} Riemann--Hilbert problem for a piecewise holomorphic function having jumps on the support of $\mu_{\e{eq}}$. 
Such Riemann--Hilbert problems can be solved explicitly, leading to a \textit{one-fold} integral representation for the density of $\mu_{\e{eq}}$. 
This property was first observed by Carleman \cite{CarlemanRHPSolutionTransfoHankelAxeFini}.

On a heuristic level of rigour, the leading asymptotics of $\ln \mc{Z}_{N}^{(\be)}$ issue from a saddle-point like estimation of the integral \eqref{int1}. This 
statement has been made precise within the framework of large deviations by Ben Arous and Guionnet \cite{BenArousGuionnetLargeDeviationForWignerLawLEadingAsymptOneMatrixIntegral}. 
The calculation of the sub-leading corrections to \eqref{ecriture large N asymptotics beta ensembles} 
is usually based on the use of loop equations. This name refers to a tower of equations which relate multi-point expectation values of test functions 
versus the probability measure induced by $\mc{Z}_N^{(\be)}$.
First calculations of sub-leading terms were carried out in the seminal papers of Ambj\o{}rn, Chekhov and Makeenko \cite{AmbjornChekovMakeenkoFirstEffectiveFormulationLoopEqns}
and of Ambj\o{}rn, Chekhov, Kristjansen  and Makeenko  \cite{AmbjornChekovKristiansenMakeenkoAsymptoticSaddlePointMatrixIntegrals}. 
These papers developed a formal\symbolfootnote[2]{Namely based, among other things, on the assumption of the very existence of the 
asymptotic expansion.} approach allowing one for an order-by-order computation of the large-$N$ asymptotic behaviour of $\mc{Z}_N^{(2)}$.  
However due to its combinatorial intricacy, the approach was quite complicated to set in practice.  
In \cite{EynardTopologicalExpansions1MatrixIntegralsFirstIntro}, Eynard proposed a rewriting of the solutions of loop equations
in a geometrically intrinsic form that strongly simplified the task. Chekhov and Eynard then described the corresponding diagrammatics \cite{CheckovEynardArgumentsAndSomeCalculationsTopoExpGnrlBetaEns}, 
what led to the emergence of the so-called topological recursion. The concept was then fully developed by Eynard and Orantin in
\cite{EynardOrantinTopologicalExpansions2MatrixIntegrals,EynardOrantinTopologicalExpansionsGeneralForm}. The latter allows one,
in its present setting, for a formal yet quite systematic order-by-order calculation of the large-$N$ expansion of $\ln \mc{Z}_N^{(\be)}$.
When the equilibrium measure is supported on a single interval, this expansion takes the form 
\beq
\ln \mc{Z}^{(\be)}_N  \; =  \; \sul{k = 0 }{K} N^{2-k} F_k^{(\be)}[V] + O(N^{-K})
 \label{ecriture DA fct part beta ens une coupure}
\enq
for any $K \geq 0$ and with coefficients being some $\be$-dependent functionals of the potential $V$. 
 When $\be=2$, the existence and form of the expansion up to $\e{o}(1)$ was proven by Johansson \cite{JohanssonEigenvaluesRandomMatricesLeadingAsympt}
for polynomial $V$ such that $\e{supp}(\mu_{\e{eq}})$ is a segment. An important input to Johansson's analysis was the 
\textit{a priori} bounds for the expectation values which were first obtained by Boutet de Monvel, Pastur et Shcherbina
\cite{BoutetdeMonvelPasturShcherbinaAPrioriBoundsOnFluctuationsAroundEqMeas}. 
The existence of the all-order asymptotic expansion at $\be=2$ was proven by Albeverio, Pastur and Shcherbina \cite{AlbeverioPasturShcherbinaProof1overNExpansionForGeneratingFunctionsAtBeta=2}. 
 Finally, Borot and Guionnet  \cite{BorotGuionnetAsymptExpBetaEnsOneCutRegime} systematised and extended to all $\beta > 0$
the approach of \cite{AlbeverioPasturShcherbinaProof1overNExpansionForGeneratingFunctionsAtBeta=2}, hence 
establishing the existence of the all-order large-$N$ asymptotic expansion of $\mc{Z}_N^{(\be)}$ at arbitrary 
$\be$ and for convex real analytic potentials.
When $\mu_{\e{eq}}$ is supported on several segments, the form 
\eqref{ecriture DA fct part beta ens une coupure} of the asymptotic expansion is not valid anymore. In particular, some additional oscillatory terms in $N$ arise. 
For real-analytic off-critical potentials and general $\beta > 0$, the all-order asymptotic expansion was conjectured in \cite{EynardConjectureMultiCutPartFct} and established in \cite{BorotGuionnetAsymptExpBetaEnsMultiCutRegime}. 
Also the asymptotic expansion has been obtained, in the two-cut regime at $\be=2$ by Claeys, Grava and McLaughlin \cite{ClaeysGravaMcLaughlinAERMT2CutsBeta2} by means of Riemann-Hilbert problems. 

So far, I only discussed the case of varying interactions, namely when the confining potential $V$ is preceded by a power of $N$. 
This scaling ensures that, for typical configurations of the $\la_a$'s, the logarithmic repulsion is of the same order of magnitude in $N$ than the confining potential.
The case of non-varying weights (\textit{i.e.} when $NV \hookrightarrow W$ in \eqref{int1}) is, in a sense, closer to the type of potentials that arise in the class of  multiple integrals associated with
the quantum separation of variables method. However, non-varying confining potentials  were much less studied. 
When $W$ is a polynomial, one can restore the varying nature of the potential by a proper rescaling of the integration variables. 
Such an analysis has been carried out in \cite{DeiftKriechMcLaughVenakZhouOrthogonalPlyExponWeights}.  
Still, the polynomial case is by far not representative of the complexity represented by working with non-varying weights. Indeed, the genuinely hard part of the analysis stems form the fact that, in this case, 
it is the large-variable asymptotics of the confining potential which drive the large-$N$ behaviour of the integral. 
Basically, one has to rescale the integration variables $\la_a=T_N y_a$ with $T_N\tend + \infty$ in such a way that the resulting varying potential $V_N(\la)= \tf{W(T_N\la)}{N}$
produces, at large-$N$, a typical contribution of the same order of magnitude in $N$ that the two-body Vandermonde interaction. 
The main problems are then related with the fact that: \vspace{1mm}
\begin{itemize}
 \item[$\bullet$] the rescaled potential may not have a well defined large-$N$ behaviour; \vspace{1mm}
\item[$\bullet$] the non-varying potential $W$ may have singularities in the complex plane. 
Then,  the singularities of the rescaled potential $V_N(\la)= \tf{W(T_N\la)}{N}$ 
  will collapse, with a $N$-dependent rate, on the integration domain. \vspace{1mm}
\end{itemize}
In this situation, the usual scheme for obtaining sub-leading corrections breaks down. 
So far, the large-$N$ asymptotic analysis of a "non-trivial" $\be$-ensemble multiple integral with non-varying interactions 
has been carried out only when $\be=2$ by Bleher and Fokin \cite{BleherFokinAsymptoticsSixVertexFreeEnergyDWBCDisorderedRegime}. 
The most delicate point of their analysis was to absorb 
the contribution of the sequence of poles $\zeta_n/N$, $n=1,2,\dots$, of the rescaled potential that were collapsing on $\R$. \textit{In fine}, they
obtained the asymptotic expansion of the logarithm of the integral up to $\e{o}(1)$ corrections.

 Taken all this into account, one can  raise the questions
\begin{itemize}
\item[$\bullet$] is it possible to generalise the existing results for $\be$-ensembles with varying interaction of the type \eqref{int1} to more complex cases of multiple integrals with varying interactions? 
\item[$\bullet$] is it possible to adapt the techniques that appeared fruitful in the random matrix context so as to carry out the large-$N$ asymptotic analysis of the class of multiple integrals that 
is of interest to the  quantum separation of variables method ?  
\end{itemize}
I will provide a positive  answer to the first question and report on a substantial progress relative to solving the second issue in Chapter \ref{Chapitre AA des integrales multiples}.

\section{List of publications}

This habilitation thesis is based on the works below, listed thematically and in reverse chronological order of appearance. 

\subsubsection{Large-volume behaviour of form factors and related topics}

\begin{itemize}

\item[$\bs{A1}$] M. Dugave, F. G\"{o}hmann, K. K. Kozlowski et J. Suzuki, { \it "On form factor expansions for the XXZ chain in the massive regime."}, J. Stat. Mech. (2015), P05037. \vspace{1mm}

\item[$\bs{A2}$] K. K. Kozlowski, {\it "Low-T asymptotic expansion of the solution to the Yang-Yang equation."},
Lett. Math. Phys., {\bf 104}, 55-74, (2014). \vspace{1mm}

\item[$\bs{A3}$] K. K. Kozlowski et E. K. Sklyanin, { \it "Combinatorics of generalized Bethe equations."}, Lett. Math. Phys. {\bf 103}, 1047-1077 (2013). \vspace{1mm}

\item[$\bs{A4}$] K. K. Kozlowski, {\it "On Form Factors of the conjugated field in the non-linear Schr\"{o}dinger model."}, 
J. Math. Phys. {\bf 52} , 083302 (2011). \vspace{1mm}

\item[$\bs{A5}$] N. Kitanine, K. K. Kozlowski, J.-M. Maillet,  N. A. Slavnov et V. Terras, { \it
"Thermodynamic limit of particle-hole form factors in the massless XXZ Heisenberg chain."}, J. Stat. Mech., P05028, 
(2011). \vspace{1mm}

\item[$\bs{A6}$] N. Kitanine, K. K. Kozlowski, J.-M. Maillet,  N. A. Slavnov et V. Terras, { \it
"On the thermodynamic limit of form factors in the massless XXZ Heisenberg chain."}, J. Math. Phys. {\bf 50}, 095209, (2009). \vspace{1mm}

 \end{itemize}

\subsubsection{Multidimensional deformation flow approach}

\begin{itemize}

\item[$\bs{A7}$] K. K. Kozlowski, {\it "Large-distance and long-time asymptotic behavior of the reduced density matrix in the non-linear Schr\"{o}dinger model."}, Ann. H. Poincar\'e , {\bf  16} , Issue 2,  437-534, (2015).  \vspace{1mm}

\item[$\bs{A8}$] K. K. Kozlowski, {\it "Riemann--Hilbert approach to the time-dependent generalized sine kernel."}, Adv. Theor. Math. Phys. 
{\bf 15}-6, 1655-1743, (2011). \vspace{1mm}

\item[$\bs{A9}$] K. K. Kozlowski et V. Terras {\it "Long-time and large-distance asymptotic behavior of the current-current correlators
 in the non-linear Schr\"{o}dinger model."}, J. Stat. Mech., P09013, (2011). \vspace{1mm}

\item[$\bs{A10}$] K. K. Kozlowski, J.-M. Maillet et  N. A. Slavnov, {\it "Correlation functions of one-dimensional bosons at low temperature."}, 
J. Stat. Mech., P03019, (2011). \vspace{1mm}

\end{itemize}

\subsubsection{Riemann--Hilbert approach to $c$-shifted integrable integral operators}

\begin{itemize}
\item[$\bs{A11}$] A. R. Its et K. K. Kozlowski, { \it "Large-x analysis of an operator valued Riemann--Hilbert problem."},  Int. Math. Res. Not., doi: 10.1093/imrn/rnv188, (2015). \vspace{1mm}
\item[$\bs{A12}$] K. K. Kozlowski, { \it "On lacunary Toeplitz determinants."}, J. Asympt. Analysis {\bf 88}, 1-16, (2014).   \vspace{1mm}
\item[$\bs{A13}$] A.R. Its et K. K. Kozlowski, { \it "On determinants of integrable operators with shifts."}, Int. Math. Res. Not., doi: 10.1093/imrn/rnt191, (2013). \vspace{1mm}
\end{itemize}

\subsubsection{Form factor approach to the critical regime}

\begin{itemize}

\item[$\bs{A14}$] K. K. Kozlowski et J. M. Maillet, {\it "Microscopic approach to a class of 1D quantum critical models."}, math-ph.:1501.07711. \vspace{1mm}

\item[$\bs{A15}$]  N. Kitanine, K. K. Kozlowski, J.-M. Maillet et V. Terras, { \it
"Long-distance asymptotic behaviour of multi-point correlation functions in massless quantum models."}, J. Stat. Mech., P05011, (2014). \vspace{1mm}

\item[$\bs{A16}$]  N. Kitanine, K. K. Kozlowski, J.-M. Maillet,  N. A. Slavnov et V. Terras, { \it
"Form factor approach to dynamical correlation functions in critical models."},
J. Stat. Mech., P09001, (2012). \vspace{1mm}

\item[$\bs{A17}$] N. Kitanine, K. K. Kozlowski, J.-M. Maillet,  N. A. Slavnov et V. Terras, { \it
"A form factor approach to the asymptotic behavior of correlation functions in critical models."},
J. Stat. Mech., P12010, (2011).  \vspace{1mm}

\end{itemize}

\subsubsection{The quantum Toda chain}

\begin{itemize}

\item[$\bs{A18}$] K. K. Kozlowski, { \it "Unitarity of the SoV transform for the Toda chain."}, Comm. Math. Phys. {\bf 334}, 223-273, (2015). \vspace{1mm}

\item[$\bs{A19}$] K. K. Kozlowski, { \it "Aspects of the inverse problem for the Toda chain."}, J. Math. Phys. {\bf 54}, 121902, (2013).  \vspace{1mm}

 \item[$\bs{A20}$] K. K. Kozlowski et J. Teschner, {\it "TBA for the Toda chain."}, Festschrift volume for Tetsuji Miwa, "Infinite 
Analysis 09: New Trends in Quantum Integrable Systems".

\end{itemize}

\subsubsection{Asymptotic analysis of multiple integrals}

\begin{itemize}
 
\item[$\bs{A21}$]  G. Borot, A. Guionnet, K. K. Kozlowski, { \it "Asymptotic expansion of a partition function related to the sinh-model."},  math-ph: 1412.7721. \vspace{1mm}

\item[$\bs{A22}$]  G. Borot, A. Guionnet, K. K. Kozlowski, { \it
"Large-N asymptotic expansion for mean field models with Coulomb gas interaction."}, Int. Math. Res. Not., doi: 10.1093/imrn/rnu260, (2015).

\end{itemize}

\subsubsection{Quantum integrable models at finite temperature}

\begin{itemize}
 
\item[$\bs{A23}$] M. Dugave, F. G\"{o}hmann et K. K. Kozlowski, { \it "Functions characterizing the ground state of the XXZ spin-1/2 chain in the thermodynamic limit."},  SIGMA {\bf 10}, 043, 18 pages, (2014). \vspace{1mm}

\item[$\bs{A24}$] M. Dugave, F. G\"{o}hmann et K. K. Kozlowski, { \it "Low-temperature large-distance asymptotics of the transversal two-point functions of the XXZ chain."}, J. Stat. Mech., P04012, (2014). \vspace{1mm}

\item[$\bs{A25}$] M. Dugave, F. G\"{o}hmann et K. K. Kozlowski, { \it "Thermal form factors of the XXZ chain and the large-distance asymptotics of its
temperature dependent correlation functions."}, J. Stat. Mech., P07010,  (2013).   \vspace{1mm}

\item[$\bs{A26}$] K. K. Kozlowski et B. Pozsgay, { \it "Surface free energy of the open XXZ spin-1/2 chain."}, J. Stat. Mech., P05021, (2012). \vspace{1mm}

\item[$\bs{A27}$] K. K. Kozlowski, J.-M. Maillet et  N. A. Slavnov, {\it "Long-distance behavior of temperature correlation functions
  of the one-dimensional Bose gas."}, J. Stat. Mech., P03018, (2011).

\end{itemize}

\section{Plan of the habilitation thesis}

In Chapter \ref{Chapitre AB grand volume FF dans modeles integrables}, I will describe methods allowing one to extract the 
large-volume asymptotic behaviour of form factors of local operators in quantum integrable models solvable by the algebraic Bethe Ansatz. 
I  shall illustrate the method on the example of the non-linear Schr\"{o}dinger model although many other models can be treated by the method. 
This chapter also contains results relative to making rigorous the passage from the Izergin-Korepin lattice discretisation of the non-linear Schr\"{o}dinger model
towards the Hamiltonian $\op{H}_{NLS}$ introduced earlier on. 
The content of this chapter is based on the works [$\bs{A1}, \bs{A2}, \bs{A3}, \bs{A4}, \bs{A5}, \bs{A6}$].

In Chapter \ref{Chapitre AB grand tps et distance via series de Natte}, I will introduce the concept of multidimensional Fredholm series on the example of 
the series of multiple integrals representing the field-conjugated field time and space dependent two-point function in the non-linear Schr\"{o}dinger model. 
I will explain how certain expansions of Fredholm determinants that can be obtained within the setting of Riemann--Hilbert problems allow one, upon introducing the 
concept of multidimensional deformation flow, to extract the long-distance and large-time asymptotic behaviour of the aforementioned series. 
The approach builds on the hypothesis of convergence of auxiliary series and is rigorous on other aspects.
The content of this chapter is based on the works [$\bs{A7},\bs{A8},\bs{A9},\bs{A10}$]. 

In Chapter \ref{Chapitre Solving some c-shifted RHP}, I will briefly comment on the problems related to extending 
the multi-deformation flow method to the asymptotic analysis of correlators represented by so-called "critical"
multidimensional Fredholm series. I will provide an account of the first steps taken in the direction of obtaining these asymptotics
and shall as well briefly describe the spin-off that these have generated in respect to characterising the 
large-size asymptotic behaviour of so-called lacunary Toeplitz determinants. The content of this chapter is based on the works [$\bs{A11},\bs{A12},\bs{A13}$]. 

In Chapter \ref{Chapitre approche des FF aux asymptotiques des correlateurs}, I will present another method allowing one to extract the 
asymptotic behaviour of correlation functions. This method is based on a direct analysis, in the long-distance and/or large-time regime, of a correlator's 
form factor expansion. It allows one to extract the long-distance and large-time asymptotic behaviour of two and multi-point correlation functions
in massless models as well as to characterise the critical behaviour of the dynamical response functions in the vicinity of the particle-hole 
excitation thresholds. It is, by far, less rigorous than the one described in Chapter \ref{Chapitre AB grand tps et distance via series de Natte} in that it completely
ignores issues of convergence, exchanges of limits and control on remainders. Still, it has the tremendous advantage of remaining very close, at each of its steps, to the objects usually
used in theoretical physics. This method culminated with the construction of a microscopic setting allowing one to argue, starting from first principles,
the appearance of an effective description of the large-distance regime of correlators in massless one-dimensional quantum models -not necessarily integrable- in terms 
of a free boson conformal field theory. The content of this chapter is based on the works [$\bs{A14},\bs{A15},\bs{A16},\bs{A17}$]. 

In Chapter \ref{Chapitre qSoV pour Toda}, I will discuss the progress I made in respect to the various aspects
 of the quantum separation of variables method for the quantum Toda chain. 
First of all,  I will discuss the proof I gave of the Nekrasov-Shatashvili conjecture on the quantisation conditions for the closed Toda chain. This result strongly simplifies the 
description of  the spectrum of the quantum Toda chain, in that it reduces the latter to a resolution of non-linear integral equations.
Then, I shall present a method allowing to prove the unitarity of the separation of variables transform. 
This method builds solely on objects that are natural to the quantum inverse scattering method and thus appears to be 
generalisable to other models. Finally,  I shall discuss some progress I made in respect to solving the quantum inverse problem for the Toda
chain. The content of this chapter is based on the works [$\bs{A18},\bs{A19},\bs{A20}$]. 

In Chapter \ref{Chapitre AA des integrales multiples}, I will report on the progress I made relatively to extracting the 
asymptotic behaviour in respect to the large-number of integrations  $N$ in an $N$-fold multiple integral. 
I shall first discuss the case of certain generalisations of $\be$-ensemble integrals with varying weights and, subsequently, 
describe the large-$N$ analysis of a toy model integral which is at the root of understanding the large-$N$ behaviour of certain classes of quantum separation of variables issued multiple
integrals. The content of this chapter is based on the works [$\bs{A21},\bs{A22}$]. 

Finally, in Chapter \ref{Chapitre QIS a tempe finie et cptmt grde distance de leurs correlateurs}, I will present several results relative to the 
characterisation of the correlation functions in quantum integrable models at finite temperature. 
I shall start by presenting the calculation of the so-called thermal form factors for the local spin operators in the 
XXZ spin-$\tf{1}{2}$ chain. Then, I will briefly comment on the technique that allows one to extract the low-temperature behaviour out of the obtained results.  
I shall conclude this chapter by discussing the computation of the finite-temperature surface free energy of the XXZ chain subject to 
diagonal boundary fields, which constitutes the first step towards characterising finite temperature correlation functions in quantum integrable models
subject to so-called open boundary conditions. The content of this chapter 
is based on the works [$\bs{A2},\bs{A23},\bs{A24},\bs{A25},\bs{A26},\bs{A27}$].

\chapter{Large-volume asymptotic behaviour of form factors in integrable models}
\label{Chapitre AB grand volume FF dans modeles integrables}

The investigation of the large-volume behaviour of form factors of local operators has been made possible thanks to determinant-based representations for 
these objects. The first study has been carried out by Slavnov \cite{SlavnovFormFactorsNLSE}  in 1990 and concerned  
the form factors of the density operator in the non-linear Schr\"{o}dinger model. Slavnov obtained the leading in $L$ asymptotic behaviour 
of the form factors taken between the ground state and an excited stated consisting of $n$ particle-hole excitations. He found that the form factor was decaying like a non-integer 
 power of the volume. 
Slavnov used his results to study the large-volume behaviour of the form factor expansion of the time and space dependent density-density correlation function at zero temperature. He was able to show,
to the first order in perturbation around $c=+\infty$, that the non-integer behaviour in the volume cancels out once that the summation over all the appropriate 
excited states is done. 
Then, in 2006, Arikawa, Kabrach, M\"{u}ller and Wiele \cite{ArikawaKabrachMullerWieleXXAsymptoticsofFFs} calculated the large-volume behaviour of form factors
involving two particle-hole excitations at the free fermion point of the XXZ chain, also observing the presence of a non-integer power-law decay in the volume. 
The presence of such a power-law behaviour in the volume is typical for massless models and is, in fact, the main obstruction to writing down  
form factor expansion for these models directly in the infinite volume limit. This contrast strongly with the case of massive models
where expressions for the form factors could have been obtained directly in the continuum, this as much for certain lattice models \cite{JimboMiwaFormFactorsInMassiveXXZ}
then for numerous cases of massive integrable field theories \cite{KarowskiWeiszFormFactorsFromSymetryAndSMatrices,KirillovSmirnovFirstCompleteSetBootstrapAxiomsForQIFT,SmirnovFormFactors}. 
Still, until my recent work, the only investigation, starting from a finite lattice, of the large-volume behaviour of form factors of local operators in massive models has been carried out in 1999 
by Izergin, Kitanine, Maillet and Terras \cite{IzerginKitMailTerSpontaneousMagnetizationMassiveXXZ}. They were able to extract 
the large-volume asymptotic behaviour of a particular form factor of the $\sg^{z}$ operator in the massive phase of the XXZ chain corresponding to
the staggered magnetisation. In that special case, the finite-volume form factor approaches, exponentially fast in $L$, Baxter's formula \cite{BaxterStaggeredPolarisationInFModel} for the staggered magnetisation.

In the present chapter, I will describe the progress I achieved in the extraction of the large-volume $L$ asymptotic behaviour of the form factors of local 
operators in massless quantum integrable models such as the XXZ chain at anisotropy $-1 \leq \De \leq 1$ and the non-linear Schr\"{o}dinger model. 
This analysis takes in roots in the 1990 work of  Slavnov \cite{SlavnovFormFactorsNLSE}. Among other things, it brings several improvements to Slavnov's method,
renders it systematic and rigorous. In its present state of the art, the method of asymptotic analysis solely builds on the hypothesis of existence of a large-$L$ asymptotic expansion of the 
so-called counting function associated with the model. The existence of an asymptotic expansion for the counting function can be proven
in certain cases (the non-linear Schr\"{o}dinger model and 
the XXZ chain\symbolfootnote[2]{After this thesis has been defended, I proved the existence of the asymptotic expansion for any $\De>-1$ in \cite{KozProofOfDensityOfBetheRoots}.} at $-1<\De\leq 0$) hence raising the obtained large-$L$ behaviour of the form factors to the level of theorem. 
To be more specific, the problem of the large-$L$ analysis of form factors corresponds to extracting the large-$L$ asymptotic behaviour of an $N\times N$ 
determinant $\det_N\big[ M(\la_{a},\la_{b}) \big]$ in the case where $M$ is some specific function of two variables, $N/L\tend cst$ when $L\tend +\infty$, and the 
parameters $\{\la_j\}_1^N$ densify on an interval $\intff{-q}{q}$. I will discuss the method of large-$L$ analysis on the example of the non-linear Schr\"{o}dinger model. 
The analysis of the $XXZ$-chain case can be found in the three articles that I have co-authored [$\bs{A1}, \bs{A5}, \bs{A6}$].  Two [ $\bs{A5}, \bs{A6}$] deal with the massless regime of the chain 
and have been written in collaboration with Kitanine, Maillet, Slavnov and Terras. The most recent one [$\bs{A1}$] deals with the massive regime and has been written 
in collaboration with Dugave, G\"{o}hmann and Suzuki.

This chapter is organised as follows. Section \ref{Section NLSE discret} is devoted to a presentation of a lattice discretisation of the non-linear Schr\"{o}dinger model. 
In Section \ref{Section FF resultat principal}, I will  build on the latter so as to prove [$\bs{A4}$] the determinant representation for the form factors of the 
conjugated field operator in the model.   In Section \ref{Section Thermo limit FF}, I will discuss the large-$L$ behaviour of this form factor. 
Finally, in Section \ref{Section Cpt large L des FF XXZ massif}, I will shortly discuss the structure of the large-volume asymptotic behaviour of form factors of local operators in the massive regime of 
the XXZ chain. I have studied this behaviour in  [$\bs{A1}$]. This short section will permit me to illustrate the fundamental difference between
the form factors in massive and massless models.

\section{The lattice discretisation of the non-linear Schr\"{o}dinger model}

\label{Section NLSE discret}

\subsection{The Lax matrix}

The lattice discretisation of the non-linear Schr\"{o}dinger model acts on the Hilbert space  
\beq
\mf{h}_{LNS} \; = \; \bigotimes\limits_{n=1}^{M} \mf{h}_{LNS}^{n} \qquad  \e{with} \qquad \mf{h}_{LNS}^{n} \simeq L^{2}(\R) \quad \e{and} \; \; M \in 2 \mathbb{Z}\;. 
\enq
With each local space $\mf{h}_{LNS}^{n}$, one associates the Lax matrix found by Izergin and Korepin \cite{IzerginKorepinLatticeVersionsofQFTModelsABANLSEandSineGordon}
\beq
\op{L}_{0 n}(\la) \; = \;  \left( \ba{cc}     - \i \f{\la}{2}\De + Z_n +c \tf{\chi_n^{*}\chi_n}{2} &  - \i\sqrt{c} \chi_n^{*} \cdot \rho_{Z_n}  \\
                           \i \sqrt{c} \rho_{Z_n} \cdot \chi_n  & \i \f{\la}{2}\De + Z_n +c \tf{\chi_n^{*}\chi_n}{2}     \ea \right) , \qquad \e{where}
\quad Z_{n}=1+ (-1)^{n} \f{c\De}{4} \;.                           
\label{ecriture Matrice de Lax}
\enq
$\op{L}_{0 n}(\la)$ is a $2\times 2$ matrix on the auxiliary space $V_0\simeq \Cx^2 $ with operator-valued entries whose joint domain 
corresponds to some dense subspace 
of $\mf{h}^n_{LNS}$. The operators $\chi_n$, $\chi_m^{*}$ correspond to the canonical bosonic creation and annihilation operators satisfying to the commutation relations 
$\big[ \chi_n , \chi_m^{*} \big] \, = \,  \De \de_{n,m}$. The operator $\chi_n^{*}$ is the adjoint of $\chi_n$ and $\rho_{Z_n}=\sqrt{Z_n+c\tf{\chi^{*}_n\chi_n}{4}}$
is expressed in terms of the "number of particles" operator at site $n$. Finally, the parameter $\De$ corresponds to the lattice spacing. It is related to the model's volume $L$
as $L=\De M$. 

The Lax matrix \eqref{ecriture Matrice de Lax} satisfies the Yang-Baxter equation
\beq
R_{00^{\prime}}\pa{\la-\mu} \op{L}_{0n}\pa{\la} \op{L}_{0^{\prime}n}\pa{\mu} 
		=   \op{L}_{0^{\prime}n}\pa{\mu} \op{L}_{0n}\pa{\la}    R_{00^{\prime}}\pa{\la-\mu} \; ,
\label{ecriture YBE}		
\enq
driven by the rational $R$-matrix $R_{00^{\prime}}(\la) \,  =  \, \la -\i c \mc{P}_{00^{\prime}}$, with $\mc{P}_{00^{\prime}}$
being the permutation operator in $V_0\otimes V_{0^{\prime}}$. The $R$ matrix reduces to a one-dimensional projector at $\la = \i c$ which ensures that the Lax matrix $\op{L}_{0n}\pa{\la}$ satisfies
the quantum determinant relation
\beq
\op{L}_{0n}(\la)\sg_{0}^{y} \op{L}_{0n}^{t_0}(\la+\i c) \sg_{0}^{y}  \;  = \;
\f{\De^2}{4} ( \la-\nu_n) (\la-\nu^*_n+\i c) \quad \e{with} \quad \nu_n=-\f{2 \i Z_n}{\De} \; .
\enq
I remind that $z^{*}$ stands for the complex conjugate of $z$. 

Izergin and Korepin \cite{IzerginKorepinLatticeVersionsofQFTModelsABANLSEandSineGordon} observed that the zeroes of the quantum determinant correspond to the values of the spectral parameter
where the Lax matrix is of rank one. In other words, the Lax matrix \eqref{ecriture Matrice de Lax}  reduces to a direct projector at the points $\nu_n, \nu^*_n - \i c$, \textit{e}.\textit{g}.:
\beq
\pac{ \op{L}_{0n}\pa{\nu_n} }_{ab}=\a_a^{\pa{+}}\!\pa{n} \, \be_b^{\pa{+}}\!\pa{n} \; , \quad \e{with} \qquad  
\a^{\pa{+}}\!\pa{n}=\pa{\ba{c} \sqrt{c} \, \chi_n^{*} \\ 2 \i \rho_{Z_n} \ea} \; , \quad
\be^{\pa{+}}\!\pa{n}=\f{1}{2}\pa{\ba{c} \sqrt{c} \, \chi_n \\ -2 \i \rho_{Z_n} \ea} \;.
\label{definition des projecteurs beta plus et alpha plus}
\enq
Furthermore, the Lax matrix becomes a reverse projector  at the points $\nu_n+\i c, \nu^*_n$, \textit{e}.\textit{g}.:
\beq
\pac{ \op{L}_{0n}\pa{\nu_n+ \i c}}_{ab}= \de_b^{\pa{+}}\!\pa{n} \, \ga_a^{\pa{+}}\!\pa{n} \; , \quad \e{with} \qquad   
 \de^{\pa{+}}\!\pa{n} = \f{1}{2}\pa{\ba{c} \sqrt{c} \, \chi_n \\ -2 \i \rho_{Z_n-\tf{\De c}{4}} \ea} \; , \quad
\ga^{\pa{+}}\!\pa{n} = \pa{\ba{c} \sqrt{c} \, \chi_n^{*} \\  2 \i \rho_{Z_n-\tf{\De c}{4}}  \ea} \;.
\label{definition des projecteurs delta plus et gamma plus}
\enq
These properties allow one to build a quasi-local Hamiltonian for the model.  The latter is constructed out of the monodromy matrix: 
\beq
\op{T}_{0;1\dots M}(\la) \, \equiv \,  \op{T}_{0}(\la) \,  = \,  \op{L}_{0M}(\la) \dots \op{L}_{01}(\la) \;  = \;  
\left( \ba{cc} \op{A}(\la) & \op{B}(\la) \\ \op{C}(\la) & \op{D}(\la) \ea \right)
\quad \e{with} \quad \;  M \in 2 \mathbb{Z} \;.
\enq
The monodromy matrix is a $2\times 2$ matrix on the auxiliary space $V_0$
with entries being operators acting on the quantum space $\mf{h}_{LNS}$.
The fact that Lax matrices reduce to direct (at odd sites) or reverse (at even sites) projectors when the spectral parameter is set to 
\beq
\nu  \equiv \nu_{2n-1} = \nu_{2n} + \i c = -\f{2 \i}{\De} + \i \f{c}{2} 
\label{definition nu}
\enq
allows one to construct the quasi-local local Hamiltonian out of the transfer matrix $\op{t}_{LNS}(\la) = \e{tr}_0\pac{\op{T}_0(\la)}$:
\bem
\op{H}_{LNS}\; \equiv  \; \op{t}_{LNS}^{-1}(\nu) \cdot \op{t}_{LNS}^{\prime}(\nu) \;  =  \; \sul{k=1}{M/2} 
\paa{ \,  \big[\be^{\pa{+}}\!\pa{2k+1}\!\big]^{\bs{T}_0} \op{L}_{02k}\!\pa{\nu} \op{L}_{02k-1}\!\pa{\nu} \ga^{\pa{+}}\!\pa{2k-2} }^{-1}  \\
\cdot
\, \big[\be^{\pa{+}}\!\pa{2k+1}\!\big]^{\bs{T}_0} 
\f{\Dp{} }{ \Dp{} \la }\pac{ \op{L}_{02k}\!\pa{\la} \op{L}_{02k-1}\!\pa{\la}}_{\mid\la=\nu } \ga^{\pa{+}}\!\pa{2k-2} \;. 
\nonumber
\end{multline}
Above, $^{\bs{T}_0} $ refers to the operation of transposition of the operator valued vector $\be^{\pa{+}}\!\pa{2k+1}$. 
Note that there exists an alternative way of constructing the local integrals of motion for the lattice non-linear Schr\"{o}dinger model
by means of trace identities at infinity \cite{IzerginKorepinSmirnovTraceIdentitiesForNLSMWithReducedLaxMatrix}. 

According to Izergin and Korepin \cite{IzerginKorepinLatticeVersionsofQFTModelsABANLSEandSineGordon}, in the continuum limit:
\beq
\De \tend 0 \quad \e{with} \quad L=\De M \quad \e{fixed}
\enq
 $\op{H}_{LNS}$  approaches, formally, to the Hamiltonian $\op{H}_{NLS}$ of the non-linear Schr\"{o}dinger model introduced earlier on. 
In this  continuous limit, one can think of the $k^{\e{th}}$ site of the lattice model as contributing to the degrees of freedom attached to the "continuous coordinate" 
$x_k=k\De$. Then, the discrete fields $\chi_n$ are expected to be related to the canonical Bose fields $\Phi\pa{x}$ as
\beq
\chi_n=\Int{ n\De }{\pa{n+1}\De} \!\!  \Phi\pa{x}\;  \dd x \; . 
\enq
However, such an identification can only be given a formal sense in as much as, strictly speaking, the \textit{rhs} 
does not have a precise mathematical meaning. On the other hand, the \textit{lhs} is perfectly well defined since 
the local operators $\chi_n$ and $\chi^{*}_n$ can be constructed explicitly in terms of the creation/annihilation operators
for the one-dimensional harmonic oscillator. 

\subsection{The spectrum and eigenvectors}
\label{SousSectionSpectrumEigenvectors}

The transfer matrix $\la  \mapsto \op{t}_{LNS}(\la)$ is diagonalised by means of standard considerations of the  algebraic
Bethe Ansatz.  One introduces the so-called pseudo-vacuum state $\ket{0}=\ket{0}_1 \otimes \dots \otimes \ket{0}_M $
where $\ket{0}_n$ is uniquely defined by the condition $\chi_n \ket{0}_n=0$ for all $n$. The state
\beq
\big| \, \psi \, \big( \big\{ \check{\la}_a \big\}_1^{N_\kappa}  \big) \big> \, = \, \op{B}( \check{\la}_1) \dots \op{B}( \check{\la}_{N_\kappa}) \ket{0}
\enq
is an eigenstate of the transfer matrix $\op{t}_{LNS}(\la)$  associated with the eigenvalue
\beq
t_{LNS}\pa{\la\mid \big\{ \check{\la}_a \big\}_1^{N_\kappa}} \, = \, a\pa{\la} \cdot \pl{p=1}{N_\kappa} \f{\la - \check{\la}_p + \i c}{ \la - \check{\la}_p } \; + \;  d\pa{\la} \cdot \pl{p=1}{N_\kappa} \f{ \la-\check{\la}_p - \i c}{ \la - \check{\la}_p}
\enq
where
\beq
a\pa{\la} = \paa{ -\i \f{ \la \De}{2} +1 + \f{c\De}{4}  }^{\f{M}{2}} \hspace{-2mm} \cdot \, \paa{ - \i \f{ \la \De}{2} +1 - \f{c\De}{4}  }^{\f{M}{2}}
\quad \e{and} \quad
d\pa{\la} = \paa{ \i \f{ \la \De}{2} +1 + \f{c\De}{4}  }^{\f{M}{2}} \hspace{-2mm}  \cdot \, \paa{ \i \f{ \la \De}{2} +1 - \f{c\De}{4}  }^{\f{M}{2}} \; 
\enq
provided that the parameters $\big\{ \check{\la}_a \big\}_1^{N_\kappa}$ solve the Bethe Ansatz equations (BAE)
\beq
\f{ d(\check{\la}_r )  }{  a (\check{\la}_r)  } = \pl{ \substack{p=1 \\ p \not= r} }{N_\kappa} \f{ \check{\la}_r-\check{\la}_p + \i c  }{ \check{\la}_r-\check{\la}_p - \i c } \;\;\; , \qquad 
r=1,\dots, N_\kappa \;.
\label{ecriture BAE modele discret}
\enq
The solutions to \eqref{ecriture BAE modele discret} are real valued, satisfy the so-called repulsion principle:
\beq
 \e{if} \; a \not= b \quad \e{then} \quad \check{\la}_a \not= \check{\la}_b \;, 
\label{repulsion principle}
\enq
and are in a one-to-one correspondence with a certain subset $\mathbb{S}_{\De;L}$ (depending on $\De$ and $L$ for $\De M=L$ fixed) 
of the sets of all ordered integers $\ell_1 < \dots < \ell_{N_\kappa}$, $\ell_a \in \mathbb{Z}$. 
This statement follows, \textit{e}.\textit{g}. from the analysis developed in [$\bs{A3}$], but can be readily
established by a generalisation of the ideas presented in \cite{BogoliubiovIzerginKorepinBookCorrFctAndABA}. 

The subset $\mathbb{S}_{\De;L}$ goes inductively to the set of all ordered integers $\ell_1 < \dots < \ell_{N_\kappa}$ in the $M\tend + \infty$ limit at $L$ fixed. 
More precisely, given any choice of integers $\ell_1<\dots <\ell_{N_\kappa}$, 
there exists a $\wt{\De}_{\bs{\ell}_{N_{\kappa}}}$ such that, for $\De<\wt{\De}_{\bs{\ell}_{N_{\kappa}}}$ (with $\De M=L$ fixed) there exists a unique solution $\{ \check{\mu}_{\ell_a} \}_1^{N_\kappa}$ to the below set of logarithmic Bethe equations
\beq
- \i \ln\paf{ d( \check{\mu}_{\ell_r} )  }{  a( \check{\mu}_{\ell_r} )  } \; +  \;
\sul{p=1}{N_\kappa} \th \big( \check{\mu}_{\ell_r} - \check{\mu}_{\ell_p} \big) \;  =  \; 2\pi \pa{ \ell_r - \f{N_{\kappa}+1}{2} } \;\;\; , \quad 
r=1,\dots, N_\kappa \; \; \; \e{with} \;\;\th\pa{\la} = \i  \ln  \paf{ \i c + \la  }{ \i c - \la  } \;.
\label{ecriture log BAE}
\enq
Finally, using elementary properties of \eqref{ecriture log BAE}, it can be shown that, 
given a fixed product $\De M =L$ and any choice of integers $\ell_1<\dots <\ell_{N_\kappa}$, there exists a $\De_0>0$ such that the parameters 
$\check{\mu}_{\ell_a} \equiv \check{\mu}_{\ell_a}\pa{\De}$ are continuous in $\De\in \intff{0}{\De_0}$. 

In fact, the $\De \tend 0$, $M\De =L$ limit of such a solution $\wh{\mu}_{\ell_a} = \lim_{\De \tend 0} \check{\mu}_{\ell_a}\pa{\De} $
gives rise to the set of parameters solving the logarithmic Bethe equations arising in the $N_{\kappa}$ quasi-particle sector of the continuous model described by 
the non-linear Schr\"{o}dinger Hamiltonian $\op{H}_{NLS}$:
\beq
L \wh{\mu}_{\ell_r} \; +  \;
\sul{p=1}{N_\kappa} \th \big( \wh{\mu}_{\ell_r} - \wh{\mu}_{\ell_p} \big)  \; = \;  2\pi \pa{ \ell_r - \f{N_{\kappa}+1}{2} } \;\;\;, \qquad 
r=1,\dots, N_\kappa \;.
\label{ecriture Log BAE continuous model}
\enq
In the present chapter, a set of Bethe roots $\{ \check{\mu}_{\ell_a} \}_1^{N_\kappa}$ with a $\check{} $ will refer to the solutions of the Bethe Ansatz equations for the 
model at finite lattice spacing $\De$. For clarity purposes, I omit writing down explicitly this dependence on $\De$. 
When the $\check{} $ is replaced by a $\, \wh{}\,$, the set of Bethe roots $\{ \wh{\mu}_{\ell_a}\}_1^{N_\kappa}$ is to be understood as one built up from the solutions to the Bethe Ansatz 
equations \eqref{ecriture Log BAE continuous model} for the model in the continuum and at \textit{finite} volume $L$.

It has been shown by Dorlas \cite{DorlasOrthogonalityAndCompletenessNLSE} in 1993 that the vectors $\ket{ \psi \, \big( \big\{ \check{\mu}_{\ell_a} \big\}_1^{N_\kappa}  \big) }$
converge, in some suitable sense, to the eigenfunctions 
\beq
\big| \, \Psi\big(  \{ \mu_{\ell_a} \}_{a=1}^{N_\kappa} \, \big)  \big> = \f{1}{N_{\kappa} !}\Int{0}{L} 
\vp\big( x_1,\dots, x_{N_\kappa} \mid \{ \wh{\mu}_{\ell_a} \}_1^{N_\kappa} \big)  \; 
\Phi^{\dagger}\big(x_1\big) \dots  \Phi^{\dagger}\big(x_{N_\kappa}\big) \ket{0}  \,  \dd^{N_\kappa}\!x \;\;  
\enq
of the non-linear Schr\"{o}dinger Hamiltonian $\op{H}_{NLS}$ in the ${N_\kappa}$ quasi-particle sector. 
The function $\vp\big( x_1,\dots, x_{N_\kappa}\! \mid \!\{  \wh{\mu}_{\ell_a} \}_1^{N_\kappa} \big)$ are, in fact, the eigenfunctions of the $\de$-function $N_{\kappa}$ particle Bose gas. They
can be constructed through the coordinate Bethe Ansatz \cite{BrezinPohilFinkelbergFirstIntroBoseGas,LiebLinigerCBAForDeltaBoseGas} and, for a generic set of 
variables $\{ \la_a \}_1^{ N_{\kappa} }$, take the form 
\beq
\vp\big( x_1, \dots, x_{ N_{\kappa} }  \mid \{ \la_a \}_1^{ N_{\kappa} } \big) = \pa{- \i \sqrt c}^{{ N_{\kappa} }}
\sul{ \sg \in \mf{S}_{ N_{\kappa} } }{} \pl{ a<b }{ { N_{\kappa} } } \paa{ \f{ \la_{\sg\pa{a}} - \la_{\sg\pa{b}} - \i c \e{sgn}\pa{x_a-x_b}  }
{  \la_{\sg\pa{a}} - \la_{\sg\pa{b}} } } \cdot 
\pl{a=1}{ N_{\kappa} } \ex{ \i \la_{\sg\pa{a} } \cdot \big( x_a -\f{L}{2}  \big) }  \;.
\label{ecriture fonction propre model continu}
\enq
Above, I have adopted the below definition for the sign function:
\beq
\e{sgn}\pa{x} =  1 \quad \e{for} \; x >0 \qquad , \quad \e{sgn}\pa{x} =  0 \quad \e{for} \; x =0 \qquad , \quad 
\e{sgn}\pa{x} =  -1 \quad \e{for} \; x <0 \;.
\label{definition fonction signe}
\enq

\subsection{Structure of the space of states}

As mentioned in the introduction, the machinery of the algebraic Bethe Ansatz provides one with determinant representations 
for the norms \cite{KorepinNormBetheStates6-Vertex} of Bethe eigenvectors as well as with those for the scalar products \cite{SlavnovScalarProductsXXZ} between 
Bethe vectors and Bethe eigenvectors. The form of these representations for the lattice non-linear Schr\"{o}dinger model is given  below.  
\begin{prop} \cite{KorepinNormBetheStates6-Vertex}
\label{Proposition normes etat Bethe}
Let $\big\{ \check{\mu}_{\ell_a} \big\}_1^{N_{\kappa}+1}$ be any solution to the Bethe Ansatz equations \eqref{ecriture log BAE},  then the norm of the 
associated Bethe state admits the  determinant representation
\beq
\big|\big| \psi\big( \big\{ \check{\mu}_{\ell_a} \big\}_1^{N_{\kappa}+1} \big) \big|\big|^2  \; = \;  
\pl{a=1}{N_{\kappa}+1} \paa{ 2 \i \pi L \check{\xi}^{\prime}_{\paa{\ell_a}} \big( \check{\mu}_{\ell_a} \big) a\big( \check{\mu}_{\ell_a} \big) d\big( \check{\mu}_{\ell_a} \big) } \cdot 
\f{ \pl{a,b=1}{N_{\kappa}+1} \pa{\check{\mu}_{\ell_a} - \check{\mu}_{\ell_b}- \i c }  }{ \pl{ \substack{ a,b=1 \\ a\not= b} }{N_{\kappa}+1} \pa{ \check{\mu}_{\ell_a} - \check{\mu}_{\ell_b}  }  } 
\cdot \det_{N_{\kappa}+1}\Big[ \Xi^{(\check{\mu})} \Big] \;.
\enq
The entries of the matrix $\Xi^{(\check{\mu})}$ read 
\beq
\Xi^{(\check{\mu})}_{ab} = \de_{ab} -\f{  K\pa{\check{\mu}_{\ell_a}-\check{\mu}_{\ell_b}} }{ 2\pi L \check{\xi}_{\paa{\ell_a}}^{\prime}\big( \check{\mu}_{\ell_b} \big) }
\quad with \quad 
\check{\xi}_{\{\ell_a\}}\pa{\om} = -\f{ \i}{2\pi L} \ln\paf{ d \pa{ \om }  }{ a\pa{ \om }  } \; +  \;
\f{1}{2\pi L}\sul{p=1}{N_{\kappa}+1} \th \big(\om - \check{\mu}_{\ell_p} \big)   \;+\; \f{N_{\kappa}+2}{2 L} \; ,
\enq
and we have agreed upon $K\pa{\la}=\th^{\prime}\!\pa{\la}$. 
\end{prop}

\begin{theorem} \cite{SlavnovScalarProductsXXZ}
\label{Theorem Nikita Scalar Products}
Let $\{ \check{\la}_{\ell_a} \}_1^{N_{\kappa}+1}$ be a solution to the logarithmic Bethe equations \eqref{ecriture log BAE}
and $\paa{ \mu_a }_{1}^{N_{\kappa}+1}$ a generic set of parameters. Then, the below scalar product reads
\beq
\braket{  \psi\Big( \{ \check{\la}_{\ell_a} \}_{1}^{N_{\kappa}+1} \Big)  }{ \psi\pa{ \{\mu_a \}_{1}^{N_{\kappa}+1} } } = 
\f{ \pl{a=1}{N_{\kappa}+1} d\pa{ \check{\la}_{\ell_a} } }{ \pl{a>b}{N_{\kappa}+1} \Big\{ \big( \check{\la}_{\ell_a} - \check{\la}_{\ell_b} \big) \cdot \pa{ \mu_{b}-\mu_{a}} \Big\} }
\cdot \det_{N_{\kappa}+1}\Big[ \Om_{\De}\pa{  \{ \check{\la}_{\ell_a} \}_{1}^{N_{\kappa}+1} , \{ \mu_a \}_{1}^{N_{\kappa}+1}  }  \Big] \;,
\label{ecriture produits scalaires}
\enq
where
\beq
\pac{ \Om_{\De} \pa{  \{ \la_a \}_{1}^{N_{\kappa}+1} , \{ \mu_a \}_{1}^{N_{\kappa}+1}  } }_{jk}=
a\pa{\mu_k} t\big(\la_j,\mu_k\big) \pl{a=1}{N_{\kappa}+1} (\la_a-\mu_k - \i c ) \; -\; d\pa{\mu_k} t\big(\mu_k,\la_j\big) \pl{a=1}{N_{\kappa}+1} (\la_a - \mu_k + \i c)
\enq
and
\beq
t\pa{\la,\mu}  = \f{ - \i c  }{ \pa{\la-\mu}\pa{ \la - \mu - \i c}} \;.
\enq
\end{theorem}

 Oota \cite{OotaInverseProblemForFieldTheoriesIntegrability} observed that one can build on the reduction of the 
Lax matrix to projectors and reverse projectors so as to reconstruct certain local operators in the lattice discretised model in terms of 
the entries of the model's monodromy matrix. The reconstruction identity which will be of interest takes the form 
\beq
\op{t}_{LNS}^{-1}(\nu) \cdot  \op{B}(\nu)  \; = \;  \bigg\{ \sul{r=1}{2} \ga_r^{\pa{+}}\!\pa{M}\be_r^{\pa{+}}\!\pa{1}  \bigg\}^{-1} \hspace{-2mm} \cdot  
\, \ga_{1}^{\pa{+}}\!\pa{M} \be_2^{\pa{+}}\!\pa{1} \; .
\label{ecriture reconstruction Oota}
\enq
Using the explicit formulae for $\ga^{\pa{+}}\pa{k}$  and $\be^{\pa{+}}\pa{k}$, \textit{c.f.} \eqref{definition des projecteurs beta plus et alpha plus}-\eqref{definition des projecteurs delta plus et gamma plus},
one gets 
\beq
\sul{r=1}{2}\ga_r^{\pa{+}}\!\pa{M} \be_r^{\pa{+}}\!\pa{1} = \f{c}{2} \, \chi_M^{*} \, \chi_1 \; + \;  2 \rho_{Z_M-\f{\De c}{4}} \, \rho_{Z_1}
\quad \e{and} \quad \ga_1^{\pa{+}}\!\pa{M}\be_2^{\pa{+}}\!\pa{1}  =  
- \i \sqrt{c} \, \chi_M^{*} \, \rho_{Z_1} \;.
\enq
Thus, by taking the formal $\De\tend 0$ expansion of the \textit{rhs} of \eqref{ecriture reconstruction Oota}, one recovers the conjugated field operator 
\beq
\op{t}_{LNS}^{-1}(\nu) \cdot  \op{B}(\nu) \;  =  \;  -\f{ \i \sqrt{c} }{ 2 } \De \, \Phi^{\dagger}\! \pa{0} +\e{O}\,\big( \De^2 \big) \;.
\label{ecriture identification formelle ac champ continu}
\enq

\section{Form factors of the conjugated field operator}
\label{Section FF resultat principal}

The main goal of  paper [$\bs{A4}$] was to make the formal identification \eqref{ecriture identification formelle ac champ continu} rigorous. 
This result closes the gaps that are necessary so as to prove the determinant representations for the matrix elements of the conjugated field $\Phi^{\dagger}(0)$ and density $\Phi^{\dagger}(0)\Phi(0)$
operators in the non-linear Schr\"{o}dinger model. The proof of the theorems given below can be found in Appendix A of [$\bs{A4}$] and builds on 
the determinant representations in the lattice discretised model obtained in Section 2.1 of that paper. 
I remind that the matrix elements of the density operator can be deduced from the scalar products between a Bethe eigenvector an a Bethe eigenvector in the so-called $\be$-twisted 
model, see \textit{e}.\textit{g}. \cite{KozKitMailSlaTer6VertexRMatrixMasterEquation} for the reconstruction and \eqref{ecriture CB modele beta twiste} for a definition of the $\be$-twist. 
Also, a determinant representation for the form factors of the conjugated field operator in the non-linear Schr\"{o}dinger model 
has been obtained by Korepin and Slavnov \cite{KorepinSlavnovFormFactorsNLSEasDeterminants} in 1999 through the use of the two-site model. This determinant representation
was then reproduced by Oota \cite{OotaInverseProblemForFieldTheoriesIntegrability} in 2004 on the basis of the inverse problem discussed earlier. 
However, the mentioned results all relied implicitly on the hypothesis of the convergence of the lattice discretisation to the continuous model, hypothesis 
that I have proven in [$\bs{A3}$], \textit{c.f}. Theorems \ref{Theorem cvgence produits scalaires}-\ref{Theorem cvgce lattice discreization} below. 
\begin{theorem}
\label{Theorem cvgence produits scalaires}
Let $\{ \check{\la}_{\ell_a} \}_1^{ N_{\kappa} }$ be a solution of the logarithmic Bethe equations  \eqref{ecriture log BAE}
in the $N_{\kappa}$ particle sector and  $\{\mu_a\}_1^{ N_{\kappa} }$ a set of generic, pairwise distinct, complex numbers.  Then 
the below scalar product in the lattice model converges, in the $\De \tend 0$ continuum  limit, to the scalar product in the continuous model
\beq
\braket{  \psi\big( \{ \mu_{a} \}_{1}^{ N_{\kappa} } \big)  }{ \psi\big( \{ \check{\la}_{\ell_a} \}_{1}^{ N_{\kappa} } \big) } \limit{ \De }{ 0 }
 \Int{0}{L} \f{ \dd^{ N_{\kappa} } x }{ N_{\kappa} ! } \;  \ov{ \vp\big( x_1, \dots, x_{ N_{\kappa} } \mid \{ \mu_{a} \}_1^{ N_{\kappa} } \big) } \; 
\vp\big( x_1,\dots, x_{ N_{\kappa} } \mid  \{ \wh{\la}_{\ell_a} \}_1^{ N_{\kappa} } \big) \;.
\label{equation convergence PS discret vers integrale}
\enq
As a consequence, one has the below determinant representation for the scalar products in the continuous model:
\beq
\big<  \, \Psi\big(  \{ \mu_{a} \}_{a=1}^{ N_{\kappa} } \, \big)  \big| \, \Psi\big(  \{ \wh{\la}_{\ell_a} \}_{a=1}^{ N_{\kappa} } \, \big)  \big>   \; = \;
\f{ \pl{a=1}{ N_{\kappa} } \Big\{ \ex{  \f{ \i L}{2} \wh{\la}_{\ell_a} }  \Big\}   }
{ \pl{a>b}{ N_{\kappa} } \big( \wh{\la}_{\ell_a}-\wh{\la}_{\ell_b}\big)\pa{\mu_{b}-\mu_{a}} } 
\cdot \det_{ N_{\kappa} }\Big[ \Om\big(  \{ \wh{\la}_{\ell_a} \}_1^{ N_{\kappa} } , \{ \mu_a \}_1^{ N_{\kappa} }  \big)  \Big]  
\enq
where $\Om$ is the $\De \tend 0$ limit of $\Om_{\De}$ and, for generic parameters $\{\la_a\}$ and $\{\mu_a\}$ takes the form 
\beq
\Big[   \Om \pa{  \{ \la_a \}_{1}^{ N_{\kappa} } , \{ \mu_a \}_{1}^{ N_{\kappa} }  } \Big]_{jk}=
\ex{  -\f{ \i L}{2} \mu_{k} } t\big(\la_j,\mu_k\big) \pl{a=1}{ N_{\kappa} } \pa{\la_a-\mu_k- \i c} \; -\; \ex{ \f{ \i L}{2} \mu_{k} } t\big(\mu_k,\la_j\big) \pl{a=1}{ N_{\kappa} } \pa{\la_a - \mu_k + \i c} \;. 
\enq
\end{theorem}
\begin{theorem}
\label{Theorem cvgce lattice discreization}
Let $\{ \check{\mu}_{\ell_a} \}_1^{N_{\kappa}+1}$ and $\{ \check{\la}_{r_a}\}_1^{ N_{\kappa} }$ be any two solution of the logarithmic Bethe equations  \eqref{ecriture log BAE}
in the $N_{\kappa}+1$ and $N_{\kappa}$ particle sectors respectively. Then, the expectation value 
\beq
F_{\Phi^{\dagger}}^{\pa{\De}}\pa{  \{ \check{\mu}_{\ell_a} \}_1^{ N_{\kappa} +1} ;  \{ \check{\la}_{r_a} \}_1^{ N_{\kappa} } } =  \f{ 2 \i }{\De \sqrt{c}}
\cdot  \bra{ \psi\pa{ \{ \check{\mu}_{\ell_a} \}_1^{ N_{\kappa} +1} } }  \, \op{t}_{LNS}^{-1} (\nu) \cdot  \op{B}(\nu) \, \ket{ \psi\pa{ \{ \check{\la}_{r_a} \}_1^{ N_{\kappa} }} } \;, 
\label{definition valeur moyenne discrete pour FF champ conj}
\enq
converges, when $\De \tend 0^+$, to the below form factor of the field operator in the continuous model defined as 
\beq
F_{\Phi^{\dagger}}\Big(  \{ \wh{\mu}_{\ell_a} \}_1^{ N_{\kappa} +1} ;  \{ \wh{\la}_{r_a} \}_1^{ N_{\kappa} } \Big) =
 \Int{0}{L} \f{ \dd^{ N_{\kappa} } x }{ N_{\kappa}! } \;   \ov{ \vp\big( 0,x_1, \dots, x_{ N_{\kappa} } \mid \{ \wh{\mu}_{\ell_a} \}_1^{ N_{\kappa} +1} \big) } \cdot 
\vp\big( x_1,\dots, x_{ N_{\kappa} } \mid  \{ \wh{\la}_{r_a} \}_1^{ N_{\kappa} } \big) \;.
\label{ecriture rep int pour FF chmp}
\enq
The latter admits the below determinant representation 
\beq
F_{\Phi^{\dagger}}\pa{ \{ \wh{\mu}_{\ell_a} \}_1^{ N_{\kappa} +1}; \{ \wh{\la}_{r_a} \}_1^{ N_{\kappa} }  } \; = \;  
\i \sqrt{c} \pl{a=1}{ N_{\kappa} +1} \ex{\f{ \i L }{2} \wh{\mu}_{\ell_a} } \cdot
\pl{k=1}{ N_{\kappa} } \paa{ \ex{-\f{ \i L}{2} \wh{\la}_{r_k} }\pac{1-\ex{-2 \i \pi \wh{F}_{\{\ell_a\} }^{ \{r_a\} }( \wh{\la}_{r_k} )  }} 
\pl{b=1}{ N_{\kappa} +1} \f{ \wh{\mu}_{\ell_b} - \wh{\la}_{r_k} - \i c }{ \wh{\mu}_{\ell_b} - \wh{\la}_{r_k}  } }
\det_{ N_{\kappa} }\big[ \de_{jk} + U_{jk}  \big]
\label{ecriture representation FF en determinant prepare}
\enq
with
%
%
%
\beq
U_{jk} = - \i  \pl{a=1}{ N_{\kappa} + 1}\f{ \wh{\la}_{r_j} - \wh{\mu}_{\ell_a}  }{ \wh{\la}_{r_j} - \wh{\mu}_{\ell_a} + \i c} \cdot 
 \f{ \pl{a=1}{ N_{\kappa} } \big( \wh{\la}_{r_j} - \wh{\la}_{r_a} + \i c \big) }{ \pl{ \substack{a=1 \\ \not=j } }{ N_{\kappa} } \big( \wh{\la}_{r_j} - \wh{\la}_{r_a}  \big) }  
 \cdot \f{ K \big(  \wh{\la}_{r_j} - \wh{\la}_{r_k} \big)  }{ \ex{-2 \i \pi \wh{F}_{\{\ell_a\} }^{ \{r_a\} }( \wh{\la}_{r_j} ) } -1 } \;.
\enq
The function $\wh{F}_{\{\ell_a\} }^{ \{r_a\} }$  appearing in the above expressions is the so-called discrete shift function for the continuous model:
\beq
\ex{ - 2 \i \pi \wh{F}_{\{\ell_a\} }^{ \{r_a\} } (\om) } \, = \,  \pl{a=1}{ N_{\kappa} +1} \f{ \wh{\mu}_{\ell_a}-\om + \i c }{ \wh{\mu}_{\ell_a}-\om - \i c } \cdot
\pl{a=1}{ N_{\kappa} } \f{ \wh{\la}_{r_a}-\om - \i c }{ \wh{\la}_{r_a}-\om + \i c } \;.
\enq
\end{theorem}

Theorem \ref{Theorem cvgce lattice discreization} already presents the determinant in a "good" form by explicitly factoring out their most singular behaviour $(\mu_{\ell_a}-\la_b)^{-1}$. 
Such a factorisation can be achieved by starting from the representation in terms of $\det_{ N_{\kappa} }[\Om]$ obtained from  \eqref{definition valeur moyenne discrete pour FF champ conj}-\eqref{ecriture rep int pour FF chmp},
 multiplying and dividing it by a Cauchy determinant and then taking the matrix products explicitly. In fact, recasting the form factor in the form 
\eqref{ecriture representation FF en determinant prepare} is the first step which allows one to study their large-$L$ asymptotic behaviour. 
Such Cauchy determinant factorisation trick was used, for the first time, in \cite{SlavnovFormFactorsNLSE}. 
Independently of the Cauchy determinant factorisation, Theorem \ref{Theorem cvgce lattice discreization}  provides one with a slightly different, in respect to the works
\cite{KorepinSlavnovFormFactorsNLSEasDeterminants,OotaInverseProblemForFieldTheoriesIntegrability}, determinant representation for $F_{\Phi^{\dagger}}\pa{  \{ \wh{\mu}_{\ell_a} \}_1^{ N_{\kappa} +1 } ;  \{ \wh{\la}_{r_a} \}_1^{ N_{\kappa} } } $. 
The equivalence of my representation with those obtained previously can be checked with the help of 
determinant identities analogous to those established in 
\cite{KozKitMailSlaTerXXZsgZsgZAsymptotics}-[$\bs{A5}$].

\section{The large-volume behaviour of the form factors of the conjugated field}
\label{Section Thermo limit FF}

In this section, I will describe the structure of the large volume asymptotic behaviour of the form factors $F_{\Phi^{\dagger}}$
associated with a specific class of excited states.  Namely, I shall consider the form factors taken between an excited state $\{\mu_{\ell_a}\}_1^{N+1}$
corresponding to an $n$-particle/hole excitation, $n$-independent of $L$, above the lowest energy state in the $N+1$-quasi particle sector and the ground-state of the model 
$\{ \wh{\la}_{a} \}_1^N$. These states and, in particular, the particle-hole terminology  will be all described below. 
After having presented the results, I will shortly review the methods that I have developed so as to prove such a asymptotic expansion. 
The details can be found in the papers
that I co-authored with Kitanine, Maillet, Slavnov and Terras [$\bs{A5},\bs{A6}$]. 
As I mentioned, \textit{per se}, various elements of the method take their roots in the pioneering work of Slavnov \cite{SlavnovFormFactorsNLSE}.
Although I will not discuss this case here, I do stress that it is not hard to generalise these results to the case of form factors involving two excited states of particle-hole types.

\subsection{Some general facts about the thermodynamic limit}

As already discussed, the eigenstates of the non-linear Schr\"{o}dinger model in the $N_{\kappa}$ quasi-particle sector, \textit{viz}. those of the system of 
$N_{\kappa}$-bosons in $\de$ two-body interactions, are parametrised by the solutions to the logarithmic Bethe Ansatz equations \eqref{ecriture Log BAE continuous model}. 
In fact, by slightly changing the boundary conditions from periodic to $\be$-twisted ones:
\beq
\vp\big(0,x_2,\dots,x_{N_{\kappa}} \mid \{ \wh{\mu}_{\ell_a} \}_1^{N_{\kappa}} \big)  \; = \; \ex{- 2 \i \pi  \be }  \vp\big(x_2,\dots,x_{N_{\kappa}}, L\mid \{ \wh{\mu}_{\ell_a} \}_1^{N_{\kappa}} \big)
\label{ecriture CB modele beta twiste}
\enq
one obtains a parametrisation of the eigenstates in terms of solutions to $\be$-twisted logarithmic Bethe Ansatz equations:
\beq
L \wh{\mu}_{\ell_r} \; +  \;
\sul{p=1}{N_{\kappa}} \th \big( \wh{\mu}_{\ell_r} - \wh{\mu}_{\ell_p} \big) \; = \;  2\pi \Big(  \ell_r  - \f{ N_{\kappa} + 1 }{ 2 }  \Big) \, + \, 2\i \pi \be \;\;\;, \qquad 
r=1,\dots,N_{\kappa} \;.
\label{definition eqn Bethe log}
\enq
Since the parameter $\be$ will appear very useful in the following, I will focus on describing the solutions to the $\be$-twisted equations even though this might appear artificial at this stage. 
For any $\ell_1<\dots<\ell_{ N_{\kappa} }$, the system of equations \eqref{definition eqn Bethe log} can be interpreted\symbolfootnote[4]{It is enough to change variables  $\mu \hookrightarrow \mu-2\i\pi\be/L$.}
as conditions for the minimum of a strictly convex function -the so-called Yang-Yang function- on $\R^{ N_{\kappa} }$ that diverges at infinity. 
 They thus admit a unique solution.

 The energies, \textit{i}.\textit{e}. eigenvalues of $\op{H}_{NLS}$ associated with a set of roots $\{ \wh{\mu}_{\ell_a} \}_1^{N_{\kappa}}$, take at $\be=0$, the form
\beq
\mc{E}\big( \{ \wh{\mu}_{\ell_a} \}_1^{N_{\kappa}} \big) \; = \; \sul{a=1}{ N_{\kappa} } \wh{\mu}_{\ell_a}^2 \; - \; h\cdot N_{\kappa} \; . 
\enq
The set of Bethe roots giving rise to the lowest energy
in the sector with a fixed  number $N_{\kappa}$ of quasi-particles 
is conjectured\symbolfootnote[2]{Although this fact has been thoroughly checked, there is still no proof of this statement to the best of my knowledge. 
The statement is easy to see at $c=+\infty$ and $c=0$ through explicit calculations. It thus also holds perturbatively 
around these points for $c\in \R^+$ as ensured by the continuity in $c$
of the solutions to \eqref{definition eqn Bethe log} and the point-wise nature of the spectrum. For generic $c$ it can be checked that the choice of consecutive integers 
does give rise to the lowest value of the energy among the class of bounded in $L$ particle/hole excitation integers $\ell_a$.}  to be given by the choice of consecutive integers $\ell_a=a$, $a=1,\dots,N_{\kappa}$. 
When $\be\not=0$, I will still continue calling the solution subordinate to this choice of integers the ground state in the $N_{\kappa}$ quasi-particle sector, even though it might no longer lead 
to  the lowest energy state in this sector. 

In their turn, the integers $ \{ \ell_j \}$ associated with the excited state above the $N_{\kappa}$ quasi-particle ground state are most conveniently presented in terms of holes
in the consecutive distribution of integers for the ground state in the given sector and of "particle" integers which correspond to values taken by the $\ell_a$ that are outside of $\intn{1}{N_{\kappa}}$:
\beq
\ell_a=a \quad \e{for} \; \; a \in \intn{ 1 }{ N_{\kappa} } \setminus{ h_1 , \dots , h_n }  \quad \e{and}  \quad
\ell_{h_a}=p_a \quad \e{for} \;\;  a=1,\dots,n \; .
\label{definition correpondance entiers ella et particules-trous}
\enq
The integers\symbolfootnote[3]{This choice of a parametrisation for the $\ell_a$'s is not respecting the ordering $\ell_1<\cdot < \ell_{N_{\kappa}} $. The latter is however readily recovered upon an
appropriate permutation.} $p_1<\dots<p_n$ and $h_1<\dots<h_n$ satisfy
\beq
p_a \not \in \intn{1}{N_{\kappa}}\equiv \{ 1,\dots, N_{\kappa} \} \qquad \e{and} \qquad  h_a\in \intn{1}{N_{\kappa} } \; .
\enq

There exists a convenient way of characterising the large-volume $L$ behaviour of the solutions to the logarithmic Bethe Ansatz equations.  
 The main idea consist in studying the large-$L$ asymptotic behaviour of their counting function:
\beq
\wh{\xi}_{\paa{\ell_a}}^{\, (\be)} (\om) \equiv \wh{\xi}_{\paa{\ell_a}}^{\,(\be)} \big( \om \mid \{ \wh{\mu}_{\ell_a} \}_{1}^{N_{\kappa}} \big) \; = \;  
\f{\om}{2\pi} +\f{1}{2\pi L} \sul{ a=1}{ N_{\kappa}  } \th\big(\om - \wh{\mu}_{\ell_a} \big) + \f{N_{\kappa}+1}{2L} - \i \f{\be}{L} \;.
\label{definition counting function mu}
\enq
By construction, it is such that $\wh{\xi}_{\paa{\ell_a}}^{\,(\be)} \big( \wh{\mu}_{\ell_a} \big)=\tf{\ell_a}{L}$, for $a=1,\dots, N_{\kappa}$.
Actually, $\wh{\xi}_{\paa{\ell_a}}^{\,(\be)}\pa{\om }$ defines a set of background parameters $\{ \wh{\mu}_a \}_{a\in \mathbb{Z} }$,
as the unique solutions to $\wh{\xi}_{\paa{\ell_a}}^{\,(\be)}\pa{\wh{\mu}_a}=\tf{a}{L}$, $a\in \mathbb{Z}$. The latter allows one to define the rapidities
$\wh{\mu}_{p_a}$, resp. $\wh{\mu}_{h_a}$, of the particles, resp. holes, entering in the description of the Bethe roots for the excited state:
\beq
\big\{ \wh{\mu}_{\ell_a} \big\}_1^{N_{\kappa} } \; = \;  \bigg\{ \big\{ \wh{\mu}_{a} \big\}_1^{N_{\kappa} } \setminus \big\{ \wh{\mu}_{p_a} \big\}_1^{n } \bigg\} \cup \big\{ \wh{\mu}_{h_a} \big\}_1^{n } \; .
\enq
I do stress that the set of parameters $\{ \wh{\mu}_a \}_{a\in \mathbb{Z} }$ does depend on the choice of $ \ell_a$, \textit{viz}. is \textit{not} universal for 
all excited states. I have omitted the explicit writing of this dependence since it would induce too bulky notations. 

One can show that the counting function satisfies a non-linear integral equation that linearises in the large-$L$ limit. This allows one to prove that $\wh{\xi}_{\paa{\ell_a}}^{\, (\be)} $
admits an all order in $L^{-1}$ asymptotic expansion in some $L$-independent strip around $\R$. In order to describe this expansion, I first need to give a little more precision
on the way that the thermodynamic limit is taken.

At $\be=0$, in each sector with a fixed number of particles $N_{\kappa}$, the model has a unique ground state. 
The $N$ quasi-particle sector giving rise to the overall ground state of the model will issue from a balancing of the kinetic energy terms of an $N_{\kappa}$-particle sector ($\sum \wh{\mu}_a^2$), which grows to 
$+\infty$ with $N_{\kappa}$, and the chemical potential energy ($-hN_{\kappa} $) which decreases down to $-\infty$ with $N_{\kappa}$. 
In the following, $N$ will \textit{always} refer to the number of quasi-particles in the overall ground state. I stress that this number does depend on $h$. 
Accordingly, I will parametrise $N_{\kappa}=N+\kappa$ in the following. 
Although I will not discuss how such a mechanism arises, I do remind that, when $L\tend +\infty$, the roots $\{ \mu_{\ell_a} \}_1^N$ with $ \ell_a$ as in \eqref{definition correpondance entiers ella et particules-trous} 
and $n$ bounded in $L$ form a dense
distribution on the interval $\intff{-q}{q}$ called the Fermi zone. The endpoint $q$ of the Fermi zone is the parameter which arises in the unique solution 
$\big(q, \veps(\la\mid q) ) $ to the set of equations 
\beq
\veps(\la \mid Q )- \Int{-Q}{Q} K\pa{\la-\mu} \veps(\mu \mid Q)  \f{\dd \mu}{2\pi} \; = \;  \veps_{0}\pa{\la} \qquad \e{with} \; \; \veps_0\pa{\la} = \la^2 - h
\qquad \e{and} \;\; \veps(\pm Q \mid Q) =0 \;.
\label{definition eqn int eps}
\enq
Although the unique solvability of the system \eqref{definition eqn int eps} has been stated for many years, it was only in 2012 that I gave a proof thereof in [$\bs{A2}$]. 
The proof boils down to proving that $Q \mapsto \veps(\pm Q \mid Q)$ is continuous, strictly increasing, strictly negative at $Q=0$ and going to $+\infty$
when $Q\tend +\infty$. The strictly increasing part being the hardest to establish. 
In collaboration with Dugave and G\"{o}hmann [$\bs{A23}$],
we proved the existence of various other analogous quantities associated with the linear integral equations describing the massless regime of the XXZ chain. I stress that the spin-chain case 
the proof built on other ideas since, for certain values of the anisotropy, the XXZ analogue of the integral kernel $K$ arising in \eqref{definition eqn int eps}
is strictly negative. I refer to that paper for more details and to \cite{Yang-YangXXZStructureofGS} where part of the used techniques were pioneered.

In the following, I set $\veps(\la)\equiv\veps(\la\mid q)$. For reasons that will become clear in the following, this quantity is called the dressed energy of the excitations. 
Pursuing the discussion of the thermodynamic limit, one can show that the ratio $N/L$ admits a limit $D \in \intff{0}{+\infty}$:
\beq
D=\f{ p(q) }{\pi} \qquad \e{with} \quad  p(\la) - \Int{-q}{q} \th\pa{\la-\mu} p^{\prime}\pa{\mu} \f{\dd \mu}{2\pi} = \la \quad \e{and} \quad p(-\la)=-p(\la) \;. 
\label{definition moment habille}
\enq
The solution to the above linear integro-difference equation is called the dressed momentum of the excitations, while $p_F=p(q)$  is called the Fermi momentum. 
The model in infinite volume with a fixed value of $D$ can always be approached in such a way that $N/L-D=\e{O}(1/L^2)$; I will build on this assumption in the following.

These preliminary parameters being introduced, I am in position to describe the large-$L$ asymptotic expansion of the counting function \eqref{definition counting function mu}. 
The latter reads:
\beq
\wh{\xi}_{\paa{\ell_a}}^{\, (\be)} (\om) \; = \; \xi(\om) \;+\; \f{ 1 }{ L }  \xi_{-1}( \om ) \; + \; \f{1}{L} F_{\be}^{(\kappa) }\pabb{\om}{ \{ \mu_{p_a}  \} } { \{ \mu_{h_a} \} } 
\; + \; \e{O}\Big( \f{1}{L^2} \Big) \;. 
\enq
The important point is that the remainder is uniform and holomorphic in some strip around $\R$. In this expansion, $\xi$ is the thermodynamic limit of the counting function given by 
\beq
\xi(\om)\, = \, \f{ p(\om) }{ 2\pi } \, + \, \f{D}{2} \, .
\label{ecriture fct cptge thermo}
\enq
The thermodynamic limit of the counting function defines  particles' $\{ \mu_{p_a} \}$ and holes' $\{\mu_{h_a}\}$ asymptotic rapidities. These correspond to  
 the unique solutions to 
\beq
\xi ( \mu_{p_a} ) =\f{p_a}{L} \qquad \e{and} \qquad  \xi\pa{\mu_{h_a}}=\f{h_a}{L} \; . 
\label{definition particle trou thermo lim}
\enq
$\xi_{-1}$ represents an overall correction to the asymptotic behaviour of the counting function. It is the one that is responsible for the $1/L$ corrections to the ground state energy
whose form can be predicted by conformal field theoretic arguments. 
It takes the same form for all excited states. Finally, $F_{\be}^{(\kappa)}$ is the so-called shift function. It gathers all the $1/L$ corrections to the
counting function issuing from  the particular excited state. The shift function has another interpretation. It measures the spacing between the ground state roots $\wh{\la}_a$ and the background 
parameters $\wh{\mu}_a$ defined by $\wh{\xi}_{\paa{\ell_a}}^{\, (\be)}$: 
\beq
\wh{\mu}_a-\wh{\la}_a \, = \,  F_{\be}^{(\kappa)}\big( \wh{\la}_a \big)\cdot \pac{L \xi^{\prime}\big( \wh{\la}_a \big) }^{-1}  \big( 1+ \e{O}\big(L^{-1}\big) \big) \; .
\enq
The shift function is expressed in terms of two auxiliary functions: the dressed phase $\phi\pa{\la,\mu}$ and the dressed charge $Z\pa{\la}$ which correspond to the solutions to  the linear integral 
equations\symbolfootnote[2]{The invertibility of $I-\tf{K}{2\pi}$ on $L^2(\intff{-q}{q})$ can be readily checked. See Lemma 2.2 of [$\bs{A2}$].}
\beq
\phi\pa{\la,\mu}- \Int{-q}{q} K\pa{\la-\tau} \phi\pa{\tau,\mu} \f{\dd \tau}{2\pi} =  \f{1}{2\pi}\th\pa{\la-\mu} \qquad \e{and} \qquad
Z\pa{\la}- \Int{-q}{q} K\pa{\la-\tau} Z\pa{\tau} \f{\dd \tau}{2\pi} =  1 \;.
\label{definition eqn int Z et phi}
\enq
The explicit expression for the shift function reads 
\beq
 F_{\be}^{(\kappa)}\pabb{\la}{ \{ \mu_{p_a} \} } {\paa{\mu_{h_a}} }
 =  \Big( \i \be - \f{\kappa}{2} \Big) Z\pa{\la} \; -\; \kappa \phi\pa{\la,q} \; - \; \sul{a=1}{n} \pac{\phi ( \la,\mu_{p_a}) - \phi\pa{\la,\mu_{h_a}}} \;. 
\label{ecriture limite thermo fction shift}
\enq
It will sometimes be useful, in the following, to deal with the finite-$L$ shift function 
\beq
\wh{F}_{ \{\ell_a\} }\pa{\om} = L \pac{ \wh{\xi}\pa{\om}  - \wh{\xi}_{\paa{\ell_a}}^{\,(\be)}\pa{\om} } \;,
\enq
where 
\beq
\wh{\xi}\pa{\om}  = \f{\om}{2\pi} \; + \; \f{1}{2\pi L} \sul{a=1}{ N } \th\big(  \om- \wh{\la}_a \big) \; +  \; \f{N+1}{2L} \;
\label{definition counting function la}
\enq
is the counting function associated with the model's ground state, \textit{i}.\textit{e}. $\wh{\la}_a$ solve the logarithmic Bethe Ansatz equations \eqref{definition correpondance entiers ella et particules-trous}
at $\kappa=0$, $\be=0$ and for $\ell_a=a$, with $a=1,\dots,N$. The thermodynamic limit of finite-$L$ shift function $\wh{F}_{ \{\ell_a\} }$  is precisely   $F_{\be}^{(\kappa)}$.

\subsection{The excitation energy and momentum}
\label{Soussection Excitation energy and momentum}

Given an eigenstate parametrised by the Bethe roots $\{ \wh{\mu}_{\ell_a} \}$, it holds
\beq
\op{H}_{NLS} \cdot \ket{ \Psi \big( \{ \wh{\mu}_{\ell_a} \}_1^{ N_{\kappa} }  \big)  }  \; = \; \mc{E} \big( \{ \wh{\mu}_{\ell_a} \}_1^{ N_{\kappa} }  \big) \ket{ \Psi \big( \{\wh{\mu}_{\ell_a} \}_1^{ N_{\kappa} }  \big)  }
\qquad \e{and} \qquad 
\op{P}_L \cdot \ket{ \Psi \big( \{ \wh{\mu}_{\ell_a} \}_1^{ N_{\kappa} }  \big)  }  \; = \; \mc{P} \big( \{ \wh{\mu}_{\ell_a} \}_1^{ N_{\kappa} }  \big) \ket{ \Psi \big( \{ \wh{\mu}_{\ell_a} \}_1^{ N_{\kappa} }  \big)  }
\label{ecriture action op impulsion ete energie sur etat propre}
\enq
where the eigenvalues $\mc{E} \big( \{ \wh{\mu}_{\ell_a} \}_1^{ N_{\kappa} }  \big)$ and $ \mc{P} \big( \{ \wh{\mu}_{\ell_a} \}_1^{ N_{\kappa} }  \big) $ are expressed in terms of the Bethe roots parametrising the state 
\beq
\mc{P}\big( \{ \wh{\mu}_{\ell_a} \}_1^{N_{\kappa}} \big) \; = \; \sul{a=1}{ N_{\kappa} } p_0( \wh{\mu}_{\ell_a} ) \; \; \e{with}\;\; p_0(\la) =\la 
\qquad \e{and} \qquad \mc{E}\big(\{ \wh{\mu}_{\ell_a} \}_1^{N_{\kappa}} \big) \; = \; \sul{a=1}{ N_{\kappa} } \veps_0\big( \wh{\mu}_{\ell_a}   \big) \; \; \e{with}\;\; \veps_0(\la) =\la^2-h  \;. 
\label{ecriture impulsion et energie}
\enq
The form taken by the eigenvalues of $\op{H}_{NLS} $ follows from the calculations carried out when implementing the coordinate Bethe Ansatz for the model \cite{BrezinPohilFinkelbergFirstIntroBoseGas,LiebLinigerCBAForDeltaBoseGas}.  
The one taken by the eigenvalues of the momentum operator $\op{P}_L$ has been obtained in \cite{BogoliubiovIzerginKorepinBookCorrFctAndABA}.

The counting function formalism allows one to estimate the thermodynamic limit of the momentum  and energy of an excited state. 
In fact, one can show that, for $L$ large enough
\beqa
\mc{P}_{\e{ex}} & = & \mc{P}\big(\{ \wh{\mu}_{\ell_a} \}_1^{N_{\kappa}} \big) \, - \, \mc{P}\big(\{ \wh{\la}_{a} \}_1^{N} \big) \; = \; 2\i\be p(q) \, + \,  \sul{a=1}{ n } p\big( \mu_{p_a} \big) - p\big( \mu_{h_a} \big)    \; + \; \e{O}\Big( \f{1}{L} \Big) 
\label{ecriture impulsion relative excitation} \\ 
\mc{E}_{\e{ex}} & = & \mc{E}\big(\{ \wh{\mu}_{\ell_a} \}_1^{N_{\kappa}} \big) \, - \, \mc{E}\big(\{ \wh{\la}_{a} \}_1^{N} \big) \; = \;  \sul{a=1}{ n } \veps\big( \mu_{p_a} \big) - \veps\big( \mu_{h_a} \big)    \; + \; \e{O}\Big( \f{1}{L} \Big) 
\label{ecriture energie relative excitation} \;. 
\eeqa
The remainder is uniform provided that $n$ is fixed. Furthermore, the rapidities $\{\mu_{p_a} \}$  and $\{ \mu_{h_a} \}$ appearing above are  defined according to \eqref{definition particle trou thermo lim}. 

The form taken by the relative excitation momentum and energy clarifies the particle-hole and dressed energy/ dressed momentum terminology used earlier on. 
Indeed, in the thermodynamic limit, the excitations are formed by adding particles -each carrying an energy $\veps(\mu_{p_a})$ and a momentum $p(\mu_{p_a})$- with rapidities outside
the Fermi zone and removing some of the particles with rapidities inside the Fermi zone, \textit{viz}. creating holes in the Fermi zone -each carrying an energy $-\veps(\mu_{h_a})$ and a momentum $-p(\mu_{h_a})$-

\subsubsection{The critical $\ell$ class}

When inspecting the form taken by the relative excitation energy $\mc{E}_{\e{ex}}$ \eqref{ecriture energie relative excitation}, one particular configuration singles out among all the possible values of the 
rapidites $\{ \mu_{p_a} \}$ and $\{ \mu_{h_a} \}$: 
\beq
\mu_{p_a} \; = \; \pm q \, + \, \e{O}\Big( \f{1}{L} \Big) \qquad \e{and} \qquad 
\mu_{h_a} \; = \; \pm q \, + \, \e{O}\Big( \f{1}{L} \Big) \; \; \e{for} \;\;  a=1,\dots, n \; . 
\enq
Indeed, owing to the very definition  $\veps(\pm q) =0$ of the Fermi boundary $q$, one has that, for such configurations,  
$\mc{E}_{\e{ex}} = \e{O}\big( \tf{1}{L} \big)$. As a consequence, in the thermodynamic limit, the excitation energy of these
states vanishes. In other words, at $L=+\infty$, such eigenstates span the massless part of the spectrum. 
Exited states enjoying of the above property will be called critical. 

A set of integers $\{ p_a \}$  and $\{ h_a \}$ describing an excitation in the $N_{\kappa}$-quasi particle sector is said to parametrise
a critical excited state if the particle-hole integers can be represented as
\beq
\big\{ p_a  \big\}_1^{ n } \; = \;  \big\{  N_{\kappa}  + p_{a;+} \big\}_1^{ n_{p;+} } \, \cup  \,
\big\{ 1 -  p_{a;-} \big\}_1^{n_{p;-}}  \qquad \e{and} \qquad 
 \big\{ h_a \big\}_1^{n } \; = \;  \big\{1+ N_{\kappa} - h_{a;+} \big\}_1^{n_{h;+}} \, \cup  \, 
\big\{ h_{a;-}  \big\}_1^{ n_{h;-} }   \;. 
\label{ecriture decomposition locale part-trou close Fermi zone}
\enq
In such a decomposition,  the integers $p_{a;\pm}, h_{a;\pm} $ are "small" compared to $L$, \textit{i}.\textit{e}.
\beq
\lim_{L\tend +\infty} \f{ p_{a;\pm} }{ L }  \; = \; \lim_{L\tend +\infty} \f{ h_{a;\pm} }{ L }  \; = \; 0 \;. 
\enq
 Furthermore, the integers $n_{p;+}$ or $n_{p;-}$, resp. $n_{h;+}$ or $n_{h;-}$, corresponding to the number of particles, resp. holes,
collapsing on the right or left Fermi boundary are subject to the constraint
\beq
n_{p;+} \, + \,  n_{p;-} \; = \;  n_{h;+}\, + \,  n_{h;-} \;= \; n \;. 
\enq

Although all "critical" excitations have a vanishing excitation energy, they may bear a non-zero relative excitation momentum. 
This allows one to regroup them into so-called 
$\ell$-critical classes, where $\ell$ parametrises the thermodynamic limit of their relative excitation momentum as 
$2\ell p_F$, where $p_F=p(q)$ is the so-called Fermi momentum. The integer $\ell \in \mathbb{Z}$ is defined as 
\beq
\ell \; = \;  n_{p;+}  \; - \;  n_{h;+} \; = \; n_{h;-}  \; - \;  n_{p;-} \;. 
\label{ecriture lien shift ells et differences part trous sur bords zone Fermi}
\enq
In fact, by summing up the logarithmic Bethe equations for the excited state of the critical $\ell$ class and those for the ground state, 
one recasts the relative excitation momentum \eqref{ecriture impulsion relative excitation} in the form
\beq
\mc{P}_{\e{ex}} \; =\;  2 \pi \Big( \ell  \; + \; \i  \be \Big) \f{N_{\kappa}}{L} 
\; + \; \f{2\pi}{L} \Bigg\{ \sul{a=1}{ n_{p;+} } p_{a;+} \; + \;  \sul{ a=1 }{ n_{h;+} } (h_{a;+}  -1 ) \Bigg\}
\; - \; \f{2\pi}{L} \Bigg\{ \sul{a=1}{n_{p;-} } (p_{a;-} -1)\; + \;  \sul{a=1}{n_{h;-}}  h_{a;-}  \Bigg\}  \; . 
\label{ecriture ex momentum pour etats ell shiftees}
\enq

\subsection{Thermodynamic limit of form factors}
\label{Subsection Thermo Lim FF}

\subsubsection{The smooth and discrete parts}

By applying Proposition \ref{Proposition normes etat Bethe} and Theorem \ref{Theorem cvgce lattice discreization},
it  follows that the normalised modulus squared of the ground-to-excited state form factor of the conjugated field  can be decomposed  into a products of two terms
\beq
\f{  \abs{ \bra{\Psi\pa{ \{ \wh{\mu}_{\ell_a} \}_1^{N+1}  }} \Phi^{\dagger}\pa{0} \ket{ \Psi\pa{ \{ \wh{\la}_a \}_1^N}  } }^2 }
{ \norm{\Psi\pa{ \{  \wh{\mu}_{\ell_a} \}_{1}^{N+1} }}^2 \cdot  \norm{\Psi\pa{\{ \wh{\la}_a \}_{1}^{N} }}^2     } 
= \wh{\mc{G}}_{N;1} \pab{ \! \!   \paa{p_a}_1^n \! \!  }{ \! \!  \paa{h_a}_1^n \! \!  } \pac{\wh{F}_{\paa{\ell_a}}, \wh{\xi}_{\paa{\ell_a}}^{\, (\be)}, \wh{\xi} \, }
\cdot \wh{D}_{N}\pab{ \! \!   \paa{p_a}_1^n \! \!  }{ \! \!  \paa{h_a}_1^n \! \!  } \pac{\wh{F}_{\paa{\ell_a}}, \wh{\xi}_{\paa{\ell_a}}^{\, (\be)} , \wh{\xi} \, }
%
%
%
\enq
one called the smooth part $\wh{\mc{G}}_{N;1}$  and one called the discrete part $\wh{D}_N$. 

\vspace{3mm}
The smooth  part reads:
\beq
\wh{\mc{G}}_{N;1} \pab{ \! \!   \paa{p_a}_1^n \! \!  }{ \! \!  \paa{h_a}_1^n \! \!  } \pac{\wh{F}_{\paa{\ell_a}},\wh{\xi}_{\paa{\ell_a}}^{\, (\be)}, \wh{\xi} \, } = 
W_{N}\pab{  \{  \wh{\mu}_{\ell_a} \}_{1}^{N}  }{ \{ \wh{\la}_a \}_{1}^{N} } 
\; \pl{a=1}{N}  \bigg| \f{ \wh{\la}_a - \wh{\mu}_{\ell_{N+1}} - \i c }{ \wh{\mu}_{\ell_a} - \wh{\mu}_{\ell_{N+1} } -  \i c  } \bigg|^2 
\cdot  \f{ \det_{N }\Big[ \de_{jk}+U_{jk}  \Big] \cdot  \ov{ \det_{N}\Big[ \de_{jk}+U_{jk} \Big] } }
{ \det_{N+1}\Big[ \, \Xi^{(\mu)} \big[ \, \wh{\xi}_{ \{\ell_a\} }^{\,(\be)} \,  \big]  \, \Big] \cdot \det_{N} \Big[ \,  \Xi^{(\la)}\big[ \, \wh{\xi} \, \big] \, \Big] } \;,
\label{definition partie lisse}
\enq
where $W_N$ stands for the double product 
\beq
W_{N} \pab{ \paa{z_a}_1^{N} }{ \paa{w_a}_1^{N} }  = 
\pl{a,b=1}{N} \bigg\{ \f{ \pa{z_{a}-w_b-\i c} \pa{w_{a}-z_{b}- \i c}  }{ \pa{z_{a}-z_{b}- \i c} \pa{w_{a}-w_{b}- \i c}  } \bigg\} 
\enq
while the matrices $ \Xi^{(\la)}\big[ \, \wh{\xi} \, \big] $ and $ \Xi^{(\mu)} \big[ \, \wh{\xi}_{ \{\ell_a\} }^{\,(\be)} \,  \big] $ take the form:
\beq
\Big( \Xi^{(\la)}\big[ \, \wh{\xi} \, \big] \Big)_{ab} \; = \;  \de_{ab} -\f{  K\big( \wh{\la}_{a} - \wh{\la}_{b} \big) }{ 2\pi L \,  \wh{\xi}^{\prime}\big( \wh{\la}_{b} \big) } 
\qquad \e{and} \qquad 
\Big( \Xi^{(\mu)} \big[ \, \wh{\xi}_{ \{\ell_a\} }^{\,(\be)} \, \big] \Big)_{ab}  \; = \; 
\de_{ab} -\f{  K\big( \wh{\mu}_{\ell_a} - \wh{\mu}_{\ell_b} \big) }{ 2\pi L \, \Big( \wh{\xi}_{\paa{\ell_a}}^{\,(\be)}  \Big)^{\prime}\big( \wh{\mu}_{\ell_b} \big) } \;.
\enq
In its turn, the discrete part is represented as 
\beq 
\wh{D}_{N}\pab{ \! \!   \paa{p_a}_1^n \! \!  }{ \! \!  \paa{h_a}_1^n \! \!  } \pac{\wh{F}_{\paa{\ell_a}}, \wh{\xi}_{\paa{\ell_a}}^{\, (\be)}, \wh{\xi} \, } \;  =  \;
\f{ \prod_{k=1}^{N} \paa{4 \sin^{2}\!\big[ \pi \wh{F}_{ \{\ell_a\} }( \wh{\la}_k )  \big] }  }
{ \pl{a=1}{N+1} \Big\{ 2\pi L \, \Big( \wh{\xi}_{\paa{\ell_a}}^{\,(\be)}  \Big)^{\prime}\big( \wh{\mu}_{\ell_a} \big)\Big\} 
				\pl{a=1}{N} \Big\{ 2\pi L \, \wh{\xi}^{\prime}\big( \wh{\la}_a \big)  \Big\} }
\cdot \pl{a=1}{N} \bigg\{ \f{ \wh{\mu}_{\ell_a} - \wh{\mu}_{\ell_{N+1}} }  { \wh{\la}_a - \wh{\mu}_{\ell_{N+1}} }  \bigg\}^2 
\cdot  \det_{N}^2\pac{ \f{1}{ \wh{\mu}_{\ell_a} - \wh{\la}_b } }  \;.
\label{definition partie discrete}
\enq

Although both the discrete and smooth parts are explicit functions of the Bethe roots describing the two states $\{ \wh{\mu}_{\ell_a} \}_1^{N+1} $
and  $\{ \wh{\la}_a \}_1^N$, it is more convenient, for further purposes, to interpret them as functionals of the finite-$L$ shift function $\wh{F}_{ \{ \ell_a \} }$
and of the two counting functions $\wh{\xi}_{\paa{\ell_a}}^{\, (\be)}$ and  $\wh{\xi}$ associated with the two sets of Bethe roots. 
Note that, in fact, the whole information on the integers $\{ \ell_a \}$ describing the given excited state is contained in the two collections of integers 
$\{p_a\}$ and $\{h_a\}$ as follows from \eqref{definition correpondance entiers ella et particules-trous}. 
This interpretation of the smooth and discrete parts in terms of functionals allows one to apply them on more general objects that the finite-$L$ counting functions associated with 
some solutions to the logarithmic Bethe Ansatz equations. 

To be more precise, let $h$ be a biholomorphism on some open neighbourhood of $\R$, strictly increasing on $\R$ and mapping $\R$ onto $\R$ and 
let $\nu$ be any holomorphic function defined on the same neighbourhood. Then set 
\beq
h_{\nu} (\om) \; = \; h(\om) + \f{1}{L} \nu(\om) \; . 
\enq
In this construction $L$ is assumed to be large enough for $h_{\nu}$ to be a biholomorphism as well. 

The explicit expressions for $\wh{D}_{N}\pac{\nu, \xi, \xi_{\nu}}$
(and $\wh{\mc{G}}_{N;1}\pac{\nu, \xi, \xi_{\nu}}$) are then given as above, with the obvious substitutions 
\beq
\wh{F}_{\paa{\ell_a}} \; \hookrightarrow \; \nu \;\;, \qquad \wh{\xi}_{\paa{\ell_a}}^{\, (\be)} \; \hookrightarrow \; h \; \; , \qquad \wh{\xi}   \; \hookrightarrow \;  h_{\nu} \;
\enq
whenever these are written explicitly. Also, in such a writing, one adopts the convention that two sets of parameters $\{ \la_a \}_1^N$ and $\{\mu_{\ell_a}\}_1^{N+1}$
are now \textit{defined} as follows: \vspace{1mm}
\begin{itemize}
\item[$\bullet$] $\mu_k$, $k\in \mathbb{Z}$ is the unique solution to $h\pa{\mu_{k}}=\tf{k}{L}$, \textit{i}.\textit{e}. they are defined by
the \textit{second} argument of the functionals; \vspace{1mm}
\item[$\bullet$] $\la_k$, $k \in \intn{1}{N}$ is the unique\symbolfootnote[2]{The uniqueness follows from Rouch\'{e}'s theorem when L is large enough.}
 solution to $h_{ \nu }\!\pa{\la_k}=\tf{k}{L}$, \textit{i}.\textit{e}. they are defined by
the \textit{third} argument of the functionals.\vspace{1mm}
\end{itemize}
I insist that here and in the following, whenever the parameters $\mu_k$ or $\la_p$ enter in the explicit expressions for these functionals, they are \textit{always}
to be understood in this way. Also, I remind that the integers $\ell_a$ are obtained from the integers $\{p_a\}_1^n$ and $\{h_a\}_1^n$ as explained in
\eqref{definition correpondance entiers ella et particules-trous}.

\subsubsection{The large-$L$ behaviour of the smooth and discrete parts}

In the remainder of this subsection, I will discuss the large-$L$ behaviour of the smooth and discrete parts. 
Since I will focus on a specific class of excited states belonging to the $N_{1}=N+1$ quasi-particle sector, it appears more convenient to introduce a simplified notation for the associated shift function:
\beq
F_{\be} (\la) \; = \;   F_{\be}\pabb{\la}{ \{ \mu_{p_a} \} } {\paa{\mu_{h_a}} } \equiv F_{\be}^{(1)}\pabb{\la}{ \{ \mu_{p_a} \} } {\paa{\mu_{h_a}} } 
\label{ecriture fonction shift beta def notation}
\enq
Also, whenever the dependence on the rapidities of the particles and holes is suggested by the context, I will simply drop the explicit writing of this dependence in $F_{\be}$,
exactly as it has been written in the most \textit{lhs} equality of \eqref{ecriture fonction shift beta def notation}.

\subsubsection*{ $\bullet$ The smooth part}

$\wh{\mc{G}}_{N;1}$ is called the smooth part due to the fact that its thermodynamic limit $\mc{G}_n$ only depends on the rapidities
of the particles $\{\mu_{p_a}\}_1^n$ and holes $\{\mu_{h_a}\}_1^n$ entering in the description of the thermodynamic limit of the excited state. 
Furthermore, this dependence is smooth. I remind that these two sets of rapidities are defined as in \eqref{definition particle trou thermo lim}. 
One of the consequences of smoothness is that a small (of the order $\e{O}(1)$ when $L\tend +\infty$) change in the value of the integers
parametrising the state, say $p_a \hookrightarrow  p_a +k_a$ with $k_a$ bounded in $L$, will \textit{not} change the value of the smooth part's thermodynamic limit but only 
manifest itself on the level of the $\tf{1}{L}$ corrections. 

The thermodynamic limit $\mc{G}_n$ is  expressed  [$\bf{A5}$]
in terms of  the thermodynamic limit $F_{\be}$ \eqref{ecriture limite thermo fction shift} of the shift function associated with the excited state
$\{\mu_{\ell_a}\}_1^{N+1}$:
\beq
\wh{\mc{G}}_{N;1} \pab{ \! \!   \paa{p_a}_1^n \! \!  }{ \! \!  \paa{h_a}_1^n \! \!  } \pac{\wh{F}_{\paa{\ell_a}}, \wh{\xi}_{\paa{\ell_a}}^{\, (\be)}, \wh{\xi} \, } = \mc{G}_n\pab{ \{ \mu_{p_a} \} }{ \paa{\mu_{h_a}} } \big[ F_{\be} \big] 
\times \pa{1+\e{O}\pa{L^{-1}}} \; . 
%
%
%
\enq
The remainder is uniform provided that $n$, the number of particle-hole excitations, is fixed. 
For any function $f$ that is holomorphic on an open neighbourhood of $\intff{-q}{q}$ and such that $\ex{\pm 2\i \pi f }-1$ has no zeroes there,
the functional representing the leading large-$L$ asymptotic behaviour of the smooth part takes the form:
\bem
\mc{G}_n\pab{ \{ \mu_{p_a} \} }{ \paa{\mu_{h_a}} } [f] = 
\pl{a=1}{n} \pl{\eps=\pm}{} \paa{ \f{   \mu_{h_a}-q+\eps \i c }{ \mu_{p_a}-q+\eps \i c   } 
\f{ \ex{2 \i \pi   C[f]\pa{\mu_{h_a} + \eps \i c}    }    } {  \ex{2 \i \pi   C[f]\pa{\mu_{p_a} + \eps \i c} }  } }
\cdot \f{  \ex{- 2 \i\pi  \sul{\eps=\pm}{} C[f]\pa{q + \eps \i c}    }  }{   \det^2\pac{I-\tf{K}{2\pi}}  }   \ex{C_0[f]} 
 \\
 \times W_n\pab{ \{ \mu_{p_a} \} }{ \paa{\mu_{h_a}} }
%
%
\cdot   \det_{ \msc{C}_{q} }\bigg[ \e{id}  + \op{U}[f]\pab{ \{\mu_{p_a}\}_1^n }{ \{\mu_{h_a}\}_1^n} \bigg]
				\det_{ \msc{C}_{q} }\bigg[ \e{id} + \ov{\op{U}}[f]\pab{ \{\mu_{p_a}\}_1^n }{ \{\mu_{h_a}\}_1^n} \bigg]  \; .
\label{formule explicite G+ thermo}
\end{multline}
There $C[f]$ is the Cauchy transform on $\intff{-q}{q}$ and $C_0[f]$ is given by a double integral
\beq
 C[f]\pa{\la} = \Int{-q}{q} \f{\dd \mu}{2 \i \pi} \f{ f\pa{\mu} }{\mu-\la} \qquad \e{and} \qquad 
C_0[f] = -\Int{-q}{q} \f{ f\pa{\la} f\pa{\mu} }{  \pa{\la-\mu - \i c}^2 }   \dd \la \dd \mu  \;.
\label{appendix themo FF definition transfo Cauchy et C0}
\enq
All determinants appearing in \eqref{formule explicite G+ thermo} are Fredholm determinants 
of integral operators of the type $\e{id}+\op{A}$ with $\op{A}$ a compact trace class operator. The integral operator $\e{id}-\tf{\op{K}}{2\pi}$ acts on $L^2(\intff{-q}{q})$ and has integral kernel $-\tf{ K(\la-\mu) }{ 2\pi }$. 
The integral kernels of $\op{U}$ and $\ov{\op{U}} $ are given by 
\beq
U[f]\pab{ \{\mu_{p_a}\}_1^n }{ \{\mu_{h_a}\}_1^n} \pa{\om,\om^{\prime}}  \;  = \;  \f{-1}{2\pi}\f{\om-q}{\om-q+ \i c}
\pl{a=1}{n} \Bigg\{ \f{ (\om- \mu_{p_a}) \pa{\om- \mu_{h_a} + \i c} }
{\pa{\om- \mu_{h_a}} (\om- \mu_{p_a} + \i c) }   \Bigg\}  \cdot \f{  \ex{C\pac{ 2\i\pi f }\pa{\om} } }{ \ex{ C\pac{ 2 \i \pi f}\pa{\om + \i c}  } }  \cdot
 \f{ K\pa{\om-\om^{\prime}} }{ \ex{-2 \i \pi f\pa{\om}}-1 }
\label{formule noyau integral U}
\enq
and
\beq
\ov{U}[f]\pab{ \{\mu_{p_a}\}_1^n }{ \{\mu_{h_a}\}_1^n} \pa{\om,\om^{\prime}}  \; = \; \f{1}{2\pi} \f{\om-q}{\om-q - \i c}
\pl{a=1}{n} \Bigg\{ \f{ (\om- \mu_{p_a}) \pa{\om- \mu_{h_a}- \i c} }
{\pa{\om- \mu_{h_a}} (\om- \mu_{p_a}-\i c) }   \Bigg\}  \cdot \f{ \ex{C\pac{2 \i \pi f}\pa{\om}} }{ \ex{  C\pac{2 \i \pi f}\pa{\om - \i c}  }}
\cdot  \f{ K\pa{\om-\om^{\prime}} }{ \ex{ 2 \i \pi f\pa{\om}}-1 } 
\label{formule noyau integral U bar}
\enq
and depend parametrically on the positions of the particles and holes. 
The operators $\e{id}+\op{U}[f]$ and $\e{id}+\ov{\op{U}}[f]$ should be understood as 
acting on $L^2(\msc{C}_{q})$, where $\msc{C}_{q}$ is a counterclockwise loop surrounding the interval $\intff{-q}{q}$
but not any other singularity of the integrand. This means that the poles at $\om=\mu_{h_a}$ of the integral kernels are located inside of $\msc{C}_q$
whereas  the zeroes of $\la\mapsto \ex{-2 \i \pi f\pa{\la}}-1$
are located outside of the contour. In Section \ref{SousSection propriete determinant Fredholm} below, I provide a more precise
definition of their determinants, as, in principle, the existence of such a contour is not guaranteed for all possible choices
of the parameters $\{\mu_{p_a}\}$, $\{\mu_{h_a}\}$ when its function argument $f$ is given by the shift function $F_{\be}$ associated with this
excited state.

The function $\mc{G}_n$ posses two properties worth singling out. \vspace{1mm}
\begin{itemize} 
\item[$\bullet$] $\mc{G}_n$ is invariant under any permutation of the particle (or hole)  rapidities; \vspace{1mm}
\item[$\bullet$] $\mc{G}_{n}$ satisfies the reduction properties
\beq
\mc{G}_n\pab{ \{ \mu_{p_a} \}_1^n }{ \{ \mu_{h_a} \}_1^n } \pac{f}_{\mid \mu_{p_n} = \mu_{h_n} } \; = \;  \mc{G}_{n-1}\pab{ \{ \mu_{p_a} \}_1^{n-1} }{ \{ \mu_{h_a} \}_1^{n-1} } \pac{f}
\enq
\end{itemize}

Note that these two properties still hold when $\mc{G}_n$ is evaluated on the shift function $F_{\be}$ since the latter also satisfies to the same reduction properties.

\subsubsection*{The discrete part}

$\wh{D}_N$ is called the discrete part since its leading large-$L$ behaviour not only depends on the "macroscopic" rapidities $\{ \mu_{p_a} \}$ and $\paa{\mu_{h_a}}$ entering in the description of the excited state
but also on the set of integers $\{ p_a \}$ and $\{ h_a \}$ characterising the excited state. 
This strongly contrasts with the smooth part case in that an $\e{O}(1)$ change of the integers parametrising the excited state, say $p_a \hookrightarrow p_a +k_a $, $k_a$ bounded in $L$, leads to a 
change in the value of the leading asymptotics $\wh{D}_N$. 

First elements of the technique to analyse the large-$L$ behaviour of $\wh{D}_N$ appeared in 
\cite{SlavnovFormFactorsNLSE} and were fully developed in [$\bs{A5},\bs{A6}$]. 
The large-$L$ asymptotic behaviour of $\wh{D}_N$ takes the form
\beq
\wh{D}_{N}\pab{ \! \!   \paa{p_a}_1^n \! \!  }{ \! \!  \paa{h_a}_1^n \! \!  } \pac{\wh{F}_{\paa{\ell_a}}, \wh{\xi}_{\paa{\ell_a}}^{\, (\be)}, \wh{\xi} \, }
=
D_0[F_{\be}]\cdot \mc{R}_{N,n}\pab{ \{ \mu_{p_a} \}; \paa{p_a} }{ \paa{\mu_{h_a}}; \paa{h_a}  }[F_{\be}] \times  \pa{1+\e{O}\paf{\ln L}{L}} \;. 
\label{AppendixThermolim thermolim D+}
\enq
The remainder is uniform in the integers $\{p_a\}$ and $\{h_a\}$ provided that the number of particle-hole excitations remains fixed. The functional $D_0$
takes the form, for any function $\nu$  smooth on $\R$
\bem
D_0\!\pac{\nu} =  \f{2q}{2\pi}  \cdot 
\f{\big\{ \varkappa[\nu](-q) \big\}^{\nu(-q)} }{ \big\{ \varkappa[\nu](q) \big\}^{\nu(q) + 2} }
  \pl{a=1}{n} \paf{\la_{N+1}-\mu_{p_a}}{\la_{N+1}-\mu_{h_a}}^2
 \f{ G^2\big( 1-\nu(-q) \big) \cdot G^2\big( 2+\nu(q) \big)  }{  \pa{2\pi}^{\nu(q) - \nu(-q) } \cdot \pac{ 2q L \xi^{\prime}(q) }^{\pa{\nu(q) + 1}^2 + \nu^2(-q)}  }   \\
\times \exp\bigg\{  \Int{-q}{q} \f{\nu^{\prime}\!\pa{\la}\nu\!\pa{\mu}-\nu^{\prime}\!\pa{\mu}\nu\!\pa{\la} }{2(\la-\mu) } \dd \la \dd \mu \bigg\}
 \; .
\label{AppendixThermoLimD+zero}
\end{multline}
The parameter $\la_{N+1}$ appearing above is defined as the unique solution to $L \wh{\xi}(\la_{N+1}) \; = \; N+1$, $G$ is the Barnes function and
\beq
\varkappa\pac{\nu}\pa{\la} = \exp\Bigg\{  -\Int{-q}{q} \f{\nu\pa{\la}-\nu\pa{\mu}}{\la-\mu} \dd \mu \Bigg\} \;.
\label{definition fonction kappa}
\enq
Finally, the part of $\wh{D}_N$'s asymptotics depending on the excited state reads 
\bem
\mc{R}_{N,n}\pab{ \{ \mu_{p_a} \}; \paa{p_a} }{ \paa{\mu_{h_a}}; \paa{h_a}  } [\nu] \, = \, 
\pl{a=1}{n} \Bigg\{ \f{ \vp\big(\mu_{h_a},\mu_{h_a}\big)  \vp\big(\mu_{p_a},\mu_{p_a}\big)  \vp\big(q,\mu_{p_a}\big)  \ex{\al[\nu]\big(\mu_{p_a}\big) } }
 { \vp\big(\mu_{p_a},\mu_{h_a}\big)  \vp\big(\mu_{h_a},\mu_{p_a}\big) \vp\big(q,\mu_{h_a}\big) \ex{\al[\nu]\big(\mu_{h_a}\big)  }  }  \Bigg\}^2
\\
\times \f{ \pl{a<b}{n} \vp^2\big( \mu_{p_a},\mu_{p_b} \big) \vp^2\big(\mu_{h_a},\mu_{h_b}\big) }
                                        { \pl{a\not= b }{n} \vp^2\big(\mu_{p_a},\mu_{h_b}\big) }\cdot \det^{2}_{n}\pac{ \f{ 1}{ h_a-p_b} } \\
\times \pl{a=1}{n} \Bigg\{    \bigg( \f{ \sin \pac{ \pi \nu(\mu_{h_a}) } }{\pi   } \bigg)^2  \cdot 
\Ga^{2}\pab{ \{ p_a-N+\nu(\mu_{p_a}) \} , \paa{p_a} +\i 0^+ , \paa{N+1-h_a -\nu\pa{\mu_{h_a}}} , \paa{h_a +\nu\pa{\mu_{h_a}}} }
{ \{ p_a-N-1 +\i 0^+  \} , \{ p_a+\nu(\mu_{p_a}) \}, \paa{N+2-h_a} , \paa{h_a}  } \Bigg\} \;.
\label{definition fonctionelle RNn}
\end{multline}
There
\beq
\al[\nu]\pa{\om} =  \nu\pa{\om} \ln \paf{\vp\pa{\om,q} }{ \vp\pa{\om,-q} }
+ \Int{-q}{q}  \f{\nu\pa{\la}-\nu\pa{\om}}{\la-\om} \dd \la  \; \qquad  \e{and} \quad
\vp\pa{\la,\mu}= 2\pi \f{ \la-\mu }{ p\pa{\la} - p\pa{\mu} }  \; .
\label{definition fonction aleph et varphi}
\enq
Above, I have introduced the hypergeometric-like representation for products of $\Ga$-functions:
\beq
\Ga\pab{ \paa{a_k} }{  \paa{ b_k } }= \pl{k=1}{n} \f{ \Ga\pa{a_k} }{ \Ga\pa{b_k} } \; .
\enq

Note that the $\i 0^+ $ regularisations in the expression for $\mc{R}_{N,n}$ are necessary so as to give rise to a meaningful expression when some of the integers $p_a$ take negative values.

Formula \eqref{AppendixThermolim thermolim D+} provides one with the most general form of the asymptotic behaviour of the discrete part, namely one that is valid for all possible 
choices of the integers $\{p_a\}$ and $\{h_a\}$. It can be further simplified if one provides additional information on the specific choice of the particle and hole integers, see Section \ref{SousSection FF dans la classe critique ell}.

\subsubsection{ Some steps of the proofs }   

{\bf $\bullet$ The smooth part }

The key results for proving  of the leading asymptotics of the smooth part is a repetitive application of  Lemma \ref{Lemme comportement asymptotique produit simple} given below,
or its generalisation to the case of multiple products. 

\begin{lemme}
\label{Lemme comportement asymptotique produit simple}
 
 Let $f$ be holomorphic in some open neighbourhood of $\intff{-q}{q}\times O$, $O$ open in $\Cx$ and such that it does not admit zeroes in this neighbourhood. Then, it holds
\beq
\f{ \pl{a=1}{N+1} f\big( \mu_{\ell_a},\om \big) }{ \pl{a=1}{N} f(\la_a,\om) } \; = \; f(q,\om) \cdot \pl{a=1 }{n} \bigg\{ \f{ f\big(\mu_{p_a},\om\big) }{  f\big(\mu_{h_a},\om\big) } \bigg\} \cdot 
\exp\Bigg\{ \Int{-q}{q} \Dp{\la}\ln f(\la,\om) \cdot F_{\be}\pabb{\la}{ \{ \mu_{p_a} \} } {\paa{\mu_{h_a}} }  \cdot \dd \la  \Bigg\} \cdot \bigg( 1+\e{O}\Big( \f{1}{L} \Big)  \bigg)\;. 
\enq
\end{lemme}

\begin{figure}[h]

\begin{pspicture}(5,7)

\psline[linestyle=dashed, dash=3pt 2pt]{->}(2.2,4)(9,4)

\rput(3.1,3.7){$-\wh{q}$}
\psdots(3.5,4)

\pscurve[linestyle=dashed](4.2,4)(4.5,4.2)(3.5,4.8)(3,4.4)(2.7,4)(3.1,3.5)(3.5,3.2)(3.9,3.2)(4.2,4)
\rput(7.9,3.7){$\wh{q}$}
\pscurve[linestyle=dashed](6.9,4)(7,4.1)(7.2,4.5)(7.4,4.6)(7.6,4.9)(7.9,4.5)(8.5,4)(7.6,3.1)(7.2,3.4)(6.9,4)
\psdots(7.5,4)

\pscurve(3.5,4)(3.5,3.8)(3.7,3)(4.5,2.4)(5,1.8)(6,1.3)(6.3,1.7)(6.6,2)(7,3)(7.5,3.8)(7.5,4)
(7.5,4.2)(7.7,4.4)(7.2,4.7)(6.4,5.5)(6.2,5.7)(5.7,6.3)(5.3,6)(5,6.5)(4.5,6)(4.2,5)(3.7,4.3)(3.5,4.2)(3.5,4)

\psline[linewidth=2pt]{->}(4.45,2.45)(4.55,2.35)
\rput(4.3,2.2){$\msc{C}^{-}$}
\psline[linewidth=2pt]{->}(6.45,5.45)(6.35,5.55)
\rput(6.6,5.7){$\msc{C}^{+}$}

\pscircle[linestyle=dashed](12,4){1.5}
\psline(12,2.5)(12,5.5)
\psdots(12,2.5)(12,5.5)(12,4)
\rput(12,2 ){\tiny $ \f{N+\tf{1}{2}}{L} -i\eps $}
\rput(12,6 ){\tiny $ \f{N+\tf{1}{2}}{L} +i\eps $}
\rput(12.5,4 ){\tiny $ \f{N+\tf{1}{2}}{L} $}

\pscurve{->}(8.2,3)(9.3,2.5)(10.3,3)
\rput(9.3,2){$\wh{\xi}$}

\end{pspicture}

\caption{Contours arising in the definition of the operator $I-\tf{\wh{K}}{2\pi}$. The two endpoints  $\wh{q}$ and $-\wh{q}$ satisfy to 
$\wh{\xi}\,(\, \wh{q} \, )=\f{N+\tf{1}{2}}{L}$ and $\wh{\xi}(-\wh{q} \, )=\f{1}{2L}$. \label{Figure Contour pour preuve cvgce dets I-K} }
\end{figure}
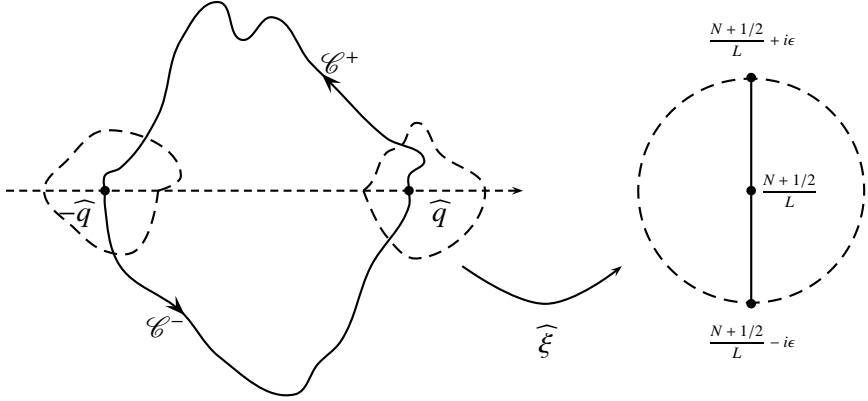

The convergence of the finite-size determinants in \eqref{definition partie lisse} to appropriate Fredholm determinants can be seen as follows. 
One first recasts the determinants in the numerator as
\beq
\det_{ N }\Big[ \de_{jk}+U_{jk}  \Big]  \; = \;  \det_{ \msc{C}_q }  \Big[ \e{id} + \wh{\op{U}}\big[ \wh{F}_{ \{ \ell_a \} } \big] \Big]
\enq
where the contour $\msc{C}_q$ is as defined previously while the integral kernel reads
\beq
\wh{U}\big[ \wh{F}_{ \{ \ell_a \} } \big] (\om,\om^{\prime}) \; = \; \f{-1}{2\pi}   \pl{a=1}{N+1}\f{ \om-\mu_{\ell_a} }{ \om - \mu_{\ell_a} + \i c} \cdot 
\pl{a=1}{N} \f{ \om - \la_{r_a} + \i c }{  \om- \la_{r_a}  }  
 \cdot \f{ K \pa{  \om - \om^{\prime} }  }{ \ex{-2 \i \pi \wh{F}_{ \{ \ell_a \} }( \om ) } -1 } \;. 
\enq
The contour $\msc{C}_q$ is supposed to satisfy the aforementioned hypothesis. If these do not hold, one should proceed as in the proof of Proposition \ref{Proposition Holomorphie en beta et rapidite part-trou det fred},
\textit{c}.\textit{f}. Section \ref{SousSection propriete determinant Fredholm} and Section 3.3 of [$\bs{A4}$].
I will not venture in such technicalities here. It is readily seen, by using Lemma \ref{Lemme comportement asymptotique produit simple} that 
\beq
\Norm{ \wh{\op{U}}\big[ \wh{F}_{ \{ \ell_a \} } \big] - \op{U}\big[ F_{ \be} \big] \pab{ \{\mu_{p_a}\}_1^n }{ \{\mu_{h_a}\}_1^n} }_{ HS }  \; + \; 
\Oint{ \msc{C}_{q} }{} \bigg|  \wh{\op{U}}\big[ \wh{F}_{ \{ \ell_a \} } \big](\om,\om) -  \op{U}\big[ F_{ \be} \big] \pab{ \{\mu_{p_a}\}_1^n }{ \{\mu_{h_a}\}_1^n} (\om,\om)   \bigg| \cdot |\dd \om |
\; = \; \e{O}\Big( \f{1}{L} \Big)   \;. 
\label{ecriture bornes sur cvgence det I + U}
\enq
The first term corresponds to the Hilbert-Schmidt norm of the operators.  Therefore, by using the bounds issuing from the Lipschitz continuity of $2$-determinants in respect to the 
Hilbert-Schmidt norm and the  bound on the second term in \eqref{ecriture bornes sur cvgence det I + U} which ensures that the traces converge to each other, one gets  the convergence of the Fredholm determinants. 
The strategy is similar in what concerns the determinants arising in the denominator. It builds on the equality, at finite $L$:
\beq
\det_N\big[ \Xi^{(\la)} \big] \; = \; \det\Big[ I-\f{\wh{K}}{2\pi}\Big] \quad \e{with} \quad \wh{K}(\om,\om^{\prime}) \; = \; K(\om-\om^{\prime})\bs{1}_{\intff{-\wh{q}}{\wh{q}} }
+\sul{\eps=\pm}{}\bs{1}_{ \wh{\msc{C}}^{\pm} }(\om^{\prime})  \f{ \eps K(\om-\om^{\prime}) }{ \ex{-2\i \pi \eps L \wh{\xi}(\om^{\prime})} -1 }
\enq
where the Fredholm determinant on the \textit{rhs} corresponds to an operator acting on  $L^{2}\big( \intff{-\wh{q} }{ \wh{q} } \cup \wh{\msc{C}}^{+} \cup \wh{\msc{C}}^{-} \big)$. These
 three auxiliary contours are defined in Figure \ref{Figure Contour pour preuve cvgce dets I-K}. The rest follows from similar bounds to \eqref{ecriture bornes sur cvgence det I + U} 
 along with the fact that $\wh{q}-q=\e{O}\big( L^{-1} \big)$. I also remind on this occasion that given a set $A$, $\bs{1}_{A}$ stands for its indicator function.

\vspace{2mm}
{\bf $\bullet$ The discrete part }
\vspace{2mm}

The main problem in the analysis stems from the fact that the parameters $\{\mu_{\ell_a} \}$ and $\{\la_a\}$ densify, when $L\tend + \infty$, on the interval $\intff{-q}{q}$. 
This generates a certain amount of divergences issuing from products involving $\mu_{\ell_a}-\la_b$ with $b-\ell_a = \e{O}(1)$. Some of these divergences will then be cancelled by 
a similar effect issuing from the Vandermonde part of $\wh{D}_N$. 

The first step of the analysis consists in separating, by hand, the contribution of the holes and particles from the one of the "back-ground" parameters that will condensate
on the Fermi zone $\intff{-q}{q}$. In other words, one recasts $\wh{D}_N$ as
\beq
\wh{D}_N \; = \; \wh{D}_{ \Phi^{\dagger} }^{ (0) } \cdot \wh{D}_{ \Phi^{\dagger} }^{ (\e{ex}) } \; 
\enq
where $ \wh{D}_{ \Phi^{\dagger} }^{ (0) } $ represents the contribution of the condensing "back-ground" parameters:
\beq
 \wh{D}_{ \Phi^{\dagger} }^{ (0) } \; = \; 
\f{ \prod_{k=1}^{N} \paa{4 \sin^{2}\!\big[ \pi \wh{F}_{ \{\ell_a\} }(\wh{\la}_k )  \big] }  }
{ \pl{a=1}{N+1} \Big\{ 2\pi L \, \Big( \wh{\xi}_{\paa{\ell_a}}^{\,(\be)}  \Big)^{\prime}\big( \wh{\mu}_{a} \big)\Big\} 
				\pl{a=1}{N} \Big\{ 2\pi L \, \wh{\xi}^{\prime}\big( \wh{\la}_a \big)  \Big\} }
\cdot \f{ \prod_{a<b}^{N}(\wh{\la}_a-\wh{\la}_b)^2  \cdot  \prod_{a<b}^{N+1}(\wh{\mu}_a-\wh{\mu}_b)^2  }{ \prod_{a=1}^{N} \prod_{b=1}^{N+1} (\wh{\mu}_a-\wh{\la}_b)^2}
\enq
while $\wh{D}_{ \Phi^{\dagger} }^{ (\e{ex}) } $ grasps the most important part of the contributions to the asymptotic behaviour of the form factor, namely those issuing from the 
specific kind of excitations present in the excited state:
\beq
 \wh{D}_{ \Phi^{\dagger} }^{ (\e{ex}) } \; = \; 
 \pl{a=1}{n} \Bigg\{ \f{ \Big( \wh{\xi}_{\paa{\ell_a}}^{\,(\be)}\Big)^{\prime}\big( \wh{\mu}_{h_a} \big)  }{  \Big( \wh{\xi}_{\paa{\ell_a}}^{\,(\be)}\Big)^{\prime}\big( \wh{\mu}_{p_a} \big) } \Bigg\}
\cdot \det_n^2 \bigg[ \f{1}{ \wh{\mu}_{p_a} - \wh{\mu}_{h_a}  } \bigg]
\pl{b=1}{n} \Bigg\{   \f{  \prod_{a=1}^{N+1} (\wh{\mu}_a-\wh{\mu}_{p_b})^2 \cdot   \prod_{a=1}^{N} (\wh{\la}_a-\wh{\mu}_{h_b})^2  }
{\prod_{ \substack{ a=1  \\ a\not= h_b} }^{N+1} (\wh{\mu}_a-\wh{\mu}_{h_b})^2 \cdot   \prod_{a=1}^{N} (\wh{\la}_a-\wh{\mu}_{p_b})^2   }  \Bigg\} \;. 
\enq
Then, main idea consists in regularising the singular factors involving differences of variables that condense on $\intff{-q}{q}$ in the following way 
\beq
\f{ 1 }{ \wh{\mu}_a - \wh{\la}_b } \; = \; \f{1}{ \wh{\vp}\big( \wh{\mu}_a , \wh{\la}_b \big) } \cdot \f{ L }{ a - b   + \wh{F}_{ \{ \ell_a \} } \big( \wh{\la}_a \big)   } \qquad \e{with} \quad 
 \wh{\vp}(\mu,\la) \; = \; \f{ \mu - \la }{ \wh{\xi}_{ \{\ell_a\} }(\mu)  - \wh{\xi}_{ \{\ell_a\} }(\la)    } \;. 
\enq
%
%

The first term appearing in the \textit{rhs} is regular when $\mu_a\tend \la_b$. Furthermore, $ \wh{\vp} \tend \vp$ when $L \tend +\infty$. 
This allows to treat all the parts containing $\wh{\vp}$ exactly as it was done for the smooth part. The second term generates an explicit dependence on $L$. 
Furthermore, the dependence on $a$ and $b$ is rather explicit what allows one to take explicitly part of the products generated by this substitution. 
The simplest situation arises in the case of $ \wh{D}_{ \Phi^{\dagger} }^{ (\e{ex}) }$ so that I describe this factor first. 
After some algebra, one can recast it as 
\beq
\wh{D}_{ \Phi^{\dagger} }^{ (\e{ex}) } \; = \; \wh{E}_0 \cdot \wh{J}_0 \cdot \wh{\vartheta}_0 \cdot \det_n^2\Big[ \f{1}{p_a-h_b} \Big]\;. 
\enq
There, I have made use of the following abbreviations
\beq
 \wh{E}_0 \; = \; \f{ \pl{a<b}{n} \wh{\vp}^2\big( \wh{\mu}_{p_a},\wh{\mu}_{p_b} \big) \wh{\vp}^2\big( \wh{\mu}_{h_a}, \wh{\mu}_{h_b}\big) }
                                        { \pl{a , b }{n} \wh{\vp}^2\big( \wh{\mu}_{p_a},\wh{\mu}_{h_b}\big) 
                                        \cdot \pl{a=1}{n} \Big\{  \Big( \wh{\xi}_{\paa{\ell_a}}^{\,(\be)}\Big)^{\prime}\big( \wh{\mu}_{h_a} \big)  \Big( \wh{\xi}_{\paa{\ell_a}}^{\,(\be)}\Big)^{\prime}\big( \wh{\mu}_{p_a} \big)  \Big\} } 
\pl{b=1}{n}\Bigg\{ \f{ \wh{\vp}\big(\wh{\la}_{N+1},\wh{\mu}_{p_b}\big) }{ \wh{\vp}\big(\wh{\la}_{N+1},\wh{\mu}_{h_b}\big) }  
\cdot  \pl{a=1}{N+1} \bigg[ \f{ \wh{\vp}\big(\wh{\mu}_{a},\wh{\mu}_{p_b}\big)  \wh{\vp}\big(\wh{\mu}_{a},\wh{\mu}_{h_b}\big)  }
 {\wh{\vp}\big(\wh{\la}_{a},\wh{\mu}_{p_b}\big)  \wh{\vp}\big(\wh{\la}_{a},\wh{\mu}_{h_b}\big)  }  \bigg] \Bigg\}^2  \;, 
\enq
as well as 
\beq
\wh{J}_0 \; = \; \pl{b=1}{n} \pl{a=1}{N+1} \Bigg\{  \f{ \big[ a-h_b-\wh{F}_{\{\ell_a\}}\big(\wh{\la}_a\big) \big] \cdot \big[ a - p_b - \wh{F}_{\{\ell_a\}}\big(\wh{\la}_{p_b} \big) \big] }
{ \big[ a-h_b-\wh{F}_{\{\ell_a\}}\big(\wh{\la}_{h_b} \big) \big] \cdot \big[ a - p_b - \wh{F}_{\{\ell_a\}}\big(\wh{\la}_{a} \big) \big]   } \Bigg\}^2
\enq
and
\beq
  \wh{\vartheta}_0 \; = \; \pl{a=1}{n} \Bigg\{ \bigg( \f{\sin \big[ \pi \wh{F}_{\{\ell_a\}}\big(\wh{\la}_{p_a} \big)  \big]  }{  \pi   } \bigg)^2  \cdot 
\Ga^{2}\pab{ \{ p_a-N+  \wh{F}_{\{\ell_a\}}\big(\wh{\la}_{p_a} \big) \} , \{ p_a +\i 0^+ \} , \{ N+1-h_a - \wh{F}_{\{\ell_a\}}\big(\wh{\la}_{h_a} \big) \} , \{ h_a + \wh{F}_{\{\ell_a\}}\big(\wh{\la}_{h_a} \big) \} }
{ \{ p_a-N-1 +\i 0^+  \} , \{ p_a+  \wh{F}_{\{\ell_a\}}\big(\wh{\la}_{p_a} \big)  \}, \{ N+2-h_a \} , \{ h_a \}  } \Bigg\} \; . 
\enq
Having such a representation for $\wh{D}_{ \Phi^{\dagger} }^{ (\e{ex}) }$ at one's disposal, one can already take the thermodynamic limit of $\wh{E}_0$ and $\wh{\vartheta}_0$
 by using Lemma \ref{Lemme comportement asymptotique produit simple} and the uniform in $\R^2$ convergence
\beq
\wh{\vp}(\la,\mu) \; \limit{L}{+\infty} \; \vp(\la,\mu) \;. 
\enq
The only problematic factor is the one involving the products over the continuous integers in $\wh{J}_0$. The latter can be deal with, basically  
by applying Proposition \ref{Proposition contribution singular continuous products} below.

\begin{prop}\label{Proposition contribution singular continuous products}

Let $f \in \mc{C}^{1}\big(\R\big)$, $N,L$ be integers such that $\tf{N}{L}\to D \in \R^+$ and $h_L$ be a sequence of integers such that $\tf{h_L}{L}$ tends to some finite value $x\in \R$ when $L\to +\infty$.
Let $\xi(\la)$ be a strictly monotonous smooth function on $\R$
such that $\xi^{-1}(0)=-q$ and $\xi(D)=q$. Define
$\la_k=\xi^{-1}(k/L)$, $k \in \mathbb{Z}$. Then, it holds
 \begin{equation}\label{prod-sing}
 \lim_{\substack{ N,L\to\infty \\ N/L \to D }} \bigg\{  \prod_{k=1}^N \frac{k-h_L+f(\lambda_k)}{k-h_L+f(\lambda_{h_L})} \bigg\}
 =
 \exp\Bigg\{ \int\limits_{-q}^q \frac{ f(\lambda)-f\big(\xi^{-1}(x)\big) }{ \xi(\lambda)-x}  \xi^{\prime}(\lambda)\, \dd \lambda  \Bigg\}.
 \end{equation}
\end{prop}

The proof can be found in Appendix A of [$\bs{A5}$].

\noindent It thus remains to discuss $\wh{D}_{ \Phi^{\dagger} }^{ (0) }$. After some algebra, it can be recast as 
\beq
\wh{D}_{ \Phi^{\dagger} }^{ (0) } \; = \;  \f{ \pl{a<b}{N} \wh{\vp}^2\big( \wh{\la}_{a},\wh{\la}_{b} \big)  \pl{a<b}{N+1} \wh{\vp}^2\big( \wh{\mu}_{a}, \wh{\mu}_{b}\big) }
                                        { \pl{a =1 }{N} \pl{b =1 }{N+1} \wh{\vp}^2\big( \wh{\mu}_{b},\wh{\la}_{a}\big)  }
\pl{a=1}{N} \Bigg\{  \f{ \Big( \wh{\xi}_{\paa{\ell_a}}^{\,(\be)}\Big)^{\prime}\big( \wh{\la}_{a} \big)  }{ \wh{\xi}^{\prime}\big( \wh{\la}_{a} \big)  }  \Bigg\}
\cdot \wh{H}_0
\label{ecriture representation pratique de hat d0 N}
\enq
where 
\bem
 \wh{H}_0 \; = \; \f{1}{2\pi L}\Ga^{2}\pab{ N+1 ,1 + \wh{F}_{\{\ell_a\}}\big(\wh{\la}_{N+1} \big) }{  N+1 + \wh{F}_{\{\ell_a\}}\big(\wh{\la}_{N+1} \big)  }
 \pl{a=1}{N} \bigg( \f{\sin \big[ \pi \wh{F}_{\{\ell_a\}}\big(\wh{\la}_{a} \big)  \big]  }{  \pi   \wh{F}_{\{\ell_a\}}\big(\wh{\la}_{a} \big)   } \bigg)^2 \\
\times \pl{a<b}{N}  \bigg( 1  - \f{ \wh{F}_{\{\ell_a\}}\big(\wh{\la}_{a} \big) - \wh{F}_{\{\ell_a\}}\big(\wh{\la}_{b} \big)   }{ a - b } \bigg)^2
\cdot \pl{ \substack{a,b=1 \\ a\not= b } }{ N }  \bigg( 1  - \f{ \wh{F}_{\{\ell_a\}}\big(\wh{\la}_{a} \big)   }{ a - b } \bigg)^2 \;. 
\end{multline}

 Again the first set of products in \eqref{ecriture representation pratique de hat d0 N} is readily taken care of by means of Lemma \ref{Lemme comportement asymptotique produit simple}. 
It remains to deal with $\wh{H}_0$. The idea consists in separating the double products  between those pairs $(a,b)$ such that $a-b$ is "large" 
and those where the difference is "small". In the former case, one can treat, in a sense, the product perturbatively in that the difference of shift functions will be "small".
In the latter case, one can, effectively speaking, treat the difference of shift functions as a constant. 
More details can be found in Propositions 5.2 and 5.3 of  [$\bs{A6}$].

\subsection{The Fredholm determinants}
\label{SousSection propriete determinant Fredholm}

I will now present the result which provides one with the correct way of understanding the Fredholm determinants appearing in \eqref{formule explicite G+ thermo}
this in the case where the contour $\msc{C}_q$, as it has been described previously, does not exist. 
The below definition holds, as well, in the case of complex valued rapidities. 
Prior to stating the result, I need to introduce some new notations.
Given $\de>0$ and $a>0$, let
\beq
U_{\de} = \Big\{ z \in \Cx \; : \; \abs{\Im\pa{z}} < 2 \de \Big\} \qquad  , \qquad 
K_{a}= \Big\{ z \in \Cx  \; : \;  \abs{\Im \pa{z} } \leq \de  \; \e{and} \; \abs{\Re \pa{z} } \leq a \Big\} \;,
\label{definition strip U delta et compact KA}
\enq
and, given $\be_0 \in \Cx$, set
\beq
\bs{U}_{\be_0} = \paa{  z \in \Cx  \; : \;  10 \Re\pa{\be_0} \geq \Re\pa{z}\geq \Re\pa{\be_0} \;\; \e{and} \;\; \abs{\Im\pa{z}} \leq \Im\pa{\be_0}  } \;. 
\enq
Let, also,  $\mc{D}_{0,\eps}$ stands for the open disk of radius $\eps$ that is centred at $0$. I remind that, given a set $S$, $\ov{S}$ stands for its closure
and $\e{Int}(S)$ for its interior.

\begin{prop}
\label{Proposition Holomorphie en beta et rapidite part-trou det fred}

Let $m\in \mathbb{N}$ be fixed and $\eps, \de>0$ be small enough. Assume that one is given 
two  holomorphic function $\nu$ and $h$  on $U_{2\de}$,  such that 
\beq
h\pa{ U_{2\de} }\subset \paa{ z \; : \; \Re\pa{z} >0 } \quad and  \quad z \mapsto \Im \pa{h\pa{z}} \; is \; bounded \; on \; U_{2\de} \,.
\enq
\noindent Then, there exists
\begin{itemize}

\item[$\bullet$] $\be_0 \in \Cx$ with $\Re\pa{\be_0}>0$ large enough  and $\Im\pa{\be_0}>0$ small enough ; \vspace{1mm}
\item[$\bullet$] $\ga_0 > 0$ but small enough ; \vspace{1mm}
\item[$\bullet$]  a small counterclockwise loop $\msc{C}_q$ around $K_{q+\eps}$ and in $U_{2\de}$; \vspace{1mm}
\end{itemize}
such that, given $\nu_{\be}\pa{\la}=\nu\pa{\la}+ \i \be h\pa{\la} $, one has 
\beq
 \ex{-2 \i \pi \ga \nu_{\be}(\la) }-1 \not= 0  \qquad for\,  all \;\;  \la \quad on \; and \; inside \; \msc{C}_{q} 
 \quad and \; uniformly \; in \quad  ( \be , \ga )  \in \bs{U}_{\be_0} \times \mc{D}_{0,\ga_0} \; .
\enq
Moreover, given the integral kernel 
\beq
U\big[  \ga \nu_{\be} \big] \pab{  \{ \mu_{p_a} \}_1^n }{ \paa{\mu_{h_a}}_1^n  }(\om,\om^{\prime})
\nonumber
\enq
as defined by \eqref{formule noyau integral U}, there exists $\be_{0} \in \mathbb{H}^+$, $\Re(\be_0)>0$, such that the function
\beq
\mf{D}\big( \bs{z} \big)  = 
G\big(1- \ga \nu_{\be}\pa{-q} \big) G\big(2+  \ga \nu_{\be} \pa{q}\big)  \pl{a=1}{n} \pa{ \ex{-2 \i \pi \ga \nu_{\be}\pa{\mu_{h_a}}} - 1 } \; \cdot \;
\det_{ \msc{C}_q }\Big[ \e{id} + \ga \op{U}\big[  \ga \nu_{\be} \big]\pab{  \{ \mu_{p_a} \}_1^n }{ \paa{\mu_{h_a}}_1^n  }  \Big] \;
\label{appendix thermo lim FF fonction avec limite themor reg}
\enq
is holomorphic in $\bs{z} = \big( \{ \mu_{p_a} \}_1^n,\paa{\mu_{h_a}}_1^n,\be, \ga \big)$
belonging to $U_{\de}^{n} \times \e{Int}\big( K_{q+\eps}  \big)^n \times   \bs{U}_{\be_0} \times \mc{D}_{0,\ga_0}$, 
this uniformly in $0\leq n\leq m$.

$\mc{D}$ admits a (unique) analytic continuation to $ U_{\de}^{n} \times  \e{Int}\big( K_{q+\eps} \big)^n\times 
\paa{ z  \in \Cx \; : \; \Re\pa{z} \geq -\eps} \times \mc{D}_{0,1+\eps} $.

\end{prop}

The proof can be found in Section 3.3 of paper [$\bs{A4}$].

\subsection{Form factors of the critical $\ell$-classes}
\label{SousSection FF dans la classe critique ell}

I have mentioned earlier that the expressions for the large-$L$ behaviour of the form factors can be further simplified if one has more information on the 
integers $\{p_a\}$ and $\{h_a\}$ and, in particular, on their magnitude in respect to $N$. Due to the prominent role the $\ell$-critical classes play in the large-distance and long-time asymptotic
analysis of two and multi-point functions, I shall now discuss the reduction of the large-$L$ asymptotic behaviour 
of the conjugated field form factor in the case when the particle-hole integers describe an excitation belonging to an $\ell$-critical class, \textit{c}.\textit{f}. \eqref{ecriture decomposition locale part-trou close Fermi zone}. 
Effectively, such form factors are parametrised by the integer $\ell$, the values of the thermodynamic limit of the counting function associated to an 
$\ell$-critical class on the left or right Fermi boundary:
\beq
\mf{F}^{\pm}_{\ell} \; = \; \ell (Z(q)-1) \pm \f{1}{2} Z^{-1}(q) \;, 
\label{definition fct shift sur les bords}
\enq
and the sets of "local" left/right integers:
\beq
\{ p_{a;+} \}_1^{ n_{p;+} } \cup  \{ p_{a;-} \}_1^{ n_{p;-} }   \quad \e{and} \quad 
\{ h_{a;+} \}_1^{ n_{h;+} } \cup  \{ h_{a;-} \}_1^{ n_{h;-} }   
\label{ecriture ensembles locaux part trous}
\enq
where $p_{a;\pm}$ and $h_{a;\pm}$ appear in  \eqref{ecriture decomposition locale part-trou close Fermi zone}. 

Note that the expression for the values of the shift function on the Fermi boundaries follows from the identities 
\beq
Z(\la) \; = \; 1 + \phi(\la,-q) \, - \, \phi(\la,q) \quad \e{and} \quad
 Z^{-1}(q) \; = \; 1 + \phi(-q,q) \, - \, \phi(q,q)  \;.
\label{ecriture relation Z moins 1 et phi}
\enq
The first one is straightforward while the second one has been established in
\cite{KorepinSlavnovNonlinearIdentityScattPhase,SlavnovNonlinearIdentityScattPhase}.

\begin{cor}
\label{Corollaire cptmt FF sur etat critique ell}

Given an excited state belonging to the $\ell$-critical class and parametrised by the sets of local particle-hole integers \eqref{ecriture ensembles locaux part trous}, the form factor of the conjugate field admits the large-$L$ asymptotic expansion:
\beq
\f{  \abs{ \bra{\Psi\pa{ \{\mu_{\ell_a} \}_1^{N+1}  }} \Phi^{\dagger}\pa{0} \ket{ \Psi\pa{ \{ \la_a \}_1^N}  } }^2 }
{ \norm{\Psi\pa{ \{ \mu_{\ell_a} \}_{1}^{N+1} }}^2 \cdot  \norm{\Psi\pa{\{ \la_a \}_{1}^{N} }}^2     }  \; = \; \mc{A}^{(\ell)}\big( \{ p_{a;\pm} \} ; \{ h_{a;\pm} \}  \big) 
\cdot \Bigg( 1 + \e{O}\bigg( \f{\ln L + \e{max} ( \{ p_{a;\pm} ; h_{a;\pm} \} ) }{ L } \bigg) \Bigg)
\enq
where the coefficient $\mc{A}^{(\ell)}\big( \{ p_{a;\pm} \} ; \{ h_{a;\pm} \}  \big) $ reads 
\bem
\mc{A}^{(\ell)}\big( \{ p_{a;\pm} \} ; \{ h_{a;\pm} \}  \big)  \; = \; \f{ \big| \mc{F}_{\ell}\big(  \Phi^{\dagger} \big) \big|^2 }{ L^{ \big( \mf{F}^{+}_{\ell} +\ell +1 \big)^2 +   \big( \mf{F}^{-}_{\ell} +\ell \big)^2 } }
\cdot \,\frac{ G^2 \big( 2 + \mf{F}^{+}_{\ell} \big) \cdot G^2\big( 1 - \mf{F}^{-}_{\ell} \big) }
  {  G^2 \big( 2 + \ell + \mf{F}^{+}_{\ell} \big) \cdot G^2\big( 1 - \ell - \mf{F}^{-}_{\ell} \big) } \\
 \times
R_{n_{p;+},n_{h;+}}\big( \{p_{a;+} \} ; \{ h_{a;+} \} \mid  \mf{F}^{+}_{\ell} +1\big) \;
R_{n_{p;-},n_{h;-}}\big( \{p_{a;-} \} ; \{ h_{a;-} \} \mid  - \mf{F}^{-}_{\ell} \big)
\label{ecriture limite sectel ell critique FF}
 \end{multline}
and
\begin{equation}\label{def-R}
 R_{n_p,n_h}\big( \{ p_a \} , \{ h_a \} \mid \mf{F} \big)= \left(\frac{\sin\big[\pi \mf{F} \big] }{\pi}\right)^{ 2 n_{h} } 
\frac{\prod\limits_{a>b}^{n_p}(p_a-p_b)^2 \cdot \prod\limits_{a>b}^{n_h} (h_a-h_b)^2 }
 {\prod\limits_{a=1}^{n_p} \prod\limits_{b=1}^{n_h}(p_a+h_b-1)^2} \;
\prod_{a=1}^{n_p}\frac{ \Gamma^2\big( p_a+\mf{F} \big) }{ \Gamma^2\big( p_a \big) }
\prod_{a=1}^{n_h}\frac{\Gamma^2\big( h_a - \mf{F} \big) }{ \Gamma^2 \big( h_a \big)}.
  \end{equation}

\end{cor}

The explicit expression for the amplitude $\mc{F}_{\ell}\big(  \Phi^{\dagger} \big)$ can be readily inferred from the expression for the smooth and discrete parts. 
Since it will play no role in the following, I refer to Section 3 of \cite{KozKitMailSlaTerThermoLimPartHoleFormFactorsForXXZ}. 
Note that $\msc{F}_{\ell}\big(  \Phi^{\dagger} \big)$ can be interpreted as the properly normalised in the volume $L$ thermodynamic limit of the
form factor of the conjugated field operator taken between the ground state and the fundamental representative of the excited state in the $\ell$ critical class. 
More precisely, let $ \{ \ell_a \}_1^{N+1}$ be a set of integers giving rise to an excited state in the  $\ell$ critical class  parametrised by the integers:
\beq
\left\{   \ba{cc} \Big\{  \{ p_{a;+}  = a \}_1^{\ell} \; ;  \; \{ \emptyset \}  \Big\} \bigcup
\Big\{  \{ \emptyset \} \; ; \;  \{ h_{a;-} = a \}_1^{\ell}  \Big\}  &
				\e{if} \; \ell \geq 0   \vspace{2mm} \\ 
		\Big\{  \{ \emptyset \} \; ; \;  \{ h_{a;+}  = a \}_1^{-\ell}   \Big\} \bigcup
\Big\{    \{ p_{a;-} = a \}_1^{-\ell}  \; ; \; \{ \emptyset \}  \Big\}   &
				\e{if} \; \ell \leq 0 \ea \right. 		\; . 
\enq
Then, it follows from \eqref{ecriture limite sectel ell critique FF}, that 
\beq
\lim_{L\tend + \infty} \Bigg\{  L^{ \big( \mf{F}^{+}_{\ell} +\ell \big)^2 +   \big( \mf{F}^{-}_{\ell} +\ell \big)^2 } \cdot 
\f{  \abs{ \bra{\Psi\pa{ \{\mu_{\ell_a} \}_1^{N+1}  }} \Phi^{\dagger}\pa{0} \ket{ \Psi\pa{ \{ \la_a \}_1^N}  } }^2 }
{ \norm{\Psi\pa{ \{ \mu_{\ell_a} \}_{1}^{N+1} }}^2 \cdot  \norm{\Psi\pa{\{ \la_a \}_{1}^{N} }}^2     }  \Bigg\}  \; = \; 
 \big| \mc{F}_{\ell}\big(  \Phi^{\dagger} \big)  \big|^2 \;. 
\enq

\section{Large volume behaviour of the form factors in the massive regime of the XXZ chain}
\label{Section Cpt large L des FF XXZ massif}

I have mentioned in the introduction that the XXZ Hamiltonian is a massive model as soon as the anisotropy exceeds one. 
Actually, the model in its massive phase posses a two-fold degenerated ground state. 
Starting from the model in finite-volume, one can obtain \cite{KMTFormfactorsperiodicXXZ} determinant representations for the form factors of its local operators, the spin
operators $\sg_a^{x}$, $\sg_a^{y}$ and $\sg_a^{z}$ acting non-trivially on the $a^{\e{th}}$ site of the chain as the corresponding Pauli matrices. 
Building on these expressions, Izergin, Kitanine, Maillet and Terras computed the infinite volume limit of $| \bra{\Psi_{G;1} } \sg_1^{z} \ket{\Psi_{G;2} }|^2$,
where $\Psi_{G;\iota}$, $\iota=1,2$ are the two ground states. In the joint work with M. Dugave, F. G\"{o}hmann and J. Suzuki [$\bs{A1}$], 
we have pushed these results much further and obtained the large-volume asymptotic behaviour of all form factors of the form $| \bra{\Psi_{G;\iota} } \sg_1^{z} \ket{\Psi_{ \e{ex} } }|^2$, $\iota =1,2$
where  $\ket{\Psi_{ \e{ex} } }$ refers to the excited states. 
These form factors exhibit a integer power-law decay $L^{-n_h}$ in the volume. The exponent $n_h$ corresponds to the number of continuous modes -the holes- 
that describe the given excited state in the $L\tend +\infty$ limit. 

To be more precise about the structure of the answer, I would like to recall that the two ground states $\ket{\Psi_{G;\iota} } $ of the XXZ spin-$\tf{1}{2}$
chain in the massive regime belongs to the zero magnetisation sector
\cite{Yang-YangXXZproofofBetheHypothesis}. Under very reasonable hypothesis, on can show that, in the large-$L$ limit, the
excited states $\ket{\Psi_{ \e{ex} } } \, = \, \ket{ \{\chi_a\}_1^{n_{\chi}} ; \{\nu_{h_a}\}_1^{n_h}  }$
having a \textit{finite} excitation energy above the ground state 
are parametrised by hole-rapidities $\{\nu_{h_a}\}_1^{n_h}$ and
complex roots $\{\chi_a\}_1^{n_{\chi}}$. 
In the thermodynamic limit, the hole-rapidities become free parameters belonging to $\intff{-\tf{\pi}{2}}{\tf{\pi}{2}}$
while the value of the complex roots are discrete and given as solutions to a set of higher-level
Bethe Ansatz equations \cite{BabelondeVegaVialletStringHypothesisWrongXXZ,%
VirosztekWoynarovichStudyofExcitedStatesinXXZHigherLevelBAECalculations} which depend on the set of hole parameters $\{\nu_{h_a}\}_1^{n_h}$
that are associated with the state:
\beq
\mc{Y}_{0}\big(\, \chi_a\mid  \{\chi_a\}_1^{n_{\chi}} ;  \{\nu_{h_a}\}_1^{n_h} \big) \, = \, 0 \; , \;  a=1,\dots, n_{\chi} \; . 
\label{ecriture HLBAE massive XXZ}
\enq
As a result of the asymptotic analysis of the determinant expressions for the form factors,  it was shown in  [$\bs{A1}$]
that the properly normalised ground to excited state  form factor of the $\sg_1^z$ operator 
admits an interpretation in terms of a form-factor density
$\mc{F}^{(z)}\big( \{\nu_{h_a}\}_1^{n_h} ; \{\chi_a\}_1^{n_{\chi}} \big)$:
\beq
\Big|\bra{\Psi_{G;\iota} } \sg^z \ket{ \{\chi_a\}_1^{n_{\chi}} ;  \{\nu_{h_a}\}_1^{n_h}  } \Big|^2 \, = \, \pl{a=1}{n_h}\bigg\{ \f{1}{L \rho(\nu_{h_a} ) } \bigg\}
\cdot \f{ \Big( \mc{F}^{(z)}\big( \{\nu_{h_a}\}_1^{n_h} ; \{\chi_a\}_1^{n_{\chi}} \big)\, \Big)^2  }
{ \det_{n_{\chi}} \Big[  \f{ \Dp{} }{ \Dp{} u_b } \mc{Y}_{0}\big( u_a\mid  \{u_c\}_1^{n_{\chi}} ;  \{\nu_{h_a}\}_1^{n_h} \big)  \Big]_{\mid u_a=\chi_a}  } \cdot \Big(1\, + \, \e{O}\big( L^{-1} \big)  \Big)\;. 
\enq
Here the function $\rho$ represents the density of real  Bethe roots in the interval $\intff{-\tf{\pi}{2}}{\tf{\pi}{2}}$. 
The Jacobian of the function $\mc{Y}_{0}$ generating the higher-level
Bethe Ansatz equations can be thought of as a higher-level expression
for the norm of an excited state. It provides one with the proper normalisation factor that is intimately attached to the discrete measure  
corresponding to a summation over all the solutions to the higher level Bethe Ansatz equations \eqref{ecriture HLBAE massive XXZ}.

The clear interpretation of the formula in terms of a form factor density allows one to write down, under appropriate hypothesis of convergence, form factor expansions for the model in infinite volume. 
This strongly contrasts with the model in its massless phase where the non-integer power-law decay of the form factors of local operators does not allow
one to write down, in any meaningful way, the form factor expansion directly in the infinite volume limit of the model.

\section*{Conclusion}

The main result discussed in this chapter consists in providing the typical form of the large-$L$ behaviour of form factors in massless quantum integrable models. 
I have discussed this form on the example of the conjugated field form factor in the non-linear Schr\"{o}dinger model but the 
overall feature of the asymptotic behaviour are actually true for other local operators and also other quantum integrable models. 
The general feature is that the form factors decay as a non-integer power in the volume. This renders the treatment of such models
complicated directly in the infinite volume limit. As it will become clear later on, the expressions for the large-$L$
behaviour of the form factors are at the root of understanding the large-distance and long-time asymptotic behaviour of multi-point correlation functions in massless models.

Very briefly, in the last Section, I have discussed the form taken by the large-$L$ asymptotic behaviour of  form factors of local operators in massive models. 
These decay as an integer power of the volume. 
This is in strong contrast with the massless case and explains why it was possible to treat massive models directly in the continuum.

\chapter{Large-distance and time asymptotics of two-point functions: the multidimensional Natte series approach}
\label{Chapitre AB grand tps et distance via series de Natte}

As follows from the discussion carried out in the last chapter, one is able to represent the form factors of local operators
in a wide class of quantum integrable models solvable by the algebraic Bethe Ansatz as finite-size determinants  \cite{KMTFormfactorsperiodicXXZ,OotaInverseProblemForFieldTheoriesIntegrability}.
It has been shown by Korepin and Slavnov \cite{KorepinSlavnovTimeDepCorrImpBoseGas} in 1990 that, for free fermion equivalent models,
it is possible to build on such a determinant structure so as to explicitly sum-up the form factor expansion of two-point correlation functions.
 Slavnov  generalised \cite{SlavnovPDE4MultiPtsFreeNLSM} the setting in 1996 
to the case of a specific multi-point correlation function arising in the  free fermion limit of the non-linear Schr\"{o}dinger model. 
Within the free fermionic form factor summation approach, once the infinite volume limit is taken, a two-point function is represented by a Fredholm determinant (or some finite rank perturbation thereof) of an integrable
integral operator $\e{id}+ \op{V}_{x}$ acting on some contour $\msc{C}_{\op{V}}$ that is determined by the properties of the model.
For models subject to periodic boundary conditions, the operator $\op{V}_{x}$ depends on the distance separating the two operators as well
as on the difference of time evolution between them. In fact, for a general free-fermion equivalent model at zero temperature, the 
integral operator  $\e{id}+ \op{V}_{x}$ associated with the form factor expansion of the time and space dependent two-point functions
acts on a compact interval $\intff{-q}{q}$ of $\R$ while its integral kernel $V_x(\la,\mu)$  fits into the class of kernels 
\beq
V_x(\la,\mu) \;  =  \; 4  \f{ \sin [ \pi\nu(\la) ]  \sin [\pi\nu(\mu) ] }{ 2 \i \pi (\la-\mu) } e(\la) e(\mu)
\Big\{ O_{\msc{C}_{ \op{V} } }[\nu,e^{-2}](\la) - O_{\msc{C}_{ \op{V} } }[\nu,e^{-2}](\mu)  \Big\} \; \; ,  \quad e(\la) \, =\,  \ex{ -\i \f{ x u(\la) }{2} - \f{ g(\la) }{2}} \; .
\label{definition noyau GSK FF}
\enq
The above kernel is expressed in terms of two auxiliary functions $\nu$ and $e$. 
The function $\nu$ encodes the fine structure of the excitations above the ground state, $e$ is some oscillating function while  
 $O_{ \msc{C}_{ \op{V} } } $ stands for the integral transform:
\beq
 O_{\msc{C}_{ \op{V} } }[\nu,E](\la) \; = \;  \i  \Int{ \msc{C}_{ \op{V} } }{}  \f{ E\pa{s} }{ s-\la } \cdot \f{ \dd s }{ 2\pi }
 \; + \;   \f{ E\pa{\la} }{  \ex{-2\i \pi \nu(\la) } -1 }   \;.
\label{definition fonction E+}
\enq
The functions $\nu$, $u$ and $g$ just as the integration curve $\msc{C}_{ \op{V} }$ appearing in \eqref{definition fonction E+} depend on the specific model 
that one considers. In fact, for free-fermion equivalent models, $\nu$  and $g$ are constants while $u$ takes a simple form. The contour $\msc{C}_{ \op{V} }$ corresponds, in this case,
to a deformation of the real axis. 
The asymptotic analysis of Fredholm determinants $\det\big[\e{id} + \op{V}_x \big]$ for various cases of interest to free fermion equivalent models has been carried out, in particular,
in the works \cite{CheianovZvonarevZeroTempforFreeFermAndPureSine,ItsIzerginKorepinVarguzinTimeSpaceAsymptImpBoseGaz}.

When considering quantum integrable models that are away of their free-fermion point, it is not possible to represent correlation functions 
in terms of Fredholm determinants  or finite rank perturbations thereof any more. However, as first observed in the 
work \cite{KozKitMailSlaTerXXZsgZsgZAsymptotics} and later developed in  [$\bs{A7},\bs{A9},\bs{A10}$] 
one can represent the correlation functions through series of multiple integrals of the type 
(or some small variation this form)
\beq
 \sul{n \geq 0}{} \f{1}{n!}  \Int{-q}{q} \pl{a=1}{ n } \f{\dd \la_a}{2 \i \pi}  
\Oint{ \msc{C}_{\bs{z}} }{} \hspace{-1mm} \pl{a=1}{ n } \f{\dd z_a}{2 \i \pi}   \Oint{ \msc{C}_{\bs{y}} }{} \hspace{-1mm} \pl{a=1}{  n } \f{\dd y_a}{2 \i \pi}   
 \pl{a=1}{n } \bigg\{ \f{   \ex{\i x [ u(y_a) -  u(\la_a)] }  }{ z_a -\la_a } \bigg\} \cdot \det_{n}\Big[ \f{ 1 }{ z_a-\la_b } \bigg]
\cdot \mc{F}_n \pab{ \la_1,\dots,\la_n }{ y_1,\dots, y_n } \;. 
\label{definition serie de Fredholm multidimensionelle}
\enq
There, $\msc{C}_{\bs{z}}$ and  $\msc{C}_{\bs{y}}$   are some contours in $\Cx$ encircling  $\intff{-q}{q}$, 
$u$ is a holomorphic function in a neighbourhood of $\intff{-q}{q}\cup \msc{C}_{\bs{y}}$ and is real valued on $\intff{-q}{q}$
while $\mc{F}_n$ is some  holomorphic function in a neighbourhood of the integration contour satisfying to some technical hypothesis. 
I call such series multidimensional Fredholm series. This name originates from the fact that, at the free fermion point,
these series degenerate to the Fredholm series associated with some integral operator. 
There are quite a few open question relatively to multidimensional Fredholm series, 
a general theory allowing one to set sharp convergence criterion in particular. 
 However convergence is a question that can be considered on separate
grounds from the one of the large-$x$ asymptotic expansion of the object that the series represent. Still, in order to carry a more concrete
discussion, it is easier to think in terms of the series than in terms of the object it represents. Thus, for the time being, I will assume the
series to be convergent.

I showed in the works  [$\bs{A7},\bs{A9}$]  that 
the analysis of the large-$x$ behaviour of such series boils down to one of an effective generalised free fermion model. 
 As a consequence, the large-$x$ analysis boils down to one of Fredholm determinants of operators $\e{id}+\op{V}_x$ with $\op{V}_x$ given by \eqref{definition noyau GSK FF}. 
However, the price is that one has to consider functions $\nu$ and $e$ that are quite general in contrast to the free fermion point.
The approach of [$\bs{A7},\bs{A9}$]  shows that kernels of the type 
\eqref{definition noyau GSK FF} appear as a natural basis of special functions
allowing one to represent series of the type \eqref{definition serie de Fredholm multidimensionelle} and thus the correlation functions of a wide class of interacting (\textit{i.e.} away from their free fermion point)
integrable models as the action of a certain multidimensional flow on their associated Fredholm determinants. The main motivation for my study [$\bs{A8}$] of the time-dependent generalised sine 
kernel  \eqref{definition noyau GSK FF} was to obtain a convenient and effective representation - that I called the Natte series\symbolfootnote[3]{The origin of this name issues from
the so-called pig-tail (or braid) hairstyle that is called Natte in French. A braid is a specifically
ordered reorganization of the loose hair-do style. Similarly, the Natte series reorganizes the Fredholm series in a very specific way, so that
the resulting representation is perfectly fit for carrying out an asymptotic expansion.}- 
for the associated Fredholm determinant. This series is built in such a way that one can \textit{read off} the large-$x$ asymptotic behaviour of the
determinant. Besides, it also allows one to access to the structure of the all order large-$x$ asymptotic expansion of the determinant. 
The Natte series for the Fredholm determinant  allows one to recast
the representation of \eqref{definition serie de Fredholm multidimensionelle} in terms of the action of a multidimensional flow 
 as some explicit new series of multiple integrals. The latter provides one with a new type of
representation for two-point functions.  The main feature of this representation is that it is built in such a way that one is able to \textit{read-off} the asymptotic behaviour of the
correlators out of it.
Therefore, the results established in my paper [$\bs{A8}$] provide one of the fundamental tools that are necessary for carrying out
the long-distance and large-time asymptotic analysis of the two-point functions in massless integrable models
which was carried out in [$\bs{A7},\bs{A9}$].

I will discuss the method of multidimensional flows and the results it begets in this chapter. 
In Section \ref{soussection hypotheses}, I will present results relative to the large-$x$ asymptotic behaviour of the Fredholm determinant of the operator $\e{id}+\op{V}_x$ with $\op{V}_x$ given by \eqref{definition noyau GSK FF}. 
On the one hand, I will provide the full structure  of its large-$x$ asymptotic expansion and, on the other hand, I will present its Natte series expansion. 
Finally, I will also discuss the Natte series representation for the Fredholm determinant associated with a rank one perturbation of the operator  $\e{id}+\op{V}_x$. 
This last result will be of main interest to the study of the large-distance and long-time asymptotic behaviour of the two-point function
considered in this chapter. In Sections \ref{Section method and main results} and \ref{Section Facteurs de Formes series}, I will focus on  
discussing the method of multidimensional flows on the example of the zero temperature reduced density matrix in the non-linear Schr\"{o}dinger model. This correlator is defined, for the model in finite volume $L$, as
\beq
\hspace{2cm} \rho_N \! \pa{x,t}\equiv \bra{\Psi\big( \{ \la_a \}_1^N \big) }\Phi\pa{x,t}\Phi^{\dagger}\!\pa{0,0}\ket{ \Psi\big( \{ \la_a \}_1^N \big) }
\cdot  \Norm{ \Psi\Big( \{ \la_a \}_1^{N} \Big)  }^{-2}  \;.
\label{definition reduced density matrix}
\enq
Above, the parameters $\{ \la_a \}_1^N$ correspond to the set of Bethe roots giving rise to the model's ground state. 

In Section \ref{Section method and main results}, I will provide a quick sketch of the main ideas behind the multidimensional flow method. 
Then, in Subsection \ref{Theorem asymptotique rho},  I will present the results for the large-distance and long-time asymptotic behaviour of the field-conjugated field
correlator. In Section \ref{Section Facteurs de Formes series}, I will explain how, starting from the form factor series expansion of $\rho_N(x,t)$, one can construct, in the infinite-volume limit,
the multidimensional Fredholm series representation for the field-conjugated field correlator. This construction builds on certain technical hypothesis of convergence of auxiliary series and relies 
on the \textit{per se} multidimensional deformation flow method. I will then provide the multidimensional Natte series representation for $\rho_N(x,t)$, directly in the infinite volume.


\section{Large-$x$ asymptotic behaviour of an auxiliary Fredholm determinant}
\label{soussection hypotheses}

\subsubsection{Hypothesis}

The characterisation of the large-$x$ asymptotic behaviour of $\det\big[\e{id} +\op{V}_x\big]$, or of the determinant of some small rank perturbations of $\op{V}_x$, 
holds provided that the functions $u$, $g$ and $\nu$ entering in the description of
the integrable kernel \eqref{definition noyau GSK FF} satisfy to a certain amount of hypotheses: \vspace{1mm}
\begin{itemize}
\item[$\bullet$] There  exists $\de>0$ such that $u$ and $g$ are both holomorphic on strip $U_{ \tf{\de}{2} }$, \textit{c.f.} \eqref{definition strip U delta et compact KA}.

\vspace{2mm}
\item[$\bullet$] The function $u$ is real valued on $\R$ and has a unique saddle-point in $U_{ \tf{\de}{2} }$ located at $\la_0 \in \R\setminus \{ \pm q \}$. This saddle-point
is a zero of $u^{\prime}$ with multiplicity one, that is to say $ \exists \, ! \; \la_0 \in \R \; : \; u^{\prime}\!\pa{\la_0}=0$ and $u^{\prime\prime}\!\pa{\la_0}<0$. \vspace{2mm}
\item[$\bullet$] $u$ is such that, given any $\eta>0$, $\ex{ \i \eta u(\la) }$ decays exponentially fast in $\la$ when $\pm \Im\pa{\la} >\de>0$
for any fixed $\de>0$,  and $\Re(\la) \tend  \mp \infty$ with $ \la \in U_{ \tf{\de}{2} } $. \vspace{2mm}
\item[$\bullet$] The function $\nu$ is holomorphic in an open neighbourhood $\mc{N}_q \subset U_{\tf{\de}{2}}$ of $\intff{-q}{q}$. It is also such that $\sin \pac{\pi \nu\pa{\la}}$ has no zeroes in some open neighbourhood of $\intff{-q}{q}$
lying in $U_{ \tf{\de}{2} }$. \vspace{2mm}
\item [$\bullet$]The function $\nu$ has a "sufficiently" small real part at $ \pm q$, \textit{i}.\textit{e}. $\abs{\Re\pac{\nu\pa{\pm q}} } <\tf{1}{2}$. \vspace{2mm}
\item[$\bullet$] The contour $\msc{C}_{V}$ is as depicted in Figure \ref{Figure contour CV}. 
\end{itemize}

\begin{figure}[h]
\begin{center}

\begin{pspicture}(6,4)

\psline[linestyle=dotted, dash=3pt 2pt]{->}(-1,2)(7.8,2)
\psline[linestyle=dotted, dash=3pt 2pt]{->}(-1,0.5)(7.8,0.5)
\psline[linestyle=dotted, dash=3pt 2pt]{->}(-1,3.5)(7.8,3.5)

\rput(7.2,2.2){$\R$}
\rput(7.2,3.8){$\R + \i \de$}
\rput(-0.7,0.8){$\R - \i \de$}


\psdots(2,2)(4,2)

\rput(2.1,1.8){$-q$}
\rput(3.8,1.8){$q$}

\pscurve(-1,3.3)(-0.5,3.3)(0,3.3)(0.7,3)(0.9,2.9)(1,2.8)(1.2,2)(1.5,2)(2,2.6)(4,2.8)(4.5,2.6)(5,2)(5.5,2)(6,1.4)(6.5,1)(7,0.8)(7.5,0.9)

\psline[linewidth=2pt]{->}(4,2.8)(4.1,2.8)

\rput(7.3,1.2){$\msc{C}_{ \op{V} }$}

\end{pspicture}

\caption{The contour $\msc{C}_{ \op{V} } \subset U_{ \tf{\de}{2} }$.}
\label{Figure contour CV}
\end{center}
\end{figure}
%
%
%


\subsection{The Natte series representation of the Fredholm determinant}

In order to state the result, I first need to introduce a few auxiliary functionals that arise in the description of the Natte series. 
The functional 
\beq
\mc{V}^{\pa{0}}_x \pac{\nu,u,g} \; = \;  \f{ \mc{B}\pac{\nu,u} }{ x^{ \nu^2(q) + \nu^2(-q) } } \cdot
\exp\Bigg\{ \int_{-q}^{\,q} \pac{ \i x u^{\prime}\!\pa{\la} + g^{\prime}\!\pa{\la} } \nu\pa{\la} \dd \la \Bigg\}
\; ,
\enq
describes an overall pre-factor of the Natte series. It is expressed by means of the 
auxiliary functional $\mc{B}\pac{\nu,u}$ which takes the form
\beq
\mc{B}\pac{\nu,u} \; = \;   
\f{ G^2\big( 1+\nu\pa{q} \big) \cdot G^2\big( 1-\nu\pa{-q} \big)  \cdot \ex{ \i \f{\pi}{2} \pa{\nu^2\!\pa{q}-\nu^2\!\pa{-q}}}  \cdot   \ex{C_1\pac{\nu}} }
{ \pac{2q \big(u^{\prime}\!\pa{q}+ \i 0^+ \big) }^{\nu^2\!\pa{q}} \cdot \pac{2q \big(u^{\prime}\!\pa{-q} + \i 0^+ \big) }^{ \nu^2\!\pa{-q} }     } \cdot  \pa{2\pi}^{ \nu\pa{-q} - \nu\pa{q}}  \; ,
\label{definition fonctionnelle B}
\enq
where $G$ is the Barnes function \cite{BarnesDoubleGaFctn2} and 
\beq
C_1\pac{\nu}= \f{1}{2} \Int{-q}{q}  \f{\nu^{\prime}\pa{\la} \nu\pa{\mu}-\nu^{\prime}\pa{\mu} \nu\pa{\la} }{\la-\mu} \cdot \dd \la \dd \mu
+\nu\pa{q} \Int{-q}{q} \f{\nu\pa{q} - \nu\pa{\la}}{q-\la} \dd \la  +\nu\pa{-q} \Int{-q}{q} \f{\nu\pa{-q} - \nu\pa{\la}}{q+\la}  \dd \la \; .
\label{definition fonctionnelle C1}
\enq

Note that the $\i 0^+$ regularization is important only when  $u^{\prime}(q)<0$ or  $u^{\prime}(-q)<0$ and  allows one for a non-ambiguous definition
of the power-laws appearing above. I do stress that I always understand powers as defined through the principal branch of the logarithm, \textit{viz}.
$\e{arg} \in \intoo{-\pi}{\pi}$.

\begin{theorem}
\label{theorem representation serie de Natte}

Under the mentioned hypothesis, the Fredholm determinant $\det\big[ \e{id}  + \op{V}_x \big]$  admits the Natte series expansion:
\beq
\det\big[ \e{id}  + \op{V}_x \big]  \; = \;  \mc{V}^{\pa{0}}_x\pac{\nu,u,g} \cdot 
\paa{1 + \sul{n \geq 1}{} \mc{H}_n\pac{\nu, \ex{g}, u}  }\; .
\label{ecriture serie de Natte intro}
\enq
The coefficients of the Natte series expansion  $\mc{H}_n\pac{\nu, \ex{g}, u }$ are certain functional of $\nu, \ex{g}$ and $u $.
Their explicit characterisation, in terms of combinatorial sums, can be found in equation (7.10) of  [$\bs{A8}$]. 
These functionals satisfy to the estimates  
\beq
\abs{ \mc{H}_n\pac{ \nu,\ex{g},  u} } \leq  \big[ m(x) \big]^n \; , \qquad 
\enq
with the control function $ m\pa{x} $ being $n$-independent and growing as $m(x)=\e{O}\pa{x^{-w}} $ with 
\beq
\hspace{-5mm} w=\f{3}{4} \e{min}\pa{\tf{1}{2}, 1-\wt{w}-2 \max_{\eps=\pm} \abs{ \Re\nu\pa{\eps q} }} \quad and \quad 
\wt{w} = 2 \sup\paa{ \abs{\Re \pac{ \nu\pa{\la}-\nu\pa{\eps q}}} \; : \; \abs{\la - \eps q}=\de \;, \; \eps=\pm } \;, 
\enq
where  $\de>0$ is taken small enough. These bounds ensure the absolute convergence of the Natte series provided that $x$ is large enough.

The functionals $\mc{H}_n\pac{\nu,\ex{g}, u }$ take the form
\bem
\mc{H}_n\pac{\nu,\ex{g}, u }  \; = \;  \mc{H}_{n}^{\pa{\infty}}\pac{\nu,\ex{g}, u}
\; + \;  \sul{ m=-\pac{\f{n}{2}} }{ \pac{\f{n}{2}} }
\f{ \ex{ \i x m \pac{u\pa{q}-u\pa{-q}}} }{ x^{2 m \pac{\nu\pa{q}+\nu\pa{-q}}} } \mc{H}^{\pa{m}}_n\pac{\nu, \ex{g},u} \\
+ \; \sul{ b=1 }{ \pac{\f{n}{2}} } \sul{p=0}{b} \sul{m= b - \pac{\f{n}{2}} }{\pac{\f{n}{2}}-b}
 \f{ \ex{ \i x m \pac{u\pa{q}-u\pa{-q}}} }{ x^{2 m \pac{\nu\pa{q}+\nu\pa{-q}}} }
 \cdot x^{\f{b}{2}} \f{ \ex{ \i x \bs{\eta} \pac{b u\pa{\la_0}-p u\pa{q}-\pa{b-p}u\pa{-q}  }   } }
{ x^{ 2\bs{\eta} \pa{b-p}\nu\pa{-q} - 2\bs{\eta} p  \nu\pa{q}} }
\cdot \mc{H}^{\pa{m,b,p}}_n\pac{\nu,\ex{g}, u} \;.
\label{ecriture serie Natte detaille pour chaque Hn Intro}
\end{multline}
There $[ \cdot ]$ stands for the floor function and I have adopted the convention that 
\beq
\bs{\eta}=1 \quad in \, the \, space-like \, regime \qquad \bs{\eta}=-1 \quad  in \, the\, time-like \, regime \; .
\label{definition parametre eta space et time like}
\enq
In this decomposition, one has that $\mc{H}_{n}^{\pa{\infty}}\pac{\nu,\ex{g}, u}=\e{O}\pa{x^{-\infty}}$
only produces exponentially small in $x$ corrections to $\mc{H}_n\pac{\nu,\ex{g}, u } $ while the functionals $\mc{H}^{(m)}_n\pac{\nu, \ex{g},u}$ and $\mc{H}^{\pa{m,p,b}}_n\pac{\nu,\ex{g}, u}$ admit the asymptotc expansions
\beqa
\mc{H}_n^{\pa{m}}\pac{\nu, \ex{g},u} & \thicksim & \sul{r \geq 0}{} \mc{H}_{n; r}^{\pa{m}}\pac{\nu, \ex{g},u}  \qquad with \qquad
\mc{H}_{n; r}^{\pa{m}}\pac{\nu, \ex{g},u} = \e{O}\paf{ \pa{\log x }^{n+r -2m} }{ x^{n+r} } \; , \nonumber \\
\mc{H}^{\pa{m,b,p}}_n\pac{\nu,\ex{g}, u} & \thicksim & \sul{r \geq 0}{} \mc{H}_{n;r}^{\pa{m,b,p}}\pac{\nu,\ex{g}, u} \qquad with \qquad
\mc{H}_{n;r}^{\pa{m,b,p}}\pac{\nu,\ex{g}, u} = \e{O}\paf{ \pa{\log x }^{n+r -2\pa{m+b} } }{ x^{n+r} } .
\label{ecriture forme DA des fonctionnelle Hn}
\eeqa

\end{theorem}

This Natte series expansion is proven in Section 7 of [$\bs{A8}$]. 
The various estimates of the building blocks of the Natte series provide one with the structure of the asymptotic expansion of the Fredholm determinant.
In do stress that this asymptotic expansion goes beyond the type usually encountered for higher transcendental functions. Indeed, 
the large-$x$ asymptotic expansion contains a tower of different fractional powers of $x$, each appearing with its own oscillating pre-factor.
Once that one has focused attention on a given phase factor and associated decay as a fractional power in $x$, then
the corresponding functional coefficients $\mc{H}^{\pa{m}}_n[\nu,\ex{g}, u]$ or $\mc{H}^{\pa{m,b,p}}_n[\nu,\ex{g}, u]$ admit an asymptotic 
expansion in the more-or-less standard sense. That is to say, they admit an asymptotic expansion
into a series whose $r^{\e{th}}$ term can be written as $\tf{P_{r+n}\pa{\log x}}{x^{n+r}}$ with $P_{r+n}$ being a polynomial of degree at most $r+n$.
One of the consequences of such a structure is that an oscillating term that appears in a sense "farther" (large values of $n$) in the asymptotic series 
might be dominant in respect to a non-oscillating term present in the "lower" orders of the Natte series.
This structure of the asymptotic expansion proves the conjectures raised in 
\cite{MullerShrockDynamicCorrFnctsTIandXXAsymptTimeAndFourier,TracyVaidyaRedDensityMatrixSpaceAsymptImpBosonsT=0}
for certain specialisations of this kernel. Also, upon specialisation, one gets in this way the general structure of the asymptotic expansion of the fifth 
Painlev\'{e} transcendent  associated with the pure sine kernel \cite{JimMiwaMoriSatoSineKernelPVForBoseGaz}. First few terms of the asymptotic expansion have been obtained 
by McCoy and Tang \cite{McCoyTangSineKernelSubleadingFromPainleveV}. The presence of logarthms $\ln x$ in the expansion was first observed in \cite{KozKitMailSlaTerRHPapproachtoSuperSineKernel}.

I will now present the first few explicit terms of the large-$x$ asymptotic expansion of the Fredholm determinant. 
The subleading terms in \eqref{ecriture serie de Natte intro} are written somewhat implicitly. 
So as to be explicit, one has to distinguish between two situations depending on whether the saddle-point $\la_0$ is inside of $\intoo{-q}{q}$
or outside of it. The first case ($-q<\la_0<q$) will be called the time-like regime and to the second one ($\abs{\la_0}>q$) the space-like regime.

\begin{cor}
\label{theorem DA determinant}

Under the mentioned hypothesis, the Fredholm determinant $\det{}\big[ \e{id} + \op{V}_x \big]$ admits the leading
$x\tend +\infty$  asymptotic behaviour:
\beq
\det{}\big[ \e{id} + \op{V}_x \big]  \; =  \; \sul{ \eps \in \{ \pm 1, 0 \} }{} \mc{V}^{\pa{0}}_x\pac{\nu + \eps ,u,g} \cdot \bigg( 1+ \e{O}\Big(\f{\log x}{x} \Big) \bigg)  
\; + \; \f{b_1\pac{\nu,u,g} }{x^{\f{3}{2}}} \mc{V}^{\pa{0}}_x\pac{\nu  ,u ,g}   \cdot \bigg( 1+ \e{O}\Big(\f{\log x}{x} \Big) \bigg)    \; .
\label{ecriture DA det GSK time dpdt}
\enq
The functional $b_1\pac{\nu,u,g}$ takes different forms depending on the space-like or time-like regime:
\beq
b_1\pac{\nu,u,g}= \f{  1  }{\sqrt{-2\pi u^{\prime \prime}\!\!\pa{\la_0}} }  \left\{ \ba{cc}
        \f{\nu\pa{-q}}{u^{\prime}\!\pa{-q} \pa{\la_0+q}^2} \f{\mc{S}_0}{\mc{S}_-}
        - \f{\nu\pa{q}}{u^{\prime}\!\pa{q} \pa{\la_0-q}^2} \f{\mc{S}_0}{\mc{S}_+}  \qquad & time-like \vspace{2mm}\\
         \f{\nu\pa{-q}}{u^{\prime}\!\pa{-q} \pa{\la_0+q}^2} \f{\mc{S}_-}{\mc{S}_0}
         - \f{\nu\pa{q}}{u^{\prime}\!\pa{q} \pa{\la_0-q}^2} \f{\mc{S}_+}{\mc{S}_0} \qquad& space-like  \ea\right.
\; .
\enq
The constants $\mc{S}_{\pm}$ and $\mc{S}_0$ are defined as 
\beqa
\mc{S}_+ &=& \pac{2q x u^{\prime}\!\pa{q} + \i 0^+}^{2\nu\pa{q}} e^2\!\pa{q}  \pa{\ex{-2 \i \pi \nu\pa{q}}-1} 
\f{ \Ga\pa{1-\nu\pa{q}} }{ \Ga\pa{1+\nu\pa{q}}    }
\exp\Bigg\{-2 \int_{-q}^{\,q} \f{\nu\pa{q}-\nu\pa{\mu} }{q-\mu} \dd\mu \Bigg\}
\; , \\
\mc{S}_- &=& \f{  \pa{\ex{-2 \i \pi \nu\pa{-q}}-1}   } { \pac{2q x u^{\prime}\pa{-q} + \i 0^+ }^{2\nu\pa{-q}} } e^2\pa{-q}
 \f{ \Ga\pa{1+\nu\pa{-q}} } { \Ga\pa{1-\nu\pa{-q}} }  \exp\Bigg\{2 \int_{-q}^{\,q} \f{\nu\pa{-q}-\nu\pa{\mu} }{q+\mu} \dd\mu \Bigg\}   \;,
\label{definition fonction S plus et moins}
\eeqa
and
\beq
\mc{S}_0= e^2(\la_0) \, \ex{ \i \f{\pi}{4} } \paf{\la_0+q}{\la_0-q-\i 0^+}^{ 2 \nu\pa{\la_0} } 
\exp\Bigg\{-2 \int_{-q}^{\,q} \f{\nu\pa{\la_0}-\nu\pa{\mu} }{\la_0-\mu} \dd\mu \Bigg\}
\; \times \; \left\{ \ba{cc}  \pa{\ex{-2 \i \pi \nu\pa{\la_0}}-1}^2  & time-like \vspace{2mm} \\ 
								1 						&  space-like   \ea \right.  
\; .
\label{definition fonction S zero et kappa}
\enq
%
%

\end{cor}

The proof of this expansion can be found in Section 6.2 of  [$\bs{A8}$] and heavily relies on the asymptotic
analysis of the Riemann--Hilbert problem associated with $\op{V}_x$ that is carried out in Sections 3 to 5 of the same paper.
I will not venture into the details of this Riemann--Hilbert problem here and, rather, refer to  [$\bs{A8}$] for more precisions. 

\vspace{1mm}

The specific case of the kernel $V_x(\la,\mu)$ corresponding to 
\beq
u(\la) = \la-\tf{t\la^2}{x} \, , \quad  g=0 \; , \quad q=1 \quad \e{and} \quad \; \nu= cst 
\enq
has been studied in the literature in the context of its relation with the impenetrable fermion gas \cite{CheianovZvonarevZeroTempforFreeFermAndPureSine}. 
Upon such a specialisation, one indeed recovers the coefficients of the asymptotic expansion obtained in that paper.

The first terms in the expansion \eqref{ecriture DA det GSK time dpdt} correspond to the first non-oscillating and the first two oscillating
corrections that are not related with the presence of the saddle-point at $\la_0$ in the sense that they are purely due to 
the effects induced by the presence of the boundaries $\pm q$. These corrections can be recast in a slightly more explicit form as
\bem
\sul{ \eps \in \{ \pm 1, 0 \} }{} \mc{V}^{\pa{0}}_x\pac{\nu + \eps ,u,g} \cdot \bigg( 1+ \e{O}\Big(\f{\log x}{x} \Big) \bigg)  \; = \; 
 \exp\bigg\{ \Int{-q}{q} \pac{ \i x u^{\prime}\pa{\la} + g^{\prime}\pa{\la}  } \nu\pa{\la} \dd \la \bigg\} \times \bigg\{  \f{ \mc{B}\pac{\nu,u} }{ x^{ \nu^2(q) + \nu^2(-q) } }  \cdot \bigg( 1+ \e{O}\Big(\f{\log x}{x} \Big) \bigg) \\ 
 \, + \, \mc{B}\pac{\nu +1 ,u}  \cdot   \f{ \ex{ \i x \pac{u\pa{q}-u\pa{-q}} + g\pa{q}-g\pa{-q} }  }{ x^{ (\nu(q)+1)^2 + (\nu(-q)+1)^2 }  } \cdot \bigg( 1+ \e{O}\Big(\f{\log x}{x} \Big) \bigg)
    \, + \,  \mc{B}\pac{\nu - 1 ,u} \cdot \f{ \ex{ \i x \pac{u\pa{-q}-u\pa{q}} + g\pa{-q}-g\pa{q} }  }{ x^{ (\nu(q)-1)^2 + (\nu(-q)-1)^2 } } \cdot \bigg( 1+ \e{O}\Big(\f{\log x}{x} \Big) \bigg) \bigg\} \;. 
\nonumber
\end{multline}

\subsection{The Natte series representation for a rank one perturbation}
\label{SousSection Serie de Natte pour perturbation de rang un}

I now provide a more extensive description of the Natte series representation for the Fredholm determinant of a rank $1$ perturbation of the operator 
$\e{id} + \op{V}_x$. More details can by found in Proposition 7.2 of [$\bs{A8}$].

The object of interest $ X_{ \msc{C}_{ \op{V}} }[ \nu,e^{2} ]$ is defined as:
\beq
X_{ \msc{C}_{ \op{V}} }[ \nu,e^{2} ]  \; = \;  \bigg(  \int_{ \msc{C}_{ \op{V}} } e^{-2}(\la) \cdot \f{\dd \la}{2\pi}  \;  +  \; \f{  \Dp{}  }{ \Dp{}\a }  \bigg)_{\mid \a=0}  \cdot 
\det\big[ \e{id} + \op{V}_x  + \a \op{P} \big]
\label{definition fonctionnelle X du contour mathcal CV}
\enq
where $\op{P}$ is the one-dimensional projector acting on $L^2\big( \msc{C}_{ \op{V} } \big) $ with the integral kernel
\beq
P(\la,\mu) \; = \;  \f{2}{\pi} \sin\pac{\pi\nu\pa{\la} }  \sin\pac{\pi\nu\pa{\mu} } \, e(\la) \,  e(\mu) \, 
O_{ \msc{C}_{ \op{V}} }[\nu, e^{-2} ] (\la)  \, O_{ \msc{C}_{ \op{V}} }[\nu,e^{-2} ](\mu) \;.
\enq
$X_{ \msc{C}_{ \op{V}} }[ \nu,e^{2} ]$ can be recast in terms of the unique solution $f$ to the linear integral equation 
\beq
\sin \pac{\pi \nu\pa{\la}} f(\la) + \Int{-q}{q} V_x\pa{\la,\mu} \sin \pac{\pi \nu\pa{\mu}} f(\mu) \dd \mu
= \sin \pac{\pi \nu\pa{\la}} \, e(\la) \, O_{ \msc{C}_{ \op{V}} }[\nu, e^{-2} ] (\la)  \;.
\label{definition fonction F+}
\enq
This alternative expression reads 
\beq
X_{ \msc{C}_{ \op{V}} }[ \nu,e^{2} ] \;  = \; 
\bigg(  \int_{ \msc{C}_{ \op{V}} } e^{-2}(\la) \cdot \f{\dd \la}{2\pi}  \;  +  \;  2 \int_{-q}^{q}  \sin^2\!\pac{\pi \nu\pa{\la}} f(\la) e(\la)
O_{  \msc{C}_{ \op{V}} }\big[\nu,e^{-2} \big](\la) \cdot \f{\dd \la}{\pi}  \bigg) \cdot \det\big[ \e{id} + \op{V}_x \big] \;.
\label{appendix ecriture limite thermo XN}
\enq

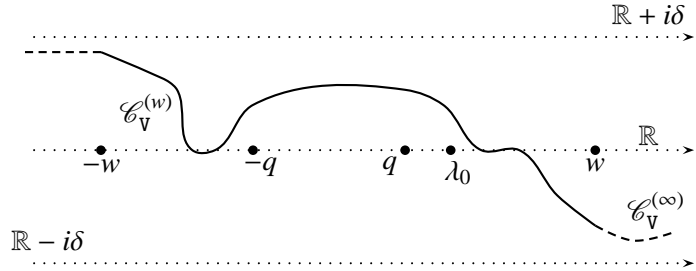
\begin{figure}[h]
\begin{center}

\begin{pspicture}(6,4)

\psline[linestyle=dotted, dash=3pt 2pt]{->}(-1,2)(7.8,2)
\psline[linestyle=dotted, dash=3pt 2pt]{->}(-1,0.5)(7.8,0.5)
\psline[linestyle=dotted, dash=3pt 2pt]{->}(-1,3.5)(7.8,3.5)

\rput(7.2,2.2){$\R$}
\rput(7.2,3.8){$\R + i\de$}
\rput(-0.7,0.8){$\R - i\de$}


\psdots(2,2)(4,2)(4.6,2)(0,2)(6.5,2)

\rput(2.1,1.8){$-q$}
\rput(3.8,1.8){$q$}
\rput(4.7,1.7){$\la_0$}
\rput(0,1.8){$-w$}
\rput(6.5,1.8){$w$}

\pscurve(0,3.3)(0.7,3)(0.9,2.9)(1,2.8)(1.2,2)(1.5,2)(2,2.6)(4,2.8)(4.5,2.6)(5,2)(5.5,2)(6,1.4)(6.5,1)


\pscurve[linestyle=dashed, dash=3pt 2pt](-1,3.3)(-0.5,3.3)(0,3.3)

\pscurve[linestyle=dashed, dash=3pt 2pt](6.5,1)(7,0.8)(7.5,0.9)

\rput(0.6,2.5){$\msc{C}_{ \op{V} }^{\pa{w}}$}

\rput(7.3,1.2){$\msc{C}_{ \op{V} }^{\pa{\infty}}$}

\end{pspicture}

\caption{The contour $\msc{C}_{ \op{V} }^{\pa{w}}$ consists of the solid line. The contour $\msc{C}_{ \op{V} }^{\pa{\infty}}$
corresponds to the union of the solid and dotted lines. The localisation of the saddle-point $\la_0$ corresponds to the space-like regime.
Both contours lie in $U_{\tf{\de}{2}}$.
\label{contour CE et sa restriction CEw}}
\end{center}
\end{figure}

Finally, I need to introduce a finite truncation of the contour $\msc{C}_{\op{V}}$:  $\msc{C}_{\op{V}}^{\pa{w}} =\msc{C}_{ \op{V} }^{\pa{\infty}} \cap \paa{ z  \in \Cx \; : \; \abs{\Re z}< w}$.
$\msc{C}_{ \op{V} }^{ (\infty) }$ and $\msc{C}_{ \op{V} }^{ (w) }$ have been depicted in  Fig.~\ref{contour CE et sa restriction CEw}.

The Natte series for the  perturbed determinant $X_{ \msc{C}_{ \op{V} }^{(w)} }\big[\nu, e^{2} \big]$ contains the auxiliary functionals:
\beq
\mc{A}_+\!\pac{\nu,p} = \f{ - 2q \cdot \varkappa^{-2}\!\pac{\nu}\pa{q} }{ \pac{ 2q p^{\prime}\!\pa{q} }^{2\nu\pa{q} + 1} }
\Ga\pab{ 1+\nu\pa{q} }{ - \nu\pa{q} }  \f{1}{\ex{-2 \i \pi \nu\pa{q}}-1}
\quad , \qquad \varkappa\pac{\nu}\pa{\la}=\exp\paa{-\int_{-q}^{\,q} \f{ \nu\pa{\la}-\nu\pa{\mu} }{ \la-\mu } \dd \mu}   \; ,
\label{definition fonctionnelle A+ et kappa}
\enq
and
\beq
\hspace{-5mm}\mc{A}_-\pac{\nu,p} = \f{ - 2q}{ \varkappa^{2}\!\pac{\nu}\!\pa{-q} }  \Ga\pab{\!\! 1-\nu\pa{-q} \!\! }{\nu\pa{-q}}
\f{\pac{2q p^{\prime}\!\pa{-q} }^{2\nu\pa{-q} - 1} }{\ex{-2 \i \pi \nu\pa{-q}}-1}
\quad \e{and} \quad
\mc{A}_0\pac{\nu} = \ex{- \i \f{\pi}{4} } \varkappa^{-2}\!\pac{\nu}\!\pa{\la_0}
\paf{ \la_{0}-q }{ \la_{0}+q }^{2\nu\pa{\la_0} } \hspace{-3mm}.
\label{definition fonctionnelle A- et A0}
\enq

\begin{theorem}
\label{Theorem series de Natte pour determinant I+V perturbe rang 1}

Let  $w> \abs{\la_0} +q >0$ and $x$ be large enough. Then, the perturbed determinant $X_{ \msc{C}_{ \op{V} }^{(w)} }\big[\nu, e^{2} \big]$ given by \eqref{appendix ecriture limite thermo XN}
admits the Natte series representation
\bem
X_{ \msc{C}_{ \op{V} }^{(w)} }\big[\nu, e^{2} \big] \; = \; \f{\mc{B}\pac{ \nu,u+ \i 0^+}}{x^{ \nu^2\pa{q} + \nu^2\pa{-q}}}
\ex{\,  \Int{ -q }{ q } \pac{ \i x u^{\prime}\! \pa{\la} + g^{\prime}\! \pa{\la}  }  \nu\pa{\la}  \dd \la }
\Bigg\{  \f{\mc{A}_0\!\pac{\nu} \bs{1}_{\intoo{q}{+\infty}}\pa{\la_0} }{\sqrt{-2\pi x u^{\prime \prime}\!\pa{\la_{0}}} }
\ex{ \i x u\pa{\la_{0}}+g\pa{\la_{0}} }
+ \f{\mc{A}_+\!\pac{ \nu,u+ \i 0^{+}} }{ x^{1+2\nu\pa{q}} } \ex{ \i x u\pa{q}+g\pa{q} }     \\
 + \; \f{\mc{A}_-\!\pac{\nu,u+\i0^+} }{ x^{1-2\nu\pa{-q}} } \ex{ \i x u\pa{-q}+g\pa{-q} } \; + \;
\sul{n \geq 1}{} \sul{ \vec{k}\in  \mc{K}_n  }{}
\sul{  \mc{E}_n(\vec{k})  }{}  \int_{ \msc{C}_{ \op{V}; \eps_{\bs{t}} }^{\pa{w}} }{}
  H_{n;x}^{\pa{\paa{\eps_{\bs{t}}}}}\big(  \paa{u\pa{z_{\bs{t}}} } ; \paa{z_{\bs{t}}}  \big) \pac{ \nu }
 \pl{ \bs{t} \in \J{ k } }{} \ex{ \eps_{\bs{t}} g\pa{ z_{\bs{t}} } }      \f{ \dd^n z_{\bs{t}} }{ \pa{2i\pi}^n }
 \Bigg\}   \; .
\label{equation developpement det serie de Natte}
\end{multline}
The second sum appearing in the last line of \eqref{ecriture serie Natte multDim coeff taylor rho eff} runs through
all the elements $\vec{k}$ belonging to
\beq
\mc{K}_n  =  \Bigg\{\vec{k} = \pa{k_1,\dots, k_{n+1}} \; : \; k_{n+1} \in \mathbb{N}^*  \; \; and \;\; k_a \in \mathbb{N} \; , a=1,\dots, n
\quad such \; that \quad  \sul{a=1}{n} a k_a  + k_{n+1}  =n \Bigg\} \;.
\enq
Once that an element of $\mc{K}_n$ has been fixed, one defines the associated set of triplets $\J{ \vec{k} }$:
\beq
\J{ \vec{k} }=\bigg\{ \pa{\bs{t}_1,\bs{t}_2,\bs{t}_3 } \, , \, \bs{t}_1\in \intn{1}{n+1} \, , \, \bs{t}_2 \in \intn{1}{k_{\bs{t}_1}} \, ,
\, \bs{t}_3 \in \intn{1}{\bs{t}_1 - n\de_{\bs{t}_1,n+1}} \bigg\} \; .
\enq
The third sum runs through all the elements $\{ \eps_{\bs{t}} \}_{\bs{t} \in \J{ \vec{k} } }$  belonging to the set
\beq
\hspace{-3mm} \mc{E}_n ( \vec{k} \, ) = \Bigg\{  \paa{\eps_{\bs{t}}}_{\bs{t} \in \J{ \vec{k} } } \: : \;
\eps_{\bs{t}} \in \paa{\pm 1, 0} \; \forall \bs{t} \in \J{ \vec{k} }
 \quad with \quad \sul{\bs{t}_3=1}{\bs{t}_1}\eps_{\bs{t}} =0 \quad for\;  \bs{t}_1=1,\dots, n \quad and \quad
\sul{p=1}{k_{n+1}}\eps_{n+1,p,1} =1 \Bigg\} \;.
\nonumber
\enq
In other words, $\mc{E}_n (\vec{k}\, )$ consists of $n$-uples of parameters $\eps_{\bs{t}}$ labelled by triplets
$\bs{t}=\pa{\bs{t}_1,\bs{t}_2,\bs{t}_3}$ belonging to $\J{ \vec{k} }$. Each element of such an $n$-uple takes its values in
$\paa{\pm 1,0}$. In addition, the components of this $n$-uple are subject to summation constraints.
These hold for any value of $\bs{t}_1$ or $\bs{t}_{2}$ and are different whether one deals with $\bs{t}_1=1,\dots,n$  or with $\bs{t}_{1}=n+1$.

The integral appearing in the $n^{\e{th}}$ summand occurring in the third line of 
\eqref{ecriture serie Natte multDim coeff taylor rho eff} is $n$-fold.
The contours of integration $\msc{C}_{ \op{V} ; \eps_{\bs{t}}}^{\pa{w}}$ depend on the choices of elements in $\mc{E}_n( \vec{k}\, )$
and are realised as $n$-fold Cartesian products of one-dimensional compact curves that correspond to various deformations of
the base curve $\msc{C}^{\pa{w}}_{\op{V}}$  depicted in Fig.~\ref{contour CE et sa restriction CEw}. In the $w\tend +\infty$ limit, these curves go to analogous
deformations of the base curve $\msc{C}_{\op{V}}^{\pa{\infty}}$. All these contours lie  in $U_{\tf{\de}{2}}$.
The detailed description of the curves $\msc{C}_{ \op{V} ; \eps_{\bs{t}}}^{\pa{w}}$ can be found in the last paragraph of Theorem 7.1, below of equation (7.27),
and in Lemma 7.2 of [$\bs{A8}$].

\end{theorem}

The above Natte series is convergent for $x$ large enough in that one has the bounds
\beq
\sul{\mc{K}_n}{}\sul{\mc{E}_n(\vec{k}) }{} \bigg| \bigg| H_{n;x}^{\pa{\paa{\eps_{\bs{t}}}}} \pac{\nu}
\pl{ \bs{t} \in \J{ k } }{} \ex{ \eps_{\bs{t}} g }  \bigg|\bigg|
_{L^{1}\big(  \msc{C}^{\pa{\infty}}_{\op{V} ; \eps_{\bs{t}}} \big) }   \;    \leq \;\;
 c_2\paf{c_1}{x}^{ n c_3 } \;.
\label{ecriture estimation de norme L1 Hn}
\enq
There $c_1$ and $c_2$ are some $n$-independent constants. These constants only depend on the values taken by  $u$, and $g$ in some small neighbourhood of the base curve
$\msc{C}_{\op{V}}^{\pa{\infty}}$ and by $\nu $ on a small neighbourhood of $\intff{-q}{q}$, whereas
\beq
c_3= \f{3}{4} \min\pa{\tf{1}{2},1-2 \max_{\tau=\pm }\abs{\Re\pac{\nu\pa{\tau q}}} -\Ups_{\eps}}  \quad
 \e{where}  \quad \Ups_{\eps} =2 \sup\bigg\{ \abs{ \Re\pac{ \nu\pa{z} - \nu\pa{\tau q}} }^{}_{} \; : \; \abs{z-\tau q} \leq \eps \;, \;  \tau=\pm   \bigg\}  \;.
\nonumber
\enq
Here $\eps>0$ is sufficiently small but arbitrary otherwise. I stress that, should these norms change, then so would change the constants $c_1$, $c_2$ and $c_3$ but the overall structure of
the estimates in $x$ would remain.

The Natte series expansion \eqref{equation developpement det serie de Natte}  has a well defined $w\tend +\infty$ limit:
all the concerned integrals are convergent as the functions $H_{n;x}^{\pa{ \{ \eps_{\bs{t}} \} }}$
approach zero exponentially fast in respect to any variable that runs to $\infty$ along $\msc{C}^{\pa{\infty}}_{ \op{V} ; \eps_{\bs{t}}}$.
In fact, one even has 
\beq
\sul{\mc{K}_n}{}\sul{\mc{E}_n(\vec{k}) }{} \bigg| \bigg| H_{n;x}^{\pa{\paa{\eps_{\bs{t}}}}} \pac{\nu}
\pl{ \bs{t} \in \J{ k } }{} \ex{ \eps_{\bs{t}} g }  \bigg|\bigg|
_{L^{1}\big(  \msc{C}^{\pa{\infty}}_{\op{V} ; \eps_{\bs{t}}} \setminus  \msc{C}^{\pa{w}}_{\op{V} ; \eps_{\bs{t}}} \big)  }     \;    \leq \;\;
 c_2\paf{c_1}{x}^{ n c_3 } \cdot \ex{-c_4 x w }
\label{ecriture estimation de norme L1 Hn}
\enq
for a $x$ and $n$-independent constant $c_4>0$.

To close this section, I list several properties of the functions $H_{n;x}^{\pa{\paa{\eps_{\bs{t}}}}}$ which will be of use later on:
\begin{enumerate}
\item[i)] $H_{n;x}^{\pa{\paa{\eps_{\bs{t}}}}}\pa{ \paa{u\pa{z_{\bs{t}}} } ; \paa{z_{\bs{t}}}  }\pac{\nu}$
is a function of $\paa{u\pa{z_{\bs{t}}}}$ and $\paa{z_{\bs{t}}}$.
It is also a regular functional\symbolfootnote[2]{See Definition \ref{Definition Fonctionelle reguliere} for a discussion of this property} of $\nu$; \vspace{2mm}
\item[ii)] $ H_{n;x}^{\pa{\paa{\eps_{\bs{t}}}}}\pa{ \paa{u\pa{z_{\bs{t}}} } ; \paa{z_{\bs{t}}}  }\pac{\ga \nu} = \e{O}\pa{\ga^n}$
and the O  holds in the $\big( L^{1}\cap L^{\infty} \big) \big( \msc{C}_{\op{V}; \eps_{\bs{t}}}^{\pa{\infty}} \big)$ sense ; \vspace{2mm}
\item[iii)]  $H_{n;x}^{\pa{\paa{\eps_{\bs{t}}}}}$ can be represented as:
\beq
H_{n;x}^{\pa{\paa{\eps_{\bs{t}}}}}\pa{ \paa{u\pa{z_{\bs{t}}} } ; \paa{z_{\bs{t}}}  } \pac{ \nu } =
\wt{H}_{n;x}^{\pa{\paa{\eps_{\bs{t}}}}} \pa{  \paa{\nu\pa{z_{\bs{t}}} }  ; \paa{u\pa{z_{\bs{t}}} } ; \paa{z_{\bs{t}}}  }
\pl{ \bs{t} \in \J{k} }{} \pa{  \varkappa\pac{\nu}\pa{z_{\bs{t}}} }^{-2\eps_{\bs{t}}}
\times
\pl{ \substack{ \bs{t} \in \J{k} \\ \eps_{\bs{t}}=1 } }{} \pa{\ex{-2\i\pi \nu\pa{z_{\bs{t}}}} -1}^2 
\label{ecriture dependence serie Natte en nu}
\enq
where $\wt{H}_{n;x}^{\pa{\paa{\eps_{\bs{t}}}}}$ is a holomorphic function for $ \abs{\Re\pa{\nu}} \leq \tf{1}{2}$ and $\varkappa\pac{\nu}\pa{\la}$
is given by \eqref{definition fonctionnelle A+ et kappa} ; \vspace{2mm}

\item[iv)]  $H_{1;x}^{\pa{\paa{\eps_{\bs{t}}}}}=\e{O}\pa{x^{-\infty}}$ and for $n\geq 2$
\beq
\hspace{-5mm}  H_{n;x}^{\pa{\paa{\eps_{\bs{t}}}}} = \e{O}\pa{x^{-\infty}}  +
\sul{b=0}{ \pac{\tf{n}{2}} } \sul{p=0}{b} \sul{ m=b-\pac{\f{n}{2}} }{ \pac{\tf{n}{2}}-b}
%
%
 \paf{ \ex{ \i x \pac{u\pa{\la_0}-u\pa{-q}} } }{ x^{2 \nu\pa{-q}   }  }^{\bs{\eta} b} \hspace{-1mm}\cdot  \,
\paf{ \ex{ \i x \pac{u\pa{q}-u\pa{-q}} } }{ x^{2\pac{ \nu\pa{q}+\nu\pa{-q}}}  }^{m-\bs{\eta} p}\hspace{-2mm}\cdot  \;
\sul{\tau \in \paa{\pm 1; 0} }{}  \f{ \mf{e}_{\tau} }{ x^{n-\f{b}{2}}} \, \cdot  \,
\pac{ H_{n;x}^{\pa{\paa{\eps_{\bs{t}}}}} }_{m,p,b,\tau} \;.
\label{ecriture forme detaille fonction Hn}
\enq
\end{enumerate}

The $\e{O}\pa{x^{-\infty}}$ appearing above holds in the $ \big( L^{1}\cap L^{\infty} \big)\big(\msc{C}^{\pa{\infty}}_{\op{V}; \eps_{ \bs{t} } } \big)$ sense.
In order to lighten the formula, I dropped the argument-dependent part. However, I do stress that the $\e{O}\pa{x^{-\infty}}$
as well as $\big[ H_{n;x}^{\pa{\paa{\eps_{\bs{t}}}}} \big]_{m,p,b,\tau} $ depend on the same set of variables as $H_{n;x}^{\pa{\eps_{\bs{t}}}}$.
Also, $\bs{\eta}$ is as defined by \eqref{definition parametre eta space et time like}, while \eqref{ecriture forme detaille fonction Hn}
 makes use of the shorthand notation
\beq
\mf{e}_{+}= \ex{\i x u\pa{q}} x^{-2\nu\pa{q}}\; ,  \mf{e}_{-}= \ex{ \i x u\pa{-q}} x^{2\nu\pa{-q}}
\qquad \e{and} \quad \mf{e}_0=\pa{1+\bs{\eta}}\ex{ \i x u\pa{\la_0}} \; .
\enq
Finally, the functions  $\big[ H_{n;x}^{\pa{\paa{\eps_{\bs{t}}}}} \big]_{m,p,b,\tau} $  are only supported on a small
vicinity of the points $\pm q$ and $\la_0$. When integrated along $\msc{C}^{(\infty)}_{\op{V}; \eps_{ \bs{t} } }$
the integration reduces for each variable $z_{\bs{t}}$, to one over a small circle $\Dp{}\mc{D}_{0;q_{\tau}}$ around $q_{\tau}$ ($q_{\pm} =\pm q$, $q_0=\la_0$). 
Their dependence on $x$ is as follows. If a variable $z_{\bs{t}}$ is integrated in a vicinity of $q_{\tau}$, 
the function  $\big[ H_{n;x}^{\pa{\paa{\eps_{\bs{t}}}}} \big]_{m,p,b,\tau} $ contains a fractional power of 
$x^{ \pm \pac{ 2\nu\pa{z_{\bs{t}}} -\nu\pa{q_{\tau}} } }$, multiplied by a function of 
$z_{\bs{t}}$ which has a uniform on  $\Dp{}\mc{D}_{0;q_{\tau}}$ asymptotic expansion into inverse powers of $x$. 
 The coefficients in this asymptotic expansions contain higher and higher order poles at $z_{\bs{t}}=q_{\tau}$.
Evaluating the integrals associated to the various terms of the asymptotic expansion by residues at the poles at $z_{\bs{t}}=q_{\tau}$
provides one with the pre-factor in front of the $x^{-r}$ terms. Since derivatives may hit on the fractional powers  
$x^{ \pm \pac{ 2\nu\pa{z_{\bs{t}}} -\nu\pa{q_{\tau}} } }$, one ends up with a polynomial in $\ln x$ of degree $r$. 
Hence, \textit{in fine}, one ends up with a contribution that is a $\e{O}\big( \pa{\tf{\ln x}{x}}^r \big)$.
The structure of the integrands is such that, upon computing all the partial derivatives and for any holomorphic 
function $h$ in the vicinity of the points $\pm q$, $\la_0$, one should make the replacement:
\beq
\sul{\bs{t} \in \J{ \, \vec{k} } }{} \eps_{\bs{t}} h\pa{z_{\bs{t}}} \hookrightarrow  \bs{\eta} b \pa{ h\pa{\la_0}-h\pa{-q}} + \pa{m-\bs{\eta}p}\pa{h\pa{q}-h\pa{-q}}
+ \pa{\de_{\tau;1}+\de_{\tau;-1} + \pa{1+\bs{\eta}} \de_{\tau;0}/2 } h\pa{q_{\tau}}\;.
\enq

\subsection{The main steps of the proof}

The Natte series is a consequence of the non-linear steepest descent based representation for the solution $\chi$ to the $2\times 2$ Riemann--Hilbert problem associated with
the integrable integral operator $\op{V}_x$. Indeed, the non-linear steepest descent analysis allows one to put the Riemann--Hilbert problem for $\chi$ in correspondence with 
the Riemann--Hilbert problem for an auxiliary, piecewise analytic, matrix $\Pi$. The jump matrices arising in the Riemann--Hilbert problem for 
$\Pi$ are uniformly and $L^1\cap L^2$ close to the identity matrix, what allows one to represent $\Pi$ in terms of a Neumann series. 
Further, one has the differential identity (\textit{c.f.} Proposition 3.1 of [$\bs{A8}$])
\beq
\Dp{x} \log \det{}\big[I+V\big]  \; = \;  - \i \f{\Dp{}}{\Dp{}{\eta}} \Bigg\{ \;   \Oint{ \Ga\pa{ \msc{C}_{\op{V} } } }{} \f{ \dd z }{ 4\pi }  \ex{ \i \eta u(z) }
\e{tr}\Bigg[ \pa{\Dp{z}\chi}(z)  \bigg( \sg^z \, +\,  2 \Int{\msc{C}_{\op{V}} }{} \f{ e^{-2}(s) }{s-z}  \cdot \f{\dd s }{ 2\i \pi} E^{12} \bigg) \chi^{-1}(z)  \Bigg]   \Bigg\}_{ \mid \eta=0^+}
\label{ecriture identite differentielle log det dp x pour GSK time dependent}
\enq
where $E^{12}$ is the $2\times 2$ matrix having $1$ in its upper corner and $0$ everywhere else while $\Ga\pa{\msc{C}_{E}}$ is a loop in $U_{\tf{\de}{2}}$ enlacing counterclockwise the contour $\msc{C}_{ \op{V} }$.
This loop is such that it goes to infinity in the regions where $\ex{ \i\eta u(z)}$, $\eta>0$ is decaying exponentially fast.
The Natte series is then obtained by taking the ante-derivative over $x$ of \eqref{ecriture identite differentielle log det dp x pour GSK time dependent}
after having inserted the expression for $\chi$ in terms of $\Pi$ and represented the latter by means of its Neumann series representation. 
Various technical details related with such handlings can be found in Section 7 and Appendices B \& C of [$\bs{A8}$].

\section{The large-distance and long-time asymptotic expansion of the reduced density matrix}
\label{Section method and main results}

\subsection{A bird's view of the method}

In the following, $\rho\!\pa{x,t} = \lim_{N,L \tend +\infty} \rho_{N}\!\pa{x,t}$ will stand for the, presumably existing, thermodynamic limit of the one-particle reduced density matrix 
$\rho_N\!\pa{x,t}$ given by \eqref{definition reduced density matrix}.
I will not develop  on the existence of this limit, and take this as a quite reasonable working hypothesis, although some arguments in favour thereof can be given.

The analysis starts with the non-linear Schr\"{o}dinger model in finite volume.
I will first provide certain re-summation formulae for $\rho_{N}\!\pa{x,t}$ starting from its form factor expansion.
Unfortunately, the very intricate structure of the summation over all the excited states in a form factor expansion prevented me, so far, from analysing its thermodynamic limit rigorously from the
very beginning. I therefore introduce  simplifying hypothesis. Namely, denoting the relative excitation energy by $\mc{E}_{\e{ex}}$, \textit{c}.\textit{f}. 
\eqref{ecriture energie relative excitation}, I will argue that all contributions issued from excited states such that
$\mc{E}_{\e{ex}}$ scales with  $L$ do not contribute to the thermodynamic limit of $\rho_N \! \pa{x,t}$. 
This hypothesis then reduces the problem to the analysis of an effective form factor series $\rho_{N;\e{eff}}\!\pa{x,t}$. I then introduce 
a certain $\ga$-deformation $\rho_{N;\e{eff}}\!\pa{x,t\mid \ga}$ thereof.
My conjecture states that $\rho_{N;\e{eff}}\!\pa{x,t\mid \ga=1}=\rho_{N;\e{eff}}\!\pa{x,t}$ has the \textit{same} thermodynamic limit as $\rho_N\!\pa{x,t}$.

Then, I will focus on the analysis of the Taylor coefficients at $\ga=0$ of the $\ga$-deformation:
\vspace{-2mm}
\beq
\hspace{1cm}\rho_{N;\e{eff}}^{\pa{m}}\!\pa{x,t} \equiv   \f{ \Dp{}^m }{ \Dp{}\ga^m } \left. \rho_{N;\e{eff}}\pa{x,t\mid \ga} \right|_{\ga=0} \;.
\enq

\vspace{-2mm} All rigorous, conjecture-free, results of this analysis are relative to these Taylor coefficients.
I will show  that  $\rho_{N;\e{eff}}^{\pa{m}}\!\pa{x,t}$ admits a well-defined thermodynamic
limit $\rho_{\e{eff}}^{\pa{m}}\!\pa{x,t}$. One can provide two different representations for this limit, each being given in terms of a  
 finite sum of multiple integrals. \vspace{1mm}
\begin{itemize}
\item[$\bullet$]  The first representation is in the spirit of the ones obtained in
\cite{KozKitMailSlaTerXXZsgZsgZAsymptotics} and [$\bs{A10}$].
It corresponds to a truncation of a multidimensional deformation of the Fredholm series for the Fredholm determinant \eqref{definition fonctionnelle X du contour mathcal CV}. \vspace{1mm}
\item[$\bullet$] The second representation is structured in such a way that it allows one to  \textit{read off} straightforwardly
the first few terms of the asymptotic expansion of $\rho_{\e{eff}}^{\pa{m}}\!\pa{x,t}$.
 The various terms appearing in this representation
are organised in such a way that the identification of those that are negligible in the $x\tend +\infty$ limit (\textit{e}.\textit{g}. exponentially small in $x$) is
trivial. The second series corresponds to a truncation of an object that can be thought of as a multidimensional deformation of a Natte series. \vspace{1mm}
%
%
\end{itemize}

The above two results are derived rigorously without any  additional conjecture.
However, in order to push the results further and compute the large-distance and long-time asymptotic behaviour of the 
reduced density matrix in infinite volume $\rho(x,t)$, I need, again, to rely on additional conjectures.
Namely that
\begin{enumerate}
\item  the series of multiple integrals that arises upon summing up the thermodynamic limits of the Taylor coefficients
$\sum_{m=0}^{+\infty} \tf{ \rho_{\e{eff}}^{\pa{m}}\!\pa{x,t} }{m!}$  is convergent; \vspace{1mm}
\item  this series coincides with the thermodynamic limit of $\rho_{N;\e{eff}}\!\pa{x,t \mid \ga=1}$  and hence, due to the first
conjecture, with $ \rho\pa{x,t}$.
\end{enumerate}
On the basis of these conjectures one can write down two types of series of multiple integral representations for the thermodynamic limit
of the reduced density matrix. One of the series is a multidimensional  Fredholm series, \textit{viz}. is of the form
\eqref{definition serie de Fredholm multidimensionelle}. The second series issues from a multidimensional deformation flow  of the  Natte series expansion for the rank one perturbation of the Fredholm determinant 
described in Section \ref{SousSection Serie de Natte pour perturbation de rang un}, Theorem \ref{Theorem series de Natte pour determinant I+V perturbe rang 1}. Just as 
the original Natte series, the multidimensional one, \textit{viz}. the one resulting from the deformation flow, has all the virtues in respect to 
the computation of the long-time and large-distance asymptotic behaviour of $\rho\!\pa{x,t}$: it is structured in
such a way that one readily \textit{reads off} from its very form, the sub-leading and the first few leading terms of the asymptoics.

Before going a bit deeper into the details, it seems important to emphasise that all of the conjectures stated above 
are supported by the fact that they can be proven to hold in the limiting case of a generalised free fermion model [$\bs{A8}$].
Unfortunately, in the general case $+\infty > c > 0$ considered here, the highly coupled nature of the integrands involved in the mentioned representations does not allow
one for any simple check of the convergence properties.
Doing so demands to push the large deviation based approach to extracting the large number of integration
asymptotic behaviour of multiple integrals far beyond its present boarders, at least in my opinion. The progress I obtained recently in this field, see Chapter \ref{Chapitre AA des integrales multiples},
can be seen as a first step toward obtaining such estimates.

\subsection{Asymptotic behaviour of the reduced density matrix}
\label{Theorem asymptotique rho}

I now describe the first few terms of the large-distance and long-time asymptotic behaviour of $\rho(x,t)$, this provided that the hypothesis and conjectures 
described above do hold. 
Recall that the relative excitation momentum \eqref{ecriture impulsion relative excitation} and energy \eqref{ecriture energie relative excitation} of an excited state is described 
in terms of the dressed momentum $p$ \eqref{definition moment habille} and dressed energy $\veps$ \eqref{definition eqn int eps}. 
One constructs out of these functions the function 
\beq
u(\la)=p(\la)-\tf{t \veps(\la)}{x}
\enq
which plays an essential role in the large-distance and long-time asymptotic behaviour. 
It can be seen as a dressed counterpart of the bare function $u_0(\la) = p_0(\la)-\tf{t \veps_0(\la)}{x}$. 
It is readily seen that the bare function $u_0$ is such that $u_0^{\prime}$ admits a unique zero on $\R$ that is, furthermore, simple. It can be readily  seen through a perturbative 
argument in respect to  $c^{-1}$ -that is however non-uniform in the ratio $t/x$- that $u^{\prime}$ admits a unique zero $\la_0$ on $\R$ , and that this zero is simple. 
This also seems to be confirmed by numerical simulations at lower values of $c$. I shall assume \symbolfootnote[2]{ The analysis would still work, with a few minor modifications,
should this hypothesis break. Indeed, in the worst case scenario one would then deal with a functions $u$
such that $u^{\prime}$ has at most a finite number of zeros on $\R$, each of them being of finite order.} that this still holds true for any value of $c$. 
I will also make the additional assumption that $\la_0$
is uniformly away from the endpoints of the Fermi zone, \textit{viz}. $\la_0 \not= \pm q$.

\vspace{2mm}

Let $x>0$ be large, the ratio $\tf{x}{t}$ be fixed and such that $\la_0$ is uniformly away from $\pm q$. 
Then, \textit{under the validity of the aforestated conjectures}, the thermodynamic limit of the zero-temperature one-particle reduced density matrix
$\rho\pa{x,t}$ admits the large-$x$ asymptotic expansion
\bem
\rho\pa{x,t} = \sqrt{   \f{ -2 \i \pi }{ t \veps^{\prime\prime}\!\pa{\la_0} -x p^{\prime\prime}\!\pa{\la_0}  }   }
 \f{ p^{\prime}\!\pa{\la_0} \; \ex{ \i x  \pac{ u\pa{\la_0} - u\pa{q} } } \;  \abs{ \mc{F}_q^{\la_0} }^2  }
 {  \pac{ \i \pa{x+v_F t}}^{ \big[ F^{\la_0}_q\pa{-q}  \big]^2  }   \pac{ - \i \pa{x-v_F t}}^{ \big[ F^{\la_0}_q\pa{q}  \big]^2  } }
\pa{ \bs{1}_{\R \setminus \intff{-q}{q} }(\la_0)  +\e{o}\pa{1} } \\
\hspace{5cm} +\;   \f{ \ex{-2 \i x p_F} \abs{ \mc{F}_q^{-q} }^2   }
{  \pac{ \i \pa{x+v_F t}}^{ \big[ F^{-q}_q\pa{-q} -1 \big]^2  }   \pac{ - \i \pa{x-v_F t}}^{ \big[ F^{-q}_q\pa{q} \big]^2  } } \big( 1+\e{o}\pa{1}  \big)   \\
+ \; \f{ \abs{ \mc{F}_{\emptyset}^{\emptyset} }^2    }
{  \pac{ \i \pa{x+v_F t}}^{ \big[ F_{\emptyset}^{\emptyset}\pa{-q} \big]^2  }
\pac{ - \i \pa{x-v_F t}}^{ \big[ F_{\emptyset}^{\emptyset}\pa{q} +1\big]^2  } } \big( 1+\e{o}\pa{1}  \big) 
\;\;  + \hspace{-2mm} \sul{ \substack{ \ell^{+}, \ell^{-} \in \mathbb{Z} \\ \bs{\eta} (\ell^+ + \ell^-) \geq 0}  }{ }  \hspace{-5mm} ^{*} \; 
 C_{\ell^+;\ell^-} \f{ \ex{ \i x \vp_{\ell^+;\ell^-} } }{ x^{\De_{\ell^+,\ell^-} } } \big( 1+\e{o}\pa{1}  \big) \;. 
\label{equation fondamentale asymptotiques rho x et t}
\end{multline}

In this formula, $\pm v_F$ corresponds to the velocity of the
excitations on the right/left Fermi boundary. I remind that it is expressed as $v_F= \tf{ \veps^{\prime}\!\pa{q} }{ p^{\prime}\!\pa{q} }$.

The large $x$ (with $\tf{x}{t}$ fixed)  asymptotic expansion has been organised with respect to the various oscillating 
phases it contains. Each phase appears with its own critical exponent driving the power-law decay in $x$. The method, in its present setting, allows one to 
determine the leading (\textit{i}.\textit{e}. up to $\e{o}\pa{1}$ corrections) behaviour of each harmonic. Just as it was the case of the asymptotics of the 
Fredholm determinant of the generalised sine kernel, the $\e{o}\pa{1}$ remainder terms stemming from one of the harmonics may be dominant even in respect to the leading terms coming from another harmonic.

\vspace{1mm}

Also, $ \bs{1}_{ \R \setminus \intff{-q}{q}}$ stands for the characteristic function of the interval $\R \setminus \intff{-q}{q}$.
It is there so as to indicate that, to the leading order, the contribution stemming from the saddle-point only appears in the
space-like regime $\la_0> q$. I do stress however that there also exists saddle-point contributions in the time-like regime where $\la_0 \in \intoo{-q}{q}$. 
These appear in the terms present in the sum over $\ell^{+}, \ell^{-}$.

\subsubsection*{$\bullet$ The critical exponents}

\noindent There are two types of algebraic decay present in \eqref{equation fondamentale asymptotiques rho x et t}:\vspace{1mm}

\begin{itemize}
 
 \item[i)] the square root power-law decay $\pa{t\veps^{\prime\prime}\!\pa{\la_0}- x p^{\prime\prime}\!\pa{\la_0} }^{-\f{1}{2}}$. Its stems from the presence of a Gaussian saddle-point at $\la_0$ and has a characteristic
 critical exponent $ - 1/2$. \vspace{1mm}
 
 \item[ii)] the decay involving the relativistic combinations $x\pm v_F t$ which  exhibits non-trivial critical exponents expressed in terms of the thermodynamic limit of the shift function associated with 
 the specific class excitations generating the contribution.  \vspace{1mm}

\end{itemize}

The power-law behaviour in $x \pm v_F t$ arising in the second and third term in \eqref{equation fondamentale asymptotiques rho x et t}
corresponds to an extrapolation of the conformal field theoretic predictions
for the equal-time correlation functions  to the $t\not=0$ case. However, the first line goes out of the scope of the type of asymptotic behaviour that can be predicted on the basis of 
conformal field theory. In particular, it was absent in the predictions proposed in \cite{BerkovichMurthyWrongCFTBasedPredictionTimeMultiCorrNLSE}. 
The contribution containing the Gaussian term stems from excitations that are far from the linear regime of the spectrum which is the only one that can be grasped by a conformal field theory
and which corresponds to particle/hole excitations with rapidities collapsing on the Fermi boundaries $\pm q$. The full form of the asymptotic behaviour  \eqref{equation fondamentale asymptotiques rho x et t} can be 
argued on the basis of the non-linear Luttinger liquid approach, see \textit{e}.\textit{g}. the excellent review \cite{GlazmanImambekovSchmidtReviewOnNLLuttingerTheory}. 
My result can thus be seen as a strong check of this heuristic method.  The result has also a strong structural resemblance with the non-linear Luttinger liquid based predictions for the 
edge exponents \cite{ImambekovGlazmanEdgeSingInDSFBoseGas} and amplitudes \cite{CauxGlazmanImambekovShashiAsymptoticsStaticDynamicTwoPtFct1DBoseGas}
arising in the behaviour of the spectral 
function\symbolfootnote[2]{The latter corresponds to the space and time Fourier transform of 
$\moy{\Phi\pa{x,t}\Phi^{\dagger}\pa{0,0}} \bs{1}_{\intoo{0}{+\infty}}\pa{t}  + 
\moy{\Phi^{\dagger}\pa{0,0}\Phi\pa{x,t}} \bs{1}_{\intoo{-\infty}{0}}\pa{t} $.} close to the particle or hole excitation thresholds. In particular the definitions
of the amplitudes in terms of form factors coincide. This topic will be given some more attention in Chapter \ref{Chapitre approche des FF aux asymptotiques des correlateurs}.

The critical exponents governing the algebraic decay in the distance of separation are expressed in terms of the thermodynamic limit 
of the shift function (at $\be=0$) associated with certain explicit excited states, namely \vspace{1mm}
\begin{itemize}
\item[$\bullet$] $F_{\emptyset}^{\emptyset}\pa{\la} = -\frac{1 }{ 2}  Z(\la)  - \phi\pa{\la, q} $ is the shift function associated with the excited state having no particle-hole excitations in the $N+1$ quasi-particle sector and 
given by the $N+1$ Bethe roots $\{ \mu^{\emptyset}_{\emptyset} \} $. I remind that $Z$ is the dressed charge while $\phi$ is the dressed phase, \textit{c.f.} \eqref{definition eqn int Z et phi}. \vspace{2mm}
\item[$\bullet$] $F^{-q}_{q}\pa{\la} = -\frac{1 }{ 2}  Z(\la) - \phi\pa{\la, -q}$  is the shift function associated with the excited state corresponding to a single particle-hole excitation such that $p_1=0$ and
$h_1=N+1$ and given by the $N+1$ Bethe roots $\{ \mu_q^{-q} \}  $. \vspace{2mm}
\item[$\bullet$] $F^{ \la_0 }_{ q }\pa{\la} = -\frac{1 }{ 2}  Z(\la)  - \phi\pa{\la, \la_0}$  is the shift function associated with the excited state corresponding to a single particle-hole excitation such that
$h_1=N+1$ and $\mu_{p_a}=\la_{0}$  and given by the $N+1$ Bethe roots  $\{ \mu_q^{\la_0} \} $. \vspace{2mm}
\end{itemize}

The last term in \eqref{equation fondamentale asymptotiques rho x et t} corresponding to a summation over the integers $\ell^+, \ell^-$ is slightly less explicit.   
The sum runs over all integers $\ell^{\pm}$  subject to the  constraint $\bs{\eta}(\ell^+ + \ell^-)\geq 0$.  The parameter $\bs{\eta}$
is as defined in \eqref{definition parametre eta space et time like}. Finally, the $*$ in the sums indicates that one should not sum up over those integers $\ell^+, \ell^-$ giving rise to 
the frequencies that are present in the first three lines of \eqref{equation fondamentale asymptotiques rho x et t}. This sum represents the contributions of the harmonics oscillating with a bigger phase than the first few terms and given by 
\beq
\vp_{\ell^+  ; \ell^-} = \ell^+ u\pa{q} + \ell^- u\pa{-q} \; - \;  (\ell^+ + \ell^-) u(\la_0) \;. 
\enq
Each term is characterised by its own critical exponents 
\beq
\De_{\ell^+;\ell^-} = \pa{ 1 + \ell^+ + \De^+ }^2 + \pa{ \De^- - \ell^-}^2  + \f{ |\ell^+ + \ell^- | }{2} \;,
\enq
where, 
\beq
\De^{ \eps } = - \f{Z\pa{\eps q}}{2} \; - \; \ell^- \phi\pa{\eps q ,-q } \; - \; (\ell^+ + 1) \phi\pa{\eps q ,q }
 \; + \; (\ell^+ + \ell^-) \phi(\eps q , \la_0) \;. 
\enq
The critical exponents can, in fact, be recast in terms of the values taken by appropriate shift functions on the right ($+$)
or left ($-$) Fermi boundaries. I will not discuss this interpretation here as this will be done in Chapter \ref{Chapitre approche des FF aux asymptotiques des correlateurs}
where the asymptotics expansion \eqref{equation fondamentale asymptotiques rho x et t} will be re-derived through a less rigorous but physically more transparent method. 

The method of analysis discussed here allows one to prove that the only harmonics present in the asymptotics are those 
oscillating with one of the frequencies $\vp_{\ell^+,\ell^-}$ and that they decay, to the leading order (\textit{i.e.}
up to $\e{o}\pa{1}$ terms), with the critical exponent $\De_{\ell^+,\ell^-}$. Due to the lack of a more precise information on the 
structure of the sub-leading terms in the large-$x$ asymptotic behaviour of the Fredholm determinant discussed previously, 
the method does not allow one to provide an explicit prediction for the amplitudes $C_{\ell^+,\ell^-}$.

\subsubsection*{$\bullet$ The amplitudes}

The amplitudes ($\big| \mc{F}_q^{\la_0} \big|^2$,
$\big|  \mc{F}_{\emptyset}^{\emptyset} \big|^2$ or $\big| \mc{F}_q^{-q} \big|^2$) appearing in front of the three first terms of the series have a clear physical interpretation. 
They correspond to properly normalised in the volume $L$  moduli squared of form factors of the conjugated field $\Phi^{\dagger}$.
More precisely, \vspace{1mm}

\begin{itemize}

\item[$\bullet$] $\abs{\mc{F}_{\emptyset}^{\emptyset}}^2$ corresponds to the case where the form factor  of $\Phi^{\dagger}$ is taken between
the ground state represented by the set of Bethe roots $\{\la\}$ and the excited state in the $N+1$ quasi-particle sector  with no holes and particles 
that is parametrised  by the Bethe roots $\big\{ \mu^{\emptyset}_{\emptyset} \big\}$:
\beq
\abs{ \mc{F}_{\emptyset}^{\emptyset} }^2=  \lim_{N,L\tend +\infty}
\paf{L}{2\pi}^{ \big[ F_{\emptyset}^{\emptyset}\pa{q} + 1 \big]^2 + \big[ F_{\emptyset}^{\emptyset} \pa{-q} \big]^2 }
\Bigg|  \f{ \bra{ \Psi\big( \big\{\mu^{\emptyset}_{\emptyset} \big\} \big) }\Phi^{\dagger}\!\pa{0,0}\ket{ \Psi\big(\big\{\la \big\} \big)  }  }
{ \norm{  \Psi\big(\big\{\la \big\} \big)   } \cdot \norm{ \Psi\big( \big\{ \mu^{\emptyset}_{\emptyset} \big\} \big) } } \Bigg|^2  \;.
\enq

\item[$\bullet$]  $\abs{\mc{F}_q^{ \la_0 } }^2$ involves the form factor of $\Phi^{\dagger}$
taken between the ground state and an excited state in  the $N+1$ quasi-particle sector with one particle at $\la_{0}$ and one hole at $q$ given by the Bethe roots $\big\{\mu^{\la_0}_{q} \big\} $:
\beq
\abs{ \mc{F}_{q}^{\la_0} }^2=  \lim_{N,L\tend +\infty} \paf{L}{2\pi}^{ \big[ F_{q}^{\la_0}\pa{q} \big]^2 + \big[ F_{q}^{\la_0}\pa{-q} \big]^2 +1 }
\Bigg|  \f{ \bra{ \Psi\big( \big\{\mu^{\la_0}_{q} \big\} \big)  }\Phi^{\dagger}\!\pa{0,0}\ket{  \Psi\big( \big\{\la \big\} \big)  } }
{ \Norm{  \Psi\big( \big\{\la \big\} \big)  } \cdot \Norm{ \Psi\big( \big\{\mu^{\la_0}_{q} \big\} \big)   }    } \Bigg|^2  \;.
\enq
\item[$\bullet$] $\abs{\mc{F}_q^{-q}}^2$  involves the form factor of $\Phi^{\dagger}$
taken between the ground state and an excited state in the  $N+1$ quasi-particle sector with one 
particle at $-q$ and one hole at $q$ given by the Bethe roots $\big\{\mu^{-q}_{q} \big\}$:
\beq
\abs{ \mc{F}_{q}^{-q} }^2=  \lim_{N,L\tend +\infty}
\paf{L}{2\pi}^{ \big[ F_{q}^{-q} \pa{q} \big]^2 + \big[ F_{q}^{-q}\pa{-q} -1 \big]^2 }
\Bigg|  \f{ \bra{ \Psi\big( \big\{\mu^{-q}_{q} \big\} \big)  }\Phi^{\dagger}\!\pa{0,0}\ket{ \Psi\big( \big\{\la \big\} \big) } }
{ \norm{ \Psi\big( \big\{\la \big\} \big) }   \cdot \norm{ \Psi\big( \big\{\mu^{-q}_{q} \big\} \big)  }  } \Bigg|^2 \;.
\enq

\end{itemize}

\vspace{1mm}

The very existence of the $L\tend + \infty$ limit is ensured by the results presented in Subsection \ref{Subsection Thermo Lim FF}. 
The explicit (but a bit cumbersome) expressions for the amplitudes can be found in Appendix A.3 of [$\bs{A7}$]. 
It is important to remark that the  exponents governing the large-volume $L$ decay of the form factors associated with the 
amplitudes ($| \mc{F}_q^{\la_0} |^2$, $|  \mc{F}_{\emptyset}^{\emptyset} |^2$ or $| \mc{F}_q^{-q} |^2$)
coincide precisely with the critical exponents governing the decay of the relativistic combinations $x \pm v_F t$. 
This confirms the conformal field theoretic predictions for the behaviour of such amplitudes in respect to the volume, \textit{c.f.} \eqref{ecriture prediction CFT pour definition amplitude dans fct 2 pts}.



\section{The multidimensional flow analysis of the reduced density matrix}
\label{Section Facteurs de Formes series}

Starting from \eqref{definition reduced density matrix}, one inserts the closure relation\symbolfootnote[3]{The completeness of the states built through the Bethe Ansatz has been proven in \cite{DorlasOrthogonalityAndCompletenessNLSE}} 
between the field and conjugated field operators, invokes the form of the spatial and temporal evolution of the fields 
\beq
\Phi\!\pa{x,t} = \ex{ - \i x \op{P}_L  + \i t \op{H}_{NLS} } \Phi\!\pa{0,0} \ex{ \i x \op{P}_L - \i t \bs{H}_{NLS}  } \; ,
\enq
where $\op{H}_{NLS} $ is the non-linear Schr\"{o}dinger model Hamiltonian and $ \op{P}_L $ is the momentum operator and then applying the 
formulae  \eqref{ecriture action op impulsion ete energie sur etat propre}-\eqref{ecriture impulsion et energie} for their eigenvalues, one obtains the form factor 
expansion for the zero-temperature reduced density matrix \eqref{definition reduced density matrix}:
\beq
\rho_N\pa{x,t} =
\sul{ \substack{\ell_1 < \dots < \ell_{N+1} \\ \ell_a \in \mathbb{Z} } }{}
\f{  \pl{a=1}{N+1} \ex{ \i x u_0\pa{\mu_{\ell_a}}}  }{  \pl{a=1}{N} \ex{ \i x u_0\pa{\la_a}}   }
\f{ \abs{ \bra{ \psi\pa{\paa{\mu_{\ell_a}}_1^{N+1}} } \Phi^{\dagger}\!\pa{0,0} \ket{ \psi\pa{\paa{\la_a}_1^N} } }^2 }
{ \norm{ \psi\pa{\paa{\mu_{\ell_a}}_1^{N+1} } }^2  \cdot \norm{ \psi\pa{\paa{\la_a}_1^{N} } }^2   } \quad \e{with} \quad u_0 \, = \,  p_0  \, - \, \f{t}{x} \veps_0 \;.
\label{equation developement correlateur serie FF}
\enq
The above series runs through all the possible  choices of integers $\ell_a$, $a=1, \dots, N+1$  such that $\ell_1<\dots<\ell_{N+1}$.
This series is convergent, but \textit{only} in the sense of distributions. This can be readily seen on the level of the integral representation
along with the completeness, in an appropriate subspace of $L^2_{\e{sym}}(\intff{0}{L}^{N+1})$,  of the system $ \big\{ \vp\big( \bs{x}_{N+1} \mid \{ \mu_{\ell_a} \}_1^{N+1} \big)  \big\}_{ \{ \ell_a \}_1^{N+1} }  $.

\vspace{1mm}
As announced, I will start by arguing in favour of several reasonable approximations  which reduce the above form factor series to another, effective one, 
whose structure already allows one for its rigorous analysis. However, in order to write down the effective form factor series, I need
revisit the conclusions of the large-volume behaviour of the form factors of  the conjugated field.

\subsection{The effective form factors}
\label{Subsection effective FF}

It have shown in Subsection \ref{Subsection Thermo Lim FF} of the previous chapter, that the below factorisation holds:
\beq
\f{  \abs{ \bra{\psi\big( \paa{\mu_{\ell_a}}_1^{N+1} \big) } \Phi^{\dagger}\!\pa{0,0} \ket{ \psi\big( \paa{\la_a}_1^N \big)  }   }^2 }
{ \norm{\psi\big( \paa{\mu_{\ell_a}}_{1}^{N+1} \big) }^2  \norm{ \psi\big( \paa{\la_a}_{1}^{N} \big) }^2     } =
\wh{\mc{G}}_{N;1} \pab{ \! \!   \paa{p_a}_1^n \! \!  }{ \! \!  \paa{h_a}_1^n \! \!  } \pac{\wh{F}_{\paa{\ell_a}}, \wh{\xi}_{\paa{\ell_a}}, \wh{\xi} \, }
\cdot \wh{D}_{N}\pab{ \! \!   \paa{p_a}_1^n \! \!  }{ \! \!  \paa{h_a}_1^n \! \!  } \pac{\wh{F}_{\paa{\ell_a}}, \wh{\xi}_{\paa{\ell_a}}, \wh{\xi} \, }
 \;.
\label{ecriture FF exact}
\enq
This factorisation is given in terms of two functionals $\wh{D}_N$ and $\wh{G}_{N;1}$ which, according to the results on the large-$L$ behaviour of the smooth and discrete parts, satisfy
\beq
\pa{ \wh{\mc{G}}_{N;1} \wh{D}_{N}}\pab{ \! \!   \paa{p_a}_1^n \! \!  }{ \! \!  \paa{h_a}_1^n \! \!  }
\pac{\wh{F}_{\paa{\ell_a}}, \wh{\xi}_{\paa{\ell_a}}, \wh{\xi} \, }
%
%
=
\wh{\mc{G}}_{N;1} \pab{ \! \!   \paa{p_a}_1^n \! \!  }{ \! \!  \paa{h_a}_1^n \! \!  } \pac{F_0, \xi, \xi_{F_0} }
\cdot \wh{D}_{N}\pab{ \! \!   \paa{p_a}_1^n \! \!  }{ \! \!  \paa{h_a}_1^n \! \!  } \pac{F_0, \xi, \xi_{F_0} } \cdot 
\pa{1+\e{O}\paf{\ln L}{L}} \;.
\label{ecriture comportement FF part trou a grand L}
\enq
The remainder is uniform in $\{p_a\}$ and $\{h_a\}$ provided that the number $n$ of particle-hole excitations is bounded in $L$. 
The functionals appearing on the \textit{rhs} act on \vspace{2mm}
\begin{itemize}
\item[i)]  \textit{the thermodynamic limit} $\xi\pa{\la}$ of the counting function \eqref{ecriture fct cptge thermo},\vspace{2mm}
\item[ii)]  \textit{the thermodynamic limit} $F_{0}\!\pa{\la}$ of the shift function at $\be=0$ associated with an excited state labelled by
the set of integers $\{ p_a \}_1^{n}$ and $\{ h_a \}_1^{n}$ \eqref{ecriture limite thermo fction shift}. The auxiliary arguments of $F_{0}$ are the rapidities 
$\{ \mu_{p_a} \}_1^n$ and $\paa{\mu_{h_a}}_1^n$ defined as $\mu_a = \xi^{-1}\big(a/L \big)$, $a \in \mathbb{Z}$. \vspace{2mm}
\item[iii)] The counting function associated  with $F_0$: $\xi_{F_0}\!\pa{\la} = \xi\!\pa{\la} + \tf{F_0\!\pa{\la}}{L}$.  \vspace{2mm}
\end{itemize}
 I have dropped the dependence of the shift function $F_0$ on the auxiliary arguments $\{ \mu_{p_a} \}_1^n$ and $\paa{\mu_{h_a}}_1^n$ in 
 \eqref{ecriture comportement FF part trou a grand L} since this information is undercurrent by the set of integers labelling the functionals
$\wh{D}_{N}$ and $\wh{\mc{G}}_{N,1}$.

Also, note that a straightforward analysis of the linear integral equation satisfied by the dressed momentum $p\pa{\la}$
ensures that it is a biholomorphism on some sufficiently narrow strip $U_{\de}$ around the real axis and that $p$ is an increasing bijection from $\R$
onto $\R$. Therefore, due to \eqref{ecriture fct cptge thermo}, $\xi$ enjoys these properties as well. This guarantees that the parameters $\mu_a$ are well defined.

The explicit expression for the discrete part is given by \eqref{definition partie discrete}. Below, I provide a rewriting of the original expression for 
the smooth part \eqref{definition partie lisse} which will appear to be more useful for further handlings. 

\subsubsection*{  $\bullet$ The smooth part revisited}

The functional $\wh{\mc{G}}_{N,1}$ represents the so-called smooth part of the form factor. I extend its definition to any $\ga$ as
\bem
\hspace{-1cm}\wh{\mc{G}}_{N;\ga} \pab{ \! \!   \paa{p_a} \! \!  }{ \! \!  \paa{h_a} \! \!  } \pac{F_0, \xi, \xi_{F_0} } =
\f{  V_{N;1} \!\pa{\mu_{N+1}} V_{N;-1} \!\pa{\mu_{N+1}}   }{ \det_{N+1}\big[ \Xi^{\pa{\mu}} \pac{\xi} \big] \det_{N}\big[ \Xi^{\pa{\la}}\pac{\xi_{F_0}} ] }
W_n\pab{ \{\mu_{p_a} \}_1^n }{ \{\mu_{h_a} \}_1^n }  \pl{a=1}{n} \pl{\eps=\pm}{} \paa{  \f{ V_{N;\eps} (\mu_{p_a}) }{ V_{N;\eps} (\mu_{h_a}) }
 \f{ \mu_{h_a}-\mu_{N+1} + \i \eps c }{ \mu_{p_a}-\mu_{N+1} + \i \eps c  } }
 \\
\times   W_N\pab{ \{\la_a \}_1^N }{ \{\mu_a \}_1^N }
 \det_{N}\Big[ \de_{jk}+ \ga \wh{V}_{\! jk}\!\pac{F_0}\!\pa{ \{\la_a \}_1^N ; \{\mu_{\ell_a} \}_1^{N+1} }  \Big]
 \det_{N}\Big[ \de_{jk}+ \ga \wh{\ov{V}}_{\! jk}\! \pac{F_0}\!\pa{ \{\la_a \}_1^N ; \{\mu_{\ell_a} \}_1^{N+1} }  \Big] \;.
\label{ecriture fonctionnelle hat GN gamma}
\end{multline}
 For any set of generic parameters
$\pa{ \{ z_a \}_1^n ; \{ y_a \}_1^{n} } \in U_{\de}^n\times U_{\de}^{n}$ one has
\beq
W_n\pab{ \{ z_a \}_1^n}{ \{ y_a \}_1^{n} } = \pl{a,b=1}{n} \f{ \pa{z_a-y_b- \i c}\pa{y_a-z_b- \i c} }{ \pa{y_a-y_b - \i c}\pa{z_a-z_b- \i c} }
\qquad \e{and} \qquad
V_{N;\eps} \pa{\om} = \pl{a=1}{N} \f{ \om -\la_b + \i \eps c }{ \om -\mu_b + \i \eps c } \;. 
\label{definition fonction WN et VNepsilon}
\enq
 Also, I remind the definitions 
\beq
\Xi^{\pa{\mu}}_{ab} \pac{\xi} = \de_{ab} - \f{  K \pa{\mu_{\ell_a} -\mu_{\ell_b} } }{ 2\pi L \, \xi^{\prime}\!\pa{\mu_{\ell_b}} }
\qquad \e{and} \qquad
\Xi^{\pa{\la}}_{ab} \pac{\xi_{F_0}} = \de_{ab} - \f{  K \pa{\la_{a} -\la_{b} } }{ 2\pi L \, \xi^{\prime}_{F_0}\!\!\pa{\la_{b}} }
\enq

Finally, for any set of generic parameters $\pa{ \{ z_a \}_1^n ; \{ y_a \}_1^{n+1} } \in U_{\de}^n\times U_{\de}^{n+1}$
the entries of the two determinants in the numerator read
\beqa
\wh{V}_{\! k\ell}\!\pac{\nu}\pa{ \{ z_a \}_1^n ; \{ y_a \}_1^{n+1} }  &=& - \i\,  \f{ \pl{a=1}{n+1} \pa{ z_k-y_a}}{ \pl{a\not= k }{ n} \pa{ z_k-z_a} }
\f{ \pl{a=1}{n} \pa{ z_k-z_a + \i c} }{ \pl{a=1}{n+1} \pa{z_k-y_a+ \i c}}  \f{ K\pa{z_k-z_{\ell}} }{ \ex{-2 \i\pi \nu\pa{z_k}}-1 }  \nonumber \\
\wh{\ov{V}}_{\! k\ell}\!\pac{\nu}\pa{ \{ z_a \}_1^n ; \{ y_a \}_1^{n+1} }  &=&  \i \, \f{ \pl{a=1}{n+1} \pa{z_k-y_a}}{ \pl{a\not= k }{ n} \pa{z_k-z_a} }
\f{ \pl{a=1}{n} \pa{z_k-z_a - \i c} }{ \pl{a=1}{n+1} \pa{z_k-y_a - \i c}}  \f{ K\pa{z_k-z_{\ell}} }{ \ex{2 \i \pi  \nu\pa{z_k}}-1 }
\label{appendix thermo lim FF definition entree V et Vbar chapeau}
\eeqa
I stress that the singularities at $z_k=z_j$, $j \not=k$ of the determinants associated with, either $\wh{V}$ or $\wh{\ov{V}}$, 
are, in fact, only apparent. The mechanism which leads to the cancellation of singularities  is discussed in \cite{KozKitMailSlaTer6VertexRMatrixMasterEquation,KozKitMailSlaTerXXZsgZsgZAsymptotics}.

\subsection{Arguments for the effective form factors series}
\label{Subsection effective FF series}

It is a common belief\symbolfootnote[2]{ The results obtained in Appendix B.2 and B.3 of [$\bs{A7}$]
constitute a proof of this statement in the case of a generalised free fermion model.}
that the contribution  to the form factor expansion of a two-point function issuing from the excited states whose energies differ macroscopically 
from the ground state's one (\textit{i.e.} by a quantity scaling as some positive power of $L$) disappears in the $L\tend +\infty$ limit. 
Heuristically, this vanishing issues from an extremely quick oscillation of the associated phase factors
adjoined to a sufficiently fast decay of the form factors taken between states differing by large values of their momenta and energies.

In the following, I will build on the assumption that the only part of the form factor expansion in \eqref{equation developement correlateur serie FF}
that has a non-vanishing contribution to the thermodynamic limit $\rho\pa{x,t}$ of $\rho_N\pa{x,t}$ corresponds to a summation over all the
excited states which are realised as some finite (in the sense that not scaling with $L$) number $n$, $n=0,1,\dots$  of particle-hole excitations above
the $N+1$-particle ground state. Indeed, these are the only excited states that can have a finite (\textit{i}.\textit{e}. not scaling with $L$) relative excitation 
energy.

\vspace{2mm}

Even when dealing with excited states realised as a finite number $n$ of particle-hole excitations in the $N+1$ quasi-particle sector, it is still
possible to end-up with a macroscopically large (\textit{i}.\textit{e}. diverging when $L \tend + \infty$) relative excitation energy
if the rapidities of the particles become very large (\textit{i}.\textit{e}. scale with $L$).
Such configurations of the integers are obtained by choosing the integers $p_a$ to lie at a distance larger than $\e{O}(L)$ from $\intn{1}{N+1}$.
Consequently, I will also drop the contribution of such excited states.

\vspace{2mm}

Limiting the sum over all the excited states in the $N+1$-particle sector to those having the same per-site energy
that the ground state means that one effectively neglects correcting terms in the lattice size $L$.
It thus seems reasonable to assume that solely the leading large-$L$ asymptotic behaviour
of individual form factors contributes to the thermodynamic limit of $\rho_{N}\pa{x,t}$.
The assumption can be checked to hold as long as one focuses on the contributions of all the excited states
realised as $0,1,\dots,n$- particle/hole excitations, with $n$ bounded in $L$. 
However, in principle, problems could arise when the number $n$ grows with $L$.

\vspace{2mm}
\noindent The above assumptions allow one for the following simplifications of the form factor series \eqref{equation developement correlateur serie FF}: \vspace{1mm}

\begin{itemize}

\item[i)] The summation over the excited states having a large excitation energy is dropped. To be more precise,  this means that I introduce
a "cut-off" in respect to the range of the integers entering in the description of the rapidities of the particles. Namely, I assume that
the integers $p_a$ are restricted to belong to the set\symbolfootnote[3]{One can choose $w_L$ to scale as any power $L^{1+\eps}$, where $\eps>0$ is small enough
but arbitrary otherwise. The choice $\eps=\tf{1}{4}$ has only been made for definiteness. One can find a  more explicit illustration of this property in 
Appendix B.1 of [$\bs{A7}$].}
\beq
 \mc{B}^{\e{ext}}_L \equiv \Big\{ n \in \mathbb{Z} \; : \;  -w_L < n < w_L  \Big\} \setminus \intn{1}{N+1} \qquad \e{where} \qquad
w_L \sim  L^{1+\f{1}{4}}   \;.
\enq

\item[ii)] The exact form factors are represented by the \textit{rhs} of \eqref{ecriture comportement FF part trou a grand L} while dropping  
  the $\e{O}\pa{L^{-1} \cdot \ln L }$ remainders  in the large-volume behaviour of form factors
given in \eqref{ecriture comportement FF part trou a grand L}.

\item[iii)] The oscillating exponent $\sul{a=1}{N+1}u_0\pa{\mu_{\ell_a}}- \sul{a=1}{N}u_0\pa{\la_a} $
is replaced by its thermodynamic limit:
\beq
\sul{a=1}{n} \big[ u(\mu_{p_a}) \, - \, u(\mu_{h_a}) \big] \;. 
\enq

\end{itemize}

Note that point ii) above implies that the roots $\paa{\mu_{\ell_a}}_1^{N+1}$ for an excited state whose particles' (resp. holes')
rapidities are labelled by the integers  $\paa{p_a}_{a=1}^{n}$ (resp. $\paa{h_a}_{a=1}^{n}$) are now given as the pre-image $\mu_{\ell_a}\,=\, \xi^{-1}\big( \tf{\ell_a}{L} \big)$
of $\ell_a/L$ by the asymptotic counting function. Therefore, the value taken by the positions of the rapidities \textit{does not depend} any more 
on the specific choice of the excited state one considers.
In this respect, one effectively recovers a description of the excitations that is in the spirit of a free fermionic model.

The above assumptions, put together, lead to the conjecture
\begin{conj}
\label{Conjecture Fondamentale}
The thermodynamic limit of the reduced density matrix $\rho_N\pa{x,t}$ coincides with the thermodynamic limit
of the effective reduced density matrix $\rho_{N;\e{eff}}\pa{x,t}$:
\beq
\lim_{N,L \tend +\infty} \rho_N\pa{x,t}=
\lim_{N,L \tend +\infty}   \rho_{N;\e{eff}}\pa{x,t}
\enq
where $\rho_{N;\e{eff}}\pa{x,t}$ is given by the series
\beq
\hspace{-5mm} \rho_{N;\e{eff}}\pa{x,t} =  \sul{n = 0}{ N+1 }
\sul{ \substack{p_1<\dots < p_n \\ p_a \in \mc{B}^{\e{ext}}_{L}  } }{}
\sul{ \substack{h_1<\dots < h_n \\ h_a  \in \mc{B}^{\e{int}}_{L}  } }{}
\pl{a=1}{n} \bigg\{ \f{ \ex{ -\i xu\pa{\mu_{h_a}} } }{ \ex{-\i xu\pa{\mu_{p_a}} } } \bigg\}  \cdot
\pa{ \wh{D}_{N} \; \wh{\mc{G}}_{N; 1 } } \pab{ \!\! \paa{p_a}_1^n \!\!  }{  \!\! \paa{h_a}_1^n \!\!  }
\pac{  F_{0}\pabb{ * }{ \{ \mu_{p_a} \} } {\paa{\mu_{h_a}} } \; ; \xi_{} \; ; \xi_{ F_{0}}  } \;.
\label{ecriture series FF effective pour sommation}
\enq
There $\mc{B}_L=\paa{n\in \mathbb{Z} \; : \; -w_L < n < w_L}$, \; $\mc{B}_L^{\e{ext}}=\mc{B}_L\setminus \intn{1}{N+1}$ and
$\mc{B}_L^{\e{int}}=\intn{1}{N+1}$. Also, the $*$ refers to the running variable of $F_0$ on which the two functionals act.

%
%
%
%
%
%
%
%
%

\end{conj}

The effective form factor series  \eqref{ecriture series FF effective pour sommation}
possesses several differences with respect to one that would arise in a generalised free fermion model (\textit{c}.\textit{f}.
equation \eqref{definition Fonction generatrice X_N} of Appendix B in [$\bs{A7}$]). Namely, 
\begin{itemize}
\item[$\bullet$] the shift function $F_{0}$ depends parametrically on the
rapidities of the particles and holes entering in the description of \textit{each} excited state one considers, \textit{c}.\textit{f}.
\eqref{ecriture limite thermo fction shift}.  It is thus summation \textit{dependent}.\vspace{1mm}
\item[$\bullet$]  Each summand is weighted by the factor  $\wh{\mc{G}}_{N;1}$  that takes into account the more complex structure of the scattering and of the scalar
products in the interacting model.  This introduces a strong coupling between the summation variables $\{p_a\}_1^n$ and $\{h_a\}_1^n$.
%
\end{itemize}
\vspace{2mm}

A separation of variables that would allow one for a resummation of \eqref{ecriture series FF effective pour sommation} is not
possible for precisely these two reasons. To bypass this issue, it appears convenient to follow the steps described below.
First, one introduces a $\ga$-deformation $\rho_{N;\e{eff}}\!\pa{x,t\mid \ga}$ of \eqref{ecriture series FF effective pour sommation} such that
 $\rho_{N;\e{eff}}\pa{x,t\mid \ga}_{\mid \ga=1}=\rho_{N;\e{eff}}\pa{x,t}$:
\beq
\rho_{N;\e{eff}}\!\pa{x,t\mid \ga} =  \sul{n = 0}{ N +1}
\sul{ \substack{p_1<\dots < p_n \\ p_a \in \mc{B}^{\e{ext}}_{L}  } }{}
\sul{ \substack{h_1<\dots < h_n \\ h_a  \in \mc{B}^{\e{int}}_{L}  } }{}
\pl{a=1}{n} \bigg\{ \f{ \ex{ -\i x u\pa{\mu_{h_a}} } }{ \ex{- \i x u\pa{\mu_{p_a}} } } \bigg\} \cdot 
\pa{ \wh{D}_{N} \; \wh{\mc{G}}_{N;\ga} } \pab{ \!\! \paa{p_a}_1^n \!\!  }{  \!\! \paa{h_a}_1^n \!\!  }
\pac{ \ga F_{0}\pabb{ * }{ \{ \mu_{p_a} \}_1^n } {\paa{\mu_{h_a}}_1^n } \; ; \xi_{} \; ; \xi_{\ga F_{0}}  }
%
%
\;.
\label{ecriture series FF effective gamma deformee}
\enq
For any finite $N$ and $L$, one can prove by using the explicit representations \eqref{definition partie discrete} for $\wh{D}_{N}$
and \eqref{ecriture fonctionnelle hat GN gamma} for $\wh{\mc{G}}_{N;\ga}$
that the $\ga$-deformation $\rho_{N;\e{eff}}\pa{x,t\mid \ga}$ is holomorphic in $\ga$ belonging to an open neighbourhood of the
closed unit disc\symbolfootnote[2]{The apparent singularity of the determinants at $\ex{\pm 2 \i \pi F_0\,(\la_k)}-1=0$,
\textit{c}.\textit{f}. \eqref{appendix thermo lim FF definition entree V et Vbar chapeau}, are cancelled by the pre-factors 
$\sin^{2} [ \pi F_0(\la_k) ]$ present in $\wh{D}_N$, \textit{c}.\textit{f}. \eqref{definition partie discrete}.}. 
Hence, its Taylor series around $\ga=0$ converges up to $\ga=1$.
It is show in Theorem C.1 of [$\bs{A7}$] that, given any fixed $m$, the $m^{\e{th}}$ Taylor coefficient of
$\rho_{N;\e{eff}}\!\pa{x,t\mid \ga}$ at $\ga=0$:
\beq
\rho_{N;\e{eff}}^{\pa{m}}\!\pa{x,t} = \left. \f{\Dp{}^m}{\Dp{}\ga^m} \rho_{N;\e{eff}}\!\pa{x,t\mid \ga}   \right|_{\ga=0} \;,
\label{definition rho N eff derivee m ieme}
\enq
admits a well-defined  thermodynamic limit $\rho_{\e{eff}}^{\pa{m}}\!\pa{x,t}$. This fact is absolutely not clear on the level of 
\eqref{definition rho N eff derivee m ieme} as, due to \eqref{AppendixThermoLimD+zero}-\eqref{definition fonctionelle RNn},
each individual summand vanishes as a fractional power-law in $L$ that depends on the excited state considered.
I proved the existence of this thermodynamic limit by carrying out a re-summation of the series representing 
$\rho_{N;\e{eff}}^{\pa{m}}\!\pa{x,t}$ so as to recast the object in terms of a finite amount of Riemann-sums converging
to multiple integrals in the thermodynamic limit. This approach leads to a representation for the thermodynamic limit
$\rho_{\e{eff}}^{\pa{m}}\!\pa{x,t}$ given in terms of a finite sum of multiple integrals
and corresponds to a truncation of the so-called multidimensional Fredholm series, see Appendix C of  [$\bs{A7}$].

It is then possible to build on the information provided by very existence of the thermodynamic limit so as to recast 
$\rho_{\e{eff}}^{\pa{m}}\!\pa{x,t}$ in terms of yet another finite sum of multiple integrals. This new representation corresponds 
to a truncation of the so-called multidimensional Natte series that will be discussed below. The latter gives a straightforward access to the large-$x$ asymptotic expansion
of $\rho_{\e{eff}}^{\pa{m}}\!\pa{x,t}$.

The proof of the existence of the thermodynamic limit and the construction of the truncated multidimensional Natte series for
$\rho^{\pa{m}}_{\e{eff}}\!\pa{x,t}$ constitute the rigorous and conjecture free part of the
analysis. Those results will be summarised in Theorem \ref{Theorem comportement asympt coeff Taylor limite Thermo} to come.

 Working on the level of the Taylor coefficients $\rho^{\pa{m}}_{N;\e{eff}}\!\pa{x,t}$ instead of the full function $\rho_{N;\e{eff}}\!\pa{x,t\mid \ga}$
taken at $\ga=1$ has the advantage of separating all questions of convergence of the obtained representations from the question of 
well-definiteness of the various re-summations and deformation procedures that are carried out
on $\rho_{N;\e{eff}}^{\pa{m}}\!\pa{x,t}$ (and subsequently on $\rho_{\e{eff}}^{\pa{m}}\!\pa{x,t}$ once that the thermodynamic limit is taken).
Indeed, by taking the $m^{\e{th}}$ $\ga$-derivative at $\ga=0$,  one will always end up dealing with a finite number of sums of multiple integrals.
However, if one would have carried out the re-summations directly on the level of $\rho_{\e{eff}}\!\pa{x,t}$, one would have ended up with a series of
multiple integrals instead of a finite sum. The convergence of such a series constitutes a separate question that deserves, in its own right,
another study. 

There is also a way to check, on a heuristic level of rigour, the well-foundedness of the various "truncation" hypothesis. This check can be preformed on the example
of the equal time ($t=0$) density-density correlation function. One can compute the presumably existing thermodynamic limit 
of the correlator in two ways. On the one hand, by using its form factor expansion and the hypothesis I described above. 
On the other hand, by using the master equation approach \cite{KozKitMailSlaTerXXZsgZsgZAsymptotics} which takes into account the contribution of \textit{all}
excited states and builds on totally different techniques. In the thermodynamic limit, both approaches lead to the same answer. I refer to Appendix D.3
of [$\bs{A9}$] for more details.

\subsection{An operator ordering}
\label{Subsection Operator ordering for functional translation}

Prior to carrying out the re-summation of the form factor expansion for $\rho_{N;\e{eff}}^{\pa{m}}\!\pa{x,t}$,
I will discuss a way of representing functional translations and generalisations thereof. These tools allow one
to separate the variables in the sums occurring in \eqref{definition rho N eff derivee m ieme}, and carry out the various re-summations.
A more precise analysis and discussion of these constructions is postponed to Appendix D of [$\bs{A7}$].
In the following, $\msc{O}\pa{W}$ stands for the ring of holomorphic functions in $\ell$ variables on the set $W \subset \Cx^{\ell}$.
Also, $f \in \msc{O}\pa{W}$, with $W$ non-open means that $f$ is a holomorphic function
on some open neighbourhood of $W$. Finally, given $f\in L^{\infty}(S)$,  
let $\norm{f}_{L^{\infty}(S)} = \e{supess}_{s \in S} \abs{f\!\pa{s}}$.

\vspace{2mm}
In the course of the analysis, one deals with various examples ($\wh{D}_{N}$ , $\wh{\mc{G}}_{N;\ga}$, $\dots$)
of functionals $\mc{F}\pac{\nu}$ acting on holomorphic functions $\nu$.
The function $\nu$ will always be defined on some compact subset $M$ of $\Cx$ whereas the explicit expression for $\mc{F}\pac{\nu}$
will only involve the values taken by $\nu$  on a smaller compact\symbolfootnote[3]{I remind that
$\e{Int}\pa{M}$ stands for the interior of the set M.} $K \subset \e{Int}\pa{M}$.
In fact, all the functionals that  will be encountered share the regularity property below:

\begin{defin}
\label{Definition Fonctionelle reguliere}
Let $M$, $K$ be compacts in $\Cx$ such that $K \subset \e{Int}\pa{M}$. Let $ W_{z}$ be a compact in $ \Cx^{\ell_{z}}$, $\ell_z \in \mathbb{N}$. An $\ell_z$-parameter family of functionals $\mc{F}\pac{\cdot}\pa{\bs{z}}$
depending on a set of auxiliary variables $\bs{z} \in W_{z}$
is said to be regular (in respect to the pair $\pa{M,K}$) if

\begin{enumerate}[i)]

\item there exists constants  $C_{\mc{F}}>0$ and $C^{\prime}>0$ such that for any $f,g \in \msc{O}\pa{M}$
\beq
\norm{f}_{ L^{\infty}(K) } \, + \, \norm{g}_{ L^{\infty}(K) } < C_{\mc{F}} \quad \Rightarrow \quad
\norm{ \mc{F}\pac{f}\pa{\cdot}-\mc{F}\pac{g} \pa{\cdot} }_{ L^{\infty}(W_z) } \; <  \; C^{\prime} \cdot \norm{f-g}_{ L^{\infty}(K) } \;,
\label{Appendice transl fonct Conte fnelle}
\enq
where the $\cdot$ indicates that the norm is computed in respect to the set of auxiliary variables $ \bs{z} \in W_{ z }$.

\item Given any open subset  $W_y \subset \Cx^{\ell_y}$, for some $\ell_y \in \mathbb{N}$,
if $\nu\pa{\la,\bs{y} } \in \msc{O}\pa{M \times W_y}$ is such that $\norm{\nu}_{L^{\infty}(K\times W_y)} <C_{\mc{F}}$, then
the function $\pa{\bs{y},\bs{z}} \mapsto \mc{F}\pac{\nu\pa{ *, \bs{y}}}\pa{\bs{z}}$ is holomorphic on $W_y\times W_z$.
The $*$ indicates the running variable $\la$ of $\nu\pa{\la,\bs{y}}$ on which the functional $\mc{F}\pac{\cdot}\pa{\bs{z}}$ acts.
\end{enumerate}

\noindent The constant $C_{\mc{F}}$ appearing above is called constant of regularity of the functional.

\end{defin}

This regularity property is at the heart of the aforementioned representation for the functional translation and generalisations thereof. However, 
priori to discussing these , I first need to define the discretisation of the boundary of a compact.

\begin{defin}
\label{Definition point discretisation}
Let $M$ be a compact with $n$ holes (\textit{i}.\textit{e}. $\Cx\setminus M$ has $n$ bounded connected components) and such that
$\Dp{}M$ can be realised as a disjoint union of $n+1$ smooth Jordan curves 
\beq
\ga_{a} : \intff{0}{1} \tend \Dp{}M \; , \qquad i.e. \quad \Dp{}M=\bigsqcup_{a=1}^{n+1} \ga_a\pa{ \intff{0}{1}} \; .
\nonumber
\enq
A discretisation (of order $s$) of $\Dp{}M$  corresponds to
a collection of $\pa{n+1}\pa{s+2}$ points 
\beq
t_{j,a} \, = \, \ga_a(x_j) \qquad  with  \quad j=0,\dots,s+1 \quad  and   \quad a=1,\dots,n+1 \quad 
 where  \qquad x_0=0\leq x_1<\dots<x_{s}\leq 1=x_{s+1} \;
\nonumber
\enq
is a partition of $\intff{0}{1}$ of mesh $\tf{2}{s}$ that is to say $|x_{a+1}-x_a| \leq \tf{2}{s}$.

\end{defin}

\subsubsection{Translations}

Suppose that one is given a compact $M$ in $\Cx$ without holes whose boundary is a smooth Jordan curve $\ga :  \intff{0}{1} \tend \Dp{}M$.
Let $K$ be a compact such that $K \subset \e{Int}\pa{M}$ and  $\mc{F}$ a regular functional (\textit{c}.\textit{f}. Definition
\ref{Definition Fonctionelle reguliere}) in respect to $\pa{M,K}$, for simplicity, not depending on auxiliary parameters $\bs{z}$.
Further introduce the functions $I_n$ and  $f_{s}$ 
\beq
I_n \pabb{ \la }{  \{y_a \}_1^n  } {\paa{ z_a }_1^n }   = \sul{a=1}{n} \psi ( \la , y_a  ) - \psi\pa{\la,z_a}
\quad \qquad
f_{s}\pa{\la\mid \paa{\vsg_a}_1^s}\equiv f_s\pa{\la}=  \sul{j=1}{s} \f{ \pa{t_{j+1}-t_j} }{t_j-\la}  \cdot \f{\vsg_j}{ 2 \i \pi} \; .
\label{definition fonction W et fs dans section bulk papier}
\enq
$I_n$ is defined in terms of an auxiliary function $\psi\pa{\la,\mu}$ that is holomorphic on $M \times M$ while the 
definition of $f_s$  utilises a set of $s+1$ discretisation points $t_j$ of $\Dp{}M$.

Finally, define $\wh{g}_s\pa{\la} $ as the below differential operator in respect to variables $\vsg_a$, with $a=1,\dots, s$:
\beq
\wh{g}_s\pa{\la} =  \sul{j=1}{s}  \psi\big( t_j,\la \big) \f{ \Dp{} }{ \Dp{}\vsg_j }  \;.
\enq
Again, the $t_j$ appearing above correspond to the same set of $s+1$ discretisation points of $\Dp{}M$. Proposition D.1 of [$\bs{A7}$] establishes that, 
for $\abs{\ga}$ small enough, it holds
\beq
  \mc{F}\pac{ \ga I_n \pabb{ * }{  \{y_a \}_1^n  } {\paa{ z_a }_1^n } }   =
 \lim_{s\tend +\infty} \; \paa{  \pl{a=1}{n} \ex{ \wh{g}_s\pa{ y_a }- \wh{g}_s\pa{ z_a } } \; \cdot \;
 \mc{F}\pac{\ga f_s}    }_{ \left| \vsg_k=0  \right. }   \hspace{-3mm}.
\label{ecriture action operateur translation sur fnelle generique}
\enq

\vspace{2mm}
The limit in \eqref{ecriture action operateur translation sur fnelle generique} is uniform in the parameters $y_a$ and $z_a$ belonging to $M$
and in $\abs{\ga}$ small enough. Actually, the magnitude of $\ga$ depends on
the value of the constant of regularity $C_{\mc{F}}$. If the latter is large enough, one can even set $\ga=1$.
The limit in \eqref{ecriture action operateur translation sur fnelle generique} also holds uniformly in respect to any finite
order partial derivative of the auxiliary parameters. In particular,
\beq
\pl{a=1}{n} \bigg\{ \f{\Dp{}^{p_a} }{ \Dp{}y_a^{p_a}} \f{\Dp{}^{h_a} }{ \Dp{}z_a^{h_a}}  \bigg\} \cdot   \f{ \Dp{}^m }{ \Dp{}\ga^m }
\mc{F}\pac{ \ga W_n \pabb{ *  }{  \{ y_a \}  } {\paa{ z_a } } } _{ \mid \ga=0} =
 \lim_{s\tend +\infty} \paa{  \pl{a=1}{n} \bigg\{ \f{\Dp{}^{p_a} }{ \Dp{}y_a^{p_a}} \f{\Dp{}^{h_a} }{ \Dp{}z_a^{h_a}}  \bigg\}
 \pl{a=1}{n} \ex{ \wh{g}_s\pa{y_a}- \wh{g}_s\pa{z_a} } \; \cdot \;
\f{ \Dp{}^m }{ \Dp{}\ga^m } \mc{F}\pac{\ga f_s}    }_{ \left| \substack{ \vsg_k=0 \\ \ga=0} \right. }   \;.
\enq
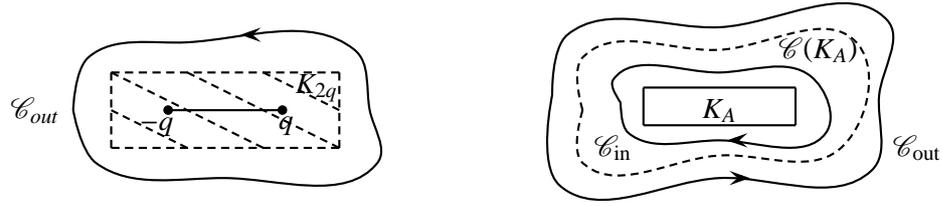
\begin{figure}[h]
\begin{center}

\begin{pspicture}(12.5,4)




\psline[linestyle=dashed, dash=3pt 2pt](2,2.5)(2,1.5)
\psline[linestyle=dashed, dash=3pt 2pt](2,1.5)(5,1.5)
\psline[linestyle=dashed, dash=3pt 2pt](5,1.5)(5,2.5)
\psline[linestyle=dashed, dash=3pt 2pt](5,2.5)(2,2.5)

\pscurve(1.5,2)(1.7,2.7)(3,2.9)(5.2,2.8)(5.4,2)(5.5,1.3)(3.5,1.1)(1.8,1.3)(1.5,2)

\psline[linestyle=dashed, dash=3pt 2pt](2,2)(3,1.5)
\psline[linestyle=dashed, dash=3pt 2pt](2,2.5)(4,1.5)
\psline[linestyle=dashed, dash=3pt 2pt](3,2.5)(5,1.5)
\psline[linestyle=dashed, dash=3pt 2pt](4,2.5)(5,2)

\psline(2.75,2)(4.25,2)
\psdots(2.75,2)(4.25,2)
\rput(2.6,1.8){$-q$}
\rput(4.3,1.8){$q$}

\rput(4.7,2.3){$K_{2q}$}
\rput(1,2){$\msc{C}_{out}$}

\psline[linewidth=2pt]{->}(3.8,3)(3.7,3)


\psline(9,2.3)(9,1.8)
\psline(9,1.8)(11,1.8)
\psline(11,1.8)(11,2.3)
\psline(11,2.3)(9,2.3)

\pscurve(8.7,2.1)(8.8,1.7)(10,1.6)(11.4,1.7)(11.2,2.45)(10,2.5)(8.6,2.4)(8.7,2.1)

\pscurve[linestyle=dashed, dash=3pt 2pt](8.2,2)(8.2,1.3)(10,1.4)(11.6,1.5)(11.8,2.9)(10,2.8)(8.2,2.7)(8.2,2)

\pscurve(7.8,2.2)(7.9,1)(10,1.1)(12,1.2)(12.1,2.2)(12.2,3.2)(10,3.1)(8,3)(7.8,2.2)

\psline[linewidth=2pt]{->}(10.3,1.1)(10.4,1.1)

\psline[linewidth=2pt]{->}(10.2,1.6)(10.1,1.6)

\rput(10,2){$K_A$}
\rput(11.3,2.8){$\msc{C}\!\pa{K_A}$}

\rput(8.6,1.5){$\msc{C}_{\e{in}}$}
\rput(12.6,1.5){$\msc{C}_{\e{out}}$}

\end{pspicture}

\caption{Example of discretised contours. In the $lhs$ the compact $M$ is located inside of its boundary $\msc{C}_{out}$
whereas the compact $K$ corresponds to $K_{2q}$ as defined in \eqref{definition compact KA}. In this case $M$ has no holes.
In the $rhs$ the compact $M$ is  delimited by the two Jordan curves $\msc{C}_{in}$ and $\msc{C}_{out}$ depicted in solid lines.
The associated compact $K$ (of Definition \ref{Definition Fonctionelle reguliere}) corresponds to the loop $\msc{C}\pa{K_A}$ depicted by dotted lines.
The compact $M$ depicted in the \textit{rhs} has one hole. This hole contains a compact $K_A$ inside.
\label{contour exemple de courbes encerclantes} }
\end{center}
\end{figure}

The proof can be found in Appendix D of [$\bs{A7}$]. Here I only describe the overall mechanism. 
By properly tuning the value of $\ga$ and invoking the regularity property of the functional $\mc{F}\pac{\ga f_s}$
one gets that, for any $s$, $\paa{\vsg_a}_1^s \mapsto \mc{F}\pac{\ga f_s}$ is holomorphic in a sufficiently large neighbourhood of $0 \in \Cx^s$.
This allows one to act with the translation operators $\prod_{b=1}^n\ex{\wh{g}_s(y_b) -\wh{g}_s(z_b) }$.
Their action replaces each variable $\vsg_a$  occurring in $f _s$ by the combination
$\sum_{b=1}^{n} \pac{ \psi\pa{t_a,y_b}-\psi (t_a,z_b) } $.
Taking the $s\tend +\infty$ limit changes the sum over $t_a$ occurring in $f_s$ into a contour integral over $\msc{C}_{out}$, \textit{c.f.} \textit{lhs} of
Fig.~\ref{contour exemple de courbes encerclantes}. Due to the presence
of a pole at $t=\la$, this contour integral exactly reproduces the function $I_n$
that appears in the \textit{rhs} of \eqref{ecriture action operateur translation sur fnelle generique}.

\vspace{2mm}

Note that such a realisation of the functional translation can also be built in the case of compacts $M$ having several holes as depicted in
the \textit{rhs} of Fig.~\ref{contour exemple de courbes encerclantes}. Also, there is no problem to consider regular functionals
$\mc{F}\pac{\cdot} \pa{\bs{z}}$ that depend on auxiliary sets of parameters $\bs{z}$.

\subsubsection{Generalization of translations}
\label{Subsubsection generalization of translations}

In the course of the analysis, in addition to dealing with functional translations as defined above,
one also has to manipulate expressions containing more involved series of partial derivatives.
Namely, assume that one is given a regular functional $\mc{F}\pac{f,g}$ of two arguments $f$ and $g$.
Then, the expression $ \bs{:} \Dp{\ga}^m \mc{F}\pac{\ga f_s, \wh{g}_s }_{\mid \ga=0} \bs{:}$ is to be understood as the left substitution of the various $\Dp{\vsg_a}$ derivatives symbols stemming from $\wh{g}_s$.

More precisely, let $\wt{g}_s$ be the below holomorphic function of $a_1,\dots, a_s$
\beq
\wt{g}_s\pa{\la}=  \sul{j=1}{s} \psi\big( t_j,\la \big) \, a_j \;.
\label{definiton fonction gs scalaire}
\enq
The regularity of the functional $\mc{F}$ ensures that the function $\{ a_p\} \mapsto \Dp{\ga}^m \mc{F}\pac{\ga f_s, \wt{g}_s}$ is holomorphic
in $a_1,\dots, a_s$ small enough. As a consequence, the  below multidimensional Taylor series is convergent for $a_j$ small enough:
\beq
\f{\Dp{}^m}{\Dp{}\ga^m} \cdot  \mc{F}\pac{\ga f_s, \wt{g}_s} _{\mid \ga=0}=
\sul{n_j \geq 0}{}   \pl{j=1}{s} \paa{ \f{ a_j^{n_j} }{n_j!} \f{ \Dp{}^{n_j} }{ \Dp{}a_j^{n_j}  }  }
\f{\Dp{}^m}{\Dp{}\ga^m} \cdot  \mc{F}\pac{\ga f_s, \wt{g}_s}_{\left| \substack{\ga=0 \\ a_j=0} \right. } \;.
\label{ecriture developpement F nu et g en puissance des aj}
\enq
%
%
%
Since $f_s$ \eqref{definition fonction W et fs dans section bulk papier} is a holomorphic function of $\vsg_1,\dots,\vsg_s$, the functional of $f_s$ coefficients of the above series give rise to a family
of holomorphic functions  in the variables $\vsg_1,\dots, \vsg_s$. Their analyticity follows, again, from the regularity of
the functional $\mc{F}\pac{f, g}$ and the smallness of $\abs{\ga}$.

The $\bs{:} \cdot \bs{:}$ ordering consists in substituting $a_j  \hookrightarrow \Dp{\vsg_j}$, $j=1,\dots,s$ in such a way that
all differential operators appear to the left. That is to say,
\beq
\bs{:} \f{\Dp{}^m}{\Dp{}\ga^m} \cdot  \mc{F}\pac{\ga f_s, \wh{g}_s} _{ \mid \ga=0} \bs{:} \; \equiv  \;
\sul{n_j \geq 0}{} \pl{j=1}{s}  \paa{  \f{\Dp{}^{n_j} }{\Dp{}\vsg_j^{n_j} } } \cdot \pl{j=1}{s} \paa{ \f{1}{n_j!} \f{ \Dp{}^{n_j} }{ \Dp{}a_j^{n_j}  }  } \cdot
\f{\Dp{}^m}{\Dp{}\ga^m} \cdot \mc{F}\pac{\ga f_s, \wt{g}_s} _{ \left| \substack{\ga=0 \\ a_j=0; \vsg_j=0} \right. } \;.
\label{definition substitution operatorielle}
\enq
Although there where no convergence issues on the level of expansion \eqref{ecriture developpement F nu et g en puissance des aj},
these can \textit{a priori} arise  on the level of the $rhs$ in \eqref{definition substitution operatorielle}.
Clearly, convergence depends on the precise form of the functional $\mc{F}$, and  should thus be
studied on a case-by-case basis. 

At this point, two observations are in order. \vspace{2mm}
\begin{itemize}
 \item[i)] \eqref{definition substitution operatorielle}
bears a strong resemblance with an $s$-dimensional Lagrange series. \vspace{2mm}
\item[ii)] The functional (of $f_s$) coefficients appearing in the $rhs$ of
\eqref{definition substitution operatorielle} are completely determined by the functional $ \mc{F}\pac{\ga f_s, \wt{g}_s} $
whose expression only involves standard (\textit{i}.\textit{e}. non-operator valued) functions. Should this functional have two (or more) equivalent representations, then
any one of them can be used as a starting point for computing the coefficients in \eqref{ecriture developpement F nu et g en puissance des aj}
and then carrying out the substitution \eqref{definition substitution operatorielle}.
\end{itemize}
\vspace{1mm}

It is important to stress that, for the class of functionals  of interest to the study, no convergence issues arise in respect to the $\bs{:} \cdot \bs{:}$. Indeed, in all of the cases,
the $m^{\e{th}}$ $\ga$-derivative at $\ga=0$ of the $\bs{:}\cdot \bs{:}$ ordered functionals of interest appears as a finite linear combinations
(or integrals thereof) of expressions of the type
\beq
\wh{\msc{E}}_{m;s}=
\bs{:} \f{ \Dp{}^m }{  \Dp{}\ga^m} \paa{ \pl{a=1}{r} \ex{ \eps_a  \wh{g}_s (\la_{\a_a}) }  \cdot \pl{b=1}{\wt{r} } \ex{ \upsilon_b  \wh{g}_s\pa{ y_b } }
\cdot \mc{F}\pac{\ga f_s}}_{\mid \ga=0} \hspace{-2mm}\bs{:}  \qquad \e{where} \;\; \a_a \in \intn{1}{N} \quad \e{and}
\quad  \eps_a, \upsilon_b \in \paa{\pm 1}\;.
\label{equation explication prise gamma derivee}
\enq
Above $y_a$ are some auxiliary and generic parameters whereas $\la_{\a_a}$ are implicit functions of $\ga$ and $\vsg_1,\dots,\vsg_s$.
For $L$-large enough, $\la_{\a_a}$ is the unique solution  to the equation $\xi_{\ga f_s}\!\pa{\la_{\a_a}}=\tf{\a_a}{L}$.

\vspace{2mm}

In the case of the specific example given in \eqref{equation explication prise gamma derivee}, the   $\bs{:} \cdot \bs{:}$ prescription  goes as follows. One first
substitutes $\wh{g}_s \hookrightarrow \wt{g}_s$ as defined in \eqref{definiton fonction gs scalaire}.  Then, one
computes the  $m^{\e{th}}$ $\ga$-derivative at $\ga=0$ of \eqref{equation explication prise gamma derivee}, this
in the presence of non-operator valued functions $\wt{g}_s$.
In the process, one has to differentiate in respect to $\ga$ \textit{both} the functional $\mc{F}\pac{\ga f_s}$ and the variables $\la_{\a_a}$ of $\wt{g}_s\pa{\la_{\a_a}}$.
Using that ${\la_{\a_a}}_{\mid \ga=0} = \mu_{\a_a}$, one arrives to
\beq
\wt{\msc{E}}_{m;s}  \equiv
\f{ \Dp{}^m }{  \Dp{}\ga^m} \paa{ \pl{b=1}{r} \ex{ \eps_b  \wt{g}_s (\la_{\a_b}) }  \pl{b=1}{\wt{r} } \ex{ \upsilon_b  \wt{g}_s\pa{ y_b } }
\cdot \mc{F}\pac{\ga f_s}}_{\mid \ga=0} =   \pl{b=1}{r} \ex{\eps_b \wt{g}_s(\mu_{\a_b})} \pl{b=1}{\wt{r} } \ex{ \upsilon_b  \wt{g}_s\pa{ y_b } }
\sul{n_1,\dots, n_s =0}{m} \pl{j=1}{s}  a_j^{n_j}    \; \cdot  \;  c_{\paa{n_j}}\pac{ f_s} \;.
\label{ecriture mise en place ordre operatoriel}
\enq
The sum is truncated at most at $n_j=m$, $j=1,\dots,m$ due to taking the $m^{\e{th}}$ $\ga$-derivative at $\ga=0$.
It is readily verified that the $\{n_j\}-$dependent coefficients $c_{\paa{n_j}}\pac{ f_s}$ are regular functionals of $f_s$
with sufficiently large constants of regularity.
It remains to impose the operator substitution $a_j \hookrightarrow \Dp{\vsg_j}$ on the level of \eqref{ecriture mise en place ordre operatoriel}
 with all differential operators $\Dp{\vsg_k}$, $k=1,\dots,s$, appearing to the left. It is clearly not a problem
to impose such an operator order on the level of the polynomial part of the above expression.
Indeed, the regularity of the functionals $c_{\paa{n_j}}\pac{f_s}$ implies that these are holomorphic in $\vsg_1,\dots,\vsg_s$
belonging to an open neighbourhood $\mc{N}_{0}$ of $0\in\Cx^s$. Hence, $\prod_{k=1}^{s} \Dp{\vsg_k}^{m_k} \cdot  c_{\paa{n_j}}\pac{f_s}_{\mid \vsg_k=0} $
is well-defined for any set of integers $\{ m_k \}$. In fact, in all the cases of interest to the analysis, the neighbourhood $\mc{N}_0$ is always large enough
so as to make the Taylor series issued from the products of translation operators
 $\prod_{a=1}^{r} \ex{\eps_a \wh{g}_s\pa{\mu_{\a_a}}} \prod_{b=1}^{\wt{r} } \ex{ \upsilon_b  \wh{g}_s\pa{ y_b } } $ convergent. Their  action can then be incorporated by a re-definition of $f_s$ leading to
\beq
\wh{\msc{E}}_{m;s} = \sul{n_1,\dots, n_s =0}{m} \pl{j=1}{s} \bigg\{  \f{ \Dp{}^{n_j} }{ \Dp{}\vsg_j^{n_j}  }  \bigg\} \cdot
c_{\paa{n_j}}\big[  \wt{f}_s \big]_{\mid \vsg_k = 0}
\enq
 \e{with}
\beq
\wt{f}_s\pa{\la} = f_s\pa{\la} \;  + \;  \sul{b=1}{s}
 \f{ \pa{t_{b+1}-t_b} }{ 2i\pi \pa{t_b-\la} }
\paa{ \sul{k=1}{r} \eps_k \psi\pa{t_b,\mu_{\a_k} }  +  \sul{k=1}{\wt{r}} \upsilon_k \psi\pa{t_b,y_{k} } }  \;.
\nonumber
\enq
 In this way, one obtains a (truncated to a finite number of terms) s-dimensional Lagrange series. The procedure for dealing with such series
and taking their $s\tend +\infty$ limits is described in Proposition D.2 of [$\bs{A7}$].

\subsection{Resummation of the finite-volume Taylor coefficients}
\label{Subsection resum effective FF series}

The functional  $\wh{\mc{G}}_{N;\ga}$ entering in the expression for the effective form factor expansion are  not regular so as to 
allow for an action of functional translations or generalisations thereof. To be able to act with these, 
one needs to regularise $\wh{\mc{G}}_{N;\ga}$ by deforming its expression with the twist parameter $\be$.
This regularisation allows one to represent it as a regular functional that, moreover, has
a form suitable for carrying out the intermediate calculations.

\subsubsection*{The $\be$-regularisation}

It is easy to see that
\beq
\pa{ \wh{D}_{N} \; \wh{\mc{G}}_{N;\ga} } \pab{ \!\! \paa{p_a}_1^n \!\!  }{  \!\! \paa{h_a}_1^n \!\!  }
\pac{ \ga F_{0} \; ; \xi_{} \; ; \xi_{\ga F_0}  }
= \lim_{\be \tend 0} \Bigg\{  \wh{D}_{N}  \pab{ \!\! \paa{p_a}_1^n \!\!  }{  \!\! \paa{h_a}_1^n \!\!  }
\pac{ \ga F_{\be} \; ; \xi_{} \; ; \xi_{\ga F_{\be}}  }
 \wh{\mc{G}}_{N;\ga} \pab{ \!\! \paa{p_a}_1^n \!\!  }{  \!\! \paa{h_a}_1^n \!\!  }
\pac{ \ga F_{\be} \; ; \xi_{} \; ; \xi_{\ga F_{\be}}  } \Bigg\}
\enq
I will now introduce  a prescription for taking the $\be \tend 0$ limit. To discuss better its purpose, I remind that 
when considered as a separate object from $\wh{D}_{N}$, the functional $\wh{\mc{G}}_{N;\ga}$ may exhibit singularities
should it happen that $ \ga^{-1}\{ \ex{2i\pi \ga F_{\be}(\la_j)}-1 \} =0$, \textit{c}.\textit{f}.
\eqref{ecriture fonctionnelle hat GN gamma}-\eqref{appendix thermo lim FF definition entree V et Vbar chapeau}.
For $\abs{\ga}$ small enough, which will always be the case of interest to the analysis, such potential zeroes correspond to solutions of $F_{\be}(\la_j)=0$. 
Now recall the set
\beq
 \bs{U}_{\be_0} = \paa{ z \in \Cx \; :\; 10\, \Re\pa{\be_0} \geq \Re\pa{z} \geq  \Re\pa{\be_0}  \;\; \e{and} \;\; \abs{\Im\pa{z}}\leq \Im\pa{\be_0}} \;. 
\enq
When $\Re\pa{\be_0}>0$ is large enough and $\Im\pa{\be_0}>0$ is small enough, it can be proved that there are no solutions of
\beq
F_\be\!\pabb{\om}{ \{ y_a \}_1^n }{ \{ z_a \}_1^n }=0 \quad \e{for} \quad \om\in U_{\de} \; ,
\;\; \e{and} \; \e{uniformly} \; \e{in} \quad
 0\leq n \leq m \; \e{and}\; \pa{\be,  \{ y_a \}_1^n, \{  z_a \}_1^n } \in \bs{U}_{\be_0} \times U_{\de}^n \times U_{\de}^n \;.
\nonumber
\enq
The optimal value of $\be_0$ preventing the existence of such solutions depends on the width $\de$ of
the strip $U_{\de}$ and on the integer $m$.

The prescription for taking the $\be \tend 0$ limit is as follows. The computations always start on the level of a representation that is holomorphic in
the half-plane $\Re(\be) \geq 0$, as for instance \eqref{ecriture series FF effective gamma deformee}-\eqref{definition rho N eff derivee m ieme}.
In the intermediate calculations whose purpose is to allow one to relate
the initial representation to another one,  it will be assumed that $ \be \in \bs{U}_{\be_0}$.
This allows one to avoid the problem of the aforementioned poles and represent $\wh{\mc{G}}_{N;\ga}$ in terms of a regular functional
that is moreover fit for carrying out the intermediate calculations.
Then, on the level of the final expression that will be obtained after several manipulations, one checks that the new representation is in fact holomorphic in the half-plane $\Re\pa{\be} \geq 0$
and has thus a unique extension from the open set $ \bs{U}_{\be_0}$ up to
$\be=0$. As the same property holds for the initial representation, both are equal at $\be=0$ and the $\be \tend 0$ limit can be readily taken on the level 
of the new representation.

\vspace{3mm}

Having the $\be$-regularisation prescription at hand,  the effective form factor expansion-based representation
for $\rho_{N; \e{eff}}^{\pa{m}}\! \pa{x,t}$ \eqref{ecriture series FF effective gamma deformee} can be simplified by the use of the below rewriting of the functionals 
$\wh{D}_N$ and $\wh{\mc{G}}_{N;\ga}$.

\subsubsection*{The functional $\wh{\mc{G}}_{N;\ga}$}

Given $A \in \R^+$, I remind the definition of the compact $K_A$ 
\beq
K_A = \paa{ z \in \Cx \; : \; \abs{\Im z} \leq \de \;, \; \abs{\Re z} \leq A} \;,
\label{definition compact KA}
\enq
and denote the open disk of radius $r$ by $\mc{D}_{0,r}= \paa{ z  \in \Cx \; : \; \abs{z} < r }$.
The compact $K_A$ is contained in $U_{\de}$. 

It is shown in Lemma A.2 of  [$\bs{A7}$] that given $A>0$ large enough and $m\in \mathbb{N}^*$ fixed, there exists \vspace{1mm}
\begin{itemize}
\item[$\bullet$] a complex number
$\be_0$ with a sufficiently large real part and an imaginary part small enough; \vspace{1mm} 
\item[$\bullet$] a positive number $\wt{\ga}_0>0$ small enough; \vspace{1mm} 
\item[$\bullet$] a regular functional $\wh{\msc{G}}_{\ga;A}^{\,\pa{\be}}$ ; \vspace{1mm} 
\end{itemize}
such that, uniformly in  $0\leq n \leq m$, $\pa{\ga,\be, \{ \mu_{p_a}\}_1^n, \{ \mu_{h_a}\}_1^n } \in \mc{D}_{0,\wt{\ga}_0} \times \bs{U}_{\be_0} \times K_A^n \times K_A^n$ one has
\beq
 \wh{\mc{G}}_{N;\ga}  \pab{ \!\! \paa{p_a}_1^n \!\!  }{  \!\! \paa{h_a}_1^n \!\!  }
\pac{ \ga F_{\be} \; ; \xi_{} \; ; \xi_{\ga F_{\be}}  }  =
\wh{\msc{G}}_{\ga; A}^{\,\pa{\be}}\pac{ H\pabb{ * }{ \{ \mu_{p_a} \}_1^n  } {  \{ \mu_{h_a} \}_1^n  } }
\quad \e{with} \quad H\pabb{ \la }{ \{ \mu_{p_a} \}_1^n  } {  \{ \mu_{h_a} \} _1^n } = \sul{a=1}{n} \f{1}{\la-\mu_{p_a}} - \f{1}{\la - \mu_{h_a}} \;.
\label{definition facteur forme lisse simple fonctionnelle}
\enq
The $*$ in the argument of $\wh{\msc{G}}_{\ga;A}^{\,\pa{\be}}$ appearing above indicates the running variable of $H$ on which this functional acts.
The explicit expression for the functional $\wh{\msc{G}}_{\ga;A}^{\,\pa{\be}}$ can be found in the statement of Lemma A.2.

\vspace{1mm}

The main advantage of such a representation is that all the dependence on the auxiliary parameters is now solely contained in the function $H$ given in
\eqref{definition facteur forme lisse simple fonctionnelle}.
The constant $\wt{\ga}_0$ is such that
\beq
 \abs{ \ga F_{\be}\pabb{\om}{ \{ y_a \}_1^n }{  \{ z_a  \}_1^n  }  } < \f{1}{2} \quad \e{uniformly} \; \e{in} \quad
\pa{\ga,\be, \{ y_a \}_1^n, \{ z_a \}_1^n } \in \mc{D}_{0,\wt{\ga}_0} \times \bs{U}_{\be_0} \times K_A^n \times K_A^n
\;\; \e{and} \;\; 0\leq n \leq m \;.
\label{ecriture condition Fbeta et gamma petit}
\enq

The functional $\wh{\msc{G}}_{\ga;A}^{\, \pa{\be}}$ is regular in respect to the to the pair $\pa{M_{\msc{G}_A}, \msc{C}\pa{K_A} }$
where $\msc{C}\pa{K_A}$ in a loop in $U_{\de}$ around $K_A$ as depicted in the $rhs$  of Fig.~\ref{contour exemple de courbes encerclantes}
and $M_{\msc{G}_A}$ corresponds to the compact with one hole that is delimited by $\msc{C}_{in}$ and $\msc{C}_{out}$.
This hole contains $K_{A}$. Finally, the parameters $\be_0 \in \Cx$ and $\wt{\ga}_0>0$ are such that the
 constant of regularity  $C_{\msc{G}_A}$ of $\wh{\msc{G}}^{\, (\be)}_{\ga;A}$ satisfies to the estimates
\beq
 C_{\msc{G}_A}  \cdot  \f{\pi \, d \!\pa{ \Dp{}M_{\msc{G}_A}, \msc{C}\!\pa{K_A} } }
 {  \abs{\Dp{}M_{\msc{G}_A}}+2\pi \, d \!\pa{ \Dp{}M_{\msc{G}_A}, \msc{C}\!\pa{K_A} }} >A \;,
\label{ecriture condition grandeur constante de regularite GAkappa}
\enq
where $\abs{\Dp{}M_{\msc{G}_A}}$ stands for the length of the boundary $\Dp{}M_{\msc{G}_A}$ and 
$d \!\pa{ \Dp{}M_{\msc{G}_A}, \msc{C}\!\pa{K_A} }>0$ stands for the distance of $\msc{C}\pa{K_A}$ to $\Dp{}M_{\msc{G}_A}$.

\vspace{2mm}
Similarly to an earlier discussion of Section \ref{Subsubsection generalization of translations} and
according to Proposition D.1 of [$\bs{A7}$], one has that, uniformly in $n,p\in \paa{0,\dots,m}$,
and $z_j$, $y_j$, $j=1,\dots, m$ belonging to $K_A$:
\beq
 \f{\Dp{}^p}{\Dp{}\ga^p }\cdot   \wh{\msc{G}}_{\ga;A}^{\, \pa{\be}} \pac{ H \pabb{ * }{  \{ z_j \}_1^{n} \vspace{1mm} }{ \{y_j\}_1^{n}  } }
_{\left| \ga=0 \right.}
=  \lim_{r \tend + \infty } \Bigg\{
\pl{j=1}{n} \ex{\wh{g}_{2,r}\pa{z_j}-\wh{g}_{2,r}\pa{y_j}}  \cdot
 \f{\Dp{}^p}{\Dp{}\ga^p } \wh{ \msc{G} }_{\ga;A}^{\, \pa{\be}}\pac{ \varpi_r}  \Bigg\}_{ \left| \substack{ \eta_{a,p}=0 \\ \ga=0 }\right. }
\hspace{-5mm} .
\label{ecriture decomposition Archeologue G+m+1}
\enq

The compact $M_{\msc{G}_A}$ has one hole. Hence, following the discussion of Section D.3 of [$\bs{A7}$]
one has to consider two sets of discretisation points  $t_{1,p}$, $p=1,\dots, r+1$
for $\msc{C}_{in}$ and $t_{2,p}$, $p=1,\dots,r+1$ for $\msc{C}_{out}$.
The function $\varpi_r$ appearing in \eqref{ecriture decomposition Archeologue G+m+1} is a linear polynomial
in the variables $\eta_{a,p}$ with $a=1,2$ and $p=1,\dots,r$:
\beq
\varpi_r \big( \la \mid \{  \eta_{a,p} \}  \big) =  \sul{p=1}{r}  \f{ t_{1,p+1}-t_{1,p} }{ 2 \i \pi \pa{t_{1,p}-\la} } \eta_{1,p}
\; + \; \sul{p=1}{r}  \f{ t_{2,p+1}-t_{2,p}  }{ 2 \i \pi \pa{t_{2,p}-\la} } \eta_{2,p} \;.
\label{definition fonction varpi r}
\enq
Finally, $\wh{g}_{2,r} \pa{\la}$ is a differential operator in respect to $\eta_{a,p}$ with $a=1,2$ and $p=1,\dots,r$:
\beq
\wh{g}_{2,r} \pa{\la} = \sul{p=1}{r} \f{ 1 }{ t_{1,p} -\la } \f{ \Dp{} }{ \Dp{}\eta_{1,p} }
\; + \; \sul{p=1}{ r } \f{ 1 }{ t_{2,p} -\la } \f{ \Dp{} }{ \Dp{}\eta_{2,p} }  \; .
\enq

\subsubsection*{The functional $\wh{D}_{N}$}

One can draw a small loop  $\msc{C}_{out}$ around $K_{2q}$ in $U_{\de}$ as depicted in the $lhs$ of Fig.~\ref{contour exemple de courbes encerclantes}.
Let $M_{\wh{D}}$ be the compact without holes whose boundary is delimited by $\msc{C}_{out}$.
Then, given $L$ large enough, the functional $\wh{D}_{N}$  \eqref{definition partie discrete} is a regular
functional (in respect to the pair $(M_{\wh{D}},K_{2q})$) of $\ga F_{\be}$ with $\be \in \bs{ U }_{\be_0}$ and $\abs{\ga} \leq \wt{\ga}_0$.
The parameters $\be_0$ and $\wt{\ga}_0$ are as defined previously.
This regularity is readily seen by writing down the integral representation:
\beq
\la_{j} = \Oint{ \Dp{}K_{2q}}{}   \f{ \xi_{\ga F_{\be} }^{\prime} \!\pa{\om} }{  \xi_{\ga F_{\be} }\!\pa{\om} - \tf{j}{L} } 
\cdot \f{\dd \om}{2 \i \pi}  \;,
\qquad j=1,\dots,N
\enq
which holds provided that $L$  is large enough (indeed then all $\la_j$'s are located in a very small vicinity of the interval 
$\intff{-q}{q}$).
Therefore, according to the results developed in Appendix D of [$\bs{A7}$] and described earlier on, one has that, uniformly in $\be \in \bs{U}_{\be_0}$
and $0\leq p,n \leq m$
\bem
 \f{\Dp{}^p}{\Dp{}\ga^p}  \Bigg\{ \wh{D}_{N} \pab{\!\! \paa{p_a}_1^n \!\! }{ \!\! \paa{h_a}_1^n \!\!  }
\pac{ \ga F_{\be}\pabb{\cdot}{  \{\mu_{p_a} \}_1^n  } {\paa{\mu_{h_a}}_1^n } \; ; \xi_{} \; ; \xi_{\ga F_{\be}}  }   \Bigg\}_{\mid\ga=0}  \\
=
 \lim_{s\tend +\infty} \Bigg[  \pl{a=1}{n} \ex{ \wh{g}_{1,s}\pa{\mu_{p_a}}- \wh{g}_{1,s}\pa{\mu_{h_a}} } \; \cdot \;
 \f{ \Dp{}^p }{ \Dp{}\ga^p } \paa{  \wh{D}_{N} \pab{ \!\! \paa{p_a}_1^n \!\! }{ \!\! \paa{h_a}_1^n \!\! }
 \pac{ \ga \nu_s \; ; \xi_{} \; ; \xi_{\ga \nu_s}  }    } _{ \left|  \substack{ \ga=0 \\ \vsg_k=0} \right.   }    \Bigg] \;.
\label{ecriture fonctionnelle Dn+ Fonct Trans}
\end{multline}

The function $\la \mapsto \nu_s\pa{\la}$ appearing above is holomorphic in some open neighbourhood of $K_{2q}$ in $M_{\wh{D}}$ and given by
\beq
\nu_{s}\pa{\la\mid \paa{\vsg_a}_1^s}\equiv \nu_s\pa{\la}= \pa{ \i \be-\tf{1}{2}}Z\pa{\la} -\phi\pa{\la,q}+ \sul{j=1}{s}
 \f{ \pa{t_{j+1}-t_j} }{t_j-\la}  \cdot \f{\vsg_j}{2 \i \pi} \; .
\label{definition fonction nus}
\enq
The parameters $t_j$, $j=1,\dots,s$ correspond to a discretisation (\textit{cf} Definition \ref{Definition point discretisation})
of the loop $\msc{C}_{out}$ around $K_{2q}$ in $U_{\de}$
that has been depicted in the $lhs$ of Fig.~\ref{contour exemple de courbes encerclantes}. $\vsg_j$ are some
sufficiently small complex numbers and $\wh{g}_{1,s}\pa{\la} $  is a differential operator in respect to $\vsg_a$:
\beq
\wh{g}_{1,s}\pa{\la} = - \sul{j=1}{s}  \phi\big(t_j,\la \big) \f{ \Dp{} }{ \Dp{}\vsg_j }  \;.
\label{definition widehat gs}
\enq

Note that the parameters $\la_a$ appearing
in the second line of \eqref{ecriture fonctionnelle Dn+ Fonct Trans} through the expression  \eqref{definition partie discrete} for
$\wh{D}_{N}$ are the unique\symbolfootnote[2]{Here, as previously, the uniqueness follows from Rouch\'{e}'s theorem. By writing down an integral
representation
for $\xi^{-1}_{\ga \nu_s}$, one readily convinces oneself that,  for $\ga$ small enough and given any fixed $s$,
the function $ \paa{\vsg_a}_1^s \mapsto  \la_a$ is holomorphic. It is also holomorphic in $\ga$ belonging to some open neighbourhood of $\ga=0$.}
 solutions to $\xi_{\ga\nu_s}\pa{\la_a}=\tf{a}{L}$.
As such, the $\la_a$'s become holomorphic functions of $\{ \vsg_a\}_1^s$ when these belong to a sufficiently small neighbourhood of the origin
in $\Cx^{s}$.


\subsubsection*{A new representation for the Taylor coefficients}

To implement the simplifications induced by the functional translations on the level of $\rho_{N;\e{eff}}^{\pa{m}}\!\pa{x,t}$, one first
observes that all rapidities $\mu_{p_a}$ and $\mu_{h_a}$ occurring in the course of summation in \eqref{ecriture series FF effective gamma deformee}
belong to the interval $\intff{-A_L}{B_L}$ with $L\xi\pa{-A_L}=-w_L-\tf{1}{2}$ and $L\xi\pa{B_L}=w_L+\tf{1}{2}$ ($A_L>B_L$).
Hence,  \textit{a fortiori}, they belong to the compact $K_{2A_L}$. This allows one to represent
the smooth part functional as $\wh{\msc{G}}_{\ga;2A_L}^{ \, \pa{\be}}$.

One can readily check that $\wh{D}_{N} \big( \{p_a\}_1^n, \{h_a\}_1^n \big) \propto \ga^{2 \pa{n-1} }$ and $\wh{\msc{G}}_{\ga;2A_L}^{ \, \pa{\be}} \pac{\varpi_r}$
has no singularities around $\ga=0$. Thus, since one computes the $m^{\e{th}}$  $\ga$-derivative of \eqref{ecriture series FF effective gamma deformee}
at $\ga=0$, all the terms issuing from  $n$ particle/hole excitations with $n\geq m$ will not contribute to the value of the derivative. Hence, one can truncate the sum over $n$ in
\eqref{ecriture series FF effective gamma deformee} at $n=m$.
Once that the sum is truncated, it remains to represent the functional $\Dp{\ga}^m \cdot \Big\{ \wh{D}_{N} \cdot \wh{\msc{G}}_{\ga;2A_L}^{\,\pa{\be}} \Big\}_{\mid \ga=0}$
by means of the identities \eqref{ecriture fonctionnelle Dn+ Fonct Trans}
and \eqref{ecriture decomposition Archeologue G+m+1}. All together, this yields 
\bem
 \rho_{N;\e{eff}}^{\pa{m}}\!\pa{x,t} =   \lim_{\be \tend 0} \lim_{s \tend +\infty} \lim_{r\tend +\infty} \Bigg[  \; \sul{n = 0}{m}
\sul{ \substack{p_1<\dots < p_n \\  p_a \in \mc{B}^{\e{ext}}_{L}  } }{}
\sul{ \substack{h_1<\dots < h_n \\ h_a  \in \mc{B}^{\e{int}}_{L}  } }{}
 \pl{a=1}{n} \f{\wh{E}^{\, 2}\pa{\mu_{h_a}}  }{ \wh{E}^{\, 2} (\mu_{p_a}) } \\
\cdot  \f{ \Dp{}^m }{ \Dp{}\ga^m } \Bigg\{ \wh{D}_{N} \pab{ \!\! \paa{p_a}_1^n \!\! }{ \!\! \paa{h_a}_1^n \!\!  }
\pac{ \ga \nu_{s}  \; ; \xi \; ; \xi_{\ga \nu_s}  }
\wh{\msc{G}}_{\ga;2A_L}^{\, \pa{\be}}\pac{\varpi_r} \;   \Bigg\}_{ \left| \substack{ \ga=0 \\ \vsg_p=0= \eta_{a,p} } \right.  } \Bigg]  \;.
\label{ecriture serie rho eff completement factorisee}
\end{multline}
Above, I have introduced 
\beq
\wh{E}^{\, 2}\pa{\la} = \ex{- \i x u\pa{\la} -\wh{g}\pa{\la}} \qquad \e{with} \qquad
\wh{g}\pa{\la} \equiv \wh{g}_{1,s}\pa{\la} +  \wh{g}_{2,r}\pa{\la} \;.
\label{definition fonction E hat chapeau}
\enq
In order to lighten the notation, the dependence of $\nu_s$ and $\varpi_r$ on the auxiliary parameters
$\vsg_p$, $\eta_{a,p}$ as well as the one of $\wh{E}\pa{\la}$ on the discretisation indices $r$ and $s$ has been dropped.
The hat has nonetheless been kept so as to insist on the operator nature of $\wh{E}$.
Equation \eqref{ecriture serie rho eff completement factorisee} is to be understood along the lines of
Section \ref{Subsection Operator ordering for functional translation}.

Starting from multiple-sum in the \textit{rhs} of \eqref{ecriture serie rho eff completement factorisee}, one can 
re-express $\rho_{N;\e{eff}}^{\pa{m}}\!\pa{x,t}$ in terms of the $m^{\e{th}}$ $\ga$-derivative of the  functional  $X_N\big[ \ga \nu_{s}, \wh{E}^{\,2}  \big]$,
where $X_N$ is defined as 

\beq
X_N\big[ \nu, e^{2} \big]  =     \sul{n = 0}{ N +1 } \;
\sul{ \substack{p_1<\dots < p_n \\ p_k \in \mc{B}_L^{\e{ext}} }  }{}\;  \sul{ \substack{h_1<\dots < h_n \\ h_k \in \mc{B}_L^{\e{int}} } }{}
\; \f{ \pl{a=1}{N} e^2\pa{\la_a} }{ \pl{a=1}{N+1} e^{2}\pa{\mu_{\ell_a}} } \;
\wh{D}_{N} \pab{ \!\! \paa{p_a}_1^n \!\! }{ \!\! \paa{h_a}_1^n \!\! } \pac{ \nu, \xi, \xi_{\nu} } \; , 
\label{definition Fonction generatrice X_N}
\enq
 $\xi$ is given by \eqref{ecriture fct cptge thermo} and
$\xi_{\nu}\pa{\la}=\xi\pa{\la}+\tf{\nu\pa{\la}}{L}$.

In order to establish the identification, one has to extend in \eqref{ecriture serie rho eff completement factorisee} the upper bound in the summation over $n$ from $m$ up to $N+1$. This does not alter the result as it corresponds to
adding up a finite amount of terms that are zero due to the presence of $\ga$-derivatives.
Then, one should use the identity
\bem
 \f{ \Dp{}^m }{ \Dp{}\ga^m } \Bigg\{ \wh{D}_{N} \pab{ \!\! \paa{p_a}_1^n \!\! }{ \!\! \paa{h_a}_1^n \!\!  }
\pac{ \ga \nu_{s}  \; ; \xi \; ; \xi_{\ga \nu_s}  }
\wh{\msc{G}}_{\ga;2A_L}^{\, \pa{\be}}\pac{\varpi_r} \;   \Bigg\}_{ \left| \substack{ \ga=0 \\ \vsg_p=0= \eta_{a,p} } \right.  }   \\
= \bs{:}  \f{ \Dp{}^m }{ \Dp{}\ga^m } \Bigg\{
\f{ \prod_{a=1}^{N+1}  \wh{E}^{\, 2} \pa{\mu_{a}}    }{  \prod_{a=1}^{N}  \wh{E}^{\, 2} \pa{\la_{a}}   } \cdot
 \f{  \prod_{a=1}^{N}  \wh{E}^{\, 2} \pa{\la_{a}}   }{  \prod_{a=1}^{N+1}  \wh{E}^{\, 2} \pa{\mu_{a}} }
\wh{D}_{N} \pab{ \!\! \paa{p_a}_1^n \!\! }{ \!\! \paa{h_a}_1^n \!\!  }
\pac{ \ga \nu_{s}  \; ; \xi \; ; \xi_{\ga \nu_s}  }
\wh{\msc{G}}_{\ga;2A_L}^{\, \pa{\be}}\pac{\varpi_r} \;   \Bigg\}_{ \left| \substack{ \ga=0 \\ \vsg_p=0= \eta_{a,p} } \right.  }
\, \hspace{-5mm} \bs{:} \; . 
\label{ecriture identie operatorielle produit egal a 1}
\end{multline}
Just as it is the case for the parameters $\la_j$ appearing in the expression for $\wh{D}_N$, the ones appearing in
the pre-factors of the \textit{rhs} in \eqref{ecriture identie operatorielle produit egal a 1} are the unique solutions to
$\xi_{\ga \nu_s}\!\pa{\la_s}=\tf{s}{L}$. Equation \eqref{ecriture identie operatorielle produit egal a 1}
is an expression of the type \eqref{equation explication prise gamma derivee}: to deal correctly with it one should implement a  $\bs{:}\cdot \bs{:}$ prescription for the way the
differential operators $\Dp{\vsg_a}$ or $\Dp{\eta_{a,p}}$ should  be substituted in the \textit{rhs} of
\eqref{ecriture identie operatorielle produit egal a 1}.

The identity \eqref{ecriture identie operatorielle produit egal a 1} forces the appearance of the product of function $\wh{E}$ whose presence is necessary for identifying  the sum
over the particle-hole type labelling of integers in \eqref{ecriture serie rho eff completement factorisee} with the functional
$\Dp{\ga}^m X_N\big[\ga \nu_s \; \wh{E}^{\,2} \big]_{ \mid_{\ga=0} }$ given in \eqref{definition Fonction generatrice X_N}.
This leads to the below representation:
\beq
\rho^{\pa{m}}_{N;\e{eff}} \!\pa{x,t} =   \; \lim_{\be \tend 0}\; \lim_{s \tend +\infty}  \; \lim_{r \tend +\infty}  \bs{: }
 \f{\Dp{}^m}{\Dp{} \ga^m } \; \Bigg\{  \f{ \prod_{a=1}^{N+1}  \wh{E}^{\,2} \pa{\mu_a} }{\prod_{a=1}^{N}   \wh{E}^{\,2} \!\pa{\la_a} }
X_N\pac{\ga \nu_{s}, \wh{E}^{\,2} } \, \wh{\msc{G}}_{\ga;2A_L}^{\, \pa{\be} }\!\pac{\varpi_r}\;
\Bigg\}_{\left| \substack{\ga=0 \\ \vsg_{p}=0=\eta_{a,p} } \right.  }  \hspace{-2mm}  \bs{ : } \;.
\label{equation exprimant Q cal en terme moyenne XN}
\enq

\subsection{Taking the thermodynamic limit}

As already mentioned,  Theorem C.1 of [$\bs{A7}$] establishes 
that $\rho^{\pa{m}}_{N;\e{eff}}\!\pa{x,t}$ admits a well-defined thermodynamic limit denoted as $\rho^{\pa{m}}_{\e{eff}}\!\pa{x,t}$.
This limit is given in terms of a truncation of a multidimensional Fredholm series. This series is close in spirit to the
type of series that have appeared in \cite{KozKitMailSlaTerXXZsgZsgZAsymptotics} and [$\bs{A10}$]. 
Proposition C.1 of [$\bs{A7}$] establishes, in its turn, that it is allowed to exchange \vspace{1mm}
\begin{itemize}
\item[$\bullet$] the thermodynamic limit $N,L \tend +\infty$, $\tf{N}{L} \tend D$ \vspace{1mm}
\end{itemize}
with
\begin{itemize}
\item[$\bullet$] the $\Dp{\ga}^m$ differentiation along with its associated operator substitution, \vspace{1mm}
\item[$\bullet$] the computation of the  translation generated by $\wh{g}_{2,r}$,  \vspace{1mm}
\item[$\bullet$] the computation of the $s$-dimensional Lagrange series associated with $\wh{g}_{1,s}$, \vspace{1mm}
\item[$\bullet$] the computation of the $r \tend +\infty$ and $s\tend +\infty$ limits,  \vspace{1mm}
\item[$\bullet$] the analytic continuation in $\be$ from $\bs{U}_{\be_0}$ up to $\be=0$. \vspace{1mm}
\end{itemize}
The result of such an exchange of symbols is that $\rho_{\e{eff}}^{\pa{m}}\!\pa{x,t}$ admits the representation
\beq
\hspace{-4mm}\rho^{\pa{m}}_{\e{eff}}\!\pa{x,t} = \lim_{w \tend +\infty}  \; \lim_{\be \tend 0} \;  \lim_{s \tend + \infty}
\; \lim_{r \tend + \infty}
\bs{:}  \f{ \Dp{}^m }{ \Dp{}\ga^m } \; \Bigg\{
\wh{E}^{\,2}\!\pa{q} \cdot \ex{- \Int{-q}{q} \pac{ \i x u^{\prime}\!\pa{\la} + \wh{g}^{\, \prime}\!\pa{\la} } \ga \nu_s\pa{\la} \dd \la }
 X_{\msc{C}_{\op{V}}^{\pa{w}}}\!\pac{\ga \nu_s,\wh{E}^{\,2}} \, \msc{G}_{\ga;2w}^{\pa{\be}}\!\pac{\varpi_r}
\Bigg\}_{\mid \ga=0}  \hspace{-5mm} \bs{: }   \;.
\label{equation exprimant rho m en terms moyenne X limit thermo}
\enq
This formula deserves a few comments. In the case of complex valued functions $e$,
the functional $X_{\msc{C}_{\op{V}}^{\pa{w}}}\!\big[\ga \nu_s, e \big]$ appearing
in \eqref{equation exprimant rho m en terms moyenne X limit thermo} corresponds precisely to the rank one perturbation \eqref{appendix ecriture limite thermo XN}  of the Fredholm determinant 
of the integrable integral operator $\e{id} + \op{V}_x$ acting on $L^{2}\pa{\intff{-q}{q}}$ with the integral kernel  \eqref{definition noyau GSK FF}
where the substitution $\nu \hookrightarrow \ga  \nu_s$ has been implemented.

The contour $\msc{C}_{\op{V}}^{\pa{w}}$ is as defined in Figure \ref{contour CE et sa restriction CEw}. The parameter $w$
delimiting the size of this contour plays the role of a regularisation.
This means, in particular, that the limit of an unbounded contour $\msc{C}_{\op{V}}^{\pa{\infty}}$ can only be taken after $r$ and $s$ are sent to infinity
and the analytic continuation up to $\be=0$ is carried out.
Finally, equation \eqref{equation exprimant rho m en terms moyenne X limit thermo} contains 
the functional $\msc{G}_{\ga;2w}^{\pa{\be}}$. The latter corresponds to  the thermodynamic limit of the functional
$\wh{\msc{G}}_{\ga;2w}^{\,\pa{\be}}$. Its explicit expression is provided in Lemma A.1 of [$\bs{A7}$]. 
It is important to point out  that the parameter $\be_0$ defining the domain $\bs{U}_{\be_0}$
from which one should carry out the analytic continuation up to $\be=0$ depends on $2w$. This dependence is chosen in such a way that
the constant of regularity $C_{\msc{G}_A}$ for the functional $\msc{G}_{\ga;2w}^{\pa{\be}}$
is large enough so as to make licit all the necessary manipulations with the translation operators and generalisations
thereof.

\vspace{2mm}

Equation \eqref{equation exprimant rho m en terms moyenne X limit thermo} provides one with a convenient representation for the thermodynamic limit
$\rho^{\pa{m}}_{\e{eff}}\!\pa{x,t}$. It constitutes  the first step towards extracting the large-distance $x$ and long-time $t$ asymptotic
behaviour of $\rho^{\pa{m}}_{\e{eff}}\!\pa{x,t}$. Although the proof of this representation for the thermodynamic limit
is quite technical and lengthy I stress that formula
\eqref{equation exprimant rho m en terms moyenne X limit thermo} can be readily obtained, on a formal level of rigour, without the use of any complicated and
long computations. It is enough to take the thermodynamic limit formally on the level of formula  \eqref{equation exprimant Q cal en terme moyenne XN}
what immediately yields \eqref{equation exprimant rho m en terms moyenne X limit thermo}.


\subsection{The multidimensional Natte series and asymptotics}
\label{subsection multidim Natte}

\begin{theorem}
\label{Theorem comportement asympt coeff Taylor limite Thermo}
The thermodynamic limit of
the Taylor coefficients $\rho_{\e{eff}}^{\pa{m}}\!\pa{x,t}$ admits the below truncated multidimensional Natte series representation
\bem
\rho_{\e{eff}}^{\pa{m}}\!\pa{x,t}   = \f{ \Dp{}^m }{ \Dp{}\ga^m }
\left\{
\f{ \bs{1}_{ \R \setminus\intff{-q}{q} }\pa{\la_0}    }{ \sqrt{-2\pi x u^{\prime \prime}\!\!\pa{\la_0} } } \times
\f{  \ex{\i x\pac{u\pa{\la_0}-u\pa{q}} }   \mc{B}\pac{\ga F^{\la_0}_{q} ; p }\mc{A}_{0} \pac{\ga F^{\la_0}_{q}  }  }
{  \pa{x-t v_F + \i 0^+ }^{ \big[ \ga F^{\la_0}_{q} \!\pa{q} \big]^2 }   \pa{x + t v_F + \i 0^+}^{  \big[ \ga F^{\la_0}_{q}\!\pa{-q} \big]^2 } }
\mc{G}_{1;\ga}^{\pa{0}}\pab{ \la_0 }{ q } \right. \vspace{2mm} \\
+ \f{ \ex{ \i x \pac{u\pa{-q}-u\pa{q}} }   \pa{\mc{B}\mc{A}_-}\pac{\ga F^{-q}_{q} ; p}
 \mc{G}_{1;\ga}^{\pa{0}}\pab{ -q }{ q } }
{  \pa{x-t v_F + \i 0^+ }^{ \big[ \ga F_{q}^{-q} \!\pa{q} \big]^2}    \pa{x + t v_F + \i 0^+}^{ \big[ \ga F^{-q}_{q}\!\pa{-q}-1\big]^2 } }
\; + \;
\f{ \pa{\mc{B}\mc{A}_+}\pac{\ga F^{\emptyset}_{\emptyset} ;p }
\mc{G}_{0;\ga}^{\pa{0}}\pab{ \emptyset }{ \emptyset } }
{  \pa{x-t v_F + \i 0^+ }^{  \pac{ \ga F^{\emptyset}_{\emptyset}\!\pa{q} +1 }^2}  \pa{x + t v_F + \i 0^+}^{\pac{\ga F^{\emptyset}_{\emptyset}\!\pa{-q}}^2 } }
\vspace{2mm} \\
\vspace{5mm}
\hspace{-1cm} + \ex{-\i x u\pa{q} }  \sul{n=1}{m} \sul{ \mc{K}_n }{} \sul{ \mc{E}_n (\vec{k}\, ) }{}
\int_{ \msc{C}_{\eps_{\bs{t}} }^{\pa{w}} }{}  \hspace{-2mm}
\left.
\f{  H_{n;x}^{\pa{\paa{\eps_{\bs{t}}}}}\pa{  \paa{u\pa{z_{\bs{t}}} } ; \paa{z_{\bs{t}}}  } \big[ \ga  F^{ z_{\bs{+}} }_{ z_{\bs{-}} } \big]
\mc{B}\big[\ga F^{ z_{\bs{+}} }_{ z_{\bs{-}} }  ; p \big]      }
{  \pa{x-t v_F + \i 0^+ }^{ \pac{ \ga F^{ z_{\bs{+}} }_{ z_{\bs{-}} } \!\pa{q}}^2}
						\pa{x + t v_F + \i 0^+  }^{ \pac{ \ga F^{ z_{\bs{+}} }_{ z_{\bs{-}} }\!\pa{-q}}^2 } }
\mc{G}_{ \abs{\paa{z_{\bs{+}} } } ; \ga }^{\pa{0}} \pab{ \paa{z_{\bs{+}}} }{ \paa{z_{\bs{-}}} \cup\paa{q} }
 \f{ \dd^n z_{\bs{t}} }{ \pa{2 \i \pi}^n }     \right\} _{\mid \ga=0}  \hspace{-3mm}.
\label{ecriture serie Natte multDim coeff taylor rho eff}
\end{multline}
There, I have introduced the notations
\beq
\big\{  z_{\bs{+}} \big\}  = \big\{ z_{\bs{t}} \; , \;  \bs{t} \in \J{ \vec{k} } \; : \; \eps_{\bs{t}} =1 \big\} \;,  \qquad
\big\{  z_{\bs{-}}  \big\}  = \big\{  z_{\bs{t}} \; , \;  \bs{t} \in \J{\vec{k}} \; : \; \eps_{\bs{t}} = - 1 \big\}  \; ,
\quad \big| \big\{  z_{\bs{+}} \big\} \big| \equiv \# \big\{ z_{\bs{t}} \; , \;  \bs{t} \in \J{ \vec{k} } \; : \; \eps_{\bs{t}} =1 \big\}   \;.
\label{definition des ensembles z plus et moins}
\enq
$F^{\emptyset}_{\emptyset}$, $F^{\la_0}_{q}$, $F^{-q}_{q}$ are as defined in the discussion that followed equation \eqref{equation fondamentale asymptotiques rho x et t}. 
More generally, one has
\beq
F^{ z_{\bs{+}} }_{ z_{\bs{-}} }\pa{\la} \equiv F\pabb{ \la  }{ \paa{z_{\bs{+}}}  }{ \paa{z_{\bs{-}}} \cup\paa{q}  }  = -\f{Z\pa{\la}}{2}
\; - \; \sul{ \substack{ \bs{t} \in \J{k}  \\   \eps_{\bs{t}}=1 } }{} \phi\pa{\la,\,z_{\bs{t}}}
 \; + \;  \sul{ \substack{ \bs{t} \in \J{k}  \\   \eps_{\bs{t}}=-1 } }{} \phi\pa{\la,\,z_{\bs{t}}} \;.
\label{definition fction shift multi-parametres z}
\enq
The function $\mc{G}_{n;\ga}^{\pa{0}}$ is related to the thermodynamic limit of the smooth part of the form factor. Its expression can be
found in \eqref{formule explicite G+ thermo}. The functionals $\mc{B}$, $\mc{A}_{\pm}$ and $\mc{A}_0$ are, respectively, defined in 
\eqref{definition fonctionnelle B}, \eqref{definition fonctionnelle A+ et kappa} and \eqref{definition fonctionnelle A- et A0}.

\end{theorem}

Note that  all the summands in \eqref{ecriture serie Natte multDim coeff taylor rho eff}  
involving the functional $\mc{G}_{n;\ga}^{\pa{\be}}$ are well defined at $\be=0$. Indeed, the potential singularities
present in $\mc{G}_{n;\ga}^{\pa{\be}}$ are cancelled by the zeroes of the pre-factors.

I refer to Section 4.6 of [$\bs{A7}$] for a proof of this theorem. 
The main idea behind its proof is as follows. In order to compute the effect of the $\bs{:} \cdot \bs{:}$
ordering in \eqref{equation exprimant rho m en terms moyenne X limit thermo}, one should replace 
the expression inside of $\bs{:} \cdot \bs{:}$ by 
\beq
\mc{F}\pac{\ga \nu_s, \wt{g},\varpi_r}\!\pa{\ga} = \wt{E}^{\,2}\!\pa{q} \cdot
\ex{- \Int{-q}{q} \pac{ \i x u^{\prime}\!\pa{\la} + \wt{g}^{\, \prime}\!\pa{\la} } \ga \nu_s\pa{\la} \dd \la }
 X_{\msc{C}_{ \op{V} } ^{\pa{w}}}\!\big[ \ga \nu_s,\wt{E}^{\,2} \big] \, \msc{G}_{\ga;2w}^{\pa{\be}}\!\pac{\varpi_r} 
\label{definition fonctionnelle F qui mime substitution operatorielle dans serie Natte}
\enq
where 
\beq
\wt{E}^{\, 2} \pa{\la} = \ex{- \i x u\pa{\la}-\wt{g}\pa{\la}} \quad  \e{with} \quad \wt{g}\pa{\la} = \wt{g}_{1,s}\pa{\la} + \wt{g}_{2,r}\pa{\la}\;,
\label{definition fction E-tilde}
\enq
and
\beq
\wt{g}_{1,s}\pa{\la} = - \sul{p=1}{s} \phi\big(t_p,\la\big) a_p \qquad  \e{while} \qquad
\wt{g}_{2,r}\pa{\la} =  \sul{p=1}{r}  \f{ b_{1,p} }{t_{1,p}-\la}  \; + \;  \sul{ p=1 }{ r } \f{ b_{2,p} }{t_{2,p}-\la} \;.
\label{definition fonctions tilde g1 et g2}
\enq
Above, $t_p$ and $t_{p,1},t_{p,2}$ are discretisation points, respectively, of the loops $\msc{C}_{\e{out}}$ or $\msc{C}_{\e{in}}$ and $\msc{C}_{\e{out}}$
that are used in the course of re-writing the functionals $\wh{D}_N$ and $\wh{\mc{G}}_{N;\ga}$.

In order to implement the operator substitution, one has to compute the Taylor coefficients of the series expansion of
$\mc{F}\pac{\ga \nu_s, \wt{g},\varpi_r}\!\pa{\ga}$ into powers of $b_{1,p}$, $b_{2,p}$ with $p=1,\dots, r$ and $a_p$ with $p=1,\dots,s$.
These Taylor coefficients are \textit{solely} determined by the functional $\mc{F}\pac{\ga \nu_s, \wt{g}, \varpi_r}\!\pa{\ga}$
depending on the \textit{classical} function $\wt{g}$ \eqref{definiton fonction gs scalaire}.
Therefore, one can use \textit{any} equivalent  representation for $\mc{F}\pac{\ga \nu_s, \wt{g},\varpi_r}\!\pa{\ga}$ as a starting point for
computing the various partial derivatives in respect to $b_{j,p}$ or $a_p$.
In other words, one can use \textit{any} equivalent series representation\symbolfootnote[2]{One natural representation that can be used as a starting point for taking the derivatives is the Fredholm series-like representation for
$X_{\msc{C}_E^{\pa{w}}}\big[\ga \nu_s, \wt{E}^{\,2} \big]$. In fact, it is this series representation that allows one to prove the 
representation \eqref{equation exprimant rho m en terms moyenne X limit thermo} in the very first place. }
 for  $X_{\msc{C}_E^{\pa{w}}}\big[\ga \nu_s, \wt{E}^{\,2} \big]$.
Clearly, different series representations for this rank one perturbation of the Fredholm determinant will lead to different types of expressions for the
Taylor coefficients. However, in virtue of the uniqueness of Taylor coefficients, their \textit{values} coincide.
As discussed in Subsection \ref{SousSection Serie de Natte pour perturbation de rang un}, $X_{\msc{C}_E^{\pa{w}}}\big[\ga \nu_s, \wt{E}^{\,2} \big]$ 
admits a Natte series representation. It remains to take this series representation as the starting point for the computation of the Taylor coefficients.
The rest is straightforward but slightly technical.

\subsection{Some more conjectures leading to the dominant asymptotics of $\rho\!\pa{x,t}$}

\noindent Under the hypothesis that
\begin{enumerate}
\item  the Taylor series $\sum_{m=0}^{+\infty} \ga^m \tf{\rho_{\e{eff}}^{\pa{m}}\!\pa{x,t}}{m!}$ is convergent up to $\ga=1$,

\item its sum gives $\rho\!\pa{x,t}$,

\item  the multidimensional Natte series \eqref{formule serie de Natte multidimensionnelle rho} is convergent;

\end{enumerate}
one gets that  $\rho\pa{x,t}$ can be obtained from
\eqref{ecriture serie Natte multDim coeff taylor rho eff} by removing the $m^{\e{th}}$ $\ga$-derivative symbol and setting $\ga=1$.
After identifying the coefficients in the first two lines with the
properly normalised thermodynamic limit of the form factors of the field ($\big| \mc{F}_q^{\la_0} \big|^2$,
$\big|  \mc{F}_{\emptyset}^{\emptyset} \big|^2$ or $\big| \mc{F}_q^{-q} \big|^2$) one obtains the below series of multiple integral representation for
the thermodynamic limit of the one-particle reduced density matrix:
\bem
\rho\pa{x,t} = \sqrt{ \f{-2 \i \pi }{ t \veps^{\prime \prime}\pa{\la_0} - x p^{\prime \prime}\pa{\la_0} } } \times
\f{ p^{\prime}\!\pa{\la_0} \ex{ \i x [u\pa{\la_0}-u\pa{q}] }   \abs{\mc{F}_q^{\la_0}}^2  }
{  \pac{-\i\pa{x-t v_F } }^{ \big[  F^{\la_0}_{q} \!\pa{q} \big]^2}   \pac{\i\pa{x + t v_F} }^{  \big[  F^{\la_0}_{q}\!\pa{-q} \big]^2 } }
\bs{1}_{ \R \setminus \intff{-q}{q} }\pa{\la_0}
 \\
+ \f{ \ex{-2\i x p_F}  \abs{\mc{F}_q^{-q}}^2 }
{  \pac{- \i \pa{x-t v_F } }^{ \big[ F_{q}^{-q} \!\pa{q} \big]^2}    \pac{ \i \pa{x + t v_F} }^{ \big[  F^{-q}_{q}\!\pa{-q}-1\big]^2 } }
\; + \;
\f{ \abs{\mc{F}_{\emptyset}^{\emptyset}}^2  }
{  \pac{- \i \pa{x-t v_F } }^{  \big[  F^{\emptyset}_{\emptyset}\pa{q}+1 ]^2}  \pac{ \i \pa{x + t v_F} }^{ \big[ F^{\emptyset}_{\emptyset}\pa{-q}\big]^2 } }  \vspace{5mm}\\
\hspace{-1cm}\; + \;  \ex{-\i x u\pa{q} } \sul{n=1}{+\infty} \sul{ \mc{K}_n }{} \sul{ \mc{E}_n ( \vec{k} ) }{}
\oint_{ \msc{C}_{\eps_{\bs{t}} }^{\pa{w}} }^{}
 \mc{G}_{ \abs{ \paa{z_{\bs{+}}} };1  }^{\pa{0}} \pab{ \paa{z_{\bs{+}}} }{ \paa{z_{\bs{-}}} }
\f{ H_{n;x}^{\pa{\paa{\eps_{\bs{t}}}}}\!\pa{ \paa{u\pa{z_{\bs{t}}} } ; \paa{z_{\bs{t}}}  } \big[ F^{z_{\bs{+}}}_{z_{\bs{-}}} \big]
\mc{B}\big[ F^{z_{\bs{+}}}_{z_{\bs{-}}}  ; p\big]    }
{  \pa{x-t v_F + \i 0^+ }^{ \pac{ F^{z_{\bs{+}}}_{z_{\bs{-}}} \pa{q}}^2}  \pa{x + t v_F +\i 0^+ }^{ \pac{ F^{z_{\bs{+}}}_{z_{\bs{-}}}\pa{-q}}^2 } }
    \cdot     \f{ \dd^n z_{\bs{t}} }{ \pa{2 \i \pi}^n }  \; .
\label{formule serie de Natte multidimensionnelle rho}
\end{multline}

The  above representation  immediately yields the

\begin{cor}

The reduced density matrix $\rho\pa{x,t}$ admits the asymptotic expansion as given in Subsection \ref{Theorem asymptotique rho}.

\end{cor}




\section{Conclusion}

This chapter reviewed the method of multidimensional flows which allows one to compute, starting from the form factor expansion of a correlator or from its multidimensional Fredholm series representation,
its long-distance and large-time asymptotic behaviour. Although I could make various steps of the method rigorous, still, a certain amount of hypothesis remains to be proven.  In its present setting, 
the method is already applicable to the study of numerous two and multi-point correlation functions. 
It was applied, in its formal setting, to the density-density correlation function of the non-linear Schr\"{o}dinger model in my joint work with Terras [$\bs{A9}$]. 
Also, in the joint work with Maillet and Slavnov [$\bs{A10}$] we argued the form of a multidimensional Fredholm series representation for the zero-time
density-density correlation function in the non-linear Schr\"{o}dinger model at \textit{finite} temperatures. 
At the time, we have built our analysis of the large-distance behaviour on an adaptation of the method introduced earlier in \cite{KozKitMailSlaTerXXZsgZsgZAsymptotics}. 
Yet, there is no problem to deal with the associated multidimensional Fredholm series within the multidimensional flow technique described in the present chapter.  
The method  should also be applicable to the study of the short distance asymptotics of 
the correlation functions in integrable massive field theories. For instance, the method seems applicable to the analysis of certain two-point functions (and their short-distance asymptotics)
in the sine/sinh-Gordon model whose form factors have been obtained in
 \cite{FringMussardoSimonettiFFSOmeLocalObsSinhGordon,KoubekMussardoFFForMoreOpInSinhGordon,SmirnovIntegralRepSolitonFFSineGordonBootstrap,SmirnovFormFactors}. 
 In these cases, I expect to deal 
 with some multidimensional deformation of a Fredholm determinant associated with \cite{BernardLeclairDiffEqnForPIII,McCoyTracyWuPIIIProofForDetRep,WidomFDetForPII} 
 a specific solution to the $3^{\e{rd}}$ Painlev\'{e} equation, a new type of special function whose description and
asymptotic behaviour is interesting in its own right.
In the sinh-Gordon case, the form of the kernel of the Fredholm determinant should be close to the one proposed by Korepin and Slavnov \cite{KorepinSlavnovDetRepDualFieldSinhGordon}
in their dual field based analysis of form factor expansions in this model.

\chapter{Towards an understanding of "critical" asymptotics}
\label{Chapitre Solving some c-shifted RHP}

The  multidimensional flow method that I described in the previous chapter is a way of recasting a correlation function in an interacting integrable  model
in terms of a $\bs{:} \cdot \bs{:}$ average of a Fredholm determinant depending on some operator symbol. Various possible representations for a 
Fredholm determinant lead to as many representations for the correlation function. In the previous chapter, I considered the example of a correlation function 
whose $\bs{:} \cdot \bs{:}$ representation involves the integral operator $\e{id}+\op{V}_x$ where $\op{V}_x$ is characterised by the integral kernel \eqref{definition noyau GSK FF}. 
It is natural to ask what other types of kernels can arise inside a $\bs{:} \cdot \bs{:}$ average representation for correlation functions. 
So far, I only focused on correlation functions involving products of local operators each of them separated by some spatial distance. 
 Yet, one can also consider correlation functions of diffuse operators. 
The simplest of them is the so-called emptiness formation probability correlation functions $\tau(x)$. In the context of the non-linear Schr\"{o}dinger model, 
this correlator represents the probability that, in the ground state of the model, there will be no quasi-particles located in a region of size $x$.  
Without going too deep into the details, under appropriate hypothesis, this correlator can be recast in terms of an average of the type
\beq
\tau(x) \; = \; \e{LIM} \bs{:} Y[\wh{E}^2 ]  \msc{T}\big[ \varpi \big] \bs{:} \qquad \e{where} \quad Y[ e^2 ] \; = \;  \det\Big[ \e{id} + \op{W}_x^{(\e{c})}  \Big] \; . 
\label{ecriture representation EFP NLSM comme valeur moyenne dot dot}
\enq
There, $\e{LIM}$ stands for a certain amount of limits that have to be taken, in the spirit of   
\eqref{equation exprimant Q cal en terme moyenne XN} or \eqref{equation exprimant rho m en terms moyenne X limit thermo},  after the $\bs{:}\cdot \bs{:}$ average is performed. 
$\msc{T}$ is a functional that plays an analogous role to the one played by functional $\msc{G}^{(\be)}_{\ga;2w}$ in the analysis of the last chapter. 
For the purpose of the discussion that I wish to carry, there is no need to provide more informations on the precise
for of $\msc{T}$ or of the $\e{LIM}$ symbol. The main point is that the $\bs{:} \cdot \bs{:}$ average involves the integral operator   $\e{id}+\op{W}_x^{(\e{c})}$ on $L^2(\intff{-q}{q})$
whose integral kernel is of $c$-shifted type:
\beq
W_x^{(\e{c})}(\la,\mu)  \; = \; 
\f{ \i c }{ 2\i \pi (\la-\mu) } \cdot \bigg\{ \f{ e(\mu)e^{-1}(\la)  }{ (\la-\mu) + \i c } 
\, + \, \f{ e(\la)e^{-1}(\mu) }{ (\la-\mu) - \i c } \bigg\} \qquad \e{with} \qquad e(\la) \; = \; \ex{ - \i \f{x}{2}  p(\la) - g(\la) } \;. 
\label{definition noyau integrable V dans regime critique}
\enq
The above kernel is integrable since it can be recast in the form:
\beq
W_x^{(\e{c})} (\la,\mu) \; = \;  \f{1}{(\la-\mu)}  \Int{0}{+\infty} \f{ \ex{-c s} }{2 \i \pi} 
\big[  e(\mu) \cdot e^{-1}(\la)   \cdot \ex{is(\la-\mu)} \, - \,  e(\la) \cdot e^{-1}(\mu) \cdot  \ex{is(\mu-\la)}  \big] \cdot \dd s \;. 
\enq
This seems to offer the possibility to construct the Natte series representation for $\det\Big[ \e{id} + \op{W}_x^{(\e{c})}  \Big] $ by means of a non-linear steepest descent analysis 
of the associated Riemann--Hilbert problem. Unfortunately, the fact that $W_x^{(\e{c})} (\la,\mu) $ is an integrable integral kernel described by a continuous collection of functions has its price!
Recall that an integrable integral operator $\e{id} \, + \, \op{O}$ on $L^2(J)$ with an integral kernel
\beq
O(\la,\mu) \, = \, \f{ \sum_{a=1}^{N} e_a(\la)f_a(\mu)  }{ \la - \mu } \qquad \e{and} \qquad \sul{a=1}{N} e_a(\la)f_a(\la) \; = \; 0
\enq
is associated with a Riemann--Hilbert problem for a holomorphic $N\times N$ matrix on $\Cx \setminus J$.  It is thus not astonishing that the use of a continuously labelled family of functions $e_a, f_a$
so as to describe the kernel results in an operator-valued Riemann--Hilbert problem. 
The operator valuedness constitutes an important technical complication with respect to the matrix valued case. 

In fact, Fredholm determinants of $c$-shifted integrable integral operators and their associated operator valued Riemann--Hilbert problems arose 
already  in the early days of  exploring the correlation functions in quantum integrable models out of their free fermion point. 
The Riemann--Hilbert machinery allowed to construct systems of partial differential equations satisfied by specific
examples of such Fredholm detereminants \cite{ItsIzerginKorepinSlavnovDifferentialeqnsforCorrelationfunctions,KojimaKorepinSlavnovNLSESystemPartialDiffEqnsDualFieldTempeAndTime,
KorepinSlavnovTempeNLSELongDistCompMoreTermsOldSeries,KorepinSlavnovRHPForNLSMOpertorValuedSettingInto}. 
However, at the time, not much progress has been achieved in respect to the asymptotic analysis of operator valued Riemann--Hilbert problems.
 The sole serious work on the subject has been carried out by Its and Slavnov \cite{ItsSlavnovNLSTimeAndSpaceCorrDualFields}.
These authors performed a formal non-linear steepest descent-based analysis of an operator valued Riemann--Hilbert problem depending oscillatorily on a large parameter $x$. 
This allowed them to extract the leading asymptotic behaviour in $x$ out of the logarithm of the Fredholm determinant
which gave rise to the Riemann--Hilbert problem on the first place. However,  numerous technical difficulties (the operator nature of the scalar Riemann--Hilbert problem which 
arises in the very the first step of the analysis, construction of parametrices in terms of special functions with operator index,...) 
which could not have been overcome stopped, for almost 15 years, any activity related to an asymptotic analysis of operator valued Riemann--Hilbert problems.

Yet, the \textit{per se} operator valued setting of the Riemann--Hilbert problem is not the sole difficulty to be dealt with so as to extract the large-$x$ asymptotic behaviour out of $\det\big[ \e{id}+\op{W}_{x}^{(\e{c})} \big]$. 
On top of all issues related to handling piecewise holomorphic functions taking values in some space of operators on some appropriate functional space, 
one has to deal with the fact that the operator valued Riemann--Hilbert problem of interest posses a "critical" structure that already manifested itself on the level of 
the Riemann--Hilbert problem associated with the pure sine kernel. 
While in the latter case the $2\times2$ jump matrix has one of its diagonal entries equal to zero what renders the use of a $LU$  factorisation approach to the 
non-linear steepest descent impossible, in the former case, the  $2\times2$ operator-valued jump matrix is such that its $22$ diagonal operator entry has a zero eigenvalue. 
In the pure sine kernel case, as proposed in \cite{DeiftItsZhouSineKernelOnUnionOfIntervals}, a way out of the problem consists in implementing  a $g$-function transformation.    
Thus, it seems reasonable to expect that, in the operator valued setting, one will have to build an operator valued analogue of the $g$-function transformation. However, the hard part 
is that such a transformation would have to provide a distinct treatment on the one hand of the zero eigenvalue eigenspace of the operator entry and 
on the other hand of its orthogonal complement. 
The detail that makes it difficult is that the zero eigenvalue eigenspace of the $22$ operator valued entry of the jump matrix depends on the position on the jump contour. 
How to effectively construct such $g$-functions seems unclear, even in the most trivial possible generalisation of the $2\times 2$ case, \textit{i}.\textit{e}.
a critical $3\times 3$ matrix Riemann--Hilbert problem. Some progress in constructing the $g$-functions for larger than $2\times 2$ Riemann--Hilbert problems,
although of a slightly different nature than the ones of interest here, has been achieved in \cite{DuitsKuijlaarsRHPAnalysisforMOPS2MatrixModel}. 

To summarise, there are two distinct issues that should be understood prior to addressing the large-$x$ analysis of the emptiness formation probability: \vspace{1mm}

\begin{itemize}

 \item[$\bullet$]  settle all of the problematic issues associated with the very fact of dealing with operator valued Riemann--Hilbert problems this in the off-critical case, \textit{i.e.} when there are 
 no zero eigenvalue eigenspaces;\vspace{1mm}
 
 \item[$\bullet$]  extend the procedure  to the critical case  by constructing operator valued analogues of the $g$-function transformation. \vspace{1mm}
 
\end{itemize}

In the present chapter, I shall review the progress I made relatively to the first point, namely obtaining a better understanding of $c$-shifted integrable integral operators and of their associated
operator valued Riemann--Hilbert problems. As indicated, the focus will be on the case of off-critical integral operators whose integral kernel takes the form 
\beq
W_x(\la,\mu)  \; = \; 
\f{ \i c F(\la) }{ 2\i \pi (\la-\mu) } \cdot \bigg\{ \f{ \ex{ \f{\i x}{2}[p(\la)-p(\mu)] } }{ (\la-\mu) + \i c } 
\, + \, \f{ \ex{ \f{\i x}{2}[p(\mu)-p(\la)] } }{ (\la-\mu) - \i c } \bigg\} \;,
\label{definition noyau integrable V dans regime critique}
\enq
with 
\begin{itemize}
 \item[$\bullet$]   $p(\intff{a}{b})\subset \R$ and such that $p$ is a biholomorphism from an open neighbourhood $U$ 
 of $\intff{a}{b}$ in $\Cx$ onto some open neighbourhood of $\intff{p(a)}{p(b)}$ in $\Cx$ which furthermore satisfies $p^{\prime}_{\mid \intff{a}{b}}>0$ ; \vspace{1mm}
\item[$\bullet$] $F$ is holomorphic on $U$ and satisfying $\big| \e{arg}\big(1+F(\la) \big) \big|<\pi$ uniformly in $ U$ . \vspace{1mm}
\end{itemize}

\noindent In the following, I will discuss two methods allowing one to extract the large-$x$ asymptotic behaviour out of $\det\big[\e{id} + \op{W}_x\big]$
with $W_x(\la,\mu)$ as given by \eqref{definition noyau integrable V dans regime critique}.

The first method utilises $2\times 2$ Riemann--Hilbert problems
and an appropriate rewriting of the integral kernel as a perturbation of the generalised sine kernel defined as 
\beq
\wt{S}_x(\la,\mu)  \; = \; \f{  F(\la)  }{  \pi (\la-\mu) } \cdot  \sin \Big( \f{ x}{2}[p(\la)-p(\mu)]  \Big) \;. 
\label{definition noyau integrable GSK}
\enq
On top of providing an access to the large-$x$ behaviour of determinants of $c$-shifted operators such as \eqref{definition noyau integrable V dans regime critique},
the method also allows one to extract the large-$x$ behaviour out of so-called lacunary Toeplitz determinants. \vspace{1mm}

The second method consists in a \textit{bona fide} non-linear steepest descent analysis of the operator valued  Riemann--Hilbert problem associated with the $c$-shifted kernel \eqref{definition noyau integrable V dans regime critique}.   
Here, the two main achievements consist in overcoming the technical difficulties that arose previously in the analysis 
of operator-valued Riemann--Hilbert problems, namely:
\begin{itemize}
 
 \item[i)]  the construction of  solutions to  operator valued scalar Riemann--Hilbert problem with jump on $\intff{-q}{q}$ is reduced to the one of inverting 
an integral operator acting on $L^{2}\big( \Ga(\intff{-q}{q}), \dd z \big)$, where $\Ga(\intff{-q}{q})$ denotes a small counterclockwise loop around $\intff{-q}{q}$. \vspace{1mm}  

 \item[ii)] The construction of local parametrices is strongly simplified and made rigorous. In the present case, the parametrices are given in terms of 
 special function (confluent hypergeometric functions) whose auxiliary parameters are scalar-valued holomorphic functions and \textit{not} 
 holomorphic functions taking values in some infinite dimensional Banach spaces,  as it was the case in \cite{ItsSlavnovNLSTimeAndSpaceCorrDualFields}. \vspace{1mm}
 
\end{itemize}

I do stress that both methods mentioned above allow one to establish the asymptotic expansion for $\det\big[ \e{id}\, + \, \op{W}_x \big]$.
\begin{theorem}
\label{Theorem Cptm Asympt det op int avec shift}
Let $p$ and $F$ be as described above and $\wt{\op{S}}_x$ denote the integral operator on $L^{2}\big( \intff{-q}{q}\big)$ whose integral kernel has been defined in \eqref{definition noyau integrable GSK}.
Then, the below ratio of Fredholm determinants admits the large-$x$ asymptotic behaviour
\beq
\f{ \det\big[ \e{id}\, + \, \op{W}_x \big] }{ \det\big[ \e{id}\, + \, \wt{\op{S}}_x \big]  } \; = \; 
\det \big[ \e{id}\, + \, \op{U}_+ \big] \cdot \det\big[\e{id}\, + \, \op{U}_- \big] \cdot 
\Big( 1\,+\,\e{o}(1) \Big)
\enq
where $\op{U}_{\pm}$ are integral operators on $L^{2}\big( \Ga(\intff{-q}{q}) \big)$, with $\Ga$ being a small counterclockwise loop 
around the interval $\intff{-q}{q}$. The integral kernels of $\op{U}_{\pm}$ read
\beq
U_{\pm} (\la,\mu)\, = \, \f{ \a(\la) \cdot \a^{-1}(\mu \mp \i c) }{ 2\i \pi (\la-\mu \pm \i c) } \quad with \quad
\a(\la) \, = \, \exp\bigg\{ \Int{-q}{q} \f{ \ln\big[1+F(\mu) \big] }{ \la - \mu } \cdot \f{ \dd \mu }{ 2\i \pi } \bigg\}\;. 
\label{ecriture noyau integral U pm et definition fct alpha}
\enq
\end{theorem}
I do stress that Theorem \ref{Theorem Cptm Asympt det op int avec shift} characterises the leading large-$x$ asymptotic behaviour of $\det\big[ \e{id}\, + \, \op{W}_x \big]$ in that the one of 
$\det\big[ \e{id}\, + \,  \wt{\op{S}}_x  \big]$ has been obtained in \cite{KozKitMailSlaTerRHPapproachtoSuperSineKernel}.

\section{The factorisation method}

\subsection{The $c$-shifted kernels}

The factorisation method [$\bs{A13}$] builds on the observation that the $c$-shifted kernel $W_x$ can be decomposed as 
\beq
W_x( \la , \mu) \; = \; \wt{S}_x(\la,\mu) \; + \; \wt{W}_x(\la,\mu) \qquad \e{with} \qquad 
\wt{W}_x(\la,\mu) \; = \;  - \; \f{ F(\la) }{ 2\i \pi } \bigg\{ \f{ \ex{ \f{\i x}{2}[p(\la)-p(\mu)] } }{ (\la-\mu) + \i c } 
\, - \, \f{ \ex{ \f{\i x}{2}[p(\mu)-p(\la)] } }{ (\la-\mu) - \i c } \bigg\}  \;.  
\enq

The integral operator $ \e{id}\, + \,  \wt{\op{S}}_x $ is of integrable type and has been extensively studied in \cite{KozKitMailSlaTerRHPapproachtoSuperSineKernel}. 
This operator is associated with a $2\times 2$ Riemann--Hilbert problem for a  matrix $\Omega$ that is holomorphic on $\Cx\setminus \intff{-q}{q}$. 
Under the hypothesis of the present chapter, this Riemann--Hilbert problem is uniquely solvable,  at least for $x$ large enough. 
Furthermore, for such large-$x$ at least, $ \e{id}\, + \,  \wt{\op{S}}_x $ is invertible. It is a standard fact that the resolvent kernel for 
$\e{id} \, + \, \wt{\op{S}}_x$ can be constructed explicitly in terms of $\Omega$.

The idea at the root of the factorisation method consists in factoring out, explicitly, the operator $ \e{id}\, + \,  \wt{\op{S}}_x $  out of $\det[ \e{id}\, + \,  \op{W}_x ]$
and using the properties of the solution $\Omega$ so as to recast $  \big( \e{id}\, + \,  \wt{\op{S}}_x  \big)^{-1} \cdot \wt{\op{W}}_x$
in a way that is appropriate for the large-$x$ analysis of the associated determinant. 
All-in-all, such handlings lead to

\begin{prop}
\label{Proposition factorisation Det W en terme RHP data det M}
The Fredholm determinant of the integral operator $\e{id} + \op{W}_x$ can be factorised as 
\beq
\det \big[ \e{id} + \op{W}_x \big] \; = \; \det \big[ \e{id} + \wt{\op{S}}_x \big] \cdot  \det \big[ \e{id} + \op{M}_x \big] \;. 
\enq
$\e{id} + \op{M}_x$ appearing above is an integral operator on $L^2\big(  \Ga( \intff{-q}{q} ) \big) \oplus L^2\big(  \Ga( \intff{-q}{q} ) \big)$ with $\Ga( \intff{-q}{q} )$ any counterclockwise 
compact loop around $\intff{-q}{q}$ located in the strip $|\Im (z)|<\tf{c}{2}$. This operator is characterised by its matrix integral kernel 
\beq
M_x(\la,\mu) \; = \;  \pa{  \ba{cc}  
  \f{ \big(  \bs{e}_1 \, ,  \Omega^{-1}(\la) \cdot \Omega(\mu + \i c) \cdot  \bs{e}_1 \big) }{ 2 \i \pi ( \la-\mu - \i c ) }  & 
   \f{ \big(  \bs{e}_1 \, , \,  \Omega^{-1}(\la) \cdot\Omega(\mu - \i c) \cdot \bs{e}_2 \big)  }{ 2 \i \pi( \la-\mu + \i c) } \vspace{2mm} \\
 \f{  \big(  \bs{e}_2 \, , \,  \Omega^{-1}(\la ) \cdot \Omega(\mu+ \i c) \cdot \bs{e}_1 \big)  }{ 2 \i \pi (\la-\mu - \i c) }   &
  \f{ \big(  \bs{e}_2 \, , \, \Omega^{-1}(\la) \cdot \Omega( \mu - \i c) \cdot  \bs{e}_2 \big)  }{  2 \i \pi(\la-\mu  + \i c) }      \ea } \;. 
\label{ecriture matrice M pour GSK}
\enq
Above, 
\beq
\bs{e}_1   \; = \; \left( \ba{c} 1 \\ 0 \ea \right) \qquad and \qquad \bs{e}_2   \; = \; \left( \ba{c}  0 \\ 1 \ea \right)
\enq
stand for the canonical basis on $\R^2$ and given two vectors $\bs{v}, \bs{w} \in \Cx^2$, $( \bs{v}, \bs{w} )= \sul{a=1}{2} \bs{v}_a \bs{w}_a $
represents the canonical bilinear pairing induced by the scalar product on $\R^2$. 
\end{prop}

Theorem \ref{Theorem Cptm Asympt det op int avec shift} then appears as an immediate corollary of this proposition. Indeed, it has been established in \cite{KozKitMailSlaTerRHPapproachtoSuperSineKernel} that, when $x\tend +\infty$,
given any open and relatively compact neighbourhood $O$ of $\intff{-q}{q}$, 
the unique solution $\Omega$ to the Riemann--Hilbert problem associated with the 
operator $\wt{\op{S}}_x$ \eqref{definition noyau integrable GSK}  takes on  $\Cx \setminus \ov{O}$, the form 
\beq
\Omega(\la)  \; =  \; \Om_{\infty}(\la) \cdot \a^{-\sg_3}(\la) \qquad \e{with} \qquad 
\norm{ \Om_{\infty} -I_2}_{L^{\infty}\big( \Cx \setminus \ov{O} \big) } \; = \; \e{o}\big( 1 \big) \;. 
\enq
Here, $\Om_{\infty}$ is some holomorphic matrix in $\Cx \setminus \ov{O}$ and the function $\a$
is as defined by \eqref{ecriture noyau integral U pm et definition fct alpha}. The existence of the $L^{\infty}$ bound on $\Om_{\infty}-I_2$ is 
all the information that is needed  on the matrix $\Om_{\infty}$ so as to get to the result. 

Having this information at hand, it is enough to pick some relatively compact neighbourhood $O$ of $\intff{-q}{q}$ that is small enough 
and take $\Ga(\intff{-q}{q})$  to be a small counterclockwise loop around $\intff{-q}{q}$, lying entirely in 
$\big\{\Cx \setminus \ov{O} \big\} \cap \big\{ | \Im(z) | < \tf{c}{2} \big\}$. 
Then, it is readily checked that the integral operator $ \e{id} \; + \; \op{M}^{(0)}_x$ on $L^2\big( \Ga(\intff{-q}{q}) \big) \oplus L^2\big( \Ga(\intff{-q}{q}) \big)$ characterised by the matrix kernel 

\beq
M^{(0)}_x(\la,\mu) \; = \;  \pa{  \ba{cc}  
  \f{ \a(\la) \cdot \a^{-1}(\mu +  \i c) }{ 2 \i \pi \cdot ( \la-\mu - \i c ) }  &    0   \\
0   &
 \f{ \a^{-1}(\la) \cdot \a(\mu - \i c) }{  2 \i \pi \cdot ( \la-\mu  + \i  c )  }      \ea } 
\label{ecriture approximant M0 pour GSK}
\enq
is a good approximant to $ \e{id}  +  \op{M}_x$. 
The integral kernels $M_x(\la,\mu)$ and $M^{(0)}(\la,\mu)$ are smooth functions on the compact $\Ga(\intff{-q}{q}) \times \Ga(\intff{-q}{q})$. 
Hence, the integral operators $\op{M}_x$ and $\op{M}^{(0)}$ are trace class in virtue of the criterion obtained in \cite{DudleyGonzalesBarriosMetricConditionForOpToBeTraceClass}. 
Their Fredholm determinant and trace is thus well defined. Furthermore, one has
\beq
\e{tr} \Big[ \op{M}_x \; - \; \op{M}^{(0)} \Big] \; = \;  \Oint{ \Ga\big( \intff{-q}{q}\big) }{} \big[  M_x(\la,\la)\, - \,  M_x^{(0)}(\la,\la)   \big] \cdot \dd \la  \;. 
\label{ecriture trace de la difference}
\enq
Also standard estimates for 2-Fredholm determinants (\textit{cf} \cite{GohbergGoldbargKrupnikTracesAndDeterminants,SimonsInfiniteDimensionalDeterminants})
ensure that, for some universal constant $C$, 
\beq
\Big| \det_2\big[ \e{id} + \op{M}^{(0)} \big] \; - \; \det_2 \big[ \e{id} + \op{M}_x \big]   \Big|  \; \leq \; C
\norm{ \op{M}^{(0)} - \op{M}_x }_{HS} \;, 
\enq
with $\norm{ \cdot }_{HS}$ being the Hilbert-Schmidt norm. Both, the Hilbert-Schmidt norm $\norm{ \op{M}^{(0)} - \op{M}_x }_{HS}$ and the integral in the \textit{rhs} of 
\eqref{ecriture trace de la difference} can be controlled in terms of the uniform bounds for $\Om_{\infty} - I_2$ on $\Ga(\intff{-q}{q})$ and thus approach $0$ when $x \tend +\infty$. 
Therefore, 
\beq
 \det \big[ \e{id} + \op{M}_x \big] \; = \;  \det\big[ \e{id} + \op{M}^{(0)}_x \big]  \cdot   \Big\{ 1 \; + \; \e{o} \big( 1  \big)  \Big\} \; ,
\enq
 what, in its turn, entails Theorem  \ref{Theorem Cptm Asympt det op int avec shift} in virtue of Proposition \ref{Proposition factorisation Det W en terme RHP data det M}. \qed

\subsection{Lacunary Toeplitz determinants}

I now give a short description of the results that can be obtained, within the factorisation method, relatively to the large-$N$ asymptotic behaviour of so-called lacunary Toeplitz 
determinants generated by a symbol $f$:
\beq
\det_{N}\Big[ c_{\ell_a-m_b}[f] \Big] \qquad \e{where} \qquad 
c_n[f] \; = \; \Oint{ \Dp{}\mc{D}_{0,1} }{} \f{ f(z)}{ z^{n+1} } \cdot  \f{ \dd z  }{2  \i \pi} \;. 
\label{ecriture detereminant lacunaire general}
\enq
The sequences $\ell_a$, $m_b$ appearing in \eqref{ecriture detereminant lacunaire general} are such that 
\beqa
\ell_a & = & a \quad \e{for} \quad a\in \big\{ 1, \dots, N \big\} \setminus \big\{h_1,\dots, h_n \big\} 
\qquad \e{and} \qquad 
\ell_{h_a} \; = \; p_a \quad a=1,\dots, n \;  \label{definition de la suite ella}\\
m_a & = & a \quad \e{for} \quad a\in \big\{ 1, \dots, N \big\} \setminus \big\{t_1,\dots, t_r \big\} 
\qquad \e{and} \qquad 
m_{t_a} \; = \; k_a \quad a=1,\dots, r \; .  \label{definition de la suite ma}
\eeqa
The integers $h_a \in \intn{1}{N}$ and $p_a \in \mathbb{Z} \setminus \intn{1}{N}$, $a=1,\dots, n$ 
(resp. $t_a \in \intn{1}{N}$ and $k_a \in \mathbb{Z} \setminus \intn{1}{N}$, $a=1,\dots, r$)
are assumed to be pairwise distinct.

  Tracy and Widom \cite{TracyWidomAsymptoticExpansionLacunaryToeplitz} and Bump and Diaconis \cite{BumpDiaconisLacunaryToeplitzThrougSumsSymFctsAndYoungTableaux} were
the first to focus their attention on the large-$N$ asymptotic behaviour of lacunary Toeplitz determinants. 
These authors have obtained, in 2002, two formulae of a very different kind for these large-$N$ asymptotics. 
In both cases, the large-$N$ behaviour of the lacunary Toeplitz determinant
was expressed  in terms of the unperturbed determinant $\det_{N}\Big[ c_{a - b}[f] \Big]$  times 
an extra term. 
The expression for the extra term proposed by Bump and Diaconis was based on characters of the symmetric group 
associated with the partitions $ \la $ and $\mu$ that can be naturally associated 
with the sequences $\ell_a$ and $m_b$. The answer involved the sum over the symmetric groups of $|\la|$
and $|\mu|$ elements. In their turn, Tracy and Widom obtained a determinant representation of the type
\beq
\det_{N}\big[ c_{\ell_a-m_b}[f] \big] \; = \; \det_{N-q}\big[ c_{j-k}[f] \big] \cdot 
\det_{ q }\big[ W_{jk} \big] \cdot  \Big(  1+ \e{o}(1) \Big)    \qquad q\, = \, \max \big\{ t_1,\dots , t_r , h_1,\dots, h_n   \big\} 
\label{formule asymptotique Widom}
\enq
 where $W_{jk} $ was an explicit $q \times q$ sized matrix depending on the symbol $f$ and the numbers
$h_1, \dots, h_n$, $p_1,\dots, p_n$, $t_1,\dots, t_r$ and $k_1,\dots, k_r$.
The two expressions were of a very different nature and it was unclear how to connect the two formulae directly. 
It was only in 2011  that  Dehaye \cite{DehayeProofIdentityBumpDiaconisTracyWidomLacunatyToeplitz}
proved, by a direct method, their equivalence. Some generalisations of lacunary Toeplitz determinants have been studied 
by Lions \cite{LionsToeplitzLacunaires}. 

The main problems of the mentioned asymptotic expansions was that the answer 
depended on the magnitude of the lacunary parameters $p_a, k_b, h_a, t_b$. As soon as these
parameters were also growing with $N$, the form of the answer did not allow for an easy access to the \textit{per} \textit{se} large-$N$
asymptotic behaviour of the lacunary determinant. Indeed, in Bump--Diaconis' case, the number of terms 
that were being summed over was growing as
\newline $\sum( |p_a|+ h_a )  + \sum( |k_a|+t_a)$ whereas in Tracy--Widom's case, the non-trivial determinant part 
involved a matrix of size $\max \big\{ t_1,\dots , t_r , h_1,\dots, h_n   \big\} $. 

I  now describe the expression for the large-$N$ behaviour of lacunary Toeplitz determinants
that I obtained in  [$\bs{A12}$]. Within my approach, 
the extra term in respect to $\det_{N}\Big[ c_{a - b}[f] \Big]$ is given by a determinant of a $(n+r)\times (n+r)$ matrix. The large-size asymptotic expansion 
of the determinant I obtained is thus free from the problems related to the growth in $N$ of the integers parametrising the lacunary lines and columns. 
In fact, I showed in  [$\bs{A12}$] that the setting is enough so as to treat certain cases of lacunary parameters $p_a, k_b, h_a, t_b$ going to infinity. 
The structure of the asymptotics when $r\not=0$ (\textit{i}.\textit{e}. $m_a\not=a$ as defined in \eqref{definition de la suite ma}) is slightly more complex, so that 
I refer to Corollary 2.1 and Theorem 2.1 of [$\bs{A12}$] for the details. Below, I will solely present the form taken by the asymptotics in the case of the line-lacunary Toeplitz determinants 
\begin{theorem}
\label{Theorem Asymptotiques Toeplitz lacunaire a lignes}
Let $f$ be a non-vanishing function on $\Dp{}\mc{D}_{0,1}$ such that $f$ and $\ln f$
are holomorphic on some open neighbourhood of $\Dp{}\mc{D}_{0,1}$. 
Let $\ell_a$ be defined as in \eqref{definition de la suite ella} and $\ga$
be the piecewise analytic function 
\beq
\ga(z) \; = \;   \exp\bigg\{ -  \sul{n \geq 0}{} c_{ n}\big[ \ln f \big] \cdot z^{n}  \bigg\} 
\quad  for\;  z \in \mc{D}_{0,1}  \qquad and \qquad 
			\ga(z) \; = \;	  \exp\bigg\{  \sul{n \geq 1}{} c_{- n}\big[ \ln f \big] \cdot z^{-n}  \bigg\}  
								  \quad  for \; z \in \Cx\setminus \ov{\mc{D}}_{0,1}   \;  . 
\nonumber 
\enq
Then, provided that the matrix $M$ given below is non-singular, the line-lacunary Toeplitz determinant 
$\det_{N}\big[ c_{\ell_a-b}[f] \big]$ admits the representation 
\beq
\det_{N}\big[ c_{\ell_a-b}[f] \big] \; = \; \det_{N}\big[ c_{a-b}[f] \big] \cdot  \det_{n}\big[ M_{ab} \big] \cdot  
\Big(  1+ \e{O}\big( N^{-\infty} \big) \Big) \;, 
\label{Theorem intro ecriture forme DA}
\enq
where the $n\times n$ matrix $M$ reads 
\bem
M_{ab} \; = \; - \bs{1}_{\mathbb{N}}(p_a) 
\Oint{  \Dp{}\mc{D}_{0,\eta_z}    }{} \hspace{-2mm} \f{ \dd z }{ 2 \i \pi} \cdot 
\Oint{  \Dp{}\mc{D}_{0,\eta_s}    }{} \hspace{-2mm} \f{ \dd s }{ 2 \i \pi} 
\cdot \f{ \ga(z) }{ \ga(s) } \cdot \f{ s^{N-p_a} \cdot z^{h_b-N-1}  }{ z- s }    \\
\; + \; \bs{1}_{\mathbb{N}}(-p_a) \Oint{  \Dp{}\mc{D}_{0, \eta_z^{-1} }    }{} \hspace{-2mm} \f{ \dd z }{ 2 \i \pi} \cdot 
\Oint{  \Dp{}\mc{D}_{ 0, \eta_s^{-1} }    }{} \hspace{-2mm} \f{ \dd s }{ 2 \i \pi} 
\cdot \f{ \ga(s) }{ \ga(z) } \cdot \f{ s^{-p_a} \cdot z^{h_b-1}  }{ z- s }  \;, 
\label{ecriture formule asymptotique matrice M cas lacunaire a ligne}
\end{multline}
 where $1>\eta_{z} > \eta_{s} >0 $ and $\bs{1}_{\mathbb{N}}$ stands for the indicator function of $\mathbb{N}$\;. 

\end{theorem}

The theorem above allows one to obtain the large $N$-asymptotic expansion of the line-lacunary Toeplitz determinants
independently of the magnitude (in respect to $N$) of the lacunary parameters $\{h_a\}$ and $\{ p_a \}$. 
Indeed, since the size of the matrix $M$ does not depend on the integers $\{h_a\}$ or $\{p_a\}$, the problem boils down to a \textit{classical}
asymptotic analysis of one-dimensional integrals defining its entries.

The idea of the proof of Theorem \ref{Theorem Asymptotiques Toeplitz lacunaire a lignes} is the following. One represents the Toeplitz determinant in terms of the Fredholm determinant of the operator $\e{id}\,+\, \op{V}_0 \, + \, \op{V}_1$
on $L^2\big( \Dp{} \mc{D}_1 \big)$, where 
\beq
V_0\big( z, s \big)  \; = \; \big( f(z) - 1 \big) \cdot 
\f{  z^{\f{N}{2}}\cdot s^{-\f{N}{2}} \; - \; z^{-\f{N}{2}}\cdot s^{\f{N}{2}}  }
{ 2 \i \pi \big( z - s \big) }
\label{definition noyau integral V0}
\enq
is an integrable kernel and $\op{V}_1$ is a finite rank perturbation. The results follows from a factorisation of the operator $\e{id}\,+\, \op{V}_0 $ out of the determinant
followed by some reductions of the remaining determinant of a finite rank operator. 

More details can be found in [$\bs{A12}$].

\section{The operator valued Riemann--Hilbert problem}

In the present section, I will provide an overview of the steepest descent analysis of the operator valued Riemann-Hilbert problem related with the integrable kernel 
 $W_x(\la,\mu)$ defined in \eqref{definition noyau integrable V dans regime critique}. However, first, I need to provide a few definitions.

\subsection{Some preliminary definitions}

 $\bullet $ The superscript $^{\bs{T}}$ will denote the transposition of vectors, \textit{viz}.
\beq
\e{if} \quad \vec{v} \, = \,  \left( \ba{c} v_1 \\ \vdots \\ v_N \ea \right) \qquad \e{then} \quad 
\vec{v}^{\bs{T}} \, = \,  \left(  v_1 \, \dots \,  v_N  \right) \;. 
\enq

\vspace{2mm} $\bullet$ The space $\mc{M}_p(\Cx)$ of $p\times p$ matrices over $\Cx$ is endowed with the norm 
$\big| \big| M \big| \big|=\max_{a,b}|M_{a,b}|$.

\vspace{2mm} $\bullet$ The space $\mc{M}_p\Big(L^2\big( X , \dd \nu \big)\Big)$ denotes the space of $p\times p$ matrix valued
functions on $X$ whose matrix entries belong to $L^2\big( X , \dd \nu \big)$. This space is endowed with the norm 
\beq
\big| \big| M \big| \big|^2_{\mc{M}_p\Big(L^2\big( X , \dd \nu \big)\Big)} \; = \; \Int{X}{} \e{tr}\big[ M^{\dagger}(x)\cdot M(x) \big] \cdot \dd \nu(x) \quad \e{with} \quad 
\Big(M^{\dagger}\Big)_{ab} \, = \, M_{ba}^{*}
\enq
and $^*$ refers to the complex conjugation of scalars.

\vspace{2mm} $\bullet$ $\e{id}$ refers to the identity operator on $L^2(\R^+,\dd s)$, $I_p\otimes \e{id}$ refers to the matrix integral operator on 
$\oplus_{a=1}^{p}L^2\big(\R^+,\dd s \big)$ which has the identity operator on its diagonal and zero everywhere else.

\vspace{2mm} $\bullet$ Given a vector $\vec{\bs{E}}$ of functions $\bs{E}_a \in L^2(\R^+,\dd s)$
\beq
\vec{\bs{E}}\; = \; \left( \ba{c} \bs{E}_1\\ \vdots\\ \bs{E}_p \ea \right) \quad \e{and} \; \e{a}\; \e{vector} \; \e{of} \; \; 1-\e{forms} \quad 
\vec{\bs{\kappa}} \; = \; \left( \ba{c} \bs{\kappa}_1\\ \vdots\\ \bs{\kappa}_p \ea \right)
\enq
on $L^2(\R^+,\dd s)$, their scalar product refers to the below sum
\beq
\big( \vec{\bs{\kappa}} , \vec{\bs{E}}  \big) \; = \; \sul{a=1}{p} \bs{\kappa}_a[\bs{E}_a]
\enq
in which one evaluates the one-form -appearing to the left- on the function -appearing to the right-. 
Furthermore, the notation $\vec{\bs{E}}\otimes \big(\vec{\bs{\kappa}}\big)^{\bs{T}}$ refers to the matrix operator on 
$\oplus_{a=1}^{p}L^2\big(\R^+,\dd s \big)$ given as 
\beq
\vec{\bs{E}}\otimes \big(\vec{\bs{\kappa}}\big)^{\bs{T}} \; = \; \Big( \bs{E}_{q} \otimes \bs{\kappa}_r \Big)_{q,r=1,\dots,p}
\enq
where $\bs{E}_q\otimes \bs{\kappa}_r$ is the operator on $L^2(\R^+,\dd s)$ acting as
\beq
\big(\bs{E}_q\otimes \bs{\kappa}_r \big)[g] \; = \; \bs{E}_q \times \bs{\kappa}_r[g]     \qquad \e{for} \; \e{any} \; g \in L^2(\R^+,\dd s) \;. 
\enq

\begin{defin}

Let $\wh{\Phi}(\la)$ be an integral operator on $\oplus_{a=1}^{p}L^2\big(\R^+,\dd s \big)$ parametrised
 by an auxiliary variable $\la$. Let $\wh{\Phi}(\la\mid s,s^{\prime} )$ denote its $p\times p$ matrix integral kernel.
Given  $\mc{D}$ an open subset of $\Cx$,  we say that  $\wh{\Phi}(\la)$ is a holomorphic in $\la \in \mc{D}$ integral
operator on $\oplus_{a=1}^{p}L^2\big(\R^+,\dd s \big)$ if \vspace{1mm}
\begin{itemize}
\item[$\bullet$] pointwise in 
$\big(  s,s^{\prime} \big) \in  \big( \R^+ \big)^2$, the $p\times p$ matrix-valued function 
$\la \mapsto \wh{\Phi}(\la\mid s,s^{\prime} )$ is holomorphic in $\mc{D}$ ; \vspace{1mm}
\item[$\bullet$]  pointwise in $\la \in \mc{D}$, $\big(  s,s^{\prime} \big) \mapsto \wh{\Phi}(\la\mid s,s^{\prime} ) \in 
\mc{M}_p\Big(L^2\big( \R^+\times \R^+ , \dd s\otimes \dd s^{\prime} \big) \Big)$. 
\end{itemize}

\end{defin}

I also need to define what is meant by $\pm$ boundary values of a holomorphic integral operator. 
There are two kinds of notions that seem of main interest to operator valued Riemann-Hilbert problems. On the one hand $L^2$
and on the other hand continuous boundary values. 

\begin{defin}

 Let $\mc{D}$ be an open subset of $\Cx$ and $\Sg_{\Phi}$ an oriented smooth curve in $\Cx$. Let $n(\la)$ be the orthogonal to 
 $\Sg_{\Phi}$ at the point $\la \in \Sg_{\Phi}$ and directed towards the $+$ side of $\Sg_{\Phi}$.  \vspace{2mm}
 
\noindent A holomorphic in $\la \in \mc{D}\setminus \Sg_{\Phi}$ integral operator $\wh{\Phi}(\la)$ on $\oplus_{a=1}^{p}L^2\big(\R^+,\dd s \big)$ 
is said to admit $L^2$ $\pm$-boundary values $\wh{\Phi}_{\pm}(\la)$ on $ \Sg_{\Phi}$ if \vspace{1mm}
\begin{itemize} 
\item[$\bullet$]   there exists a matrix valued function $(\la, s, s^{\prime}) \mapsto  \wh{\Phi}_{\pm}(\la\mid s,s^{\prime} )$ belonging to 
$L^2\big( \Sg_{\Phi}\times \R^+\times \R^+ \big)$ and such that 
\beq
\lim_{\eps \tend 0^+}
\big| \big| \wh{\Phi}^{(\pm \eps)} -\wh{\Phi}_{ \pm} \big| \big|_{ \mc{M}_p\Big(L^2\big(  \Sg_{\Phi}\times \R^+\times \R^+ \big)\Big)}  \; = \; 0
\qquad where \qquad \wh{\Phi}^{( \eps)}(\la\mid s, s^{\prime} ) \; = \; \wh{\Phi}(\la+\eps n(\la) \mid s, s^{\prime} ) \;. 
\nonumber
\enq
The operators $\wh{\Phi}_{\pm}(\la)$ are then defined as the integral operators on  $\oplus_{a=1}^{p}L^2\big(\R^+,\dd s \big)$ 
characterised by the matrix integral kernel $\wh{\Phi}_{\pm}(\la\mid s,s^{\prime} )$.  \vspace{2mm}
\end{itemize}

\noindent A holomorphic in $\la \in \mc{D}\setminus \Sg_{\Phi}$ integral operator $\wh{\Phi}(\la)$ on $\oplus_{a=1}^{p}L^2\big(\R^+,\dd s \big)$ 
is said to admit continuous boundary values $\wh{\Phi}_{\pm}(\la)$ on $\Sg_{\Phi}^{\prime} \subset \Sg_{\Phi}$ if  \vspace{1mm}
\begin{itemize}
 \item[$\bullet$]  pointwise in  $\big(  s,s^{\prime} \big) \in  \big( \R^+ \big)^2$ the non-tangential limit
 $\wh{\Phi}(\la\mid s, s^{\prime}) \limit{\la  }{t} \wh{\Phi}_{\pm}(t\mid s,s^{\prime} )$ when $\la$
 approaches $t \in \Sg_{\Phi}^{\prime}$ from the $\pm$ side exists and that 
 the map $t \mapsto \wh{\Phi}_{\pm}(t\mid s,s^{\prime} )$ is continuous on $\Sg_{\Phi}^{\prime}$. 
The operators $\wh{\Phi}_{\pm}(\la)$ are then defined as the integral operators on  $\oplus_{a=1}^{p}L^2\big(\R^+,\dd s \big)$ 
characterised by the matrix integral kernel $\wh{\Phi}_{\pm}(\la\mid s,s^{\prime} )$. 

 \end{itemize}

\end{defin}

Let $\la \mapsto \bs{m}_{k}(\la)$ be the below one parameter $t$ family of functions taking values in the space of functions on $\R^+$:
\beq
\bs{m}_{1}(\la)(s) \; \equiv  \, \bs{m}_1(\la;s)\; = \; \sqrt{c} \,\ex{-\f{c s}{2}} \ex{ \i s t \la}   \qquad \e{and} \qquad
\bs{m}_{2}(\la)(s) \; \equiv  \, \bs{m}_2(\la;s) \; = \; \sqrt{c} \, \ex{-\f{c s}{2}} \ex{- \i s t \la}  \;.
\enq
Let $\la \mapsto \bs{\kappa}_{k}(\la)$ be the below one-parameter $t$ family of functions taking values in the space of one-forms on $L^2(\R^+,\dd s)$:
\beq
\bs{\kappa}_1(\la)[f]\; = \; \sqrt{c} \Int{0}{+\infty} \ex{-\f{c s}{2}} \ex{ -\i s t \la} f(s) \cdot \dd s   \qquad \e{and} \qquad
\bs{\kappa}_2(\la)[f]\; = \; \sqrt{c} \Int{0}{+\infty} \ex{-\f{c s}{2}} \ex{ \i s t \la } f(s) \cdot \dd s    \; . 
\enq
Note that, uniformly in $\la\in\intff{-q}{q}$, the function $s \mapsto \bs{m}_{k}(\la;s)$ belongs to $\big(L^1\cap L^{\infty}\big)(\R^+,\dd s)$. The one-forms and functions introduced above satisfy to 
\beq
\bs{\kappa}_{k}(\la)[\bs{m}_k(\mu)]  \; = \; \f{ \i c \eps_k }{ t(\la-\mu) + \i \eps_k c } \qquad \e{where} \quad  k=1,2 \; \quad 
\e{and} \quad \left\{ \ba{c}   \eps_1 = -1 \\
				    \eps_2 = 1 \ea \right. .
\label{definition epsilon k}
\enq
Now  introduce the one-form on $L^2(\R^+,\dd s)$ valued  vector $\vec{\bs{E}}_L(\mu)$ and the $L^2(\R^+,\dd s)$-valued vector $\vec{\bs{E}}_R(\mu)$:
\beq
\vec{\bs{E}}_L(\mu) \; = \;  F(\mu) \left( \ba{c}   \ex{ - \f{\i x}{2} p(\mu)  } \cdot \bs{\kappa}_{1}(\mu)  \\  
			   -  \ex{  \f{\i x}{2} p(\mu)  }  \cdot \bs{\kappa}_{2}(\mu)  \ea \right)
\qquad \e{and} \qquad 
\vec{\bs{E}}_R(\mu) \; = \; \f{ -1 }{ 2 \i \pi } \left( \ba{c}   \ex{ \f{\i x}{2} p(\mu)  } \cdot \bs{m}_{1}(\mu)  \\ 
								  \ex{ - \f{\i x}{2} p(\mu)  } \cdot \bs{m}_{2}(\mu)   \ea \right) \;. 
\enq
These allow one to construct a $t$-deformation $W_{x;t}(\la,\mu)$ of the integrable integral kernel  \eqref{definition noyau integrable V dans regime critique}
\beq
W_{x;t}(\la,\mu) \; = \; \f{\Big( \vec{\bs{E}}_L(\la), \vec{\bs{E}}_R(\mu) \Big)    }{\la-\mu}  \; = \; 
\f{ \i c F(\la)   }{ 2\i \pi (\la-\mu) } \cdot \bigg\{ \f{ \ex{ \frac{\i x}{2} [p(\la)-p(\mu)]  } }{ t(\la-\mu) + \i c }
\, + \, \f{  \ex{ \frac{\i x}{2} [p(\mu)-p(\la)]  }  }{ t(\la-\mu) - \i c } \bigg\} \; . 
\enq
The reason for studying the $t$-deformation instead of the kernel $W_{x}$ itself is that the parameter $t$
allows one to interpolate between the generalised sine kernel $W_{x;0}(\la,\mu) = \wt{S}_x(\la,\mu)$ and the kernel of
interest $W_{x;1}(\la,\mu) = W_x(\la,\mu)$. For technical reasons, it is more convenient  to interpolate between 
$\wt{\op{S}}_x$ and $\op{W}_x$ by varying the extra parameter $t$ then by means of deforming the apparently "more natural"
parameter $c$. 

The strategy for proving Theorem \ref{Theorem Cptm Asympt det op int avec shift} is classical. 
One recasts $\Dp{t}\ln \det\big[\e{id} \, + \, \op{W}_{x;t}  \big]$ in terms of the solution of the associated Riemann-Hilbert problem, see
Lemma \ref{Lemme derivee partielle en t det fred de I + Wt}. 
The fine bounds on the large-$x$ asymptotic behaviour of the solution to this Riemann--Hilbert problem then allow one to 
integrate $\Dp{t}\ln \det\big[\e{id} \, + \, \op{W}_{x;t}  \big]$  over a path joining $t=0$ to $t=1$  and asymptotically in $x$.

 \subsection{The initial operator-valued Riemann--Hilbert problem}

\noindent The integral  kernel $W_{x;t}(\la,\mu)$ is associated with the Riemann--Hilbert problem for a $2\times 2$  operator-valued 
matrix $\chi(\la) = I_2\otimes \e{id} + \wh{\chi}(\la) $ : \vspace{2mm}
\begin{itemize}
 \item[$\bullet$] $\wh{\chi}(\la)$ is a holomorphic in  $\la \in \Cx \setminus \intff{-q}{q}$ integral operator on 
 $L^{2}\big( \R^+, \dd s \big) \oplus L^{2}\big( \R^+, \dd s \big) $; \vspace{2mm}
 \item[$\bullet$] $\wh{\chi}(\la)$ admits continuous $\pm$-boundary values $\wh{\chi}_{\pm}(\la)$ on $\intoo{-q}{q}$; \vspace{2mm}
\item[$\bullet$] uniformly in $(s,s^{\prime}) \in \R^+\times \R^+$ and for any compact $K$ such that $\pm q \in \e{Int}(K) $, there exist a constant $C_K>0$
such that 
\beq
\big| \big| \wh{\chi}\big( \la \mid s,s^{\prime}\big) \big| \big|\; \leq \; \f{ C_K }{1+|\la|} \cdot \ex{-\f{c}{4} (s+s^{\prime})} \quad \e{on} \quad \Cx \setminus \Big\{ K \cup \intff{-q}{q} \Big\} \;. 
\label{ecriture bornes chi a l'infini}
\enq

\item[$\bullet$] there exists $\la$-independent vectors $\vec{\bs{N}}_{\varsigma}$, $\varsigma \in \{-q,q\}$ whose entries are functions in 
$\big(L^1\cap L^{\infty}\big)(\R^+,\dd s)$  and an integral operator 
$ \wh{\chi}_{\e{reg}}^{(\vsg)}( \la )$ on $L^{2}\big( \R^+, \dd s \big) \oplus L^{2}\big( \R^+, \dd s \big) $ such that 
\beq
\chi( \la)\; = \; I_2\otimes \e{id} \; + \; \ln \big[ w(\la) \big] \cdot \vec{\bs{N}}_{\varsigma}  \otimes \Big(\vec{\bs{E}}_{L}(\varsigma)\Big)^{\bs{T}}
\; + \; \wh{\chi}_{\e{reg}}^{(\vsg)}( \la ) \qquad \e{where} \quad  w(\la) \, = \,  \f{\la-b}{\la-a} \;. 
\label{ecriture coptmt local chi en a ou b} 
\enq
The integral kernel $\wh{\chi}_{\e{reg}}^{(\vsg)}\big( \la \mid s,s^{\prime}\big)$ satisfies the bound
\beq
\big| \big|  \wh{\chi}_{\e{reg}}^{(\vsg)}\big( \la \mid s,s^{\prime}\big)  \big| \big| 
\; \leq  \; C \ex{-\f{c}{4} (s+s^{\prime})} (s+1)(s^{\prime}+1) 
\qquad \e{uniformly} \; \e{in} \; \; \la \in U_{\vsg} \;\; \e{and} \; \;  (s,s^{\prime}) \in \R^+\times \R^+
\label{ecriture borne sur chi reg}
\enq
for some open neighbourhood $U_{\vsg}$  of $ \vsg \in \{-q,q\}$. \vspace{2mm}

\item[$\bullet$] the $\pm$ boundary values satisfy $\chi_+(\la) \cdot G_{\chi}(\la) \; = \; \chi_-(\la)$ where the jump matrix corresponds to the 
matrix operator on $L^{2}\big( \R^+, \dd s \big) \oplus L^{2}\big( \R^+, \dd s \big) $ :
\beq
G_{\chi}(\la) \; = \; \left( \ba{cc} \e{id} - F(\la) \cdot  \bs{m}_1(\la)\otimes \bs{\kappa}_{1}(\la)  
									& F(\la) \, \ex{\i x p(\la)} \cdot \bs{m}_1(\la)\otimes \bs{\kappa}_{2}(\la) \\
 -F(\la) \, \ex{- \i x p(\la)} \cdot \bs{m}_2(\la)\otimes \bs{\kappa}_{1}(\la)  
						  & \e{id} + F(\la) \cdot \bs{m}_2(\la)\otimes \bs{\kappa}_{2}(\la)   \ea \right) \;. 
\enq
\end{itemize}

\begin{prop}

The Riemann--Hilbert problem for $\chi$ admits, at most, a unique solution. 
Furthermore, there exists $\de>0$ and small enough such that for any $t$ such that $|\Im(t)| < \de$
and $\det\big[\e{id} + \e{W}_{x;t} ] \not=0$, this unique solution exists and takes the explicit form
\beq
\chi(\la) \;  =  \; 
I_2\otimes \e{id} - \Int{-q}{q}  \f{   \vec{\bs{F}}_{R}(\mu) \otimes \Big( \vec{\bs{E}}_{L}(\mu) \Big)^{\bs{T}} }{ \mu- \la }  \cdot \dd \mu    \qquad and \qquad
\chi^{-1}(\la) \; =  \; I_2\otimes \e{id} + \Int{-q}{q}   \f{  \vec{\bs{E}}_{R} (\mu)\otimes \Big( \vec{\bs{F}}_{L}(\mu) \Big)^{\bs{T}} }{ \mu - \la }  \cdot \dd \mu 
\label{formules reconstruction chi chi-1 en terms F R et FL}
\enq
where $\vec{\bs{F}}_{R}(\la) $ and $\vec{\bs{F}}_{L}(\la) $ correspond to the solutions to the below linear integral equations
\beq
\vec{\bs{F}}_{R}(\la) \; + \; \Int{-q}{q} W_{x;t}(\mu,\la) \cdot \vec{\bs{F}}_{R}(\mu)  \cdot \dd \mu \; =\;   \vec{\bs{E}}_{R}(\la)
\qquad and \qquad 
\vec{\bs{F}}_{L}(\la)  \; + \; \Int{-q}{q} W_{x;t}(\la,\mu) \cdot \vec{\bs{F}}_{L}(\mu)  \cdot \dd \mu  \; =  \;  \vec{\bs{E}}_{L}(\la) \;. 
\enq
Reciprocally, the solutions $\vec{\bs{F}}_{R/L}(\la)$ can be constructed in terms of $\chi$ as
\beq
\vec{\bs{F}}_{R}(\mu) \; = \;  \chi(\mu) \cdot \vec{\bs{E}}_{R}(\mu) \qquad and \qquad 
\Big( \vec{\bs{F}}_{L}(\mu) \Big)^{\bs{T}} \; = \;  \Big( \vec{\bs{E}}_{L}(\mu) \Big)^{\bs{T}} \cdot \chi^{-1}(\mu)
\qquad with \quad \mu \in \intoo{a}{b} \;. 
\label{reconstruction vecto ops FR et FL}
\enq

\end{prop}

The proof is close in spirit to the matrix Riemann--Hilbert problem case, although some additional care is needed due to the handling of operators. 
I refer for more details to the proof given in [$\bs{A11}$] after the statement of Proposition 2.1.

\subsection{Auxiliary operator-valued  scalar Riemann--Hilbert problems}
\label{SoussectionAuxiliaryScalarRHP}

I now present a way of solving the operator valued scalar Riemann--Hilbert problems that arise when implementing the first step of the 
non-linear steepest descent method. Let
\beq
\tau_{1}(\la) \; = \;  -\f{ F(\la) }{ 1 + F(\la) }  \qquad \e{and} \qquad 
\tau_{2}(\la) \; = \; F(\la) \;. 
\enq
The Riemann--Hilbert problem for $\be_{k}=\e{id}+\wh{\be}_{k}$ with $k=1,2$ reads: \vspace{1mm}
\begin{itemize}
 \item[$\bullet$] $\wh{\be}_k(\la)$ is a holomorphic in $\la \in \Cx \setminus \intff{-q}{q}$ integral operator on $L^{2}\big( \R^+, \dd s \big)$; \vspace{1mm}
 \item[$\bullet$] $\wh{\be}_k(\la)$ admits continuous $\pm$-boundary values $\wh{\be}_{k;\pm}$ on $\intoo{-q}{q}$; \vspace{1mm}
\item[$\bullet$] uniformly in $(s,s^{\prime}) \in \R^+\times \R^+$ and for any compact $K$ such that $\pm q \in \e{Int}(K)$, there exist a constant $C>0$
such that 
\beq
\big| \wh{\be}_k\big( \la \mid s,s^{\prime}\big) \big| \; \leq \; \f{ C }{1+|\la|} \cdot \ex{-\f{c}{4} (s+s^{\prime})} \quad \e{for} \quad \Cx \setminus K \;. 
\label{ecriture bornage a infini noyau beta k}
\enq
 \vspace{1mm}
\item[$\bullet$] There exists a function $\bs{n}_{k;\varsigma} \in \big(L^1\cap L^{\infty}\big) \big(\R^+,\dd s \big)$ 
and a neighbourhood $U_{\vsg}$ of $\varsigma \in \{-q,q\}$ such that for $\la \in U_{\zeta}$
\beq
\wh{\be}_k(\la) \; = \; \big[ w(\la) \big]^{ - \nu_k (\vsg)  } \cdot \bs{n}_{k;\varsigma}\otimes \bs{\kappa}_{k}(\varsigma) 
\;+ \; \wh{\be}_{k;\e{reg}}^{(\vsg)}(\la) \quad \e{with} \quad \nu_{k}(\mu) \; = \; \f{-1}{2\i \pi} \ln \big[ 1+\tau_k(\mu) \big]
\label{defintion nuk}
\enq
where $w(\la)$ is as given in \eqref{ecriture coptmt local chi en a ou b} while, for any $\la \in U_{\vsg}$,
\beq
\big| \wh{\be}_{k;\e{reg}}^{(\vsg)}\big( \la \mid s, s^{\prime} \big) \big| \; \leq \; C \ex{-\f{c}{4}(s+s^{\prime})}(s+1)(s^{\prime}+1) \quad \e{for} \; \e{some} \quad C>0 \;. 
\enq
\item[$\bullet$] the boundary values satisfy $\be_{k;+}(\la) \cdot \Big( \e{id} \, + \, \tau_{k}(\la) \cdot \bs{m}_{k}(\la)\otimes \bs{\kappa}_{k}(\la) \Big) \; = \; \be_{k;-}(\la)$. 
\end{itemize}

\begin{prop}
 
There exists $\de>0$ small enough such that the Riemann--Hilbert problem for $\be_{k}$ admits a unique solution provided that $1+\tau_{k}(\la) \not=0$ on $\intff{-q}{q}$
and $|\Im(t)|<\de$. Furthermore, the solution exists as soon as 
\beq
|\Im(t)|<\de \quad and \quad   \det_{\Ga(\intff{-q}{q}) } \big[ \e{id} \, + \, \mc{U}_{k;t} \big] \; \not= \; 0
\enq
where the integral kernel $U_{k;t}(\la,\mu)$ of the integral operator $\mc{U}_{k;t}$ acting on $L^2\big( \Ga(\intff{-q}{q}) \big)$ reads 
\beq
U_{k;t}(\la,\mu) \; = \;   -  t \f{  \a_{k}(\la) \cdot \a_k^{-1}(\mu + \i \eps_k \tf{c}{t})  }{ 2 \i \pi \cdot \big[ t (\mu-\la) + \i \eps_k c \big]  } \; 
\qquad with \quad \a_{k}(\la)\; = \; \exp\bigg\{  \Int{-q}{q} \f{ \nu_k(\mu)  }{ \mu - \la } \cdot \dd \mu  \bigg\} 
\label{definition noyau integral U k et t}
\enq
$\nu_k$ as in \eqref{defintion nuk} and $\eps_k$ as in \eqref{definition epsilon k}. The solution $\be_k$ can be represented as 
\beq
\be_k(\la) \; = \; \e{id} \, - \, \Int{-q}{q} \f{ \tau_k(\mu) \, \bs{\rho}_k(\mu)\otimes \bs{\kappa}_k(\mu) }{ \mu - \la } \cdot \f{ \dd \mu }{ 2\i \pi } \;. 
\enq
Above $\bs{\rho}_{k}(\la)$ denotes the function $\big(\bs{\rho}_{k}(\la)\big)(s) \, = \, \rho_{k}(\la;s)$  which is
defined as the unique solution to the linear integral equation 
\beq
\Big( \e{id} \, + \, \mc{K}_{k;t} \Big)[\rho_k(*;s)](\la) \; = \; \a_{k;+}(\la) 
\Oint{ \Ga\big( \intff{-q}{q}\big)  }{}  \f{ \sqrt{c} \cdot  \ex{-\f{c s }{2} - \i \eps_k t s \mu} }
{ \a_k(\mu) \cdot ( \mu-\la ) } \cdot \f{ \dd \mu }{ 2 \i \pi } \;. 
\label{eqn integrale pour fct rhok} 
\enq
There $*$ denotes the variable on which the integral operator acts. 
The integral kernel $K_{k;t}(\la,\mu)$ of the integral operator $\mc{K}_{k;t}$ on $L^2\big( \intff{-q}{q} \big)$ driving this linear integral equation reads
\beq
K_{k;t}(\la,\mu) \; = \; -    t \f{ \a_{k;+}(\la) \cdot \a_k^{-1}(\mu + \i \eps_k \tf{c}{t} )  }{ 2 \i \pi \cdot \big( t(\mu-\la) + \i \eps_k c \big)  } \cdot \tau_k(\mu) \;.
\enq

\end{prop}

\subsection{First transformation of the Riemann--Hilbert problem}

The jump matrix $G_{\chi}$ admits a factorisation
\beq
G_{\chi}(\la) \; = \;  \left( \ba{cc} \be_{1;+}^{-1}(\la) & 0  \\ 
						0  &  \be_{2;+}^{-1}(\la) 	\ea \right)
\cdot M_{\ua;+}(\la) \cdot M_{\da;-}(\la) \cdot \left( \ba{cc} \be_{1;-}(\la) & 0  \\ 
						0  &  \be_{2;-}(\la) 	\ea \right)
\enq
in terms of matrices $M_{\ua/\da}$:
\beq
M_{\ua}(\la) \; = \; \left( \ba{cc} \e{id}  &  \bs{P}(\la)  \ex{\i x p(\la)} \\
		      0  & \e{id}  \ea \right)   \qquad  \e{and} \qquad 
M_{\da}(\la) \; = \; \left( \ba{cc} \e{id}  &  0 \\
		      \bs{Q}(\la)  \ex{- \i x p(\la)}  & \e{id}  \ea \right) \;. 
\enq
Their definition builds on the two operators 
\beq
\bs{P}(\la) \; = \;   \f{ F(\la) }{ 1+F(\la)  }  \be_{1}(\la) \cdot \bs{m}_1(\la)\otimes \bs{\kappa}_{2}(\la) \cdot \be_{2}^{-1}(\la)
 \qquad \e{and} \qquad 
\bs{Q}(\la) \; = \;  - \f{ F(\la) }{ 1+F(\la)  }  \be_{2}(\la) \cdot  \bs{m}_2(\la)\otimes \bs{\kappa}_{1}(\la) \cdot \be_{1}^{-1}(\la) \;. 
\enq
Note that the operators $\bs{P}$ and $\bs{Q}$ can be recast as 
\beq
\bs{P}(\la) \; = \; -2 \i \ex{ \i\pi \nu(\la) } \f{ \sin\big[ \pi \nu(\la) \big] }{ \a^2(\la)  } \cdot \bs{O}_{12}(\la) \qquad \e{and} \qquad
\bs{Q}(\la) \; = \; 2 \i \ex{ \i \pi \nu(\la) } \sin\big[ \pi \nu(\la) \big]  \a^2(\la)  \cdot   \bs{O}_{21}(\la)
\label{ecriture represebtation avec sing expl pour P et Q}
\enq
where $ \bs{O}(\la)$ is the  integral operator on $L^2\big(\R^+, \dd s ) \oplus L^2\big(\R^+, \dd s )$ defined as 
\beq
\bs{O}(\la) \; = \; \left( \ba{cc}  \be_{1}(\la) \cdot \bs{m}_1(\la) \otimes   \bs{\kappa}_1(\la) \cdot \be_{1}^{-1}(\la)   & 
					    \a^2(\la)  \be_{1}(\la)\cdot  \bs{m}_1(\la) \otimes   \bs{\kappa}_2(\la) \cdot \be_{2}^{-1}(\la)  \\
		\a^{-2}(\la)  \be_{2}(\la) \cdot  \bs{m}_2(\la) \otimes   \bs{\kappa}_1(\la) \cdot  \be_{1}^{-1}(\la)   &
					\be_{2}(\la) \cdot  \bs{m}_2(\la) \otimes   \bs{\kappa}_2(\la) \cdot  \be_{2}^{-1}(\la)   \ea \right) \;. 
\label{definition operateur regulier O}
\enq
Even though the individual operators appearing in the above matrix elements have cuts, the operator $\bs{O}(\la)$, taken as a whole, 
is regular. More precisely, one has the

\begin{lemme}
\label{Lemme regularite certains op a trace}
There exists an open neighbourhood $V$ of the segment $\intff{-q}{q}$ such that the integral operator $\bs{O}(\la)$ on $L^2\big(\R^+, \dd s ) \oplus L^2\big(\R^+, \dd s )$ 
defined in \eqref{definition operateur regulier O} is holomorphic on $V$. 
\end{lemme}

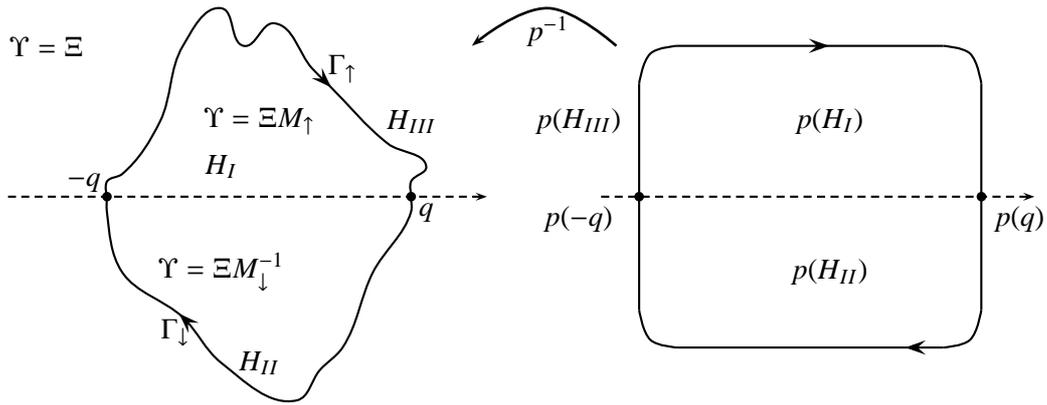
\begin{figure}[h]
\begin{center}

\begin{pspicture}(14.5,7)

\psline[linestyle=dashed, dash=3pt 2pt]{->}(0.2,4)(6.5,4)

\rput(1.2,4.2){$-q$}%
\psdots(1.5,4) \rput(5.7,3.8){$q$} \psdots(5.5,4)
\pscurve(1.5,4)(1.5,3.8)(1.7,3)(2.5,2.4)(3,1.8)(4,1.3)(4.3,1.7)(4.6,2)(5,3)(5.5,3.8)(5.5,4)
(5.5,4.2)(5.7,4.4)(5.2,4.7)(4.4,5.5)(4.2,5.7)(3.7,6.3)(3.3,6)(3,6.5)(2.5,6)(2.2,5)(1.7,4.3)(1.5,4.2)(1.5,4)

\psline[linewidth=2pt]{<-}(2.45,2.45)(2.55,2.35)
\rput(2.4,2.2){$\Gamma_{\da}$}
\psline[linewidth=2pt]{<-}(4.45,5.45)(4.35,5.55)
\rput(4.6,5.7){$\Gamma_{\ua}$}

\rput(0.7,6){$\Upsilon=\Xi$} \rput(3.5,5){$\Upsilon=\Xi M_{\ua}$}
\rput(3,3){$\Upsilon=\Xi M_{\da}^{-1}$}

\pscurve[linewidth=1pt]{->}(8.2,6)(7.3,6.5)(6.3,6)
\rput(7.3,6.2){$p^{-1}$}

\psline[linestyle=dashed, dash=3pt 2pt]{->}(8,4)(13.7,4)
\psdots(8.5,4)(13,4) \rput(7.7,3.7){$p(-q)$}
\rput(13.5,3.7){$p(q)$}

\psline{-}(8.5,2.5)(8.5,5.5)
\pscurve{-}(8.5,5.5)(8.65,5.9)(9,6)

\psline{-}(9,6)(12.5,6)
\pscurve{-}(12.5,6)(12.85,5.9)(13,5.5)

\psline{-}(13,5.5)(13,2.5)
\pscurve{-}(13,2.5)(12.85,2.1)(12.5,2)

\psline{-}(12.5,2)(9,2)
\pscurve{-}(9,2)(8.65,2.1)(8.5,2.5)


\rput(11,5){$p(H_I)$}

\rput(11,3){$p(H_{II})$}
\rput(7.7,5){$p(H_{III})$}


\rput(3,4.4){$H_I$}

\rput(3.5,1.8){$H_{II}$}
\rput(5.5,5){$H_{III}$}

\psline[linewidth=2pt]{<-}(12,2)(12.1,2)
\psline[linewidth=2pt]{<-}(11,6)(10.9,6)

\end{pspicture}
\caption{Contours $\Gamma_{\ua}$ and
$\Gamma_{\da}$ associated with the RHP for $\Upsilon$. The second figure depicts how $p$ maps the contours
$\Ga_{\da}$ and $\Ga_{\ua}$. \label{contour pour le RHP de Y}}
\end{center}
\end{figure}

To implement the first step of the non-linear steepest descent, one defines the matrix $\Ups$ and the contour $\Sg_{\Ups}$ according to Fig.~\ref{contour pour le RHP de Y} 
where 
\beq
\Xi(\la) \; = \; \chi(\la) \cdot \left( \ba{cc}   \be_1^{-1}(\la) &  0  \\ 
						      0   & \be_2^{-1}(\la) \ea \right) \;. 
\enq

\noindent $\Ups(\la) = I_2\otimes \e{id} + \wh{\Ups}(\la) $ solves the Riemann--Hilbert problem: \vspace{2mm}
\begin{itemize}
 \item[$\bullet$] $\wh{\Ups}(\la)$ is a holomorphic in  $\la \in \Cx \setminus \Sg_{\Ups}$ integral  operator on $L^{2}\big( \R^+, \dd s \big) \oplus L^{2}\big( \R^+, \dd s \big) $;\vspace{2mm}
 \item[$\bullet$] $\wh{\Ups}(\la)$ admits continuous $\pm$-boundary values $\wh{\Ups}_{\pm}(\la)$  on $\Sg_{\Ups}\setminus \{-q,q\}$;\vspace{2mm}
 \item[$\bullet$] uniformly in $(s,s^{\prime}) \in \R^+\times \R^+$ and for any compact $K$ such that $\pm q \in \e{Int}(K)$, 
there exist a constant $C>0$ such that 
\beq
\big| \wh{\Ups}\big( \la \mid s,s^{\prime}\big) \big| \; \leq \; \f{ C }{1+|\la|} \cdot \ex{-\f{c}{4} (s+s^{\prime})} \quad \e{for} \quad \Cx \setminus K \;. 
\enq
\item[$\bullet$] there exists an open neighbourhood $U_{\vsg}$ of $\vsg \in \{-q,q\}$,   
vector valued functions $\vec{\bs{N}}_{\varsigma}$ as well as functions $\wt{\bs{n}}_{k;\varsigma}$, $k=1,2$,  
all belonging to $\big(L^{1}\cap L^{\infty}\big)\big( \R^+, \dd s\big)$ such that, for $\la \in U_{\varsigma} \cap H_{III}$ 
one has $\Ups(\la) \; = \; \Ups_{H_{III}}(\la)$ where 
\bem
\Ups_{H_{III}}(\la) \; = \; \Big( I_2 \otimes \e{id} \; + \; \ln\big[ w(\la) \big] \cdot \vec{\bs{N}}_{\varsigma}\otimes \big( \vec{\bs{E}}_{L}(\varsigma) \big)^{\bs{T}} \; + \; 
\wh{R}_{ \Ups}^{(\varsigma)}(\la) \Big)    \\ 
\times \; \left(\ba{cc} 
\e{id} + \big[ w(\la) \big]^{\nu_{1}(\la)}  \wt{\bs{n}}_{1;\varsigma}\otimes \bs{\kappa}_1(\varsigma)  \; + \; r_{1;\Ups}^{(\varsigma)}(\la) & 0  \\
0 & \e{id} + \big[ w(\la) \big]^{\nu_{2}(\la)}  \wt{\bs{n}}_{2;\varsigma}\otimes \bs{\kappa}_1(\varsigma)  \; + \; r_{2;\Ups}^{(\varsigma)}(\la) \ea \right) \;,
\nonumber
\end{multline}
$w(\la)$ is as defined in \eqref{ecriture coptmt local chi en a ou b}, and $\wh{R}_{ \Ups}^{(\varsigma)}(\la)$, resp. $r_{k;\Ups}^{(\varsigma)} $, 
is an integral operator on $L^2(\R^+,\dd s)\oplus L^2(\R^+,\dd s) $, 
resp. $L^2(\R^+,\dd s)$, such that for any $\la \in U_{\vsg}$ 
\beq
\big| \big| \wh{R}_{ \Ups}^{(\varsigma)}(\la\mid s, s^{\prime})  \big| \big| \; \leq  \;  C
\ex{-\f{c}{4}(s+s^{\prime})} (s+1)(s^{\prime}+1)
\qquad \e{resp.}  \qquad 
\big| r_{k;\Ups}^{(\varsigma)}(\la \mid s, s^{\prime}) \big| \; \leq  \; C \ex{-\f{c}{4}(s+s^{\prime})}(s+1)(s^{\prime}+1)  
\enq
for some constant $C>0$. Furthermore, one has that 
\beqa
\Ups(\la) & = &  \Ups_{H_{III}}(\la) \cdot \left( \ba{cc} \e{id} & \big[ w(\la) \big]^{-2\nu(\la) } \bs{P}_{\e{reg}}(\la) \\ 
						  0     & \e{id}  \ea \right)   \quad \e{where} \quad  \la \tend \varsigma \in \{-q,q\}
\quad \e{with} \quad \la \in U_{\varsigma} \cap H_{I}  \nonumber\\ 
\Ups(\la) & = &  \Ups_{H_{III}}(\la) \cdot \left( \ba{cc} \e{id} & 0 \\ 
						  \big[ w(\la) \big]^{2\nu(\la) } \bs{Q}_{\e{reg}}(\la)     & \e{id}  \ea \right)   \qquad \e{where} \quad  \la \tend \varsigma \in \{-q,q\}
\quad \e{with} \quad \la \in U_{\varsigma} \cap \in H_{II} \nonumber 
\eeqa
where $\bs{P}_{\e{reg}}(\la)$ and $\bs{Q}_{\e{reg}}(\la) $ are integral operators on $L^2(\R^+,\dd s)$ such that, 
\beq
\big| \bs{P}_{\e{reg}}(\la\mid s, s^{\prime} ) \big|  \; \leq \;  C \ex{-\f{c}{4}(s+s^{\prime})} (s+1)(s^{\prime}+1) 
\qquad \e{and} \qquad
\big|  \bs{Q}_{\e{reg}}(\la\mid s, s^{\prime} ) \big|  \; \leq  \;  C \ex{-\f{c}{4}(s+s^{\prime})} (s+1)(s^{\prime}+1) 
\label{ecriture proprietes operateurs P et Q reg}
\enq
for some constant $C>0$ and any $\la \in U_{\vsg}$.  \vspace{2mm}

\item[$\bullet$] the boundary values satisfy $\Ups_+(\la) \cdot  G_{\Ups}(\la) \; = \; \Ups_-(\la)$ where the jump matrix reads
\beq
G_{\Ups}(\la) \; = \; M_{\ua}(\la) \quad \e{for} \quad \la \in \Ga_{\ua} \qquad \e{and} \qquad 
G_{\Ups}(\la) \; = \; M_{\da}^{-1}(\la) \quad \e{for} \quad \la \in \Ga_{\da} \; . 
\enq
\end{itemize}

This Riemann--Hilbert problem is uniquely solvable and hence, its solution is in one-to-one correspondence with the one 
to the Riemann--Hilbert problem for $\chi$. The fact that the operators $\bs{P}_{\e{reg}}(\la)$ and $\bs{Q}_{\e{reg}}(\la)$
satisfy \eqref{ecriture proprietes operateurs P et Q reg} follows from \eqref{ecriture represebtation avec sing expl pour P et Q}, 
Lemma \ref{Lemme regularite certains op a trace} as well as from the local 
behaviour of $\a$ around $\la=\vsg \in \{-q,q\}$. Finally, the local behaviour 
of $\Ups$ around $\vsg\in \{-q,q\}$ is inferred from the one of $\chi$, \textit{c}.\textit{f}.
Fig. \ref{contour pour le RHP de Y}.

\subsection{The parametrix around $-q$}
\label{Section parametrices}

\begin{figure}[h]
\begin{center}

\begin{pspicture}(8.2,5)
\pscurve{-}(3,0.8)(3.5,1)(4,2.35)(4,2.4)(4,2.5)(4,2.6)(4,2.65)(4.6,4.4)
\pscircle(4,2.5){2}
\psline[linewidth=2pt]{->}(6,2.5)(6,2.6)

\rput(4.3,2.5){$-q$}
\psdots(4,2.5)

\rput(3.7,3.5){$\Gamma_{\ua}$}
\psline[linewidth=2pt]{->}(4.15,3.2)(4.2,3.32)

\rput(4.4,1.5){$\Gamma_{\da}$}
\psline[linewidth=2pt]{->}(3.89,1.7)(3.94,1.85)

\psline{->}(7.1,2.5)(8,2.5) \rput(8,2.2){$\Re (\la)$}

\psline{->}(7.1,2.5)(7.1,3.4) \rput(6.7,3.2){$\Im (\la)$}

\end{pspicture}

\caption{Contours in the RHP for $\mc{P}_{-q}$.\label{contours for the RHP for P}}
\end{center}
\end{figure}
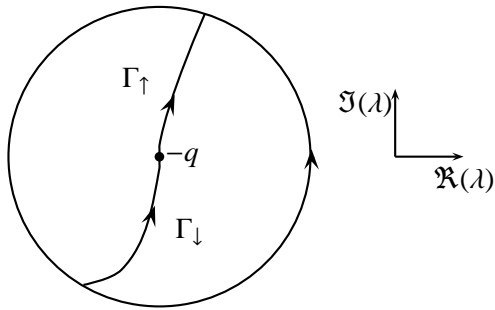

The local parametrix $\mc{P}_{-q}=\e{id} + \wh{\mc{P}}_{-q}$ on a small disk
$\mc{D}_{-q,\delta}\subset U$ of radius $\delta$ and centred at $-q$,
 is an exact solution of the Riemann--Hilbert problem: \vspace{2mm}
\begin{itemize}
\item[$\bullet$] $\wh{\mc{P}}_{-q}(\la)$ is a holomorphic in $\la \in\mc{D}_{-q,\delta}\setminus \big\{ \Gamma_{\ua} \cup \Gamma_{\da} \big\} $
integral operator on $L^{2}\big( \R^+, \dd s \big) \oplus L^{2}\big( \R^+, \dd s \big) $; \vspace{1mm}
\item[$\bullet$]  $\wh{\mc{P}}_{-q}(\la)$ admits continuous $\pm$-boundary values $\big(\wh{\mc{P}}_{-q}\big)_{\pm} (\la)$  on 
$ \big\{ \Gamma_{\ua} \cup \Gamma_{\da} \setminus\{ -q\}  \big\} \cap \mc{D}_{-q,\delta}$; \vspace{1mm}
\item[$\bullet$] $\wh{\mc{P}}_{-q} (\la)$ has the same singular structure as $\Ups$ around $\la = -q $;\vspace{1mm}
\item[$\bullet$] uniformly in $(s,s^{\prime}) \in \R^+\times \R^+$ and $\la \in \partial \mc{D}_{-q,\de} $, one has 
\beq
\big| \big| \wh{\mc{P}}_{-q} \big( \la \mid s,s^{\prime}\big) \big| \big| \; \leq \; \f{ C }{ x^{1-\veps_{-q}}  } \cdot \ex{ - \f{c}{4} (s+s^{\prime}) } \quad \e{for} \; \e{some} 
\quad C>0 \; ; 
\label{ecriture RHP param borne sur bord disque}
\enq
\vspace{1mm}
\item[$\bullet$] $\left\{ \ba{ll}
   \mc{P}_{-q;+}(\la) \cdot M_{\ua}(\la) \; = \; \mc{P}_{-q;-}(\la) \quad
   &\text{for }\la \in \Gamma_{\ua} \cap \mc{D}_{-q,\delta}, \vspace{2mm} \\
   \mc{P}_{-q;+}(\la) \cdot  M_{\da}^{-1}(\la) \; = \; \mc{P}_{-q;-}(\la) \quad
   &\text{for }\la \in \Gamma_{\da} \cap \mc{D}_{-q,\delta}
                \ea \right.$ \; . 
\end{itemize}
Here $\veps_{-q}= 2 \!\underset{ \la \in \partial \mc{D}_{-q,\delta}}{\sup}\!\!\abs{\Re \big(\nu(\la) \big)}< 1$. The canonically
oriented contour $\partial \mc{D}_{-q,\delta}$ is depicted in
Fig.~\ref{contours for the RHP for P}.

\noindent 
\noindent Let $\zeta_{-q}(\la)=x\big( p(\la)-p(-q) \big)$ and set 
\begin{equation}
\mc{P}_{-q}(\la) \; = \; \Psi_{-q}(\la) \cdot L_{-q}(\la) \cdot \big[ \zeta_{-q}(\la) \big]^{-\nu(\la)\sg_3} \cdot \ex{ \f{ \i \pi\nu(\la) }{2} }\; + \;  
\left( \ba{cc}   \e{id}-\bs{O}_{11}(\la) &  0  \\  
			  0    		& \e{id}-\bs{O}_{22}(\la) \ea \right) \; . 
\label{parametrice en -q1}
\end{equation}
Above, $\Psi_{-q}(\la)$ denotes the matrix integral operator
\begin{equation}
\Psi_{-q}(\la)= 
              \begin{pmatrix}
                        \Psi\big(-\nu(\la),1;- \i\zeta_{-q}(\la) \big)  \cdot \bs{O}_{11}(\la)  
				& \i b_{12}(\la) \cdot \Psi\big(1+\nu(\la),1; \i\zeta_{-q}(\la) \big) \cdot \bs{O}_{12}(\la) \\
  -\i b_{21}(\la)\cdot \Psi\big( 1-\nu(\la),1;-\i \zeta_{-q}(\la) \big) \cdot \bs{O}_{21}(\la) & \Psi\big(\nu(\la),1; \i \zeta_{-q}(\la) \big)  \cdot \bs{O}_{22}(\la) 
              \end{pmatrix} \; ,
\label{ecriture parametrice en a}
\end{equation}
with
\beqa
b_{12}(\la)&=& - \i \f{ \sin \big[ \pi \nu(\la) \big] \cdot \Gamma^{2}\big( 1 + \nu(\la) \big) }
				      { \pi \a^2_{\e{reg}}(\la) \cdot \big[ x\big( p(q)-p(\la)\big) \big]^{ 2\nu(\la) }  }   \cdot \ex{ ixp(-q) }  \,  , \\
 b_{21}(\la) & = & -\i \f{ \pi \a^2_{\e{reg}}(\la) \cdot  \big[ x \big( p(q)-p(\la)\big)  \big]^{ 2\nu(\la) }  }
					  { \sin\big[ \pi \nu(\la) \big] \cdot  \Gamma^{2}\big( \nu(\la) \big)  }  \cdot \ex{-\i x p(-q) }  \, .
\eeqa
In \eqref{ecriture parametrice en a}, $\Psi(a,c;z)$ denotes the Tricomi confluent hypergeometric function
(CHF) of the second kind (see equations \eqref{cut-Psi-2}-\eqref{asy-Psi} of the appendix) with the convention of choosing the cut along $\R^-$. 
 Note that this choice for the cut of $\Psi$ implies the use of the principal branch of the logarithm: $-\pi < \e{arg}(z) < \pi$.
Also, the definition of $b_{12}$ and $b_{21}$ makes use of $\a_{\e{reg}}$ the regular part of $\a$ 
\beq
\a_{\e{reg}}(\la) \; = \; \exp\bigg\{ \Int{-q}{q} \nu(\mu) \Big[ \f{1}{\mu-\la} \,-\, \f{ p^{\prime}(\mu) }{ p(\mu)-p(\la) } \Big] \cdot \dd \mu \bigg\}
\enq
which is a holomorphic function on $U$. 
Finally, the expression for the piecewise holomorphic constant matrix $L_{-q}(\la)$ depends on the
region of the complex plane. Namely,
\begin{equation}
L_{-q}(\la) \; = \;  \left\{\ba{cc}  I_2 \otimes \e{id}                           & -\tf{\pi}{2}<\e{arg}\big[p(\la)-p(-q) \big] < \tf{\pi}{2} \vspace{4mm},\\
\left( \ba{cc} \e{id}  &0 \\
                                0& \ex{-2 \i \pi \nu(\la) }\cdot \e{id} 
                                    \ea \right)  & \tf{\pi}{2}<\e{arg}\big[ p(\la)-p(-q) \big] < \pi \vspace{4mm},\\
\left( \ba{cc} \ex{-2\i \pi \nu(\la) } \cdot \e{id} &0 \\
                                     0& \e{id}
                                    \ea \right)  & -\pi<\e{arg}\big[ p(\la) - p(-q) \big] < -\tf{\pi}{2}.
        \ea \right.
\label{matrice L constante}
\end{equation}

The fact that $\mc{P}_{-q}$ proposed in \eqref{parametrice en -q1} is indeed a solution to the Riemann--Hilbert problem around $-q$ 
can be checked by using standard properties of the confluent hypergeometric function along with the relation 
\beq
\bs{O}_{jl}(\la) \cdot \bs{O}_{lk}(\la) \, = \, \bs{O}_{jk}(\la)  \; . 
\enq

The parametrix $\mc{P}_{q}$ around $q$ is constructed in a very similar way. 

\subsection{A determinant identity and asymptotics}

In order to finish with the non-linear steepest descent, it is enough to carry out the last transformation towards 
the integral operator $\Pi(\la) = \e{id}\otimes I_2\, + \, \wh{\Pi}(\la)$ defined according to 
\begin{equation}
\Pi(\la) \; = \; \left\{  \ba{ll}
                    \Upsilon(\la) \cdot \mc{P}^{-1}_{q}(\la) &\text{for } \la\in \mc{D}_{q,\delta} \, ,  \vspace{1mm} \\
                    \Upsilon(\la) \cdot \mc{P}^{-1}_{-q}(\la) &\text{for } \la \in \mc{D}_{-q,\delta} \, , \vspace{1mm} \\
                    \Upsilon(\la) &\text{for } \la \in \mathbb{C}\setminus\big\{ \ov{\mc{D}}_{-q,\delta}\cup \ov{\mc{D}}_{q,\delta} \big\} \, .
            \ea \right.
\end{equation}
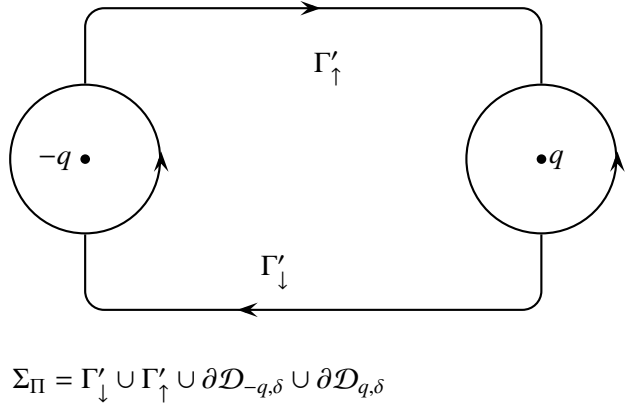
\begin{figure}[h]
\begin{center}

\begin{pspicture}(10,6.5)

\pscircle(2,4){1}
\psdots(2,4)
\rput(1.6,4){$-q$}
\psline[linewidth=2pt]{->}(3,4)(3,4.1)

\pscircle(8,4){1}
\psdots(8,4)
\rput(8.2,4){$q$}
\psline[linewidth=2pt]{->}(9,4)(9,4.1)

\psline[linearc=.25]{-}(2,5)(2,6)(8,6)(8,5)
\psline[linewidth=2pt]{->}(5,6)(5.1,6)
\rput(5.2,5.2){$\Gamma_{\ua}'$}

\psline[linearc=.25]{-}(2,3)(2,2)(8,2)(8,3)
\psline[linewidth=2pt]{<-}(4,2)(4.1,2)
\rput(4.5,2.5){$\Gamma_{\da}'$}



\rput(3.5,1){$\Sg_\Pi=\Ga_{\da}'\cup\Ga_{\ua}'\cup\Dp{}\mc{D}_{-q,\de} \cup \Dp{}\mc{D}_{q,\de}$}
\end{pspicture}

\caption{Contour $\Sg_\Pi$ appearing in the RHP for $\Pi$.\label{Contour for RHP for R}}
\end{center}
\end{figure}

\noindent $\Pi=I_2\otimes \e{id} \; + \; \wh{\Pi} $  satisfies the Riemann--Hilbert problem : \vspace{2mm}

\begin{itemize}
\item[$\bullet$] $ \wh{\Pi}(\la)$ is a holomorphic in $\la \in\Cx \setminus \Sg_{\Pi} $ 
integral operator on $L^{2}\big( \R^+, \dd s \big) \oplus L^{2}\big( \R^+, \dd s \big)$; \vspace{1mm}
\item[$\bullet$] $\wh{\Pi}(\la)$  admits continuous $\pm$-boundary values $\big( \wh{\Pi} \big)_{\pm}(\la)$ on 
$ \Sg_{\Pi} $; \vspace{1mm}
\item[$\bullet$] uniformly in $(s,s^{\prime}) \in \R^+\times \R^+$ and for any compact $K$ such that $\Sg_{\Pi}\subset \e{Int}(K)$,
one has 
\beq
\big| \big| \wh{\Pi}\big( \la \mid s,s^{\prime}\big) \big| \big| \; \leq \; \f{ C }{ x^{1-\veps }  } \cdot \ex{ - \f{c}{4} (s+s^{\prime}) } 
\quad \ba{l} \e{for} \; \e{some} \quad C>0  \; , \; \; \e{any} \; \la \in \Cx \setminus K \vspace{2mm} \\
\e{and} \; \e{for} \; \veps =  2 \!\underset{ \la \in \partial \mc{D}_{\pm q,\delta}}{\sup}\!\!\abs{\Re \big(\nu(\la) \big)}< 1 \ea \; ; 
\label{ecriture bornes uniformes sur noyau integral de Pi}
\enq
\vspace{1mm}
\item[$\bullet$]  $\Pi_+(\la) \cdot G_{\Pi}(\la) \, =  \,   \Pi_-(\la) $ \qquad  where \qquad  $ G_{\Pi}(\la) \, = \, \left\{ \ba{ll}
M_{\ua}(\la) \;\;  \text{for } \; \la \in \Gamma_{\ua}^{\prime}  \; ;  \vspace{1mm} \\ 
M_{\da}^{-1}(\la) \; \;  \text{for }  \; \la \in \Gamma_{\da}^{\prime} \;  ;     \vspace{1mm} \\
   \mc{P}_{\vsg} (\la) \; \e{for} \; \la \in \Dp{}\mc{D}_{ \vsg,\de}  \; \;\e{with} \;  \vsg \in \big\{ -q,q \big\}  \; . 
                \ea \right.$  
\end{itemize}

\vspace{2mm}

The Riemann--Hilbert problem for $\Pi$ is already in good form so as to solve it through a Neumann series expansion of the matrix and operator valued singular integral equation that is 
equivalent to this Riemann--Hilbert problem \cite{BealsCoifmanScatteringInFirstOrderSystemsEquivalenceRHPSingIntEqnMention}. 

\begin{prop}
 
 The solution to the Riemann--Hilbert problem for $\Pi$ exists and is unique, provided that $x$ is large enough
 and $|\Im(t)| <\de$, with $\de>0$ but small enough.

\end{prop}

In order to establish the large-$x$ asymptotic expansion of $\det\big[ \e{id} \, + \,  \op{W}_x\big]$ as given by Theorem \ref{Theorem Cptm Asympt det op int avec shift}, it enough to 
trace back the transformations described above, express $\chi$ in terms of $\Pi$ and insert these expressions in the differential identity provided by the below  lemma, what allows for
its asymptotic evaluation in $x$ 

\begin{lemme}
 \label{Lemme derivee partielle en t det fred de I + Wt}
 The following holds
\beq
\Dp{t}\ln\det\big[\e{id} \, + \,  \op{W}_{x;t} \big]  \; = \; \Oint{\Ga\big( \intff{-q}{q} \big) }{} \hspace{-2mm}
y \cdot \e{tr}\Big[\Dp{y}\chi(y) \cdot \sg^z \cdot \bs{\mf{s}} \cdot  \chi^{-1}(y)  \Big] \cdot \f{\dd y}{2\pi}
\qquad where \quad \sg^z \; = \; \left(\ba{cc} 1 & 0 \\ 0 & -1 \ea \right)
\enq
and $\bs{\mf{s}}$ is the operator of multiplication by $s$, \textit{viz}. $\big(\bs{\mf{s}}\cdot f \big)(s) \, = \, s f(s)$. 
 Note that $\e{tr}$ appearing above refers to the matrix and operator trace. 
 
\end{lemme}

The proof and the details of the asymptotic in $x$ integration of the resulting formula can be found in Section 5 of [$\bs{A11}$].

\section*{Conclusion}

The present chapter reviewed the progress I achieved in respect to understanding Fredholm determinants of so-called $c$-shifted integrable integral operators. 
This progress consists in setting a factorisation method on the one hand and pushing the theory of operator valued Riemann--Hilbert problems and their 
non-linear steepest descent analysis on the other one. However, in order to be able to carry out the asymptotic analysis of the "diffuse" correlation function
such as the emptiness formation probability, one needs to push the technique much farther. In particular, it is necessary to develop a non-linear steepest descent analysis 
of the critical $c$-shifted integrable integral operators such as $\op{W}^{(\e{c})}_x$.

\chapter{The form factor approach to the asymptotic behaviour of correlation functions in massless models}
\label{Chapitre approche des FF aux asymptotiques des correlateurs}

I have described, so far, methods allowing one to study the large-distance and long-time asymptotic behaviour of correlation functions that were based on
the existence of specific series of multiple integral representations for these quantities. Although these representations arise quite naturally in quantum integrable systems, they are not very natural from 
the point of view of the tools used in theoretical physics. In fact, form factor expansions are the most widely used objects
due to their close connection with a model's spectrum. Hence, in order to extend the method of asymptotic analysis to a wider class of models, 
and also so as to gain a deeper physical insight into the structure of the large-distance and/or long-time asymptotic behaviour, it is important to be able to extract the asymptotic
behaviour of a correlation function by carrying out all the analysis \textit{solely} on the level of its form factor expansion. 

In the present chapter, I will develop the form factor based approach to the large-distance and long-time asymptotic behaviour of two and multi-point correlation functions. 
The method is not rigorous but very easy to implement. Furthermore, it works, under reasonable hypothesis on the structure of the model, for a large class of one-dimensional quantum models belonging to the Luttinger liquid universality class. 
On top of providing an efficient way for computing the time and space asymptotic expansion of correlation functions it also allows one to characterise the singular structure
of the response functions in the vicinity of the particle or hole excitation thresholds. 
In fact, the form factor approach that I will describe allows one to go much farther then that: it provides one with a microscopic justification, namely one that is carried out directly on the level of 
a model at finite $L$ where no heurtistic approximation by a field theory has been made,  of an effective description of the large-distance regime of correlation functions
by a free boson $c=1$ conformal field theory. 

The method is based on an astonishing decomposition of the function $z \mapsto (1-z)^{-\nu^2}$ and generalisations thereof, see 
\eqref{definition somme restreinte}-\eqref{ecriture resultat sommation somme restreinte}. 
Within the context of quantum integrable models, in collaboration with Maillet and Slavnov [$\bs{A27}$], I discovered the identity 
in the context of studying the large-distance asymptotic behaviour of the density-density correlation functions in the 
Bose gas at low-temperatures. Its use has been subsequently systematised and improved in the series of works which were carried out in collaboration with 
Kitanine, Maillet, Slavnov and Terras [$\bs{A16},\bs{A17}$] and then with 
Kitanine, Maillet, Slavnov and Terras [$\bs{A15}$]. 
The elucidation of the mechanism which gives rise to an effective description by a conformal field theory was obtained in a joint work with Maillet [$\bs{A14}$]. 
Therefore this last work, although based on formal manipulations, constitutes an important step in understanding the emergence of universality in one-dimensional massless quantum models.

This chapter is organised as follows. In Section \ref{Section Form factor approach to AB of correlation fcts}, I will describe the main ideas of the method on the example of 
the equal-time reduced density matrix in the non-linear Schr\"{o}dinger model. 
I will also describe the various other types of expansions that can be obtained by the method. 
In Section \ref{Section Universal microscopic model},  I will provide a general microscopic setting, generalising the one provided by the Bethe Ansatz, which allows one to 
handle the large-distance or time asymptotic behaviour of correlation function in any model where such a behaviour holds. 
I will then explain, within such a microscopic setting, the mechanism at the root of an effective description by a conformal field theory.

\section{A form factor approach to the asymptotics}
\label{Section Form factor approach to AB of correlation fcts}

 \subsection{Large-distance behaviour of the reduced density matrix}

Recall the form factor expansion of the zero-temperature reduced density matrix 
\beq
 \rho_N(x,0) \; = \; \f{\bra{ \psi\big( \paa{\la_a}_1^N \big) } \Phi(x,0) \Phi^{\dagger}(0,0)  \ket{ \psi\big( \paa{\la_a}_1^N \big) } }{ \norm{ \psi\big( \paa{\la_a}_1^{N} \big) }^2  }\; = \; 
\sul{ \substack{\ell_1 < \dots < \ell_{N+1} \\ \ell_a \in \mathbb{Z} } }{}
\ex{\i x \mc{P}_{\e{ex}}} \cdot \Big| \mc{F}_{\bs{\ell}_{N+1}}\big( \Phi^{\dagger} \big) \Big|^2 
\enq
where $\bs{\ell}_{N+1}\, = \, \big(\ell_1,\dots,\ell_{N+1}\big)$ and
\beq
\Big| \mc{F}_{\bs{\ell}_{N+1}}\big( \Phi^{\dagger} \big) \Big|^2 \; = \; \f{ \abs{ \bra{ \psi\big( \paa{\mu_{\ell_a}}_1^{N+1} \big) } \Phi^{\dagger}\!\pa{0,0} \ket{ \psi\big( \paa{\la_a}_1^N \big) } }^2 }
{ \norm{ \psi\big( \paa{\mu_{\ell_a}}_1^{N+1} \big) }^2  \cdot \norm{ \psi\big( \paa{\la_a}_1^{N} \big) }^2   }  \;. 
\enq
The excitation momentum $\mc{P}_{\e{ex}}$ is as defined by \eqref{ecriture impulsion relative excitation}. By taking the sum of the logarithmic Bethe equations at zero twist \eqref{definition eqn Bethe log}
for the ground state and those describing  an excited state parametrised in terms of  the particle-hole integers \eqref{definition correpondance entiers ella et particules-trous},
one recasts the excitation momentum as
\beq
\mc{P}_{ \e{ex} } \; = \; \f{2\pi}{L} \sul{a=1}{n} (p_a-h_a)  \;. 
\enq
It was shown in Chapter \ref{Chapitre AB grand volume FF dans modeles integrables} that the  conjugated field operator's form factors involving the excited states
parametrised by the integers $\ell_1,\dots,\ell_{N+1}$ defined as in \eqref{definition correpondance entiers ella et particules-trous} admits the large-$L$ asymptotic behaviour 
\beq
\Big| \mc{F}_{\bs{\ell}_{N+1}}\big( \Phi^{\dagger} \big) \Big|^2  \; = \; \f{1}{ L^{\th( \{ \mu_{p_a} \}; \{ \mu_{h_a} \} )  }  } \cdot \mc{A}\big( \{\mu_{p_a} \} ; \{ \mu_{h_a} \} \mid  \{ p_a \} ; \{ h_a \} \big)
\cdot \bigg( 1+\e{O} \Big( \f{ \ln L }{ L } \Big) \bigg)
\enq
where $\mc{A}\big( \{\mu_{p_a} \}, \{ \mu_{h_a} \} \mid  \{ p_a \} , \{ h_a \} \big)$ is a finite amplitude that depends smoothly on the rapidities of the particles and holes
which are defined in terms of the asymptotic counting function $\mu_{a} = \xi^{-1}\big( \tf{a}{L} \big)$. For a generic excited state, the amplitude has, as well, an explicit
dependence on the particle-hole integers. Disregarding the corrections in the volume, the form factor expansion can be recast as 
\beq
\rho_N(x,0) \; =  \;   \sul{n = 0}{ N+1}
\sul{ \substack{p_1<\dots < p_n \\ p_a \in \mathbb{Z}\setminus \intn{1}{N+1} } }{}
\sul{ \substack{h_1<\dots < h_n \\ h_a  \in \intn{1}{N+1}  } }{}
 \f{ \pl{a=1}{n}\Big\{  \ex{ \f{2\i \pi}{L} (p_a-h_a)  }\Big\}  }{ L^{\th( \{ \mu_{p_a} \}; \{ \mu_{h_a} \} )  }  } \cdot \mc{A}\big( \{\mu_{p_a} \} ; \{ \mu_{h_a} \} \mid  \{ p_a \} ; \{ h_a \} \big)  \;. 
\label{ecriture effective series FF approchee}
\enq
One is interested in extracting the large-$L$ and large-$x$ behaviour of the sum, this in the limit where $\tf{x}{L} \ll 1$. 
I will now argue that, in this limit, the sum localises around the critical $\ell$ class, \textit{c}.\textit{f}.  Subsection \ref{Soussection Excitation energy and momentum}. 
For this purpose, it is useful to make an analogy with an oscillating integral. 
Suppose that one is interested in the large-$x$ behaviour of the integral 
 \begin{equation}\label{int-example}
I_n(x) \; = \; \int\limits_{ \big( \mathbb{R}\setminus [-q;q] \big)^n }\hspace{-5mm} \dd^n\mu_{p_a}  \int\limits_{-q}^{q} \hspace{-2mm} \dd^n\mu_{h_a}\; 
f\big(\{ \mu_{p_a} \} ; \{ \mu_{h_a} \} \big) \pl{a=1}{n}  \ex{ \i x\big[ \upsilon(\mu_{p_a}) - \upsilon(\mu_{h_a})\big] } \; ,
\end{equation}
where $f\big(\{ \mu_{p_a} \};\{ \mu_{h_a} \} \big)$ is a holomorphic function  in a neighbourhood of $\R^{2n}$  with a sufficiently fast decay at $\infty$ 
while $\upsilon(\mu)$ is holomorphic in a neighbourhood of $\R$ and such that $\upsilon^{\prime}_{\mid \R}>0$. 
The large-$x$ behaviour of such integrals can be extracted by deforming the integration contours into the complex plane. Such a deformation leads to the conclusion that 
all parts of the integration contours produce exponentially small corrections with the exception of an immediate vicinity of the endpoints $\pm q$.
As a consequence, the large-$x$ asymptotic analysis reduces to an evaluation of the integral in a small vicinity of the
endpoints. Furthermore, in first approximation, the function $f\big(\{ \mu_{p_a} \};\{ \mu_{h_a} \} \big)$ can be replaced by the value it takes on the endpoints of interest.
In the case where $f\big(\{ \mu_{p_a} \};\{ \mu_{h_a} \} \big)$ has integrable singularities at $\pm q$,
for example 
\beq
f\big( \{ \mu_{p_a} \} ; \{ \mu_{h_a} \} \big) \, = \,  ( q - \mu_{h_1} )^{ \nu_+ } \cdot ( \mu_{h_1} + q )^{ \nu_- } \cdot f_{ \e{reg} } \big(\{ \mu_{p_a} \};\{ \mu_{h_a} \} \big) \; ,
\enq
with 
$f_{ \e{reg} }$ smooth on $\R$, then, when carrying out the asymptotic analysis in the vicinities of $\pm q$, one has to keep the singular factors
$(q\mp\mu_{h_1})^{\nu_\pm}$ as they are, but, in what concerns the leading order, can replace the regular part $f_{ \e{reg} }\big(\{ \mu_{p_a} \};\{ \mu_{h_a} \} \big)$ by 
its value at the endpoints of interest.

By analogy with the case of multiple integrals one may expect that, since the oscillatory factor has no saddle-point, the sum  \eqref{ecriture effective series FF approchee} 
will localise around the boundaries of summation for the integers $p_a$ and $h_a$. This corresponds, in the rapidity picture,  to a localisation on one of the two endpoints of
the Fermi zone. In other words, the sum will localise on the critical excited states that were discussed in Chapter \ref{Chapitre AB grand volume FF dans modeles integrables}, Section \ref{Soussection Excitation energy and momentum}. 
The part of the amplitude $\mc{A}\big( \{\mu_{p_a} \} ; \{ \mu_{h_a} \} \mid  \{ p_a \} ; \{ h_a \} \big) $  smoothly depending  on the rapidities  
plays the role of the regular part $f_{ \e{reg} }$ in the multiple integrals \eqref{int-example}. Thus
replacing $ \{\mu_{p_a} \} $ and  $ \{ \mu_{h_a} \} $ by their values in a given critical $\ell$ class:
\beq
\mc{A}\big( \{\mu_{p_a} \} ; \{ \mu_{h_a} \} \mid  \{ p_a \} ; \{ h_a \} \big) \; \longrightarrow \; 
\mc{A}\big(  \{ q \}^{ n_p^+ } \cup \{ -q \}^{ n_p^- } ; \{ q \}^{ n_h^+ } \cup \{ -q \}^{ n_h^- }\mid  \{ p_{a;\pm} \} ; \{ h_{a;\pm} \} \big) \; \equiv \; 
\mc{A}^{(\ell)}\big( \{ p_{a;\pm} \} ; \{ h_{a;\pm} \}  \big) 
\label{restriqt-q}
\enq
should not change the leading large-$x$ asymptotic behaviour of the form factor series
The remaining coefficient $\mc{A}^{(\ell)}\big( \{ p_{a;\pm} \} ; \{ h_{a;\pm} \} \big)$  depends  explicitly on $\ell$ and on the quantum numbers $\{ p_a \}$ and  $\{ h_a \}$. 
It plays an analogous role to the singular factors   $(q\mp\mu_{h_1})^{\nu_\pm}$ in the  integral \eqref{int-example}
since, for large $L$, it varies quickly  in the vicinity of the Fermi boundaries.
This means that, upon localising the sum to a given $\ell$ critical class, one should sum up  $\mc{A}^{(\ell)}\big( \{ p_{a;\pm} \} ; \{ h_{a;\pm} \} \big)$ over all the excited states 
(namely over all the possible numbers $\{ p_a \}$ ,  $\{ h_a \} $), while keeping $L$ large but finite. The limit $L\tend +\infty$ should only be taken at the very
end of the calculations. 

Note that, one can, on the same basis, replace the critical exponent governing the 
power-law decay of the form factors as 
\beq
\th( \{ \mu_{p_a} \}; \{ \mu_{h_a} \} ) \hookrightarrow \th \big(  \{ q \}^{ n_p^+ } \cup \{ -q \}^{ n_p^- } ; \{ q \}^{ n_h^+ } \cup \{ -q \}^{ n_h^- } \big) \; \equiv \;
\big( \mf{F}^{+}_{\ell} +\ell \big)^2  \; + \;   \big( \mf{F}^{-}_{\ell} +\ell \big)^2 \;. 
\enq

The expressions for the factor $\mc{A}^{(\ell)}\big( \{ p_{a;\pm} \} ; \{ h_{a;\pm} \} \big)$ and for the exponents $ \mf{F}^{\pm}_{\ell} $ have been already given in Corollary \ref{Corollaire cptmt FF sur etat critique ell}, equation 
\eqref{ecriture limite sectel ell critique FF}. Inserting the expression for the excitation momentum \eqref{ecriture ex momentum pour etats ell shiftees}, carrying out
the replacements issuing from a localisation on the Fermi boundaries, and, finally, taking the partial thermodynamic limit based on $\lim_{L\tend +\infty}\big( \tf{N}{L} \big) \, = \,  2\pi p_F $, with $p_F$ being 
the Fermi momentum expressed in terms of the dressed momentum as  $p_F =p(q)$, one gets that 
\beq
\rho(x,0) \; \sim \; \lim_{L \tend + \infty} \Bigg\{ \sul{\ell \in \mathbb{Z} }{}   \ex{2 \i \ell p_F}   \f{ \big| \mc{F}_{\ell}\big(  \Phi^{\dagger} \big) \big|^2  }
{ L^{ \big( \mf{F}^{+}_{\ell} +\ell \big)^2 +   \big( \mf{F}^{-}_{\ell} +\ell \big)^2 } } \cdot
\frac{ G^2 \big( 1 + \mf{F}^{+}_{\ell} \big) \cdot G^2\big( 1 - \mf{F}^{-}_{\ell} \big) }
  {  G^2 \big( 1 + \ell + \mf{F}^{+}_{\ell} \big) \cdot G^2\big( 1 - \ell - \mf{F}^{-}_{\ell} \big) } 
\msc{S}_{\ell}\big( \mf{F}^{+}_{\ell} \mid  \ex{ 2\i \pi \f{ x }{ L } }  \big) 
\cdot \msc{S}_{-\ell}\big( -\mf{F}^{-}_{\ell} \mid  \ex{ - 2\i \pi \f{ x }{ L } }  \big)   \Bigg\}  \; . 
\label{ecriture densite reduite limite thermo ac somme restreintes}
\enq
The coefficients $\mf{F}^{\pm}_{\ell}$ are as defined in \eqref{definition fct shift sur les bords}, 
the properly normalised in the volume thermodynamic limit of form factors $\big| \mc{F}_{\ell}\big(  \Phi^{\dagger} \big) \big|^2 $ appears in \eqref{ecriture limite sectel ell critique FF}
while $\msc{S}_{\ell}\big( \nu \mid  z   \big)$ stands for the infinite sum
\bem
\msc{S}_{\ell}\big( \nu \mid  z  \big) \; = \; 
\sul{ n_p - n_h = \ell }{\infty}  \sul{ \substack{ p_1<\cdots< p_{n_p} \\ p_a \in \mathbb{N}^*}   }{}
\sul{ \substack{ h_1< \cdots< h_{n_h} \\ h_a \in \mathbb{N}^*}   }{}
\pl{a=1}{n_p} \Big\{ z^{p_a-1} \Big\} \pl{a=1}{n_h} \Big\{ z^{h_a}  \Big\} \cdot \left(\frac{\sin\pi \nu }{\pi} \right)^{ 2 n_h }  \\
\times \frac{\prod\limits_{a>b}^{n_p}(p_a-p_b)^2\prod\limits_{a>b}^{n_h}(h_a-h_b)^2}
{\prod\limits_{a=1}^{n_p}\prod\limits_{b=1}^{n_h}(p_a+h_b-1)^2}
\pl{a=1}{n_p}\bigg\{ \frac{ \Gamma^2(p_a+\nu)}{ \Gamma^2(p_a) } \bigg\} 
\pl{a=1}{n_h} \bigg\{ \frac{ \Gamma^2(h_a-\nu) }{ \Gamma^2(h_a) } \bigg\} \; .
\label{definition somme restreinte}
\end{multline}
The multiple sum is convergent inside of the unit disk $|z|<1$, this uniformly in $\nu$ belonging to compact subsets of $\Cx$. Hence, the functions $\msc{S}_{\ell}$ present in \eqref{ecriture densite reduite limite thermo ac somme restreintes}
should be understood in the sense of boundary values\symbolfootnote[2]{This is, in fact, consistent with the definition of correlation functions as boundary values of the time-dependent functions
when the time $t\in \mathbb{H}^+$ approaches $t\in \R$.}. 
Although, at first sight, it looks that there is little hope of recasting $\msc{S}_{\ell}\big( \nu \mid  z  \big)$ into a simpler form, 
quite surprisingly, it can actually be computed in closed form. When $\ell=0$, the multiple summation arose for the first time in the context of studying certain probability measures on Young diagrams
\cite{KerovOlshanskiVershikFirstOccurenceAbstracxtLevelMagicFormula,OlshanskiPointProcessAndInfiniteSymmetricGroup} and was computed in closed form. 
In the context of quantum integrable models, the multiple sum $\msc{S}_{\ell}\big( \nu \mid  z  \big)$ appeared, for the first time, in the paper [$\bs{A27}$]
where I studied, with Maillet and Slavnov, the low-temperature limit of the large-distance asymptotic expansion of the density-density correlation function in the non-linear Schr\"{o}dinger model. 
At the time, we were unaware of the results of \cite{KerovOlshanskiVershikFirstOccurenceAbstracxtLevelMagicFormula,OlshanskiPointProcessAndInfiniteSymmetricGroup} 
and found an independent proof thereof in [$\bs{A17}$]. The proof allowed us to deal with the case of $\nu\in \mathbb{Z}$ and general $\ell$. Also, in that paper, we argued the closed form for 
$\msc{S}_{\ell}\big( \nu \mid  z  \big)$, for any $\nu$,  on the basis of a determinant identity due to Widom \cite{WidomAsymptoticsOfBlockToeplitz}. 
However, the argument relied on certain formal manipulations and  was thus not rigorous. In fact, the closed expression for $\msc{S}_{ 0 }\big( \nu \mid  z  \big)$
can also be established as a summation formula for certain Schur functions, see \textit{e}.\textit{g}. \cite{AlexandrovZaborodinTechniquesOfFreeFermions}.  The result is, in fact, 
particularly simple, namely
\beq
\msc{S}_{\ell}\big( \nu \mid  z  \big) \; = \;  \f{ z^{ \ell \f{ \ell - 1 }{ 2 } } }{ (1-z)^{(\nu+\ell)^2} } \frac{ G^2(1+\ell+\nu) }{ G^2(1+\nu) }  \, .
\label{ecriture resultat sommation somme restreinte}
\enq
The closed form expression for $\msc{S}_{\ell}$ then recasts the reduced density matrix as 
\beq
\rho(x,0) \; \sim \;  \sul{\ell \in \mathbb{Z} }{}   \ex{2 \i \ell p_F}    \big| \mc{F}_{\ell}\big(  \Phi^{\dagger} \big) \big|^2  \cdot
\lim_{L \tend + \infty}  \Bigg\{ \bigg( \f{1}{ L \big(1-  \ex{ 2\i \pi \f{ x }{ L } } \big) }  \bigg)^{   \big( \mf{F}^{+}_{\ell} +\ell \big)^2  }
\cdot \bigg( \f{1}{ L \big(1-  \ex{ -2\i \pi \f{ x }{ L } } \big) }  \bigg)^{   \big( \mf{F}^{-}_{\ell} +\ell \big)^2  } \bigg\}    \; . 
\label{ecriture densite reduite limite thermo ac somme restreintes}
\enq
The main point of the above resummed formula is that one can take readily the large-$L$ limit. 
This then leads to the large-$x$ asymptotic expansion for the reduced density matrix
\beq
\rho(x,0) \; \sim \;  \sul{\ell \in \mathbb{Z} }{}   \ex{2 \i \ell p_F}   \f{  \big| \mc{F}_{\ell}\big(  \Phi^{\dagger} \big) \big|^2  }
{ \big(-2\i \pi x \big)^{   \big( \mf{F}^{+}_{\ell} +\ell \big)^2  } \cdot \big( 2\i \pi x \big)^{   \big( \mf{F}^{-}_{\ell} +\ell \big)^2  } }
\cdot \Big( 1+\e{o}(1) \Big) \; . 
\label{ecriture densite reduite limite thermo ac somme restreintes}
\enq
The above series provides one with the leading large-$x$ asymptotic behaviour of each oscillating harmonics that is present in the large-$x$
asymptotic expansion of the reduced density matrix. Even though this expansion has been obtained on a quite heuristic basis, 
the simplicity of the intermediate reasonings and calculations is more than appealing. Note also that \textit{per se}, the method
only allows one to argue that the remainders associated with each oscillating harmonic are a $\e{o}(1)$. However, on the basis of the results presented in Chapter
\ref{Chapitre AB grand tps et distance via series de Natte}, one can affirm that these are, in fact, a $\e{O}\big( \ln x / x \big)$. 

On top of its simplicity, the method of asymptotic analysis of the form factor expansion can be generalised so as to encompass the 
case of the large-distance and long-time asymptotic behaviour of two-point functions, extract the critical behaviour of the spectral functions on the 
particle-hole excitation tresholds or, even, treat the case of the large-distance asymptotic behaviour of multi-point correlations.  
 I shall describe the results one obtains for the first two generalisations on the example of the time-dependent reduced density matrix
 and its Fourier transform. 
To finish the discussion, I stress that even though I have only dealt with the case of the reduced density matrix, the method is very general and applies 
to many other correlation functions.

\subsubsection{The large-distance and long-time asymptotic behaviour}

The long-distance and large-time asymptotic behaviour  corresponds to sending $(x,t) \tend \infty$ with a 
prescribed value for the ratio $x/t$. The asymptotic expansions given below is expected to be uniform in the whole region $\abs{\tf{x}{t}} \not= v_F$.
As already discussed in Chapter \ref{Chapitre AB grand tps et distance via series de Natte}, there will be essentially\symbolfootnote[2]{One also expects the emergence of a third region enjoying its own regime 
and corresponding to $\abs{\tf{x}{t}}=v_F$. The extraction of the associated asymptotic behaviour goes, however, beyond the applicability  
of the method.}  two regions for these asymptotics: the space-like region $\abs{\tf{x}{t}} > v_F$,
and the time-like region $\abs{\tf{x}{t}} < v_F$. I remind that once that the value of the ratio $\tf{x}{t}$ becomes fixed, 
by hypothesis, there exists a unique saddle-point $\la_0$ of the plane-wave combination 
\beq
u(\la) \; = \; p(\la) - t \tf{ \veps(\la) }{ x } \;. 
\enq
The  space-like regime corresponds to  $\abs{\la_0}>q$ and the time-like regime to $\abs{\la_0}<q$ . 
 
In order to describe the asymptotic expansion, it appears convenient to use $\bs{\eta}$ already introduced in \eqref{definition parametre eta space et time like}
\beq
\bs{\eta}=1 \quad \e{in} \, \e{the} \, \e{space-like} \, \e{regime} \qquad \bs{\eta}=-1 \quad  \e{in} \, \e{the}\, \e{time-like} \, \e{regime} 
\enq
since then one is able to treat both regimes in a rather uniform way.

I still need to introduce compact notations for the critical exponents associated with the right and left Fermi boundaries. 
\beq
\De_{\ell_+;\ell_-}^{(+)} = \big[ F_{\ell_+;\ell_-}(q) + \ell_{+} + 1 \big]^2
\qquad \e{and} \qquad
\De_{\ell_+;\ell_-}^{(-)} = \big[ F_{\ell_+;\ell_-}(-q) - \ell_{-}  \big]^2 \;, 
\label{definition exposants critiques asympt x t}
\enq
where the shift function associated with these excitations reads 
\beq
F_{\ell_+;\ell_-}(\om) =  -\, \Big( \f{Z\pa{\om}}{2} + \phi\pa{\om,q} \Big)  - \ell_+ \phi\pa{\om,q} - \ell_- \phi\pa{\om,-q} + 
 \pa{\ell_+ + \ell_-} \phi\pa{\om,\la_0} \;.
\label{definition fction shift pour asympt x t}
\enq

The asymptotic behaviour of the reduced density matrix takes the form 
\bem
\rho(x,t)  \; \sim  \;   \sul{ \substack{ \ell_+; \ell_- \in \mathbb{Z} \\ \bs{\eta}\pa{\ell_+ + \ell_-} \leq 0 } }{} 
 \big|\mc{F}_{\ell_+,\ell_-}\big(\Phi^{\dagger} \big) \big|^2  \cdot \f{ ( 2\pi )^{\f{\abs{\ell_+ + \ell_-} }{2} } } {G\pa{1+\abs{\ell_+ + \ell_-}} } 
\cdot \bigg( \f{ \big( p^{\prime}(\la_0) \big)^2  }{  - 2 \i \pi  \bs{\eta}  \big[ x p^{\prime\prime}\pa{\la_0} - t \veps^{\prime\prime}\pa{\la_0} \big]   } \bigg)^{ \f{ \abs{\ell_-+\ell_+}^2}{2} } \\
\times  \f{ \ex{ \i x \pa{\ell_+-\ell_-}p_F} \ex{-\i x \pa{\ell_+ + \ell_-}u\pa{\la_0}}      }
{ \pac{- 2\i \pi ( x - v_F t )   }^{ \De_{\ell_+;\ell_-}^{(+)} } \cdot  \pac{ 2\i \pi  (x + v_F t )     }^{ \De_{\ell_+;\ell_-}^{(-)} } } \cdot  \Big( 1+\e{o}(1) \Big)\;.
\label{equation resultat pple Phi dagger Phi}
\end{multline}

Similarly to the long-distance regime, the amplitudes $ \big|\mc{F}_{\ell_+,\ell_-}\big(\Phi^{\dagger} \big) \big|^2 $
correspond to the thermodynamic limit of properly normalised in $L$ form factors of the conjugated field operator. 
More precisely, let $\paa{\mu_{\ell_a}}_{1}^{N +1 }$ be an excited state such that \vspace{1mm}
\begin{itemize}

\item[$\bullet$] for $\ell_+ > 0$ (resp. $\ell_+<0$) there are $ \abs{\ell_+}$ particles (resp. holes)  on the right Fermi boundary. 
The integers characterising their position are given by $p_a=N+1+a$, (resp. $h_a=N+2 -a $) 
with  $a=1,\dots, \abs{\ell_+}$;\vspace{1mm}

\item[$\bullet$] for $\ell_- > 0$ (resp. $\ell_-<0$) there are $ \abs{\ell_-}$ particles (resp. holes) on the left Fermi boundary. 
The integers characterising their position are given by $p_a=1-a$ (resp. $h_a= a $) with  $a=1,\dots, \abs{\ell_-}$; \vspace{1mm}

\item[$\bullet$] for $\ell_+ +\ell_-<0$ (resp. $\ell_+ +\ell_->0$) there are $\abs{\ell_+ + \ell_-}$ particles (resp. holes) 
whose rapidity goes to $\la_0$ in the thermodynamic limit. Their associated integers are given by $p_a = m_L + a$
(resp. $h_a=m_L+a$) with $a=1,\dots,\abs{\ell_+ + \ell_-}$ and $m_L$ is any sequence such\symbolfootnote[2]{I remind that $\xi$ stands for the thermodynamic limit of the counting function, \textit{c}.\textit{f}. 
\eqref{ecriture limite thermo fction shift}.}
that 
\beq
\lim_{L\tend +\infty} \xi^{-1} \Big( \f{m_L}{L} \Big) = \la_0 \;.
\nonumber
\enq

\end{itemize}
The amplitudes $ \big|\mc{F}_{\ell_+,\ell_-}\big(\Phi^{\dagger} \big) \big|^2 $ is then defined as 
\beq
\big|\mc{F}_{\ell_+,\ell_-}\big(\Phi^{\dagger} \big) \big|^2= \lim_{N,L\tend +\infty} \paa{   
L^{\De_{\ell_+;\ell_-}^{(-)} + \De_{\ell_+;\ell_-}^{(+)} + \abs{\ell_+ + \ell_-}^2 } 
\Bigg| \f{ \bra{ \psi\big( \{\mu_{\ell_a} \}_1^{ N +1 } \big)  }\Phi^{\dagger}\!\pa{0,0} \ket{ \psi\big( \{ \la_a \}_1^N \big)  } }
{\big|\big|  \psi\big( \{\mu_{\ell_a} \}_1^{N+1} \big)  \big|\big|  \cdot \big|\big| \psi\big( \{ \la_a \}_1^N \big) \big|\big|  } \Bigg|^2   } \; . 
\label{definition facteur background pour harmonique}
\enq

I stress that the shift function \eqref{definition fction shift pour asympt x t} entering in the definition of the critical exponents 
on the right/left Fermi boundaries \eqref {definition exposants critiques asympt x t} arising in \eqref{equation resultat pple Phi dagger Phi}
is precisely the one associated with the excited state entering in the definition of the properly normalised form factor  
\eqref{definition facteur background pour harmonique}.

 \subsubsection{The critical behaviour of the spectral function}

 I will now shortly describe the asymptotic behaviour of the zero-temperature spectral function in the vicinity of the particle and hole excitation spectra. I remind that 
this quantity is defined as the space and time Fourier transform of 
the time ordered reduced density matrix and can be presented\symbolfootnote[3]{Due to the form of the large $(x,t)$ asymptotic expansion of the reduced density matrix, the double integral
should be understood in the sense of an oscillatory integral.} as 
\beq
A(k,\om) \; =  \; \f{ \e{sgn}(\om) }{ 2 \pi} \Int{\R^2}{}  \ex{ \i(\om t - k x)} \Big\{ \moy{ \Phi^{\dagger}\pa{-x,-t} \Phi\pa{0,0} } + 
\moy{ \Phi\pa{x,t} \Phi^{\dagger}\pa{0,0} } \Big\} \;  \dd x \, \dd t    \;. 
\label{definition SF}
\enq
In physical terms, the spectral function measures the response of the model's ground state to the addition of an excitation with momentum $k$ and energy $\om$.

By symmetry, it is enough to focus on the behaviour of the spectral function in the first quarter of the $(k,\om)$
plane: $k\geq0\,, \om \geq0 $. The behaviour of the spectral function at generic values of the energy and momentum is 
proper to each model and hard to characterise in any manageable way, even for quantum integrable models. 
For the latter case, however, one can study this observable and its analogues on a numerical basis quite effectively 
\cite{CauxCalabreseDynamicalStructureFactoBoseGas,CauxCalabreseSlavnovSpectralFunctionBoseGas,CauxHagemansMailletDynamicalCorrFunctXXZinFieldPlots,CauxMailletDynamicalCorrFunctXXZinFieldPlots}. 
Yet, as it has been argued in the introduction, this observable 
exhibits a universal behaviour in the vicinity of the particle-hole excitation thresholds. 
 The arguments in favour of universality, at least in the case of the 
Luttinger liquid universality class, have been advanced in \cite{GlazmanImambekovGeneralRelationEdgeExpForGeneralModel,GlazmanImambekovDvPMTCompletTheoryNNLL,GlazmanKamenevKhodasPustilnikNLLLTheoryAndSpectralFunctionsFremionsFirstAnalysis,
GlazmanKamenevKhodasPustilnikDSFfor1DBosons,GlazmanKamenevKhodasPustilnikNLLLTheoryAndSpectralFunctionsFremionsBetterStudy}.

\subsubsection{Behaviour on the hole tresholds}

In this paragraph  $\la_0$ stands to the rapidity of a hole located at a finite distance
from the endpoints  $\pm q$ of the Fermi zone. In other words, there exists a $\de^{\prime}>0$ such that 
$\la_0 \in \intff{-q+\de^{\prime} }{ q  -\de^{\prime}}$. 

Let the frequency and momentum $\pa{\om,k}$ be parametrised as 
\beq
 \om=  - \veps\pa{\la_0} +\de \om  \quad \e{and} \quad k=p_F - p\pa{\la_0} \; . 
\enq
$k$ coincides with the momentum of an excited state consisting of a hole with rapidity $\la_0$ and of a particle on the right edge $q$ of the Fermi zone. 
The frequency $\om$ is however slightly detuned in respect to the energy $-\veps\pa{\la_0}$ carried by such an excited state.

 When $\de \om \tend 0$, the spectral function admits the below behaviour 
\begin{equation}\label{A-hole}
A(k,\om) \; = \;
   \frac{ \bs{1}_{\R^+}(\delta\om)\,   \big|\mc{F}_{1,0}\big(\Phi^{\dagger} \big) \big|^2 }
            {2\pi \cdot \Gamma\Big( \De^{\pa{+}}_{1;0} + \De^{\pa{-}}_{1;0} \Big) \, (v_{_F}-v)^{ \De^{\pa{+}}_{1;0} } \,(v_{_F}+v)^{\De^{\pa{-}}_{1;0}}   }     \;
      \left(\frac{\delta\om}{2\pi}\right)^{ \De^{\pa{+}}_{1;0} + \De^{\pa{-}}_{1;0} -1} \Big( 1+ \e{O}\pa{\de \om \ln (\de \om) }  \Big)  \; . 
\end{equation}
I remind that $\bs{1}_{A}$ stands for the indicator function of the set $A$.  
The asymptotic expansion is obtained on the basis of formal arguments. In particular, the $ \e{O}$ does not indicate a rigorous control on the remainder,
but rather a formal one based on  algebraic manipulations. The presence of logarithms in the remainder is reminiscent of exactly the same effect that gives rise
to $(\ln x) /x$ corrections in the large-distance asymptotic behaviour of correlation functions in quantum integrable models as discussed in 
Chapter \ref{Chapitre AB grand tps et distance via series de Natte}.

\subsubsection{Behaviour on the particle tresholds}

I now discuss the behaviour of the spectral function in the neighbourhood of the particle threshold. 
In this context,  $\la_0$ will correspond to the rapidity of a particle located at a finite distance
from the right endpoint  $q$ of the Fermi zone, namely $\la_0 \in \intfo{q+\de^{\prime} }{ +\infty }$ for some 
$\de^{\prime}>0$.

In the present case, it is convenient to parametrise the frequency and momentum  $(\om,k)$ as 
\beq
 \om \, = \,   \veps(\la_0) +\de \om  \qquad \e{and} \qquad k \, = \,  p(\la_0) - p_F  \;, 
\enq
with $\de \om$ small. In other words, in the present case,  $k$ coincides with the momentum of an excited state built up from 
a hole located on the right Fermi boundary $q$ and a particle having the rapidity $\la_0$. The frequency $\om$ is, again, slightly detuned
from the energy $\veps\pa{\la_0}$ of this excitation. 

In such a situation, the spectral function admits the below behaviour in the $\de \om \tend 0$ limit 
\bem
A(k,\om) \; = \; \Gamma\Big( 1\, - \, \De^{\pa{+}}_{-1;0} \,-\,  \De^{\pa{-}}_{-1;0} \Big) \cdot 
   \frac{ \bs{1}_{\R^+}(\delta\om)\, \sin\big[ \pi \De^{(-)}_{-1;0} \big] \, + \,    \bs{1}_{\R^+}(-\delta\om)\, \sin\big[ \pi \De^{(+)}_{-1;0} \big]  }
            {2\pi^2 \cdot  \, (v_{_F}-v)^{ \De^{(+)}_{-1;0} } \,(v_{_F}+v)^{\De^{(-)}_{-1;0}}   }  \\
\times \big|\mc{F}_{-1,0}\big(\Phi^{\dagger} \big) \big|^2  \cdot 
      \left| \frac{\delta\om}{2\pi} \right|^{ \De^{\pa{+}}_{-1;0} + \De^{\pa{-}}_{-1;0} -1} \Big( 1+ \e{O}\pa{\de \om \ln (\de \om) }  \Big)  \; . 
\label{A-particle}
\end{multline}
Here, again, one expects on the basis of algebraic handlings that the corrections to the main asymptotics should go like $\de \om \cdot \ln (\de \om)$. 

\vspace{2mm}

The critical behaviour of the so-called density structure factor, which corresponds to the space and time Fourier transforms of the 
density-density correlation function, can be treated as well within the method, see Subsection 4.2 of  [$\bs{A16}$]. 
I refer to Section 4 of  [$\bs{A16}$] where arguments in favour of the critical behaviour 
\eqref{A-hole}-\eqref{A-particle} have been developed.

For both Fourier transforms, the obtained results confirm the non-linear Luttinger liquid-based predictions 
for the edge exponents \cite{GlazmanImambekovComputationEdgeExpExact1DBose,GlazmanImambekovDvPMTCompletTheoryNNLL,GlazmanKamenevKhodasPustilnikDSFfor1DBosons}
and for the amplitudes \cite{CauxGlazmanImambekovShashiAsymptoticsStaticDynamicTwoPtFct1DBoseGas}.

 \section{A universal microscopic model}
\label{Section Universal microscopic model}

  In the present section, I will introduce a general setting, clearly mimicking the one present for quantum integrable models. In this description, 
 I follow very closely the presentation given in [$\bs{A14}$]. This setting concerns certain general properties of the spectrum, 
of the states and of the form factors of a given one-dimensional quantum Hamiltonian. The matter is that if this setting holds, then one can argue how the $c=1$ free boson field 
theory arises as an effective description of the long-distance regime of such a model's correlation functions. The derivation is not rigorous - there is absolutely 
no control of the remainders and on the exchanges of the summation symbols, just as in the method discussed in the previous section of this chapter- but provides 
a quite nice physical picture of the manifestation of universality on the level of the large-distance asymptotics. 
The setting I describe clearly holds for quantum integrable models in one-dimension, see \textit{e}.\textit{g}. \cite{BogoliubiovIzerginKorepinBookCorrFctAndABA} 
in what concerns the spectrum part and the papers  [$\bs{A5},\bs{A6}$]  in what concerns 
the form factor part. The validity of the picture has been also established, on a formal perturbative level, in \cite{CauxGlazmanImambekovShasiNonUniversalPrefactorsFromFormFactors}. 
This setting should, however, be common to all models belonging to the universality class of the Luttinger model, although in my present understanding of the situation, it is not clear to me how to argue, without even 
mentioning to prove, the validity of the picture.

\subsection{General hypothesis on the spectrum}

Below  $\op{H}_L$ corresponds to a one-dimensional quantum Hamiltonian representing a physical system in \textit{finite} volume $L$. 
Just as in the introduction, the volume $L$ represents the number of sites in the case of a lattice model or the overall volume of the 
occupied space in the case of a model already formulated in the continuum. 

I assume that the eigenstates of the Hamiltonian can be organised into sectors with a fixed pseudo-particle number $N^{\prime}$. 
In practical situations, the integer $N^{\prime}$ may be related to the total longitudinal spin of a state (in the case of spin chains
with total longitudinal spin conservation) or may simply correspond to the number of bare particles building up the 
many body eigenstate of $\op{H}_L$ (in the case of models enjoying a conservation of the total number of bare particles). 

The ground state of the model $\ket{ \Psi_{g;N} }$ is located within the sector with $N$ pseudo-particles. 
This number does depend on $L$ and is such that, in the thermodynamic limit $L\tend + \infty$, one has 
$\lim_{L\tend +\infty} \big( \tf{N}{L}  \big) \, = \, D>0$. Furthermore, taking the thermodynamic limit restricts the space of states
to the sector corresponding to excitations having a \textit{finite}, when $L\tend+\infty$,  energy relatively to the ground state. 
I assume that the eigenstates having this property are located in sectors with $N^{\prime}$ bare particles where 
$N^{\prime}$ is such that  the difference $s=N^{\prime}-N$ remains finite in the thermodynamic limit. 
Having in mind the spin-chain setting, $s$ will be called the spin of the 
excited state.  

I further assume that the excited states are only built up from particle-hole excitations. 
From the technical point of view, this means that one can label the eigenstates, within each sector built up from $N+s$ quasi-particles,
by a set of integers 
\beq
\mc{I}_{n}^{(s)} \; = \;\Big\{  \{ p_a^{(s)} \}_1^{n} \quad ;  \quad \{ h_a^{(s)} \}_1^{n} \Big\} 
\label{definition ensemble total pour parametriser etats}
\enq
containing two collections of integers which label the so-called particle $\{ p_a^{(s)} \}_1^{n}$ and hole $\{ h_a^{(s)} \}_1^{n}$ excitations. 
In this parametrisation, the integer $n$ may run through $ 0, 1, \dots N + s$ while :
\beq
p_1^{(s)} < \dots < p_n^{(s)} \qquad \e{and}  \qquad h_1^{(s)} < \dots < h_n^{(s)} \qquad \e{with} \qquad 
\left\{ \ba{c}  p_a^{(s)} \in  \intn{-M_{L}^{(1)} }{ M_{L}^{(2)} } \setminus \intn{ 1 }{ N +s }   \vspace{2mm} \\ 
				h_a^{(s)}  \in \intn{ 1 }{ N + s}   \ea \right. \;. 
\enq
The precise values of the integers $M_{L}^{(a)}$ defining the range of the $p_a$'s vary from one model to another. 
Typically for models having no upper bound on their energy, one has $M_{L}^{(a)}=+\infty$ while
for model having an upper bound, $M_{L}^{(1)}, M_{L}^{(2)}$ are both finite but such that $M_{L}^{(a)}-N$, $a=1,2$,
both go to $+\infty$ sufficiently fast with $L$. 

This assumption allows one to denote the eigenstates of the model as $\ket{ \mc{I}_{n}^{(s)} } $. 
Such eigenstate correspond to the so-called microscopic description of the model, namely a complete parametrisation of the space of states
in terms of discrete integers.

From the point of view of the physics of a model, it is however the macroscopic description that is pertinent for describing the thermodynamic limit of the observables in the model. 
In the present setting, the macroscopic description arises by means of the so-called counting function $\wh{\xi}_{ \mc{I}_{n}^{(s)} }$ associated with each given excited state $\ket{ \mc{I}_{n}^{(s)} }$. 
More precisely, the particle, resp. the hole, excitations are described by a set of rapidities $\{\wh{\mu}_{p_a}^{(s)} \}_1^n$, resp. 
 $\{\wh{\mu}_{h_a}^{(s)} \}_1^n$ corresponding to the unique solutions to 
\beq
\wh{\xi}_{ \mc{I}_{n}^{(s)} }\big( \wh{\mu}_{p_a}^{(s)} \big)  \; = \;  \f{ p_a^{(s)} }{L}  \qquad \e{and} \qquad 
\wh{\xi}_{ \mc{I}_{n}^{(s)} }\big( \wh{\mu}_{h_a}^{(s)} \big)  \; = \;  \f{ h_a^{(s)} }{L}\;. 
\label{ecriture equations definissant rap part et trous}
\enq
The counting function does depend, \textit{a priori} on the set of integers labelling the excited state 
$\ket{ \mc{I}_{n}^{(s)} }$. Therefore, the system of equations \eqref{ecriture equations definissant rap part et trous} is, in fact, extremely involved.
 This very characterisation also depends on the hypothesis that the counting function is strictly increasing on $\R$, this irrespectively of the excited state $\ket{ \mc{I}_{n}^{(s)} }$ to which it is attached.

In the following, I will build on the assumption that any counting function $\wh{\xi}_{ \mc{I}_{n}^{(s)} }$ admits, in the $L\tend + \infty$ limit, the asymptotic expansion
\beq
\wh{\xi}_{ \mc{I}_{n}^{(s)} }\big(\om \big)  \; = \; \xi(\om) \; + \; \f{ 1 }{  L   }  \xi_{-1} (\om)
\; - \;  \f{ 1 }{  L   } F_{\mc{R}_{n}^{(s)}} (\om)   \; + \; \e{O} \Big( \f{ 1 }{ L^2 } \Big) \;. 
\label{ecriture definition fonction comptage}
\enq
This asymptotic expansion involves three "macroscopic" functions.  \vspace{1mm}

\begin{itemize}
\item[$\bullet$] The function $\xi$ is the asymptotic counting function.  It is the same for \textit{all} excited states and assumed to be 
 strictly increasing. This function defines a set of "macroscopic" rapidities $\{ \mu_a \}_{a \in \mathbb{Z} } $ 
\beq
\xi( \mu_a ) \; = \;  \f{ a }{ L } \;.  
\label{ecriture equation definition rapidites macro}
\enq
These macroscopic rapidities provide one with the leading order in $L$ approximation of the rapidities $\{\wh{\mu}_{p_a}^{(s)} \}_1^n$ and 
 $\{\wh{\mu}_{h_a}^{(s)} \}_1^n$:
\beq
\wh{\mu}_{p_a}^{(s)} \; \simeq \;  \mu_{ p_a^{(s)} }  \qquad \e{and} \qquad 
\wh{\mu}_{h_a}^{(s)} \; \simeq \;  \mu_{ h_a^{(s)} }  \;. 
\label{ecriture eqn Asymp pour position reseau rapidites}
\enq
\item[$\bullet$]  The function $F_{\mc{R}_{n}^{(s)} } $ stands for the shift function (of the given excited state in respect to the model's ground state). 
It is a function of the spin $s$ and of the set macroscopic rapidities  
\beq
\mc{R}_{n}^{(s)} \; = \; \Big\{ \{ \mu_{ p_a^{(s)} } \}_1^n \; ; \; \{ \mu_{h_a^{(s)}} \}_1^n \Big\} \;. 
\label{introduction variables macroscopiques}
\enq
This function measures the small $\e{O}(L^{-1})$ drift in the position of a rapidity in the Fermi sea under the effect of interactions. \vspace{1mm}
\item[$\bullet$] Finally, the function $\xi_{-1}$ represents the potential $1/L$ corrections to the ground state's counting functions. 
 I stress that this way of decomposing the $1/L$ corrections to the counting function 
is such that the ground state's shift function vanishes, \textit{i}.\textit{e}. $F_{\mc{R}_{0}^{(0)} } =0 $.  \vspace{1mm}
\end{itemize}

In the large-$L$ limit,  the rapidities for the ground state form a dense distribution 
 on the Fermi zone $\intff{-q}{q}$ with a density $\xi^{\prime}$.
Each particle or hole excitation with rapidity $\mu$ carries a momentum $p(\mu)$ and a energy $\veps(\mu)$. 
The dressed energy $\veps$ and dressed momentum $p$ are smooth functions that satisfy to the general properties
\beq
p^{\prime}_{\mid \R } >0 \qquad \; \qquad  \veps_{\intoo{-q}{q}}<0  \qquad \e{and} \qquad \veps_{\R\setminus \intff{-q}{q}}>0 \;. 
\label{ecriture proprietes generales de p et veps}
\enq
The relative momentum and energy of an excited state are expressed in terms of the dressed momentum and energy as 
\beqa
\De\mc{E}\big( \mc{I}_{n}^{(s)}  \big) \; \equiv \; \mc{E}\big( \mc{I}_{n}^{(s)} \big) \; - \;  \mc{E}\big( \mc{I}_{0}^{(0)} \big) 
& = &  \sul{a=1}{n} \Big( \veps\big( \mu_{p_a^{(s)}} \big) \; - \;  \veps\big( \mu_{h_a^{(s)}} \big) \Big)
\; + \; \e{O}\Big( \f{1}{L} \Big)  \label{definition relative ex energy} \\
\De\mc{P}\big( \mc{I}_{n }^{(s)}  \big) \; \equiv \; \mc{P}\big( \mc{I}_{n}^{(s)} \big) \; - \;  \mc{P}\big( \mc{I}_{0}^{(0)} \big) & = & 
\sul{a=1}{n} \Big( p\big( \mu_{p_a^{(s)}} \big) \; - \;  p\big( \mu_{h_a^{(s)}} \big) \Big) 
\; + \; \e{O}\Big( \f{1}{L} \Big)\;.
\label{definition relative ex momentum}
\eeqa
Above, $\mc{I}_{0}^{(0)}=\{ \emptyset ; \emptyset \}$ refers to the set of integers which parametrises the ground state
of the model and $\mc{P}\big( \mc{I}_{n}^{(s)} \big)$ and $\mc{E}\big( \mc{I}_{n}^{(s)} \big)$ are, respectively, the momentum and energy of the state parametrised by the set of integers $\mc{I}_{n}^{(s)}$.  
The superscript $(s)$ in the sets and integers reminds the spin sector $s$ to which the excitation belongs to.




\subsubsection{The operators and their form factors}

The model is assumed to come  with a collection of local operators $\mc{O}_r$. 
These operators are characterised in terms of their form factors and are taken such that 
they connect only states differing by a fixed value of their spin, namely 
%
%
%
%
%
\beq
\mc{F}_{\mc{O}_r} \Big( \mc{I}_{ m }^{(s^{\prime})} \, ; \,   \mc{I}_{ n }^{(s)} \mid x  \Big) \; = \; 
\ex{i x [ \mc{P}( \mc{I}_{ n }^{(s)} ) -  \mc{P}( \mc{I}_{ m }^{(s^{\prime})} ) ] } 
\cdot  \bra{ \mc{I}_{ m }^{(s^{\prime})} } \mc{O}_r(0) \ket{ \mc{I}_{ n }^{(s)} }  \; \not= \; 0
\qquad \e{only}\; \e{if} \quad s \, - \, s^{\prime} \;  = \;  o_r\;. 
\enq
The large-$L$ behaviour of the form factors is parametrised by \textit{both} the set of macroscopic rapidities $\mc{R}_{ n }^{(s+o_r)}, \mc{R}_{ m }^{(s)} $ 
and the set of \textit{discrete} integers $\mc{I}_{n}^{(s+o_r)}, \mc{I}_{ m }^{ (s) }$. 
Namely, for properly normalised states $\ket{ \Psi\big( \mc{I}_{ n }^{(s+o_r)} \big) } $ and 
their duals $\bra{ \Psi\big( \mc{I}_{ m }^{(s)} \big) } $, the large-$L$ behaviour factorises as
\bem
\mc{F}_{\mc{O}_r}\big( \mc{I}_{ m }^{(s)} \, ; \,  \mc{I}_{ n }^{(s+o_r)}  \mid 0 \big) 
\; = \; 
\mc{S}^{( \mc{O}_r)} \Big(   \mc{R}_{ m }^{(s)} \, ; \,   \mc{R}_{n}^{(s+o_r)}     \Big)
     \cdot 	\mc{D}^{(s)}
 \Big(   \mc{R}_{m}^{(s)}\, ; \,  \mc{R}_{n}^{(s+o_r)}     \big| \, 
    \mc{I}_{ m }^{(s)} \, ; \,   \mc{I}_{ n }^{(s+o_r)}   	\Big)	  \cdot 
\bigg(  1+ \e{O}\Big( \f{\ln L }{ L } \Big) \bigg)      		 \;. 
\end{multline}
The smooth part $\mc{S}^{( \mc{O}_r )}$ represents the non-universal part of the model's form factors. It depends smoothly on the rapidities parametrising an excited state. 
As a consequence,  a small $\e{O}(1)$ change in the value of the integers parametrising an excited state will only affect
the value of the smooth part by producing some additional  $1/L$ corrections, exactly as it was discussed on the example of the conjugated field form factor in Section \ref{Subsection Thermo Lim FF}
of Chapter \ref{Chapitre AB grand volume FF dans modeles integrables}.
 The discrete part $\mc{D}^{(s)}$ is, however, universal and only depends on the spin $o_r$ carried by the operator $\mc{O}_r$. Its general explicit expression
corresponds to a slight generalisation of formulae \eqref{AppendixThermolim thermolim D+}, \eqref{AppendixThermoLimD+zero} and  \eqref{definition fonctionelle RNn}. 
However, since it will not play a role in the analysis, I will not develop further on the matter. 
 Below, I will however provide its simplified expression when one restricts to excited states belonging to critical $\ell_s$-classes. 
The main feature of the discrete part is that it depends explicitly on \textit{both}
types of parametrisations attached to an excited state: the macroscopic rapidities $\mc{R}_{ n }^{(s+o_r)}, \mc{R}_{ m }^{(s)} $ and 
 the sets of integers labelling the excited states $\mc{I}_{ n }^{(s+o_r)}, \mc{I}_{ m }^{(s)} $. It thus keeps track of the microscopic details of the different
excited states. It is this component of the large-$L$ behaviour of a form factor that is responsible for the ill-definiteness of 
form factor expansion in the infinite volume limit of massless quantum integrable models.

\subsection{The critical $\ell_s$ class}

A set of integers $\mc{I}^{(s)}_n$ is said to parametrise 
a critical excited state if the associated particle-hole integers $\{ p_a^{(s)} \}_1^{ n }$ and $\{ h_a^{(s)} \}_1^{ n }$  can be represented as
\beq
\big\{ p_a^{(s)}  \big\}_1^{ n } \; = \;  \big\{  N+s + p_{a;+}^{(s)} \big\}_1^{ n_{p;+}^{(s)} } \, \cup  \,
\big\{ 1 -  p_{a;-}^{(s)} \big\}_1^{ n_{p;-}^{(s)} }  \qquad \e{and} \qquad 
 \big\{ h_a^{(s)} \big\}_1^{ n } \; = \;  \big\{1+ N+ s - h_{a;+}^{(s)} \big\}_1^{ n_{h;+}^{(s)} } \, \cup  \, 
\big\{ h_{a;-}^{(s)}  \big\}_1^{ n_{h;-}^{(s)} }   
\label{ecriture decomposition locale part-trou close Fermi zone}
\enq
where the integers $p_{a;\pm}^{(s)}, h_{a;\pm}^{(s)} \in \mathbb{N}^{*}$ are "small" compared to $L$, \textit{i}.\textit{e}. 
\beq
\lim_{L\tend +\infty} \f{ p_{a;\pm}^{(s)} }{ L }  \; = \; \lim_{L\tend +\infty} \f{ h_{a;\pm}^{(s)} }{ L }  \; = \; 0 \;, 
\enq
 and the integers $n_{p/h;\pm}^{(s)}$ satisfy to the constraint
\beq
n_{p;+}^{(s)} \, + \,  n_{p;-}^{(s)} \; = \;  n_{h;+}^{(s)} \, + \,  n_{h;-}^{(s)} \;= \; n \;. 
\enq
Within this setting, one can readily check that the critical excited state described above 
will have $n_{p;+/-}^{(s)}$ particles, resp. $n_{h;+/-}^{(s)}$ holes, on the right/left end of the Fermi zone $\intff{-q}{q}$
associated with the spin $s$ sector. 

The main feature of the critical states is that they all have a vanishing as $\e{O}(1/L)$ relative excitation energy.
One can organise the critical states into $\ell_s$-critical classes corresponding to all states whose relative excitation
momentum, up to $\e{O}(1/L)$ corrections, corresponds to $2\ell_s p_F$, where $p_F=p(q)$ is the so-called Fermi momentum and 
\beq
\ell_s \; = \;  n_{p;+}^{(s)}  \; - \;  n_{h;+}^{(s)} \; = \; n_{h;-}^{(s)}  \; - \;  n_{p;-}^{(s)} \;. 
\label{ecriture lien shift ells et differences part trous sur bords zone Fermi}
\enq
The index $s$ in $\ell_s$ specifies the spin $s$ sector to which the critical state belongs to.

The relative excitation momentum associated with an excited state belonging to the $\ell_s$ critical class can be expanded as:
\bem
\De\mc{P}\big( \mc{I}_{n }^{(s)}  \big) \; =\;  2 \ell_s  p_F 
\; + \; \f{2\pi}{L} \a^{+} \Bigg\{ \sul{a=1}{ n_{p;+}^{(s)} } (p_{a;+}-1) \; + \;  \sul{ a=1 }{ n_{h;+}^{(s)} } h_{a;+}   \Bigg\}
\; - \; \f{2\pi}{L} \a^{-} \Bigg\{ \sul{a=1}{n_{p;-}^{(s)} } (p_{a;-} -1)\; + \;  \sul{a=1}{ n_{h;-}^{(s)} }  h_{a;-}  \Bigg\} \\
\; + \; \f{2\pi}{L} \Big\{ \a^{-} \ell_s \f{(\ell_s+1)}{2} \, - \, \a^{+} \ell_s \f{(\ell_s-1)}{2} \Big\} \; + \; \dots
\label{ecriture ex momentum pour etats ell shiftees}
\end{multline}
where 
\beq
\a^{\pm} \; = \; \f{1}{2\pi} \f{ p^{\prime}(\pm q) }{ \xi^{\prime}(\pm q)  } \;. 
\label{ecriture formule pour alpha pm}
\enq
The terms that have been included in the $\dots $ \vspace{1mm}
\begin{itemize}
 \item[$\bullet$]  are of the order of $\e{O}(1/L)$  \textit{but} do \textit{not} depend on the integers $p_{a;\pm}$ and $h_{a;\pm}$ nor on $n_{p/h;\pm}$ (they can, nonetheless depend on $\ell_s$); \vspace{1mm}
\item[$\bullet$]  depend on these integers \textit{but} are of the order of $\e{O}(1/L^2)$. \vspace{1mm}
\end{itemize}
According to the description of what the $\dots$ are, the second line of  \eqref{ecriture ex momentum pour etats ell shiftees} could just have been incorporated there. 
Those contributions have nevertheless been kept for further convenience of normalisations: equation \eqref{ecriture ex momentum pour etats ell shiftees} with the $\dots$ being dropped allows one for a
more straightforward correspondence with the free boson model. 

Due to \eqref{ecriture decomposition locale part-trou close Fermi zone}, it appears convenient to parametrise excited states belonging to an $\ell_s$ critical class directly
in terms of the two subsets $\mc{J}^{(s)}_{ n_{p;+}^{(s)} ; n_{h;+}^{(s)} } \cup \mc{J}^{(s)}_{ n_{p;-}^{(s)} ; n_{h;-}^{(s)} } $ where 
\beq
  \mc{J}^{(s)}_{  n ; m  }  \; = \; \Big\{ \{ p_{a}^{(s)} \}_1^{ n } 
\; ; \;  \{ h_{a}^{(s)} \}_1^{ m } \Big\}  \;. 
\enq
Thus, for any state belonging to a critical $\ell_s$ class, I will make the identification between 
\beq
\mc{I}^{(s)}_n \qquad  \e{and} \qquad 
\mc{J}^{(s)}_{ n_{p;+}^{(s)} ; n_{h;+}^{(s)} } \cup \mc{J}^{(s)}_{ n_{p;-}^{(s)} ; n_{h;-}^{(s)} } \;.
\enq
Due to \eqref{ecriture decomposition locale part-trou close Fermi zone}, the value of the spin $s$ does play a role in the correspondence between 
$\mc{I}^{(s)}_n$ and $\mc{J}^{(s)}_{ n_{p;+}^{(s)} ; n_{h;+}^{(s)} } \cup \mc{J}^{(s)}_{ n_{p;-}^{(s)} ; n_{h;-}^{(s)} }$. 
I also draw the reader's attention to the fact that the value of $\ell_s$ is encoded in the very notation $\mc{J}^{(s)}_{ n_{p;+}^{(s)} ; n_{h;+}^{(s)} } 
\cup \mc{J}^{(s)}_{ n_{p;-}^{(s)} ; n_{h;-}^{(s)} } $, see \eqref{ecriture lien shift ells et differences part trous sur bords zone Fermi}.

\subsubsection{Large-$L$ expansion of form factors connecting critical states}
\label{Section dvpt a grand L des FF}

It is precisely the form factors of local operators $\mc{O}_r$ taken between two eigenstates belonging to critical classes
that are responsible for the emergence of a effective field theory description at large distances of separation between the operators. 
To describe these form factors, let 
\beq
\mc{I}^{(s)}_m \; \equiv \; \mc{J}_{ m_{p;+}; m_{h;+} } \cup \mc{J}_{ m_{p;-}; m_{h;-} } 
\quad \e{and} \quad 
\mc{I}^{(s+o_r)}_n \; \equiv \; \mc{J}_{ n_{p;+}; n_{h;+} } \cup \mc{J}_{ n_{p;-}; n_{h;-} } 
\enq
be two sets of integers\symbolfootnote[2]{Since the 
context of the setting is clear, I have omitted the spin sector label so as to lighten the formulae.} parametrising excited states belonging to the 
\beq
 \ell_{\e{out}}\; = \; m_{p;+} -  m_{h;+} \; = \; m_{h;-} -  m_{p;-}   \qquad \e{and} \qquad
 \ell_{\e{in}}\; = \; n_{p;+} -  n_{h;+} \; = \; n_{h;-} -  n_{p;-}
\enq
critical classes. The form factors of local operators taken between two excited states belonging to the critical classes 
introduced above take the universal form:  
\bem
\mc{F}_{\mc{O}_r}\bigg(  \mc{I}^{(s)}_m  ;  \mc{I}^{(s+o_r)}_n \mid x_r \bigg)  \; = \; \Big\{ \ex{2 \i p_F x_r }  \Big\}^{ \ell_{ \e{out} }- \ell_{\e{in}}  }
 \cdot C^{(\ell_{ \e{out} } - \ell_{\e{in}})} \big( \nu_r^+, \nu_r^- \big) \cdot \mc{F}_{\ell_{ \e{out} } - \ell_{\e{in}}}\big( \mc{O}_r \big)
\cdot \bigg( \f{ 2\pi }{ L } \bigg)^{ \rho\big(\nu_r^{+}+\ell_{ \e{out} }- \ell_{\e{in}} \big) + \rho\big(\nu_r^-+\ell_{ \e{out} }- \ell_{\e{in}}\big) }   \\
\times   \, 
\msc{F}\Big[  \mc{J}_{ m_{p;+}; m_{h;+}  } ;  \mc{J}_{ n_{p;+} ; n_{h;+} }     \mid   \nu_r^{+}  ,   \om_r^{+} \Big] \cdot 
 \msc{F}\Big[ \mc{J}_{ m_{p;-} ; m_{h;-} } ;  \mc{J}_{ n_{p;-} ; n_{h;-} }  \mid - \nu_r^{-}  , \om_r^{-} \Big] \\
\times \big(\om_r^+ \big)^{ \ell_{\e{in}}\f{\ell_{\e{in}}-1}{2} - \ell_{\e{out}}\f{\ell_{\e{out}}-1}{2} } \cdot  %
\big(\om_r^- \big)^{ \ell_{\e{in}}\f{\ell_{\e{in}}+1}{2} - \ell_{\e{out}}\f{\ell_{\e{out}}+1}{2} } 
\cdot \bigg( 1 \, + \, \e{O}\Big( \f{ \ln L }{ L } \Big) \bigg)\;. 
\label{ecriture conjecture form general facteur de forme deux etats excites}
\end{multline}
The constituents of the above formula are parametrised by the values 
\beq
\nu_{r}^{+} \; =  \; \nu_r(  q )  \, - \, o_r \qquad \e{and} \qquad
\nu_{r}^{-} \; =  \; \nu_r(  -q )
\enq
that the relative shift function between the $\ell_{\e{in}}, \ell_{\e{out}}$ critical states 
\beq
\nu_{r}(\la) \; = \;  F_{s}(\la) \; - \; F_{s+o_r}(\la)  
\label{definition fction shift relative etats gauche et droit}
\enq
takes on the right/left endpoints of the Fermi zone, up to subtracting the "spin" $o_r$ of the operator $\mc{O}_r$
in the case of the right endpoint. They also depend on 
\beq
\om_r^{+} \; = \; \ex{ -2\i \pi  \f{x_r}{L} \a^+ }  \qquad \e{and} \qquad  \om_r^{-} \; = \; \ex{  2\i \pi \f{x_r}{L}  \a^- }   
\enq
representing the exponent of the individual momentum brought by a particle excitation on the left or right Fermi boundary. 
In the above large-$L$ asymptotics, the quantity $\mc{F}_{\ell_{ \e{out} }- \ell_{ \e{in} } }\big( \mc{O}_r \big)$ 
represents the properly normalised finite and non-universal (\textit{i}.\textit{e}. model and operator dependent) part of the large-$L$ behaviour of the 
form factor of the operator $\mc{O}_r$ taken between fundamental representatives of the $\ell_{ \e{in} }$ and $\ell_{ \e{out} }$ critical classes. 
More precisely, it is defined as
\beq
\mc{F}_{\ell_{ \e{out} }- \ell_{ \e{in} } }\big( \mc{O}_r \big) \; = \; \lim_{L\tend +\infty}\Bigg\{
 \bigg( \f{ L }{ 2\pi } \bigg)^{ \rho\big(\nu_r^{+}+\ell_{ \e{out} }- \ell_{\e{in}} \big) + \rho\big(\nu_r^-+\ell_{ \e{out} }- \ell_{\e{in}}\big)  }
\bra{  \mc{L}_{ \ell_{ \e{out} } }^{(s)}  } \mc{O}_r(0)   \ket{  \mc{L}_{ \ell_{ \e{in} } }^{(s+o_r)}  } \Bigg\} \;. 
\label{ecriture definition facteur de forme proprement normalise macroscopique}
\enq
Above, the sets of integers $\mc{L}_{ \ell_{ \e{out} } }^{(s)}$  and $ \mc{L}_{ \ell_{ \e{in} } }^{(s+o_r)} $ 
parametrise the lowest possible excited state belonging to the $\ell_{ \e{out} }$ and $\ell_{ \e{in} }$ critical class in the spin $s$ sector 
\beq
 \mc{L}_{ \ell }^{(s)}  \; = \; 
\left\{   \ba{cc} \Big\{  \{ p_{a;+}^{(s)}  = a \}_1^{\ell} \; ;  \; \{ \emptyset \}  \Big\} \bigcup
\Big\{  \{ \emptyset \} \; ; \;  \{ h_{a;-}^{(s)} = a \}_1^{\ell}  \Big\}  &
				\e{if} \; \ell \geq 0   \vspace{2mm} \\ 
		\Big\{  \{ \emptyset \} \; ; \;  \{ h_{a;+}^{(s)}  = a \}_1^{-\ell}   \Big\} \bigcup
\Big\{    \{ p_{a;-}^{(s)} = a \}_1^{-\ell}  \; ; \; \{ \emptyset \}  \Big\}   &
				\e{if} \; \ell \leq 0 \ea \right. 		\; . 
\enq
Note that this form factor \textit{solely} depends on the \textit{difference}  $\ell_{ \e{out} }- \ell_{ \e{in} }$. The sole dependence on the difference $\ell_{ \e{out} }- \ell_{ \e{in} }$ issues from the fact that, \textit{a priori},
in the thermodynamic limit, any fundamental representative can be chose to be the reference ground state. Yet, relatively to the fundamental representative of the $\ell_{\e{out}}$-class,
the other excited state corresponds to the fundamental representative of the $\ell_{ \e{out} }- \ell_{ \e{in} }$ class.

The power of the volume $L$ arising in \eqref{ecriture conjecture form general facteur de forme deux etats excites} and 
\eqref{ecriture definition facteur de forme proprement normalise macroscopique} involves
the right $\rho(\nu_r^++\ell_{ \e{out} }- \ell_{\e{in}})$ and left $\rho(\nu_r^-+\ell_{ \e{out} }- \ell_{\e{in}})$ scaling dimensions whose generic expression reads  
\beq
 \rho(\nu) \; = \;  \f{\nu^2}{2}  \;. 
\label{definition scaling dimension}
\enq

The local microscopic form factor $\msc{F}$ represents the universal part of the form factor's large-$L$ asymptotics. 
It contains  all the "microscopic" contributions issuing from excitations 
localised on a given Fermi boundary. The local microscopic form factor depends on the 
value taken on the right or left Fermi boundary by the relative shift function $\nu_r$ associated with the critical excited states of interest
and on the position of the operator. Finally, it depends as well on the sets of integers $\mc{J}_{ m_{p;\pm}; m_{h;\pm} } $ and  
$\mc{J}_{ n_{p; \pm }; n_{h; \pm } } $ parametrising the excitations on the boundary of the Fermi zone to which they are associated. 
The explicit expression for the local microscopic form factor is a bit bulky 
but its representation theoretic interpretation is crystal clear, \textit{c}.\textit{f}. Theorem \ref{Theorem FF Vop entre etats generaux}. 
It is convenient to represent it as a product of two fundamental building blocks $\varpi$ and $\msc{D}$ which are set functions.  
 Given two sets of integers 
\beq
\mc{J}_{n_p,n_h} \; = \; \Big\{  \{p_a\}_1^{n_p} \; ; \;  \{h_a\}_1^{n_h}  \Big\}  \qquad \e{and} \qquad 
\mc{J}_{n_k,n_t} \; = \; \Big\{  \{k_a\}_1^{n_k} \; ; \;  \{t_a\}_1^{n_t}  \Big\} 
\label{definition ensemble Jnpnh et Jnknt}
\enq
 one has 
\beq
 \varpi\Big( \mc{J}_{ n_{p}; n_{h} } ;  \mc{J}_{ n_{k}; n_{t} }    \mid \nu \Big)  
			\; = \;  \pl{a=1}{n_h  }   \Bigg\{  \f{  \pl{b=1}{ n_k } \big( 1-k_b-h_a+\nu  \big)    }
{  \pl{b=1}{n_t } \big( t_b-h_a+\nu \big)  }  \Bigg\}  \cdot 
\pl{a=1}{n_p}  
\Bigg\{  \f{ \pl{b=1}{n_t} \big(p_a+t_b+\nu  -1 \big)    }
{  \pl{b=1}{n_k} \big( p_a - k_b + \nu \big)  }  \Bigg\}		\;,  
\label{definition de la fonction  varpi}
\enq
and 
\bem
 \msc{D}\Big( \mc{J}_{ n_{p}; n_{h} }   \mid \nu , \om \Big)  \; = \; \bigg( \f{\sin [\pi \nu] }{ \pi } \bigg)^{  n_{h}  }    
\cdot \pl{a=1}{n_p} \Bigg\{ \om^{p_a-1} \,  \Ga\Bigg( \ba{c}  p_{a}+\nu    \\
						    p_{a} \ea \Bigg) \Bigg\}  \cdot 
\pl{a=1}{n_h} \Bigg\{ \om^{h_a} \, \Ga\Bigg( \ba{c}  h_{a}-\nu    \\
						    h_{a} \ea \Bigg) \Bigg\}  \\
\times \f{ \pl{a>b}{ n_{p} } (p_{b} - p_{a} ) \cdot \pl{a>b}{ n_{h}  } ( h_{b} - h_{a} ) }
{ \pl{a=1}{ n_{p} } \pl{b=1}{ n_{h} } (p_{a} + h_{b} - 1) } \;. 
\end{multline}
Then the local microscopic form factor reads 
\bem
\msc{F}\Big( \mc{J}_{ n_{p}; n_{h} } ;  \mc{J}_{ n_{k}; n_{t} }  \mid \nu ,\om \Big)
\; = \;  (-1)^{ n_{p} + n_{t} }\cdot (-1)^{ \f{(n_p-n_h)(n_p-n_h+1)}{2} } \cdot \bigg( \f{\sin [\pi \nu] }{ \pi } \bigg)^{ n_{p} \, - \, n_{h}  }   
\cdot  \msc{D}\Big( \mc{J}_{ n_{p}; n_{h} }   \mid \nu , \om \Big)  \\
 \times \msc{D}\Big( \mc{J}_{ n_{k}; n_{t} }   \mid -\nu , \om^{-1} \Big) \cdot  \varpi\Big( \mc{J}_{ n_{p}; n_{h} } ;  \mc{J}_{ n_{k}; n_{t} }    \mid \nu \Big)  \;.   
\label{definition microscopic form factor}
\end{multline}

 I remind that \eqref{definition microscopic form factor} is expressed by using hypergeometric-like notations for ratios of products of $\Ga$-functions introduced in Chapter \ref{Chapitre AB grand volume FF dans modeles integrables}.

Finally, formula \eqref{ecriture conjecture form general facteur de forme deux etats excites} contains  the normalisation constant  $C^{ (\ell_{ \e{out}} - \ell_{\e{in}} ) } \big( \nu_r^+, \nu_r^- \big)$.
The latter is chosen in such a way that it cancels out the $L$-independent contributions of the right and left critical form factors
when particularising the expression to the lowest excited states of the $\ell_{\e{out}}$ and $\ell_{ \e{in} }$ critical classes. These special excited states are 
parametrised by the sets of integers $\mc{L}_{ \ell_{ \e{out} } }^{(s)}$  and $ \mc{L}_{ \ell_{ \e{in} } }^{(s+o_r)} $. 
The explicit form of the normalisation factor can be computed in terms of ratios of Barnes $G$-function \cite{BarnesDoubleGaFctn1}:
\beq
C^{(\ell_{ \e{out} } - \ell_{\e{in}} )} \big( \nu_r^+, \nu_r^- \big) \; = \; 
  G\bigg( \ba{cc} 1 + \nu_r^- , 1-\nu_r^+   \\
1  + \nu_r^- + \ell_{ \e{out} } - \ell_{\e{in}}  ,   1-\nu_r^+ - \ell_{ \e{out} } + \ell_{\e{in}}    \ea \bigg)   
	\;. 	
\enq

\subsubsection{General remarks on the scope of applicability of the model}

The general structure of the spectrum and form factors that I discussed throughout this section was first introduced in my work on multi-point correlation functions in massless one-dimensional 
quantum models [$\bs{A15}$].

 The formula for the form factors of local operators taken between excited states belonging to critical classes
can be proven within the framework of the algebraic Bethe Ansatz for various quantum integrable models on the basis of determinant representations for their
form factors \cite{KMTFormfactorsperiodicXXZ,KorepinSlavnovApplicationDualFieldsFredDets,OotaInverseProblemForFieldTheoriesIntegrability,SlavnovFormFactorsNLSE} and [$\bs{A4}$]. 
The corresponding calculations are a straightforward generalisation of the method developed in \cite{SlavnovFormFactorsNLSE} and  [$\bs{A5},\bs{A6}$]. 
However, I do trust that the decomposition \eqref{ecriture conjecture form general facteur de forme deux etats excites} is, in fact, universal. 
More precisely, the properly normalised form factor $\mc{F}_{  \ell_{ \e{out} }- \ell_{ \e{in} }  }\big( \mc{O}_r \big)$ is definitely model dependent
and thus can only be obtained on the basis of exact computations. Its explicit expression is available for many quantum integrable models and can be found in the 
aforementioned works. 
I do trust that, in fact, the local microscopic pre-factors and the leading power-law behaviour in $L$ is universal: namely that they take the same form for models belonging to the Luttinger liquid universality class. 
Part of this has been confirmed by formal perturbative calculations around a free model \cite{CauxGlazmanImambekovShasiNonUniversalPrefactorsFromFormFactors}.



 \subsection{The free boson space of states and effective description of correlation functions}

I will recall in the present section the free fermion based construction of the space of states for the free boson model. The presentation 
 basically follows the notations and conventions that can be found in the excellent review paper \cite{AlexandrovZaborodinTechniquesOfFreeFermions}. 
The various results found in this review originate from a long series of developments which started with the cornerstone work of Kyoto's school 
(Jimbo, Miwa and Sato) in the late '70's on holonomic quantum fields \cite{JimMiwaSatoQuantumFieldsI}.

\subsection{Overall definitions and generalities} 

\subsubsection{The space of states}

The whole construction of the space of states builds on a collection of fermionic operators $\{\psi_n\}_{n \in \mathbb{Z} }$ and their $*$ associates $\{\psi_n^*\}_{n \in \mathbb{Z}}$
satisfying to the anti-commutation relations
\beq
\{ \psi_n,\psi_m \} \; = \; \{ \psi_n^*,\psi_m^* \} \; = \; 0 \qquad \e{and} \qquad
\{ \psi_n,\psi_m^* \}\;= \; \de_{n,m} \;, 
\enq
where $\de_{n,m}$ is the Kronecker symbol. One assumes the existence of a vacuum vector $\ket{0}$ characterised by 
\beq
\psi_n \ket{0} \; = \; 0 \; \; \e{for} \; \; n<0  \qquad \e{and} \qquad   \psi_n^{*} \ket{0} \; = \; 0 \; \; \e{for} \; \; n \geq 0  \;. 
\enq
The dual vacuum $\bra{0}$ fulfils the analogous properties  
\beq
\bra{0} \psi_n^{*}  \; = \; 0 \; \; \e{for} \; \; n<0 \qquad \e{and} \qquad  \bra{0} \psi_n  \; = \; 0 \; \; \e{for} \; \; n \geq 0 \;. 
\enq
The vacuum (resp. the dual vacuum) allows one to construct other vectors (resp. dual vectors) through a repetitive action of the fermion operators.  
All vectors built in this way are parametrised by sets 
\beq
\mc{J}_{n_p;n_h} \; = \; \Big\{ \{p_a\}_1^{n_p} \; ; \; \{h_a\}_1^{n_h} \Big\}
\enq
built up of two collections of ordered integers $1\leq p_1 < \dots < p_{n_p}$ and $1\leq h_1 < \dots < h_{n_h}$. For later convenience, 
one can think of the $p_a$'s as particles and of the $h_a$'s as holes. To each set $\mc{J}_{n_p;n_h}$,  one associates the vector 
\beq
\ket{ \mc{J}_{n_p;n_h} } \; = \; \psi_{-h_1}^{*}\cdots \psi_{-h_{n_h}}^{*} \cdot \psi_{p_{n_p}-1} \cdots \psi_{p_1-1} \ket{0}
\enq
and the dual vector 
\beq
\bra{ \mc{J}_{n_p;n_h} } \; = \;  \bra{0}\psi_{p_1-1}^{*} \cdots \psi_{p_{n_p}-1}^{*} \cdot \psi_{-h_{n_h}} \cdots \psi_{-h_1} \;. 
\enq
The Hilbert space $\mf{h}_{FB}$ of the model is then defined as the span of the vectors introduced above
\beq
\mf{h}_{FB} \; = \; \e{span} \bigg\{ \ket{ \mc{J}_{n_p;n_h} } \; \e{with} \; n_p, n_h\in \mathbb{N} \; \; \e{and} \; \; 
\ba{c} 1\leq p_1 < \dots < p_{n_p} \\  1\leq h_1 < \dots < h_{n_h} \ea  \; \; 
p_a, h_a \in \mathbb{N}^{*} \bigg\}. 
\label{definition espace de Hilbert boson lilbre}
\enq

Note that, when the number of particle and hole-like integers coincide, \textit{i}.\textit{e}. $n_p=n_h=n$, one can 
identify the set $\mc{J}_{n_p;n_h}$ with a Young diagram. The one-to-one map is obtained by interpreting the integers 
$\Big\{ \{p_a\}_1^{n} \; ; \; \{h_a\}_1^{n} \Big\}$  as the Frobenius coordinates of the Young diagram, this within
the slightly unusual convention that the Frobenius coordinates include the diagonal and hence always start from $1$.

One can also provide another basis $\ket{ \mathbb{Y} ; \ell}$ of $\mf{h}_{FB}$ which is directly connected with Young
diagrams $\mathbb{Y}$. First of all, this alternative basis  takes, as its starting point, the so-called fixed charge $\ell$ pseudo-vacuum and dual pseudo-vacuum:
\beq
\ket{ \ell } \; = \; \left\{ \ba{cc} \psi_{\ell-1}\cdots \psi_{0}\ket{0} &  \quad \ell >0 \vspace{1mm} \\ 
					\psi_{\ell}^{*}\cdots \psi_{-1}^{*}\ket{0} & \quad  \ell < 0   \ea \right.  \qquad \qquad \e{and} \qquad \qquad 
\bra{ \ell } \; = \; \left\{ \ba{cc} \bra{0} \psi_{0}^{*}\cdots \psi_{\ell-1}^{*} & \quad \ell >0 \vspace{1mm} \\ 
					\bra{0} \psi_{-1}\cdots \psi_{\ell} & \quad \ell < 0   \ea \right.  \;. 
\enq
The states $\ket{ \mathbb{Y} ; \ell}$ are build as equal in number particle-hole excitations over the vaccum $\ket{\ell}$, resp.  its dual vaccum $\bra{\ell}$. 
Let 
\beq
\mathbb{Y} \; = \; \Big\{ \{\a_a\}_1^d \; : \; \{\be_a\}_1^d  \Big\} \quad \e{with} \qquad 
\left\{  \ba{c} 		
				  1\leq \a_1 \, < \, \cdots \, <\; \a_d  \vspace{1mm}\\ 
			     1\leq \be_1 \, < \, \cdots \, <\; \be_d  \ea \right. 
\enq
be the Frobenius coordinates of the Young diagram $\mathbb{Y}$, then
\beq
\ba{ccc}
\ket{\mathbb{Y};\ell} & = & \psi^{*}_{\ell - \be_1 } \cdots  \psi^{*}_{ \ell - \be_d } \, \psi_{\ell + \a_d - 1 } \cdots  \psi_{ \ell + \a_1 - 1} \, \ket{\ell}  \vspace{2mm} \\
\bra{\ell ; \mathbb{Y} } & = & \bra{\ell} \, \psi^{*}_{\ell + \a_1 -1 } \cdots  \psi^{*}_{ \ell + \a_d -1} \, \psi_{\ell - \be_d  } \cdots  \psi_{ \ell - \be_1} 
    \ea \; .   
\label{ecriture bases labelle par les diagrammes de Young}
\enq
In fact, one can also consider mixtures of the basis $\ket{\mathbb{Y};\ell}$ and $\ket{ \mc{J}_{n_p;n_h} }$, namely the basis 
\beq
\ket{ \mc{J}_{n_p;n_h} ;\ell  } \; = \; \psi_{\ell - h_1}^{*}\cdots \psi_{ \ell - h_{n_h}}^{*} \cdot \psi_{p_{n_p}+\ell - 1} \cdots \psi_{p_1+\ell-1} \ket{\ell} \;. 
\enq
Clearly
\bem
\e{Span}\Big\{ \ket{\mathbb{Y};\ell+r} \; : \; \mathbb{Y}\; \;  \e{Young} \; \e{diagram} \Big\} \; = \; 
\e{Span}\Big\{  \ket{ \mc{J}_{n_p;n_h}  } \; : \; \e{sets} \; \mc{J}_{n_p;n_h} \; \;\e{with} \; \; \;  n_p-n_h=\ell+r  \Big\}  \\
\; = \; \e{Span}\Big\{  \ket{ \mc{J}_{n_p;n_h} ; \ell } \; : \; \e{sets} \; \mc{J}_{n_p;n_h} \; \;\e{with} \; \; \; n_p-n_h=r  \Big\}   \;. 
\end{multline}

\subsubsection{The space of operators}

Several important operators on $\mf{h}_{FB}$ can be built explicitly using the fermionic operators $\psi_j$ and $\psi_j^*$. 
The operators that will be of interest to the problem are the current operators. 
The latter are built by using current operator modes $J_k$, $k \in \mathbb{Z}$.
The most fundamental mode $J_0$ is the so-called charge operator which takes the explicit form
\beq
J_0 \; = \; \sul{k\geq 0}{} \psi_k \, \psi^{*}_k \; - \; \sul{k<0}{} \psi_k^{*} \, \psi_k \;. 
\enq
The definition might appear formal due to convergence issues. However, the infinite sums truncate to finite ones as soon as one computes matrix elements taken between
the fundamental system of basis vectors labelled by the sets $\mc{J}_{n_p;n_h}$. It is in this sense that the above and all of the following expressions should be understood. 
The vector $\ket{ \mc{J}_{n_p;n_h} } $ is associated with the eigenvalue $n_p-n_h$ of the charge operator $J_0$.
Thus, one says that the vector $\ket{ \mc{J}_{n_p;n_h} } $ has charge  $n_p-n_h$. 
As a consequence, the charge operator induces a grading of the Hilbert space $\mf{h}_{FB}$
 into the direct sum of spaces $\mf{h}_{FB;\ell}$ having a fixed charge $\ell$
\beq
\mf{h}_{FB} \; = \; \bigoplus_{\ell \in \mathbb{Z} } \mf{h}_{FB;\ell}  \; \qquad \e{with}  \qquad
\mf{h}_{FB; \ell} \; = \; \e{span} \Big\{ \ket{ \mc{J}_{n_p;n_h} } \; : \;  n_p-n_h\, = \, \ell \Big\}. 
\enq
The current operators modes are defined as
\beq
J_k \; = \; \sul{j\in \mathbb{Z} }{} \psi_j \psi^*_{j+k} \qquad  \;  \e{for} \; \;  k \not= 0 \; 
\qquad \e{and} \; \e{satisfy} \quad \big[ J_k, J_{\ell} \big] \; = \; k \de_{k,-\ell} \;.
\enq
Furthermore, for $k\in \mathbb{N}^{*}$, one has
\beq
J_k \ket{0} \; = \; 0 \qquad \e{and} \qquad \bra{0} J_{-k}  \; = \; 0 \;. 
\enq
All this allows one to interpret $\{J_k\}_{k\in \mathbb{N}^*}$ as bosonic creation operators while $\{J_{-k}\}_{k\in \mathbb{N}^*}$ as  bosonic annihilation operators. 
It is on the basis of such an observation that the construction of the free boson field theory can be obtained by using the picture provided by the free fermion model. 
The main advantage for using the free fermion model is that the fermionic structure allows one to simplify many intermediate calculations, \textit{e}.\textit{g}. by using some version of Wick's theorem. 

The bosonic operator modes allow one to define a class of very important operators, the current operators which correspond to the series 
\beq
\msc{J}_{\pm}(\nu,\om) \; \; = \; \mp \nu \sul{ k \geq 1}{} \f{  \om^{\mp k} }{ k }  \cdot J_{\pm k } \;. 
\enq

There is also another operator that will play an important role in the construction: the so-called shift operator $\ex{P}$ 
which maps $\mf{h}_{\ell}$ onto $\mf{h}_{\ell+1}$. One can interpret the operator $P$ arising in the exponent as the conjugate operator to $J_0$. Although the shift operator has no 
simple expression in terms of the fermions, it takes a particularly simple form in the basis of $\mf{h}$
subordinate to Young diagrams, \textit{cf}. \eqref{ecriture bases labelle par les diagrammes de Young}. 
Indeed, any integer power $\ex{r P}$ of the shift operator satisfies $\ex{r P }\ket{ \mathbb{Y}  ; \ell} \, = \, \ket{ \mathbb{Y} ; \ell + r}$, 
\textit{viz}. 
\beq
\ex{r P } \; = \; \sul{ \mathbb{Y} \, , \, \ell }{} \ket{ \mathbb{Y} ; \ell + r}\bra{ \ell ; \mathbb{Y} }
\label{ecriture operateur translation dans base des coord Frob a n parts}
\enq
where the sum runs over all integers $\ell$ and all Young diagramms $\mathbb{Y}$ (see \cite{AlexandrovZaborodinTechniquesOfFreeFermions} for more details). 

I am finally in position to provide the definition of the $r$-shifted bosonic vertex operators
\beq
\msc{V}(\nu, r \mid \om) \; = \; \ex{ \msc{J}_- (\nu + r,\om)  } \cdot \ex{ \msc{J}_+ (\nu + r,\om)  } \cdot \ex{r P} \;. 
\label{definition  r shifted vertex operators}
\enq
It is this operator that plays a central role in the effective description of the large-distance asymptotic behaviour of multi-point functions in massless models 
by means of the free boson model. 
In fact, the main tool in the construction of this correspondence is the below theorem

\begin{theorem}
\label{Theorem FF Vop entre etats generaux}

The form factor of the $r$-shifted vertex operator $\msc{V}(\nu, r\mid \om)$  reads
\beq
\bra{ \mc{J}_{ n_{p}; n_{h} } } \msc{V}(\nu, r\mid \om) \ket{  \mc{J}_{ n_{k}; n_{t} } }
\; = \; (-1)^{\frac{r(r+1)}{2}} \cdot  \f{ \de_{n_p-n_h, n_k-n_t+r} }{  \big(\om\big)^{\frac{r(r-1)}{2}+r(n_k-n_t)} }\cdot G\bigg( \ba{c} 1-\nu \\ 1-\nu -r \ea\bigg)
\cdot  \msc{F}\Big( \mc{J}_{ n_{p}; n_{h} } ;  \mc{J}_{ n_{k}; n_{t} }  \mid \nu,\om \Big) \;. 
\label{ecriture representation VO pour la valeur moyenne a shift r general}
\enq
The symbol $G$ appearing above stands for the hypergeometric-like notation for the ratio of two Barnes functions. 
\end{theorem}

The Kronecker symbol in \eqref{ecriture representation VO pour la valeur moyenne a shift r general} arises since
the form factor is zero unless $n_p-n_h=n_k-n_t+r$. This is due to the fact  that the exponents of current operators preserve the charge of a state
while the shift operator changes the charge of the state by $r$. 

The theorem is well known in the case $n_p=n_h=r=0$. Indeed, then, the representation \eqref{ecriture representation VO pour la valeur moyenne a shift r general}
follows, for instance, from the Giambelli determinant representation for Schur functions along 
with the possibility to compute explicitly the Schur functions $s_{\a,\be}(\bs{t}_{\pm})$ associated
with hook diagrams $(\a,\be)$ in Frobenius notations (\textit{cf}. Appendix A of \cite{AlexandrovZaborodinTechniquesOfFreeFermions} for more
details). However, to the best of my knowledge,  the general case given above is new and its proof 
relies on new ideas that are not related to the theory of Schur functions. I also emphasise that Theorem \ref{Theorem FF Vop entre etats generaux}
above provides one with a new type of explicit representation for skew Schur functions associated with the parameters $\bs{t}_{\pm}$.

 The proof of the above theorem is the main technical result of [$\bs{A14}$]. It constitutes a synthesis 
of Lemma 1.1 and Proposition 1.1 of Section 1.2 of that paper. The proof of Proposition 1.1 can be found in Appendix C of the same paper.




\subsection{The effective free boson field theory model}

The effective Hilbert space is defined as the tensor product of two copies $\mf{h}_L$ and $\mf{h}_R$ of the free boson Hilbert space $\mf{h}_{FB}$
introduced previously:
\beq
\mf{h}_{\e{eff}} \; = \;   \mf{h}_{L}  \otimes \mf{h}_{ R } \;. 
\enq
The first space (resp. second) arising in the tensor product models the effects of the left (resp. right) Fermi surface
of the physical model.

Given a local operator $\mc{O}_r(x_r)$ in the physical model, one builds the below operator on $\mf{h}_{ \e{eff} }$ 
\beq
\msc{O}_r(\om_r) \; = \; \sul{ \kappa \in \mathbb{Z} }{} \mc{F}_{\kappa}\big(\mc{O}_r \big)\cdot \Big( \f{2\pi}{L} \Big)^{\rho(\nu_r(q) -o_r + \kappa)+\rho(\nu_r(-q)+\kappa)} \cdot 
\ex{2\i p_F \kappa x_r} \cdot  \msc{V}_{L}\big(-\nu_r(-q)  , -\kappa \mid \om_r^{-}  \big) \cdot  \msc{V}_{R}\big(\nu_r(q)-o_r,\kappa - o_r \mid \om_r^{+}  \big)  \;. 
\enq
In this formula, 
\begin{itemize}
 \item[$\bullet$] $\msc{V}_{L/R}(\nu,\kappa;\om)$ stands for the operator acting non-trivially on the $L/R$ copy of the original 
Hilbert space as the vertex operator $\msc{V}(\nu,\kappa;\om)$ defined in \eqref{definition  r shifted vertex operators}, \textit{viz}. 
\beq
\msc{V}_{L}(\nu,\kappa \mid \om) \; = \; \msc{V}(\nu,\kappa \mid  \om) \otimes \e{id} \qquad \e{and} \qquad 
\msc{V}_{R}(\nu,\kappa \mid  \om) \; = \;  \e{id} \otimes \msc{V}(\nu,\kappa \mid  \om)  \;. 
\enq
\vspace{1mm} 
\item[$\bullet$] $\rho(\nu)$ are the scaling dimensions introduced in \eqref{definition scaling dimension} . \vspace{1mm} 
\item[$\bullet$] $\mc{F}_{\kappa}\big( \mc{O}_r \big)$ is the thermodynamic limit of the properly normalised in the volume amplitude defined in 
\eqref{ecriture definition facteur de forme proprement normalise macroscopique} and taken between excited states satisfying $\ell_{\e{out}}-\ell_{\e{in}}=\kappa$. \vspace{1mm} 
\item[$\bullet$] $\om_s^{\pm}$ is a phase factor that reads
\beq
\om_s^{\pm} \; = \;  \ex{ \mp 2\i \pi   \f{ x_s }{ L } \a^{\pm} } \;. 
\enq

\end{itemize}

 The crucial observation is that given two sets 
\beq
\mc{I}^{(s)}_m \; \equiv \; \mc{J}_{ m_{p;+}; m_{h;+} } \cup \mc{J}_{ m_{p;-}; m_{h;-} }
\quad \e{and} \quad 
\mc{I}^{(s+o_s)}_n \; \equiv \; \mc{J}_{ n_{p;+}; n_{h;+} } \cup \mc{J}_{ n_{p;-}; n_{h;-} }
\label{ecriture correspondance etats hilbert initial et hilbert effectif}
\enq
 parametrising critical excited states in the physical Hilbert space $\mf{h}_{\e{phys}}$, one 
has an equality, up to $\e{O}\big( \ln L / L \big)$ corrections between matrix elements:
\bem
\mc{F}_{ \mc{O}_r }\Big(  \mc{I}^{(s)}_m  ;  \mc{I}^{(s+o_r)}_n \mid x_r \Big)  \; =  \; (-1)^{\varkappa}
\Big\{ \bra{ s; \mc{J}_{ m_{p;-}; m_{h;-} }  } \otimes \bra{ s;  \mc{J}_{ m_{p;+}; m_{h;+} } }  \Big\}
\msc{O}_r(\om_r) \\ 
\Big\{ \ket{ s ; \mc{J}_{ n_{p;-}; n_{h;-} } }\otimes \ket{ s+o_r;  \mc{J}_{ n_{p;+}; n_{h;+} }  } \Big\}  \cdot  \bigg( 1 + \e{O}\Big( \f{ \ln L }{ L }\Big) \bigg) 
\label{ecriture identification asymptotique des FF}
\end{multline}
where
\beq
\varkappa \; = \; m_{p;+}-m_{h;+}-n_{p;+}+n_{h;+} \;. 
\enq

This identity allows one then to argue that , when $|x_a-x_b| p_F \gg 1$, 
\beq
\big< \mc{O}_1(x_1)\cdots \mc{O}_{r}(x_r)  \big>_{ \mf{h}_{\e{phys}} } \; \simeq  \; 
\big< \msc{O}_1(\om_1)\cdots \msc{O}_{r}(\om_r) \big>_{ \mf{h}_{\e{eff}} } \;, 
\enq
where the $\simeq$ sign is to be understood as an equality up to the first leading correction arising in each oscillating harmonics in the difference of positions $|x_a-x_{a+1}|$. 
Above, the subscripts $\mf{h}_{\e{phys}}$ and $\mf{h}_{\e{eff}}$ are present so as to insist on the Hilbert space where the operators "live" and hence where 
the expectation values are computed. 
The asymptotic equivalence is argued on the basis of the form factor series decomposition for the two objects. See Section 3.2 of [$\bs{A14}$] for more details.

\section{Conclusion}

In the present chapter, I gave a account of the form factor series expansion approach to the asymptotic behaviour of multi-point correlation functions in quantum integrable models. 
Although heuristic, the method is extremely powerful and allows one to reproduce all the results that can be obtained though the use of a correspondence with a 
conformal field theory at $c=1$ or its more evolved version -the non-linear Luttinger liquid- allowing one to grasp the effects of time dependence. 
The main advantage of the method in respect to the field theoretic approach is that it is constructive. True it builds on certain hypothesis on the structure of the space of a model. 
However, for the remaining part of the analysis, the method stays coherent with itself in that it doesn't require numerous heuristics that are necessary to deal with in the 
field theoretical framework. Furthermore, the method is, \textit{de facto}, relatively simple provided that one takes for granted summation identities such as the multiple sums 
\eqref{definition somme restreinte}-\eqref{ecriture resultat sommation somme restreinte}. Last but not least, the method is extremely effective. Once that we have understood 
its basics in the first paper of the series [$\bs{A17}$], it was a question of a couple of months to generalise the method to the case of time and space dependent
asymptotics as well as to the case of the critical behaviour of the dynamical response functions in  [$\bs{A16}$]. 
It is worth reminding that it took around twenty-five years starting from the early days of Luttinger liquid approach developed by Luther and Peschel \cite{LutherPeschelCriticalExponentsXXZZeroFieldLuttLiquid}
and Haldane \cite{HaldaneCritExponentsAndSpectralPropXXZ,HaldaneStudyofLuttingerLiquid} so as to grasp all the details that were necessary to set the non-linear Luttinger liquid 
approach \cite{GlazmanImambekovDvPMTCompletTheoryNNLL,GlazmanImambekovSchmidtReviewOnNLLuttingerTheory,ImambekovGlazmanEdgeSingInDSFBoseGas,GlazmanKamenevKhodasPustilnikNLLLTheoryAndSpectralFunctionsFremionsFirstAnalysis} which allows one to deal with the time dependent case, \textit{viz}. take into account the non-linear part of the spectrum.  
True, the non-linear Luttinger liquid based approach results did help to guide the research developed in  [$\bs{A16}$]. However, due to the constructive setting, without knowing much 
of the structure or type of the answer in the time-dependent case, one is still able to derive it after a few "experiments".

\chapter{The quantum separation of variables for the Toda chain}
\label{Chapitre qSoV pour Toda}

The first description of the quantisation conditions for the Toda chain goes back to the works of Gutzwiller \cite{GutzwillerResolutionTodaChainSmallNPaper2}, 
Gaudin-Pasquier \cite{GaudinPasquierQOpConstructionForTodaChain} and Goodmann-Wallach \cite{GoodmannWallachQuantumTodaIII}. 
The first two collaborations obtained a description in terms of the Baxter equation while the last one proposed a quite different 
form for the quantisation conditions. These characterisation of the spectrum of the Toda chain were somewhat complicated in that 
they involved finding zeroes or poles of infinite determinants of Hill type. Recently, Nekrasov and Shatashvili \cite{NekrasovShatashviliConjectureTBADescriptionSpectrumIntModels} 
conjectured a way of describing the spectrum of the Toda chain through non-linear integral equations. 
Its relation to the quantum separation of variables issued description of the spectrum was, however, unclear. 

As mentioned in the introduction,  the implementation of the quantum separation of variables builds on the construction of an
integral transform that is supposed to map unitarily the original Hilbert space where the model is defined onto
some model $L^2$ space where the quantum separation of variables takes place. Although there exist quantum inverse scattering based methods allowing one to construct 
and fully characterise the integral kernel of this transform, the proof of its unitarity was only available for the
Toda chain case and relied on earlier results issuing from the group theoretic based approach to the model. 
Since the group theoretic interpretation is missing for more complex models solvable by the quantum separation of variables, it is more than desirable
to find a way of proving the unitarity solely by using structures that are natural to the quantum inverse scattering framework. 
Indeed, then, one may expect that, with minor modifications, the proof can also be applicable to other models.

The quantum separation of variables provides one with an utterly simple description of the Toda chain eigenfunctions on the space $L^2(\R^N\times \R,\dd \mu(\bs{y}_N)\otimes \dd \veps)$ where the 
quantum separation of variables takes place. 
In order to bring the calculations of the correlation functions to the sole handling of objects living on this space one needs 
to find how the local operators of the model realise as operators on  $L^2(\R^N\times \R,\dd \mu(\bs{y}_N)\otimes \dd \veps)$. 
A very effective way for accessing to such an information is to obtain, as proposed by Babelon  \cite{BabelonActionPositionOpsWhittakerFctions}, 
equations in dual variables: 
\beq
\mc{O} \cdot \Psi_{\bs{y}_{N};\veps} (\bs{x}_{N+1}) \; = \; \wh{\mc{O}} \cdot  \Psi_{\bs{y}_{N};\veps} (\bs{x}_{N+1})
\enq
in which $\mc{O} $ represented a certain class of operators acting on the variables  $\bs{x}_{N+1}$ attached to the original space $\mf{h}_{\e{Td}}$ 
whereas its dual operator $\wh{\mc{O}}$ acts on the variables attached to the space where the separation of variables occurs, \textit{viz}. the dual variables $(\bs{y}_N; \veps)$. 
Since $\Psi_{\bs{y}_{N};\veps} (\bs{x}_{N+1})$ corresponds to the integral kernel of the separation of variables transform, such equations
are indeed enough so as to solve the inverse problem.

In this chapter, I will describe the progress I made relatively to three features related with the quantum Toda chain:     \vspace{1mm}
\begin{itemize} 
\item[$\bullet$] proving the non-linear integral based description of the quantisation conditions that was conjectured by Nekrasov and Shatashvili; \vspace{1mm}
\item[$\bullet$] developing a proof of the unitarity of the quantum separation of variables transform $\mc{U}_N$ which solely builds on objects that are "natural" to the quantum inverse scattering method;\vspace{1mm} 
\item[$\bullet$] pushing further the resolution of the spectral problem by following the approach of Babelon \cite{BabelonActionPositionOpsWhittakerFctions}. \vspace{1mm}
\end{itemize}

This chapter is organised as follows. In Section \ref{Section spectrum}, I give a short description of the quantum inverse scattering approach to the Toda chain and of Gutzwiller's 
solution of the scalar $\op{t}-\op{Q}$ equation for the Toda chain as well as of the quantisation conditions that result from this construction. 
I will then build on this setting so as to prove the Nekrasov-Shatashvili description of the quantisation conditions. 
In Section \ref{Section unitarite SoV transform}, I will describe the quantum inverse scattering based method that I developed so as to prove the unitarity of the 
separation of variables transform. In particular, I will describe the two possible ways of constructing of the integral kernel of the transform, leading to its 
Gauss-Givental and Mellin-Barnes multiple integral representations. I will remind how the Gauss-Givental representation allows one to easily prove, on a formal level of 
rigour, the isometricity of the separation of variables transform. I will then discuss my contributions. 
Finally, in Section \ref{Section probleme inverse Toda}, I will present the equations in dual variables for a class of local operators associated to the quantum Toda chain. 
In the simplest case, I will explain how these allow one to access to multiple integral representation for the form factors of the model.

\section{Spectrum of the Toda chain}
\label{Section spectrum}

\subsection{Integrability of the quantum Toda chain and Gutzwiller's quantisation}

The Lax matrix for the Toda chain takes the form  
\beq
\op{L}_{0n}(\la) \; = \; \pa{\ba{cc} \la-\op{p}_n & \ex{- \op{x}_n} \\ - \ex{ \op{x}_n} & 0 \ea}_{\pac{0}} \quad \e{with} \quad 
 \pac{ \op{x}_k , \op{p}_{\ell} } \, = \,   \i \hbar \de_{k\ell} 
\label{ecriture matrice de Lax Toda}
\enq
and gives rise to the monodromy matrix of the model by the standard construction of the quantum inverse scattering method:
\beq
\op{T}_{0;1,\dots,N+1}(\la)  \; = \, \op{L}_{01}(\la) \dots \op{L}_{0N+1}(\la) \;  =  \; 
	\pa{ \ba{cc} \op{A}_{1, N+1} (\la) & \op{B}_{1, N+1} (\la) \\ 
	\op{C}_{1, N+1} (\la) & \op{D}_{1, N+1}(\la) \ea}_{\pac{0}}\;.
\enq
The transfer matrix $ \op{t}_{\e{Td}}(\la) \; = \; \e{tr}_0\big[ \op{T}_{0;1,N+1}(\la)   \big]$  is a monoic operator valued polynomial in $\la$ of degree $N+1$
such that $\big[ \op{t}_{\e{Td}}(\la) ,\op{t}_{\e{Td}}(\mu) \big]=0$ for any $\la,\mu$. Hence, it allows one to construct $N+1$ charges in involution by considering its 
expansion  into powers of  $\la$:
\beq
 \op{t}_{\e{Td}}(\la) \; = \; \la^{N+1}  \; + \; \sul{k=1}{N+1}(-1)^k \la^{N-k+1}  \op{t}_{\e{Td};k} \qquad \e{with} \qquad 
 \op{t}_{\e{Td};1} \; = \; \sul{a=1}{N+1} \op{p}_a \qquad \e{and}  \qquad \op{t}_{\e{Td};2} \; = \; \op{t}_{\e{Td};1}^2 - \op{H}_{\e{Td} \mid \kappa=1} \;. 
\enq
Observe that $N+1$ Hamiltonians in involution are self-adjoint. Hence, the eigenvalues $t_{\e{Td}}(\la)$ of the transfer matrix $ \op{t}_{\e{Td}}(\la)$ are polynomials 
$t_{\e{Td}}(\la) \, = \, \prod_{k=1}^{N+1}(\la-\tau_k)$ such that the set $\{ \tau_a \}_1^{N+1} $ of their zeroes is necessarily self-conjugated $\paa{\tau_a}=\paa{ \tau_a^* }$.

As already stated in the introduction, the quantum separation of variables reduces the spectral problem for the closed quantum Toda chain  
to the problem of finding all solutions $\big( t_{ \e{Td} }(\la) \, , \,   q_{t_{ \e{Td} }}(\la) \big)$ 
to the Baxter equation 
\beq
t_{ \e{Td} }(\la) \cdot  q_{t_{ \e{Td} }}(\la) \; = \; (\i)^{N+1} q_{t_{ \e{Td} }}(\la + \i \hbar) \; + \;  (-\i)^{N+1} q_{t_{ \e{Td} }}(\la-\i\hbar) 
\label{ecriture eqn TQ scalaire Toda coeur du Chapitre SoV Toda}
\enq
that, furthermore, satisfy the conditions: \vspace{1mm}
\begin{itemize}
 \item[i)] $t_{\e{Td}}(\la)$ is a polynomial of the form $t_{\e{Td}}(\la) \, = \, \prod_{k=1}^{N+1}(\la-\tau_k)$ with $\big\{ \tau_a \big\} = \big\{ \tau_a^* \big\}$; \vspace{1mm}
 \item[ii)]  $q_{t_{ \e{Td} }}$ is entire and satisfies, for some $N$-dependent $C>0$, to the bound
\beq
 \abs{ q_{t_{ \e{Td} }}(\la) }\, \leq  \, C \cdot \ex{-(N+1)\f{\pi}{2\hbar}\abs{\Re\pa{\la}}}  \abs{\la}^{\frac{N+1}{2\hbar}\pa{ 2\abs{\Im\pa{\la}}-\hbar } } \qquad \e{uniformly}\;\e{in}\quad 
\la \in \paa{ z\; : \; \abs{\Im\pa{z} } \leq \tf{\hbar}{2}}  \;; 
\label{ecriture borne sur fction q Baxter Toda}
\enq
\item[iii)] the roots $\{\tau_k\}_1^{N+1}$ satisfy  to $\sum_{a=1}^{N+1}\tau_a = \veps$. \vspace{1mm}
\end{itemize}
The third condition iii) relates the $\tau_a$'s to the eigenvalue $\veps$ of the total momentum operator $\sum_{a=1}^{N+1}\op{p}_a$. 
This constraint issues from the fact that the Toda chain Hamiltonian is translationally invariant, hence making it more convenient
to describe the spectrum of the chain directly in a sector corresponding to a fixed eigenvalue $\veps$ of the total momentum operator $\sum_{a=1}^{N+1}\op{p}_a$.
After such a reduction, $q_{t_{ \e{Td} }}$ represents the "normalisable" part of the Toda chain eigenfunction, \textit{c.f.} \eqref{definition transfo integrale U super cal} and \eqref{definition forme fct propre Toda dans espace separe}.

\vspace{2mm}

As already observed by Gutzwiller \cite{GutzwillerResolutionTodaChainSmallNPaper1,GutzwillerResolutionTodaChainSmallNPaper2}, given any polynomial $t_{\e{Td}}$ satisfying condition i), 
one can fairly easily find two  meromorphic linearly independent solutions of the Baxter equation that have the required decay at $\Re(\la) \tend + \infty$, \textit{c.f.}
\eqref{ecriture borne sur fction q Baxter Toda}.  
For this purpose, given a polynomial $t_{\e{Td}}$ as in condition i), one introduces its associated Hill determinant \cite{GohbergGoldbargKrupnikTracesAndDeterminants}
\beq
\mc{H}_{ t_{\e{Td}} }\pa{\la} = \det \pac{ \ba{cccccc}  \dots &  \ddots & \ddots& \ddots & \dots & \dots\\
                                        \dots & \f{ 1 }{ t_{\e{Td}}\pa{\la-\i\hbar} } & 1 & \f{1}{ t_{\e{Td}}\pa{\la - \i \hbar} } &  0 & \dots \\
        \dots  &0 & \f{ 1 }{t_{\e{Td}}\pa{\la} } & 1 & \f{1}{ t_{\e{Td}}\pa{\la} }& 0 \dots \\
        \ddots & \ddots  & \ddots & \ddots & \ddots  &\ddots
  \ea} \; .
\label{definition determinant infini Hill}
\enq
The Hill determinant can be shown to admit the representation \cite{GaudinPasquierQOpConstructionForTodaChain}
\beq
\mc{H}_{ t_{\e{Td}} } (\la) \;  =  \; \pl{a=1}{N+1} \f{  \sinh\f{\pi}{\hbar}\pa{\la-\sg_a}  }{  \sinh\f{\pi}{\hbar}\pa{\la-\tau_a} } \; . 
\label{equation factorisation H}
\enq
One can always chose the $\sg_a$ such that  $-\tf{\hbar}{2} \, < \, \big| \Im\pa{\sg_k} \big| \, \leq  \, \tf{\hbar}{2}$ and $\{\sg_a^{*}\}=\{\sg_a\}$. Doing so defines unambiguously the set $\{\sg_a\}_1^{N+1}$ in terms of $t_{ \e{Td} }(\la)$.
The two linearly independent  meromorphic solutions $q_{ t_{\e{Td}} }^\pm$ are then defined as
\begin{equation}\label{qfromQ}
q_{ t_{\e{Td}} }^\pm(\la)\,=\,\frac{ \mf{q}_{ t_{\e{Td}} }^\pm(\la)}{\prod_{a=1}^{N+1} \big\{\ex{-\f{\pi \la}{\hbar}}\sinh\frac{\pi}{\hbar}(\la-\sg_a) \big\}}\,,
\end{equation}
where 
\beq\label{Qpmdefinition}
\mf{q}_{  t_{\e{Td}}  }^+(\la)\,=\, \f{ K^+_{  t_{\e{Td}}  }(\la) \cdot  \ex{-(N+1)\f{\pi}{\hbar}\la} }
	    {  \pl{k=1}{N+1} \Big\{ \hbar^{-\i\frac{\la}{\hbar}} \Ga\Big( 1 -  \i \f{\la-\tau_k}{\hbar} \Big)   \Big\} } \qquad \e{and} \quad  
\quad
\mf{q}_{  t_{\e{Td}}  }^-(\la)\,=\,\f{  K^-_{  t_{\e{Td}}  }(\la)  \cdot  \ex{-(N+1)\f{\pi}{\hbar}\la}  }
	    {   \pl{k=1}{N+1} \Big\{ \hbar^{\i\frac{\la}{\hbar}} \Ga\Big( 1 +  \i \f{\la-\tau_k}{\hbar} \Big)   \Big\}   }   \,. 
\enq
The functions $K^{\pm}_{  t_{\e{Td}}  }(\la)$ are expressed as half-infinite determinants:
\beq
K^+_{  t_{\e{Td}}  }(\la) \;  =  \; \det \pac{ \ba{cccccc}  1  & t^{-1}_{  t_{\e{Td}}  } (\la + \i \hbar) & 0 & \cdots    \\
            t^{-1}_{  t_{\e{Td}}  }(\la+2 \i \hbar)  & 1 & t^{-1}_{  t_{\e{Td}}  }(\la+2 \i \hbar) & 0 & \cdots \\
            0 & \ddots & \ddots & \ddots  &   \ddots  & \cdots
                                    \ea }
\label{definition determinant K+}
\enq
while  $K^-_{  t_{\e{Td}}  }(\la) \, = \, \big( K^+_{  t_{\e{Td}}  } (\la^{*}) \big)^*$ (recall that $\{ \tau_k \} = \{ \tau_k^* \} $). 
The proof that the two functions $q_{ _{  t_{\e{Td}}  } }^{\pm}$ do satisfy to all the requirements can be found in Appendix A of [$\bs{A20}$]. 

The functions $\mf{q}_{  t_{\e{Td}}  }^{\pm}$ are linearly independent entire functions whose discrete Wronskian can be evaluated explicitly, \textit{cf}. Lemma 1 of  [$\bs{A20}$] : 
\beq
\mf{q}_{  t_{\e{Td}}  }^+\!\pa{\la} \mf{q}_{  t_{\e{Td}}  }^-\!\pa{\la + \i \hbar} \, - \,  \mf{q}_{  t_{\e{Td}}  }^+\!\pa{\la + \i \hbar} \mf{q}_{  t_{\e{Td}}  }^-\!\pa{\la} \, = \, 
\ex{-2(N+1)\f{\pi}{\hbar}\la} \pl{a=1}{N+1} \Big\{ \f{ \hbar }{ \i \pi } \sinh\f{\pi}{\hbar}\pa{\la-\tau_a}  \Big\} \cdot \mc{H}_{  t_{\e{Td}}  }\pa{\la} \;.
\label{ecriture Wronskien determinant de Hill}
\enq

The general solution of \eqref{ecriture eqn TQ scalaire Toda coeur du Chapitre SoV Toda} with a polynomial subject to condition i) can be presented in the form 
\beq
q_{  t_{\e{Td}}  }(\la) \; = \; P^+ \mf{q}_{  t_{\e{Td}}  }^{+}(\la) \; + \;  P^- \mf{q}_{  t_{\e{Td}}  }^{-}(\la) \quad \e{with} \quad P^{\pm}\in \Cx \;. 
\enq

Thus, in order to obtain solutions $q_{  t_{\e{Td}}  }$ to the Baxter equation under the requirements i), ii) and iii), one has to fine tune the roots $\{\tau_a\}$ of the polynomial $t_{\e{Td}}$ in such a 
way that the parameters $\{\sg_a\}_{1}^{N+1}$ defining the roots of the associated Hill determinant $\mc{H}_{t_{\e{Td}}}$ satisfy to the condition 
\begin{equation}\label{q-cond}
\mf{q}_{t_{\e{Td}}}^+(\sg_a)-\zeta\, \mf{q}_{t_{\e{Td}}}^-(\sg_a)\,=\,0\,,\quad\text{for}\quad
a=1,\dots,{N+1}\, \quad \text{and} \; \text{some} \; \zeta \in \Cx, \;\; \abs{\zeta}=1\;.
\end{equation}
These $N+1$ equations should be supplemented by the condition fixing the total momentum $\sum_{k=1}^{N+1}\tau_k \,= \, \veps$.
This way of formulating the quantisation conditions for the Toda chain has been developed in \cite{GaudinPasquierQOpConstructionForTodaChain,GutzwillerResolutionTodaChainSmallNPaper1}. 
Clearly, it is a quite involved system of equations
on zeroes $\tau_k$ of $t_{\e{Td}}(\la)$, unreasonably exceeding in complexity  
the systems of Bethe Ansatz equations usually encountered for other models. Indeed, the equations on the $\tau_a$ not only involve 
some higher transcendental functions but also functions defined through infinite determinants. The worst, however,
is that the equations involve the parameters $\sg_1,\dots,\sg_{N+1}$ which have to be related to the roots $\tau_k$
by constructing the zeroes of the Hill determinant $\mc{H}_{t_{\e{Td}}}$. Without even mentioning any attempt to 
deal with these issues in some abstract way, dealing with the quantisation conditions on a numerical level already seems hopeless.

\subsection{The Nekrasov-Shatashvili quantisation}

In 2009, Nekrasov and Shatashvili \cite{NekrasovShatashviliConjectureTBADescriptionSpectrumIntModels} 
argued a way of recasting the  quantisation conditions for the Toda chain in terms of a function $Y_{ \{ \sg_a \} } $ satisfying to the
non-linear integral equation:
\begin{equation}\label{TBA}
\log Y_{ \{ \sg_a \} } (\la)\,=\,\int_{\R} \dd  \mu\;K(\la-\mu)\,\log\left( 1 +  \frac{ Y_{ \{ \sg_a \} }(\mu) }{ \vth(\mu-\i \hbar/2)\vth(\mu+ \i \hbar/2) }\right)\,,
\end{equation}
where
\begin{equation}\label{kerneldef}
K\pa{\la} = \f{ \hbar } { \pi ( \la^2+\hbar^2 ) }  \qquad \e{and} \qquad 
\vth(\la)=\prod_{k=1}^{N+1}(\la-\sg_k) \;.
\end{equation}
Building on the solution $Y_{ \{\sg_a\} }$, which they assumed to exists and be unique for each $\{ \sg_a \}$,  Nekrasov and Shatashvili defined a function, the so-called Yang potential,
$\mc{W}_{ \{n_a\} }\big( \{ \sg_a \} \big) \; = \; \mc{W}^{\e{inst}}_{ \{ n_a\} }\big( \{ \sg_a \} \big) \, + \,  \mc{W}^{\e{pert}}\big( \{ \sg_a \} \big)$. 
The two building blocks of Yang's potential are defined as
\beq
 \mc{W}^{\e{pert}}\big( \{ \sg_a \} \big)  \; = \; \i \sul{k=1}{N+1} \f{\sg_k^2}{2} \log\pa{\hbar^{\tf{(N+1)}{\hbar}}} \, - \,  \log \zeta \sul{k=1}{N+1} \sg_k
\, + \, \sul{j,k=1}{N+1} \ga\pa{\sg_k-\sg_j}  \,  -  \, 2 \i \pi \sul{k=1}{N+1} n_k \sg_k \,,
\enq
where $\ga^{\prime}\pa{\la}= \log \Ga\pa{1+ \i \tf{\la}{\hbar}}$ and 
\beq
\mc{W}^{\e{inst}}\big( \{ \sg_a \} \big) \; = \;  - \Int{\R}{} \paa{ \f{\log Y_{\{ \sg_a\} } (\mu) }{2} \log \pa{ 1+ \f{ Y_{\{ \sg_a\} }(\mu) }{ \abs{\vth\pa{\mu- \i\tf{\hbar}{2}}}^2 } }
 \, +  \, \e{Li}_2\pa{ \f{ - Y_{\{ \sg_a\} }(\mu) }{ \abs{\vth\pa{\mu-\i \tf{\hbar}{2}}}^2 } } } \f{ \dd \mu }{ 2 \i \pi } \;,
\enq
with
\beq
\e{Li}_2\pa{z} = \Int{z}{0} \f{ \log\pa{1-t}}{t} \dd t \;.
\enq
The Yang potential $\mc{W}_{ \{n_a\} }\big( \{ \sg_a \} \big)$ depends on a set of $N+1$ auxiliary integers $\{n_a\}$. In the Nekrasov-Shatashvili approach, the equations defining extrema of the Yang potential give rise to quantisation
conditions on the parameters $\{\sg_a\}_1^{N+1}$. Nekrasov and Shatashvili also argued that a set of parameters $\{ \sg_a \}_1^{N+1}$ corresponding to an extremum of the Yang potential
gives rise to $N+1$ eigenvalues $\mf{E}_k$, $k=1,\dots,N+1$ of N+1 Hamiltonians in involution associated with the model through the formula:
\beq
\mf{E}_k \; = \; \sul{p=1}{N+1} \sg_p^k \; - \; 
k \Int{\R}{} \f{\dd \tau}{2 \i \pi}  \paa{ \pa{\tau + \i \tf{\hbar}{2}}^{k-1}- \pa{\tau - \i \tf{\hbar}{2}}^{k-1} } \cdot \log \pa{  1+ \f{  Y_{\{ \sg_a\}}\pa{\tau} } { \vth\pa{\tau-\i \tf{\hbar}{2}} \vth\pa{\tau + \i \tf{\hbar}{2}}  }   }  \; .
\enq
They did not provide any reasonable characterisation of these Hamiltonians, though. 
Within the Nekrasov-Shatashvili approach, since the positions of the extrema of the Yang potential depend on the choice of the integers $\{n_a\}$, 
it is enough to vary the integers $\{n_a\}$ so as to obtain more eigenvalues.

The   Nekrasov-Shatashvili \cite{NekrasovShatashviliConjectureTBADescriptionSpectrumIntModels} proposal of quantisation conditions
was based on rather indirect arguments coming from the study of supersymmetric gauge theories. 
With Teschner, I managed to prove the above reformulation of the quantisation conditions 
by building on certain properties of the Baxter equation for the Toda chain, and, in particular,
on the Wronskian relation \eqref{ecriture Wronskien determinant de Hill}. The very fact that there  exits a connection between the Baxter 
equation and nonlinear integral equations has been observed in various works, see \textit{e}.\textit{g}.
\cite{BatchelorKlumperFirstIntoNLIEForFiniteSizeCorrectionSpin1XXZAlternativeToRootDensityMethod,BazhanovLukyanovZamolodchikovIntStructureofCFTQOpAndDDVEqns,DestriDeVegaAsymptoticAnalysisCountingFunctionAndFiniteSizeCorrectionsinTBAFirstpaper,
TeschnerSpectrumSinhGFiniteVolume,ZalmolodchikovTBAForSinhGordon}. 
Still, the precise form and structure of the NLIE heavily depends  on the
analytic properties imposed on the solutions to the Baxter equation of interest to a given model.
Further, the Toda chain particle system brings a new interesting feature into the game: the quantisation conditions \eqref{q-cond} are 
not formulated as some equations on the zeros of the solution to the Baxter equation, but instead, they 
are equations on the zeroes of the Wronskian formed out of the two linearly independent solutions of the 
Baxter equation. The locii $\{ \sg_1,\dots,\sg_{N+1} \}$ 
of these zeroes correspond precisely to the variables which arise in the Yang potential $\mc{W}_{ \{ n_a \} }\big( \{ \sg_a \} \big)$.

The first result we obtained with Teschner corresponds to the unique solvability of the non-linear integral equation 

\begin{theorem}
\label{Proposition Unicite et existence solutions}
 Let $\vth\pa{\la}$ be the polynomial $\vth\pa{\la}=\prod_{k=1}^{N+1}\pa{\la-\sg_k}$. Assume that its zeroes $\{\sg_k\}$,  $|\Im(\sg_k)| < \tf{\hbar}{2}$, 
are given by the $N+1$ zeroes of the Hill determinant built out of some polynomial $t(\la)$ subject to condition $i)$ and whose zeroes satisfy $|\Im(\tau_a)|< \tf{\hbar}{2}$. Then, there 
exists a function $Y_{\{ \sg_a\} } $
that is meromorphic on $\Cx$ with poles accumulating in the direction $\abs{\arg\pa{\la}}=\tf{\pi}{2}$. 
The function $Y_{\{ \sg_a\} } $ is bounded for $\la\tend \infty$
uniformly away from the set of its poles  and $Y_{\{ \sg_a\} } \tend 1$ for $\la \rightarrow \infty$
in any sector $\abs{\e{arg}\pa{\la}-\tf{\pi}{2}}\abs{\e{arg}\pa{\la}+\tf{\pi}{2}}>\eps$ for some fixed $\eps>0$. 

Finally, the function $\log Y_{\{ \sg_a\} } $ is continuous, positive and bounded on $\R$. 
It is the unique solution, in this class of functions, to the 
non-linear integral equation \eqref{TBA}.
\end{theorem}

This theorem is a consequence of Lemma 3 and Proposition 3  of  [$\bs{A20}$]. Although  Theorem \ref{Proposition Unicite et existence solutions} 
does not guarantee, \textit{a priori} the existence of solution for all possible choices of parameters $\{\sg_a\}_1^{N+1}$, it does however guarantee
the existence of solutions for all cases of interest to the study of solutions to the scalar $\op{t}-\op{Q}$ equation for the quantum Toda chain. 
Thus, in respect to applications to the Toda chain case, this potential lack of solution is not a restriction.

\vspace{2mm}

\noindent The properties of $Y_{\{ \sg_a\} } $ allow one to define two auxiliary functions:
\beq
\log v_{\uparrow}\pa{\la}  = - \Int{\R}{}  \f{\dd \mu}{2\i\pi}  
\f{1}{ \la-\mu +   \i \tf{\hbar}{2} } \cdot \log\left(1+ \f{  Y_{\{ \sg_a\} } \pa{\mu} }{ \vth\pa{\mu - \i\tf{\hbar}{2}}  \vth\pa{\mu + \i\tf{\hbar}{2}}  }\right)\,,
\enq
and
\beq
\log v_{\downarrow}\pa{\la- \i \hbar}  =  \Int{\R}{}  \f{\dd \mu}{2 \i \pi}  \f{1}{ \la-\mu - \i \tf{\hbar}{2}} \log\left(1+ \f{ Y_{\{ \sg_a\} } \pa{\mu} }{ 
\vth\pa{\mu - \i\tf{\hbar}{2}}  \vth\pa{\mu + \i\tf{\hbar}{2}}  }\right) \;.
\label{definition v up-down section vth}
\enq
The auxiliary functions $v_{\ua/\da}$ then gives rise to the functions 
\begin{equation}
\mf{q}_{\{ \sg_a\} }^+(\la)\, =\,    \f{  \hbar^{ \i\f{(N+1)\la}{\hbar} } \ex{-\f{(N+1)\pi}{\hbar}\la} \cdot v_{\uparrow}\pa{\la}}
{ \pl{k=1}{N+1} \Big\{ \Ga \Big(1- \i \f{\la-\sg_k}{\hbar} \Big) \Big\}  } 
\; \; ,\qquad
\mf{q}_{\{ \sg_a\} }^-(\la)\, =\,  \f{ \hbar^{-\i\f{(N+1)\la}{\hbar} }  \ex{-\f{(N+1)\pi}{\hbar}\la} \cdot v_{\downarrow}\pa{\la-i\hbar} }
{\pl{k=1}{N+1} \Big\{ \Ga \Big(1 +  \i \f{\la-\sg_k}{\hbar} \Big) \Big\}   }\,.
\label{definition Q vth plus/moins} 
\end{equation}
 Lemma 4 of Appendix C of [$\bs{A20}$] establishes that the functions $\mf{q}_{\{ \sg_a\} }^{\pm}$ are entire. 
Building on the Wronskian relation satisfied by $\mf{q}_{\{ \sg_a\} }^{\pm}$:
\beq
\mf{q}_{\{ \sg_a\}}^+(\la) \mf{q}_{\{ \sg_a\}}^-(\la+ \i\hbar) \,  - \, \mf{q}_{\{ \sg_a\}}^-(\la) \, \mf{q}_{\{ \sg_a\}}^+(\la+ \i \hbar) \,  = \, 
\bigg(\frac{\hbar \ex{-\f{2\pi\la}{\hbar}}}{ \i \pi}\bigg)^{N+1}  \pl{k=1}{N+1} \Big\{ \sinh\f{\pi}{\hbar}\pa{\la-\sg_k} \Big\} \; .
\label{q-Wronski}
\enq
Proposition 4 of that Appendix establishes that 
\beq
t_{\{ \sg_a\}} (\la) \; =  \;
\f{ \mf{q}_{\{ \sg_a\}}^{+}(\la-\i \hbar) \mf{q}_{\{ \sg_a\}}^{-}(\la +  \i \hbar)  \, - \,  \mf{q}_{\{ \sg_a\}}^{+}(\la + \i \hbar) \mf{q}_{\{ \sg_a\}}^{-}(\la - \i \hbar)  }
{\mf{q}_{\{ \sg_a\}}^+(\la) \mf{q}_{\{ \sg_a\}}^-(\la+i\hbar) \, - \, \mf{q}_{\{ \sg_a\}}^+(\la + \i \hbar) \mf{q}_{\{ \sg_a\}}^-(\la)  }  
\label{t-defn}
\enq
is a monoic polynomial in $\la$  of degree $N+1$ that has, furthermore, a self-conjugated set of roots.

 The main point is that the functions $q_{\{ \sg_a\}}^\pm(\la)$ defined from $\mf{q}_{\{ \sg_a\}}^\pm(\la)$ by analogues of 
 \eqref{qfromQ} are meromorphic solutions to the Baxter equation 
\eqref{ecriture eqn TQ scalaire Toda coeur du Chapitre SoV Toda} associated with the polynomial $t_{\{ \sg_a\}}$. Furthermore, 
the functions $q_{\{ \sg_a\}}^\pm(\la)$ satisfy to the bounds of condition ii) above.

The conclusion is that the solutions of the non-linear integral equation \eqref{TBA} 
provide one with an alternative construction of the fundamental system of solutions to the Baxter equations.
These are \textit{only} parametrised by the locii of their poles $\{ \sg_a\}$. 
One may therefore use directly the function $q_{\{ \sg_a\}}^\pm(\la)$ to construct Gutzwiller's solutions.
This has the advantage of recasting the quantization conditions \eqref{q-cond}  in the form
\begin{align}\label{ecriture conditions de quantification'}
2\pi n_k & = \f{ (N+1) \sg_k}{\hbar} \log \hbar   + \i \log \zeta
-\i \sul{p=1}{N+1} \log \frac{ \Gamma\big( 1+\i \tf{ (\sg_k-\sg_p) }{\hbar} \big) }{ \Gamma \big( 1 - \i \tf{ (\sg_k-\sg_p) }{\hbar} \big) } \\
& \quad+\Int{\R}{} \f{\dd \tau}{2\pi} \paa{ \f{1}{ \sg_k-\tau + \i \tf{\hbar}{2}}  +\f{1}{ \sg_k-\tau - \i \tf{\hbar}{2}}  }
\log \pa{  1+ \f{  Y_{\{ \sg_a\}}\pa{\tau} }{ \vth \pa{\tau - \i \tf{\hbar}{2} }  \vth \pa{ \tau + \i\tf{\hbar}{2} }  }   }  \;,
\nonumber\end{align}
see Appendix A of  [$\bs{A20}$] for some more details. These are precisely the quantisation conditions in the form argued by  Nekrasov and Shatashvili.

To close this section, I do stress that even though this way of representing the solutions to the $\op{t}-\op{Q}$ equations 
solely builds on the poles $\{\sg_a\}$ of the Hill determinant, it still does allow one to compute the associated set $\{\tau_a\}$ of zeroes of $t_{\e{Td}}$. 
More precisely, I have shown with Teschner that one has the below representation for the Newton polynomials in the zeroes $\tau_a$ : 

\beq
\label{Newtonreconstr}
\sul{p=1}{N+1} \tau^{k}_p \; = \; \sul{p=1}{N+1} \sg_p^k \; - \; 
k \Int{\R}{} \f{\dd \tau}{2 \i \pi}  \paa{ \pa{\tau + \i \tf{\hbar}{2}}^{k-1}- \pa{\tau - \i \tf{\hbar}{2}}^{k-1} } \cdot \log \pa{  1+ \f{  Y_{\{ \sg_a\}}\pa{\tau} } { \vth\pa{\tau-\i \tf{\hbar}{2}} \vth\pa{\tau + \i \tf{\hbar}{2}}  }   }  \; .
\nonumber
\enq
Our relation thus proves the Nekrasov-Shatashvili conjecture on the description of the spectrum and, on top of it, provides a clear identification of the commuting
family of Hamiltonians whose joint spectrum in obtained by this procedure.








\section{Unitarity of the separation of variables transform}
\label{Section unitarite SoV transform}

\subsection{The integral representations of the integral kernel}

I have recalled in the introductory chapter that the eigenfunctions of the quantum Toda chain, in a sector corresponding to the fixed value $\veps$ of the
total momentum, can be recast as
\beq
\Phi_{ \veps; t_{\e{Td}} }\big( \bs{x}_{N+1} \big) \; = \; \mc{U}_{N}\big[  \wt{\Phi}_{\veps; t_{\e{Td}} }(*,x_{N+1}) \big](\bs{x}_N) \qquad \e{where} \quad
\wt{\Phi}_{\veps; t_{\e{Td}} }\big( \bs{y}_N , x_{N+1} \big) \; = \; \ex{\f{\i}{\hbar}(\veps-\ov{\bs{y}}_N)x_{N+1}}  \pl{a=1}{N} \Big\{  q_{t_{\e{Td}}}(y_a)  \Big\} \;. 
\label{ecriture fctions propres Toda dans SoV}
\enq
I do remind that $\bs{x}_N=(x_1,\dots,x_N)$ stands for an $N$-dimensional vector while $\ov{\bs{x}}_N=\sum_{a=1}^{N}x_a$.  The
function $  q_{t_{\e{Td}}}(y_a) $ appearing in \eqref{ecriture fctions propres Toda dans SoV} solves the scalar $\op{t}-\op{Q}$ equation \eqref{ecriture eqn TQ scalaire Toda coeur du Chapitre SoV Toda} associated with the eigenvalue $t_{\e{Td}}$
of the transfer matrix. Furthermore, the transform $\mc{U}_{N}$ acts on the $\bs{y}_N$ variables of $\wt{\Phi}_{\veps; t_{\e{Td}} }$ and, given any 
$F \in L^{1}_{\e{sym}}\big(\R^N , \dd \mu(\bs{y}_N)\big)$,  takes the form 
\beq
\mc{U}_N[F] (\bs{x}_N) \; = \;\f{1}{ \sqrt{N!} }  \Int{ \R^N }{}  \vp_{\bs{y}_N}(\bs{x}_N) \cdot F(\bs{y}_N)  \cdot \dd \mu(\bs{y}_N)  \;. 
\label{definition transfo U_N chapitre SoV Toda Intro} 
\enq
The integration measure in \eqref{definition transfo U_N chapitre SoV Toda Intro}  is Lebesgue-continuous and reads
\beq
\dd \mu(\bs{y}_N) \; = \; \mu(\bs{y}_N) \cdot \dd^N y 
\quad \e{with} \quad  \mu(\bs{y}_N)  \; = \; 
\f{1}{(2\pi \hbar)^{N}}  \pl{ k \not= p }{ N } \Ga^{-1}\Big( \f{y_k-y_p}{ \i \hbar}  \Big) \;. 
\label{definition mesure Sklyanin}
\enq
The integral kernel $ \vp_{\bs{y}_N}(\bs{x}_N) $ can be interpreted as a Whittaker 
function associated with $GL(N,\R)$. Although this interpretation might appear interesting from the point of 
representation theory, it is not so advantageous form the point of view of a general development of the quantum separation of variables simply because
it only holds for the quantum Toda chain. If some properties of the $\mc{U}_N$ transform can be deduced from the general theory of Whittaker functions,
it will not be possible to generalise the result straightforwardly to more complex models solvable by the quantum separation of variables since the Whittaker function interpretation is absent.
As I have pointed out earlier on, one can construct two kinds of multiple integral representation for the integral
kernel by methods that solely rely on general algebraic features of the quantum inverse scattering method.
As a consequence, although the details of the construction may differ from one model to another, the overall structure of the construction will be common 
to numerous other quantum integrable models. 

\subsubsection{The Mellin-Bernes multiple integral representation}

The Mellin-Barnes multiple-integral representation issues from the resolution of a recurrence relation provided by the recursive structure
of the model's monodromy matrix
\beq
\op{T}_{0;1,\dots,N+1}(\la)  \; = \ \op{T}_{0;1,\dots,N}(\la) \cdot  \op{L}_{0N+1}(\la) \;. 
\label{ecriture equation recursive pour Mellin Barnes  niveau matrice de monodromie} 
\enq
The recursive scheme leading to Mellin-Barnes multiple integral representations has been proposed by Sklyanin in \cite{SklyaninSoVGeneralOverviewAndConstrRecVectPofB}
and has been applied to the Toda chain case by Kharchev and Lebedev \cite{KharchevLebedevIntRepEigenfctsPeriodicTodaFromRecConstrofEigenFctOfB}. 
For the model of interest, the main building block of this representation 
is the integral kernel:
\beq
\varpi( \bs{w}_{N} \mid \bs{y}_{N+1} ) \; = \; 
 \pl{a=1}{N} \pl{b=1}{N+1} \Big\{ \hbar^{\f{ \i }{\hbar}(w_a-y_b) } \Ga\Big( \f{y_b-w_a}{ \i \hbar} \Big) \Big\} 
 \cdot \pl{a \not= b}{N}  \Ga^{-1}\Big( \f{w_b-w_a}{ \i \hbar } \Big)  \;.
\enq
The Mellin-Barnes representation is an encased integral which couples the neighbouring sets of variables through the $\varpi$ kernel:
\beq
\vp_{\bs{y}_N}(\bs{x}_N) \; = \; \ex{ \f{ \i }{ \hbar }  \ov{\bs{y}}_{N} x_{N} }   \pl{s=1}{N-1} \Int{ (\R - \i \a_s)^{N-s} }{  } \hspace{-3mm} 
\f{ \dd^{N-s} w^{(s)}  }{ (N-s)! (2\pi \hbar)^{N-s} } 
\pl{s=1}{N-1}  \ex{ \f{ \i }{\hbar} \ov{\bs{w}}_{N-s}^{(s)} \big( x_{N-s} - x_{N-s+1} \big) } 
\pl{s=1}{N-1}\varpi( \bs{w}_{N-s}^{(s)} \mid \bs{w}_{N-s+1}^{(s-1)} )  \;,
\label{definition fonction vp apres resolution MellinBarnes}
\enq
where $0 < \a_1 < \dots < \a_{N-1}$ and  $\bs{w}_{N}^{(0)} = \bs{y}_N $. Also, this formula makes use of the previously introduced notations for $k$-dimensional vectors. 

One can prove that the multiple integral \eqref{definition fonction vp apres resolution MellinBarnes} converges strongly (exponentially fast), see \textit{e}.\textit{g}. 
\cite{GerasimovKharchevLebedevRepThandQISM}, and that it defines a smooth function
\beq
(\bs{y}_N,\bs{x}_N) \; \mapsto \;  \vp_{\bs{y}_N}(\bs{x}_N)  \; \in \; L^{\infty}(\R^N\times \R^N, \dd^N y \otimes \dd^N x) \;.  
\enq
The boundedness of $ \vp_{\bs{y}_N}(\bs{x}_N)$ ensures that the $\mc{U}_N$ transform is well-defined for any $F\in L^1\big( \R^N, \dd \mu(\bs{y}_N) \big)$. 
One can be even more specific.

\begin{prop}
\label{Proposition charactere de transfo UN}

Given any $F \in \msc{C}^{\infty}_{\e{c}}(\R^N)$, the integral transform $\mc{U}_N[F]$ is well 
defined and belongs to the Schwartz class $\mc{S}(\R^N)$. In particular,  for such functions $F$,
one has that $\mc{U}_N[F] \in L^{2}\big( \R^N, \dd^N x \big)$.  

\end{prop}

The proof of the statement can be found in Appendix A of [$\bs{A18}$].

\subsubsection{The Gauss-Givental representation}

The fundamental building blocks of the Gauss-Givental representation for $\vp_{\bs{y}_N}(\bs{x}_N)$ are extracted from the integral kernel of 
the chain's $\op{Q}$-operator which has been constructed by Gaudin-Pasquier  \cite{GaudinPasquierQOpConstructionForTodaChain}. This way of
constructing  the integral kernel of the separation of variables transform was proposed, for the first time, 
in \cite{DerkachovKorchemskyManashovXXXSoVandQopNewConstEigenfctsBOp} for the XXX $\mf{sl}(2,\Cx)$ non-compact chain. 

The integral kernel $Q_{\la}(\bs{x}_N,\bs{x}^{\prime}_N) $ of the $\op{Q}$-operator for the Toda chain takes the form  
\beq
Q_{\la}(\bs{x}_N,\bs{x}^{\prime}_N)   \; =  \; 
\pl{n=1}{N} \bigg\{ V_{\la;-}\big( x_n-x^{\prime}_n \big) V_{\la;+}\big( x_n-x^{\prime}_{n-1} \big)  \bigg\} \quad \e{where} \quad 
V_{\la ; \pm }(x) \; = \; \exp\Big\{ -\f{1}{\hbar} \ex{\pm x} \; + \; \i \f{\la x }{2 \hbar }  \Big\} \;. 
\label{ecriture noyau integral Q}
\enq
The building block of the Gauss-Givental representation is the function
\beq
\La_{\la}^{(N)} \big( \bs{x}_N \mid \bs{x}^{\prime}_{N-1} \big)  \; = \;  \ex{ \f{ \i \la }{ 2 \hbar } ( x_1 + x_N )  }
\pl{n=1}{N-1} \bigg\{ V_{\la;-}\big( x_n-x^{\prime}_n \big) \bigg\} \pl{n=2}{N} \bigg\{ V_{\la;+}\big( x_n-x^{\prime}_{n-1} \big)  \bigg\} \;. 
\enq
It provides one with the integral kernel of the integral operator

\beqa
\La^{(N)}_{\la} \quad  : \quad L^{\infty}(\R^{N-1})  &\tend &   L^{\infty}(\R^{N})  \nonumber \\
\phantom{ \La^{(N}_{\la}\quad  : \quad}  f  & \mapsto & \Int{ \R^{N-1} }{  } 
\La^{(N)}_{\la}(\bs{x}_N \mid \bs{z}_{N-1} ) \, f(\bs{z}_{N-1})  \; \pl{a=1}{N-1}\dd z_a \;. \nonumber
\eeqa
The integral kernel $\La_{\la}^{(N)} \big( \bs{x}_N \mid \bs{x}^{\prime}_{N-1} \big) $ arises as the non-trivial part of the $x_N^{\prime}\tend +\infty$ 
and point-wise in the other variables $(\bs{x}_N,\bs{x}_{N-1}^{\prime})$ behaviour of the integral kernel of the $\op{Q}$-operator:
\beq
Q_{\la}\big( \bs{x}_N,\bs{x}^{\prime}_N \big)  \; =  \; \La_{\la}^{(N)} \big( \bs{x}_N \mid \bs{x}^{\prime}_{N-1} \big) 
\,  \ex{-\i \f{\la}{\hbar} x^{\prime}_N } \, 
\exp\Big\{  - \f{ 1 }{ \hbar } \ex{x^{\prime}_N-x_N^{} }  \Big\}  \cdot \big( 1+\e{o}(1) \big) \;. 
\enq
The Gauss-Givental representation for the integral kernel is then realised as the encased action of the operators $\La^{(N)}_y$, namely
\beq
\vp_{\bs{y}_N}(\bs{x}_N) \; = \; \big( \La^{(N)}_{y_1} \cdots  \La^{(1)}_{y_N} \big)(\bs{x}_N)   \; .  
\label{ecriture forme pyramidale fct vp}
\enq
One can show that this encased action is well-defined and that it defines a symmetric function of $y_1,\dots, y_N$. This last statement 
follows from the operator identity $\La_{\la}^{(N)} \La_{\mu}^{(N-1)} = \La_{\mu}^{(N)} \La_{\la}^{(N-1)}  $, \textit{c}.\textit{f}. Proposition 1.2 of [$\bs{A18}$]. 
The first statement follows, in its turn, from Lemma 1.2 of the same paper.

It is useful to introduce the operator $\ov{\La}^{\pa{N}}_{\la}$ conjugated to $\La_{\la}^{(N)}$ :
\beqa
\ov{\La}^{\pa{N}}_{\la} \quad  : \quad  \mc{S}(\R^N)  &\tend &  L^{\infty}(\R^{N-1}) \\
\phantom{ \ov{\La}^{\, \pa{N}}_{\la} \quad  : \quad}  f  & \mapsto & \Int{\R^N}{} 
 {\ov{\La} }^{(N)}_{\la} \big( \bs{z}_{N-1} \mid \bs{x}_{N} \big)   f(\bs{x}_N)  \,  \pl{a=1}{N}\dd  x_a \;,
\eeqa
where $\mc{S}(\R^N)$ refers to Schwartz functions on $\R^N$ and 
\beq
\ov{\La}_{\la}^{ (N) } \big( \bs{z}_{N-1} \mid \bs{x}_{N}  \big)  \;  = \; \ex{ - \f{ \i \la }{ 2 \hbar } (x_1+x_N) }
\pl{n=1}{N-1} \bigg\{ V_{- \la; - }\big( x_n-z_n \big) \bigg\} \cdot \pl{n=2}{N} \bigg\{  V_{- \la;+ }\big( x_n-z_{n-1} \big)\bigg\}  \;.
\enq

Provided a proper regularisation of the operator products is introduced, the operators $ \ov{\La}_{ y^{\prime} }^{(N)}$ and $ \La_{ y }^{(N)} $ satisfy the identity 
\beq
 \ov{\La}_{ y^{\prime} }^{(N)} \cdot \La_{ y }^{(N)}   
 \; =  \; 
 \Ga\Big( \f{ y - y^{\prime} }{ \i \hbar } \Big)  \Ga\Big(  \f{ y^{\prime} - y  }{ \i \hbar } \Big) \cdot   \La_{ y }^{(N-1)} \cdot \ov{\La}_{ y^{\prime} }^{(N-1)}  
\label{ecriture relation d'echange entre operateurs Lambda et Lambda bar}
\enq
when $y\not=y^{\prime}$. See Lemma 1.3 of [$\bs{A18}$] for a precise statement.

\vspace{3mm}

I have described two quite different types of multiple integral representations for the functions $\vp_{\bs{y}_N}(\bs{x}_N)$. 
\textit{A priori}, it is not at all clear if these define the same function. This has been established in \cite{GerasimovLebedevOblezinBAxtOpMixedRepsForTodaWhittakerAndMore}
by using the results on completeness and orthogonality of the Whittaker functions $\vp_{\bs{y}_N}(\bs{x}_N)$. 
In Appendix B of  [$\bs{A18}$] I proved, without using any information on completeness or orthogonality, 
that the functions defined by the \textit{rhs} of \eqref{definition fonction vp apres resolution MellinBarnes} and of \eqref{ecriture forme pyramidale fct vp} coincide.  
See Proposition B.1 of that paper for more details.

\subsection{Heuristics of the proof}

The main result of [$\bs{A18}$] was to establish the below theorem \textit{solely}
by relying on properties and objects that arise naturally within the quantum inverse scattering method. 
My work thus opens the possibility of establishing analogous results for more complex models solvable by the quantum separation of variables
method where the group theoretic interpretation of the Toda chain is missing, and hence where one cannot rely on some generalisation 
of the results of \cite{WallachRealReductiveGroupsII}. I also stress that my proof is completely independent from the previous scheme of works 
that require a group theoretical interpretation of the model in the spirit of \cite{GoodmannWallachQuantumTodaI,
GoodmannWallachQuantumTodaII,GoodmannWallachQuantumTodaIII,KostantIdentificationOfEigenfunctionsOpenTodaAndWhittakerVectors}. 
Moreover, the proof I proposed is conceptually simple and relatively short: on a formal level of rigour, it is almost immediate to implement.
Finally, its steps do not rely on any property specifically associated with the Toda chain
but rather on some algebraic properties of the Mellin-Barnes and Gauss-Givental integral representations. 
As a consequence, it appear likely that the method can be adapted so as to prove the unitarity of separation of variables transforms which appears in other quantum integrable models solvable by the method.

\begin{theorem}
 The integral transform $\mc{U}_N$ defined by \eqref{definition transfo U_N chapitre SoV Toda Intro}  for functions 
$ F \in \msc{C}^{\infty}_{\e{c};\e{sym}}\big(\R^N \big)$ extends to  
a unitary map $\mc{U}_N \; : \; L^2_{\e{sym}}\big(\R^N, \dd \mu(\bs{y}_N) \big)  \;  \tend  \; L^2\big(\R^N, \dd^N x \big)  $. 
\end{theorem}

The proof goes in two steps. One first shows the isometric nature of $\mc{U}_N$ and then the isometric nature of its formal adjoint $\ov{\mc{V}}_N$. 
The isometricity then ensures that $\ov{\mc{V}}_N$ is, in fact, the adjoint of $\mc{U}_N$ hence proving the above theorem. 
I will comment on the main steps leading to these results below. 

\subsubsection{Isometric nature of $\mc{U}_N$}

This property is resumed by the 
\begin{theorem}
\label{Theorem completude en x de la transfo SoV}
The map $\mc{U}_N$ defined for $\big(L^1\cap L^2\big)_{\e{sym}}\big(\R^N, \dd\mu(\bs{y}_N) \big)$ functions by the integral
transform \eqref{definition transfo U_N chapitre SoV Toda Intro}  extends into an isometric linear map 
\beq
 \mc{U}_N \; : \;  L^{2}_{\e{sym}} \big( \R^N, \dd \mu(\bs{y}_N) \big) \;  \tend \;   L^{2}\big( \R^N, \dd^N x \big)  \;. 
\enq
In other words, one has the equality 
\beq
\norm{ \mc{U}_N[F] }_{ L^{2}\big( \R^N, \dd^N x\big) }   \; = \; 
\norm{ F }_{ L^{2}_{\e{sym}}\big( \R^N,\dd \mu(\bs{y}_N) \big) } \;. 
\enq

\end{theorem}

The proof of this theorem can be found in Section 2 of [$\bs{A18}$]. Below, I will only discuss the formal handlings leading to the result. 

On a formal level of rigour, the statement boils down to the relation 
\beq
 \Int{ \R^{N} }{} \Big( \vp_{\bs{y}_{N}^{\prime}} (\bs{x}_{N}) \Big)^{*}   \cdot 
 \vp_{\bs{y}_{N}} (\bs{x}_{N}) \cdot \dd^{N} x  \; = \; 
 \big[  \mu(\bs{y}_N) \big]^{-1}  
\sul{ \sg \in \mf{S}_N }{}  \pl{a=1}{N} \de\big( y_a - y^{\prime}_{\sg(a)} \big) \;, 
\label{ecriture relation completude par rapport espace originel}
\enq
Assume that $ y_1< \dots < y_N$, $ y_1^{\prime}< \dots < y_N^{\prime}$ and that $y_N^{\prime} \not= y_1,\dots, y_{N-1}$. 
Denote the integral on the \textit{lhs} as $\Pi_N\big(\bs{y}_N^{\prime}, \bs{y}_N \big) $. Then, by \eqref{ecriture forme pyramidale fct vp}, one can recast this quantity as the below product of operators. 
\beq
 \Pi_N\big(\bs{y}_N^{\prime}, \bs{y}_N \big)  \; = \; \ov{\La}^{(1)}_{ y_1^{\prime} } \dots  \ov{\La}^{(N)}_{ y_N^{\prime} }  \La^{(N)}_{y_1} \dots  \La^{(1)}_{y_N} \;. 
\enq
One moves the operator $\ov{\La}^{(N)}_{ y_N^{\prime} } $ through the string of $\La$ operators by means of a repetitive application of \eqref{ecriture relation d'echange entre operateurs Lambda et Lambda bar} and, 
at the very last stage, applies the identity $  \ov{\La}^{(1)}_{ y_N^{\prime} } \La^{(1)}_{y_N} \; = \; 2\pi \hbar \de (y_N^{\prime} - y_N)$. 
This leads to the induction 
\beq
 \Pi_N\big(\bs{y}_N^{\prime}, \bs{y}_N \big)  \; = \; 2\pi \hbar \de (y_N^{\prime} - y_N) \pl{a=1}{N-1} \bigg\{ \Ga\Big( \f{ y_a - y^{\prime}_N }{ \i \hbar } \Big)  \Ga\Big(  \f{ y^{\prime}_N - y_a  }{ \i \hbar } \Big) \bigg\} 
 \cdot \Pi_{N-1}\big(\bs{y}_{N-1}^{\prime}, \bs{y}_{N-1} \big)  \; , 
\enq
which is straightforward to solve. The formal identity \eqref{ecriture relation completude par rapport espace originel} then follows upon symmetrising the solution of the induction. 
Although formal, the simplicity of the handlings is striking. The formal method was first developed in \cite{DerkachovKorchemskyManashovXXXSoVandQopNewConstEigenfctsBOp} 
and applied to the case of the Toda chain in \cite{SilantyevScalarProductFormulaTodaChain}. This paper improved the rigour of certain formal steps.
In Section 2 of [$\bs{A18}$], I managed to get all the steps rigorous.

\subsubsection{Isometric nature of $ \mc{U}_N^{\dagger} $}

The formal adjoint of $\mc{U}_N$ is given by the map $\ov{\mc{V}}_N$ defined on $L^1\big( \R^N, \dd^N x \big)$ through 
\beq
\ov{\mc{V}}_N[F](\bs{y}_N) \; = \; \f{ 1 }{ \sqrt{N!} } \Int{ \R^N }{} \Big( \vp_{\bs{y}_N}(\bs{x}_N) \Big)^* \cdot F(\bs{x}_N) \cdot \dd^{N} x \; .
\enq
Using that $\Big( \vp_{\bs{y}_N}(\bs{x}_N) \Big)^*  \; = \;  \vp_{-\bs{y}_N}(\bs{x}_N)  $, the isometricity 
of $\ov{\mc{V}}_N$ is equivalent to the one of  the operator $\mc{V}_N $
whose action on $\big(L^1 \cap L^2 \big)\big( \R^N, \dd^N x \big)$ is given by the integral transform
\beq
\mc{V}_N[F](\bs{y}_N) \; = \; \f{ 1 }{ \sqrt{N!} } \Int{ \R^N }{} \vp_{\bs{y}_N}(\bs{x}_N) F(\bs{x}_N) \cdot \dd^{N} x \;. 
\label{definition transformation VN}
\enq
This isometricity is guaranteed by the 

\begin{theorem}

The transform $\mc{V}_N$ defined through \eqref{definition transformation VN} is such that 
given  any
\beq
F\in \msc{C}^{\infty}_{\e{c}}\big( \R^N \big) \; ,  \qquad 
\mc{V}_N\big[ F\big]  \in \mc{S}(\R^N )\cap L^{2}_{\e{sym}}\big(\R^N, \dd \mu(\bs{y}_N) \big) \;. 
\enq
$\mc{V}_N$ extends to an isometric operator $\mc{V}_N \; : \; L^2\big(\R^N, \dd^N x \big) \; \tend \;  L^{2}_{\e{sym}}\big(\R^N, \dd\mu(\bs{y}_N) \big)$:
\beq
\big|\big|  \mc{V}_N\big[ F\big]     \big|\big|_{L^{2}_{\e{sym}}\big( \R^N, \dd \mu(\bs{y}_N) \big) } \; = \; 
 || F ||_{L^{2}\big(\R^N, \dd^N x\big)} \;. 
\enq

\end{theorem}

The method which allowed me to prove the isometric nature of $\mc{V}_N$ 
has never been proposed, even on a formal level of rigour, previously. 
The handlings leading to the result are from the point of view of formal manipulations, quite simple. 
The main idea of the proof is that the orthogonality of the system $ \{ \vp_{\bs{y}_N}(\bs{x}_N) \}$ obtained by means of the Gauss-Givental representation
translates, when expressing it though the Mellin-Barnes multiple integral representation for $\vp_{\bs{y}_N}$, into 
a set of integral identities depending on free external parameters $\bs{y}_N$ and $\bs{y}_N^{\prime}$. 
Sending some of the variables in $\bs{y}_N$ and $\bs{y}_N^{\prime}$ to infinity in a specific way then yeilds the precise integral identity that is 
necessary for proving completeness. Below, I will explain how the procedure works on a formal level of rigour. 
The proof can be found in Section 3 of [$\bs{A18}$]. 
The main point is that there seems to exist a sort of hidden duality between the isometric character of the 
operator $\mc{U}_N$ and $\mc{V}_N$. As soon as one isometry is proven,
the second one follows from it.

Formally, the isometricity of the $\mc{V}_N$ transform boils down to the completeness of the system $ \{ \vp_{\bs{y}_N}(\bs{x}_N) \}$
\beq
\chi_N\big(\bs{x}_N^{\prime},\bs{x}_N\big) \; = \; \Int{ \R^{N} }{} \Big( \vp_{\bs{y}_{N}} (\bs{x}_{N}^{\prime}) \Big)^{*}  \vp_{\bs{y}_{N}} (\bs{x}_{N})\; 
 \cdot   \dd \mu(\bs{y}_N)  \; = \; 
\pl{a=1}{N+1} \de(x_a - x^{\prime}_{a}) \;, 
\label{ecriture relation completude par rapport espace SoV}
\enq
By using the recursive form of the Mellin-Barnes integral representation 
\beq
\vp_{\bs{y}_{N}}(\bs{x}_{N}) \; = \; \Int{ (\R - \i \a)^{N} }{  } \hspace{-3mm} 
\ex{ \f{ \i }{ \hbar } (\ov{\bs{y}}_{N} - \ov{\bs{w}}_{N-1}) x_{N} } 
\vp_{\bs{w}_{N-1}}(\bs{x}_{N-1}) \varpi( \bs{w}_{N-1} \mid \bs{y}_{N} ) \cdot \f{ \dd^{N-1} w  }{ (N-1)! (2\pi \hbar)^{N-1} } \qquad \e{with} \; \; \a >0 
\label{definition fct Whittaker par Mellin-Barnes}
\enq
one is able to recast $\chi_N\big(\bs{x}_N^{\prime},\bs{x}_N\big)$ as
\beq
\chi_N\big(\bs{x}_N^{\prime},\bs{x}_N\big) \; = \hspace{-2mm}  \Int{ \R^{N-1} \times \R^{N-1} }{  } \hspace{-4mm} 
\f{  \ex{ \f{\i}{\hbar} ( \ov{\bs{w}}_{N-1}^{\prime} x_N^{\prime} - \ov{\bs{w}}_{N-1} x_N) }  }
{  \hbar^{ \tf{\i ( \ov{\bs{w}}_{N-1}^{\prime} - \ov{\bs{w}}_{N-1}) }{\hbar}  }  }\cdot 
\Big( \vp_{ \bs{w}_{N-1}^{\prime} } (\bs{x}_{N-1}^{\prime}) \Big)^{*}  \vp_{\bs{w}_{N-1}} (\bs{x}_{N-1}) 
\cdot  
\psi\big(\bs{w}_{N-1},\bs{w}_{N-1}^{\prime}; x_N-x_N^{\prime} \big)  \cdot \f{ \dd\mu(\bs{w}_{N-1}) \dd \mu(\bs{w}_{N-1}^{\prime} ) }{ (N-1)! }
\nonumber
\enq
where $\psi$ is a distribution defined in terms of the formal integral representation
\beq
\psi\big(\bs{w}_{N-1},\bs{w}_{N-1}^{\prime}; x_N-x_N^{\prime} \big)   \; = \; 
\Int{ \R^{N} }{}  \ex{ \f{\i}{\hbar} \ov{\bs{y}}_N (x_N-x^{\prime}_N) }
\cdot     \pl{a=1}{N-1} \pl{b=1}{N} \bigg\{ \Ga \Big(  \f{ y_b -w_a + \i 0^+}{ \i \hbar } ,  \f{ w_a^{\prime} - y_b + \i 0^+}{ \i \hbar}  \Big) \bigg\} \cdot  \f{ \dd\mu(\bs{y}_{N}) }{ (N-1)! } \;. 
\enq
The proof of orthogonality boils down to finding an alternative representation for the distribution $\psi$. I will argue below that the latter can be recast as 
\beq
\psi\big(\bs{w}_{N-1},\bs{w}_{N-1}^{\prime}; x_N-x_N^{\prime} \big) \; = \; \f{ \de\big( x_N \, - \,  x_N^{\prime} \big) }{  \mu( \bs{w}_{N-1} ) } 
\sul{ \sg\in \mf{S}_{N-1} }{}    \pl{a=1}{N-1} \de\big( w_a - w^{\prime}_{\sg(a)} \big) \;. 
\label{ecriture representation distribution psi dans unitarite U}
\enq
This being settled, one obtains the recurrence
\beq
\chi_N\big(\bs{x}_N^{\prime},\bs{x}_N\big) \; = \; \de\big( x_N \, - \,  x_N^{\prime} \big)  \cdot \chi_{N-1}\big( \bs{x}_{N-1}^{\prime} , \bs{x}_{N-1} \big) \;. 
\enq
what readily allows one to establish completeness by induction on $N$.

In order to establish \eqref{ecriture representation distribution psi dans unitarite U}, one rewrites the orthogonality condition obtained by means of the 
Gauss-Givental representation on the level of the inductive construction of the Mellin-Barnes integral representation. This yields the identity 
\beq
\sul{ \sg\in \mf{S}_{N} }{}   \f{ \prod_{a=1}^{N} \de\big( w_a - w^{\prime}_{\sg(a)} \big) }{ \mu( \bs{w}_{N} ) }
\;= \; 2\pi \hbar \cdot \de\big( \ov{\bs{w}}_N - \ov{\bs{w}}^{\prime}_{N} \big)  \Int{ \R^{N-1} }{} 
   \pl{a=1}{N-1} \pl{b=1}{N} \bigg\{ \Ga \Big(  \f{ y_a -w_b^{\prime} + \i 0^+}{ \i \hbar } ,  \f{ w_b - y_a + \i 0^+}{ \i \hbar}  \Big) \bigg\} \, \cdot \,  \f{ \dd\mu(\bs{y}_{N-1}) }{ (N-1)! } \;. 
\label{premiere identite pour completude}
\enq
Upon multiplying both sides by the function
\beq
\ex{ \f{\pi}{2\hbar} (N-1)(\ov{\bs{w}}_{N-1} -\ov{\bs{w}}_{N-1}^{\prime} ) }\cdot  \bigg( \f{K}{\hbar} \bigg)^{ \f{\pi}{\i\hbar} (N-1)(\ov{\bs{w}}_{N-1} -\ov{\bs{w}}_{N-1}^{\prime} ) } \; , 
\enq
substituting $w_N \hookrightarrow w_N + K $,  $w_N^{\prime} \hookrightarrow w_N^{\prime} + K $ and then sending $K\tend+ \infty$ one observes that the sole terms
in the sum over permutations in the \textit{lhs} of \eqref{premiere identite pour completude} which survives to the limit correspond to permutations $\sg \in \mf{S}_N$
stabilising $N$, \textit{viz}. $\sg(N)=N$. Since $\de\big( \ov{\bs{w}}_N - \ov{\bs{w}}^{\prime}_{N} \big)$ appears in both sides of \eqref{premiere identite pour completude}, 
one can simplify this factor provided a proper substitution of the variable $w^{\prime}_N$ is made. Further, moving all the dependence 
on $w_N$ to the \textit{rhs}, one gets the identity
\bem
\sul{ \sg\in \mf{S}_{N-1} }{}   \f{ \prod_{a=1}^{N-1} \de\big( w_a - w^{\prime}_{\sg(a)} \big) }{ \mu( \bs{w}_{N-1} ) }
\;= \;  \Int{ \R^{N-1} }{}  \pl{a,b=1}{N-1} \bigg\{ \Ga \Big(  \f{ y_a -w_b^{\prime} + \i 0^+}{ \i \hbar } ,  \f{ w_b - y_a + \i 0^+}{ \i \hbar}  \Big) \bigg\} \cdot \ex{ \f{\pi}{\hbar}(\ov{\bs{y}}_{N-1}-\ov{\bs{w}}_{N-1} )(1-0^+)  } \\
\times \lim_{K\tend + \infty} \Big\{ \mc{T}_K\big( \bs{y}_{N-1},\bs{w}_{N-1},\bs{w}_{N-1}^{\prime} \big) \Big\}  \cdot \f{ \dd\mu(\bs{y}_{N-1}) }{ (N-1)! }  \;. 
\label{deuième identite pour completude}
\end{multline}
There, one has
\bem
 \mc{T}_K\big( \bs{y}_{N-1},\bs{w}_{N-1},\bs{w}_{N-1}^{\prime} \big) \; = \;    \pl{a=1}{N-1} \bigg\{ \Ga \Big(  \f{ y_a -\ov{\bs{w}}_{N}+\ov{\bs{w}}_{N-1}^{\prime}  + \i 0^+}{ \i \hbar } ,  \f{ w_N + K- y_a + \i 0^+}{ \i \hbar}  \Big) \bigg\} 
 \cdot \ex{ - \f{\pi}{\hbar}(\ov{\bs{y}}_{N-1}-\ov{\bs{w}}_{N-1} )(1-0^+)  } \\
\times   \pl{a=1}{N-1}\bigg\{ \Ga \Big(  \f{ w_N -w_a +K }{ \i \hbar } ,  \f{ w_a - K - w_N}{ \i \hbar}  \Big) \bigg\}^{-1}
\cdot \ex{ \f{\pi}{2\hbar} (N-1)(\ov{\bs{w}}_{N-1} -\ov{\bs{w}}_{N-1}^{\prime} ) } \cdot \Big( \f{K}{\hbar} \Big)^{ \f{\pi}{\i\hbar} (N-1)(\ov{\bs{w}}_{N-1} -\ov{\bs{w}}_{N-1}^{\prime} ) } \;. 
\end{multline}
The $(1-0^+)$ regularisation is present so as to ensure the convergence of the integral at infinity.
It is readily seen that 
\beq
\lim_{K\tend + \infty} \Big\{ \mc{T}_K\big( \bs{y}_{N-1},\bs{w}_{N-1},\bs{w}_{N-1}^{\prime} \big)  \Big\}  \; = \; 1
\enq
hence providing one with an auxiliary multiple integral representation for a symmetrised multi-variable $\de$-function. Starting from \eqref{deuième identite pour completude}
with the $K\tend + \infty$ limit replaced by 1, one multiples both sides by 
\beq
\Ga\Big(  \f{ K \ex{s^{\prime} } }{ \i \hbar } \Big) \cdot \Ga^{-1}\Big(  \f{ K \ex{s } }{ \i \hbar } \Big) \qquad \e{and} \; \e{substitues} \quad
w_{N-1} \hookrightarrow -K\ex{s} \quad w_{N-1}^{\prime} \hookrightarrow  - K \ex{ s^{\prime} } \quad 
\enq
in terms of new variables $s, s^{\prime}$ and, finally, moves all the $K$ and $s$ dependence to the \textit{rhs}. Then, sending $K\tend +\infty$, yields
\bem
\de(s-s^{\prime}) \sul{ \sg\in \mf{S}_{N-2} }{}   \f{ \prod_{a=1}^{N-2} \de\big( w_a - w^{\prime}_{\sg(a)} \big) }{ \mu( \bs{w}_{N-2} ) }
\;= \;  \Int{ \R^{N-1} }{} 
\pl{a=1}{N-1}\pl{b=1}{N-2}  \bigg\{ \Ga \Big(  \f{ y_a -w_b^{\prime} + \i 0^+}{ \i \hbar } ,  \f{ w_b - y_a + \i 0^+}{ \i \hbar}  \Big) \bigg\} \cdot \ex{\f{\i}{\hbar}(s-s^{\prime})\ov{\bs{y}}_{N-1} }   \\
\times \lim_{K\tend + \infty} \Big\{ \mc{R}_K\big( \bs{y}_{N-1},\bs{w}_{N-2},\bs{w}_{N-2}^{\prime} ; s, s^{\prime} \big) \Big\}  \cdot \f{ \dd\mu(\bs{y}_{N-1}) }{ (N-1)! }  \;. 
\label{deuième identite pour completude}
\end{multline}
The function under the limit takes the form 
\bem
\mc{R}_K\big( \bs{y}_{N-1},\bs{w}_{N-2},\bs{w}_{N-2}^{\prime} ; s, s^{\prime} \big) \; = \;  \ex{ \f{\pi}{\hbar}(\ov{\bs{y}}_{N-1}-\ov{\bs{w}}_{N-1} )(1-0^+)  }
\cdot \ex{- \f{\i}{\hbar}(s-s^{\prime})\ov{\bs{y}}_{N-1} } \cdot \Ga\bigg(  \f{ K \ex{s^{\prime} } }{ \i \hbar } \bigg) \cdot \Ga^{-1}\bigg(  \f{ K \ex{s } }{ \i \hbar } \bigg) \cdot 
  \\
\times  \pl{a=1}{N-1} \bigg\{ \Ga \Big(  \f{ y_a + K \ex{s^{\prime}}   + \i 0^+}{ \i \hbar } ,  \f{ -K \ex{s }- y_a + \i 0^+}{ \i \hbar}  \Big) \bigg\}  \cdot 
\pl{a=1}{N-1}\bigg\{ \Ga \Big(  \f{K \ex{s } +w_a }{ \i \hbar } ,  \f{ -w_a - K \ex{s }}{ \i \hbar}  \Big) \bigg\}^{-1}  
\end{multline}
and is such that 
\beq
\lim_{K\tend + \infty} \Big\{ \mc{R}_K\big( \bs{y}_{N-1},\bs{w}_{N-2},\bs{w}_{N-2}^{\prime} ; s, s^{\prime} \big) \Big\}  \; = \; 1 \;. 
\enq
This yields the sought identity.





\section{Inverse problem for the Toda chain}
\label{Section probleme inverse Toda}

\subsection{Equations in dual variables}

It follows from the discussion carried out at the beginning of Section \ref{Section unitarite SoV transform} that, in a sector corresponding to 
a fixed total momentum $\veps$, the eigenfunctions of the quantum Toda chain take the generic form 
\beq
\Phi_{ \veps; t_{\e{Td}} }\big( \bs{x}_{N+1} \big) \; = \;  \Int{ \R^N }{} \Psi_{\bs{y}_N,\veps }\big( \bs{x}_{N+1} \big)  \cdot \pl{a=1}{N} \Big\{  q_{t_{\e{Td}}}(y_a)  \Big\} \cdot \f{ \dd \mu(\bs{y}_N)  }{ \sqrt{N!} }
\quad \e{where} \quad
\Psi_{\bs{y}_N,\veps }\big( \bs{x}_{N+1} \big) \; = \; \ex{\f{\i}{\hbar}(\veps-\ov{\bs{y}}_N)x_{N+1}}   \vp_{\bs{y}_N}(\bs{x}_N) \; . 
\label{ecriture fct propre Toda} 
\enq

This means that in order to obtain the action of a local operator $\op{O}$ on the eigenfunction-part of the above integral representation it is enough to obtain an equation in dual variables as described in the introduction
to the chapter
\beq
\op{O} \cdot \Psi_{\bs{y}_N,\veps }\big( \bs{x}_{N+1} \big)  \; = \; \wh{\op{O}}\cdot \Psi_{\bs{y}_N,\veps }\big( \bs{x}_{N+1} \big) 
\label{ecriture equation aux variables duales}
\enq
where  $\wh{\op{O}}$ acts on the dual variables $( \bs{y}_N,\veps )$. Once that one disposes of such an equation, it solely remains 
to use contour deformations or integration by parts so as to move the action of the operator $\wh{\op{O}}$ on the eigenfunction part $\prod_{a=1}^{N} \big\{  q_{t_{\e{Td}}}(y_a)  \big\}$.

In order to establish equations in dual variables \eqref{ecriture equation aux variables duales}, it is important to have well-suited representation 
for the integral kernel $\Psi_{\bs{y}_N,\veps }\big( \bs{x}_{N+1} \big)$ at one's disposal. In \cite{BabelonActionPositionOpsWhittakerFctions}, Babelon managed to determine
such equations for the local operators
\beq
\pl{a=1}{r} \Big\{ \ex{ \op{x}_a - \op{x}_{N+1} } \Big\}
\enq
by building on the Mellin-Barnes integral representation \eqref{definition fonction vp apres resolution MellinBarnes} for the functions $\Psi_{\bs{y}_N,\veps }\big( \bs{x}_{N+1} \big)$. 
As I will show in the following, it is indeed the one that is most adapted for such calculations. 

However, first, I would like to remind that there exists a privileged family of operators on $\mf{h}_{\e{Td}}$ for which it is easy to build systems of equations in dual variables: 
the entries of the monodromy matrix of the model.

\begin{prop}
\label{propositon action operateurs ABD}

The function $\Psi_{\bs{y}_N,\veps }\big( \bs{x}_{N+1} \big)$ satisfies to the equations in dual variables: 
\beq
\big[ \op{D}_{1,N+1}(\la) \Psi_{\bs{y}_N, \veps} \big] (\bs{x}_{N+1}) \; = \; 
\sul{p=1}{N} (-\i)^{N+1} \pl{ \substack{ s=1 \\ \not=p} }{ N } \Big( \f{\la-y_s}{y_p-y_s} \Big) 
\cdot  \Psi_{\bs{y}_N- \i \hbar \bs{e}_p, \veps} (\bs{x}_{N+1}) \;, 
\label{ecriture action D sur Psi}
\enq
as well as
\beq
\big[ \op{A}_{1,N+1}(\la) \Psi_{\bs{y}_N, \veps} \big] (\bs{x}_{N+1}) 
\; = \; \big( \la-\veps + \ov{\bs{y}}_N \big)  \pl{\ell=1}{N}(\la-y_{\ell}) \Psi_{\bs{y}_N, \veps} (\bs{x}_{N+1})
\; + \; \sul{p=1}{N} (\i)^{N+1} \pl{ \substack{ s=1 \\ \not=p} }{ N } \Big( \f{\la-y_s}{y_p-y_s} \Big) 
\cdot  \Psi_{\bs{y}_N+\i\hbar \bs{e}_p, \veps} (\bs{x}_{N+1}) \;. 
\label{ecriture action A sur Psi}
\enq
There $\bs{e}_p$ stands for the unit vector in $\R^N$ with a $1$ solely in its $p^{\e{th}}$ entry:
\beq
\bs{e}_p \; = \; \big( \underbrace{0, \dots, 0}_{p-1 \;  \e{terms} } , 1, 0,\dots, 0  \big) \;. 
\enq
Finally, one also has the identities
\beq
\pl{a=1}{N} \ex{x_a} \cdot \Psi_{\bs{y}_N,\veps}(\bs{x}_{N+1}) \; = \; 
 \Psi_{\bs{y}_N-\i \hbar \bs{e},\veps- i \hbar N}(\bs{x}_{N+1}) \qquad and  \qquad
\ex{-\ell x_{N+1}} \cdot \Psi_{\bs{y}_N,\veps}(\bs{x}_{N+1}) \; = \; \Psi_{\bs{y}_N,\veps+  \i \ell \hbar}(\bs{x}_{N+1}) \;. 
\label{ecriture proprietes action ex x sur fct propre B}
\enq
The vector $\bs{e}$ introduced above takes the form $\bs{e} = \sul{p=1}{N} \bs{e}_p$ .

\end{prop}

The proof of the proposition  can be found in Appendix B of [$\bs{A19}$] although I stress that it is very standard. 
The main ideas in the proof should be attributed to Sklyanin \cite{SklyaninSoVGeneralOverviewFuncBA,SklyaninSoVGeneralOverviewAndConstrRecVectPofB}. 
I refer to Corollary 1.1 of [$\bs{A19}$] for the equations in dual variables associated with the operator $\op{C}_{1,N+1}(\la)$.

The main observation one can make on the level of these formulae is that it is fairly easy to compute the action on $\Psi_{\bs{y}_N,\veps}(\bs{x}_{N+1})$ of operators acting non-trivially on the "full"
Hilbert space $\mf{h}_{\e{Td}}$ as opposed to a sub-set of its local spaces. Therefore, my main idea for obtaining equations in dual variables consisted in using a slightly different variant of the Mellin-Barnes multiple integral representation. 
Indeed, the representation presented earlier in the text \eqref{definition fonction vp apres resolution MellinBarnes} was obtained by 
an inductive procedure based on the splitting of the chain of $N+1$ sites into a concatenation of an $N$ site chain and a single site one, \textit{c}.\textit{f}. \eqref{ecriture equation recursive pour Mellin Barnes  niveau matrice de monodromie}.
One can establish analogous representations corresponding to any splitting of the chain into sub-chains of respective lengths $r$ and $N+1-r$.

\begin{defin}
\label{Definition procedure recursive construction fcts propre B}
Let $\bs{y}_N \in \Cx^N$ and $\veps \in \Cx$ be given. Then, define 
the function $\Psi_{\bs{y}_N,\veps}(\bs{x}_{N+1})$ inductively as follows. 
First set 
\beq
\Psi_{\emptyset , \veps} (x)  \; = \; \ex{ \i \f{\veps}{\hbar} x}
\enq
and then define the collection of functions $\Psi_{\bs{y}_N, \veps} (\bs{x}_{N+1})$ by 
\beq
\Psi_{\bs{y}_N, \veps} (\bs{x}_{N+1})  \; = \; \Int{ \msc{C}_{r-1; N-r} }{}  \Psi_{ \bs{w} , \ov{\bs{y}}_N-\ov{\bs{z}} }(\bs{x}_1) 
\Psi_{\bs{z} , \veps-\ov{\bs{y}}_N+\ov{\bs{z}} }(\bs{x}_2) \;  
\varpi(\bs{w},  \bs{z} \mid \bs{y}_N) \cdot \pl{a=1}{r-1}\dd w_a  \cdot \pl{a=r}{N-1}\dd z_a  \;. 
\label{ecriture rep int fct propre B}
\enq
There the integration runs through two sets of variables $w_a$ and $z_b$ which are collected in a vector notation
\beq
\bs{w} \; = \; (\underbrace{w_1,\dots, w_{r-1}, 0, \dots, 0}_{N-1})  \qquad and \qquad 
\bs{z} \; = \; (\underbrace{0,\dots, 0, z_r,\dots, z_{N-1}}_{N-1}) \; .
\label{definition des deux ensembles de variables d'integration}
\enq
The vectors 
\beq
\bs{x}_1 = (x_1,\dots, x_{r} )  \qquad and \qquad 
			\bs{x}_2 = (x_{r+1},\dots, x_{N+1} )
\enq
correspond to a splitting of the coordinates of the position vector $\bs{x}_{N+1} = (x_1,\dots, x_{N+1} )$. 
Further, the integration runs through the domain
\beq
\msc{C}_{r-1; N-r} \; = \; \big(\R - \i \a\big)^{r-1}\!\! \times \big(\R + \i \a \big)^{N-r} \quad where \; \a 
\; is \; such \; that \quad 
 \a > \max_{k\in \intn{1}{N}} \big( |\Im(y_k)| \big) \;. 
\label{definition du contour d'intégration pour fcton Whittaker}
\enq
Finally, the weight $\varpi(\bs{w},  \bs{z} \mid \bs{y}_N)$ arising under the integral sign is given by  
\beq
\varpi(\bs{w},  \bs{z} \mid \bs{y}_N) \; = \; \f{ (2\pi \hbar)^{1-N} }{(r-1)! (N-r)! } \cdot 
\f{ \pl{a=1}{r-1} \pl{b=1}{N} \bigg\{  \Ga\Big( \f{y_b-w_a}{ \i \hbar} \Big) \, \hbar^{\f{ \i }{\hbar} (w_a-y_b)}  \bigg\}  
\cdot \pl{a=r}{N-1} \pl{b=1}{N} \bigg\{  \Ga\Big( \f{z_a-y_b}{ \i \hbar} \Big) \,  \hbar^{\f{ \i }{\hbar} (y_b-z_a)}  \bigg\}  }
{  \pl{ \substack{a, b =1 \\a \not= b} }{ r-1} \Ga\Big( \f{w_a-w_b}{ \i \hbar} \Big) \cdot 
\pl{ \substack{a,b=r \\ a \not= b} }{ N-1} \Ga\Big( \f{z_a-z_b}{ \i \hbar} \Big) 
\cdot \pl{a=1}{r-1} \pl{b=r}{N-1} \bigg\{  \Ga\Big( \f{z_b-w_a}{ \i \hbar} \Big) \,  \hbar^{ \f{ \i }{\hbar} (w_a-z_b)}  \bigg\}  } \;. 
\enq
\end{defin}

\noindent It is shown in Appendices A and B of [$\bs{A19}$] that this definition \vspace{1mm}
\begin{itemize}
\item[$\bullet$] is clean-cut, \textit{i}.\textit{e}. that the integrals \eqref{ecriture rep int fct propre B} are 
indeed convergent;\vspace{1mm}
\item[$\bullet$] is consistent, \textit{i}.\textit{e}. the functions 
$\Psi_{\bs{y}_N,\veps}(\bs{x}_{N+1})$ do not depend on the value of the integer $r$ used in the splitting of the 
integration variables \eqref{definition des deux ensembles de variables d'integration}.\vspace{1mm}
\end{itemize}

The recursive form of the Mellin-Barnes integral representation \eqref{definition fct Whittaker par Mellin-Barnes} corresponds to setting $r=N$ in \eqref{ecriture rep int fct propre B}
and then simplifying the exponential prefactors so as to recover $ \vp_{\bs{y}_N}(\bs{x}_N) $ according to \eqref{ecriture fct propre Toda}.

The $r$-split representation for $\Psi_{\bs{y}_N,\veps}(\bs{x}_{N+1})$ is very effective for obtaining equations in dual variables
for composite operators. The result is resumed in the theorem below whose proof is given in Appendix C and in Proposition 2.1 of [$\bs{A19}$].

\begin{theorem}
\label{proposition action operateur Or sur fct propres B}
The operator 
\beq
\msc{D}_r(\la) \; = \; \pl{a=1}{r} \Big\{  \ex{ \op{x}_a - \op{x}_{N+1} }  \Big\} \; \cdot  \; \op{D}_{r+1,N+1}(\la) 
\enq
 satisfies the equation in dual variables
\beq
\msc{D}_r(\la) \cdot \Psi_{\bs{y}_N,\veps}(\bs{x}_{N+1}) \; = \; 
- \sul{ \substack{  \mc{I}_N =  \sg \cup  \ov{\sg} \\ \#\sg = r+1} }{}  \pl{ \substack{ a\in \sg \\ b\not\in \sg }}{} \bigg\{ \f{-  \i}{y_a-y_b} \bigg\}
\cdot  \pl{ b\not\in \sg }{} (\la-y_b)  \cdot \Psi_{\bs{y}_N- \i \hbar \sul{a \in \sg }{} \bs{e}_a,\veps}(\bs{x}_{N+1}) \;, 
\label{ecriture action operateur O r sur Psi}
\enq
where $\mc{I}_N = \intn{1}{N}$ and the sum runs through all partitions $\sg \cup \ov{\sg}$ of $\mc{I}_N$ under the constraint
$\# \sg = r+1$.

Similarly,  the operator 
\beq
\msc{A}_r(\la) \; = \; \pl{a=1}{r} \Big\{  \ex{ \op{x}_a - \op{x}_{N+1} }  \Big\} \; \cdot  \; \op{A}_{1,r}(\la) 
\enq
gives rise to the equation in dual variables
\beq
\msc{A}_r(\la) \cdot \Psi_{\bs{y}_N,\veps}(\bs{x}_{N+1}) \; = \; 
 \sul{ \substack{  \mc{I}_N =  \sg \cup  \ov{\sg} \\ \#\sg = r} }{}  
\pl{ \substack{ a\in \sg \\ b\not\in \sg }}{} \bigg\{ \f{- \i }{ y_a - y_b } \bigg\}
\cdot  \pl{ b\in \sg }{} (\la-y_b)  \cdot \Psi_{\bs{y}_N - \i \hbar \sul{a \in \sg }{} \bs{e}_a,\veps}(\bs{x}_{N+1}) \;. 
\label{ecriture action operateur Ar cal sur Psi}
\enq
\end{theorem}

One can, in principle, construct equations in dual variables for other operators associated with the Toda chain, see \textit{e}.\textit{g}. Corollary 2.2  of [$\bs{A19}$].
Also, Theorem \eqref{proposition action operateur Or sur fct propres B} is already enough so as to recover the equations in dual variables
obtained by Babelon \cite{BabelonActionPositionOpsWhittakerFctions,BabelonQuantumInverseProblemConjClosedToda} and Sklyanin \cite{SklyaninResolutionIPFromQDet}. Indeed, one has the 
\begin{cor}
The below equations in dual variables hold 
\beq
\pl{a=1}{r} \Big\{  \ex{ \op{x}_a - \op{x}_{N+1} }  \Big\} \cdot \Psi_{\bs{y}_N,\veps}(\bs{x}_{N+1}) \; = \; 
 \sul{ \substack{ \mc{I}_N=\sg \cup \ov{\sg} \\ \#\sg = r} }{}  \pl{ \substack{ a\in \sg \\ b\not\in \sg }}{} 
 \bigg\{ \f{- \i }{y_a-y_b} \bigg\}
\cdot \Psi_{\bs{y}_N - \i \hbar \sul{a \in \sg }{} \bs{e}_a,\veps}(\bs{x}_{N+1})
\label{ecriture action operateur position 1 vers r sur Psi}
\enq
and 
\beq
\pl{\ell=1}{r} \Big\{ \ex{ \op{x}_{\ell} - \op{x}_{N+1}} \Big\} \cdot \Big( \sul{\ell=1}{r} \op{p}_{\ell} \Big) \cdot
 \Psi_{\bs{y}_N, \veps}(\bs{x}_{N+1}) \; = \;
 \sul{ \substack{  \mc{I}_N=\sg \cup \ov{\sg} \\ \#\sg = r} }{} \; 
 \pl{ \substack{ a\in \sg \\ b\not\in \sg }}{} \bigg\{ \f{- \i }{y_a-y_b} \bigg\}
\, \cdot   \,\Big( \sul{a \in \sg}{} y_a \Big)  \cdot 
\Psi_{\bs{y}_N - \i\hbar \sul{a \in \sg }{} \bs{e}_a,\veps}(\bs{x}_{N+1}) \;. 
\label{ecriture action operateurs impuslion sur fonction propre chaine ouverte}
\enq
\end{cor}

\Proof

This is a straightforward consequence  of formula \eqref{ecriture action operateur O r sur Psi}
as soon as one observes that 
\beq
\op{D}_{r,N+1}(\la)  \; = \; - \la^{N-r} \ex{ \op{x}_{r} - \op{x}_{N+1}} \; + \; \e{O}\big( \la^{N-r-1} \big)  \; .
\enq
Likewise, equation \eqref{ecriture action operateurs impuslion sur fonction propre chaine ouverte} is 
deduced from \eqref{ecriture action operateur Ar cal sur Psi} upon observing that 
\beq
\op{A}_{1,r}(\la)  \; = \; \la^{r} \, - \, \la^{r-1} \cdot \Big( \sul{\ell=1}{r} \op{p}_{\ell} \Big)  \; + \;  \e{O}\big( \la^{r-2} \big)  \; .
\enq

\qed

\subsection{Multiple integral representations for certain form factors of the model}

Equations in dual variables allow one to obtain $N$-fold multiple integral representations for the
form factors of various local operators. I will describe the result in the case of the operators
$\prod_{a=1}^{r} \big\{  \ex{ \op{x}_a - \op{x}_{N+1} }  \big\} $. Since the model is translation invariant, one has to fix
the centre of coordinates in order to obtain a finite answer for the form factor, hence waving off the infinite
contribution of the continuous modes of the total momentum operator. Below I adopt a coordinate system where  $x_{N+1}=0$.

\begin{prop}
 
 Let $\Phi_{ \veps ; t_{\e{Td}} }$ and $\Phi_{ \veps ; t_{\e{Td}}^{\prime} }$ be two eigenfunctions of the Toda chain in the  sector characterised by the total momentum 
$\veps$.  The associated finite part of the form factor of the operator $\prod_{a=1}^{r} \big\{  \ex{ \op{x}_a - \op{x}_{N+1} }  \big\} $
takes the form 
\bem
\Big( \Phi_{ \veps ; t_{\e{Td}}^{\prime} } , \prod_{a=1}^{r} \Big\{  \ex{ \op{x}_a - \op{x}_{N+1} }  \Big\}  \cdot \Phi_{ \veps ; t_{\e{Td}} }  \Big)_{ \mid x_{N+1}=0 } \; = \; 
\f{ N! }{r! (N-r)! } \Int{ \R^N }{} \pl{a<b}{N} \Big\{ \f{y_a-y_b}{\pi \hbar} \sinh\big[ \f{\pi}{\hbar}(y_a-y_b) \big]  \Big\}
\cdot \pl{a=1}{N} \Big\{ \big( q_{ t_{\e{Td}}^{\prime} }(y_a) \big)^* q_{ t_{\e{Td}} }(y_a) \Big\} \\
\times \pl{a=1}{r} \bigg\{ \f{ q_{ t_{\e{Td}} }(y_a+\i \hbar) }{ q_{ t_{\e{Td}} }(y_a) } \cdot \pl{b=r+1}{N} \Big[ \f{ \i }{ y_a - y_b } \Big]    \bigg\} \cdot  \dd^N y \;. 
\label{ecriture rep int mult correlateurs dans chaine Toda}
\end{multline}
The index $\mid x_{N+1}=0$ refers to the fact that the coordinates are chosen so that $x_{N+1}=0$. 

\end{prop}

\Proof

Starting from the representation \eqref{ecriture fct propre Toda} for the eigenfunction, one acts with the operator 
$\prod_{a=1}^{r} \big\{  \ex{ \op{x}_a - \op{x}_{N+1} }  \big\} $ under the integral sign by using \eqref{ecriture action operateur position 1 vers r sur Psi}. 
One can then shift the integration contours in the variables labelled by the set $\sg$ in \eqref{ecriture action operateur position 1 vers r sur Psi} in virtue of the analyticity of the integrand, leading to  
\bem
 \prod_{a=1}^{r} \Big\{  \ex{ \op{x}_a - \op{x}_{N+1} }  \Big\}  \cdot  \Phi_{ \veps; t_{\e{Td}} }\big( \bs{x}_{N+1} \big) \; = \;   \sul{ \substack{  \mc{I}_N =  \sg \cup  \ov{\sg} \\ \#\sg = r} }{}  
\Int{ \R^N }{} \Psi_{\bs{y}_N,\veps }\big( \bs{x}_{N+1} \big)  \cdot
 \pl{ \substack{ a\in \sg \\ b\not\in \sg }}{} \bigg\{ \f{- \i}{y_a-y_b +\i \hbar} \bigg\} \\
\pl{ a \in  \sg }{} \Big\{  q_{t_{\e{Td}}}(y_a+\i \hbar)  \Big\} \cdot \pl{ a \in  \ov{\sg} }{} \Big\{  q_{t_{\e{Td}}}(y_a)  \Big\}  
\cdot \mu\big(\bs{y}_N+\i \hbar \sum_{a \in \sg} \bs{e}_a  )  \cdot \f{ \dd^N y }{ \sqrt{N!} }  \;. 
\label{ecriture fct propre Toda} 
\end{multline}
It remains to apply the finite difference equation satisfied by the measure $\mu$:
\beq
  \f{ \mu\big(\bs{y}_N+\i \hbar \sum_{a \in \sg} \bs{e}_a  )  }{ \mu(\bs{y}_N) } \; = \; 
 \pl{ \substack{ a\in \sg \\ b\not\in \sg }}{} \bigg\{ \f{y_a-y_b +\i \hbar}{y_b-y_a} \bigg\}   
\enq
and then invoke the symmetry of the integrand so as to recast the action as 
\bem
\prod_{a=1}^{r} \Big\{  \ex{ \op{x}_a - \op{x}_{N+1} }  \Big\}  \cdot \Phi_{ \veps; t_{\e{Td}} }\big( \bs{x}_{N+1} \big) \; = \;   \f{ N! }{r! (N-r)! }
\Int{ \R^N }{} \Psi_{\bs{y}_N,\veps }\big( \bs{x}_{N+1} \big)  \cdot
 \pl{ a=1}{r} \pl{b=1}{r+1} \bigg\{ \f{ \i }{y_a-y_b } \bigg\} \\
\pl{ a =1 }{r} \Big\{  q_{t_{\e{Td}}}(y_a+\i \hbar)  \Big\} \cdot \pl{ a = r +1 }{ N }  \Big\{  q_{t_{\e{Td}}}(y_a)  \Big\}  
\cdot  \f{ \dd \mu(\bs{y}_N) }{ \sqrt{N!} }  \;. 
\label{ecriture fct propre Toda} 
\end{multline}
The rest follows by virtue of the unitarity of the $\mc{U}_N$ transform.  \qed 

\vspace{1mm}

\section*{Conclusion}

This chapter covered the progress I achieved in understanding various aspects of the spectral problem for the Toda chain. 
I have proved the validity of the Nekrasov-Shattashvili description of the model's spectrum,
I developed a new approach allowing one to establish the unitarity of the separation of variables transform, and finally, I pushed 
forward the resolution of the spectral problem. All this leads to $N$-fold multiple integral representations for the form factors of various
operators of the Toda chain, as shown in Proposition \ref{ecriture rep int mult correlateurs dans chaine Toda}. These representations
fit, upon taking specific limits, to the sinh-model multiple integrals that have been mentioned in the introduction.

\chapter{Towards the asymptotic analysis of quantum separation of variables issued multiple integrals}
\label{Chapitre AA des integrales multiples}

I have briefly reviewed, in the introduction chapter, the state of the art of the techniques allowing one to extract the large-$N$ asymptotic behaviour out of $N$-fold multiple integrals
corresponding to partition functions of $\be$-ensembles with varying weights. One of the most straightforward generalisations of these integrals, already proposed by Boutet de Monvel, Pastur and Shcherbina 
\cite{BoutetdeMonvelPasturShcherbinaAPrioriBoundsOnFluctuationsAroundEqMeas}, consists in replacing the 
varying one-body confining potential $N\cdot V$ by a varying multi-particle potential, hence leading to integrals of the type 
\beq
\mc{Z}_N^{(\be;r)}\big[ \{V_p\}_1^r \big]\; = \; \Int{ \mathsf{A}^N }{} \pl{a<b}{N} |\la_a-\la_b|^{\be} \cdot 
 \exp\bigg\{ - \sul{p=1}{r} \frac{N^{2-p}}{p!} \sul{ \substack{ i_1<\dots < i_p \\ i_a=1,\dots, N} }{} V_{p}\big( \la_{i_1}, \dots, \la_{i_p} \big) \bigg\} \cdot \dd^N \la  \;. 
\label{label ecriture int mult pour r interacting pot}
\enq
There $\mathsf{A}$ is a finite union of intervals of $\R$ and $V_p$ are some sufficiently regular $p$-particle interactions. The scaling $N^{2-p}$ in front of $V_p$ 
is such that all interactions between the integration variables have, effectively, the same order of magnitude in $N$.
 When $\mathsf{A}=\R$, G\"{o}tze, Venker \cite{GotzeVenkerLocalUniversalityGeneral2bdyBetaRepInteraction} and Venker \cite{VenkerParticleSystemsWithRepBeta} have shown that 
the bulk universality properties of the probability distribution associated with the model at $r=2$ corresponds to the universality class of $\beta$-ensembles. 
Furthermore, when  $r=2$, $\be=2$ and 
\beq
V_{2}(\la_1,\la_2) = \ln\Bigg(\frac{f(\la_2) - f(\la_1)}{\la_2 - \la_1}\Bigg) \,  , 
\enq
the multiple integral has a determinantal structure \cite{BorodinBiOrthogonalEnsIntroAnStudyOfExamples,ClaeysRomanoStudyForGeneralPeriodsOfSinh-BasedPlys,ClaeysWangIntroRHPForBioRthSinhLikePlys}
what allows for its study through bi-orthogonal polynomials \cite{KonhauserFirstintoBiOrthPlys}. Some of these polynomials - for specific choices of $f$'s- can be characterised by means of Riemann--Hilbert problems 
\cite{ClaeysRomanoStudyForGeneralPeriodsOfSinh-BasedPlys,ClaeysWangIntroRHPForBioRthSinhLikePlys}. 
The formal asymptotic expansion of the partition function of multiple integrals of the type  \eqref{label ecriture int mult pour r interacting pot}
has been argued by Borot \cite{BorotFormalLargeNAEForMultiPtInteractionsMI} on the basis of a generalisation of the topological recursion. 
In [$\bs{A22}$], in collaboration with Borot and Guionnet, I proved the existence of the all-order asymptotic expansion 
under certain regularity assumptions on the multi-particle interactions which are, for instance, satisfied for perturbations of the Gaussian one-body potential
or for various examples of two-body (\textit{viz}. $r=2$ in \eqref{label ecriture int mult pour r interacting pot}) interactions depending on the difference of arguments.

Even though interesting from the point of view of its applications, dealing with this class of multiple integrals was merely a technical issue. 
Indeed, upon minor embellishments, the overall analysis fits quite well into the scheme developed for $\be$-ensembles. 
However, developing an approach allowing one to extract the large-$N$ asymptotic expansion out of the class of multiple integrals that are of central interest to the quantum separation of variables appeared to be a quite different issue. 
Generically, such multiple integrals can be written as: 
\beq
\mf{z}_N[W_N] \; = \; \Int{\R^N}{} \pl{a<b}{N} \Big\{ \sinh\big[ \f{\pi}{\om_1} (y_a-y_b) \big]  \sinh \big[ \f{\pi}{\om_2} (y_a-y_b) \big] \Big\} 
\cdot \pl{a=1}{N}\ex{-W_N(y_a)} \cdot \dd^N y \;. 
\label{ecriture MI qSoV comme exemple}
\enq
There $W_N$ is a model dependent confining potential while $\om_1, \om_2$ are related to coupling constants of the given model. 
Here, the integration runs over $\R^N$ and this case will be of prime interest in the following. 
Nevertheless, in some models, one should trade the integration contour from $\R^N$ to $\msc{C}^N$, with $\msc{C}$ a curve in $\Cx$. 
Although such results have seldom been written down explicitly, it is more or less clear that scalar products of states 
\cite{DerkachovKorchemskyManashovXXXSoVandQopNewConstEigenfctsBOp,DerkachovKorchemskyManashovXXXreelSoVandQopABAcomparaison} and [$\bs{A18}$], certain partition functions \cite{KazamaKomatsuNishimuraSoVLikeMIRepForPartFctRat6Vertex}
and also form factors \cite{BabelonQuantumInverseProblemConjClosedToda,GrosjeanMailletNiccoliFFofLatticeSineG,NiccoliCompleteSpectrumAndSomeFormFactorsAntiPeriodicXXZ,NiccoliCompleteSpectrumAndSomeFormFactorsXXXHigherSpin,SklyaninResolutionIPFromQDet} 
and [$\bs{A19}$] of numerous local operators
in models solvable by this method are either given directly as integrals of the type \eqref{ecriture MI qSoV comme exemple}
or by certain perturbations thereof as in the case of the Toda chain issued form factors \eqref{ecriture rep int mult correlateurs dans chaine Toda}. 
The large-$N$ asymptotic behaviour of these integrals is of main interest to the physics of the models in which such multiple integrals arise. 
For instance, in the case of the Toda chain, sending $N\tend +\infty$ corresponds to taking the limit of 
infinitely many quantum particles in interaction. Then, extracting the $N\tend + \infty$ asymptotic behaviour of the multiple integrals \eqref{ecriture rep int mult correlateurs dans chaine Toda}
is the \textit{sin equa non} condition for being able to describe the dynamical properties of such a system of infinitely many interacting particles. 
In this respect, it is also clear that it will be more or less direct to understand the large-$N$ behaviour of perturbations, such as \eqref{ecriture rep int mult correlateurs dans chaine Toda}, 
of \eqref{ecriture MI qSoV comme exemple} once that it becomes clear how to deal with the large-$N$ behaviour of $\mf{z}_N[W_N]$.

Independently of its applications to the physics of quantum integrable models, the multiple integrals \eqref{ecriture MI qSoV comme exemple} are interesting in their own right:
they naturally constitute  the next, after $\be$-ensembles, class of $N$-fold integrals to investigate and understand 
when having in mind developing a general, effective, theory of the  large-$N$ asymptotic analysis of $N$-fold multiple integral. 
On the one hand, the integrand in \eqref{ecriture MI qSoV comme exemple}  bears certain structural similarities with the one arising in $\be$-ensembles.
On the other hand, it brings two new features into the game hence providing one with a good playground for pushing forward the methods of 
asymptotic analysis. 
To be more precise, the main features of the integrand in $\mf{z}_N[W]$ which render  the use of the already established methods  problematic 
stem from the presence of \vspace{1mm}
\begin{itemize}
\item[1°)] either a non-varying confining one-body potential $W_N=W$ \textit{or} a potential \symbolfootnote[2]{This is precisely the case of potentials that would be of interest to the Toda chain} 
$W_N$ such that, when $\la \tend  \infty$, $W_N(\la)=N \pi \big( \om_1^{-1} + \om_2^{-1} \big) |\la| \big( 1 + \e{o}(1) \big)$, hence making the integral \textit{minimally} convergent, and 
such that $W_N$ converges uniformly on compact subsets of $\R$ to a non-varying potential;\vspace{1mm}
\item[2°)]  a two-body interaction that has the same local (\textit{viz}. when $\la_a\tend \la_b$)
singularity structure as in the $\be$-ensemble case, while breaking other properties of the Vandermonde interaction such as 
its effective invariance under a re-scaling of all the integration variables. \vspace{1mm}
\end{itemize}

The present chapter is devoted to describing the progress I achieved in understanding certain aspects of the large-$N$ asymptotic expansion
of the multiple integrals $\mc{Z}_N^{(\be;r)}\big[ \{V_p\}_1^r \big]$ and, above all, $\mf{z}_N[W_N]$. This chapter is organised as follows. 
Section \ref{Section Multi particle potential Chapter AA SoV Int} provides a short description of the 
large-$N$ expansion of the partition function $\mc{Z}_N^{(\be;r)}\big[ \{V_p\}_1^r \big]$ in the so-called one-cut regime. 
This section is based on the work [$\bs{A22}$]. The characterisation of the asymptotic expansion in the multi-cut regime
is also possible and can be found in the mentioned paper. I will start Section \ref{Section AA Baby SoV integrals Chapter AA SoV In} by 
arguing that, in order to tackle the large-$N$ asymptotic expansion of multiple integrals of the type $\mf{z}_N[W_N]$, 
one first needs to be able to extract, in a efficient way, the large-$N$ asymptotic expansion of the below sinh-model partition functions 
with varying interactions 
\beq
Z_N[V] \; = \; 
\Int{\R^N}{}  \pl{a<b}{N} \Big\{ \sinh\big[ \frac{\pi}{\om_1} T_N (\la_a-\la_b)\big] 
				\sinh\big[ \frac{\pi}{\om_2} T_N (\la_a-\la_b)\big]  \Big\}
\cdot \pl{a=1}{N} \Big\{  \ex{-N T_N V(\la_a)}  \Big\} \cdot  \dd^N \la
\enq
where $T_N \tend +\infty$ when $N\tend + \infty$. In Subsection \ref{SousSection Resultat Pour model Rescale}, I will present the form for the large-$N$ behaviour of $Z_N[V] $
that I proved with Borot and Guionnet in [$\bs{A21}$]. In Subsection \ref{SousSection Resultats Auxiliaires},
I will discuss some auxiliary results of importance to the analysis of [$\bs{A21}$]. I will, in 
particular, describe briefly the structure of the equilibrium measure associated with the sinh varying interactions. 
In will conclude this section by discussing, in Subsection \ref{SousSection Strategi de la preuce modele sinh rescale}, the overall strategy allowing one
to prove the mentioned asymptotic expansion of $Z_N[V] $.

\section{Mean field models with Coulomb interactions } 
\label{Section Multi particle potential Chapter AA SoV Int}

\subsection{The overall strategy}

The main idea for attacking the problem of estimating the partition function at large-N consists in constructing a one parameter $t$ family of potentials 
$\{ V_{p;t}\}_1^r$ such that $V_{p;1}=V_{p}$ and $V_{p;0}=0$ for any $p \geq 2$. 
Given such a family, one has that the logarithmic $t$-derivative of the partition function 
can be recast as
\beq
\Dp{t} \ln \mc{Z}_N^{(\be;r)}\big[ \{V_{p;t} \}_1^r \big] \; = \;
- N^2 \sul{p=1}{r} \f{ 1 }{ p! } \Oint{ \Ga( \mathsf{A} )^p }{} \Dp{t} V_{p;t}(\xi_1,\dots,\xi_p) \cdot \wt{W}_p^{ \{V_{p;t} \}_1^r }\big(\xi_1,\cdots,\xi_p \big) \cdot \f{ \dd^p \xi }{ (2\i \pi)^p } \;. 
\label{equation derivee partielle de fct part multi particules}
\enq
Above, the quantities $\wt{W}_p^{ \{V_{p;t} \}_1^r }$ correspond to the Stieltjes transforms of the $n$-th order moments of the empirical measure 
\beq
\label{discon}\widetilde{W}_n^{\{V_{p} \}_1^r}(x_1,\ldots, x_n) \; = \; 
\mathbb{P}_{ \{V_{p} \}_1^r  }\Big[\prod_{a = 1}^{n}\Int{}{} \f {\dd L_N^{(\bs{\la})}(\xi_a)}{x_a - \xi_a}\Big] \qquad \e{with} \quad \dd L_N^{(\bs{\la})} \;= \; \f{1}{N} \sul{a=1}{N} \de_{\la_a} \;. 
\enq
Above $\mathbb{P}_{ \{V_{p} \}_1^r  }$ stands for the probability measure on $\mathsf{A}^N$ that is induced by the partition function $\mc{Z}_N^{(\be;r)}\big[ \{V_{p} \}_1^r \big]$ and $\de_x$
is the Dirac mass centred at $x$. 

Integrating the \textit{rhs} of \eqref{equation derivee partielle de fct part multi particules} over $t$ yields
\beq
\Int{0}{1}\Dp{t} \ln \mc{Z}_N^{(\be;r)}\big[ \{V_{p;t} \}_1^r \big] \dd t \; = \; \ln \bigg(  \f{ \mc{Z}_N^{(\be;r)}\big[ \{V_{p;1} \}_1^r \big] }{ \mc{Z}_N^{(\be;r)}\big[ \{V_{p;0} \}_1^r \big] }  \bigg) \;. 
\enq
The term in the denominator is a one-body confining potential $\be$-ensemble partition function; its large-$N$ asymptotic behaviour has been determined in the works 
\cite{BorotGuionnetAsymptExpBetaEnsMultiCutRegime,BorotGuionnetAsymptExpBetaEnsOneCutRegime}. Thus, to access to the large-$N$ expansion of the partition function of interest,
it remains to find a uniform in $t \in \intff{0}{1}$ asymptotic expansion of the \textit{rhs} of \eqref{equation derivee partielle de fct part multi particules}
up to the desired order of precision in $N$ and then integrate it out over $t$. Once that I provide a description of the main result of  [$\bs{A22}$] in the one-cut case, 
I will describe the overall strategy on which the proof is based.

\subsection{A brief description of the result}

In order to describe the main result of  [$\bs{A22}$], I first need to introduce a few objects
and state the hypothesis under which the analysis holds. First of all, the $p$-body potentials $V_p$, $p=1,\dots,r$, are assumed to be real analytic. 
This hypothesis allows one to rely on the tools of complex analysis so as to carry out the large-$N$ analysis of the loop equations subordinate to this model.
Although this is not primordial, see [$\bs{A21}$], it allows one for an important simplification of the analysis  

For further purpose, it is convenient to introduce  a rewriting of the $p$-particle potentials $V_p$ as
\beq
T(x_1,\ldots,x_r) \; = \;   \sul{ J \subseteq \intn{1}{r} }{}   \big( r - |J| \big)! \, V_{|J|} \big( \bs{x}_{J} \big)\quad \text{with} \quad \left\{  \ba{c}  J=\{j_1,\dots , j_{|J|}\} \\
\bs{x}_{J} = \big(x_{j_1},\dots , x_{j_{|J|}} \big) \ea \right. \;. 
\enq
In the case where $\mathsf{A}$ has some connected component going to infinity, one needs to ensure the convergence of the integral.  This can be shown to hold provided 
that  $T$ is bounded from below as 
\beq
T(x_1,\ldots,x_r) \geq  C \; + \; (r - 1)! \sum_{a = 1}^r \big( \beta + \eps \big) \ln |x_a| \qquad \e{for} \; \e{some} \;  \eps>0 \;\;\e{and} \; C \in \R \;. 
\label{borne inf pour T}
\enq
This will be taken as a working hypothesis.

Quite analogously to the case of $\be$-ensemble integrals, the large-$N$ asymptotic expansion of the partition function is
driven by the properties of the so-called energy functional $\mathcal{E}_{MF}$. The latter is a functional on $\mc{M}^{1}(\R)$, which, for the case of interest takes the form
\beq
\label{defE}\mathcal{E}_{MF}[\mu] \; = \;   \Int{\mathsf{A}^r}{} \bigg\{ \frac{T(x_1,\ldots,x_r)}{r!} - \frac{\beta}{r(r - 1)}\sum_{1 \leq i \neq j \leq r} \ln|x_i - x_j|  \bigg\} \prod_{a = 1}^r \dd\mu(x_a)\;.
\enq
Under the above working hypothesis, one can  show following standard arguments \cite{AndersonGuionnetZeitouniIntroRandomMatrices,DeiftOrthPlyAndRandomMatrixRHP}  that 
$\mathcal{E}_{MF}$ has compact level sets and that there exists a measure $\mu_{\e{min}} \in \mc{M}^1(\R) $ 
minimising $\mc{E}_{MF}$ :
\beq
\mc{E}_{MF}\big[ \mu_{\e{min}} \big] \; = \; \inf_{\mu \in \mc{M}^{1}(\R) } \Big\{ \mc{E}_{MF}\big[ \mu \big] \Big\} <+\infty \; .
\enq
The uniqueness of the minimum is a different issue and depends on the specific form of the $p$-body potentials $V_p$. 
This uniqueness is a necessary ingredient for the analysis that has been developed in [$\bs{A22}$] to hold. 
I will thus take it as a working hypothesis and refer to the unique minimiser as the equilibrium measure $\mu_{\e{eq}}$. 

\vspace{1mm}

Exactly as for $\be$-ensembles, the equilibrium measure can be characterised in terms of a set of variational equations. Under the hypothesis of real analytic $p$-body interactions 
and uniqueness of the minimiser, one can establish other properties of $\mu_{\e{eq}}$, analogous to the $\be$-ensemble case: 
 the existence of a density, the fact that its support $\mathsf{S}$ consists of a finite union of intervals or characterise
the behaviour of its density at the edges of the support $\Dp{}\mathsf{S}$, this as much for the hard and soft edge case. 
I remind that \vspace{1mm}
\begin{itemize}
\item[$\bullet$] $a \in \Dp{}\mathsf{S}$ is a hard edge if $a \in \Dp{}\mathsf{S} \cap \Dp{}\mathsf{A}$;\vspace{1mm}
\item[$\bullet$] $a \in \Dp{}\mathsf{S}$ is a soft edge if $a \in \overset{\circ}{\mathsf{A}}$. \vspace{1mm}
\end{itemize}
The bounds on the behaviour of the equilibrium measure in a neighbourhood of the edges depend on the nature thereof. One has
\beq
\f{ \dd \mu_{ \e{eq} } (\xi) }{ \dd \xi } \; = \; \e{O}\Big( \sqrt{ |\xi-a| } \Big) \; , \; a \; \e{soft} \; \e{edge} \qquad \e{and} \qquad
\f{ \dd \mu_{\e{eq}} (\xi) }{ \dd \xi } \; = \; \e{O}\bigg( \f{1}{ \sqrt{ |\xi-a| } } \bigg) \; , \; a \; \e{hard} \; \e{edge} \; . 
\enq
The equilibrium measure is said to be off-critical if it vanishes precisely as a square root at the soft edges and as an inverse square root at the hard edges. 
These various results relative to the characterisation of the equilibrium can be found in Theorems 2.2-2.3 and Lemmata 2.4-2.5 of [$\bs{A22}$]. 

Prior to stating the result, I still need to introduce the functional $\mathcal{Q}_{\e{eq}}$ defined on the space of bounded signed measures on $\mathsf{A}$ as
\beq
\mathcal{Q}_{\e{eq}}[\nu] = - \beta  \,\Int{\mathsf{A}^2}{} \ln|\xi - \eta|\dd\nu(\xi)\dd\nu(\eta) \; + \;  \Int{\mathsf{A}^r}{} \frac{T(\xi_1,\ldots,\xi_r)}{(r - 2)!}\,\dd\nu(\xi_1)\dd\nu(\xi_2)\prod_{a = 3}^r \dd\mu_{\mathrm{\rm eq }}(\xi_a) \;. 
\label{definition fonctionnelle Q}
\enq

I will state the main result of  [$\bs{A22}$] in the simplest case when 
the support of the equilibrium measure consists of a single interval. 

\begin{theorem}
\label{cc1} Assume that  \vspace{1mm}
\begin{itemize}
\item[$\bullet$]the $p$-body potentials are real-analytic; \vspace{1mm}
\item[$\bullet$] $\mc{E}_{MF}$ admits a unique minimiser $\mu_{\mathrm{eq}}$ whose support consists of a single interval; \vspace{1mm}
\item[$\bullet$] the density of equilibrium measure is off-critical;  \vspace{1mm}
\item[$\bullet$]  the functional $\mc{Q}_{\e{eq}}$ is positive definite, \textit{viz}. $\mc{Q}_{\e{eq}}[\nu] \geq 0$
and $\mc{Q}_{\e{eq}}[\nu]=0$ implies $\nu=0$. \vspace{1mm}
\end{itemize}
Then, the partition function $\mc{Z}_N^{(\be;r)}\big[ \{V_{p;t} \}_1^r \big]$  admits the large-$N$ asymptotic expansion 
\beq
\mc{Z}_N^{(\be;r)}\big[ \{V_{p;t} \}_1^r \big] \; = \; N^{N \f{\be}{2} + \tau} \exp\bigg\{ \sum_{k = -2}^{k_0} N^{-k}\,F^{[k]} + o(N^{-k_0}) \bigg\} \qquad for \;any \; k_0 \in \mathbb{N}  \; . \nonumber
\enq
$\tau$ is a universal exponent depending only on $\beta$ and the nature of the edges: \vspace{1mm}
\begin{itemize}
\item[$\bullet$] $\tau \, = \,  \frac{1}{12} \Big( 3 + \frac{\beta}{2} + \frac{2}{\beta} \Big) $  \quad \;  if the support  $\mathsf{S}$ has two soft edges ; \vspace{1mm}
\item[$\bullet$] $\tau \, = \,  \frac{1}{6} \Big(  \frac{\beta}{2} + \frac{2}{\beta} \Big) $ \qquad if it has one soft edge and one hard edge ;  \vspace{1mm}
\item[$\bullet$] $\tau  \, = \,  \frac{1}{4} \Big(  \frac{\beta}{2} + \frac{2}{\beta} -1 \Big)$  \;\; if it has two hard edges.\vspace{1mm}
\end{itemize}
$F^{[k]}$ are functionals of the $p$-particle potentials that are computable order-by-order. The leading term takes the form  $F^{[-2]}=-\mc{E}_{MF}\big[ \mu_{ \e{eq} } \big]$.

\end{theorem}

I do stress that the form of the asymptotic expansion can also be established in the multi-cut case, \textit{viz}. when $\e{supp}[\mu_{\e{eq}}]$ consists of several disjoint segments.
I refer to Theorem 1.1 and Sections 7-8 of  [$\bs{A22}$]  for a precise statement. 

It is legitimate to wonder about the applicability of the general setting to concrete situations and, in particular, to ask about the possibility to prove, in concrete 
examples, the hypothesis on which the analysis relied. One can provide numerous examples of models with two-body interactions $r=2$ where the strict convexity of the 
energy functional $\mc{E}_{MF}$ can be established, see Section 1.2 of  [$\bs{A22}$] for explicit examples. For such models, the one-body confining potential can be taken arbitrary up to some reasonable 
hypothesis such as ensuring the convergence of the integral.

\subsection{The main steps of the proof}
\label{SousSection Main Steps of Proof of Mean Field Part Fct}

In order to extract the asymptotic expansion of the Stieltjes transform $\wt{W}_n$ in \eqref{equation derivee partielle de fct part multi particules}
one first obtains the one of a related object the so-called connected correlators $W_n$. These are defined as 
\beq
\label{defcow} W_n(x_1,\ldots, x_n) \; = \; \bigg( \f{-1}{N^2} \bigg)^{n}
 \f{ \Dp{}  }{ \Dp{}\tau_1 }\cdots  \f{ \Dp{}  }{ \Dp{}\tau_n } \cdot \ln  \mc{Z}_N^{(\be;r)}\big[ \{  V_{p}^{(\bs{\tau})} \}_1^r \big]\big]_{ |_{\tau_a = 0}}   
\enq
where the $ V_p^{(\bs{\tau})} $ correspond to the below deformations of the original potentials
\beq
 V_p^{(\bs{\tau})} \big( \xi_1, \dots,\xi_p \big) \; =\; V_p\big( \xi_1, \dots , \xi_p \big) \, + \,  \f{ \de_{p,1} }{ N} \sul{s=1}{n} \f{ \tau_s}{ x_s-\xi_1 } 
\enq
 and $\de_{a,b}$ is the Kronecker symbol. 
The connected correlators reconstruct $\wt{W}_n$ as 
\beq
\widetilde{W}_n(x_1,\ldots,x_n) \; = \;  
\sum_{s = 1}^n \sum_{\substack{ \intn{1}{n}=\\ J_1 \sqcup \cdots \sqcup J_s }} 
\prod_{i = 1}^s W_{|J_i|}( \bs{x}_{J_i} ) \quad \text{where}\,  \quad \left\{  \ba{c}  J=\{j_1,\dots , j_{|J|}\} \\
\bs{x}_{J} = \big(x_{j_1},\dots , x_{j_{|J|}} \big) \ea \right. \;.
\label{ecriture reconstruction correlateur en terms correlateurs connexes}
\enq
Above, the sum runs through all partitions of $\intn{1}{n}$ into $s$ disjoint sets $J_1,\dots, J_s$. 
It follows from \eqref{ecriture reconstruction correlateur en terms correlateurs connexes} that if the asymptotic expansion of the connected correlators $W_n$ is known, so is the one of the $\wt{W}_n$'s.  
The main reason for studying the connected correlators instead of the Stieltjes transforms of the $n^{\e{th}}$-order moments is that the former 
have a much better structure of the large-$N$ asymptotic expansion that the $\wt{W}_n$'s, hence making it easier to study.

The first step towards extracting the large-$N$ behaviour of the connected correlators consists in obtaining some \textit{a priori} bounds on their magnitude. 
These can be deduced from concentration of measure arguments which are obtained in Section 3.3 of  [$\bs{A22}$].
Their derivation is very analogous to the method employed by Maurel-Segala and Ma\"{i}da \cite{MaidaMaurelSegalaInegalitesPourConcentrationDeMesures} in the
$\be$-ensemble case. The main idea behind the concentration of measure is to control, in some specific way, the fluctuations of the empirical measure
\beq
L_N^{(\bs{\la})}= \f{1}{N} \sul{a=1}{N} \de_{\la_a}
\enq
around the equilibrium measure. Since the empirical measure has atoms, it is necessary to formulate the concentration of measure in terms of fluctuations of an
appropriate regularisation $\wt{L}^{(\bs{\la})}_{N}$ of the empirical measure. This regularisation is  Lebesgue continuous and close ($\e{O}(N^{-2})$)
to $L_N^{(\bs{\la})}$ in the Wasserstein metric. Now, in terms of this regularisation, one has, for some constants $c_0,c^{\prime}>0$,
\beq
\mathbb{P}_{ \{ V_p\}_1^r } \Big[ \bs{1}_{ \mf{D}_N } \Big] \; \leq \; 
\ex{  -c^{\prime} N \ln N  }  \quad \e{where} \quad  \mf{D}_N  \; = \; \Big\{\bs{\la}\in \R^N  \; : \; \mc{Q}_{\e{eq}}\big[ \wt{L}^{(\bs{\la})}_{N} - \mu_{\e{eq}}  \big] \geq c_0 \f{\ln N}{N}  \Big\} \;. 
\label{ecriture estimation mesure proba domaine avec Q fluctuation grandes}
\enq
This result means that, relatively to the probability measure $\mathbb{P}_{ \{ V_p\}_1^r }$,  the contribution of domains in $\R^N$ where the regularised empirical measure  deviates from 
the equilibrium measure is quite small. The control on the magnitude of this deviation is ensured by the functional  $\mc{Q}_{\e{eq}}$. 
One would like to use this information to control the magnitude of the connected correlators for $z \in \Cx \setminus \mathsf{A}$. In the case of $W_1$, one has to recentre around the equilibrium measure
\beq
\Big| W_1(z) \, - \, \Int{ \mathsf{A} }{} \f{ \dd \mu_{\e{eq}}(\xi)   }{z-\xi}  \Big|  \; = \; 
\mathbb{P}_{ \{ V_p\}_1^r } \bigg[  \Int{}{}  \psi_z(\xi) \cdot \dd\big(L_N^{(\bs{\la})}-\mu_{\e{eq}} \big)(\xi) \bigg] \quad \e{where} \quad 
\psi_z(\xi)=\f{ \bs{1}_{ \mathsf{A} }(\xi) }{  (z-\xi) } \;. 
\enq
In order to make the most of the concentration bounds the idea is to decompose 
\beq
 \Int{}{}  \psi_z(\xi) \cdot \dd\big(L_N^{(\bs{\la})}-\mu_{\e{eq}} \big)(\xi) \; = \;  \Int{}{}  \psi_z(\xi) \cdot \dd\big(L_N^{(\bs{\la})}-\wt{L}^{(\bs{\la})}_{N} \big)(\xi)
\; + \;  \Int{}{}  \psi_z(\xi) \cdot \dd\big(\wt{L}^{(\bs{\la})}_{N}-\mu_{\e{eq}} \big)(\xi) \; . 
\enq
The first term can be bounded easily in terms of the control - uniform in $\bs{\la}\in \R^N$- on the Wasserstein distance of $\wt{L}^{(\bs{\la})}_{N}$ to $L_N^{(\bs{\la})}$. 
However, in order to make the most of the bounds \eqref{ecriture estimation mesure proba domaine avec Q fluctuation grandes}, one should find a norm $\norm{\cdot }_{\mc{H}}$ on 
an appropriate space of functions $\mc{H} \subseteq \mc{C}^{0}\big(\mathbb{R})$ such that for any  $\varphi \in \mathcal{H}$ and bounded measure $\nu$ of total mass $0$:
\beq
 \Big|\int_{\mathsf{A}} \varphi(x)\,\dd\nu(x)\Big| \leq c_0\,\mathcal{Q}^{1/2}[\nu]\, \norm{ \vp } _{\mathcal{H}}\;.
\label{ecriture norme sur espace special H}
\enq
The space $\mc{H}$ should also contain a convenient family of functions (see Definition 3.2 and Lemma 3.3 of [$\bs{A22}$]). 
Once the bound \eqref{ecriture norme sur espace special H} is established, it is enough to split the integral versus $\mathbb{P}_{ \{ V_p\}_1^r }$ in one over $\mf{D}_N$ and one over its complement $\mf{D}_N^{\e{c}} $. 
The integration over $\mf{D}_N$ leads to exponentially small contributions due to \eqref{ecriture estimation mesure proba domaine avec Q fluctuation grandes} and the use of $L^{\infty}$
bounds on $\psi_z(\xi)$. Then, applying the bound \eqref{ecriture norme sur espace special H} to the measure $ \nu = \wt{L}^{(\bs{\la})}_{N}-\mu_{\e{eq}} $ allows one to estimate
in $N$ the integral over $\mf{D}_N^{\e{c}} $.

The space $\mc{H}$ is constructed in Section 4 and Appendix A. It is shown there that one can take $\mc{H}$ to be the Sobolev space $W^{q}_1(\mathsf{A})$ with $q>2$
where I remind that this is the space of functions on $\mathsf{A}$ that are bounded in respect to the norm
\beq
\label{defWninfnorm} \norm{f}_{ W^{\infty}_{k} (\mathsf{A},\dd \mu) } \; = \; \max
\bigg\{  \norm{ \Dp{x_1}^{a_1}\dots \Dp{x_n}^{a_n}f  }_{  L^{\infty}(\mathsf{A},\dd \mu) }  \; : \; 
  a_{\ell} \in \mathbb{N}, \; \ell=1,\dots, n,   \; \;  \e{and}\; \e{satisfying}  \; \sul{\ell=1}{n}  a_{\ell} \leq k   \bigg\}\;.
\enq
It is easy to check that the functions $\psi_z(\xi)=\tf{ \bs{1}_{ \mathsf{A} }(\xi) }{  (z-\xi) }$ which are the building blocks of the correlators,   do belong to this space. 
All these informations lead to the \textit{a priori} bounds on the correlators $W_n$ 
\beq
\Big| W_1(z) \, - \, \Int{ \mathsf{A} }{} \f{ \dd \mu_{\e{eq}}(\xi)   }{z-\xi}  \Big| \; \leq \;  \frac{C}{ N^2 \cdot d^2(z,\mathsf{A}) } + C^{\prime} \norm{ \psi_{z} }_{ W^{q}_1(\mathsf{A}) } \sqrt{ \f{\ln N }{ N} }
\label{aprioriW1}
\enq
for some constants $C,C^{\prime}$ and where $d(z,\mathsf{A})$ represents the distance of $z$ to $\mathsf{A}$.  
Similarly, for any $n \geq 2$ and $N$ large enough, one shows that there exists $c_n > 0$ such that the $n^{\e{th}}$ connected correlator satisfies:
\beq\label{aprioriWn}
\big|W_{n}(z_1,\ldots,z_n)\big| \leq c_n \prod_{a = 1}^n\bigg\{   \frac{C}{ N^2 \cdot d(z_a,\mathsf{A})^2 } + C^{\prime} \norm{ \psi_{z_a} }_{ W^{q}_1(\mathsf{A}) } \sqrt{ \f{\ln N }{ N} }  \bigg\} \; . 
\enq

I stress that these bounds are not optimal. They solely constitute the starting point of the analysis. The tool which allows one to study the large-$N$ expansion of the correlators is the
system of loop equations satisfied by the correlators. It is a tower of non-linear integral equations relating the $W_n$'s among them. 
See Section 5.1 of [$\bs{A22}$] for a precise statement. Inserting the bounds \eqref{aprioriW1}-\eqref{aprioriWn} 
into the loop equations allows one to improve them up to the optimal level by using a bootstrap argument. 
Once that the bounds are optimal, one is able to access to the large-$N$ asymptotic expansion of the correlators by some perturbative
handlings of the loop equations. This asymptotic expansion can be obtained to any order in $\tf{1}{N}$. The details associated with these steps 
can be found in Sections 5 and 6 of [$\bs{A22}$].

The large-$N$ asymptotic expansion of the Stieltjes transform obtained in this way is already enough so as to access to the large-$N$ expansion of $\mc{Z}_N^{(\be;r)}\big[ \{V_p\}_1^r \big]$,  
this both in the one and multi-cut regimes, as demonstrated  in Sections 7 and 8 of [$\bs{A22}$].







\section{The re-scaled quantum separation of variables integral}
\label{Section AA Baby SoV integrals Chapter AA SoV In}

\subsection{Motivation of the varying-interaction model}
\label{SousSection Motivation pour etude model rescale}

In order to explain the main difference between the asymptotic analysis of $\mf{z}_N[W_N]$ and $\be$-ensemble multiple integrals subject to varying potentials it is instructive 
to discuss the example of a non-varying monomial potential $W_{\e{ply}}(y) = c |y|^q$. 
First consider, at large $N$, the  contribution of a bounded domain in $\R^N$ to the integral $\mf{z}_N[W_{\e{ply}}]$. 
When restricted to such domains, the $\sinh$ two-body interactions are dominant in respect to the confinement induced by $W_{\e{ply}}$, this by one order in $N$. 
Likewise, for such domains, the Lebesgue measure should contribute to the integral by, at most, generating an exponential growth in $N$. 
Thus, the dominant contribution of bounded domains in $\R^N$ is obtained by spacing the $y_{a}$'s as far apart as possible.
Clearly,  increasing the size of the bounded domain also increases the value of the dominant contribution of this domain ... at least until the confining nature
of the potential kicks in. In fact, the confining nature of the potential also implies that if some variable is sufficiently large -on a $N$-dependent scale- the presence of the 
confining potential will result in dumping completely its contribution. This means that, in order to identify the configuration maximising the value of the integral, one should change the scale
of integration into one that depends on $N$: $y_a = T_N \la_{a}$ with  $T_N \rightarrow \infty$. 
\beq
\mf{z}_N[W_{\e{ply}}] \; = \; \Big(T_N\Big)^N
\Int{\R^N}{}  \pl{a<b}{N} \Big\{ \sinh\big[ \frac{\pi}{\om_1} T_N (\la_a-\la_b)\big] 
				\sinh\big[ \frac{\pi}{\om_2}  T_N (\la_a-\la_b)\big]  \Big\}
\cdot \pl{a=1}{N} \Big\{  \ex{- W_{\e{ply}}(T_N \la_a)}  \Big\} \cdot  \dd^N \la \;. 
\label{ecriture fct part W ply rescalee}
\enq
The sequence $T_N$  should be chosen in such a way that, for a typical distribution of the integration variables $\la_a$,  both the two-body interaction and the confinement ensured by 
the potential produce contributions of the same order of magnitude in $N$. 
For a "typical" distribution of the variables $\{\la_a\}_1^N$, the leading in $N$
asymptotic behaviour of the sum involving the potential in the \textit{rhs} below should be of the order of the \textit{lhs}. 
\beq
\sul{a=1}{N} W_{\e{ply}}\big( T_N \la_a \big) \; \sim \; T_N^{q} \,  N  \cdot C \;  . 
\enq
 Similarly, assuming a typical distribution of the variables $\{\la_a\}_1^N$ such that, for most of the variables 
$T_N|\la_a-\la_b|\gg 1$, one will have 
\beq
\sul{a<b}{N}  \ln \Big[ \sinh\big[\frac{\pi}{\om_1} T_N (\la_a-\la_b)\big] \sinh\big[ \frac{\pi}{\om_2} T_N (\la_a-\la_b)\big]  \Big] \; \sim \;
N^2 \, T_N \cdot  \wt{C} \;. 
\enq
Hence, it appears that both terms will generate the same order of magnitude in $N$ if
\beq
N^2 \cdot T_N \; = \; T_N^q \cdot N \qquad i.e. \qquad 
T_N \; = \; N^{ \f{1}{q-1} }  \;. 
\enq
Since such a rescaling tunes the interactions to the same order of magnitude in $N$, it appears plausible that the integration variables in \eqref{ecriture fct part W ply rescalee}
will condensate, analogously to what happens in the one-cut case of $\be$-ensemble, on some interval with a density $\rho_{\e{eq}}$.
These heuristics can be proven within the large-deviations approach, see Theorem 2.1 and Appendix B of [$\bs{A21}$]. 

Although I only discussed a specific example, the situation is roughly similar in the general case be it for non-varying potential $W$
or for  weakly confining potentials $W_N$ which converge, on compact subsets of $\R$ to a non-varying potential. 
In the case of non-varying potentials, the conclusion is that it is the leading large-variable asymptotic behaviour of the potential at $\pm \infty$
that drives the large-$N$ behaviour of $\ln \mf{z}_N[W]$. I should stress that, then, the analysis becomes a case-by-case study basically because this
leading asymptotic behaviour might be quite peculiar. Still, for various concrete examples of potentials $W$,  it is possible to mimic  the large-deviation approach to $\be$-ensemble integrals so as to obtain the 
leading asymptotic behaviour of $\ln \mf{z}_N[W]$. I expect that also in the case of weakly confining potentials the leading asymptotics of $\ln \mf{z}_N[W_N]$  should be tractable 
within the large deviation principle by generalising the approach of Hardy \cite{HardyBetaEnsIntWithWeakConfinment}, although I did not investigate
this issue in much details. The main point, though, is that, in order to go beyond the leading asymptotic behaviour of the logarithm, one has to alter the picture and work directly 
at the level of the rescaled multiple integral $Z_N[V_N]$, where 
\beq
Z_N[V] \; = \; 
\Int{\R^N}{}  \pl{a<b}{N} \Big\{ \sinh\big[ \frac{\pi}{\om_1} T_N (\la_a-\la_b)\big] 
				\sinh\big[ \frac{\pi}{\om_2} T_N (\la_a-\la_b)\big]  \Big\}
\cdot \pl{a=1}{N} \Big\{  \ex{-N T_N V(\la_a)}  \Big\} \cdot  \dd^N \la
\label{ecriture premiere intro dans texte de fct part rescale}
\enq
and the potential $V_N$ is given by $V_N(\la) = W(T_N \la)/(T_N N )$. 

\vspace{2mm}

It is on the level of \eqref{ecriture premiere intro dans texte de fct part rescale} that  point $2°)$ of the list given in the introduction to this chapter
introduces a striking difference in respect to $\be$-ensembles.  
Indeed, the problem of rescaling a non-varying potential may also appear in the context of $\be$-ensemble, see the works of Bleher and Fokin \cite{BleherFokinAsymptoticsSixVertexFreeEnergyDWBCDisorderedRegime}. 
However, there, the rescaling does not affect the form of the two-body Vandermonde like interactions. Hence, the situation boils down to 
the study of a $\be$-ensemble with an $N$-dependent potential. Although bringing new technicalities into the game, the problem remains globally unchanged
mainly due to the fact that all the asymptotic expansion is driven by properties and details of the equilibrium measure. The latter 
can be \textit{explicitly} constructed by means of the solution to a scalar Riemann--Hilbert problem, hence tremendously simplifying the analysis. 
However, in the case of present interest \eqref{ecriture premiere intro dans texte de fct part rescale}, the two-body interaction explicitly involves 
an $N$ dependence. In order to grasp all the relevant contributions to the large-distance asymptotic expansion one now has to incorporate 
the $N$-dependence of the two-body interaction into the analysis. Furthermore, due to the more involved form of the two-body interactions, the 
description of the equilibrium measure becomes more involved in that it is done by means of auxiliary $2\times 2$ Riemann--Hilbert problems.

The analysis of the integral \eqref{ecriture premiere intro dans texte de fct part rescale} in the presence of scaling with $N$ potentials $V_N$ can
bring several additional technical complications into the game -such as a tower of poles
squeezing onto the integration domain or a quickly oscillatory asymptotic behaviour- without really helping to clarify how the
new features brought in by this class of multiple integrals should be treated. Thus, in order to understand the large-$N$ asymptotic behaviour 
of partition functions $\mf{z}_N[W_N]$ that are of interest to quantum integrable models, it seems reasonable to start by developing the method of asymptotic analysis of \eqref{ecriture premiere intro dans texte de fct part rescale}
on the example of $N$-independent potentials $V$ satisfying to as mild hypothesis as necessary. 
This allows one to keep the complexity of the problem to a minimum and concentrate oneself solely on 
developing the framework that would allow one to deal with the varying nature of the two-body interaction. 
Doing so was the main goal of my joint work with Borot and Guionnet [$\bs{A21}$]. 
In the three subsections that follow,  I will describe the main features of the method that was developed in [$\bs{A21}$]. 

\subsection{The asymptotic expansion of $Z_N[V]$}
\label{SousSection Resultat Pour model Rescale}

The analysis of [$\bs{A21}$] builds on four hypothesis that should be satisfies  by the potential $V$ and the sequence $T_N$:

\begin{itemize}

\item[$\bullet$] the potential $V$ is confining, \textit{viz}. there exists $\eps>0$ such that 
\beq
\limsup_{ |\xi| \tend + \infty }\bigg\{ \f{  V(\xi) }{ |\xi|^{1 + \eps } }  \bigg\} \; = \; + \infty \;; 
\label{Hypothesis: confinment of the potential}
\enq
\vspace{1mm}
\item[$\bullet$]  the potential $V$ is smooth and strictly convex on $\R$ ; \vspace{1mm}
\item[$\bullet$] the potential is sub-exponential, namely there exists $\eps>0$ and $C_V>0$  such that
\beq
\label{subexpV} \forall \xi \in \mathbb{R},\qquad \sup_{\eta\in \intff{0}{\eps}} \big| V^{\prime}(\xi + \eta)\big| \, \leq \, C_V\big( |V(\xi)| +1 \big)\;,
\enq
and  given any $\kappa > 0 $ and $p\in \mathbb{N }$, there exists $C_{\kappa,p}$  such that 
\beq
 \forall \xi \in \mathbb{R},\qquad \big| V^{(p)}(\xi) \big|\,\ex{-\kappa V(\xi) } \; \leq \;  C_{\kappa,p}\;.
\label{Hypothese sous exponetialite des derivees}
\enq
\item[$\bullet$] the sequence $T_N$ grows to $+\infty$ neither too fast nor too slow:
\beq
(\ln N )^2  < T_N  < N^{\f{1}{6} }\;.
\enq
\end{itemize}

\vspace{1mm}



\subsubsection{An explicitly computable case}

Prior to stating the theorem describing the large-$N$ asymptotic expansion of $\ln Z_N[V]$ up to a $\e{o}(1)$ remainder, 
it is worthy to stress that one can compute explicitly the partition function associated with a Gaussian 
potential. Namely, one has the

\begin{prop}
\label{Proposition calcul explicite fct part Gaussienne}
Let $V_{G}(\la) =  g  \la^2 +  t \la$ be a Gaussian potential and $T_N$ be arbitrary. 
The associated partition function $Z_N[V_G]$ admits the explicit representation
\beq
\label{ecriture expression explicite fct part Gaussienne beta=1}
Z_N[V_G] \; = \;  \f{ N! }{ 2^{N(N-1)} } \,\Bigg(\frac{\pi}{g N T_N}\Bigg)^{N/2}\,
\exp\bigg\{\frac{ N^2 T_N t^2}{4g} + \frac{\pi^2(\omega_1^{-1} + \omega_2^{-1})^2}{ 12 g }T_N(N^2 - 1)\bigg\}\,
\pl{j = 1}{N } \big(  1 -  \ex{- j \frac{2 T_N \pi^2 }{ N g  \omega_1\omega_2 }   }\big)^{N - j} \;. 
\enq
Furthermore, provided that $(\ln N)^2< T_N<N^{1-\eps}$ for some $ \eps >0$, the $N\tend + \infty$ asymptotic expansion holds:
\bem
\ln Z_{N}[V_G]\;  =  \; N^{2 } T_N  \cdot  \Big[  \f{ t^2 }{4 g } \, + \,  \frac{ \pi^2(\omega_1^{-1} + \omega_2^{-1} )^2 }{12 g }  \Big]  
\; - \;  N^2 \cdot \ln 2 \; - \;  \f{N^{2}}{T_N} \cdot  \frac{g }{12  } \om_1 \om_2  \\
\; + \;   \f{ N^{2} }{ T_N^2 } \cdot  \frac{ (g \om_1 \om_2)^2 \,\zeta(3) }{ \big( 2\pi^2  \big)^2 } 
\; +\;  N   \ln\Big[ N \cdot T_N^{-1} \Big]  \;  +  \; N \cdot \ln \Big( \f{2}{\e{e}} \sqrt{\om_1\om_2} \Big)  \\
\; - \;  T_N \cdot \f{ \pi^2 (\om_1^{-1} + \om_2^{-1})^2 }{ 12 g }   
 \; + \;  \f{ \ln \big[ N^{5} \cdot T_N^{-1} \big] }{12}
\;  + \;  \f{1}{12}\ln\Bigg( \frac{128\pi^{8} }{ g \omega_1 \omega_2 } \Bigg) \, + \,  \zeta^{\prime}(-1)  \; + \; \e{o}(1)\;. 
\label{ecriture asymptotique fct part gaussienne a beta egal 1}
\end{multline}
Above, $\zeta$ refers to the Riemann $\zeta$ function. 
\end{prop}

\subsubsection{The main result}

Priori to stating the theorem giving the explicit form of the large-$N$ expansion of $Z_N[V]$ up to $\e{o}(1)$, I need to introduce a few auxiliary objects.
The function
\beq
S(\xi) \; = \; \sul{p=1}{2} \f{\pi}{\om_p}  \coth\big[ \frac{\pi}{ \om_p} \xi \big] 
\label{ecriture definition fct S}
\enq
defines the integral kernel of the singular integral operator that arises in most of the important equations of the analysis. 
Let $\mc{F}[S]$ be the principal value regularised Fourier transform of $S$. Below, I will make use of $J(u)$, the regularised Fourier transform of $1/\mc{F}[S]$
defined, for $u>0$, as 
\beq
J(u) \; = \; \Int{ \Ga_+(\i \R ^+ )  }{}  \f{ \ex{\i \la u}  }{  \mc{F}[S](\la) }  \cdot \dd \la  \qquad \e{where} \quad  \mc{F}[S](\la) \; = \;  
\f{  \i \pi \sinh\big[\tf{\la(\om_1+\om_2)}{ 2 } \big]  }{  \sinh\big[\tf{\om_1 \la}{2} \big]  \sinh\big[\tf{ \om_2 \la}{2} \big] } \;. 
\label{definition fonction j de u et ctr Gamma plus de i R}
\enq
The contour $\Ga_+(\i \R ^+ )$ appearing above is a counter-clockwise loop around $ \i \R^+ + \i \eps$, for some $\eps>0$ small enough. 
Let $R_{\da}$ denote  the Wiener-Hopf factor of $ \tf{1}{ \mc{F}[S](\la) }$
on $\mathbb{H}^-$. $R_{\da}$ admits the explicit form:
\beq
R_{\da}(\la)  \; = \; \f{ \la }{ 2\pi } \sqrt{\f{\om_1 \om_2}{\om_1+\om_2}  } \cdot  \Big( \f{\om_1}{\om_1+\om_2} \Big)^{- \f{ \i \la}{2\pi \om_1} } 
 \cdot \Big( \f{\om_2}{\om_1+\om_2} \Big)^{- \f{\i \la}{2\pi \om_2} }     
 \cdot \f{ \Ga\bigg( \f{ \i \la \om_1 }{ 2\pi } \bigg) \cdot \Ga\bigg( \f{ \i \la \om_2 }{ 2\pi  } \bigg) }
 {\Ga\bigg( \f{ \i \la }{ 2\pi  } (\om_1+\om_2) \bigg)  } \;. 
\enq

\begin{theorem}

For any potential $V$ satisfying to the four hypothesis, the partition function $Z_N[V]$ admits the asymptotic expansion
\bem
 \ln \Bigg( \f{ Z_N[V] }{ Z_N[V_{G;N}] } \Bigg)  \; = \; -N^{2} T_N \sul{p=0}{ [ 2/\alpha] +1} \f{ \capricornus_p[V] }{ T_N^p } 
\; + \; T_N \cdot \gimel_0\cdot\Big( \leo[V,V_{G;N}](b_N) -  \leo[V,V_{G;N}](a_N)  \Big) \\ 
\; + \;  \aleph_0\cdot  \Big( \leo[V,V_{G;N}]^{\prime}(b_N) +  \leo[V,V_{G;N}]^{\prime}(a_N)  \Big)  \, + \, \e{o}(1) \;. 
\label{Ecriture DA ZN dans partie Intro resultats}
\end{multline}
 $\gimel_0$ and $\aleph_0$ appearing in this expansion are numerical coefficients given, resp., in terms of a single and four-fold integral. 
Also, the answer involves the Gaussian potential 
\beq
V_{G;N}(\xi) \; = \; \f{ \pi (\om_1^{-1}+\om_2^{-1}) \cdot \Big[ \xi^2 - \big( a_N+b_N \, + \,  \e{O}(N^{-\infty}) \big) \xi \Big]   }
{ b_N-a_N \, + \, \f{1}{ T_N } \sul{p=1}{2} \f{ \om_p }{\pi } \ln \Big( \f{ \om_p }{ \om_1+\om_2 }\Big)   \; + \;  \e{O}(N^{-\infty}) } \; . 
\enq
The whole $V$-dependence of the expansion \eqref{Ecriture DA ZN dans partie Intro resultats} is encoded in the coefficients $\capricornus_p[V]$ given below, 
the function 
\beq
\leo[V,V_{G;N} ](\xi) \; = \; \f{V^{\prime}(\xi)- V_{G;N}^{\prime}(\xi) }{ V^{\prime\prime}(\xi)- V_{G;N}^{\prime \prime}(\xi)   }\,
\cdot \ln \Bigg( \f{  V^{\prime\prime}(\xi) }{ V_{G;N}^{\prime \prime}(\xi)  } \Bigg)  
\label{leoDef}
\enq
or its derivative $ \leo[V,V_{G;N}]^{\prime}(\xi)$ and in the sequences $a_N$ and $b_N$. Note that $a_N$ and $b_N$
both depend implicitly on $V$.  They will be discussed in more details in Theorem \ref{Proposition caracterisation rudimentaire mesure equilibre} to come. 

When $p=0$, one has 
\beq
\capricornus_0[V] \; = \; \f{ - \om_1 \om_2}{ 4 \pi  (\om_1+\om_2) }\Int{ a_N }{ b_N } V^-_{N}(\xi) \cdot (V_{N}^-)^{\prime\prime}(\xi)\,\dd \xi
\qquad where \quad V_{N}^{\pm} = V \pm V_{N;G} \;. 
\enq
For any $p\geq 1$, the coefficients $\capricornus_p[V]$ read :
\begin{eqnarray}
\capricornus_p[V] & = &    u_{ p+1 } \Int{a_N}{b_N} V^{-}_{N}(\xi)\cdot V^{(p+2)}(\xi)\,\dd \xi  \\
& & +  \sul{ \substack{ s+\ell=p-1 \\ s,\ell \geq 0} }{} \f{\daleth_{s,\ell}}{ s! }\Bigg\{(-1)^{\ell}\,(V^{-}_{N})^{(\ell+1)}(a_N)\cdot (V^+_{N})^{(s + 1)}(a_N) \; + \; (-1)^{s}\,(V_{N}^{-})^{(\ell+1)}(b_N)\cdot (V^+_{N})^{(s+1)}(b_N) \Bigg\}\; .  \nonumber
\end{eqnarray}
The coefficients $\daleth_{s,\ell}$ are expressed as:  
\beq
\daleth_{s,\ell} \; = \; \f{ \i^{s+\ell+1} }{2\pi } \sul{r=1}{\ell+1} \f{ s!  }{r! (s+\ell+1-r)!  } \cdot 
\f{\Dp{}^r}{\Dp{}\mu^r } \Big( \f{\mu}{ R_{\da}(-\mu) } \Big)_{\mid \mu=0} \cdot 
\f{\Dp{}^{s+1+\ell-r} }{\Dp{}\mu^{s+1+\ell-r} } \Big( \f{1}{ R_{\da}(\mu) } \Big)_{\mid \mu=0}  \;.
\label{definition constantes dialeth s ellbis}
\enq
The $V$-independent coefficient $\gimel_0$ appearing in the penultimate term in \eqref{Ecriture DA ZN dans partie Intro resultats} is defined as 
\beq
\gimel_0 \; = \; \Int{0}{+\infty} J(u) \cdot \big( uS^{\prime}(u) \, + \,  S(u)\big)  \cdot \frac{\dd u}{2\pi}
\label{gimel0def}\enq
Finally, the numerical prefactor $\aleph_0$ is expressed in terms of the four-fold integral
\bem
\aleph_0 \; = \; - \f{\om_1+\om_2}{2 \om_1 \om_2 }{\displaystyle \Int{ \R }{} }  \f{ \dd u}{2\pi} J(u)   {\displaystyle \Int{ |u| }{ + \infty }  } \dd v\,\partial_{u} 
\Big\{ S(u)\cdot\Big( \mf{r}\big[\f{v - u}{2}\big] - \mf{r}\big[\f{v + u}{2}\big]  \Big) \Big\}   \\
\; + \;  \Int{ \Ga_+(\i \R ^+ ) }{}  \f{\dd \la\,\dd \mu }{ (2 \i \pi)^2 }  \f{ \mu \cdot (\om_1^{-1}+\om_2^{-1}) }{ (\la+\mu)R_{\da}(\la)R_{\da}(\mu)   }  
{\displaystyle  \Int{0}{+\infty} } \dd x\,\dd y\,\ex{ \i \la x  + \i \mu y}\,  \partial_{x}\Bigg\{
 S(x-y)\,\Bigg( \mf{r}(x) - \mf{r}(y) - \f{ (x-y) \om_1 \om_2 }{2\pi (\om_1+\om_2)} \Bigg)\Bigg\} \;. 
\label{aleph0def} %
\end{multline}
The contour $\Ga_+(\i \R ^+ )$ is as defined in \eqref{definition fonction j de u et ctr Gamma plus de i R}. 
The integrand of $\aleph_0$ involves the function $\mf{r}$ which is given by 
\beq
\mf{r}(x) \; = \;  \frac{\mf{c}_1(x) + \mf{c}_0(x)\Bigg[\sul{p=1}{2} \f{ \om_p }{2\pi} \ln \Big( \f{ \om_p }{ \om_1+\om_2 }\Big)\Bigg]}{1+2\pi  (\om_1^{-1}+\om_2^{-1})  \mf{c}_0(x)}
\enq
with
\beq
\mf{c}_p(x) \, = \, \f{\i^{p-1} }{ 2\pi} \cdot  \sqrt{ \f{\om_1 \om_2}{\om_1+\om_2} }  \cdot
\Int{ \Ga_+(\i \R ^+ ) }{} \f{ \ex{\i \la x} }{ \la } \f{\Dp{}^p }{ \Dp{}\la^p}\Big( \f{1}{ R_{\da}(\la)} \Big) \cdot \f{ \, \dd \la }{2\i \pi  }  \;. 
\enq
\end{theorem}

With the asymptotic expansion \eqref{Ecriture DA ZN dans partie Intro resultats} at hand, it suffices to invoke Proposition \ref{Proposition calcul explicite fct part Gaussienne}
so as to access to the \textit{per se} asymptotic expansion of $\ln Z_N[V]$.

\subsection{Intermediate results: the $N$-dependent equilibrium measure and the master operator}
\label{SousSection Resultats Auxiliaires}

It is not hard to generalise the large deviation principle based method for extracting the leading asymptotics of $\be$-ensemble integrals
to the case of the varying sinh-interaction partition function $Z_N[V]$.

\begin{prop}
 \label{Proposition LDP order dominant fct part rescalee ZN}

Let $\mc{E}_{\infty}$ be the lower semi-continuous good rate function 
\beq
\mc{E}_{\infty}[ \mu ] \; =\frac{1}{2} \; \Int{}{}\Big( V(\eta)+
V(\xi) \; - \;  {\pi(\om_1^{-1}+\om_2^{-1}) } |\xi-\eta|\Big) \,\dd \mu(\xi)\dd \mu(\eta) \;.
\label{definition rate fct pour fct part ZN rescalee}
\enq
Then, under the four hypothesis stated earlier on, one has that 
\beq
\lim_{N \tend +\infty}  \bigg\{ \f{\ln Z_N[V] }{N^{2} T_N }  \bigg\} \; = \; - \inf_{\mu \in \mc{M}^{1}(\R)} \mc{E}_{\infty}\big[\mu \big] \;. 
\enq
The infimum of $\mc{E}_{\infty}$ is attained at a unique probability measure $\mu_{\e{eq}}$. This measure is continuous with respect to the Lebesgue measure and has density
\beq
\rho_{\e{eq}}(\xi) \; = \; \f{ V^{\prime \prime}(\xi) }{ 2\pi  }\cdot \f{ \om_1 \, \om_2 }{ \om_1+\om_2}\cdot\bs{1}_{\intff{a}{b}}(\xi) 
\label{denspot} 
\enq
supported on the interval $\intff{a}{b}$. The endpoints $a$ and $b$ correspond to the unique solutions to the set of equations
\beq
V^{\prime}(b) \; = \; - V^{\prime}(a) \; = \;  \pi  (\om_1^{-1}+\om_2^{-1}) \;. 
\label{ecriture equation definissant couple a b asympt}
\enq
One has, explicitly, 
\beq
\lim_{N \tend +\infty} \bigg\{ \f{ \ln Z_N[V]}{N^{2}T_N } \bigg\} \; = \; -\f{ V(a)+V(b) }{2}
\, + \, \f{ \big( V^{\prime}(b) \big)^2(b-a) \, + \,  \int_{a}^{b} \big( V^{\prime}(\xi) \big)^2\,\dd \xi   }{ 4\pi   (\om_1^{-1}+\om_2^{-1} ) }     \;.
\enq

\end{prop} 

The strict convexity of $V$ guarantees that the density  $\rho_{\e{eq}}$ given by \eqref{denspot}  is positive and that it reaches a non-zero limit at the endpoints of the support. 
This behaviour contrasts with the situation usually\symbolfootnote[2]{In some very special, non-generic, cases the square root behaviour can soften since the vanishing may occur
as some odd power of a square root.} encountered in $\beta$ ensembles with real analytic potentials; in the latter case
one deals with a square root (or inverse square root) vanishing (or divergence) of the equilibrium density at the edges of its support.

A slight refinement of Proposition~\ref{Proposition LDP order dominant fct part rescalee ZN} leads to the more precise estimates
\beq
 \ln Z_N[V]  \; = \; - N^2 T_N \cdot \mc{E}_{\infty}\big[\mu_{\e{eq}} \big] \; + \; \e{O}(N^2) \;. 
\label{ecriture DA naif fct part}
\enq
It is on the level of these estimates  that problems start to arise. Indeed, when comparing to the case of $\be$-ensembles with
varying weight, one observes a much weaker relative control on the remainder. With respect to the leading asymptotics, the remained  in \eqref{ecriture DA naif fct part} is 
smaller by the order of magnitude $T_N^{-1}$ whereas, in the $\be$-ensembe case, the remainder is smaller  by the order of magnitude $N^{-1}$. 
This loss of relative precision takes its origin in the fact that the purely asymptotic rate function $\mc{E}_{\infty}\big[\mu_{\e{eq}} \big]$ does not absorb
enough of the fine structure of the dominant configurations, \textit{viz}. the configurations of integration variables which produce the dominant contribution to the integral.  
 As a consequence, the  $\e{O}(N^2)$ remainder in \eqref{ecriture DA naif fct part} mixes \textit{both} types of contributions: \vspace{1mm}
\begin{itemize}
 \item[$\bullet$] the deviation of the dominant configurations with respect to its asymptotic position governed by $\mc{E}_{\infty}$ ;\vspace{1mm}
\item[$\bullet$] the fluctuation of the integration variables around the dominant configurations. \vspace{2mm}
\end{itemize}

It turns out that the fine, $N$-dependent, structure of the dominant configurations is  captured by the $N$-dependent
deformation\symbolfootnote[4]{The property of lower semi-continuity along with the fact that $\mc{E}_N$ has compact level sets is verified exactly as in the case of $\be$-ensembles, 
so we do not repeat the proof here.} of the rate functions $\mc{E}_{\infty}$:
\beq
\mc{E}_N [ \mu  ] \; = \f{1}{2} \; \Int{}{} \Big( V(\xi)\,+\,V(\eta) \,-\, \f{ 1 }{  T_N } 
\ln\bigg\{ \pl{p=1}{2} \sinh\big[ \f{\pi}{ \om_p } T_N (\xi-\eta) \big] \bigg\} \;  \Big)\,\dd \mu(\xi) \otimes \dd \mu(\eta) \;.
\label{definition fnelle a minimiser N dpdte}
\enq
This  $N$-dependent rate function is, in fact, naturally dictated by the very form of the integrand in \eqref{ecriture premiere intro dans texte de fct part rescale}
and turns out to be extremely effective for the purpose of the large-$N$ analysis of $Z_N[V]$. 
Namely, it allows one to re-sum, into a single term,  a whole tower of contributions to the remainders arising at various steps of the analysis.  
The improvement provided by the use of the finite-$N$ minimiser of $\mc{E}_N$, instead of the one of $\mc{E}_{\infty}$, leads to :
\beq
 \ln Z_N[V]  \; = \; - N^{2}T_N \inf_{\mu\in \mc{M}^1(\R) } \mc{E}_{N}\big[\mu \big] \; + \; \e{O}(N T_N) \;. 
\enq
 I do stress that the use of $\mc{E}_N$ 
should not be considered as a mere technical simplification of the intermediate steps. Its use is of prime importance. 
The use of the more "classical" object $\mc{E_{\infty} }$ would render the large-$N$ analysis of the Schwinger-Dyson equations impossible.

Exactly as in the case of $\be$-ensembles, the minimiser of $\mc{E}_N$ admits a  characterisation in terms of a variational problem. 
It so happens that one can still construct the solution to this variational problem by means of Riemann--Hilbert problems. 
However, in the present case one has to deal with a $2\times2$ matrix Riemann--Hilbert problem instead of a scalar one. 

\begin{theorem}
\label{Proposition caracterisation rudimentaire mesure equilibre}
 For any strictly convex potential $V$, the $N$-dependent rate function $\mc{E}_N$ admits a unique minimum on $\mc{M}^{1}(\R)$ 
 which occurs at the equilibrium  measure  $\mu_{\e{eq}}^{(N)}$. This equilibrium measure is supported on a segment $\intff{a_N}{b_N}$ and
corresponds to the unique solution to the integral equations
\beqa
&&V(\xi) \; - \; \f{ 1 }{ T_N } \Int{}{} 
\ln\bigg\{ \pl{p=1}{2} \sinh\big[ \frac{\pi}{\om_p}  T_N (\xi-\eta) \big] \bigg\}\,\dd \mu_{\e{eq}}^{(N)}(\eta) \; = \; C_{\e{eq}}^{(N)}  
\qquad  \e{on}  \; \; \intff{a_N}{b_N} 
\label{definition de la cste Ceq par eqn int eq meas} \\
&& V(\xi) \; - \; \f{ 1  }{  T_N } \Int{}{} 
\ln\bigg\{ \pl{p=1}{2} \sinh\big[ \frac{\pi}{\om_p}  T_N (\xi-\eta) \big] \bigg\}\,\dd \mu_{\e{eq}}^{(N)}(\eta) \; >  \; C_{\e{eq}}^{(N)} 
\qquad  \e{on} \;\; \R \setminus \intff{a_N}{b_N}  \;, 
\label{ecriture condition negativite dehors support mu eq}
\eeqa
with $C_{\e{eq}}^{(N)}$ a constant whose determination is part of the problem
\eqref{definition de la cste Ceq par eqn int eq meas}-\eqref{ecriture condition negativite dehors support mu eq}.
\vspace{2mm}

\noindent There exists $N_0$ such that for any integer $N\geq N_0$, the equilibrium measure $\mu_{\e{eq}}^{(N)}$: \vspace{1mm}

\begin{itemize}
\item[$\bullet$] is supported on the single interval $\intff{a_N}{b_N}$ whose endpoints admit the asymptotic expansion
\beq
a_N \; =  \; a \, +\, \sul{\ell=1}{k}  \f{ a_{N;\ell}  }{ T_N^{\ell}  }  \, + \, \e{O}\Big(  T_N^{-(k+1)}  \Big)
\quad and \quad b_N  \; = \;  b \, +\, \sul{\ell=1}{k}  \f{ b_{N;\ell}  }{ T_N^{\ell} }  \, + \, \e{O}\Big(  T_N^{-(k+1)}  \Big)  \;,
\label{ecriture DA grand N point bord du support de mu eq N}
\enq
where $k\in \mathbb{N}^*$ is arbitrary, $(a,b)$ are as defined in \eqref{ecriture equation definissant couple a b asympt} while
\beq
\left( \ba{c} b_{N;1} \\ 
	      a_{N;1} \ea \right)  \;=\;\bigg\{ \sul{p=1}{2} \f{\om_p }{2\pi } \ln \Big( \f{ \om_p }{ \om_1+\om_2 } \Big) \bigg\}  \cdot 
	      \left( \ba{c}  V^{\prime\prime}(a) \cdot \big\{ V^{\prime\prime}(b) \big\}^{-1}    \\ 
	      -  V^{\prime\prime}(b) \cdot \big\{ V^{\prime\prime}(a) \big\}^{-1}  \ea \right) \; ; 
\label{blbalb}
\enq

\item[$\bullet$] is continuous with respect to Lebesgue's measure. Its density $\rho_{\e{eq}}^{(N)}$ is smooth on $\intoo{a_N}{b_N}$  and vanishes like a square-root at the edges:
\beq
\rho_{\e{eq}}^{(N)}(\xi) \underset{\xi \tend a_N^{+} }{\sim} 
\Big( V^{\prime\prime}(a_N) \, + \, \e{O}( T_N^{-1} )   \Big)\, \sqrt{ \f{\om_1 \, \om_2 (\xi-a_N) }{ \pi^{3}(\om_1+\om_2) }  }\;,\qquad
  \rho_{\e{eq}}^{(N)}(\xi) \underset{\xi \tend b_N^{-} }{\sim} 
\Big( V^{\prime\prime}(b_N) \, + \, \e{O}( T_N^{-1} )   \Big)\, \sqrt{ \f{\om_1 \, \om_2 (b_N-\xi) }{ \pi^{3}(\om_1+\om_2) }  } \;.
\nonumber
\enq
Furthermore, there exists a constant $C > 0$ independent of $N$ such that:
\beq
\norm{\rho_{\e{eq}}^{(N)}}_{L^{\infty}(\intff{a_N}{b_N})}  \; \leq  \;  C\,\norm{V^{\prime\prime} }_{ L^{\infty}(\intff{a_N}{b_N}) }\;.
\label{ecriture borne uniforme sur la densite}
\enq
\end{itemize}
\noindent The density of the equilibrium measure has the explicit expression $\rho_{\e{eq}}^{(N)}=\mc{W}_N[V^{\prime}]$ which is given in terms of the $\mc{W}_N$ transform, \textit{c}.\textit{f}. \eqref{definition operateur WN} below,
of the potential.

\end{theorem}

The proof of the first part of the above theorem is rather classical. The continuity with respect to the Lebesgue measure 
and the at least square root vanishing at the endpoints follows from arguments that can be found, say in Section 2.3 of [$\bs{A22}$]. 
 The connectedness of the support and the strict inequality in \eqref{ecriture condition negativite dehors support mu eq} follows from 
a convexity argument that first appeared in \cite{MhaskarSaffLocumOfSupNormWeightedPly}. Elements of proof of these properties are gathered in Appendix C of [$\bs{A21}$]. 
However, the proof of the second part which gives a  quite explicit control on the various properties of $\rho_{\e{eq}}^{(N)}$ for $N$ large enough is much more involved. 
Those results are obtained through a characterisation of the density $\rho_{\e{eq}}^{(N)}$ as the solution to the singular integral equation $\mc{S}_N\big[ \rho_{ \e{eq} }^{(N)} \big](\xi) = V^{\prime}(\xi)$
on $\intff{a_N}{b_N}$, where 
\beq
\mc{S}_N\big[ \phi \big](\xi) \; = \; \Fint{a_N}{b_N} S\big[T_N(\xi-\eta)\big] \phi(\eta)\,\dd \eta  \;. 
\label{ecriture eqn int sing de depart}
\enq
The integral kernel of this singular integral operator involves the function $S$ which has been defined in \eqref{ecriture definition fct S}. 
The unknowns in this equation $(\rho_{ \e{eq} }^{(N)}, a_N,b_N)$ should be picked in such a way that 
$\rho_{ \e{eq} }^{(N)}$ has mass 1 on $\intff{a_N}{b_N}$ and that it has, at least, a square root behaviour at the endpoints $a_N,b_N$. 
Hence, the construction of  the equilibrium measure basically boils down to an inversion of the singular integral operator $\mc{S}_N$.
I stress that, on top of being involved in the characterisation of the equilibrium measure, 
the singular integral operator $\mc{S}_N$ plays a prominent role in the large-$N$ analysis of the Schwinger-Dyson equations.  
As a consequence, constructing its inverse $\mc{W}_N$ and then obtaining precise estimates on $\mc{W}_N[\phi]$  is of prime importance to the large-$N$ analysis. 
One of the main achievements of  [$\bs{A21}$] consisted precisely in 
obtaining a sharp control on its inverse $\mc{W}_N$, defined between appropriate functional spaces.  

When $N$ is large enough, the inversion of $\mc{S}_N$ can be accomplished by means of a non-linear  steepest descent analysis of a $2\times 2$ matrix Riemann--Hilbert problem 
followed by some additional asymptotic analysis.  Indeed, the operator $\mc{S}_N$ is of truncated Wiener--Hopf 
type so that its inversion is equivalent to solving a $2\times 2$ matrix valued Riemann--Hilbert problem for a piecewise holomorphic matrix $\chi$, see Section 4.1 of [$\bs{A21}$]
for more details. The Riemann--Hilbert problem for this $\chi$ is investigated, for $N$-large enough, in Sections 4.3-4.5 of that paper.
The obtained characterisation of its solution is then used in Section 5 of [$\bs{A21}$] so as to describe, quite explicitly, the inverse $\mc{W}_N$
of $\mc{S}_N$ understood as the map $\mc{S}_N \; : \; H_{s}(\intff{a_N}{b_N}) \mapsto \mc{S}_N\big[ H_{s}(\intff{a_N}{b_N}) \big]$.
One of the difficulties that arises at that point is that  the Riemann--Hilbert approach to the inversion, as developed by Novoksenov in \cite{NovokshenovSingIntEqnsIntervalGeneral}, provides one 
with a "raw" pseudo-inverse $\wt{\mc{W}}_N$ (\textit{viz}. modulo fixing the elements of the kernel) of the operator $\mc{S}_N$ when the latter is understood as an operator on $ H_{s}(\R) $ with $s<0$. 
In order to construct the inverse in the case when $\mc{S}_N$ acts on more regular Sobolev spaces $H_{s}(\R)$ with $0 <s <\tf{1}{2}$, 
it becomes necessary to provide, first, an effective  description of the image space $\mf{X}_s(\R)= \mc{S}_N\big[ H_{s}(\intff{a_N}{b_N}) \big]$ with $0<s<\tf{1}{2}$. 
The characterisation of $\mf{X}_s(\R)$ has to be manageable enough so as to allow one for an easy implementation of the 
restriction of the "raw" inverse $\wt{\mc{W}}_N$ to $\mf{X}_s(\R)$. 
By "easy", I mean one which produces an operator $\mc{W}_N$ on $\mf{X}_s(\R)$ that can be \textit{straightforwardly} checked to be 
continuous as an operator $ \mf{X}_{s}(\R) \mapsto H_{s}(\intff{a_N}{b_N})$ and which can also be checked, through relatively simple calculations, to be the inverse of $\mc{S}_N$. 

In order to state the conclusion of this analysis, I need to precise that $\chi_{11}$, the $(1,1)$ matrix entry of $\chi$, is such that 
$\mu  \mapsto \mu^{1/2} \cdot \chi_{11}(\mu) \in L^{\infty}(\R)$.

\begin{theorem}
Let $0 < s <1/2$ and $\chi$ be the unique solution to the Riemann--Hilbert problem discussed in Section 4.3 of [$\bs{A21}$]. 
The operator $\mc{S}_N \; : \; H_{s}\big( \intff{a_N}{b_N} \big) \tend 
\mf{X}_{s}\big( \R  \big)$ is continuous and invertible. Here, for any closed $A \subseteq \R$,  
\beq
\mf{X}_{s}\big(  A \big) \; = \; \Big\{  H \in H_{s}\big( A  \big) \; :  \;  
\Int{\R+\i \eps }{} \chi_{11}(\mu) \mc{F}[H](T_N\mu)\ex{- \i T_N \mu b_N}   \f{ \dd \mu }{ 2 \i \pi }  \; = \; 0\Big\} \; 
\enq
is a closed subspace of $ H_{s}\big(  A \big) $ such that $\mc{S}_N\big( H_{s}\big( \intff{a_N}{b_N} \big) \big) \, =  \, \mf{X}_s(\R)$.  
The inverse of $\mc{S}_N \; : \; H_{s}\big( \intff{a_N}{b_N} \big) \tend \mf{X}_{s}\big( \R  \big)$ is provided by the operator 
$\mc{W}_N $ which takes, for $H\in \mc{C}^1(\intff{a_N}{b_N})\cap\mf{X}_s(\R)$, the form of an integral transform 
\beq
\mc{W}_N[H](\xi) \; = \; \f{T_N^2 }{ 2\pi }  
\Int{ \R + 2 \i \eps }{} \f{ \dd \la }{ 2 \i \pi } \Int{ \R + \i \eps }{} \f{ \dd \mu }{ 2 \i \pi } 
\f{ \ex{- \i T_N (\xi-a_N) \la}  }{ \mu- \la }
\bigg\{   \chi_{11}(\la) \chi_{12}(\mu)  - \f{ \mu }{ \la }\cdot\chi_{11}(\mu) \chi_{12}(\la) \bigg\} 
\cdot \Int{a_N}{b_N} \!\! \dd \eta  \ex{ \i T_N \mu (\eta-b_N) } H(\eta)  \;. 
\label{definition operateur WN}
\enq
In the above integral representations the parameter $\eps >0$ is small enough but arbitrary otherwise. 
Furthermore, for any $H \in \mc{C}^{1}\big( \intff{a_N}{b_N} \big)$, the transform $\mc{W}_N$ 
 exhibits the local behaviour 
\beq
\mc{W}_N[H] (\xi) \underset{ \xi \tend a_N^+ }{ \sim  } C_L H^{\prime}(a_N) \sqrt{ \xi - a_N} \qquad and \qquad
\mc{W}_N[H] (\xi) \underset{ \xi \tend b_N^- }{ \sim  } C_R H^{\prime}(b_N) \sqrt{ b_N - \xi } \;. 
\enq
where $C_{L/R}$ are some $H$-independent constants. 
\end{theorem}

The characterisation of the inverse $\mc{W}_N$ provided by the above theorem is, however, only the beginning of the analysis. 
Starting from there, it is necessary to extract fine informations of the large-$N$ behaviour of $\mc{W}_N$ this to any order in $N$ 
and in such a way that, when acting on sufficiently regular functions, the remainders are controlled by   $W_{k}^{\infty}(\R)$ norms. I remind
that these norms are defined as 
\beq
 \norm{f}_{ W^{\infty}_{k} (A,\dd \mu) } \; = \; \max
\bigg\{  \norm{ \Dp{x_1}^{a_1}\dots \Dp{x_n}^{a_n}f  }_{  L^{\infty}(A,\dd \mu) }  \; : \; 
  a_{\ell} \in \mathbb{N}, \; \ell=1,\dots, n,   \; \;  \e{and}\; \e{satisfying}  \; \sul{\ell=1}{n}  a_{\ell} \leq k   \bigg\}\;.
\enq
Such a fine control is achieved in Section 6  of [$\bs{A21}$]. In refer to that section,  Propositions 6.4, 6.6
and Lemma 6.11, for a more precise description of the results. 

\vspace{1mm}

Having developed all this machinery, one can finally characterise the density of the equilibrium measure $\mu_{\e{eq}}^{(N)}$ completely. 
The latter is expressed in terms of the inverse as $\rho_{\e{eq}}^{(N)}=\mc{W}_N\big[ V^{\prime} \big]$ although, at this stage,
the endpoints $(a_N, b_N)$ of its supports are still unknown. As follows from the arguments advocated earlier on,  these endpoints  
corresponds to the unique solution to the system of equations 
\beq
\Int{a_N}{b_N} \mc{W}_N\big[ V^{\prime} \big](\xi)\,\dd  \xi \; = \; 1 \qquad \e{and} \qquad 
\Int{\R + \i \eps }{}  \f{ \dd \mu}{ 2 \i \pi }  \chi_{11}(\mu) 
 \Int{a_N}{b_N} \ex{\i\mu T_N (\eta- b_N)} V^{\prime}(\eta) \,\dd \eta   \; = \; 0 \;. 
\enq
The first condition guarantees that $\mu_{\e{eq}}^{(N)}$ has indeed mass 1, while the second one ensures 
that its density vanishes as a square root at the edges $a_N,b_N$. Using the control in $N$  on the inverse obtained in Section 6, 
one can obtain the all-order in $T_N^{-1 } $large-$N$ asymptotic expansion of these constraints, see Lemma 7.2 and Proposition 7.6 of [$\bs{A21}$]. 
These expansions, in their turn, allow one to prove the existence of $a_N$ and $b_N$
as well as the fact that these admit an all order asymptotic expansion in $T^{-1}_N$, see Lemma 7.9 and Corollary 7.10  of [$\bs{A21}$] for more details. 
All-in-all, one then arrives to the characterisation of the equilibrium measure as announced in Theorem \ref{Proposition caracterisation rudimentaire mesure equilibre} above.

\subsection{The overall strategy of the proof}
\label{SousSection Strategi de la preuce modele sinh rescale}

In the following, I denote by $p_N^{V} ( \bs{\la} ) $ the probability density on $ \R^N$ associated with the partition function $Z_N[V]$
defined in \eqref{ecriture premiere intro dans texte de fct part rescale}:
\beq
p_N^{V} ( \bs{\la} ) \;  =  \;  \f{1 }{Z_N[V]} 
\pl{a<b}{N} \Big\{ \sinh\big[ \f{ \pi }{ \om_1 }  T_N(\la_a-\la_b)\big] \sinh\big[ \f{ \pi }{ \om_2 }  T_N (\la_a-\la_b)\big]  \Big\}
\,\pl{a=1}{N} \ex{- N T_N V(\la_a) }   \;. 
 \label{definition mesure proba rescalee avec potentiel arbitraire}
\enq
$p_N^{V} ( \bs{\la} )$ gives rise to the probability measure $\mathbb{P}_N^{V}$ on $\R^N$. 

Prior to going deeper into the details, I need to provide the definition of the expectation value of a function in $\ell$ variables. 
Those expectation values play an important role in the analysis. 

\begin{defin}
\label{Definition moyenne contre mes stochastique}
Let $\nu_1,\dots, \nu_{\ell}$ be any (possibly depending on the stochastic vector  $\bs{\la}$) 
measures and $\psi$ a function in $\ell$ variables. 
Then 
\beq
\Big<   \psi \Big>_{\nu_1 \otimes \dots \otimes \nu_{\ell} }^{V}  \equiv   
 \Big<   \psi(\xi_1,\dots,\xi_{\ell} ) \Big>_{\nu_1 \otimes \dots \otimes \nu_{\ell} }^{V}    \equiv   
\mathbb{P}_N^{V}  \Big[     \Int{\R^{\ell} }{}    \psi(\xi_1,\dots,\xi_{\ell})\,\dd \nu_1 \otimes \dots \otimes  \dd \nu_{\ell} \Big]
\label{definition moyenne contre mesures stochastiques}
\enq
whenever it makes sense. 

\end{defin}

\noindent Note that if none of the measures $\nu_1,\dots, \nu_{\ell}$ is stochastic, \textit{viz}. does \textit{not} depend on the integration variables $\bs{\la}$, then the expectation versus $\mbb{P}_N^{V}$ 
in \eqref{definition moyenne contre mesures stochastiques} can be omitted.

\vspace{1mm}

The expectation values allow one to compute the logarithmic derivatives of the partition function. For instance, if $\{V_t\}_{t}$ is a smooth, one-parameter t, family of potentials, then 
\beq
\f{ \Dp{} }{ \Dp{} t } \ln Z_{N}[V_t] \; = \; -N^{2} T_N \cdot \Big<  \f{ \Dp{} }{ \Dp{} t } V_t   \Big>_{ L_N^{(\bs{\la})} }^{ V_t }   \qquad \e{where} \qquad L_N^{ (\bs{\la}) } \; = \; \f{1}{N} \sul{a=1}{N} \de_{\la_a}   
\label{equation reliant derivee log fct part et corr derivee pot}
\enq
is the empirical measure associated with the stochastic vector $\bs{\la}$.  I stress that the expectation value in \eqref{equation reliant derivee log fct part et corr derivee pot}
is computed in respect to the probability measure subordinate to the $t$-dependent potential $V_t$. 

\vspace{1mm}

Given any potential $V$ satisfying to the four hypothesis and giving rise to an equilibrium measure supported on $\intff{a_N}{b_N}$, 
it is shown in Lemma D.1 of [$\bs{A21}$] that there exists a quadratic potential $V_{G;N}$ whose equilibrium measure is precisely 
supported on $\intff{a_N}{b_N}$. 
Taking $V_{t} = t V + (1-t)V_{G;N}$ in \eqref{equation reliant derivee log fct part et corr derivee pot} readily allows one to integrate
the \textit{lhs} over $t$. Then, since the asymptotic expansion of Gaussian partition functions $Z_N\big[ V_{G;N}\big] $ is provided in Proposition \ref{Proposition calcul explicite fct part Gaussienne},
the extraction the large-$N$ behaviour of $Z_{N}[V]$
boils down to obtaining a sufficiently precise and uniform in $t \in \intff{0}{1}$ control on the large-$N$ behaviour
of the one-point expectation values $\moy{\cdot}_{ L_N^{(\bs{\la})} }^{ V_t } $. I do stress that since both $V$ and $V_{G;N}$ satisfy to the four hypothesis stated 
in the beginning of this section, so does $V_{t}$. 

\vspace{1mm}

Just as for the connected correlators, one can write down a set of loop equations satisfied by the expectation values. In the present context, 
these constitute a tower of equations which relate expectation values -in the sense of Definition \ref{Definition moyenne contre mes stochastique}- 
of functions in many, not necessarily fixed, variables which are integrated versus
\beq
\mc{L}^{(\bs{\la})}_N=L^{(\bs{\la})}_N-\mu_{\e{eq}} \;, 
\enq
\textit{i.e.} versus the empirical measure centred around $\mu_{\e{eq}}$. 
Below, I will just provide the first one of these equations
and refer to Proposition 3.13 of [$\bs{A21}$] for a general presentation:
\beq
\label{SDlevel1}
- \big<   \phi \big>_{  \mc{L}_N^{(\bs{\la})} }^{V}
 \; + \;  \f{1}{2 } 
 \Big<  \mc{D}_N\circ{\cal U}_{N}^{-1}[\phi] \Big>_{  \mc{L}_N^{(\bs{\la})}    \otimes   \mc{L}_N^{(\bs{\la})} }^{V}  \; = \; 0  \;. 
\enq
This equation holds for any function  $\phi$ such that $\phi \ex{-\kappa V} \in W^{\infty}_{1}(\R) $ for some $\kappa \in \intfo{0}{NT_N}$.   
The operator $\mc{D}_N$ arising in the loop equation at level $1$ corresponds to the non-commutative derivative
\beq
\label{noncommet}\mc{D}_N[\phi](\xi,\eta) \; = \;  \Bigg\{\sul{p=1}{2} \f{ \pi }{ \om_p }\coth\big[ \frac{\pi}{ \om_p } T_N (\xi-\eta) \big] \Bigg\}\cdot\big[  \phi(\xi) -\phi(\eta) \big] \;.
\enq
Furthermore, this loop equations involves the so-called master operator, $\mc{U}_N$, which is defined as 
\beq
\mc{U}_N[\phi](\xi) \; = \; \phi(\xi)\cdot \Big\{  V^{\prime}( \xi) \, -  \, \mc{S}_N[ \rho_{\e{eq}}^{(N)} ](\xi)   \Big\}
 + \mc{S}_N[\phi \cdot \rho_{\e{eq}}^{(N)} ](\xi) \;.  
\label{definition noyau integral operateur S driven by mu eq}
\enq
The invertibility of $\mc{U}_N$ is established in Proposition 8.1 while various fine properties of its inverse, in particular $N$-dependent estimates of its $W_k^{\infty}(\R)$ norms,
 are obtained in Proposition 8.2 of [$\bs{A21}$]. I refer to these for more details.

\vspace{1mm}

 Similarly to the arguments described in Section \ref{SousSection Main Steps of Proof of Mean Field Part Fct},  one can obtain concentration of measure 
 \textit{a priori} bounds on the expectation values of smooth compactly supported functions. These, adjoined to 
the above-mentioned fine bounds on the master operator allow one to
\begin{itemize}
\item[$\bullet$] first,  carry out a bootstrap analysis of the loop equations that results in an improvement of the \textit{a priori} bounds, \textit{c}.\textit{f}. Proposition 3.17 of [$\bs{A21}$];\vspace{1mm}
\item[$\bullet$] second, build on the improved bounds so as to obtain the uniform in $V$ and $H$ asymptotic expansion of the one-point expectation value $ \moy{ H }_{L_N^{(\bs{\la})}}^{V}$, this 
up to a $\e{o}(N^{-2}T_N^{-1})$ remainder, \textit{c}.\textit{f}. Proposition 3.19 of [$\bs{A21}$]. \vspace{1mm}
\end{itemize}
All-in-all, for any $V$ satisfying to the four hypothesis and $H$ sufficiently regular, one obtains the below expansion 
\beq
 -N^{2} T_N \moy{ H }_{L_N^{(\bs{\la})}}^{V}
\; = \; -N^{2} T_N\Int{a_N}{b_N} \hspace{-1mm} H(\xi)\cdot\mc{W}_N\big[V^{\prime}\big](\xi)\,\dd \xi \;+   \; 
\f{1}{2} \mf{I}_{\e{d}}\big[ H , V  \big] \; + \; \de_N[H,V] \;. 
\label{ecriture DA de ln ZV a partir eqns de boucles}
\enq
Several explanations are in order. The first integral term corresponds to an integration of $H$ \textit{versus} the $N$-dependent equilibrium measure of the problem. 
The second term stems from the leading asymptotic behaviour of the two-point expectation value in \eqref{SDlevel1} and takes the form 
\beq
\mf{I}_{\e{d}}[H,V] \; = \; \Int{a_N}{b_N} \mc{W}_N\Big[ 
\partial_{\xi}\big\{ S\big(T_N(\xi-*)\big) \cdot \mc{G}_N\big[H,V\big](\xi,*)     \big\}   \Big](\xi) \, \dd \xi\;,
\label{definition integrale double du DA correlateur a un point0}
\enq
where $*$ stands for the running variable on which $\mc{W}_N$ acts while 
\beq
 \mc{G}_N\big[H,V\big](\xi,\eta) \; = \; \f{ \mc{W}_N[H](\xi) }{ \mc{W}_N[V^{\prime}](\xi)  }
 \; - \; \f{ \mc{W}_N[H](\eta) }{ \mc{W}_N[V^{\prime}](\eta)  } \;. 
\enq
Finally $\de_N[H,V]$ appearing in \eqref{ecriture DA de ln ZV a partir eqns de boucles} is the remainder. Without venturing into technical details 
relative to the form of the bounds, I will only stress that it is controlled in such a way that given the $t$-dependent potential $V_{t} = t V + (1-t)V_{G;N}$, 
one has
\beq
\big| \de_N\big[ \Dp{t} V_{t} , V_t \big] \big| \; = \; \e{o}(1) \quad \e{uniformly} \, \e{in} \quad t \in \intff{0}{1}
\enq
what is already enough so as to use this information for an asymptotic in $N$ integration of \eqref{equation reliant derivee log fct part et corr derivee pot} 
by means of the expansion \eqref{ecriture DA de ln ZV a partir eqns de boucles}. 

At that stage, all is almost set so as to access to the large-$N$ asymptotic expansion of $Z_N[V]$. 
It remains to obtain a uniform in $t$ large-$N$ asymptotic expansion of the one-fold integral
\beq
\Int{a_N}{b_N} \hspace{-1mm} \Dp{t} V_t(\xi)\cdot\mc{W}_N\big[V_t^{\prime}\big](\xi)\,\dd \xi 
\enq
and of the functional $ \mf{I}_{\e{d}}[\Dp{t} V_t , V_t]$. This can be achieved by means of "classical" asymptotic analysis. 
The results relative to the one-fold integral can be found in Proposition 7.6 of [$\bs{A21}$] while the ones relative to the functional $\mf{I}_{\e{d}}[H,V] $
can be found in Proposition 9.10 of [$\bs{A21}$]. Having the uniform in $t\in \intff{0}{1}$, large-$N$ expansion at one's disposal,
the $t$-integration is trivial.

\subsection{Some comments}

It seems worthy to shortly comment on the form of the loop equations that has been used in the analysis. 
Usually, as it has been described for the case of the mean-field interactions generalisation of $\be$-ensembles, 
one introduces a set of auxiliary objects, the connected correlators $W_n$ which are holomorphic on $\big(\Cx\setminus \mathsf{A}\big)^n$
if the integration runs through $\mathsf{A}^N$. These objects allow one to recast the loop equations
into a set of non-linear integral equations equation for the connected correlators. The operators involved
in these equation act on functions supported on small loops around $\mathsf{A}$. The size of the loops
is $N$-independent. Such a framework allows one to strongly simplify the analysis in that then, one 
can build on powerful tools of complex analysis. However, one meets two problems when trying to implement this approach to the case of 
$Z_N[V]$. First of all, the bi-periodic structure of the two-body interaction would demand to use a collection 
of two type of correlators so as to obtain an effectively closed and simple system of equations. 
Second, and this is the most problematic point, the scaling $T_N$ results in the squeezing of the poles of $S(T_N \xi)$, the logarithmic derivative of the two-body interaction in the rescaled model,
down to the real axis. Should one want to apply the complex variable techniques that appeared fruitful 
in the $\be$-ensemble case, then one would need to introduce some integration in the complex plane. 
The latter would then generate the contribution of all the poles squeezing down to the real axis. In my present understanding of the 
situation, such handling seem intractable to me.  
For those reasons, a real variable based analysis of the set of loop equations did seem better adapted to the study of the
large-$N$ expansion of the expectation values. 

\vspace{1mm}

It is legitimate to wonder whether it is possible to bypass or relax some of the hypothesis that have been used in the analysis. 

\vspace{1mm}

The first hypothesis guarantees that the integral \eqref{ecriture premiere intro dans texte de fct part rescale} is well-defined. 
It also ensures that the integration variables will remain, with overwhelming probability ($1 - \e{O}\big(N^{-\infty}\big)$ ), inside of a ball in $\R^N$ having an $N$-independent radius. 
In principle one could hope to relax the strong confinement of the potential to a weakly confining one, \textit{viz}.  $V(\xi) \simeq \pi \big( \om_1^{-1}+\om_{2}^{-1} \big)|\xi|$,
by setting up some generalisation of the approach of Hardy \cite{HardyBetaEnsIntWithWeakConfinment}. 
It is however unclear to me,  at this point, what would exactly be the complications arising in this case. 

\vspace{1mm}

One could, in principle, relax both requirements of the second hypothesis. Indeed, there is no problem to limit the hypothesis to 
$V$ of class $\mc{C}^{k}$ provided that $k$ is large enough. Still, in the present state of the art of the method, dealing with 
small values of $k$, for instance the minimal one that make sense out of \eqref{Ecriture DA ZN dans partie Intro resultats}, seems inaccessible. 
The convexity of the potential ensures that the support of the equilibrium measure is a single segment.
When the potential is not strictly convex, one may have to deal with a multi-cut support for the equilibrium measure.  
In principle, the asymptotic analysis could be addressed by importing the ideas of \cite{BorotGuionnetAsymptExpBetaEnsMultiCutRegime} to the present framework. 
In the multi-cut regime, one would also have to alter the scheme of the asymptotic analysis of the Riemann-Hilbert problem characterising the equilibrium measure and the inverse of the 
master operator. This should be doable, with some efforts, for the price of dealing with higher dimensional matrix Riemann--Hilbert problems.

\vspace{1mm}

The third hypothesis is not essential, but allows some simplification of the intermediate proofs concerning the equilibrium measure and the large deviation estimates, \textit{e}.\textit{g}. 
Theorem 3.7 and Corollary 3.9 of  [$\bs{A21}$]. 

\vspace{1mm}

In the fourth hypothesis, one can relax  the $(\ln N)^2$ lower bound on $T_N$ down to  a logarithmic one $c \ln N \leq  T_N$. However, doing so,  
has a price; depending on the value of $c$ it may become necessary of having to push further the large-$N$
asymptotic analysis of the Riemann--Hilbert problem for $\chi$. Indeed the $\e{O}(N^{-\infty})$ reminders which cloud always be waived-off 
due to their extremely fast vanishing now only vanish as some power of $N$: $\e{O}(N^{-c^{\prime} })$ . As such, these remainders may end-up contributing to the result.  
In the present state of the art, it isn't clear to me how to deal with sequences $T_N$ having a sub-logarithmic growth in $N$. 
The upper bound $T_N  < N^{1/6}$ present in the fourth hypothesis is purely of technical origin. 
The value $1/6$ of the exponent could be increased by entering deeper into the fine structure of the analysis of the Schwinger-Dyson equation and by finding more precise \textit{local} bounds
(instead of the global ones which were derived in [$\bs{A21}$])  for the large-$N$ behaviour of the inverse of 
the master operator ${\cal U}_{N}^{-1}$. Although it is rather clear how to achieve these goals, the associated analysis will be extremely technical and bulky. 
There might also exist an alternative route for improving the upper bound $N^{1/6}$, namely by establishing a local rigidity of the integration variables close to their
classical position in the spirit of \cite{BourgadeErdosYauUniversalityEdgeBetaEnsGnlPot,BourgadeErdosYauUniversalityBulkBetaEnsConvPot}.

\section{Conclusion}

This chapter reviewed the results I obtained relatively to the characterisation of the large number of integration  asymptotic behaviour  of two classes of multiple integrals. 
The first class is a direct generalisation of the $\be$-ensemble integrals. It could be studied within the framework of the existing method, with a few technical modifications. 
The second class of integrals is closely related to the $N$-fold
multiple integrals arising within the framework of the quantum separation of variables. It has varying weights \textit{and} varying two-body sinh-interactions. 
Setting forth methods allowing one to carry out the large-$N$ analysis of such integrals is the first step towards developing a systematic and efficient
approach to the large-$N$ analysis of the multiple integrals describing the form factors for models
solvable by the quantum separation of variables method. The asymptotic analysis of this class of integrals demanded some conceptually new ideas.

\chapter{Integrable models at finite temperature and the low-temperature large-distance asymptotic behaviour of their correlators}
\label{Chapitre QIS a tempe finie et cptmt grde distance de leurs correlateurs}

It was shown in \cite{GohmannKlumperSeelFinieTemperatureCorrelationFunctionsXXZ,GohmannKlumperSeelElementaryBlocksFiniteTXXZ}
that correlation functions of quantum integrable models at finite temperature can be described effectively within 
 the so-called quantum transfer matrix formalism introduced by Koma \cite{KomaIntroductionQTM6VertexForThermodynamicsOfXXX,KomaIntroductionQTM6VertexForThermodynamicsOfXXZ}. 
A key feature in this approach is the description of the quantum transfer matrix's spectrum through a single linear integral equation proposed by Kl\"{u}mper \cite{KlumperNLIEfromQTMDescrThermoRSOSOneUnknownFcton}
and the Slavnov \cite{SlavnovScalarProductsXXZ} determinant representation of the scalar product between Bethe vectors. 
So far, the question of the large-distance asymptotic behaviour of two and multi-point functions at finite temperature remained, however, open. 
Form factor expansions subordinate to the eigenbasis of the quantum transfer matrix appear as a natural approach to tackle with this issue. 
I will discuss the progress that I made, in collaboration with Dugave and G\"{o}hmann [$\bs{A23},\bs{A24},\bs{A25}$], 
in extracting the large-distance asymptotics at finite temperature. 
The method we have set to tackle this problem is exact but not rigorous. Although many open problems from the 
point of view of a satisfactory analysis remain, the results we obtained are very effective be it from the point of view of numerical
calculations or testing the conformal field theory predictions.

The understanding of the finite temperature behaviour of quantum integrable models subject to boundary conditions other than periodic is
much less developed. In the case of the XXZ chain subject to diagonal boundary fields, some progress has been achieved by Bortz, Frahm and G\"{o}hmann \cite{BortzFrahmGohmannSurfaceFreeEnergy} in respect to representing the 
finite Trotter number approximant of the so-called surface free energy. These authors have reduced the computation of this quantity
to the one of a specific correlation function associated with the quantum transfer matrix of
the periodic XXZ chain. In collaboration with Pozsgay, I managed to bring this correlation function to a form where the infinite trotter number limit 
can be taken, under a set of hypothesis that is standard to the quantum transfer matrix approach.

This chapter is organised as follows.  Section \ref{Section QTM et series FF} is devoted to the discussion of finite temperature form factors expansions. In 
Subsection \ref{SousSection free energy}, I will remind the general features of the quantum transfer matrix approach. Then, in Subsection \ref{SousSection Corr Fcts and FF expansion},
I will describe how form factor expansions look within the quantum transfer matrix. This will allow me to introduce all the ingredients related to 
the analysis of the finite temperature form factors. Finally, in Subsection \ref{SousSectionFFQTMATrotterFini}, I will review the results I obtained and the main steps of the 
analysis relatively to the characterisation of these form factors at finite temperature. In Section \ref{Section Low T analysis of FF expansions},
I will describe the main ideas allowing one to perform the low-temperature analysis of the finite temperature form factors. 
In Section \ref{Section Energie Libre de Surface}, I will describe the progress I made relatively to characterising the surface free energy of the XXZ chain 
subject to so-called diagonal boundary fields. I will introduce this model and set the problem in Subsection \ref{SousSection Intro model XXZ bdry plus Surf Free energy}. 
Then, in Subsection \ref{SousSection Surf free Energie Recriture vers limite}, I will present the method allowing one to 
rewrite the Bortz, Frahm, G\"{o}hmann expression in such a way that the infinite Trotter limit is easy to take. Finally, in 
Subsection \ref{SousSection Finite Temp Bdry Mag}, I will discuss some applications of this result: namely the characterisation of the 
boundary magnetisation at finite temperature.

\section{The quantum transfer matrix and representation for the form factors}
\label{Section QTM et series FF}

\subsection{The free energy}
\label{SousSection free energy}

The Hamiltonian
\beq
\op{H}_{XXZ} \; = \; J \sul{a=1}{L} \Big\{ \sg_a^x \sg_{a+1}^x \, + \, \sg_a^y \sg_{a+1}^y \, + \, \De \big( \sg_a^z\sg_{a+1}^z + \e{id}_L \big)   \Big\} \; - \;  \f{ h }{ 2 } \sul{a=1}{L} \sg_a^z  
\nonumber
\enq
of the XXZ spin-$1/2$ chain has already been described in the introduction. It acts on the Hilbert space $\mf{h}_{XXZ}\simeq \otimes_{k=1}^{L} \mf{h}^k_{XXZ}$
built out of the tensor product of the local spaces $\mf{h}^k_{XXZ}\simeq \Cx^2$ that are attached to the sites of the chain.  
The quantum integrability of this model is described by the so-called $6$-vertex type $R$-matrix:
\beq
R(\la) = \pa{ \ba{cccc}  \sinh(\la+\eta) &  0 &0 &0 \\
							0 & \sinh(\la) & \sinh(\eta) & 0 \\
								0 & \sinh(\eta)& \sinh(\la) & 0 \\
							0 & 0 & 0 & \sinh(\la+\eta) \ea  } \quad \e{with} \quad \De = \cosh(\eta) \;.
\enq
These allow one to build two types of monodromy matrices  $\op{T}_a(\la)$ and $\wt{\op{T}}_a(\la)$:
\beq
\op{T}_a\pa{\la} = R_{aL}\pa{\la+\xi}\dots R_{a1}\pa{\la+\xi} \qquad \e{and} \qquad 
\wt{\op{T}}_a\pa{\la} = R_{1a}^{ \bs{T}_a } \pa{-\la-\xi} \dots R_{La}^{ \bs{T}_a }\pa{-\la-\xi_L} \;. 
\label{definition matrices de monodromie pour XXZ periodique}
\enq
Above, the roman index $a$ refers to the 2 dimensional auxiliary space whereas the integer indices $1,\dots,L$ label the local spaces 
 $\mf{h}^k_{XXZ}$. Also $^{\bs{T}_a}$ represents the transposition on the auxiliary space while $\xi$ is an inhomogeneity
parameter. 

The $R$-matrix reduces to a permutation operator at zero spectral parameter $R_{ab}(0) \; = \; \sinh(\eta) \, \op{P}_{ab} $,  
where $\op{P}_{ab}$ is the permutation operator on $\mf{h}^a_{XXZ}\otimes \mf{h}^b_{XXZ}$ : $\op{P}_{ab} (x\otimes y) \,  = \,  y\otimes x$ \;.  
This reduction of the $R$-matrix allows one to relate the transfer matrices $\op{t}_{XXZ}(\la)$ and $\wt{\op{t}}_{XXZ}(\la)$ to the 
 XXZ Hamiltonian at zero magnetic field $h=0$ as
\beqa
\op{t}_{XXZ}(\la)_{\mid \xi=0} & = & \big[ \sinh(\eta) \big]^L \cdot \op{U}_L \cdot \Big(\e{id}_L \; - \; \la \f{ {\op{H}_{XXZ}}_{\mid h=0} }{ 2 J \sinh(\eta) } \, + \, \e{O}(\la^2)  \Big) \label{ecriture DA court Lambda t xxz} \\
\wt{\op{t}}_{XXZ}(\la)_{\mid \xi=0}& = &  \Big(\e{id}_L \; + \; \la \f{ {\op{H}_{XXZ}}_{\mid h=0} }{ 2 J  \sinh(\eta) } \, + \, \e{O}(\la^2)  \Big) \cdot \op{U}_L  \cdot \big[ \sinh(\eta) \big]^L \;. 
\label{ecriture DA court Lambda t tilde xxz}
\eeqa
There $\e{id}_L$ is the identity operator on $\mf{h}_{XXZ}$. The low-$\la$ expansions \eqref{ecriture DA court Lambda t xxz}-\eqref{ecriture DA court Lambda t tilde xxz} 
lead to the below Trotter-like representation for the statistical operator
\beq
\ex{ -\f{1}{T} \op{H}_{XXZ} } \; = \; \lim_{N\tend + \infty} \bigg\{ \f{  \wt{ \op{t} }_{ XXZ } \big( \tf{\be}{N} \big) \cdot  \op{t}_{XXZ}\big( -\tf{\be}{N} \big) }{  \big[ \sinh(\eta) \big]^{2L}  }\bigg\}^N_{\mid \xi=0}
\; \; \times  \; \; \pl{a=1}{L} \ex{ \f{h}{2T}\sg_a^z } 
\qquad \e{with} \quad \be \, = \, \f{ J \sinh (\eta) }{ T } \;. 
\enq
After some algebra, this limit can be recast in terms of the ordered product of quantum  monodromy matrices $\op{T}_{\mf{q};k}(\xi)$
\beq
\ex{ -\f{1}{T} \op{H}_{XXZ} } \; = \; \lim_{N\tend + \infty} \Big\{ \e{tr}_{a_1,\dots,a_{2N}} \bigg[ \op{T}_{\mf{q};1}(\xi) \cdots \op{T}_{\mf{q};L} (\xi)  \bigg]_{\mid \xi=0} \Big\} \;. 
\enq
The quantum monodromy matrix 
\bem
\op{T}_{\mf{q};k} (\xi) \, = \,   R^{ \bs{T}_{a_{2N}}}_{a_{2N} k }\!\pa{ - \xi -  \tf{\be}{N} } R_{k a_{2N-1}}\!\pa{\xi - \tf{\be}{N} } \cdots 
R_{a_{2} k}^{ \bs{T}_{a_2} }\!\pa{-\xi - \tf{\be}{N} } R_{k a_{1}}\!\pa{\xi - \tf{\be}{N} } \f{ \ex{\f{h}{2T}  \sg^z_k }  }{[\sinh(\eta)]^{2N} } \\
 \; = \; \pa{\ba{cc} \op{A}_{\mf{q}}(\la) & \op{B}_{\mf{q}}(\la) \\ \op{C}_{\mf{q}}(\la) & \op{D}_{\mf{q}}(\la) \ea  }_{ [k] }
\label{definition QTM}
\end{multline}
 has $\mf{h}_{XXZ}^{k}$ as its auxiliary space. When represented as a matrix operator on this space, its entries are operators on the space $\mf{h}_{\mf{q}}=\otimes_{a=1}^{2N} \mf{h}_{\mf{q}}^{a}$
 with $\mf{h}_{\mf{q}}^{a} \simeq \Cx^2 $. 

It follows from general considerations \cite{RuelleRigorousResultsForStatisticalMechanics} that the the thermodynamic limit of the \textit{per} site free energy of the XXZ chain
\beq
f_{XXZ}  \; = \; - T \lim_{L\tend +\infty} \bigg\{ \f{ 1 }{L}  \ln \tr_{ \mf{h}_{XXZ} }\Big[ \ex{ -\f{1}{T} \op{H}_{XXZ} } \Big] \bigg\} \; 
\enq
is well defined. Assuming the exchangeability of 
limits\symbolfootnote[3]{Some arguments in favour thereof have been given in  \cite{KomaIntroductionQTM6VertexForThermodynamicsOfXXZ,SuzukiArgumentsForInterchangeabilityTrotterAndVolumeLimitInPartFcton}}, one then gets 
\beq
- \f{ f_{XXZ} }{ T}  \; = \;  \lim_{L\tend +\infty} \bigg\{ \f{ 1 }{L} \ln \tr_{ \mf{h}_{XXZ} }\Big[ \ex{ -\f{1}{T} \op{H}_{XXZ} } \Big] \bigg\} \; = \; 
 \lim_{N\tend +\infty} \lim_{L\tend +\infty} \f{ 1 }{L}  \ln \bigg\{  \e{tr}_{a_1,\dots,a_{2N}} \Big[ \Big( \op{t}_{\mf{q}}(\xi) \Big)^L \Big]_{\mid \xi=0}   \bigg\}
\enq
where $\op{t}_{\mf{q}}(\xi) \, = \,  \e{tr}\big[ \op{T}_{\mf{q};k}(\xi)\big]$ is the the quantum transfer matrix. Hence, assuming\symbolfootnote[2]{This has been checked at least numerically and it 
appears that, in fact, the eigenspace attached with the largest eigenvalue is one dimensional} that the largest eigenvalue 
$ t_{\mf{q}}^{ (0) }(0) $ of the quantum transfer matrix  $\op{t}_{\mf{q}}(\xi)$ at zero spectral parameter is degenerated  at 
most sub-exponentially in $L$, one gets the very simple result
\beq
- \f{ f_{XXZ} }{ T}  \; = \;   \lim_{N\tend +\infty}  \ln  \big[  t_{\mf{q}}^{ (0) }(0)  \big] \;. 
\label{ecriture limite grand N de vp max de QTM}
\enq
In this way, the problem boils down to the calculation of the largest eigenvalue of the quantum transfer matrix and then taking the infinite Trotter number limit of its
logarithm. The matter is that one can construct eigenvectors of the quantum transfer matrix by means of an algebraic Bethe Ansatz \cite{KomaIntroductionQTM6VertexForThermodynamicsOfXXX,KomaIntroductionQTM6VertexForThermodynamicsOfXXZ}
and hence carry out easily this computation.

\subsection{The correlation functions and form factor expansion}
\label{SousSection Corr Fcts and FF expansion}
\subsubsection{General considerations}

Starting from the definition of the thermal correlation functions and assuming the validity of the exchange of limits, one gets \cite{GohmannKlumperSeelFinieTemperatureCorrelationFunctionsXXZ} that, 
given a product of local operators $\op{O}_1^{(k_1)}\cdots\op{O}_m^{(k_m)}$ acting on the first $m$-sites of the chain: 
\beq
\Big< \op{O}_1^{(k_1)} \cdots \op{O}_m^{(k_m)}  \Big>_T \; = \; \lim_{N \tend + \infty} \Big< \op{O}_1^{(k_1)} \cdots \op{O}_m^{(k_m)}  \Big>_{T;N} \;. 
\label{ecriture Fct tempe finie comme limit Trotter}
\enq
The finite Trotter number $N$ approximant of the thermal correlator appearing above is defined as
\beq
\Big< \op{O}_1^{(k_1)} \cdots \op{O}_m^{(k_m)}  \Big>_{T;N} \; = \; 
\f{ \bra{\Psi_0} \e{tr}_1\big[\op{O}_1^{(k_1)} \op{T}_{\mf{q};1} (0) \big] \cdots \e{tr}_m\big[\op{O}_m^{(k_m)} \op{T}_{\mf{q};m} (0) \big]  \ket{\Psi_0} }
{ \big[ t_{\mf{q}}^{ (0) }(0)  \big]^m \cdot \braket{\Psi_0 }{\Psi_0 }  } \;. 
\enq
Note that, in order to get \eqref{ecriture Fct tempe finie comme limit Trotter}, I have made the additional hypothesis that the eigenspace attached with the largest eigenvalue $t_{\mf{q}}^{ (0) }(0) $ is one-dimensional
and spanned by $\ket{ \Psi_0 }$. 
Formula \eqref{ecriture Fct tempe finie comme limit Trotter} is already enough so as to write down the form factor expansions for two-point functions. I will consider the example of the 
spin-spin correlation function. Let $\ket{\Psi_n}$ represent the complete
set of eigenvectors of the quantum transfer matrix $\op{t}_{\mf{q}}(\xi)$ and let $t_{\mf{q}}^{(n)}(\xi)$ be the associated eigenvalues ordered decreasingly in respect to their module. Then, 
one has
\beq
\Big< \sg_1^{z} \cdot \sg_{m+1}^{z}   \Big>_{T;N} \; = \;  \sul{n=0}{ 2^{2N}-1 } \Big( \rho_{N}^{(n)} \Big)^{m}\cdot \msc{A}^{(n)}_{N} \qquad \quad  \e{where} \quad  
\rho_{N}^{(n)} \; = \; \f{ t_{\mf{q}}^{(n)}(0) }{ t_{\mf{q}}^{(0)}(0)  }
\label{ecriture dvpmt series FF thermaux sgZsgZ}
\enq
is the ratio of eigenvalues of the quantum transfer matrix while the associated amplitudes takes the form
\beq
 \msc{A}^{(n)}_{N} \; = \; \f{ \bra{\Psi_0} \big( \op{A}_{\mf{q}}(0) -  \op{D}_{\mf{q}}(0) \big) \ket{\Psi_n} \bra{\Psi_n} \big( \op{A}_{\mf{q}}(0) -  \op{D}_{\mf{q}}(0) \big) \ket{\Psi_0}  } 
 { t_{\mf{q}}^{(n)}(0)  \cdot  t_{\mf{q}}^{(0)}(0)  \cdot \braket{\Psi_0}{\Psi_0} \cdot  \braket{\Psi_n}{\Psi_n} } \;. 
\label{definition amplitude  pour sgZsgZ}
\enq
Note that due to the hypothesis of non-degeneracy of the largest eigenvalue, $|\rho_{N}^{(n)} | < 1$ and by construction 
$|\rho_{ N }^{ (n) } | \,  \geq   \, | \rho_{ N }^{ (n+1) } | $. 
Therefore, at least at finite $N$, the form factor expansion is well-suited for describing the large-distance asymptotic expansion of the two-point function. 
However, so as to be able to apply these results to the case of the XXZ chain one should be able to take the large-$N$ limit
of the form factor expansion \eqref{ecriture dvpmt series FF thermaux sgZsgZ}. Doing so can be decomposed in two steps. One first 
takes the limit of the individual terms of the series, namely of the  ratios of eigenvalues $\rho_{N}^{(n)}$ and of the amplitudes $\msc{A}^{(n)}_{N}$. 
Second, one should prove that the analogous series to \eqref{ecriture dvpmt series FF thermaux sgZsgZ} built up from these limits 
converges and that, in fact, the finite but growing with $N$ sum \eqref{ecriture dvpmt series FF thermaux sgZsgZ} does converges to this series. It is still not clear to me, for the moment, 
how to achieve rigorously the second point. However, at least provided one makes a few technical hypotheses, one can basically deal with the first step. I will be more explicit about this further on. 
The Trotter limit of the ratio of eigenvalues was studied on many occasions either through numerics or by some handlings of non-linear integral equations 
\cite{KlumperNLIEfromQTMDescrThermoXYZOneUnknownFcton,KlumperMartinezScheerenShiroishiXXZforPositiveDeltaCrossoverAtTNicePictureCondRootsAtFermiPts,
SakaiShiroishiSuzukiUmenoCorrelationLengthsXXZThroughsomeotherModel,TakahashiThermoXYZInfiniteNbrRootsFromQTM}. However, the investigation of the amplitudes
was much sparser. A numerical investigation of the amplitudes $ \msc{A}^{(n)}_{N}$ has been carried out by Fabricius, Kl\"{u}mper and McCoy \cite{FabriciusKlumperMcCoyTemperatureDrivenSpaceOscillationsNumerics} 
  in 1999 for Trotter numbers $N$ up to 16. 

\subsubsection{The algebraic Bethe Ansatz approach}

Let $ \big\{ \ket{+}, \ket{-} \big\}$ be the canonical basis of $\Cx^2$. Just as for the XXZ spin-$1/2$ chain, one 
can build on the algebraic Bethe Ansatz so as to construct eigenstates of the quantum transfer matrix. In that case, the pseudo-vacuum is given by the vector 
\beq
\ket{ 0 }_{\mf{q}} \, = \, \ket{+}_{a_1}\otimes\ket{-}_{a_2}\otimes\,  \cdots \,  \otimes \ket{+}_{a_{2N-1}} \otimes \ket{-}_{a_{2N}}  \in \mf{h}_{\mf{q}} \;. 
\enq
If the pairwise distinct parameters $\{ \mu_k \}_1^M$ satisfy the Bethe Ansatz equations
\beq
-1 =  \ex{-\f{h}{T} } \pl{k=1}{M} \Bigg\{ \f{ \sinh\big(\mu_p-\mu_k + \eta \big)  }{  \sinh\big(\mu_p-\mu_k - \eta \big) }  \Bigg\} \cdot 
 \Bigg[  \f{  \sinh\big(\mu_p+\tf{\be}{N}-\eta \big)\sinh\big(\mu_p- \tf{\be}{N} \big)  }{  \sinh\big(\mu_p-\tf{\be}{N}+\eta \big)\sinh\big(\mu_p+ \tf{\be}{N} \big)  } \Bigg]^N \; \; , \quad  p=1,\dots,M
\label{ecriture Bethe Equations QTM}
\enq
then one has the relation 
\beq
\op{t}_{ \mf{q} }(\xi) \cdot   \op{B}_{\mf{q}}\pa{\mu_M} \cdots \op{B}_{\mf{q}}\pa{\mu_1} \ket{ 0 }_{\mf{q}}  \; = \;  
 t_{\mf{q};h}\big( \xi \mid \{\mu_k\}_1^M \big) \cdot \op{B}_{\mf{q}} \pa{\mu_M}\cdots \op{B}_{\mf{q}} \pa{\mu_1} \ket{0 }_{\mf{q}}   \;, 
\enq
where 
\bem
 t_{\mf{q};h}\big( \xi \mid \{\mu_k\}_1^M \big) \, = \,  \pa{-1}^N \ex{\f{h}{2T} }\pl{k=1}{M} \Bigg\{ \f{ \s{\xi-\mu_k - \eta} }{ \s{\xi-\mu_k } }  \Bigg\} \cdot 
\paf{  \s{\xi + \tf{\be}{N} } \s{\xi - \tf{\be}{N} + \eta} }{ \sinh^2(\eta) }^N   \\ 
+ \pa{-1}^N  \ex{-\f{h}{2T} } \pl{k=1}{M}  \Bigg\{ \f{ \s{\xi-\mu_k + \eta} }{ \s{\xi-\mu_k } }  \Bigg\} \cdot 
\paf{  \s{\xi + \tf{\be}{N} -\eta} \s{\xi - \tf{\be}{N}} }{ \sinh^2(\eta) }^N \;. 
\end{multline}
The index $h$ in $ t_{\mf{q};h}$ is there so as to insist on the dependence of the eigenvalues on the magnetic field.   

\vspace{3mm}

Numerical investigations, analysis at the free fermion point $\eta=-\i\tf{\pi}{2}$ and for $\be$ small all indicate  \cite{GohmannKlumperSeelFinieTemperatureCorrelationFunctionsXXZ} that the 
dominant eigenvalue is constructed in terms of a solution to the Bethe equations with $M=N$ roots, this independently of the value of the magnetic field $h$ 
and the anisotropy $\eta \in -\i\intff{0}{\pi}\cup \R^+$. 
 I will denote this solution as $\{ \la_k \}_1^N$, so that the "dominant" eigenvector of the quantum transfer matrix takes the form 
\beq
\ket{\Psi_0}= \op{B}_{\mf{q}}(\la_1) \cdots \op{B}_{\mf{q}}(\la_N) \ket{ 0 }_{\mf{q}} \; .
\label{definition Vect Propre Dominant QTM}
\enq
The \textit{per} \textit{se} distribution of roots does, however, depend on the magnetic field and anisotropy.  

As proposed in \cite{KlumperNLIEfromQTMDescrThermoXYZOneUnknownFcton}, it is convenient to introduce a function closely related to the exponent of the counting function 
\beq
\wh{\mf{a}}_{\la}\pa{\om} \, = \,  \ex{-\f{h}{T} } \pl{k=1}{N} \Bigg\{  \f{ \sinh\big(\om-\la_k + \eta \big) }{  \sinh\big(\om-\la_k - \eta \big)  }  \Bigg\}
\cdot  \Bigg[  \f{ \sinh\big(\om + \tf{\be}{N}-\eta \big) \sinh\big(\om - \tf{\be}{N} \big)    }{  \sinh\big(\om - \tf{\be}{N}+\eta \big) \sinh\big(\om + \tf{\be}{N} \big)  } \Bigg]^N \; .   
\label{definition counting function GS}
\enq
As will be shown below, the knowledge of the $\wh{\mf{a}}_{\la}$ function allows one to recast symmetric products in the 
Bethe roots as exponents of contour integrals. It can be easily checked that the $\wh{\mf{a}}_{\la}$ function is $\i \pi$-periodic, bounded when $\Re(\om) \tend \pm \infty$ and such that 
it has, in the strip of width $\pi$, \vspace{1mm}
\begin{itemize}
\item[$\bullet$] an $N^{\e{th}}$-order pole at $\om = - \tf{\be}{N}$ \, ,  \vspace{1mm}
\item[$\bullet$] an $N^{\e{th}}$-order pole at $\om = \tf{\be}{N} - \eta$ \, ,\vspace{1mm}
\item[$\bullet$] N simple poles at $\om = \la_k+\eta$, \quad $k=1,\dots,N$.  \vspace{1mm}
\end{itemize}
Note that if some $\la$'s coincide, than the number of poles is reduced but their order grows. These properties ensure that $1+\wh{\mf{a}}\pa{\om}$ has $3N$ zeroes
counted with their multiplicity. $N$ of these are, by construction, the Bethe roots, \textit{i}.\textit{e}. 
\beq
 1+\wh{\mf{a}}_{\la} (\la_p) \, = \, 0\;\; , \qquad  p=1,\dots,N \; . 
\enq
Numerical analysis and calculations at the free fermion point 
indicate that the roots for the ground  state of the quantum transfer matrix can all be encircled by a unique loop $\msc{C}$ that is  $N$ independent 
and such that any additional root to $ 1+\wh{\mf{a}}_{\la} \pa{ \om } =0$ is located outside of this loop 
\cite{GohmannKlumperSeelFinieTemperatureCorrelationFunctionsXXZ}. 
Numerical investigations also indicate that all the other roots accumulate around the points $\pm \eta$. 
The results gathered in the following will all be based on these assumptions. It would be a very interesting result to prove these
from first principles.

 The above information are enough so as to establish that the function $\wh{\mf{a}}_{\la}$ solves the non-linear integral equation \cite{KlumperNLIEfromQTMDescrThermoXYZOneUnknownFcton}
\beq
\ln \wh{\mf{a}}_{\la} \pa{\om} \,  = \,  -\f{h}{T} \; + \;  N \ln  \pac{  \f{ \s{\om + \tf{\be}{N}+\eta}  \s{\om - \tf{\be}{N}} }{  \s{ \om - \tf{\be}{N} + \eta}  \s{ \om + \tf{\be}{N}} } }
\; - \; \Oint{ \msc{C} }{}   \theta^{\prime}_{XXZ}(\om - \nu) \ln \pac{1 +  \wh{\mf{a}}_{\la} (\nu) } \cdot \f{ \dd \nu }{2\i \pi} 
\label{ecriture NLIE fonction a hat regime general}
\enq
where  $\theta_{XXZ}(\la) \,  =  \,  \ln \pac{ \tf{ \s{\eta-\la}}{ \s{\eta+\la}}  }$. 
The contour $\msc{C}$ encircles the Bethe roots $\la_1,\dots,\la_N$, the pole at $-\tf{\be}{N}$ but not any other 
singularity of the integrand. Manifestly there exist solutions to \eqref{ecriture NLIE fonction a hat regime general}. However, it is not clear whether these are unique. 
I will take it as a working hypothesis that it is so. If uniqueness indeed holds, then one can take the non-linear integral equation as a means for characterising $\wh{\mf{a}}_{\la}$
independently of its relation to some set of Bethe roots. 

The usefulness of the function $\wh{\mf{a}}_{\la}$ manifests itself already on the level of the eigenvalues of the quantum transfer matrix in that one can show that these
admit the representation:
\bem
 \ln \big[ t_{\mf{q};h}^{(0)}\big( \xi \mid \{\la_k\}_1^M \big) \big]  \, = \, \f{h}{2T} 
 \; + \; N \ln\bigg[  \f{ \sinh\big( \eta + \xi-\tf{\be}{N}  \big)  \sinh\big( \eta - \xi - \tf{\be}{N} \big)   }{ \sinh^2(\eta) } \bigg] \\
\; + \; \Oint{ \msc{C} }{}  \f{ \sinh(\eta) }{\sinh(\xi-\nu-\eta) \sinh(\xi-\nu) } \ln \pac{1 +  \wh{\mf{a}}_{\la} \pa{\nu} } \cdot \f{ \dd \nu }{2\i \pi}  \;. 
\end{multline}

The principal advantage of the non-linear integral equation based description of $\wh{\mf{a}}_{\la}$ is that the $N$-dependence only arises parametrically on the level of the driving term. 
Furthermore, when $N\tend + \infty$, the driving term converges, uniformly on $\msc{C}$, towards a well-defined function. It is thus reasonable to expect that 
$\wh{\mf{a}}_{\la}$ converges uniformly on $\msc{C}$ towards the solution $\mf{a}_{\la}$ of a non-linear integral equation similar to \eqref{ecriture NLIE fonction a hat regime general} 
but where the driving term has been replaced by its limit:
\beq
\ln \mf{a}_{\la} \pa{\om} \,  = \,  -\f{h}{T} \; - \;  \f{ 2J \sinh^2(\eta) }{ T \sinh(\om) \sinh(\om+\eta)  }  
\; - \; \Oint{ \msc{C} }{}   \theta^{\prime}_{XXZ}(\om - \nu)  \ln \pac{1 +  \mf{a}_{\la} (\nu) } \cdot \f{ \dd \nu }{2\i \pi} \;. 
\label{ecriture eqn NLIE pour fct a lambda infinite Tortter nbr}
\enq
Again, although this is an open issue, I will build on the hypothesis that \vspace{1mm}
\begin{itemize}
\item[$\bullet$] the non-linear integral equation \eqref{ecriture eqn NLIE pour fct a lambda infinite Tortter nbr} admits a unique solution; \vspace{1mm}
\item[$\bullet$] $\wh{\mf{a}}_{\la} \tend \mf{a}_{\la}$ uniformly on $\msc{C}$. \vspace{1mm}
\end{itemize}
If these hypothesis do hold, then one can easily take the large-$N$ limit in \eqref{ecriture limite grand N de vp max de QTM}, hence yielding the below representation for the \textit{per} site free energy of the XXZ chain 
\beq
f_{XXZ} \; = \; - \f{h}{2} \, + \, 2 J \cosh(\eta) \; - \; \Oint{ \msc{C} }{}  \f{  T \sinh(\eta) }{ \sinh(\nu+\eta) \sinh(\nu) } \ln \big[ 1 + \mf{a}_{\la} (\nu) \big]  \cdot \f{ \dd \nu }{2\i \pi}  \;. 
\enq

One can describe the other solutions to the Bethe Ansatz equations in a similar way.  More precisely, given a solution $\{ \mu_a( h^{\prime} ) \}_1^N$ to the Bethe equation at some 
magnetic field $h^{\prime}$, possibly different from $h$, upon similar hypothesis, one argues that the function 
\beq
\wh{\mf{a}}_{\mu}\pa{\om} \, = \,  \ex{-\f{h^{\prime}}{T} } \pl{k=1}{N} \Bigg\{  \f{ \sinh\big(\om-\mu_k(h^{\prime}) + \eta \big) }{  \sinh\big(\om-\mu_k(h^{\prime}) - \eta \big)  }  \Bigg\}
\cdot  \Bigg[  \f{ \sinh\big(\om + \tf{\be}{N}-\eta \big) \sinh\big(\om - \tf{\be}{N} \big)    }{  \sinh\big(\om - \tf{\be}{N}+\eta \big) \sinh\big(\om + \tf{\be}{N} \big)  } \Bigg]^N  
\enq
satisfies the non-linear integral equation
\bem
\ln \wh{\mf{a}}_{\mu} \pa{\om} \,  = \,  -\f{h^{\prime}}{T} \; + \;  N \ln  \pac{  \f{ \s{\om + \tf{\be}{N}+\eta}  \s{\om - \tf{\be}{N}} }{  \s{ \om - \tf{\be}{N} + \eta}  \s{ \om + \tf{\be}{N}} } } \\ 
\, + \, \sul{a=1}{n_h} \th_{XXZ}(\om - \wh{x}_{a}) \;  - \;  \sul{a=1}{n_p} \th_{XXZ}(\om - \wh{y}_{a}) 
\; - \; \Oint{ \msc{C} }{}    \theta^{\prime}_{XXZ}(\om - \nu)  \ln \big[ 1 +  \wh{\mf{a}}_{\mu} (\nu) \big] \cdot \f{ \dd \nu }{2\i \pi} \;. 
\label{ecriture NLIE fonction a hat etat excite}
\end{multline}
This equation involves an additional in respect to \eqref{ecriture NLIE fonction a hat etat excite} set of parameters $\{\wh{x}_a\}_1^{n_h}$ and $\{\wh{y}_a\}_1^{n_p}$: \vspace{1mm}
\begin{itemize}
 \item[$\bullet$] $\wh{x}_a$ are the so-called holes. They correspond to those solutions of $1+ \wh{\mf{a}}_{\mu} (\nu)=0$ which are encircled by the contour  $\msc{C}$
but which are \textit{not} belonging to $\{ \mu_a(h^{\prime})\}$.  \vspace{1mm}
\item[$\bullet$]  $\wh{y}_a$ are the so-called particles. They correspond to those roots $\mu_a(h^{\prime})$ which are \textit{not} encircled by the contour   $\msc{C}$.  \vspace{1mm}
\end{itemize}
Similarly to the case of the function $\wh{a}_{\la}$ which is associated with the largest eigenvalue, I will assume that  \vspace{1mm}
\begin{itemize}
\item[$\bullet$] the non-linear integral equation \eqref{ecriture NLIE fonction a hat etat excite} admits a unique solution; \vspace{1mm}
\item[$\bullet$] $\wh{x}_a\tend x_a$ and $\wh{y}_a\tend y_a$ in the $N\tend + \infty$ limit ; \vspace{1mm}
\item[$\bullet$] $\wh{\mf{a}}_{\mu} \tend \mf{a}_{\mu}$ uniformly on $\msc{C}$ where $\mf{a}_{\mu}$ in the unique, by hypothesis, solution to 
\bem
\ln \mf{a}_{\mu} \pa{\om} \,  = \,  -\f{h^{\prime}}{T} \; - \;  \f{ 2J \sinh^2(\eta) }{ T \sinh(\om) \sinh(\om+\eta)  }  \\
\, + \, \sul{a=1}{n_h} \th_{XXZ}(\om - x_{a}) \;  - \;  \sul{a=1}{n_p} \th_{XXZ}(\om - y_{a}) 
\; - \; \Oint{ \msc{C} }{}   \theta^{\prime}_{XXZ}(\om - \nu)\ln \big[ 1 +  \mf{a}_{\mu} (\nu) \big] \cdot \f{ \dd \nu }{2\i \pi} \;. 
\label{ecriture eqn int limite pour a etat excite mu}
\end{multline}
\end{itemize}
Given an eigenvector $\ket{\Psi_n}=\op{B}_{\mf{q}}(\mu_1) \cdots \op{B}_{\mf{q}}(\mu_N) \ket{ 0 }_{\mf{q}} $  of the quantum transfer matrix parametrised by the roots $\{\mu_a\}_1^N$,
one can build on these expressions so as to provide a one-fold integral representation for the associated ratio of the eigenvalues of the quantum transfer matrix:
\beq
\rho_N^{(n)} \; = \; \exp\Bigg\{  \sul{a=1}{n_p} \ln \Big( \f{ \sinh( \wh{y}_a+\eta ) }{ \sinh( \wh{y}_a )  } \Big) \; - \; \sul{a=1}{n_h} \ln \Big( \f{ \sinh( \wh{x}_a+\eta ) }{ \sinh( \wh{x}_a )  } \Big)
\; + \; 
\Oint{ \msc{C} }{}  \f{  \sinh(\eta) }{ \sinh(\nu+\eta) \sinh(\nu) } \wh{\mf{z}}(\nu)_{\mid h^{\prime}=h} \cdot  \dd \nu  \Bigg\}
\label{excriture rep int pour ratio VP QTM Trotter fini}
\enq
where
\beq
\wh{\mf{z}}(\nu) \; = \; \f{1 }{2\i \pi} \ln \bigg( \f{ 1 +  \wh{\mf{a}}_{\la} (\nu) }{  1 +  \wh{\mf{a}}_{\mu} (\nu) } \bigg)  \;. 
\label{definition fct widehat z}
\enq
Note that the $h^{\prime}=h$ subscript is there to indicate that one should set $h^{\prime}=h$ in the solution $\wh{a}_{\mu}$. 
Under the previous hypothesis it is straightforward to take the infinite Trotter number limit of the ratio of eigenvalues: 
it is simply enough to carry out the substitution 
\beq
\wh{\mf{z}}(\nu) \; \hookrightarrow \; \mf{z}(\nu) \; = \; \f{1 }{2\i \pi} \ln \bigg( \f{ 1 +  \mf{a}_{\la} (\nu) }{  1 +  \mf{a}_{\mu} (\nu) } \bigg) 
\label{ecriture remplacement z hat par z}
\enq
in \eqref{excriture rep int pour ratio VP QTM Trotter fini}. In \eqref{ecriture remplacement z hat par z}, 
the functions $\mf{a}_{\la}$, resp. $\mf{a}_{\mu}$, are defined as the solutions to the non-linear integral equations \eqref{ecriture eqn NLIE pour fct a lambda infinite Tortter nbr}, resp. \eqref{ecriture eqn int limite pour a etat excite mu}. 
As a consequence, one gets that $\rho_N^{(n)} \tend \rho^{(n)}$ with
\beq
\rho^{(n)} \; = \; \exp\Bigg\{   \sul{a=1}{n_p} \ln \Big( \f{ \sinh( y_a+\eta ) }{ \sinh( y_a )  } \Big) \; - \; \sul{a=1}{n_h} \ln \Big( \f{ \sinh( x_a+\eta ) }{ \sinh( x_a )  } \Big)
\; + \; 
\Oint{ \msc{C} }{}  \f{  \sinh(\eta) }{ \sinh(\nu+\eta) \sinh(\nu) } \mf{z}(\nu)_{\mid h^{\prime}=h} \cdot  \dd \nu \Bigg\} \;. 
\label{ecriture forme ration VP Trotter infini}
\enq

\subsection{The form factors at finite Trotter number}
\label{SousSectionFFQTMATrotterFini}

Let $\ket{\Psi_n}=\op{B}_{\mf{q}}(\mu_1) \cdots \op{B}_{\mf{q}}(\mu_N) \ket{ 0 }_{\mf{q}} $ be an eigenvector of the quantum transfer matrix 
parametrised by the roots $\{\mu_a\}_1^N$ solving the Bethe Ansatz equations \eqref{ecriture Bethe Equations QTM}. Then the evaluation of the amplitude
$\msc{A}^{(n)}_{N}$ defined in \eqref{definition amplitude  pour sgZsgZ} boils down to some handlings of the exchange relations provided by the Yang-Baxter algebra
along with the use of Slavnov's determinant representation for the scalar products, I refer to Section 5 \& 6 and Appendices A\& B of [$\bs{A25}$]
for more details. All-in-all, this allows one to recast the amplitudes in terms of ratios of determinants of size $N$ matrices 
adjoined to some prefactors corresponding to simple and double products. All these quantities involve both sets of roots $\{\mu_a\}_1^N$ and $\{ \la_a \}_1^N$ and the 
overall structure is similar to the one obtained for the form factors in Section \ref{Section FF resultat principal}
of Chapter \ref{Chapitre AB grand volume FF dans modeles integrables}. The main issue, then,
is to recast the answer in such a way that one can take the Trotter limit thereof. In this case, due to a completely different pattern of the Bethe roots, 
the approach which allowed one to carry out the large-L analysis of the form factors at zero temperature is not applicable. 
The main idea, then, is to rewrite all the products and finite-size determinants in terms of integrals, resp. Fredholm determinants of integral operators, 
whose integrands, resp. integral kernels, depend on the functions $\wh{\mf{a}}_{\la}$ and $\wh{\mf{a}}_{\mu}$.

In order to state the result of all these manipulation, I first need to introduce a few notations. 
 Given a sufficiently regular compact curve $\ga \subset \Cx $, $\mc{L}_{\ga}$ refers to the  $\i \pi$-periodic Cauchy transform subordinate to the contour $\msc{C}$ which, 
for $z \in \Cx \setminus \ga $ and $f\in L^1(\ga)$ takes the form 
\beq
\mc{L}_{\ga}[f](z) \; = \;  \Oint{\ga}{} f(\om)  \coth(\om - z )   \cdot \dd \om   \;. 
\enq
This Cauchy transform allows one to define two integral kernels
\beqa
 U_{\th;\ga}^{(\la)} [ \, \wh{\mf{z}} \,] ( \om , \om^{\prime}) & = &  - 
\f{ \ex{ -\mc{L}_{\ga}[ \, \wh{\mf{z}} \, ](\om) }  
				\cdot \big[  K_{h-h^{\prime}}(\om - \om^{\prime}) - K_{h-h^{\prime}}(\th - \om^{\prime})   \big] }
		{   \ex{ -\mc{L}_{\ga}[\, \wh{\mf{z}} \, ](\om-\eta) } \; - \; \ex{ \f{ h-h^{\prime}}{T} }  \ex{ -\mc{L}_{\ga}[\, \wh{\mf{z}} \,](\om+\eta) }   }
   \label{ecriture rep noyau la limite Trotter} \\
U_{\th;\ga}^{(\mu)} [ \, \wh{\mf{z}} \,](\om,\om^{\prime}) & = & 
\f{ \ex{ \mc{L}_{\ga}[ \, \wh{ \mf{z} } \, ](\om^{\prime} ) }  
				\cdot \big[  K_{h-h^{\prime}}(\om - \om^{\prime}) - K_{h-h^{\prime}}(\om- \th )   \big] }
		{   \ex{ \mc{L}_{\ga}[\, \wh{ \mf{z} } \, ](\om^{\prime} + \eta) } \; - \;
		 \ex{ \f{ h-h^{\prime}}{T} }  \ex{ \mc{L}_{\ga}[\, \wh{ \mf{z} } \,](\om^{\prime} - \eta) }   }
\label{ecriture rep noyau mu limite Trotter}
\eeqa
whose definition also involves the function 
\beq
K_{h-h^{\prime}}(\la)  \; = \; \f{  1 }{2 \i \pi} \Big\{ \coth(\la-\eta) \; - \; \ex{ \f{ h-h^{\prime}}{T} }   \coth(\la+\eta)  \Big\} \;. 
\enq
Assume that there exists some compact $C_{\ga}$ containing $\ga$ such that the only singularities of the kernels inside of this compact are the cuts on $\ga$ of the Cauchy transforms
and let $\Ga(\ga)=\Dp{}C_{\ga}$ be its boundary. Then, due to the criterion developed in \cite{DudleyGonzalesBarriosMetricConditionForOpToBeTraceClass}, the integral 
kernels \eqref{ecriture rep noyau la limite Trotter}-\eqref{ecriture rep noyau mu limite Trotter} give rise to trace class integral operators 
$ \op{U}_{\th;\ga}^{(\la/\mu)} [ \, \wh{\mf{z}} \,] $ on $L^2\big( \Ga(\ga) \big)$. As a consequence, the Fredholm determinants of $\e{id} \, + \, \op{U}_{\th;\ga}^{(\la/\mu)} [ \, \wh{\mf{z}} \,] $ 
are well-defined. In any case where such a compact does not exists, the symbols $\det_{}\big[ \e{id} \, + \, \op{U}_{\th;\ga}^{(\la/\mu)} [ \, \wh{\mf{z}} \,] \big]$
should be understood as the analytic continuations in the spirit of Proposition \ref{Proposition Holomorphie en beta et rapidite part-trou det fred}
of the Fredholm determinants $\det_{}\big[ \e{id} \, + \, \op{U}_{\th;\ga}^{(\la/\mu)} [ \, \wh{\mf{z}} \,] \big]$ arising in a situation where the contour can be constructed.

Finally, for $\nu=\la$  or $\nu=\mu$,  
\beq
\wh{K}_{\ga;\nu}(\om,\om^{\prime}) \; = \;  \f{ K_0(\om-\om^{\prime}) }{ 1 + \big[\, \wh{\mf{a}}_{\nu}(\om) \big]^{-1} }  \;. 
\label{ecriture operateir hat K gamma et racines nu}
\enq
is the integral kernel of the operator $\wh{\op{K}}_{\ga;\nu}$ on $L^2(\ga)$.

\begin{prop}
\label{Proposition Rep Int Trotter Fini pour amplitude sgZ a tempe finie}
Let $\ket{\Psi_n}=\op{B}_{\mf{q}}(\mu_1) \cdots \op{B}_{\mf{q}}(\mu_N) \ket{ 0 }_{\mf{q}} $ be an eigenvector of the quantum transfer matrix 
parametrised by the roots $\{\mu_a\}_1^N$ solving the Bethe Ansatz equations \eqref{ecriture Bethe Equations QTM} at $h^{\prime}=h$ and such that $\braket{\Psi_n}{\Psi_n} \not=0$. 
Assume furthermore that $\{\mu_a\}_1^N \not= \{\la_a\}_1^N $. 

Then, the amplitude $\msc{A}^{(n)}_{N}$ defined in \eqref{definition amplitude  pour sgZsgZ} admits the representation
\beq
\msc{A}^{(n)}_{N} \; = \; - \f{ 2 T^2 }{ \rho_N^{(n)} } \cdot \Big( \rho_N^{(n)} - 1 \Big)^2  \cdot \f{ \Dp{}^2 }{ \Dp{}{ h^{\prime}}^2} \Big\{ \wh{\mc{S}}^{(h,h^{\prime})}  \Big\}_{ \mid h^{\prime}=h}
\label{ecriture amplitude An pour finite Trotter number}
\enq
The amplitude $\wh{\mc{S}}^{(h,h^{\prime})} $ is expressed by means of the function $\wh{\mf{z}}$  \eqref{definition fct widehat z} defined 
in terms of the solution $\wh{\mf{a}}_{\la}$ to \eqref{ecriture NLIE fonction a hat regime general} at magnetic field $h$
and $\wh{\mf{a}}_{\mu}$ to \eqref{ecriture NLIE fonction a hat etat excite} at magnetic field $h^{\prime}$:
\beq
\wh{\mc{S}}^{(h,h^{\prime})}   \; = \; 
\exp\Bigg\{  \Oint{ \msc{C}_{n} }{} \dd \om \Oint{ \msc{C}^{\prime}_{n} \subset \msc{C}_{n} }{} \!\! \dd \nu
\big[ \coth^{\prime}(\om - \nu +\eta) - \coth^{\prime}(\om - \nu ) \big]
\, \wh{\mf{z}}(\om) \,  \wh{\mf{z}}(\nu ) \Bigg\} \cdot \wh{\mc{A}}^{(h,h^{\prime})} \;. 
\enq
The contour $\msc{C}^{\prime}_{n}$ is a slight deformation of $\msc{C}_{n}$ such that $\msc{C}^{\prime}_{n}$ is entirely contained in 
$\msc{C}_{n}$ but still enjoys the same properties. Finally, one has
\bem
\wh{\mc{A}}^{(h,h^{\prime})}  \; = \;   \f{  \big( 1-\ex{ \f{ h-h^{\prime}}{T} } \big)^2   }
{  \Big\{ \ex{-\mc{L}_{ \msc{C}_{n} }[\wh{\mf{z}}](\th_1-\eta)} - \ex{ \f{ h-h^{\prime}}{T} } \ex{-\mc{L}_{ \msc{C}_{n} }[\wh{\mf{z}}](\th_1+\eta)} \Big\} 
\Big\{ \ex{\mc{L}_{ \msc{C}_{n} }[\wh{\mf{z}}](\th_2+\eta)} - \ex{ \f{ h-h^{\prime}}{T} } \ex{\mc{L}_{ \msc{C}_{n} }[\wh{\mf{z}}](\th_2-\eta)} \Big\}  } \\
\times \f{   \det_{\Ga( \msc{C}_{n} ) }\Big[ \e{id} + \op{U}^{(\la)}_{\th_1}  [ \, \wh{\mf{z}} \,] \Big]  \cdot 
\det_{\Ga( \msc{C}_{n} ) } \Big[ \e{id} + \op{U}^{(\mu)}_{\th_2} [ \, \wh{\mf{z}} \,] \Big] }{  \det\big[ \e{id} + \wh{\op{K}}_{ \msc{C}_{n} ; \la} \big]  \cdot \det_{ \msc{C}_{n} }\big[ \e{id} + \wh{\op{K}}_{ \msc{C}_{n} ; \mu} \big]    } \;. 
\end{multline}

The expressions for $\wh{\mc{S}}^{(h,h^{\prime})}$ and $\wh{\mc{A}}^{(h,h^{\prime})}$  all involve a contour $\msc{C}_n$ encircling  all the 
roots $\{\la_a\}_1^N$ and $\{\mu_a\}_1^N$ and the point $-\tf{\be}{N}$ but not any other zero (or singularity)
of $1 + \wh{\mf{a}}_{\la}(\om)$ and $1 + \wh{\mf{a}}_{\mu}(\om)$. If such a contour does not exist, the answer should be 
considered as a limit of a small deformation of the roots $\{\la_a+\veps_a\}_1^N$ and $\{\mu_a+\veps_a^{\prime}\}_1^N$
where such a contour exists. 

\end{prop}

\noindent Here, two comments are in order. \vspace{1mm}

\begin{itemize}
 \item[$\bullet$]  The parameters $\th_1, \th_2$ appearing in the expression for $\wh{\mc{A}}^{(h,h^{\prime})}$ are arbitrary. 
Although individual terms of the expression do depend on the $\th_a$'s, the expression taken as a whole,  
does not depend on the $\th_a$'s, see Appendix A.3 of \cite{KozKitMailSlaTerXXZsgZsgZAsymptotics}. \vspace{1mm}

 \item[$\bullet$] The expression for the amplitude \eqref{ecriture amplitude An pour finite Trotter number} involves a partial derivative in respect to $h^{\prime}$. 
To allow doing so, one should prove the property of differentiability in $h^{\prime}$  for $\wh{a}_{\mu}$. This boils down to the differentiability of 
the roots $\{\mu_a\}_1^N$.  The parameters $\{\mu_a\}_1^N$ one starts with in Proposition \ref{Proposition Rep Int Trotter Fini pour amplitude sgZ a tempe finie}
are assumed to be a solution of the Bethe Ansatz equations at magnetic field $h$ which has the property that the associated Jacobian of the Bethe equations does not vanish 
(the latter is equivalent to $\braket{\Psi_n}{\Psi_n}\not= 0$). 
By the implicit function theorem, this guarantees the existence of a smooth deformation $h^{ \prime} \mapsto \{ \mu_a(h^{\prime}) \}_1^N$ 
of these roots.  \vspace{1mm}

\end{itemize}

Under the hypothesis stated in the previous section relatively to the functions $\wh{\mf{a}}_{\la}$ and $\wh{\mf{a}}_{\mu}$, 
it is straightforward to take the infinite Trotter number limit on the level of the representation  \eqref{ecriture amplitude An pour finite Trotter number}:
just as for the case of the correlation lengths, one should solely carry out the replacement 
\beq
\wh{\mf{z}}(\nu) \; \hookrightarrow \; \mf{z}(\nu)  \qquad \e{and} \qquad \wh{\mf{a}}_{\la/\mu} \hookrightarrow \mf{a}_{\la/\mu}
\label{ecriture equation passage nbr de Trotter fini}
\enq
exactly as it has been discussed in \eqref{ecriture remplacement z hat par z}. This gives rise to the amplitudes $\msc{A}^{(n)}$.

All-in-all, under the additional hypothesis of convergence of the obtained series and exchangeability of limits, one gets the representation:
\beq
\Big< \sg_1^{z} \cdot \sg_{m+1}^{z}   \Big>_{T} \, - \, \Big< \sg_1^{z}   \Big>_{T}^2 \; = \;  \sul{n=1}{ +\infty } \Big( \rho^{(n)} \Big)^{m} \cdot \msc{A}^{(n)} \;. 
\enq
There, the ratios of eigenvalues are as given by \eqref{ecriture forme ration VP Trotter infini}
 while the amplitudes $\msc{A}^{(n)}$ are obtained from the amplitudes $\msc{A}^{(n)}_N$ in \eqref{ecriture amplitude An pour finite Trotter number}
by means of the substitution \eqref{ecriture equation passage nbr de Trotter fini}.

The exponential decay at large-distances of the $\sg^z-\sg^z$ connected two-point function at finite temperature follows directly from the properties of the correlation lengths.

\section{The low-temperature limit}
\label{Section Low T analysis of FF expansions}

\subsection{The strategy and main result}

The series expansion \eqref{ecriture dvpmt series FF thermaux sgZsgZ} provides one with a quite effective representation for a correlation function at finite temperature. 
In particular, upon reasonable hypothesis, the leading large-distance asymptotic expansion of the two-point function can be readily extracted from it. 
It seems reasonable to ask whether this kind of expansion allows one to test the conformal field theory based predictions for the long-distance asymptotic behaviour of two point functions in models
at finite but sufficiently low temperatures. Upon some additional hypothesis and certain formal handlings of the type described in Chapter \ref{Chapitre approche des FF aux asymptotiques des correlateurs},
it is indeed possible to reproduce the conformal predictions for the large-distance asymptotic behaviour. 
To do so, one needs to process in three steps : \vspace{1mm}
\begin{itemize}
\item[i)] obtain the low-T asymptotic expansion of the (supposedly existing) solutions $\mf{a}_{\la}$ and $\mf{a}_{\mu}$ to the non-linear integral equations 
\eqref{ecriture eqn NLIE pour fct a lambda infinite Tortter nbr} and \eqref{ecriture eqn int limite pour a etat excite mu}. This can be achieved by making a certain amount of  hypothesis
on the overall properties of the functions $\mf{a}_{\la}$ and $\mf{a}_{\mu}$. Using results obtained in [$\bs{A23}$], 
such properties can be checked to hold \textit{a posteriori} on the level of the obtained answer for the low-T expansion. This provides a consistency test of the analysis. 
So far, I only managed to prove  [$\bs{A2}$] the existence of such asymptotic expansion in the much simpler case of the solution to the non-linear integral equation \eqref{ecriture eqn YY} which drives the finite
temperature behaviour of the free energy of the non-linear Schr\"{o}dinger model.  \vspace{1mm}
\item[ii)]  Extract the low-T asymptotic behaviour of the amplitudes $\msc{A}^{(n)}$ and ratios of eigenvalues $\rho^{(n)}$. Taking for granted the 
low-$T$ asymptotic expansion for $\mf{a}_{\la}$ and $\mf{a}_{\mu}$ obtained in step i), this low-$T$ asymptotic analysis can be done rigorously. \vspace{1mm}
\item[iii)] Re-sum the series resulting from the replacement in  \eqref{ecriture dvpmt series FF thermaux sgZsgZ} of the amplitudes $\msc{A}^{(n)}$ and ratios of eigenvalues $\rho^{(n)}$ by their leading low-$T$
asymptotics.  \vspace{1mm}
\end{itemize}
The techniques allowing one to carry out the low-$T$ analysis of step ii) have been first developed in my collaboration with Maillet and Slavnov [$\bs{A27}$]. 
In collaboration with Dugave and G\"{o}hmann [$\bs{A24}$], I managed to set the derivation of into a rigorous framework, provided that the low-$T$ 
asymptotic expansions of obtained in step i) are taken for granted. 
Step iii) builds from its very start on the hypothesis that it is licit to exchange the summation over $n$ in \eqref{ecriture dvpmt series FF thermaux sgZsgZ} with 
the low-$T$ asymptotic expansion of the individual summands. Further, even if this is taken for granted,  one still has to allow for various formal 
handlings in the spirit of those described in  Section \ref{Section Form factor approach to AB of correlation fcts}. Still, despite the obvious missing bricks in the analysis, the arguments developed in steps i)-iii) lead to 
the conclusion that, in the limit $m \tend + \infty$, the product $Tm$ being fixed:
\beq
\Big< \sg_1^{z} \cdot \sg_{m+1}^{z}   \Big>_{T} \, \simeq  \, \Big< \sg_1^{z}   \Big>^2  \, - \,  \f{2 }{ \pi^2 \mc{Z}^2 } \Bigg(  \f{ \tf{\pi T}{v_F} }{ \sinh\Big[ \frac{1}{v_F} \pi T m \Big] }  \Bigg)^2
\; + \; \sul{ \ell \in \mathbb{Z}^* }{}  \big| \mc{F}_{\ell}\big(\sg^z \big) \big|^2 \ex{2\i m \ell p_F } \Bigg(  \f{ \tf{\pi T}{v_F} }{ \sinh\Big[ \frac{1}{v_F} \pi T m \Big] }  \Bigg)^{2 \ell^2  \mc{Z}^2 }
\label{ecriture DA sgZsgZ en tempe XXZ massless}
\enq
This is precisely the form predicted on the basis of conformal field theory arguments. 
The symbol $\simeq$ in \eqref{ecriture DA sgZsgZ en tempe XXZ massless} indicates that the \textit{rhs} should provide one with the leading asymptotic behaviour of each oscillating harmonic  up to $\big( 1 + \e{o}(1) \big)$ corrections.  
In order to describe the various constants arising in the expansion \eqref{ecriture DA sgZsgZ en tempe XXZ massless}, I need to remind a few results relative to 
the characterisation of the thermodynamic limit (at zero temperature) of the XXZ spin-$1/2$ chain. 
The Fermi boundary $q$ defining the endpoint of the Fermi zone at a given magnetic field $h$ corresponds the the unique solution $\big( q , \veps_{XXZ}(\la\mid q) \big) $
to the below equations on the unknowns $Q$ and $\veps_{XXZ}(\la\mid Q)$:
\beq
\veps_{XXZ}(\la\mid Q) \; + \; \Int{-Q}{Q} K_{XXZ}\big(\la-\mu\mid \zeta\big)   \veps_{XXZ}(\mu\mid Q) \cdot \f{ \dd \mu }{  2\pi } \; = \; 
h- 2J \sin(\zeta) K_{XXZ}\big( \la\mid \tf{\zeta}{2}\big)   
\enq
and
\beq
\veps_{XXZ}(Q\mid Q)  \; = \; 0 \qquad \e{where} \qquad  K_{XXZ}\big( \la\mid \zeta \big)  \; =\; \f{ \sin(2\zeta) }{ \sinh(\la-\i \zeta)\sinh(\la+\i \zeta) } \;. 
\enq
With Dugave and G\"{o}hmann, we proved (Theorem 3 of [$\bs{A23}$]) that this solution indeed exists and is unique. 
We also obtained some explicit bounds on the Fermi boundary $q$. In particular one has $0 \leq q<+\infty$ whenever $h>0$. 
The function $\veps_{XXZ}(\la) \equiv \veps_{XXZ}(\la\mid q )$ corresponds to the dressed energy of the excitations in the massless regime of the XXZ chain. 
The dressed momentum and dressed charge are then defined as the unique solutions\symbolfootnote[2]{In this case, uniqueness is relatively easy to see
since one can show that the spectral radius of the integral operator $\op{K}_{XXZ}$ on $L^2(\intff{-q}{q})$ is strictly less than 1. 
See Proposition 2 of [$\bs{A23}$] for more details.} to the linear integral equations:
\beq
 \left( \ba{cc} Z_{XXZ}(\la) \\ p^{\prime}_{XXZ}(\la) \ea \right) \; + \; \Int{-q}{q}K_{XXZ}\big( \la-\mu\mid \zeta\big)  \left( \ba{cc} Z_{XXZ}(\mu) \\ p^{\prime}_{XXZ}(\mu) \ea \right) \cdot \f{ \dd \mu }{  2\pi }
\;= \;  \left( \ba{cc} 1 \\  K_{XXZ}\big( \la \mid \tf{\zeta}{2} \big)  \ea \right) \qquad \e{and} 
\quad p_{XXZ}(\la) \; = \; \Int{0}{\la} p^{\prime}_{XXZ}(\la) \cdot \dd \la \;. 
\enq
The Fermi momentum $p_F$ of the XXZ chain is defined as $p_F=p(q)$, the Fermi velocity $v_F=\veps^{\prime}_{XXZ}(q)/p^{\prime}_{XXZ}(q)$
and the constant $\mc{Z}$ corresponds to the value taken by the dressed charge of the XXZ chain on the Fermi boundary $\mc{Z}=Z_{XXZ}(q)$.
 The Fermi momentum parametrises the so-called ground state expectation value of the $\sg_1^z$ operator:
\beq
\Big< \sg_1^{z}   \Big> \; = \; 1\, - \, \f{2}{\pi} p_F \;. 
\enq
Finally, it remains to describe the amplitude $\big| \mc{F}_{\ell}\big(\sg^z \big) \big|^2$. It corresponds to the properly normalised in the volume $L$ thermodynamic limit of the modulus squared of the
form factor of the $\sg^z$ operator taken between the ground state $\ket{G.S.}$ and the fundamental representative $\ket{\ell}$ 
of the $\ell$-critical class (the terminology is analogous to the one introduced below Corollary \ref{Corollaire cptmt FF sur etat critique ell}) of the XXZ chain:
\beq
\big| \mc{F}_{\ell}\big(\sg^z \big) \big|^2 \; = \; \lim_{ L \tend + \infty} \Bigg\{  \bigg( \f{ L }{ 2\pi } \bigg)^{2 \ell^2 \mc{Z}^2 }
\cdot  \f{ |\bra{\ell} \sg_1^z \ket{G.S.}  |^2 }{ \braket{ G.S. }{ G.S. } \cdot \braket{ \ell }{ \ell }  }   \Bigg\} \;. 
\enq

\section{The surface free energy}
\label{Section Energie Libre de Surface}

\subsection{The model and the starting expression}
\label{SousSection Intro model XXZ bdry plus Surf Free energy}

I will discuss in this section the results  relative to a characterisation of the 
surface free energy of the XXZ chain subject to so-called diagonal boundary fields that I obtained in [$\bs{A26}$]. The Hamiltonian of this model embedded in an overall magnetic field $h$  is defined as 
\bem
\op{H}_{XXZ}^{(\e{bd})} \, = \,   J \sum_{a=1}^{L-1} \Big\{ \sigma^x_a \,\sigma^x_{a+1} +   \sigma^y_a\,\sigma^y_{a+1} + \cosh \eta \,(\sigma^z_a\,\sigma^z_{a+1}+1)\Big\} \\
+ J \sinh \eta \coth\xi_- \, \sg^z_1 +  J \sinh \eta \coth\xi_+ \,\sg^z_L \; + J \f{\cosh\pa{2\eta}}{\cosh\pa{\eta}}  
-\f{h}{2} \sul{a=1}{L} \sg_a^z \;. 
\label{ecriture Hamiltonien XXZ ac bord}
\end{multline}
The parameters $\xi_{\pm}$ parametrise the intensity of the two boundary fields. The quantum integrability of this model builds on the 
reflection equation introduced by Cherednik \cite{CherednikReflectionEquationFactorisabilityOfScattering}. The transfer matrix 
$\op{t}_{XXZ}^{(\e{bd})}$ is built \cite{SklyaninABAopenmodels} out of the diagonal solutions  \cite{CherednikReflectionEquationFactorisabilityOfScattering} of the reflection equations
\beq
K^{\pm}_a\pa{\la} = K_a\pa{\la +\tf{\eta}{2} \pm \tf{\eta}{2} ; \xi_{\pm}  } \qquad \e{with} \quad 
K_a\pa{\la;\xi} = \pa{  \ba{cc} \s{\la+\xi} & 0 \\ 0 & \s{\xi-\la} \ea
}_{\pac{a}} 
\label{Kdef}
\enq
and from the monodromy matrix $\op{T}_a\pa{\la}$ of the periodic XXZ chain \eqref{definition matrices de monodromie pour XXZ periodique} :
\beq
\op{t}_{XXZ}^{(\e{bd})}(\la) \;  =  \; (-1)^L \e{tr}_a\pac{  K^+_a\pa{\la} \op{T}_a\pa{\la} K^-_a\pa{\la} \sigma_a^y \, \op{T}_a^{t_a}\pa{-\la-\eta} \sigma_a^y }_{\mid \xi=0}   \;. 
\label{crossed transfer matrix}
\enq
The boundary transfer matrix satisfies 
\beq
\op{t}_{XXZ}^{(\e{bd})}(0)  \, = \,  \f{ \e{tr}_a\pac{ K^+_a\pa{0} } \e{tr}_a\pac{ K^-_a\pa{0} } }{2} \pac{ \sinh(\eta) }^{2L}  \e{id}_L
\enq
and
\beq
\op{H}_{XXZ}^{(\e{bd})} \,  =  \, J \sinh \eta 
\Big\{ \f{\dd }{\dd \la}\ln \op{t}_{XXZ}^{(\e{bd})}(\la) \Big\} _{\mid_{\la=0}} \, - \, \f{h}{2} \sul{a=1}{L} \sg_a^z  \;. 
\enq
There, $\e{id}_L$ is the identity operator on $\mf{h}_{XXZ}$. These identities allow one to represent the partition function of the 
boundary XXZ chain as the Trotter limit:
\beq
\mc{Z}_T\big[ \op{H}^{(\e{bd})}_{XXZ}\big]  \; = \;  \e{tr}_{ \mf{h}_{XXZ} } \Big[ \ex{ -\f{ 1 }{T} \op{H}_{XXZ}^{(\e{bd})}   } \Big] 
\; = \;   \lim_{N\tend +\infty}
\Bigg\{ \e{tr}_{ \mf{h}_{XXZ} } \bigg[	\bigg( \op{t}_{XXZ}^{(\e{bd})} \big(-\tf{\be}{N}\big)  \cdot \Big( \op{t}_{XXZ}^{(\e{bd})}(0) \Big)^{-1}  \bigg)^N   \cdot \pl{a=1}{L} \ex{ \f{h}{2T}\sg_a^z } \bigg] \Bigg\}\; . 
\label{ecriture fction partition comme trace}
\enq

As shown in \cite{BortzFrahmGohmannSurfaceFreeEnergy}, the representation \eqref{ecriture fction partition comme trace} can be reorganised in the form  
\beq
\mc{Z}_T\big[ \op{H}^{(\e{bd})}_{XXZ}\big] \; = \;  \lim_{N\tend +\infty} \Bigg\{  
\bigg( \f{ 2  }{ \e{tr}_a\pac{ K^+_a\pa{0} } \e{tr}_a\pac{ K_a^-\pa{0} }  } \bigg)^N
\e{tr}_{a_1,\dots,a_{2N}} \bigg[  \op{P}_{a_1 a_2}\pa{ \tf{-\be}{N} } \dots \op{P}_{a_{2N-1} a_{2N} } \pa{ \tf{-\be}{N} } 
 \Big( \op{t}_{\mf{q}}(0) \Big)^L   \bigg]  \Bigg\}\;,
\label{equation fonction partition volume fini}
\enq
where $\op{t}_{\mf{q}}(\xi)=\e{tr}_k\big[ \op{T}_{\mf{q};k}(\xi) \big] $ is the quantum transfer matrix constructed from the quantum monodromy matrix \eqref {definition QTM} 
while $\op{P}_{\! ab}(\la)$ is the one-dimensional projector
\beq
\op{P}_{\! ab}(\la) \;  = \; K_a^+(\la) \, \big[ \ket{+}_a\ket{+}_b +  \ket{-}_a\ket{-}_b  \big]  \cdot  \big[  \bra{+}_a\bra{+}_b   +    \bra{-}_a\bra{-}_b     \big] \,  K_a^{-}(\la) 
\label{definition projecteur}
\enq
on $ \mf{h}_{\mf{q}}^{a} \otimes \mf{h}_{\mf{q}}^{b}$.

The surface free energy associated with $\op{H}_{XXZ}^{(\e{bd})}$ is defined as the difference
between the free energy of the XXZ chain subject to periodic and the one subject to the "diagonal" boundary conditions. 
Assuming that a large-$L$ asymptotic behaviour of the type \eqref{ecriture DA gd L Free energy at boundary model} holds, 
the surface free energy can be related to the large-volume $L$ limit of the ratio of partition functions:
\beq
 \ex{- \f{ f_{\e{surf}} }{ T } }  \equiv \lim_{L\tend +\infty }  \bigg\{  \f{   \mc{Z}_T\big[ \op{H}^{(\e{bd})}_{XXZ}\big]   }{  \tr_{ \mf{h}_{XXZ} }\Big[ \ex{ -\f{1}{T} \op{H}_{XXZ} } \Big]  }  \bigg\} \;.
\enq
Assuming further the exchangeability of the Trotter and infinite volume limits, it is possible to represent \cite{BortzFrahmGohmannSurfaceFreeEnergy}
$ \ex{- \f{ f_{\e{surf}} }{ T } }$ in terms of the  Trotter limit: 
\beq
 \ex{- \f{ f_{\e{surf}} }{ T } }  \; =  \;  \lim_{N \tend +\infty} \Big\{  \ex{- \f{ f_{\e{surf}}^{(N)} }{ T } }  \Big\} 
\qquad \e{where} \qquad 
\ex{- \f{ f_{\e{surf}}^{(N)} }{ T } }   = 
\f{ \bra{\Psi_0} \op{P}_{a_1 a_2}\pa{ \tf{-\be}{N} } \cdots \op{P}_{a_{2N-1} a_{2N} } \pa{ \tf{-\be}{N} } \ket{\Psi_{0}} }
{  \braket{\Psi_0}{\Psi_0} \cdot \Big\{ \e{tr}_a\pac{ K^+_a\pa{0} }\cdot \e{tr}_a\pac{ K^-_a\pa{0} } /2 \Big\}^N   }   \;. 
\label{ecriture energie surface free limit trotter}
\enq
Above $\ket{\Psi}_0$ stands for the "dominant" eigenvector \eqref{definition Vect Propre Dominant QTM} of the quantum transfer matrix for the XXZ spin -$1/2$ chain
subject to \textit{periodic} boundary conditions.

\subsection{Transformations towards taking the Trotter limit}
\label{SousSection Surf free Energie Recriture vers limite}

Building on the representation for the local projector \eqref{definition projecteur}, 
it is possible to recast $\exp\Big\{ - \f{  1 }{ T } f_{\e{surf}}^{(N)}  \Big\}$ in the form
\beq
\ex{- \f{ f_{\e{surf}}^{(N)} }{ T } }  = 
 \paf{ 2 }{ \e{tr}_0\pac{ K_+\pa{0} } \e{tr}_0\pac{ K_-\pa{0} }  }^N   \cdot 
\f{  \mc{F}^+ \cdot \mc{F}^-  }{ \braket{\Psi_0}{\Psi_0}  }    \;. 
\label{ecriture rep trotter finie SFE}
\enq
$\mc{F}^+$ and $\mc{F}^- $ stand for the expectation values
\beq
\mc{F}^+ \equiv \bra{\Psi_0} K_{a_1}^+\pa{-\tf{\be}{N}} \dots K_{a_{2N-1}}^+\pa{ -\tf{\be}{N} } \ket{v}
\quad \e{and} \quad
\mc{F}^- \equiv\bra{v} K_{a_1}^-\pa{-\tf{\be}{N}} \dots K_{a_{2N-1}}^-\pa{ -\tf{\be}{N} } \ket{\Psi_0}
\label{rewrite-eq-1}
\enq
where 
\beq
\ket{ v } = \pa{ \ket{+}_{a_1}\ket{+}_{a_2} +  \ket{-}_{a_1}\ket{-}_{a_2} } \otimes \dots \otimes 
\pa{ \ket{+}_{a_{2N-1}}\ket{+}_{a_{2N}} +  \ket{-}_{a_{2N-1}}\ket{-}_{a_{2N} } } 
\enq
and $\bra{v}$ is the dual vector. 
Formula \eqref{ecriture rep trotter finie SFE} constitutes the starting point towards taking the infinite Trotter number limit of $f_{\e{surf}}^{(N)}$.
The large-$N$ analysis of this expression is made possible thanks to the observation made in [$\bs{A26}$] that
$\mc{F}^{\pm}$ both can be recast in terms of $\mc{Z}_N\Big( \{\xi_a \}_1^N  ; \{ \la_k \}_1^N ; \xi_-  \Big) $, the partition function of the six-vertex model with reflecting ends. 
\begin{prop}
 The representation holds 
\beq
\mc{F}^{-} \; = \; \ex{-\f{Nh}{2T}} \cdot \mc{Z}_N\Big( \{-\tf{\be}{N}\}_1^N  ; \{ \la_k \}_1^N ; \xi_-  \Big) \quad and \quad 
\mc{F}^{+} \; = \; \bigg(  \f{ \sinh\big( 2\eta- 2 \tf{\be}{N} \big) }{  \ex{\f{h}{2T}}  \sinh\big( - 2 \tf{\be}{N} \big) } \bigg)^N
\cdot \mc{Z}_N\Big( \{-\tf{\be}{N}\}_1^N  ; \{ \la_k \}_1^N ; \xi_+  \Big)
\enq
where the notation $\{-\tf{\be}{N}\}_1^N$ means that the $N$ auxiliary parameters ought to be set equal to $-\tf{\be}{N}$. Also, the roots $\{ \la_a \}_1^N$
are the Bethe roots which parametrise the dominant eigenvector $\ket{\Psi_0}$ of the quantum transfer matrix \eqref{definition Vect Propre Dominant QTM}. 
\end{prop}

Tsuchiya \cite{TsuchiyaPartitionFunctWithReflecEnd} showed that the partition function of the six-vertex model with reflecting ends admits a determinant representation 
of Izergin type \cite{IzerginPartitionfunction6vertexDomainWall}:
\beq
\mc{Z}_N\pa{ \{\xi_a\}_1^N  ; \{ \la_k \}_1^N ; \xi_- } =    
\f{ \pl{a=1}{N} \Big\{  \s{\xi_- + \la_a}  \s{2\xi_a} \Big\}\cdot  \pl{a,b=1}{N} \Big\{ \sd{\xi_a,\la_b} \cdot \sd{\xi_a+\eta,\la_b} \Big\}  }
{   \pl{a<b}{N} \Big\{ \sd{\la_b,\la_a}   \s{\xi_a-\xi_b} \s{\xi_a+\xi_b+\eta} \Big\} }  \cdot   
\det_{N}\big[ \mc{N} \big] \;.
\label{ecriture Tsuchiya determinant}
\enq
$\mf{s}(\la,\mu)$ and 
the entries of the matrix $\mc{N}$ are defined as 
\beq
\sd{\la,\mu} = \s{\la+\mu}\s{\la-\mu} \qquad \e{and} \qquad 
\mc{N}_{jk} = \f{  \s{\eta}  }{   \sd{\xi_j+\eta,\la_k}\sd{\xi_j,\la_k}  } \;. 
\label{ecriture entree N et def double sinus}
\enq
 It is readily checked that all the singularities present in the denominator of \eqref{ecriture Tsuchiya determinant}
correspond to zeroes of the determinant. The coefficients $\mc{F}^{\pm}$ are expressed as the semi-homogeneous
limit of the partition function where $\xi_a \tend -\tf{\be}{N}$. Taking this semi-homogeneous limit brutally
on the level of \eqref{ecriture Tsuchiya determinant} would replace $\det_{N}\big[ \mc{N} \big]$ by a Wronskian determinant of size $N$. 
Such a representation is however not adapted for taking the large-$N$ limit. All the more that, so as to take this limit
effectively, one should find a way to extract all the potential zeroes out of the determinant, namely also 
those at $\la_a= \pm \la_b$. Taking these limits can be achieved by means of Cauchy determinant factorization techniques 
\cite{SlavnovFormFactorsNLSE}
followed by some combinatorial rewriting of the answer.

\begin{theorem}
\label{Proposition representation limit homogene Tsuchiya}
Let $\{\la_a\}_1^N$  solve the system of Bethe equations associated with the quantum tranfser matrix \eqref{ecriture Bethe Equations QTM}
and $\wh{\mf{a}}_{\la}$ be as defined by \eqref{definition counting function GS}.   
Then, the homogeneous limit $\xi_a\tend -\tf{\be}{N}$, $a=1,\dots,N$ of the partition function admits the below representation:
\bem
\mc{Z}_N\pa{ \{-\tf{\be}{N}\}_1^N ; \{ \la_a \}_1^N;\xi_-  } = \pl{a=1}{N}  \bigg\{ \f{ \s{-2\tf{\be}{N} } \s{\la_a+\xi_-} }{ \s{2\la_a} } \bigg\}
\cdot  \pl{a=1}{N} \Big[ \s{\la_a-\tf{\be}{N}} \s{\eta -\la_a-\tf{\be}{N} }\Big]^{N}  \\
\times \pl{a<b}{N}  \f{\s{\la_a+\la_b+\eta}}{\s{\la_a+\la_b}}  \cdot 
\Bigg\{  \pl{p=1}{N}[1 + \kappa \wh{\mf{a}}_{\la}(-\la_p)]^{\f{1}{2}} \cdot 
\paf{ 1 + \kappa \wh{\mf{a}}_{\la}\pa{0} }{ 1 - \kappa  \wh{\mf{a}}_{\la}\pa{0} }^{\f{1}{4}} \cdot 
\ex{\mc{F}_N^{(\kappa)} \pa{\{\la_a\}_1^N}}  \Bigg\}_{\mid \kappa =1}\; . 
\label{ecriture representation limit homogene fcton partition}
\end{multline}
For $|\kappa|$ small enough, $\mc{F}_N^{(\kappa)}$ is defined by the convergent series 
\bem
\mc{F}_N^{(\kappa)} \pa{\{\la_a\}_1^N} = 
\sul{ k = 0  }{+\infty}  \kappa^{2k+1} \Oint{ \msc{C}_1\supset\dots \supset \msc{C}_{2k+1} }{}
\hspace{-2mm}  \pl{p=1}{2k+1}\Big\{ \wh{U}(\om_p,\om_{p+1}) \Big\} \; \cdot \; 
\bigg\{ \sul{ n=k }{ +\infty } \f{ \Big[ \kappa^2 \, \wh{\mf{a}}_{\la}(\om_{2k+1}) \wh{\mf{a}}_{\la}(-\om_{2k+1})  \Big]^{n-k} }{2n+1}  \; \bigg\} 
\f{ \dd^{2k+1} \om }{ (2 \i \pi)^{2k+1} }  \\
\; -\; \sul{  k = 1   }{+\infty}  \kappa^{2k}  \Oint{ \msc{C}_1\supset\dots \supset \msc{C}_{2k} }{}
 \pl{p=1}{2k}\Big\{ \wh{U}(\om_p,\om_{p+1}) \Big\} \; \cdot \; 
\bigg\{ \sul{ n=k }{ +\infty } \f{ \Big[ \kappa^2 \, \wh{\mf{a}}_{\la}(\om_{2k}) \wh{\mf{a}}_{\la}(-\om_{2k})  \Big]^{n-k} }{2n+1}  \; \bigg\} \; \f{ \dd^{2k} \om }{ (2 \i \pi)^{2k} } \; . 
\label{ecriture F cal N kappa trotter fini}
\end{multline}
The contours arising in the multiple integrals $\msc{C}_1 \supset \dots \supset \msc{C}_p$, $p\in \mathbb{N}^*$, are encased contours such that $\msc{C}_k$, 
for $k=1,\dots,p$ encircles the roots $\la_1, \dots, \la_N$ but not any other singularity of the integrand. In particular, the poles at 
$\om_k=-\om_{k+1}$ are \textit{not} encircled by $\msc{C}_k$. Also, the integrands in the multiple integrals are expressed by means of the convention  
 $\om_{n+1}\equiv \om_1$ and are built up from the function 
\beq
\wh{U}(\om,\om^{\prime})  = \f{ - \ex{ -\f{h}{T} } \s{2\om^{\prime}+\eta} }
			{   \s{\om^{\prime} + \om }\s{\om-\om^{\prime} +\eta}}  
\pl{a=1}{N} \f{ \s{\la_a + \om^{\prime} } \s{\om^{\prime} - \la_a + \eta}  }
				{\s{\om^{\prime}-\la_a} \s{\la_a + \om^{\prime} +\eta} }    \;. 				
\enq

\end{theorem}

I should comment on how formula  \eqref{ecriture representation limit homogene fcton partition} should be understood. First of all, the function appearing inside of the last bracket is an entire function of $\kappa$. It has thus
a well defined limit when $\kappa=1$. However, the individual constituents of this function might not be well defined in the $\kappa \tend 1$ limit. 
For instance the series might not be convergent at $\kappa=1$. Thus, the expression inside of the bracket is to be understood as the analytic continuation 
from $\kappa$ small up to $\kappa=1$.  Second, the formula contain some potential singularities at $\la_a=-\la_b$. 
Should two roots satisfy $\la_a+\la_b=0$, then the formula should be understood as the appropriate limit. 
This limit is well defined in that all such singularities are cancelled by the zeroes of $ \prod_{p=1}^{N}[1 + \kappa \wh{\mf{a}}(-\la_p)]^{\f{1}{2}} $. 
In the following, I will simply admit that the analytic continuation up to $\kappa=1$ can be made simply by setting $\kappa=1$ directly in each of the 
terms present in \eqref{ecriture representation limit homogene fcton partition}. 

The proof of Theorem  \ref{Proposition representation limit homogene Tsuchiya} follows from Lemma 2.1, Lemma 2.2 and Proposition 2.2 of [$\bs{A26}$].

\vspace{2mm}

Theorem \ref{Proposition representation limit homogene Tsuchiya} is already enough so as to provide an expression for the surface free energy that constitutes a good
starting point for taking the infinite Trotter number limit. Namely, one has 
\bem
\exp\Big\{ -\f{1}{T} f^{\pa{N}}_{\e{surf}} \Big\} \;  =  \;  \ex{ \f{Nh}{T} }
\pl{ a=1 }{ N } \f{ \s{\xi_- + \la_a }  \s{\xi_+ + \la_a} }{ \s{\xi_-} \s{\xi_+} }  \cdot \pl{a=1}{N} \f{ \s{\eta} }{ \s{2\la_a +\eta} } \cdot 
\pl{a=1}{N} \f{ \s{-2\tf{\be}{N}}  }{  \s{2\la_a}    }  \\
\times   \pl{a,b=1}{N} \f{ \s{\la_a + \la_b +\eta} }{ \s{\la_a-\la_b +\eta} } \; \cdot \; 
\Bigg\{ \f{ \prod_{a \not=b }^{N} \s{\la_a-\la_b} }
{  \pl{a=1}{N} \tf{ \wh{\mf{a}}^{\, \prime}(\la_a) }{ \wh{\mf{a}}(\la_a) } }  \Bigg\}
\cdot \Bigg\{ \f{ \prod_{a=1}^{N} \big[ 1+\wh{\mf{a}}(-\la_a)\big] }{ \prod_{a,b}^{N}\s{\la_a+\la_b} }  \Bigg\} 
\paf{1+ \wh{\mf{a}}(0) }{ 1-\wh{\mf{a}}(0) }^{\f{1}{2}} \\
\times \paf{  \s{2\eta - 2\tf{\be}{N} } }{ \s{2\eta} }^{N}  \cdot \f{  \ex{2\mc{F}_N^{(1)}\pa{\{\la_a\}_1^N}}  }{  \det\big[\e{id} + \wh{\op{K}}_{\msc{C};\la} \big] } \;.
\label{equation representation energie libre finite Trotter}
\end{multline}
Above, $\msc{C}$ is the contour defined below of \eqref{ecriture NLIE fonction a hat regime general} 
and $\wh{\op{K}}_{ \msc{C};\la} $ is the integral operator on $L^2(\msc{C})$ characterised by the integral kernel \eqref{ecriture operateir hat K gamma et racines nu}.

In order to take the infinite Trotter number limit, one should recast all the dependence on the $\{\la_a\}$ variables in 
\eqref{equation representation energie libre finite Trotter} in terms of contour integrals solely involving the function $\wh{\mf{a}}_{\la}$,
similarly to what has been discussed in the case of the form factors of the quantum transfer matrix. 
The necessary rewritings are described in  Section 3.1 of [$\bs{A26}$].
All-in-all, one obtains 
\beq
\exp\Big\{ -\f{1}{T} f^{\pa{N}}_{\e{surf}} \Big\} =
\paf{ \s{\eta} \s{2\eta-2 \tf{\be}{N}}   }
{ \s{\eta-2\tf{\be}{N}} \s{2\eta}    }^{N}
\f{ \ex{- \f{1}{T}  \big[  \wh{\mc{B}}(\xi_+)  \; +\;  \wh{\mc{B}}(\xi_-) \big] } }
 {\sqrt{ 1-[\wh{\mf{a}}(0)]^2}   }    \cdot 
 \f{   1\; + \; \ex{-\tf{h}{T}}   }{  \det\big[\e{id} + \wh{\op{K}}_{\msc{C};\la} \big]  }  
  \exp\bigg\{ 2 \mc{F}_{N}\pa{ \{\la_a\}_1^N }  \; + \; \wh{ \mc{I}} \, \bigg\} \;. 
\enq
The whole dependence on the parameters $\xi_{\pm}$ encoding the intensity of the boundary fields is contained in the function 
\beq
\wh{\mc{B}}(\xi) \; = \;   -  N \ln \Big(  \f{ \s{\xi - \tf{\be}{N}} }{ \s{\xi} } \Big) 
- T \de_{\xi} \ln \big[ 1+\wh{\mf{a}}(-\xi) \big] \; + \;
T \Oint{ \msc{C} }{}  \ln \big[ 1+\wh{\mf{a}}\pa{\om} \big] \cdot  \coth\pa{\om+\xi} \cdot  \f{\dd \om}{ 2 \i \pi }  
\enq
where $\de_{\xi}=1$ if $-\xi$ is surrounded by  $\msc{C}$ and $\de_{\xi}=0$ otherwise. Further, $\wh{\mc{I}}$ is defined in terms of single and two-fold integrals 
\bem
\wh{\mc{I}} \; = \; 
 2 \Oint{\msc{C}}{} \hspace{-1mm} \f{\dd \om}{2 \i \pi} 
  \ln\big[ 1+\wh{\mf{a}}\pa{\om} \big] \Big[\coth\pa{2\om} + \coth\pa{2\om +\eta} \Big]   
 \\
 -2N \Oint{ \msc{C} }{} \hspace{-1mm} \f{\dd \om }{ 2 \i \pi }  \ln\big[1+\wh{\mf{a}}\pa{\om} \big] 
\pac{ \coth\pa{\om +\tf{\be}{N} } \, - \,   \coth\pa{\om -\tf{\be}{N} } \; + \; \coth\pa{\om -\tf{\be}{N} +\eta} \, - \,   \coth\pa{\om +\tf{\be}{N} +\eta}   }  \\
\; + \;    \Oint{\msc{C}}{} \hspace{-1mm} \f{\dd \om}{2 \i \pi}   
\hspace{-1mm} \Oint{\msc{C}^{\prime} \subset \msc{C}}{} \hspace{-1mm}  \f{\dd \om^{\prime}}{2 \i \pi} 
  \ln\big[ 1+\wh{\mf{a}}\pa{\om} \big] \ln\big[ 1+\wh{\mf{a}}(\om^{\prime}) \big]
\bigg[ - \, \coth^{\prime}(\om-\om^{\prime})
 \, +\, \Dp{\om} \Dp{\om^{\prime}} \ln\Big[\f{ \s{\om+\om^{\prime}+\eta} }{ \s{\om-\om^{\prime}+\eta} }  \Big]  
  \, - \,  \coth^{\prime}(\om+\om^{\prime})  \bigg]   \;.
\nonumber
\end{multline}
The contour $\msc{C}$ arising in the first line does \textit{not}  surround the points $ \i \tf{\pi}{2} $ and $-\tf{\eta}{2}$ mod $[ \i\pi]$. 

Finally, the function $\mc{F}_N^{(1)}\pa{ \{\la_a\}_1^N}$ is as defined through  \eqref{ecriture F cal N kappa trotter fini}
with the exception that the kernel $\wh{U}$ is now defined as 
\bem
\wh{U}(\om,\om^{\prime}) = \f{ - \ex{-\f{h}{T}} \s{2\om^{\prime}+\eta} }{ \s{\om+\om^{\prime}} \s{\om-\om^{\prime}-\eta}}
\cdot \bigg[ \f{ \s{\om^{\prime}- \tf{\be }{N}} \s{\om^{\prime}+ \tf{\be }{N}+\eta} }
	{ \s{\om^{\prime}+ \tf{\be }{N}} \s{\om^{\prime} - \tf{\be }{N}+\eta} }  \bigg]^N \\ 
\exp\Bigg\{  - \Oint{ \msc{C}_U }{} \hspace{-1mm} \f{\dd \tau }{ 2i\pi }  \ln\big[ 1+\wh{\mf{a}}\pa{\tau} \big] 
\Big[ \coth(\tau +\om^{\prime} ) \, + \, \coth( \om^{\prime} -\tau )
 \, -  \, \coth(\tau +\om^{\prime} +\eta) \, - \, \coth(\om^{\prime} -\tau +\eta)  \Big] \Bigg\}	\;.
\nonumber
\end{multline}
The contour $\msc{C}_U$ arising in its definition  is such that given any $\om^{\prime}, \om \in \msc{C}_p$,
where $\msc{C}_p$ refers to any of the encased contours present in \eqref{ecriture F cal N kappa trotter fini}, the points
\beq
 \pm \om^{\prime},\quad  \pm (\om^{\prime}+\eta) \quad \pm (\om^{\prime}-\eta)
\enq
 are \textit{not} surrounded by $\msc{C}_U$. However, $\msc{C}_U$ encircles all the roots  $\la_1, \dots , \la_N$ as well as the origin.

 Upon similar assumptions to those advanced in Section \ref{SousSection Corr Fcts and FF expansion} relatively to the infinite Trotter number limit of the function $\wh{\mf{a}}_{\la}$
and the hypothesis of convergence of the series defining $\mc{F}_N^{(1)}$ in the infinite Trotter number limit, one can readily take the limit, 
basically by replacing, everywhere, the function $\wh{\mf{a}}_{\la}$ by its limit $\mf{a}_{\la}$  defined as the solution to the non-linear integral equation \eqref{ecriture eqn NLIE pour fct a lambda infinite Tortter nbr}.
The answer can be found in Subsection 3.3 of [$\bs{A26}$]. Since it is quite similar to the one at finite Trotter number, I do not reproduce it here.

\subsection{The boundary magnetisation at finite temperatures}
\label{SousSection Finite Temp Bdry Mag}

The boundary magnetization at finite length and temperature can be expressed as 
\beq
\moy{\sg_1}_{T;L} \; = \; \f{ \sinh^2(\xi_-) }{ \be }  \cdot \f{ \Dp{}  }{ \Dp{} \xi_- } \big[  \ln \mc{Z}_T\big[ \op{H}^{(\e{bd})}_{XXZ}\big]   \big] \;. 
\enq
Assuming that it is licit to exchange the infinite volume limit $L\tend \infty$ limit with the $\xi_-$-derivation, one obtains the representation
\beq
\moy{\sg_1}_{T} = 1  \; + \; \f{ T \sinh^2(\xi_-) }{ J \s{\eta} }  \cdot \f{ \de_{\xi_-} \mf{a}^{\prime}_{\la} (-\xi_-) }{ 1+ \mf{a}_{\la} (-\xi_-) }   
\; + \; T \f{\sinh^{2}(\xi_-)  }{J \sinh(\eta)  } 
\Oint{ \msc{C}}{}   \f{  \ln \big[ 1+ \mf{a}_{\la} (\om)\big]  }{ \sinh^2(\om + \xi_- ) } 
		\cdot \f{ \dd \om }{ 2 \i \pi } \;. 
\label{representation boundary spin mean value}
\enq
There $\mf{a}_{\la}$ corresponds to the solution of the non-linear integral equation \eqref{ecriture eqn NLIE pour fct a lambda infinite Tortter nbr}. 
The integral representation \eqref{representation boundary spin mean value} is perfectly fit for a numerical investigation of the 
boundary magnetisation at finite temperature, see Section 4.2 of [$\bs{A26}$] and Figures 6-10 of that paper. 
Also, under similar assumptions to those discussed in Section \ref{Section Low T analysis of FF expansions} above, one can obtain the low-$T$
expansion of the boundary magnetisation. In the massless regime $\eta=-\i \zeta$ with $0<\zeta <\pi $, it takes the form 
\beq
\big< \sg_1^z \big>_T \;  =  \;  1 + \f{ \sinh^2(\xi_-) }{J \i  \sin(\zeta) } \Int{ \mc{C}_{\xi_{-}} }{}
 \f{ \veps_0(\la) }{ \sinh^{2}(\la+\xi_- + \i \tf{\zeta}{2})} \cdot \f{\dd \la}{2 \i \pi}  \; + \; \e{O}(T^2) \; 
\enq
in which 
\beq
\ba{cc} 
\mc{C}_{\xi} = \intff{-q}{q} \cup \Dp{}\mc{D}_{-\i\tf{\zeta}{2} - \xi,\eps}  & \e{if} \;\; 0 < -\Im(\xi)< \tf{\zeta}{2} \\ 
\mc{C}_{\xi} = \intff{-q}{q}   & \e{otherwise} \ea  
\enq
and $\eps>0$ is small enough. 
The leading term of this expansion can be shown to coincide, see page 32-33 of [$\bs{A26}$], with the answer that can be deduced 
from the multiple integral representations for the elementary blocks of the open XXZ spin-1/2 chain at zero temperature obtained in 
\cite{KozKitMailNicSlaTerElemntaryblocksopenXXZ}.

\section{Conclusion}
 
 I have described in this chapter the various contributions I brought to the understanding of correlation functions of  quantum integrable models at finite temperature 
 or to obtaining closed formulae for the surface free energy at finite temperatures. The result presented in this chapter allow one to test and confirm the conformal field theory based
predictions for the large-distance asympototic behaviour at finite, but small, temperatures of the two-point functions of the XXZ chain. 
The content of this chapter was based on the joint works with Dugave, G\"{o}hmann [$\bs{A23},\bs{A24},\bs{A25}$],
with Pozsgay [$\bs{A26}$]  and  Maillet and Slavnov [$\bs{A10},\bs{A27}$]. 
 The analysis I have presented builds on numerous hypotheses. It would be extremely interesting to prove, if not all, then at least  some of these.

\chapter{Final conclusion and outlook}

This habilitation thesis gives an account of the progress I achieved over the period ranging from January 2009 to January 2015 
in pushing forward the theory of correlation functions in quantum integrable models and, in particular,  
 solving various problems in asymptotic analysis related to these objects. 
In certain cases, I was able to push the analysis to the very end on rigorous grounds. 
Sometimes, I was able to bring into a rigorous formalism the handlings that were conjectured to hold true. 
Relatively to other problems, I only managed to carry out a consistent, rigorous, analysis provided some
hypotheses are made.  Finally, in some cases, I was only able to develop a formal asymptotic analysis. 

\vspace{1mm} 

The results I obtained allowed, among other,  for \vspace{1mm}
\begin{itemize}
\item[$\bullet$]  a thorough first principle based-check of an good deal of universal behaviours for 
models belonging to the Luttinger liquid universality class;\vspace{1mm}
\item[$\bullet$] a characterisation of the structure of the 
non-conformal (sub-leading terms) part of the large-distance and long-time asymptotic expansion of a two-point function;\vspace{1mm}
\item[$\bullet$] setting forth some of the elements of the theory of multidimensional Fredholm series, in particular, the multidimensional flow method allowing one to study their large-parameter 
asymptotics;\vspace{1mm}
\item[$\bullet$] developing the Riemann--Hilbert problem approach to $c$-shifted integrable integral operators;\vspace{1mm}
\item[$\bullet$] providing a deeper understanding of the quantum separation of variables for the Toda chain; \vspace{1mm}
\item[$\bullet$] pushing further the methods for extracting the large-number of integration asymptotic behaviour of multiple integrals;  \vspace{1mm}
\item[$\bullet$] pushing further the understanding of correlation functions in quantum integrable models at finite temperature. 
\end{itemize}

The problems I investigated led to numerous  open questions. In the years to come, I would like to dedicate my attention to some of these.

The various aspects of the large-$N$ asymptotic analysis of the sinh-model I developed so far are still not enough 
so as to extract the large-$N$ limit of the multiple integrals directly of interest to the quantum separation of variables. 
It would definitely be worth developing tools allowing one to carry out the asymptotic analysis of 
the form factors of the exponent of the field in the lattice-discretised sinh-Gordon model. Here, some more input
to the method I developed would be needed due to the weakly confining nature of the potential that is associated
with the $N$-dependent $q$ functions in this model. The next case that I would like to investigate corresponds to the $N$-fold integrals which arise within the 
framework of the "discrete" separation of variables. These integrals still correspond to two-body sinh interactions 
but now the integration runs through $\msc{C}^N$ with $\msc{C}$ some curve in $\Cx$. 
Here, several difficulties will arise due to the loss of a probabilistic interpretation. 
Independently of the quantum separation of variables setting, I am also planning to develop methods allowing one to carry out the large-number of integration analysis of 
the multiple integrals  that arise as $N^{\e{th}}$ summands of multidimensional Fredholm series. 
Here, a good example to start with would be the case of the emptiness formation probability in the non-linear Schr\"{o}dinger model. 
On top of having to deal, again, with complex valued integration, an additional difficulty would issue from the 
semi-implicit nature of the integrands which makes it difficult to obtain local estimates of the growth/decay in $N$
of the integrand. Understanding the case of the emptiness formation probability should already be enough so as to be able 
dealing with the more involved multidimensional Fredholm series such as the one describing the generating function of the 
spin-spin correlation functions in the XXZ spin-$1/2$ chain or those arising from form factor expansions in 1+1
dimensional integrable quantum field theories in infinite volume.

Independently of settling the convergence issues, I would also like to reach some understanding of the large-$x$ behaviour of so-called critical 
multidimensional Fredholm series as described in Chapter \ref{Chapitre Solving some c-shifted RHP}. The various problems in asymptotic 
analysis related to this issue look extremely stimulating and interesting. On the one hand, one would have to discover how to deal with 
$g$-functions within the setting of operator valued Riemann--Hilbert problems and, on the other hand, it will be necessary to adapt the 
method of multidimensional flows to the new class of asymptotic behaviour embodied by the "base" Fredholm determinant at the 
root of the critical multidimensional Fredholm series. 

There is yet another topic to which I  would like to dedicate my research to, namely developing a more rigorous framework for dealing 
with the quantum integrable models at finite temperature. Here, the cornestone result would consist in developing the theory of the class of 
non-linear integral equations arising in the problem and, in particular, finding ways for proving the existence and uniqueness of their
solutions. The main difficulty would consist in having to step out of the real variable setting an building on some specific
properties associated with the presence of complex variables.

Finally, I plan to explore as well the large-volume behaviour of the form factors of the XXZ chain at zero magnetic field and the structure of the 
associated form factor expansions of multi-point functions. Here, it would be extremely interesting to focus on the isotropic point, the XXX chain, where one expects
new features to arise in the large-distance asymptotic expansion of two-point functions.

\appendix 

\chapter{Main notations and symbols}

\section{General symbols and sets}

\begin{itemize}

\item[$\bullet$] $\e{o}$ and $\e{O}$ refer to standard domination relations between functions. In the case of matrix function $M(z)$ and $N(z)$,
the relation $M(z)=\e{O}(N(z))$ is to be understood entry-wise, \textit{viz}. $M_{jk}(z)=\e{O}(N_{jk}(z))$. \vspace{1mm}

\item[$\bullet$] $\e{O}(N^{-\infty})$ means $\e{O}(N^{-K})$ for arbitrarily large $K$'s.\vspace{1mm}

\item[$\bullet$] Given a set $A\subseteq X$,  $\e{Int}(A)$ for its interior $\ov{A}$ for its closure and $A^{\e{c}}$ for its complement in $X$. \vspace{1mm}

\item[$\bullet$] A letter appearing in bold usually denotes a vector. An integer subscript attached to a vector, \textit{e.g.} $\bs{\la}_N$,  denotes its dimensionality 
($N$ in this case): 
\beq
\bs{\la}_N \; = \; \big( \la_1,\dots,\la_N  \big) \in \R^N \;. 
\enq
\item[$\bullet$] $\dd^{N} \la$ denotes the product of Lebesgue measures $\prod_{a = 1}^N \dd\la_{a}$.\vspace{1mm}
 \item[$\bullet$] $\mathbb{N}$ is the set of non-negative integers $\{0,1,2,\dots \}$. \vspace{1mm}
 \item[$\bullet$]  $\mc{M}_p(\Cx)$ is the space of $p\times p$ matrices over $\Cx$. It is endowed with the norm 
$\big| \big| M \big| \big|=\max_{a,b}|M_{a,b}|$. \vspace{1mm}
\item[$\bullet$] $\de_{a,b}$ stands for the Kronecker symbol $\de_{a,b}=1$ in $a=b$ and  $\de_{a,b}=0$ otherwise. 
\item[$\bullet$] The superscript $^{\bs{T}}$ will denote the transposition of vectors, \textit{viz}.
\beq
\e{if} \quad \vec{v} \, = \,  \left( \ba{c} v_1 \\ \vdots \\ v_N \ea \right) \qquad \e{then} \quad 
\vec{v}^{\bs{T}} \, = \,  \left(  v_1 \, \dots \,  v_N  \right) \;. 
\enq

\item[$\bullet$] $I_2$ is the $2\times 2$ identity matrix while $\sg^{x},\sg^{y}$ and $\sg^{z}$ stand for the Pauli matrices:
\beq
\sg^x \; = \; \left( \ba{cc}  0 & 1 \\ 1 & 0 \ea \right) \quad , \quad
\sg^y \; = \; \left( \ba{cc}  0 & -\i  \\ \i & 0 \ea \right)  \quad , \quad
\sg^{z} \; = \; \left( \ba{cc}  1 & 0 \\ 0 & -1 \ea \right) \qquad \e{and} \qquad 2 \sg^{\pm} = \sg^x \mp \i \sg^y \;.
\enq

\item[$\bullet$] $\mathbb{H}^{\pm} = \{z \in \mathbb{C}\,\,:\,\,\mathrm{Im}\,(\pm z) > 0\}$ is the upper/lower half-plane, and $\mathbb{R}^{\pm} = \{z \in \mathbb{R}\,\,:\,\,\pm z \geq 0\}$ is the closed positive/negative real axis. \vspace{1mm}

\item[$\bullet$] $\mc{D}_{a,\eta} \subset \Cx$ is the open disk of radius $\eta$ centred at $a$. $\Dp{} \mc{D}_{a,\eta} $ stands for its canonically oriented boundary.

\end{itemize}

\section{Functions}

\begin{itemize}
 
 \item[$\bullet$] Given a set $A\subseteq X$,  $\bs{1}_{A}$ stands for the indicator function of $A$.\vspace{1mm}
 
 \item[$\bullet$] For any $x \in \R$, $[x]$ denotes the integer part of $x$, namely if $n \leq x < n+1 $, $n \in \mathbb{Z}$, then 
 $[x]=n$. \vspace{1mm}
 
 \item[$\bullet$] $\e{sgn}$ is the sign function on $\R$ defined as
\beq
\e{sgn}(x) = -1 \; \e{if} \;\; x<0 \; ,  \qquad \e{sgn}(0) = 0 \;  , \qquad \e{sgn}(x) = 1 \; \e{if} \;\; x>0  \;. 
\enq

 \item[$\bullet$] Given generic parameters $(a,c)$ the Tricomi confluent hypergeometric
function $\Psi\pa{a,c;z}$ is one of the  solutions to the
differential equation $z y'' + \pa{c-z} y' -a y=0$. It enjoys the monodromy properties 
\begin{align}
 &\Psi(a,1;ze^{2i\pi})= \Psi(a,1;z)e^{-2i\pi a}+
 \frac{2\pi ie^{-i\pi a+z}}{\Gamma^2(a)}
 \Psi(1-a,1;-z),  & \e{when}  \quad \Im (z)<0 \, ,
\label{cut-Psi-1}\\
 &\Psi(a,1;ze^{-2i\pi})= \Psi(a,1;z)e^{2i\pi a}-
 \frac{2\pi ie^{i\pi a+z}}{\Gamma^2(a)}
 \Psi(1-a,1;-z),  &\e{when}  \quad  \Im (z)>0 \, ,
\label{cut-Psi-2}
\end{align}
and has the asymptotic expansion:
\begin{equation}
 \Psi(a,c;z) \sim \sum_{n=0}^\infty(-1)^n\frac{(a)_n(a-c+1)_n}{n!}z^{-a-n},
 \quad z\to\infty,
 \quad -\frac{3\pi}2<\arg(z)<\frac{3\pi}2,
\label{asy-Psi}
\end{equation}
with $(a)_n \, = \,  \tf{ \Ga(a+n) }{ \Ga(a) }$. \vspace{1mm}

 \item[$\bullet$]  The Barnes $G$-function is a generalisation of the $\Ga$ function, in the sense that it satisfies the functional relation $G(z+1)=\Ga(z) G(z)$. 
It admits the integral representation 
\beq
G(1+z) \; = \; \Big( \sqrt{2\pi} \cdot \Ga(z) \Big)^{z} \cdot  \exp\Bigg\{ \f{z(1-z)}{2} \; - \; \Int{0}{z} \ln \Ga(s) \cdot \dd s   \Bigg\}
\label{Appendix definition Barnes function}
\enq
from which one can deduce the reflection formula
\beq
\f{ G(1-z) }{ G(1+z) } \; = \; (2\pi)^{-z} \cdot \exp\Bigg\{ \Int{0}{z} \pi s \cot(\pi s) \cdot \dd s \Bigg\} \;. 
\enq
%
%
%
%
%
%
%
%
%
%
%
 \item[$\bullet$] Products of ratios of Euler $\Ga$ or Barnes $G$-function can be presented by means of the so-called hypergeometric like notation:
\beq
\Ga \Bigg(  \ba{c} \{v_a\}_1^n \\ \{w_a\}_1^m \ea \bigg) \; = \; 
\Ga \Bigg(  \ba{c} v_1, \dots , v_n  \\ w_1, \dots , w_m \ea \bigg) \; = \;  \f{\pl{a=1}{n} \Ga(v_a) }{ \pl{a=1}{m} \Ga(w_a) }
\qquad \e{and} \qquad 
G \Bigg(  \ba{c} \{v_a\}_1^n \\ \{w_a\}_1^m \ea \bigg)  \; = \;  \f{\pl{a=1}{n} G(v_a) }{ \pl{a=1}{m} G(w_a) }\;. 
\label{introduction notation produit compact gamma functions}
\enq

\end{itemize}

\section{Functional spaces}

\begin{itemize}

 \item[$\bullet$] $\mc{M}^{1}(\R)$ denotes the space of probability measures on $\R$. The weak topology on $\mc{M}^1(\mathbb{R})$ is metrised by the Wasserstein distance, defined for any two probability measure $\mu_1$ and $\mu_2$ by:
\beq
\label{Vasera}D_{V}[\mu,\nu] = \sup_{f \in {\rm Lip_{1,1}(\R)}} \Int{\R}{}  f(\xi)\,\dd(\mu_1 - \mu_2)(\xi)\;,
\enq
where ${\rm Lip}_{1,1}(\R)$ is the set of Lipschitz functions bounded by $1$ and with Lipschitz constant bounded by $1$. If $f$ is a bounded, Lipschitz function, its bounded Lipschitz norm is:
\beq
\label{BLnorm}\norm{f}_{{\rm BL}} = \norm{f}_{L^{\infty}(\R)} + \sup_{\xi \neq \eta \in \R} \Bigg|\frac{f(\xi) - f(\eta)}{\xi - \eta}\Bigg| \;.
\enq

\item[$\bullet$] Given an open subset $U$ of $ \Cx^n$, $\mc{O}(U)$ refers to the ring of holomorphic functions on $U$. 
If $f$ is a matrix of vector valued function, the notation $f\in \mc{O}(U)$ is to be understood entrywise, \textit{viz}. 
$\forall \; a,b$ one has $f_{ab} \in \mc{O}(U)$. 

\item[$\bullet$] $\mc{C}^k(A)$ refers to the space of function of class $k$ on the manifold $A$. $\mc{C}^k_{\e{c}}(A)$ refers to the spaces built out of 
functions in  $\mc{C}^{k}(A)$ that have a compact support.  \vspace{1mm}

\item[$\bullet$] $L^p(A,\dd\mu)$ refers to the space of $p^{\e{th}}$-power integrable functions on a set $A$ in respect to the measure $\mu$. $L^p(A,\dd\mu)$
is endowed with the norm 
\beq
\norm{f}_{ L^p(A,\dd \mu) } \; = \; \bigg\{ \Int{A}{} | f(x) |^{p}\,\dd \mu(x) \bigg\}^{\f{1}{p}} \;. 
\enq
\item[$\bullet$]  More generally, given an $n$-dimensional manifold $A$, $W^{p}_{k} (A,\dd \mu )$ refers  to the $p^{{\rm th}}$ Sobolev space of order $k$
defined as 
\beq
W^{p}_{k} (A,\dd \mu ) \; = \; 
\bigg\{  f\in L^{p}(A,\dd \mu) \; : \; \Dp{x_1}^{a_1}\dots \Dp{x_n}^{a_n}f \in  L^{p}(A,\dd \mu) \; ,
      \, \sul{\ell=1}{n}  a_{\ell} \leq k  \quad \e{with} \quad   a_{\ell} \in \mathbb{N} \bigg\}  \;.
\enq
This space is endowed with the norm 
\beq
\label{defWninfnorm} \norm{f}_{ W^{p}_{k} (A,\dd \mu) } \; = \; \max
\Big\{  \norm{ \Dp{x_1}^{a_1}\dots \Dp{x_n}^{a_n}f  }_{  L^{p}(A,\dd \mu) }  \; : \; 
  a_{\ell} \in \mathbb{N}, \; \ell=1,\dots, n,   \; \;  \e{and}\; \e{satisfying}  \; \sul{\ell=1}{n}  a_{\ell} \leq k   \Big\}\;.
\enq
In the following, we shall simply write $L^p(A)$, $W^p_k(A)$ unless there will arise some ambiguity on the measure
chosen on $A$. \vspace{1mm}

\item[$\bullet$] Let $A \subset \R^N$ and $\mu$ be a measure on $A$ such that both $A$ and $\mu$ are invariant under the action of the group of permutations $\mf{S}_N$ on $\R$. 
Then $L^p_{\e{sym}}(A,\dd\mu)=L^p(A,\dd\mu)\cap \mc{S}_{\e{sym}}(A)$, where $ \mc{S}_{\e{sym}}(A)$ refers to the space of measurable functions on $A$ which are 
invariant under the action of $\mf{S}_N$. \vspace{1mm}

\item[$\bullet$] The symbol $\mc{F}$ denotes the Fourier transform on $L^2(\R)$ whose expression, versus $L^1\cap L^2(\R)$
functions, takes the form
\beq
\mc{F}[\vp](\la) \; = \; \Int{ \R }{} \vp(\xi)\,\ex{ \i\xi\la}\dd \xi \;. 
\enq
Given $\mu \in \mc{M}^1(\R)$, we shall use the same symbol for denoting its Fourier transform, \textit{viz}. $\mc{F}[\mu]$.
The Fourier transform on $L^2(\R^n)$ is defined with the same normalisation. \vspace{1mm}
\item[$\bullet$] The $s^{\e{th}}$ Sobolev space on $\R^n$ is defined as  
\beq
H_s(\R^n) \; = \; \bigg\{  u \in \mc{S}^{\prime}(\R^n) \; : \;  
\norm{u}^2_{H_s(\R^n)} \; = \; \Int{\R^n}{}  \Big(1 + \big|\sul{a=1}{n} t_a^2\big|^{\f{1}{2}} \Big)^{2s} 
\big| \mc{F}[u](t_1,\dots,t_n) \big|^2 \cdot \dd^n\bs{t} \; < \; +\infty \bigg\} \; , \; 
\enq
in which $\mc{S}^{\prime}$ refers to the space of tempered distributions. 
We remind that given a closed subset $F \subseteq \R^n$, $H_s(F)$ corresponds to the subspace of 
$H_{s}(\R^n)$ of functions whose support is contained in $F$. \vspace{1mm}

\item[$\bullet$] Given a smooth curve $\Sg$ in $\Cx$, the space $\mc{M}_{p}\big( L^2(\Sg, \dd \nu) \big)$ denotes the space of $p\times p$ matrix valued
functions on $X$ whose matrix entries belong to $L^2\big( \Sg , \dd \nu \big)$. It is endowed with the norm
\beq
\norm{M}_{\mc{M}_{p}(L^2(\Sg,\dd \nu))} \; = \; 
\bigg\{ \Int{\Sg}{} \sul{a,b}{} \big[ M_{ab}(s) \big]^{*} M_{ab}(s)\,\dd \mu(s) \bigg\}^{ \f{1}{2} }\;. 
\label{definition norme L2 fonctions matricielles}
\enq
and $^*$ denotes the complex conjugation.

\end{itemize}

\section{Boundary values, operators}

\begin{itemize}

\item[$\bullet$] $\e{id}$ refers to the identity operator on $L^2(\R^+,\dd s)$, $I_p\otimes \e{id}$ refers to the matrix integral operator on 
$\oplus_{a=1}^{p}L^2\big(\R^+,\dd s \big)$ which has the identity operator on its diagonal. \vspace{1mm}

\item[$\bullet$] Given a function $f$ defined on $\Cx\setminus \Sg$, with $\Sg$
an oriented curve in $\Cx$, we denote -if these exists- by $f_{\pm}(s)$  the boundary values of $f(z)$ on $\Sg$ when the argument $z$ 
 approaches the point $s \in \Sg$ non-tangentially and from the left ($+$) or the right ($-$) side of the curve. 
 Furthermore, if one deals with vector or matrix-valued function, then this notation is to be understood entry-wise. \vspace{1mm}

\item[$\bullet$] The symbol $\mc{C}$ refers to the Cauchy transform on $\R$:
\beq
\mc{C}[f](\la) \; = \; \Int{ \R }{}  \f{ f(s) }{ s-\la } \cdot \f{ \dd s }{ 2 \i \pi }\;.
\enq
The $\pm$ boundary values $\mc{C}_{\pm}$ define continuous operators on $H_{s}(\R)$ and admit the expression
\beq
\label{Cplusmoins}\mc{C}_{\pm}[f](\la) \; = \; \f{ f(\la) }{ 2 } \; + \;\f{ 1 }{ 2 \i } \Fint{ \R }{}  \f{ f(s)\,\dd s }{\pi(s-\la) } \; .   
\enq

\item[$\bullet$] Given an operator $\op{M} \, : \, L^2(X, \dd \nu) \mapsto L^2(X, \dd \nu)$ with an integral kernel $M(\la,\mu)$, the Hilbert--Schmidt 
norm of $\op{M}$ is defined as 
\beq
\norm{ \op{M} }_{HS}  \; = \; 
\bigg\{ \int_{X \times X}^{} \big| M(\la,\mu) \big|^2\,\dd \nu(\la) \otimes \dd \nu(\mu) \bigg\}^{ \f{1}{2} }\;. 
\label{definition norme HS}
\enq

\end{itemize}

\end{document}